\documentclass[11pt]{ociamthesis}  

\usepackage{amsmath,amssymb,amsfonts}
\usepackage{natbib,upgreek,mathtools}
\usepackage{rotating, caption, subcaption}
\usepackage{graphicx}
\usepackage{verbatim}
\usepackage{url}
\usepackage{afterpage}
\usepackage{setspace}
\usepackage{moreverb}
\usepackage{lineno}
\usepackage{array}
\usepackage{geometry}
\let\hat=\widehat

\let\leq=\leqslant

\newcommand\argmin[1]{\underset{#1}{\mathrm{argmin} \text{ }}}
\newcommand\argmax[1]{\underset{#1}{\mathrm{argmax} \text{ }}}
\newcommand{\un}[1]{\boldsymbol{#1}}
\newcolumntype{P}[1]{>{\centering\arraybackslash}p{#1}}

\title{Boundary constraints on  \\[1ex]     
       the large-scale ocean circulation}   

\author{Tomas Jonathan}             
\college{Worcester College}  

\degree{Doctor of Philosophy}     
\degreedate{Michaelmas 2021}         

\begin{document}

\baselineskip=18pt plus1pt

\setcounter{secnumdepth}{3}
\setcounter{tocdepth}{3}

\pagenumbering{gobble}
 
\begin{center}
\textbf{\Large{Boundary constraints on the large-scale ocean circulation}}

\vspace{10pt}

{\Large{Tomas Jonathan}}

\vspace{10pt}

Worcester College

University of Oxford

\vspace{10pt}

\textit{A thesis submitted for the degree of Doctor of Philosophy}

Michaelmas 2021

\end{center}	
 
\singlespacing

The Meridional Overturning Circulation (MOC) is a system of surface and deep currents encompassing all ocean basins, crucial to the Earth's climate, transporting heat, carbon and nutrients around the globe. Detecting potential climatic changes in the MOC first requires a careful characterisation of its inherent variability. Key components of the MOC are the Atlantic MOC (AMOC) and the Antarctic Circumpolar Current (ACC).

In this thesis, the role of boundary properties (including density and velocity) in determining the AMOC and ACC is investigated, as a function of cross-sectional coordinate and depth, using a hierarchy of general circulation models. The AMOC is decomposed as the sum of near-surface Ekman, depth-independent bottom velocity and eastern and western boundary density components. The decomposition proves a useful low-dimensional characterisation of the full 3-D overturning circulation, providing a means to quantify the relative importance of boundary contributions on different timescales.

The estimated total basin-wide AMOC overturning streamfunction, reconstructed using only boundary information, is in good agreement with direct calculations of the overturning using meridional velocities, in terms of magnitude and spatial structure. The time-mean maximum overturning streamfunction is relatively constant with latitude, despite its underlying boundary contributions varying considerably, especially in the northern hemisphere. Regression modelling, supplemented by Correlation Adjusted coRrelation (CAR) score diagnostics, provides a natural ranking of the various boundary components in explaining total transport variability. At short timescales the bottom, western boundary and Ekman components play a dominant role. At decadal timescales, the importance of both boundary density components is revealed.

Applying a similar decomposition diagnostic to the ACC provides insight into the differences between model simulations, revealing spatial resolution dependence of ACC transport through the Drake Passage. All models exhibit a weak ACC compared to observations, the $1/4^\circ$ model showing significant weakening over the first $40$ years of simulation. The weaker ACC transport in the $1/4^\circ$ model can be attributed to fresher, lighter southern boundary densities and a strengthened Antarctic Slope Current.

Density maps along sloping ocean boundaries are produced, incorporating the western and eastern Atlantic boundaries and the Antarctic coastline. Isopycnals are flat over long portions of eastern ocean boundaries, and slope linearly on western boundaries; possible mechanisms for these features are discussed. The relationship between boundary currents and isopycnal slope around Antarctica is also considered.

Results of this thesis serve to indicate the importance of boundary information in characterising the AMOC and the ACC, and the relative simplicity of along-sloping-boundary density structure.

\doublespacing

\maketitle                  

\singlespacing
\begin{abstract} \label{TJ_Abs}
\vspace{-0.3cm}

\end{abstract}
\doublespacing'


\chapter*{Acknowledgments}

\vspace{60pt}

\begin{center}
	\textit{``Dyfal donc a dyr y garreg.''}\\
	\hspace{150pt}Welsh proverb
\end{center}

\vspace{60pt}

Thanks to my supervisors Helen, David and Mike for their support at Oxford and the Met Office over the last four years. Thanks further to Mike, Dave and Pat for their guidance and welcome at the Met Office. Diolch i fy nheulu am bopeth, yn enwedig i mam, dad a Emma am eu help, cefnogaeth ac am ddyfalbarhau tan y diwedd.

\vspace{60pt}

\begin{center}
	\textit{``Po\v{r}\'adn\'a kn\'i\v{z}ka nen\'i pro to, aby \v{c}ten\'a\v{r} l\'ip usnul, ale vysko\v{c}il z postele a rovnou v podvl\'ika\v{c}k\'ach b\v{e}\v{z}el panu spisovateli napl\'acat dr\v{z}ku.''}\\
	\hspace{150pt}Bohumil Hrabal
\end{center}

\begin{romanpages}          
\tableofcontents            
\listoffigures              
\listoftables              
\end{romanpages}            

\pagenumbering{arabic}



\chapter{Introduction}  \label{TJ_Int}

This chapter provides a brief description of the Earth's ocean overturning circulation, its key features, driving mechanisms and spatio-temporal variability. It also outlines the theory of the dynamics of large-scale flow. It discusses the global importance of the Meridional Overturning Circulation (MOC), and summarises current literature on the Atlantic MOC (AMOC) and the Antarctic Circumpolar Current (ACC). To finish, thesis aims and structure are outlined. 

\section{Overview}
\label{I_ovr}

The Earth's oceans cover more than two thirds of its surface. These reservoirs of liquid water help determine the Earth's climate, storing and transporting heat, carbon, salt and other nutrients, and maintaining a moist atmosphere compatible with life. Physical processes in the ocean and atmosphere occur on a wide range of spatial and temporal scales (from $10^{\text{-2}}$m to $10^{\text{6}}$m, and 1s to $10^8$s), including phase transitions between liquid water, ice and vapour. The oceans' capacity to store energy far exceeds that of the atmosphere (e.g. \citealt{Williams2011}).

A highly complex system of fluid circulation in the Earth's atmosphere and oceans maintains the pattern of heat flux resulting from the need to balance the budget between solar irradiation of the Earth's surface, which is mainly in the tropics, and radiative cooling at all latitudes. In general, heat is redistributed from the tropics to the poles. Overall heat transport in the Pacific and Indian Oceans are poleward, as might be expected; however, in the Atlantic Ocean, the strong overturning circulation there results in a northward overall ocean heat flux, even in the southern hemisphere (\citealt{Trenberth:2017}). This has profound effects on multiple features of global climate, including shifts in the Inter-Tropical Convergence Zone (ITCZ), Sahel and Indian Monsoons, Atlantic hurricanes, El Nino Southern Oscillation (ENSO), and European, North American and Asian climates (\citealt{Zhang2019}). Characterising the MOC is therefore crucial to our understanding of global fluid circulation, wider climate effects and resulting societal and economic impacts. MOCs in each ocean basin are connected through the Antarctic Circumpolar Current, due to the zonally continuous Southern Ocean, isolating Antarctica.

Anthropogenic release of carbon dioxide and other greenhouse gasses into the atmosphere is currently destabilising the climate system and its overturning circulations. Approximately 93\% of the energy accumulated from anthropogenic global warming has been absorbed by the oceans, and widespread surface warming has been observed in recent decades (e.g. IPCC 2021,  \citealt{Anderson2016}).  As much as two-thirds of ocean heat uptake (OHU) since 1955 has been by the ocean's upper 700m (\citealt{Levitus2012}). The overturning circulation is critical to transferring heat into the deeper ocean. To increase our confidence in climate projections, we require a detailed understanding of the processes and mechanisms active in the climate system, including those responsible for ocean uptake of heat and carbon.

Our understanding of the global overturning is relatively poor, due to sparsity of direct observations, and the complexity of the Navier-Stokes equations defining its behaviour (\citealt{ROBINSON1959}). In particular, interpreting output from general circulation models (GCMs) can be problematic, and numerical solutions do not always provide intuitive understanding of the physical mechanisms at play. This has motivated the development of many simpler diagnostic models, to describe the characteristics of the MOC, underpinned by elementary physical relationships which govern the large-scale circulation. Near coastal boundaries, near the surface and near bottom, these simple models break down. Nevertheless, diagnostic tools provide useful means of simplifying complex four-dimensional behaviour in space and time in terms of e.g.  two-dimensional cross-sectional summaries by depth and latitude, longitude or along-boundary distance. 

\subsection*{Meridional overturning circulation}
\label{I_moc}
\begin{figure}[h!]
	\centerline{\includegraphics[width=13cm]{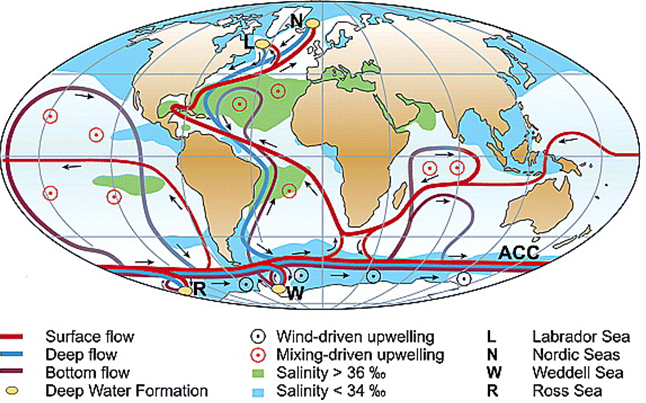}}
	\caption[Schematic of the global overturning circulation and the processes in play (reproduced from \citealt{Kuhlbrodt2007}).]{Schematic of the global overturning circulation and the processes in play (reproduced from \citealt{Kuhlbrodt2007}). }
	\label{MOC_schem}
\end{figure}
Figure \ref{MOC_schem} depicts a simplified global overturning circulation consisting of surface currents (red arrows) and deep currents (blue arrows). Overturning circulations are present in each ocean basin, and their characteristics are determined in part by the locations of boundary land masses, and bathymetry. In the Atlantic, we find warm, salty northward currents (e.g. the Gulf Stream) flowing towards high northern latitudes where large buoyancy loss to the atmosphere results in densification and downwelling (e.g. \citealt{Williams2011}); a colder, denser southward return flow is found at great depths (Deep Western Boundary Current, DWBC). To complete the cycle, the southward return flow is eventually upwelled via diffusion, diapycnal mixing and strong Ekman upwelling driven by Southern Ocean winds (e.g. \citealt{Toggweiler1995}). Overturning circulations for all ocean basins are connected via the ACC. \cite{Buckley2016} provide a review of observations, inferences and mechanisms for the AMOC. Due to the deficit of freshwater supply to the Atlantic relative to the Pacific, Atlantic salinity is higher and its sea water denser; hence North Atlantic Deep Water (NADW) formation can be observed, at high northern latitudes. This is not the case in the North Pacific (\citealt{Warren1983}) or Indian Ocean (\citealt{Tamsitt2017}). Only at a few locations at high latitudes (e.g. Nordic Seas, Weddell Sea) do the surface waters become dense enough to sink to the deep ocean (yellow discs, Figure \ref{MOC_schem}): these regions of deep water formation are few and far between, but their role in the climate system is pivotal.

\section{Driving mechanisms of the AMOC and ACC}
\label{I_drv}

In broad terms, there are two main contributors to a sustained overturning circulation: \textbf{\textit{surface buoyancy forcing}} and \textbf{\textit{mechanical forcing}}. In the past, it was believed that surface buoyancy was the sole driver of the overturning which was hence described as the ``thermohaline circulation''. Our current understanding is different. We now know the thermohaline circulation requires an input of mechanical energy to sustain a deep overturning (\citealt{Sandstrom1916}, \citealt{Munk1998}, \citealt{Yuan2005}). By now, the thermohaline circulation refers not only to flows resulting from density variations, but also to those from mechanically-driven mixing and surface water-mass transformations within the ocean. 

The density of sea water is influenced by precipitation, evaporation, atmospheric heating and cooling, and ice melting (e.g. \citealt{Marshall2008}). Density increases with depth: a water parcel of lighter density within a denser water mass therefore tends to move upwards to the level of a water mass with similar density. Sea water of the same density can vary in temperature and salinity (\citealt{Emery2001}) due to their compensatory effects.

Due to the unequal warming of the Earth by the Sun, we find warmer, less dense waters at low latitudes and colder, denser waters at high latitudes. At low latitudes, diffusion and vertical mixing due to winds distributes the heat downwards slowly. In contrast, at high latitudes, cooling of surface waters is dominated by convection (\citealt{North2014}), a much faster process resulting in denser colder waters being formed (via deep convection e.g. NADW), ultimately leading to downwelling. Formation of deep water masses occurs in narrow regions (e.g. Labrador, Nordic, Irminger, Weddell and Ross Seas) due to intense air-sea forcing (e.g. \citealt{Haine2008}). In addition, proximity to ice can drive deep water formation. Salt rejection when ice is formed increases seawater density. In North Atlantic marginal seas, regions of deep convection share common features; weak mean interior flows, doming isopycnals and cyclonic boundary currents. Air-sea interaction results in heat loss in the interior of the basin, which is balanced by lateral eddy fluxes from the cyclonic boundary current. This results in denser waters along the boundary and a reduction in vertical shear. This is followed by downwelling near the boundary, in order to maintain geostrophic balance. Therefore, lateral eddy fluxes connect the buoyancy loss by vertical transport near the boundary, and deep convection in the interior of the basin (\citealt{Johnson2019}). NADW is comprised of multiple water masses, but is typically described by two distinct layers, Lower NADW consisting of colder GIN (Greenland, Iceland and Norway) Seas water and, warmer Upper NADW formed in the Labrador Sea.

In the early 1900s, \cite{Sandstrom1916}, using only the laws of thermodynamics, came to the conclusion that a closed steady buoyancy forced circulation within the ocean could only be sustained if the heating source was located deeper than the cooling source. Therefore, an ocean interacting with the atmosphere at the surface should only exhibit a very shallow overturning. Further work by \cite{Munk1998} investigating the power ($\approx$ 2.1TW) required to support mixing in the abyssal layers and hence return deep water to the surface, and revealed the crucial role of wind and tides. The contributions of surface buoyancy forcing and geothermal heating through the sea floor were found to be negligible. \cite{Munk1998} state that an ocean, driven only by surface buoyancy forcing, would become a``stagnant pool of cold salty water with equilibrium maintained by near-surface mixing with very weak convectively driven surface-intensified circulation'' within a few thousand years. We now appreciate (e.g. \citealt{Gnanadesikan1999}) that a strong overturning circulation requires deep stratification. Stratification is deepened by sources of mechanical energy including Southern Ocean winds and tidally-driven diapycnal mixing: in the absence of winds or tidal forcing, a shallow stratification or weak overturning would be observed.     

Wind stress on the ocean's surface results in fluid transport. In approximate terms, we can think of the ocean's near-surface mixed layer itself as a sequence of horizontal layers, each layer exerting a frictional force on its neighbours. Due to the Earth's rotation and the resulting Coriolis effect, the direction of fluid flow is deflected increasingly to the right with depth in the northern hemisphere, and to the left in the southern hemisphere, forming an Ekman spiral (\citealt{Ekman1905}). Net transport due to wind stress in the surface Ekman layer is orthogonal to the direction of the wind stress. Due to equatorward coastal winds, we find coastal upwelling along eastern boundaries, bringing nutrient-rich cold water to the surface. 

\begin{figure*}[ht]
	\centerline{\includegraphics[width=13cm]{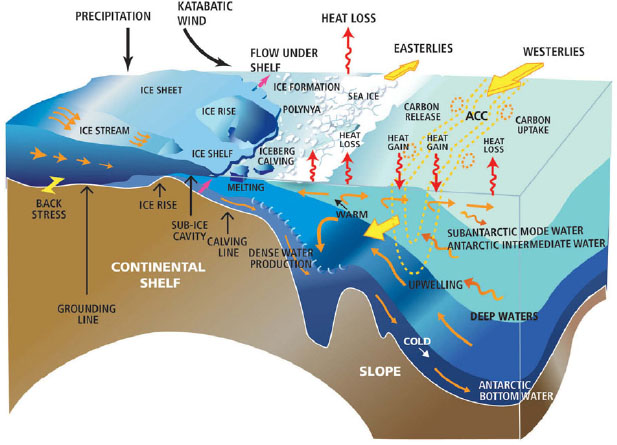}}
	\caption{ Schematic of the processes at play in the Southern Ocean and along the coastline of Antarctica (reproduced from \citealt{NationalResearchCounci2011}).}
	\label{SO_schem}
\end{figure*}

Ekman transport is particularly important for the Southern Ocean (\citealt{Gill1968}, \cite{Toggweiler1995} and Figure \ref{SO_schem}). Strong westerly winds around Antarctica cause northward surface Ekman transport, outcropping isopycnals and large-scale upwelling of NADW (formed in the high-latitude North Atlantic) before the continental shelf (\citealt{Doos1994}, \citealt{Toggweiler1995}). Due to the absence of a meridional boundary in the upper ocean at Drake Passage latitudes, a zonal pressure gradient is only formed below sill depths (\citealt{R.Rintoul2001}), maintaining a meridional geostrophic flow which can balance the northward Ekman transport in the Southern Ocean. The resulting upwelling from depth occurs mainly along isopycnals that outcrop due to the strong westerly winds (e.g. \citealt{Toggweiler1995}, \citealt{Marshall2012}). Sloping isopycnals result in an eastward ACC. The upwelling of NADW here is crucial to sustaining the AMOC (e.g. \citealt{Visbeck2007}). \cite{Munk1951a} showed that, in order to balance the input of zonal momentum from wind stresses over the Southern Ocean, a sink of momentum into the solid Earth must be present, in the form of bottom form stress (i.e. action of pressure against bathymetric barriers on the sea floor). In shallower waters, where bathymetric barriers are not present, the role of transient eddies in creating southward isopycnal flux is crucial in balancing the northward Ekman transport. To maintain the heat balance of the Southern Ocean, a poleward transfer of heat must be present (\citealt{deSzoeke1981}); \cite{Bryden1979} showed the important role of mesoscale eddies in this respect.

\begin{figure}[h!]
	\centerline{\includegraphics[width=13cm]{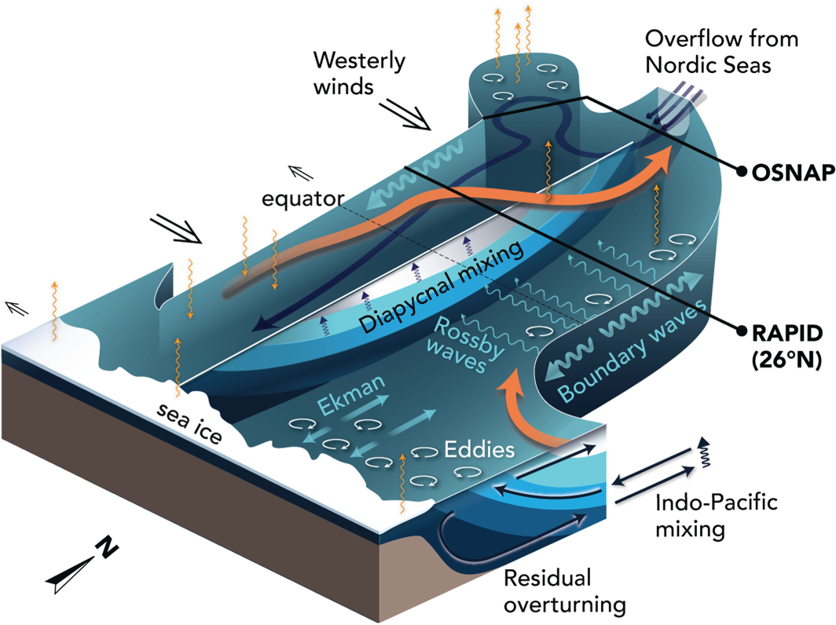}}
	\caption{Schematic illustrating the key processes which determine the strength, structure, and variability of the AMOC (reproduced from \citealt{Johnson2019}). }
	\label{MOC_Mch}
\end{figure}

\cite{Munk1998} emphasise the role of tides in returning the NADW to the surface. The gravitational pull of the Sun and moon results in stratified water being moved up and down sloping topography, producing waves on density interfaces in the ocean's interior. If a small water parcel is displaced from its equilibrium position, it will experience a returning force downwards due to gravity or upwards due to buoyancy. If the water parcel overshoots its original equilibrium position, the disturbance forms an internal gravity wave. The breaking of these internal waves causes dissipation of energy in the surrounding water parcels and thus diapycnal mixing within the interior of the ocean (as illustrated in Figure \ref{MOC_Mch}, \citealt{Marshall2012}, \citealt{Talley:2013}). Winds can also produce internal waves. 
\cite{Polzin1997} showed that within the Brazilian basin, enhanced diapycnal mixing is observed over rough bathymetry. Recently, considerable doubt has been placed on the methodology set out by \cite{Munk:1966} and \cite{Munk1998}: there is evidence that waters are found to sink rather than rise in regions where mixing rate intensifies near the ocean floor (\citealt{Gargett:1984}, \citealt{Polzin1997}, \citealt{Simmons:2004}, \citealt{Kunze:2012}). \citealt{Ferrari2016} argue instead that abyssal waters rise to the surface in ``narrow turbulent boundary layers'' which are found along abyssal ridges and continental margins. This upwelling is so intense that it overwhelms the diapycnal sinking found in the interior, resulting in net upward flow. They highlight the need for observations of these turbulent bottom boundary layers to characterise flow and mixing. Further, they question how well GCMs capture interior diapycnal downwelling, and upwelling near boundaries. In summary therefore, the role of Southern Ocean wind and diapycnal mixing is vital in returning deep water to the surface (e.g. \citealt{Badin2013}). In addition, eddies, bottom friction and turbulence influence the dynamics of large-scale overturning circulation. 

Characteristics of the overturning circulation are related to numerous local oceanic phenomena, some of which are now outlined.  

\textbf{Gyres} in subtropical and subpolar regions of ocean basins, are caused by wind-stress curl. In the northern hemisphere we find easterlies (or trade winds) at equatorial latitudes, westerlies at mid latitudes and easterlies at high latitudes. In conjunction with the Coriolis force, this wind forcing pattern leads to gyres whose structure was explained by \cite{Sverdrup1947}. For example, for the subtropical gyre found in the North Atlantic, easterlies near the equator induce a northward Ekman transport, westerlies at the mid latitudes induce a southward Ekman transport leading to Ekman convergence at intermediate latitudes. Water columns beneath the Ekman layer are squashed, and conservation of potential vorticity requires interior flow within the subtropical gyre to be equatorward. A balancing northward flow is achieved via narrow fricitional or inertial western boundary currents (\citealt{Stommel1948}, e.g. the Gulf Stream in the North Atlantic).

\textbf{Southward flow of NADW:} Greater hydrostatic pressure at high latitudes (e.g. \citealt{North2014}) results in a deep equatorward flow of polar waters. Due to the Earth's rotation, these deep currents flow along the western boundary of the basin (i.e. as the DWBC). \cite{Stommel1959, Stommel1959a} investigate western intensification of the abyssal ocean and find, for low Rossby number, that large-scale dynamics in the ocean's interior are governed by geostrophic balance (see Equation \ref{GB}). The interior of the abyssal ocean is found to flow poleward, implying the presence of the DWBC. However, analysis of float trajectories in the subpolar North Atlantic (e.g. \citealt{Lavender2000}, \citealt{Bower:2009, Bower2011}) shows a large proportion of floats are deflected into the interior of the basin, rather than following the DWBC. \cite{Bower:2009,Bower2011} show that almost 70\% of floats are deflected eastwards from the western boundary, following the path of the deep North Atlantic Current. Less than 10\% of floats were found to continue southwards with the DWBC. These interior pathways highlight the large exchange between the DWBC and the basin interior, especially on the boundary between subtropical and subpolar gyres. This can be attributed to the presence of coherent eddies, eddy-driven mean recirculation (\citealt{Lozier1997}), and mixing and stirring in the deep layers (\citealt{Lozier1997}, \citealt{Gary2011}, \citealt{Bower2019}). 

South of the $43^\circ$N, the majority of southward AMOC flow is confined to the DWBC. This persists southwards through the tropics and into the South Atlantic (\citealt{Gary2011}). Some recirculation of shallower waters is found where the DWBC flows under the Gulf Stream, and due to eddies in the tropics. Near the Victoria-Trinidade Ridge, we find two possible DWBC pathways: the main flow continues southwards along the South American coastline, with a secondary weaker flow proceeding into the interior of the South Atlantic, through the Mid Atlantic Ridge, and into the Cape Basin. We find the DWBC is eventually upwelled due to strong winds in the Southern Ocean (\citealt{Marshall2012}) and diapycnal mixing. This upwelling occurs within all basins, suggesting the AMOC and corresponding MOCs in other basins are connected via the Southern Ocean on centennial to millennial, timescales (\citealt{Buckley2016}). \cite{Bower2019} provide a review of AMOC pathways. 

\textbf{Antarctic Bottom Water} (AABW) is formed within the Weddell and Ross Seas (Figure~\ref{MOC_schem}) along Antarctica's continental shelf. The upwelling of NADW before the continental shelf results in buoyancy gain at the surface from the atmosphere, but within the Weddell and Ross Seas, ice-formation allows for shelf waters of sufficient salinity to sink to the greatest depths of the ocean. The AABW flows northwards in the abyss of all ocean basins, including into the North Atlantic, forming a second lower overturning cell. Diapycnal mixing between the AABW and NADW is important for the buoyancy budget of this abyssal cell.

\textbf{Ocean eddies} are formed mainly due to bathymetric obstacles (stationary eddies) or baroclinic instability (transient eddies, e.g. \citealt{Marshall2008}). Transient eddies are time-dependent, small-scale structures. Eddies are ubiquitous around the global ocean and are particularly evident in major ocean currents, including the Gulf Stream, the Agulhas Current, the Kuroshio Current, and the ACC. \cite{deSzoeke1981} showed the important role eddies play in the Southern Ocean, transporting heat poleward, and hence balancing the northward heat flux resulting from Ekman transport. Southern Ocean eddies are critical in opposing isopycnal steepening induced by Ekman upwelling around Antarctica. 

\section{Dynamics of large-scale flow}
\label{I_dynL}

MOCs are large-scale oceanic flows, the behaviour of which is ultimately governed by the Navier-Stokes equations (e.g. \citealt{Williams2011}, \citealt{Vallis2019}). GCMs provide ever-improving approximate numerical solutions to the Navier-Stokes equations, but obtaining these numerical solutions is computationally demanding, and interpreting the huge quantities of data they produce is challenging. On an ocean basin scale, geostrophy and hydrostatic balance provide useful approximations to the Navier-Stokes equations appropriate for synoptic-scale and larger flows. Gesotrophy and hydrostatic balance are obtained from the Navier-Stokes equations by retaining only the dominant Coriolis, gravity and pressure gradient terms. The Rossby number $R_o = {U}/{fL}$, for synoptic length $L$, speed $U$ and Coriolis parameter $f$ is $<<1$ for large-scale ocean flows, indicating the dominance of Coriolis forces. 

Adopting a local Cartesian frame of reference with horizontal axes $x$ (eastwards) and $y$ (northwards), and $z$ axis vertically upwards from the ocean surface (at $z=0$), \textbf{\textit{hydrostatic balance}} (Equation \ref{HB}) describes a force balance between the acceleration due to gravity $g$, and the upward pressure gradient force per unit mass $\frac{1}{\rho}\frac{\partial P}{\partial z}$, for pressure $P$ and density $\rho$:
\begin{equation}
	\frac{\partial P}{\partial z} = - \rho g.\label{HB}
\end{equation}

The local horizontal components of the Navier-Stokes equations simplify to a pair of equations known as \textbf{\textit{geostrophic balance}}. Horizontal pressure gradient forces are balanced by the Coriolis force on perpendicular horizontal gesotrophic flows, where the components of flow velocity in the local Cartesian frame are ($u$, $v$, $w$):
\begin{equation}
	\frac{1}{\rho} \frac{\partial P}{\partial x} = fv, 
	\label{GB}
\end{equation}
\begin{equation}
	\frac{1}{\rho} \frac{\partial P}{\partial y} = - fu. 
\end{equation}

The \textbf{\textit{thermal wind relation}} is obtained by substituting the hydrostatic balance (Equation \ref{HB}) into the vertical derivative of the gesotrophic relation (Equation \ref{GB}). The meridional thermal wind relationship (Equation \ref{TW}) relates horizontal zonal density gradients to the vertical gradient of the northward velocity,
\begin{equation}
	\frac{\partial v}{\partial z} = - \frac{g}{\rho_0 f} \frac{\partial \rho}{\partial x},
	\label{TW}
\end{equation}
where $\rho$ is the in-situ densities of sea water, and $\rho_0$ is a constant reference value having assumed Boussinesq flow.

The use of geostrophy throughout the basin to calculate the net meridional flow might seem simplistic, especially when the zonal integral must pass through a western boundary current where non-linear and agesotrophic terms may dominate. However, even in these boundary current settings, it is generally accepted that geostrophy dominates the cross-stream balance where the flow is parallel to the coastline, except in the surface Ekman layer. Issues may arise when currents are no longer along-shore, but these effects should be minor so long as they result from narrow currents (\citealt{Bingham2008}, \citealt{Bell2011}).  

\section{The Atlantic Meridional Overturning Circulation (AMOC)}
\label{I_oAMOC}
\subsection{Quantifying the AMOC}
\label{I_qtfAMOC}
The AMOC is defined in terms of the Atlantic meridional overturning streamfunction or zonally-integrated northward volume transport:
\begin{equation}
	\Psi(z;y) \equiv \int_{W(y)}^{E(y)} \int_{-H(x,y)}^{z} v(x,y,z') dz'dx,
	\label{T}
\end{equation}
for meridional velocity $v(x,y,z)$, ocean depth $H(x,y)$, and western and eastern boundaries $W(y)$ and $E(y)$. The AMOC overturning streamfunction is therefore a function of ``latitude'' or meridional coordinate $y$ and depth $z$ (see Section \ref{S_App_DD} for more grid details), with units m$^{3}\text{s}^{\text{-1}}$, where $10^6\text{m}^{\text{3}}\text{s}^{\text{-1}}$ is defined as equal to a Sverdrup (Sv).

The strength of the AMOC for a given $y$ is commonly defined as the maximum of the overturning streamfunction over depth $z$
\begin{equation}
	\Psi_{\max}(y) = \max_{z} \Psi(z;y). 
	\label{E_MOCStr}
\end{equation}

\subsection{Observing the AMOC}
\label{I_obsAMOC}

One of the main issues facing physical oceanography generally, and a major constraint on our ability to better understand oceanic systems, is the sparsity of ocean observations in space and time. Long observational datasets are limited, and confined to popular shipping routes or coastal regions. Surface properties have been routinely recorded via satellite since the 1990s. Yet only from the early 2000s has continuous monitoring of the sub-surface limbs of the overturning been performed, following hydrographic studies revealing an apparent weakening AMOC (e.g. \citealt{Bryden2005}).

\cite{Marotzke1999} showed using the adjoint of the Massachusetts Institute of Technology general circulation model (MITgcm) that the dynamical sensitivity of heat transport variability is generally greatest to density perturbations near meridional boundaries of the basin, consistent with the idea that the net thermal wind integrated across a basin is controlled by boundary density variations. This suggests the possibility of ``boundary monitoring'' the AMOC. \cite{Hirschi2003} and \cite{Baehr2004} demonstrated the practical feasibility of using an array of sparse moorings to monitor the AMOC within two eddy-permitting numerical ocean models. 

\subsubsection*{RAPID (2004-date)}

Studies such as those referenced above led to the deployment of the RAPID/MOCHA/WBTS (hereafter RAPID) mooring array in $2004$, at a latitude of 26.5$^\circ$N, to monitor zonally-integrated volume and heat transports. The main components of the estimated overturning streamfunction estimated are: (a) geostrophic interior transport, (b) Ekman transport due to wind-stresses, and (c) transport through Florida straits (\citealt{Cunningham2007}, \citealt{Kanzow2007}, \citealt{McCarthy2015, McCarthy2020}). 

RAPID consists of a sequence of moorings concentrated on the Atlantic's eastern and western boundaries and on each side of the Mid Atlantic Ridge (MAR). Estimates of the geostrophic interior flow are made using thermal wind (Equation \ref{TW})  relative to a reference level ($z = -4740$m), based on work by \cite{Lee1998} on decomposing Indian Ocean overturning to investigate its variability. Esimates for the Florida Strait transport, the surface Ekman transport and the geostrophic interior transport are summed to estimate the total overturning (e.g. \citealt{Cunningham2007}). At 26.5$^\circ$N, the contribution of the overturning through the Florida strait is estimated using a measurement of the voltage across a disused telephone cable, exploiting the fact that seawater is a conductor moving in the Earth's magnetic field, inducing a current in the cable. Usually it is assumed that the overturning streamfunction conserves volume; to achieve this, the overturning streamfunction estimate is adjusted by a uniform velocity field in the opposite direction, across the section. AMOC strength (Equation  \ref{E_MOCStr}) for the period 2004 to 2017 is estimated to be approximately $17$Sv (\citealt{Frajka-Williams2019}).

\begin{figure*}[ht!]
\centerline{\includegraphics[width=13cm]{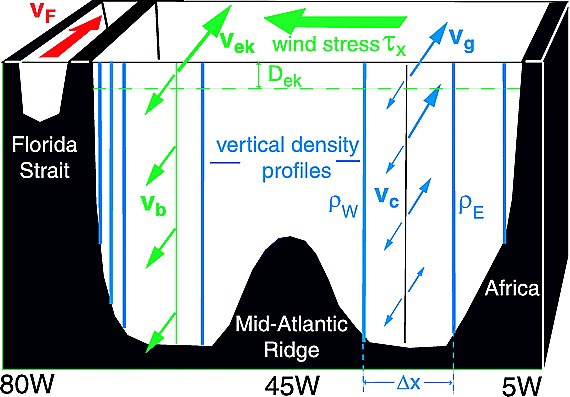}}
\caption[ Schematic of the MOC observing system across 26.5$^\circ$N in the North Atlantic (reproduced from \citealt{Hirschi2003}).]{ Schematic of the MOC observing system across 26.5$^\circ$N in the North Atlantic. The MOC is estimated from the zonal wind stress and from vertical density profiles taken at different longitudes across the basin. \cite{Hirschi2003} assumed the transport through the Florida Strait is known ($v_F$). Knowing the wind stress allows the calculation of an Ekman velocity $v_{ek}$. East of the Florida Strait a depth-dependent velocity field $v_g$ is estimated using the thermal wind relation between adjacent density profiles. The meridional transport associated with $v_g$, $v_F$ and $v_{ek}$ is compensated by constant velocity corrections $v_c$ (for $v_g$) and $v_b$ (for $v_F$ and $v_{ek}$), in order to ensure zero net meridional transport contribution. Reproduced from \cite{Hirschi2003}.}
\label{Rapid}
\end{figure*}

Figure \ref{Rapid} shows the components contributing to the overturning at RAPID. Observations suggest that the major components of the circulation are the Gulf Stream, Ekman transport, southward flow of NADW and subtropical gyre, and northward flow of Antarctic Intermediate Water. Estimates for some of these quantities are provided by the AMOC timeseries (from www.rapid.ac.uk). The Gulf Stream contribution is found to be approximately 31Sv, Ekman transport around 4Sv and the upper mid-ocean contribution is around -18Sv. 

\subsubsection*{SAMBA (2009-date)}
More recently, two additional cross-basin mooring arrays have been deployed in the South Atlantic (SAMBA) and the subpolar North Atlantic (OSNAP).

The SAMBA array is situated at approximately $34.5^\circ$S, to investigate the South Atlantic MOC, including features such as the Agulhas leakage and the Malvinas current. In contrast to RAPID, no telephone cable measurements are used; the monitoring relies on density profiles, PIES (pressure-inverted echo sounders) and CPIES (``sea'' PIES) providing baroclinic and barotropic estimates of the temporal variability of the overturning streamfunction. The use of PIES and CPIES at $1350$ dbar renders the zero net flow assumption obsolete, and hence volume compensation is not required. However, sensor drift is an issue for both PIES and CPIES measurements. Hence, capturing the time-average barotropic (depth-independent) component is difficult and estimates are unreliable. As a work-around, a time-mean reference velocity is acquired from the OFES (Ocean For the Earth Simulator) model, providing an estimate of the meridional volume transport in regions inshore of the $1,350$ dbar isobath (\citealt{Meinen2018}). A number of sensitivity studies at SAMBA have explored the temporal variability of each of the SAMBA measurement components (discussed in Section \ref{I_var}) due to a known ``leveling'' issue (\citealt{Watts1990}, \citealt{Donohue2010}) associated with the bottom pressure recorders. The monitoring array reports a maximum overturning streamfunction of the order of $15$Sv (\citealt{Frajka-Williams2019}). 

\subsubsection*{OSNAP (2014-date)}

The OSNAP observing array was set up using similar principles to RAPID, to observe and quantify the subpolar North Atlantic overturning circulation. Further, it is used to explore overflow pathways, subtropical and subpolar connectivity and relate AMOC variability to deep water mass variability. It is composed of two legs, OSNAP West across the mouth of the Labrador Sea and OSNAP East across the Irminger and Iceland basins. Each leg consists of multiple moorings along the continental boundaries and Reykjanes ridge, carrying instruments including CTDs (conductivity, temperature and depth), current meters, thermistors, acoustic Doppler current profilers and moored profilers. Away from mooring arrays, where required, geostrophic velocities are estimated from temperature and salinity fields constructed from Argo profiles, gliders etc. (\citealt{Lozier2017} and \citealt{Li2017}). 

Using density coordinates to better reflect the overturning (since much of the water mass transformation occurs in the horizontal circulation of the subpolar gyre), \cite{Lozier2019} show the importance of the regions north of the OSNAP East section to the overturning. Here, northward-flowing warm and salty Atlantic waters from the subtropics are replaced with colder, fresher southward-flowing waters moving along the western boundaries of the Iceland and Irminger basins (\citealt{Frajka-Williams2019}). The monitoring array observed a maximum overturning streamfunction of the order of $16.6$Sv (\citealt{Li2021}) over the period 2014 to 2018.    

For time-mean overturning, OSNAP monitoring (\citealt{Lozier2019}, \citealt{Li2021}) has revealed that : (a) the Labrador Sea (OSNAP East) has minimal contribution to the subpolar overturning circulation defined in density space (2.6Sv), (b) westerly wind-induced Ekman transport contributions are small (-1.7Sv), and (c) the OSNAP East contribution to the overturning at around 16.8Sv is around 7 times larger than its western counterpart. 

The sum of MOC estimates across OSNAP West and East cannot be used to estimate the total MOC, due to cancellations of northward and southward transports within the same density class; southward flow around eastern Greenland cancels with some of the northward flow along western Greenland. OSNAP East plays a dominant role for meridional heat transport, and OSNAP West dynamics are found to be important for freshwater transport, which could have a direct effect on salinity anomalies downstream. Specifically, OSNAP West is shown to provide half of the freshwater transport measured across the OSNAP array.

\subsubsection{Other monitoring}

RAPID, SAMBA and OSNAP arrays are part of a larger set of observational campaigns to investigate and quantify the ocean's overturning circulations (see \citealt{McCarthy2020} for overview of technologies and methods). The deployment of around 4000 Argo floats at any time has improved the spatial coverage of ocean hydrographic data (\citealt{Riser2016}). Each Argo float drifts around the world's oceans at around 1km depth, submerging approximately every 10 days to a depth of around 2km, before returning to the surface, to transmit a profile of temperature and salinity with depth via satellite. Argo floats leave the deep seas un-sampled. Newer floats have been tested which submerge to around $5$km, but these have issues in shallower waters. 

Bottom pressure recorders and current meters can be used to obtain further information regarding the contribution of the depth-independent flow (or external-mode) component (e.g. \citealt{Donohue2016}). This is especially useful along the Atlantic's sloping western boundary where strong boundary currents are observed. The challenge using data from bottom pressure recorders is sensor drift, leading to inaccuracies and difficulty in estimating transport associated with the depth-independent flow component (\citealt{Watts1990}, \citealt{Hughes2013}). 

\subsection{AMOC variability}
\label{I_var}
\subsubsection{Observational assessment of AMOC variability}
\label{I_var_obs}

The AMOC varies on timescales from days to millennia. The relatively short continuous timeseries available for the RAPID, SAMBA and OSNAP observing programmes have shed some light on short-term AMOC variability. For the subtropical Atlantic, RAPID has shown that wind stress on the ocean's surface dominates the seasonal variability of the AMOC (e.g. \citealt{Hirschi2007}, \citealt{Srokosz2015}) via processes such as Ekman transport, wind-stress curl impact on boundary densities, coastal upwelling and Sverdrup balance. Large seasonal fluctuations (4Sv to 34.9Sv) in the strength of the overturning circulation have been observed. Wind-stress curl on the eastern side of the basin are found to be particularly important on seasonal timescales (e.g. \citealt{Kanzow2010}, \citealt{Chidichimo:2010}). 

On interannual timescales, a weakening of the AMOC from 2004 to 2012 is attributed to a strengthening southward flow in the upper mid ocean due to a stronger recirculation of the subtropical gyre (\citealt{Smeed2014}, \citealt{Srokosz2015}). \cite{Smeed2018} show that the AMOC at RAPID has been in a reduced state since 2008 relative to 2004-2008, concurrent with a northward shift and broadening of the Gulf Stream, and altered patterns in ocean heat content and sea surface temperature (SST). In conjunction with changes in air-sea fluxes above the western boundary, these suggest that the AMOC is a major factor in decadal scale North Atlantic variability. At OSNAP, \cite{Li2021} (and \citealt{Lozier2019}) reveal that overturning variability is dominated by the overturning in the eastern subpolar gyre, measured by OSNAP East, on monthly to interannual timescales. They find that 82\% of AMOC variance is explained by OSNAP East between 2014-18.  

Low-frequency variability at RAPID is attributed to the geostrophic (boundary density) term. \cite{Elipot2014} find the variance of the western boundary density term to be significantly larger than that of its eastern counterpart, although theoretical arguments suggest eastern boundary density contribution should be largest (\citealt{Johnson2002, Johnson:2002b}). Boundary buoyancy anomalies may be fundamental to understanding mechanisms of long-timescale AMOC variability. 
 
\cite{Meinen2018} discuss the variability found at SAMBA for the South Atlantic. High frequency (short-timescale) variability dominates, with largest contributions from the geostrophic relative velocity term (boundary density contribution) with both Ekman and geostrophic reference velocity (volume compensation term, equivalent to depth-independent term) playing secondary roles. Interestingly, whereas variability at RAPID is dominated by western boundary density variations, overturning variation at SAMBA is equally dependent on both boundaries. This reflects the important role of inter-ocean exchanges between the South Atlantic and both Indian and Pacific Oceans. Seasonal variation is dominated by semi-annual density variation near the eastern boundary. Interannual variations are equally dependent on density and pressure (depth-independent, external mode) terms. In years where density variation is key, the eastern boundary is found to dominate AMOC variability, in contrast to what is found at RAPID (\citealt{McCarthy2015}). For years of large pressure or depth-independent contributions, eastern and western pressure contributions are similar. 

\subsubsection{Model-based assessment of AMOC variability}
\label{I_var_mod}

Prior to the observational data discussed in Sections \ref{I_obsAMOC} and \ref{I_var_obs} becoming available, estimates of AMOC strength and variability were mostly model-based, with the exception of limited AMOC estimates from hydrographic sections, giving snapshots in time (e.g. \citealt{Hall1982}). These estimates provide a complete description of the AMOC in space and time. However, they are highly dependent on model-specific factors, including spatial resolution, and representations of small-scale processes such as overflows, convection, ocean eddies, mixing etc. Hence, accurate quantification of even the strength of the time-mean AMOC is problematic.

On short timescales there is general consensus that the Ekman component dominates AMOC model-based variability (e.g. \citealt{Hakkinen1999}, \citealt{Dong:2002}, \citealt{Xu2014}). \cite{Zhao:2014} and \cite{Pillar2016} also show that wind forcing dominates variability at $26.5^\circ$N. 

At interannual and decadal timescales, the geostrophic component dominates variability (e.g. \citealt{Hirschi2007}, \citealt{Cabanes2008}). \cite{Bingham:2007} find that the AMOC is not coherent between subpolar and subtropical gyres; interannual variability dominates the subtropical gyre whereas decadal variability dominates the subpolar gyre (e.g. \citealt{Wunsch2013}). \cite{Buckley2016} note a lack of consensus regarding mechanisms for low frequency AMOC variability, and the concern that AMOC variability in model output is dependent on uncertain model and parameter specification. Further they observe that characterising AMOC variability on inter-annual to decadal timescales remains challenging, since observations to validate model-based estimates are lacking.  

At decadal timescales, models show meridionally coherent modes of AMOC and Atlantic ocean heat transport (e.g. \citealt{Delworth1993}, \citealt{Delworth2000}, \citealt{Knight:2005}, \citealt{Danabasoglu2012}). \cite{Zhang:2010} show that meridionally coherent AMOC anomalies emerge from subpolar regions, as a result of time-variable buoyancy forcing (\citealt{Biastoch2008}, \citealt{Robson2012a}, \citealt{Yeager2014}). 

There is no consensus from models regarding what sets the timescales dominating low frequency AMOC variability. However, numerous studies using idealised models (\citealt{Marshall2001a}, \citealt{Lee2010}) and GCMs (\citealt{Hirschi2007}, \citealt{Zanna2011,Zanna2012}, \citealt{Buckley2012}) suggest the dominant timescale of AMOC variability is determined by the time it takes for a baroclinic Rossby wave to propagate across the basin. Other studies examine the role of advective processes such as spin-up and spin-down of gyre circulation (\citealt{Delworth:1997}, \citealt{Dong2005}) and build-up of high- or low-density water in regions of deep water formation (\citealt{Dong2005}, \citealt{Msadek:2009}).

Many model studies support the idea of decadal AMOC variability being tied to regions of deep convection or deep water formation. \cite{Zhang:2010} and similar studies (e.g. \citealt{Delworth1993}, \citealt{Dong2005}, \citealt{Danabasoglu2008}, \citealt{Kwon:2012}, \citealt{Danabasoglu2012}, \citealt{Jackson2013}, \citealt{Roberts2013}) use multiple lagged correlation, to suggest that anomalies in regions of deep water formation lead AMOC anomalies. Other studies show that buoyancy forcing over regions of deep water formation dictates decadal AMOC variability (e.g. \citealt{Yeager:2012} for the Labrador Sea). \cite{Zhang:2005} and others suggest that large changes in the AMOC can be caused by disruption to deep water formation, or overflow waters (\citealt{Zhang:2011}). Hosing experiments suggest that excess freshwater input at high northern latitudes promotes AMOC weakening or collapse (\citealt{Zhang:2005}, \citealt{Zhang:2007}, \citealt{Jackson2015}). 

\cite{Biastoch2008a} show a possible role of the Agulhas leakage in decadal AMOC variability, with anomalies in thermocline depth propagated across the South Atlantic via Rossby waves, and then northward along the western boundary via Kelvin waves.

Changes in Southern Ocean westerlies, which return NADW to the surface, could  impact the ``pulling'' part of the AMOC (\citealt{Visbeck2007}). However, the local overturning response to changes in Southern Ocean wind-forcing appears to be small, due to eddy compensation (e.g. at equilibrium \citealt{Farneti2010a}, \citealt{Farneti2010a}, \citealt{Farneti2011}, \citealt{Gent2011}, \citealt{Gent2016}). The response time of the AMOC in the North Atlantic to changes in Southern Ocean winds is of the order of multiple decades or centuries, suggesting Southern Ocean winds are more likely to influence centennial rather than decadal AMOC variability (\citealt{Spence2009}, \citealt{Spooner2013}). \cite{Delworth2012} show, that in the CM2.1 model, AMOC centennial variability is tied to propagation of salinity anomalies from the Southern Ocean, northward through the basin. 

Studies such as \cite{Lozier2010} and \cite{Lozier2012} dispute the importance of deep convection for decadal AMOC variability. \cite{Lozier2019} and \cite{Li2021} show that recent OSNAP observations disagree with inferences from modelling studies (e.g. \citealt{Buckley2012}, \citealt{Buckley2016}, \citealt{Thornalley2018}), which suggest that multi-annual to decadal AMOC variability is attributed to the export and propagation of density anomalies from the Labrador Sea, due to deep convection. \cite{Menary2020} find that density anomalies, advected from the eastern subpolar North Atlantic, dominate the density variability in the western boundary of the Labrador Sea. Therefore, density anomalies found in the Labrador Sea are likely to carry a signature of upstream forcing anomalies. Furthermore, other recent studies have suggested that the linkage between Labrador Sea convection and downstream AMOC variability can be explained by shared variability in response to the North Atlantic Oscillation and other atmospheric forcings (e.g. \citealt{Zou:2019,Zou2020}) rather than by the equatorward propagation of AMOC anomalies from the Labrador Sea. 

It is widely accepted that western boundary buoyancy anomalies are central to understanding AMOC variability on decadal timescales. However, there is no consensus about driving mechanisms, as highlighted by the recent OSNAP observations. From earlier studies using idealised models (\citealt{Zanna2011}, \citealt{Buckley2012}) and GCMs (\citealt{Danabasoglu2008}, \citealt{Zhang:2008}, \citealt{Tulloch2012}), \cite{Zanna2012} suggests that the boundary between the subpolar and subtropical gyres can be thought of as the ``pacemaker'' for decadal AMOC variability, referred to by \cite{Buckley2016} as the ``transition zone''. 

\cite{Buckley2016} review the processes which could form buoyancy anomalies in this region, including (a) local atmospheric forcing (e.g. \citealt{Frankignoul:1977}, \citealt{Buckley2014, Buckley2015}), (b) advection of buoyancy anomalies by mean currents (e.g. \citealt{Tulloch2012}, \citealt{Kwon:2012}), (c) buoyancy signals communicated via baroclinic Rossby waves (e.g. \citealt{Hirschi2007}, \citealt{Cabanes2008}, \citealt{Frankcombe2009}, \citealt{Zanna2011, Zanna2012}, \citealt{Buckley2012}), (d) gyre wobbles, and shifts in Gulf Stream path (e.g. \citealt{Marshall2001}, \citealt{Zhang:2008}), (e) changes in deep convection and water mass transformation (e.g. \citealt{Curry1998}, \citealt{Pena-Molino2011}, \citealt{VanSebille2011}) and (f) the role of salinity (\citealt{Delworth1993}, \citealt{Holliday2003}, \citealt{Dong2005}, \citealt{Tulloch2012}).

\subsection{Meridional propagation of western boundary buoyancy anomalies} \label{I_adjAMOC}

For meridionally-coherent AMOC anomalies to form, buoyancy anomalies must be propagated southwards along the western boundary. Two main mechanisms have been proposed, namely fast meridional propagation via boundary waves (e.g. \citealt{Kawase1987}, \citealt{Johnson:2002b}, \citealt{Schloesser2012}, \citealt{Marshall2013}) and slow advection by the DWBC or interior pathways (\citealt{Curry1998}, \citealt{Koltermann1999}, \citealt{VanSebille2011}, \citealt{Pena-Molino2011}).

Observational studies by \cite{Curry1998}, \cite{Koltermann1999}, \cite{VanSebille2011} and \cite{Pena-Molino2011} at locations  downstream of the Labrador Sea (48$^\circ$N, 36$^\circ$N and 24$^\circ$N) generally find slow propagation of buoyancy anomalies, suggesting the advective pathway. However, the velocity inferred from the time lag corresponding to maximum anomaly correlation between locations, suggests a slow propagation of the order of centimeters per second, slower than the speed of the DWBC. The lower velocity can be attributed to eddy-driven recirculation gyres generated by instabilities of the Gulf Stream and North Atlantic Current (\citealt{Lozier1997}, \citealt{Gary2011}).  

Within a simple reduced-gravity model such as the one documented by \cite{Johnson2002}, a layer thickness anomaly produced at high latitudes due to e.g. thermohaline forcing, creates meridional velocities in the ocean's surface layers which are in geostrophic balance with a density, temperature or pressure anomaly. These anomalies are propagated southward along the western boundary via boundary waves. This process is fundamental to how the ocean adjusts to changes in its boundary conditions (\citealt{Wajsowicz1986}, \citealt{Kawase1987}). The amplitude of these waves reduces whilst nearing the equator in accordance with geostrophic balance (due to the diminishing Coriolis effect). The equator acts as a pseudo-barrier, due to the Coriolis force being zero there, forcing boundary waves to propagate along the equator and poleward along the eastern boundary, with waves distributing the anomaly evenly along that boundary. On the eastern boundary, a meridional pressure gradient can only exist if there is a geostrophic velocity into the boundary. In accordance with no normal flow through the boundary, there can therefore be no along-boundary gradient in pressure or density. This results in flat isopycnals along eastern boundaries, implying the stratification of the water column does not vary with latitude. Given that Rossby waves transmit eastern boundary information into the interior, the stratification in these simple models is quasi-uniform over much of the domain. This suggests that considerable information concerning the global overturning can be extracted from only a few stratification profiles, especially along the eastern boundary. The assumption of flat isopycnals along the eastern boundary has been widely used within diagnostic models (e.g. \citealt{Cessi2010}, \citealt{Nikurashin2011}, \citealt{Marshall2017}, \citealt{Nieves2018}). However, no attempt has been made to map the isopycnal structure along the boundary within GCMs or observationally.

Originally, Kelvin waves were thought to propagate the buoyancy anomaly. However, work by \cite{Marshall2013} showed that this is true only for short wave periods; for longer periods, the anomaly is propagated by long and short Rossby waves, both zonally and along the western and eastern boundaries respectively. 

\subsection{Wider role of the AMOC}
\label{I_AMOCrole}
\subsubsection{Atlantic Multidecadal Variability}
\label{I_role_AMV}

On seasonal to decadal timescales, AMOC variations have been shown to influence North Atlantic SST (e.g. \citealt{Duchez2016}) and sea levels on the eastern U.S. coast  (\citealt{Little2017}). Decadal variability in North Atlantic SST is quantified using the Atlantic Multidecadal Variability (AMV, or the Atlantic Multidecadal Oscillation). Studies suggest lagged correlations between low-frequency AMOC and SST (via AMV) in models (\citealt{Enfield:2001}, \citealt{Zhang2019}); however, the causal role of AMOC in creating these SST patterns is unclear. These SST changes directly influence temperatures in Northern America and Europe, as well as Atlantic hurricane activity and rainfall across Northern Africa and India. Models further suggest that the AMOC and AMV are coherent over multidecadal timescales (e.g. \citealt{Zhang:2007}). However, the AMOC's influence is not fully represented within models (\citealt{Zhang2019}), which fail to capture the observed AMV pattern under realistic external forcing (\citealt{Buckley2016}). 

\cite{Kim2020} show ocean dynamical changes are key to driving the AMV, but there is debate about the impact of the AMOC on AMV. Some argue the AMV is a result of internal AMOC variability (e.g. \citealt{Enfield:2001}), whereas others argue the AMV is a result of the SST response to changes in radiative forcing (e.g. \citealt{Booth2012}, \citealt{Bellomo:2018}) or atmospheric forcing (e.g. \citealt{Clement2015,Clement2016}, \citealt{Cane:2017}); these mechanisms have however been disputed by other authors (e.g. \citealt{Zhang2017}), \citealt{Zhang:2013}, \citealt{Zhang:2016}, \citealt{Yan:2017}, \citealt{Yan2018a}). Other studies have proposed the AMOC-AMV relationship is a coupled system with a two-way feedback (e.g. \citealt{Fraser2021}). The high number of SST-based studies is at least in part the result of the availability of relatively long records, in comparison to other observables.

\subsubsection{AMOC influence on mean climate}
\label{I_role_clm}

Numerous studies have considered the relationship between the AMOC and Altantic climate (e.g. \citealt{Cassou:2018}). The AMOC accounts for approximately 90\% of northward ocean heat transport within the Atlantic (\citealt{Johns:2011}), estimated to be approximately 1.3PW at RAPID. This has a substantial effect on winter temperatures in north-western Europe. Warm western boundary currents prevent excessive sea ice formation in the sub-polar North Atlantic, releasing heat into the atmosphere. The result is relatively warm conditions over the greater North Atlantic region compared to similar latitudes of the North Pacific (\citealt{Palter2015}). 

From spring 2009 to spring 2010, the AMOC exhibited an approximate 30\% decline, partially due to a strong negative North Atlantic Oscillation impacting the wind field. This reduction resulted in a northward heat transport of only 0.9PW at RAPID, leading to an abrupt reduction of North Atlantic SST and a noticeably colder winter for the UK (e.g. \citealt{Srokosz2015}). 

Since the AMOC transports heat across the equator, the northern hemisphere is slightly warmer than its southern counterpart. This uneven spatial distribution of heat results in the Inter Tropical Convergence Zone (ITCZ) being located to the north of the equator (e.g. \citealt{Marshall2014a}). The location of the ITCZ affects tropical rainfall and characteristic wet and dry seasons. A potential weakening in the AMOC could result in more Arctic sea ice and a colder Arctic, an equator-ward shift of the ITCZ, and weakened Asian and Indian monsoons; conversely a stronger AMOC could lead to less sea ice, warmer Arctic and a northward shift of the ITCZ (e.g. \citealt{Vellinga2002}).  

In general, the AMOC influences the rate and location at which heat is taken up and released by the ocean (e.g. \citealt{Kostov2014}), the vertical distribution of heat, and the rise in sea level due to thermal expansion. The timescale of the ocean (and climate) response to increases in atmospheric $\text{CO}_2$ depends upon the rate of ocean heat uptake and the efficiency with which the ocean transports carbon and heat into the deep ocean. With increasing greenhouse gas emissions (\citealt{HANSEN1985}, \citealt{Takahashi2009}, \citealt{Perez2013}), the criticality of the AMOC's role in sequestering heat and carbon in the deep ocean increases.

\subsubsection{AMOC weakening}
As outlined in Section \ref{I_obsAMOC}, deployment of new monitoring technologies will improve the coverage of current and future observational data for the Atlantic Ocean in space and time. Together with better data-constrained models, this will improve our understanding of the AMOC. However, at the current time, our knowledge of the inherent longer-term natural variability of the AMOC, and the processes governing this variation, is rather limited. The need to consider and quantify the effects of climate change complicate things further.

\begin{figure*}[ht!]
	\centerline{\includegraphics[width=\textwidth]{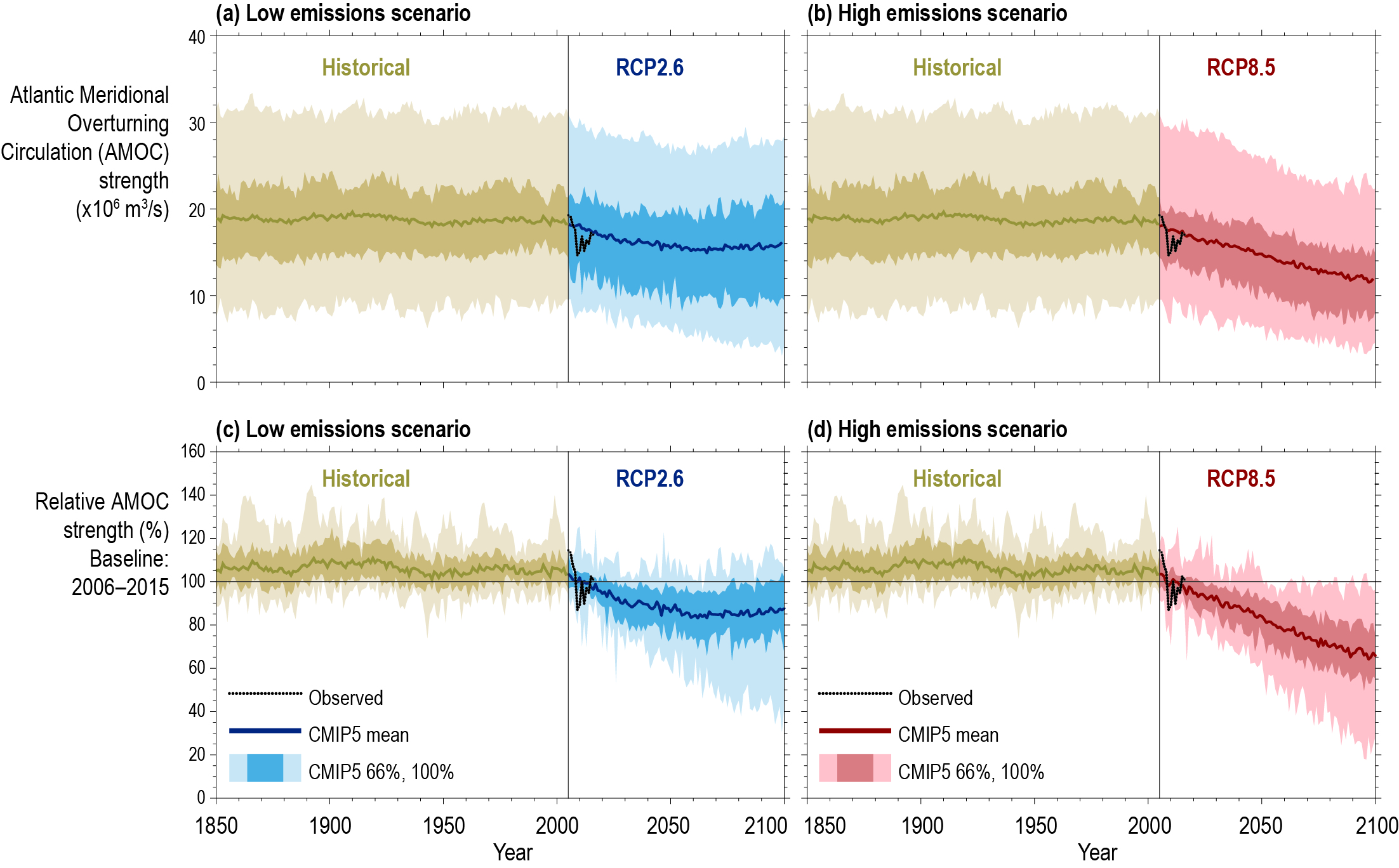}}
	\caption[AMOC changes at $26^\circ$N as simulated by 27 models (reproduced from IPCC 2021 - Figure 6.8).]{AMOC changes at $26^\circ$N as simulated by 27 models. The dotted line shows the observation-based estimate at $26^\circ$N (\citealt{McCarthy2015}) and the thick beige/blue/red lines the multi-model ensemble mean. Values of AMOC maximum at $26^\circ$N (in units $10^6 \text{m}^3\text{s}^{\text{-1}}$ ) are shown in historical simulations (most of the time 1850–2005) followed for 2006–2100 by a) Representative Concentration Pathway (RCP)2.6 simulations and b) RCP8.5 simulations. (The value following ``RCP'' indicates a specific emissions scenario. e.g. RCP8.5 results in a net top-of-atmosphere radiative imbalance of 8.5 Wm$^{\text{-2}}$ by the year 2100). In c) and d), the timeseries show the AMOC strength relative to the value during 2006–2015, a period over which observations are available. c) shows historical followed by RCP2.6 simulations and d) shows historical followed by RCP8.5 simulations. The 66\% and 100\% ranges of all-available CMIP5 simulations are shown in beige for historical, blue for RCP2.6 scenario and red for RCP8.5 scenario.. Reproduced from IPCC (2021) - Figure 6.8.}
	\label{AMOC_slow}
\end{figure*}

There is general consensus between models that the AMOC will weaken in future (see Figure \ref{AMOC_slow}). \cite{Bryden2005} claimed evidence for a weakening of $30\%$ in the AMOC, based on data from 5 hydrographic sections measured over a period of 47 years. However, their study was strongly dependent on one AMOC estimate from 1957, and moreover the observed trend is thought to be biased due to aliasing of seasonal and higher-frequency variability (\citealt{Kanzow2010}). 

The concern about a weakening overturning, and its climate and societal consequences, was one of the main motivations to establish the RAPID array. Some fear a complete shut-down in circulation is possible, but this is highly debated, and serves to emphasise the importance of improved understanding of the processes governing the ocean's response to different forcing mechanisms at all timescales. 


\cite{Rahmstorf2015}, \cite{Caesar2018} and \cite{Caesar2021} utilise proxies such as salinity, SST-based indices, sortable silt and $\delta^{\text{18}}$O to infer that the AMOC has weakened over the past century, and is currently at its weakest point in the last millennium. However, \cite{Worthington2021} reconstruct the AMOC timeseries for the past 30 years using an empirical model based on RAPID and hydrographic cruise data and find no signs of AMOC weakening. Similarly, \cite{Fraser2021} reconstruct the AMOC for the past 120 years using the Bernoulli inverse applied to the EN4.2.1 observational dataset to suggest no significant weakening trend.

Changes of the AMOC have been linked to paleo-climate shifts (\citealt{Broecker2003}), suggesting that abrupt changes in climate might have been caused by the AMOC switching on and off. Idealised models suggest the AMOC might also be bistable (e.g. \citealt{STOMMEL1961}), but this has yet to be shown with a state-of-the-art eddy-resolving GCM (\citealt{Mecking2016}, \citealt{Weijer2019a}). Whether or not the AMOC is in a bistable regime is difficult to say; possible deficiencies in the models such as inaccurate freshwater budgets, key to a bistable AMOC, have been highlighted (\citealt{Buckley2016}, \citealt{Weijer2019a}). \cite{Weijer2019a} provide a review of AMOC stability. 

With an ever warming climate, and polar amplification, increased ice melting will lead to less dense surface waters at high northern latitudes and a reduction of deep convection in sinking regions. The lighter NADW formed may become too light to outcrop in the Southern Ocean (\citealt{Wolfe2010}, \citealt{Wolfe2014}) further weakening the overturning. 

Freshwater fluxes at high latitudes had been thought to be key in understanding the slowing of the AMOC, but recent research has shown changes in air-sea heat fluxes and ocean temperature are primarily the cause (\citealt{Gregory2005}, \citealt{Weaver2007}, \citealt{Marshall2015}).  Hosing experiments within GCMs have shown a weakening of the AMOC, and if the period of hosing is sufficient, a possible shutdown (\citealt{Jackson2018}); but no shutdown has been observed under typical global warming scenarios. \cite{Jackson2015} introduce large amounts of freshwater at high latitudes to simulate the impact of a collapsed AMOC within the HadGEM3 GCM. Notable impacts on the North Atlantic include stronger storm tracks, widespread cooling, greater sea-ice coverage and large changes in precipitation. For Europe, they find less summer precipitation, more precipitation as snow and less vegetation and crop productivity.

In summary, it is difficult to estimate trends (such as possible recent slowing) in AMOC from short periods of observation, since it is not possible to distinguish between the effects of climate change and those of inherent natural variability at different timescales. Model-based assessment, using models whose credibility is established by comparison with available observations, is also crucial.

\section{The Antarctic Circumpolar Current (ACC)}
\label{I_SO} \label{I_ACC}

\subsection{The Southern Ocean}
Southern Ocean dynamics play a critical role in maintaining the AMOC. High northern latitude deep water formation is largely balanced by the Ekman upwelling caused by Southern Ocean westerly winds (\citealt{Toggweiler1995}, \citealt{Kuhlbrodt2007}). The unblocked Drake Passage latitudes allow for inter-basin exchange of waters, nutrients and biogeochemical products (e.g. \citealt{Moon:2018}). \cite{Gnanadesikan1999} highlights the Southern Ocean's critical role in determining the structure of the global pycnocline, and consequent impacts on the ocean's ability to store carbon at depth. The upwelling of NADW in the Southern Ocean, and subsequent formation of Antarctic Intermediate Water and AABW, make it an ideal region for uptake of carbon into the ocean interior.  The Southern Ocean is responsible for almost 40\% of the global ocean's anthropogenic carbon uptake (\citealt{Gruber2019}), and 75\% of the uptake of excess heat accumulated in the Earth system (\citealt{Frolicher2015}) since 1861. 

The major large-scale features of Southern Ocean dynamics are the eastward ACC, and the Weddell and Ross gyres. Along the Antarctic boundary we find a reversal of wind stresses (easterlies) and a westward flowing Antarctic Slope Current (ASC). \cite{Thompson2018} review the properties of the Antarctic shelf to determine the dynamical processes that influence the local variability of the ASC, and cross-slope transport. They classify the Antarctic shelf into three categories: (a) fresh shelf, (b) dense shelf and (c) warm shelf (see Figure 2-4 of \citealt{Thompson2018}). 

The ``warm'' shelf is characterised by a shallow layer (upper 200m) of cold surface waters, lying directly above warm waters. A warm shelf occurs typically along the West Antarctic Peninsula, where there are weak easterly winds, and no ASC (in fact, a reverse current is present). Typically, these are sites where Antarctic ice shelves are thinning rapidly (\citealt{Pritchard2012}). ``Fresh'' shelves are characterised by cold shelf waters, penetrating deeper than those on a warm shelf (over the upper 600m), adjacent to large ice shelves. East Antarctica, the western Amundsen Sea and eastern Ross Sea are regions of strong coastal easterly winds, giving rise to Ekman transport, leading to downwelling, and a barrier between cold and fresh shelf waters, and warm and salty offshore waters. In these regions, isopycnals perpendicular to the coast, slump near the coast, forming a strong ASC. ``Dense'' shelves typically correspond to regions of deep water formation, such as the Weddell and Ross Seas and near the Adelie coast, and exhibit a sideways V-shape frontal structure of dense water. Typically,  near the surface (the upper side of the V) waters are cold and fresh; the lower down-slope side consists of denser saline waters flowing down the continental shelf and slope. These regions are crucial to forming densest water masses such as the AABW. 

The ACC is the strongest oceanic current, flowing without restriction around Antarctica. Zonally unblocked latitudes are crucial for its formation and strength. An equator-to-pole temperature gradient and strong westerly winds act to steepen isopycnal gradients, leading to an eastward-flowing current. The majority of the ACC flow is concentrated into 4 fronts, namely the (a) subtropical front, (b) subantarctic front, (c) polar front, and (d) southern ACC front. The sheer size of the Southern Ocean, and its harsh conditions, make the ACC difficult to observe. The easiest monitoring location is at the Drake Passage, the point of narrowest constriction between South America and Antarctica. The driving mechanisms for the ACC, leading to tilted isopycnals with northward Ekman transport compensated by a deep geostrophic return flow and mesoscale eddies, have been discussed in Section \ref{I_drv}.

\subsection{Observing the ACC}
\label{I_obsACC}

The ACC was studied as long ago as the HMS Discovery expedition of $1929-1930$. Early estimates of the ACC transport varied greatly, in part due to the use of different reference depths for geostrophic calculations, and lack of knowledge of near-bottom current characteristics. \cite{Bryden1977} used a pilot array of 15 moorings in the Drake Passage to show a level of no-motion near the sea-floor would not be appropriate, due to bottom currents of the order of 1-2cms$^{\text{-1}}$. This work informed the International Southern Ocean Studies (ISOS) programme in the late 1970s and early 1980s (\citealt{Whitworth1985}). ISOS utilised bottom-pressure gauges, dynamic-height moorings, current-meter moorings and hydrography (\citealt{Whitworth1982}, \citealt{Whitworth1983}, \citealt{Whitworth1985}), leading to significantly improved understanding of the Southern Ocean, and estimates of ACC transport of $134\text{Sv}$ with standard deviation $11.2$Sv. \cite{Cunningham2003} revisited the ISOS data, methodology and assumptions, estimating a mean ACC transport of $134$Sv with uncertainty of between $15$Sv and $27$Sv. They cite baroclinic variability as an important contributor to net ACC variability.

\cite{Meredith2011} reviewed estimates of the baroclinic transport through the Drake Passage using $15$ hydrographic cruises along the $SR1b$ section. They showed, for the period $1993-2009$ (including data from the World Ocean Circulation Experiment), that the baroclinic transport through the Drake Passage is $136.7\pm6.9\text{Sv}$, and remained relatively constant throughout the period of observation. \cite{Firing2011} and \cite{Renault2011} provide other reviews. \cite{Koenig2014} estimate a 20-year timeseries of total ACC transport using the DRAKE mooring array, and satellite data. Moored current meter data from 2006 to 2009, and satellite altimetry data from 1993 to 2012 are combined to create a look-up table of vertical velocity profiles to estimate transport. This method relies on a dependence of the vertical velocity structure on surface velocity and latitude. They find the full‐depth transport to be $141\pm 2.7$Sv over the 20 year period, with a baroclinic component of 136Sv. 

Using data collected from the $cDrake$ experiment, a combination of CPIES and bottom current and pressure meters at a location slightly west of $SR1b$, \cite{Chidichimo2014} show that the resultant average baroclinic transport through the passage is $127.7\pm8.1\text{Sv}$ between $2007-2011$. They again emphasise the relative stability of the year-on-year time-mean transport. A further study by \cite{Donohue2016}, based on the same $cDrake$ experiment, quotes an additional depth-independent or barotropic transport component of $45.6$Sv, calculated using bottom current recorders. The updated estimate of $173.3\pm10.7$Sv for the total baroclinic plus barotropic transport through the Drake Passage is approximately $30\%$ larger than the benchmark of $130$ to $140$Sv against which ACC transport in climate models is typically compared.

\subsection{ACC variability}

Meridional density gradients across the Southern Ocean drive the ACC. Therefore, any process which acts to steepen or flatten isopycnals can influence ACC transport. For example, all of freshening, surface heating, diapycnal mixing and changes in NADW properties influence the strength of the ACC. Many studies of ACC temporal variability have concentrated on the Drake Passage. The observational studies discussed in Section \ref{I_obsACC} suggest a large uncertainty in the estimate of total transport there. This is particularly true for the barotropic or depth-independent component of the ACC transport. 

On short timescales, the barotropic component is found to be the dominant contributor to variability (\citealt{Whitworth1983}, \citealt{Hughes1999}). On sub-seasonal timescales, \cite{Hughes1999} find that ACC transport variability can be related to sea level near the Antarctic coast. They show within a numerical model of the Southern Ocean, that ACC transport is closely tied with the wind field there. Stronger westerly winds lead to a stronger northward Ekman transport, and therefore a drop in sea level near Antarctica. These changes in wind field correlate with the Southern Annual Mode (SAM) index (e.g. \citealt{Meredith2004}), and influence ACC transport variability within models over a range of timescales (\citealt{Hughes2014}). The SAM is the dominant mode of atmospheric variability in the southern hemisphere, and therefore it is not surprising that it influences ACC transport variability. Others suggest possible secondary influences of the Madden-Julian Oscillation and the El-Ni\~no Southern Oscillation (e.g. \citealt{Matthews2004}, \citealt{Sallee2008}). 

\cite{Thompson2002} observe an increasing SAM index since the 1970s in conjunction with stronger westerly winds. However, no observational evidence of a long-term trend in ACC transport is found to corroborate this (\citealt{Cunningham2003}, \citealt{Meredith2011}, \citealt{Koenig2014}, \citealt{Donohue2016}). \cite{Straub1993} and later \cite{Hallberg2001} hypothesise that this lack of dependence is due to eddy saturation. This implies that stronger wind-forcing increases mesoscale eddy energy, instead of changing the time-mean strength of the current, and hence the ACC is insensitive to changes in wind stress. The eddy saturation hypothesis is supported by model results (e.g. \citealt{Gnanadesikan2006}, \citealt{Munday2013}, \citealt{Abernathey2014}) and in line with surface observations by \cite{Hogg2015}. However, no direct ACC transport observations have yet verified this hypothesis. 

The ACC is sparsely observed, and any potential monitoring programme would require high-frequency temporal sampling, to prevent aliasing (\citealt{Meredith2005}). \citealt{Hughes2014} highlight the difficulty of resolving vertical structure which varies spatially and temporally. Once again, model-based studies would appear extremely useful in characterising ACC transport structure, and its longer-timescale variability, provided we have sufficient confidence in the model: good agreement between model output and observation (where available) is an essential first step.

\section{Thesis aims}
\label{I_aims}
This thesis seeks to explore the extent to which MOCs can be constrained by relatively easy-to-measure boundary information, and hence whether boundary information provides a basis for useful reconstruction of MOCs. The thesis aims to:
\begin{itemize}
	\item Develop and evaluate a conceptual model of basin-wide AMOC using only boundary information. This diagnostic model represents the complex three-dimensional large-scale fluid structure in two dimensions, allowing easier interpretation of spatio-temporal patterns.
	\item Quantify contributions of the various boundary components in the decomposition to the time-mean AMOC overturning streamfunction, and its variability on annual and longer timescales, as a function of latitude and depth.
	\item Investigate biases in UK Met Office HadGEM-GC3.1 models for the ACC at the Drake Passage, using a similar decomposition diagnostic tailored to zonal flows.
	\item Identify and physically interpret regions of flat and sloping along-boundary isopycnals in oceanic basins including the Atlantic and Antarctic.
\end{itemize}

\section{Thesis Structure}
\label{I_strct}
The layout of the thesis is as follows. Chapter \ref{TJ_TM} introduces a methodology to decompose the basin-wide AMOC overturning streamfunction into contributions from boundary densities, surface wind stresses and bottom velocities. The decomposition is applied to HadGEM-GC3.1 GCM output, to investigate the time-mean contributions of boundary components. Chapter \ref{TJ_Var} investigates the contributions of each boundary component to the variability of the AMOC overturning streamfunction. It includes analysis at RAPID and SAMBA locations, and quantification of the importance of the contributions from each boundary component across a range of timescales. Chapter \ref{TJ_Bdry} maps and characterises boundary density, potential temperature and salinity along sloping boundaries of the Atlantic basin and neighbouring coastlines, using a novel boundary mapping algorithm. Regions of along-boundary flat or linearly-sloping isopcynals are investigated, and possible underlying physical mechanisms discussed. Chapter \ref{TJ_ACC} extends the decomposition method of Chapter \ref{TJ_TM} to the ACC at the Drake Passage and elsewhere, to investigate biases within HadGEM-GC3.1 models at different spatial resolutions. Density structure is also investigated along the sloping boundaries of Antarctica. Chapter \ref{TJ_Smm} provides a summary of results, conclusions and opportunities for further work.

 \clearpage

\chapter{MOC decomposition and time-mean AMOC characteristics}  \label{TJ_TM}

\renewcommand{\arraystretch}{1.5}
\setlength{\arrayrulewidth}{0.5mm}

In this chapter, the theoretical basis and practical application of the decomposition diagnostic for the MOC are outlined. Using HadGEM3-GC3.1 model data, the overturning streamfunction of the AMOC for any given latitude is decomposed into its constituent boundary components. The time-mean AMOC overturning streamfunction reconstructed from these boundary components is compared to the overturning streamfunction calculated directly from time-mean meridional velocities; further, the time-mean contributions of the four boundary components are analysed. Discrepancies are investigated and possible explanations are discussed. Finally, the time-mean maximum overturning streamfunction and contributing boundary components are characterised across the full range of Atlantic latitudes. 

\section{Background}
\label{S3_Bck}

A decomposition of the overturning (introduced in Sections \ref{I_dynL} and \ref{I_oAMOC}) was first performed by \cite{Lee1998}, and formed the basis of the RAPID methodology. They decomposed the meridional velocity into Ekman, external mode and vertical shear components to highlight the relative importance of various dynamical processes within the seasonal cycle of heat transport in the Indian Ocean. 

\cite{Marotzke1999} suggested the possibility of ``boundary monitoring'' the MOC based on a study using the adjoint of the Massachusetts Institute of Technology general circulation model (MITgcm). They showed that the dynamical sensitivity of heat transport variability is generally greatest to density perturbations near meridional boundaries of the basin. Further, the net meridional transport is influenced by density anomalies only once they reach the boundary. \cite{Hirschi2003} and \cite{Baehr2004} used two eddy-permitting numerical ocean models to test the practical feasibility of the \cite{Marotzke1999} method. \cite{Baehr2004} note within the FLAME model at $26^\circ$N that the dominant contributors to the overturning streamfunction are from the boundary density terms and the Ekman transport, with sometimes significant but generally smaller contributions from other terms. The feasibility and performance of a hypothetical monitoring array in reconstructing the AMOC depends upon the strength of bottom velocities; in regions where strong currents hit sloping boundaries, the contribution of a ``depth-averaged'' velocity (or ``external mode'') term becomes significant. In such regions, the MOC cannot be reliably estimated using density and surface wind stresses alone. 

The work of these authors led to the deployment in $2004$ of the RAPID-MOCHA mooring array (Section \ref{I_obsAMOC}) at a latitude of $26.5^\circ$N to monitor overturning volume and heat transports. The main components of the overturning streamfunction estimate are: (a) the geostrophic interior flow, (b) Ekman transport due to wind-stresses and (c) transport through Florida straits, measured using a telephone cable (\citealt{Cunningham2007}, \citealt{Kanzow2007}). 

\cite{Sime2006} note the external mode (i.e. the component arising due to depth-independent flow) becomes important in regions of strong western boundary currents and overflows. They find for latitudes between $25^\circ$N and $32^\circ$N, and at $62^\circ$N that the external mode is a dominant contributor to the time-mean overturning streamfunction. 

\section{Method for decomposition of the MOC}
\label{S3_Mth}

Here we discuss theoretical principles underlying the decomposition diagnostic and the mathematical formulation of the boundary components contributing to the overturning streamfunction.

Using approximations of the Navier-Stokes equation appropriate for large-scale ocean dynamics (Section \ref{I_dynL}), we can relate the zonal pressure gradient to a meridional velocity using the geostrophic relationship (Equation \ref{GB}). It is reasonable to assume geostrophic balance holds even within the western boundary current for a flow parallel to the coastline (\citealt{Bingham2008}, \citealt{Bell2011}). In conjunction with hydrostatic balance (Equation \ref{HB}), geostrophy can be used to establish an expression for the vertical structure of the northward velocity dependent upon the zonal density gradients. This is known as the thermal wind relationship (Equation \ref{TW}). The overturning streamfunction $\Psi(z; y)$, equivalent to the cumulative meridional volume transport up to a chosen depth $z$ for a given meridional coordinate $y$, is defined by Equation \ref{T}. We partition the meridional transport into three main components: (a) depth-independent (barotropic, sea-level or surface-pressure-change-driven) flow, (b) baroclinic flow (related to density gradients) and (c) wind-stress driven (i.e. Ekman) flow. Using Equation \ref{T}, the meridional velocity $v_{N}(x,z; y)$ (with north as positive) can be decomposed as
\begin{equation}
v_{N}(x,z; y) = v_{bot}(x; y) + v_{th}(x,z;y) + v_{Ekm}(x,z;y),
\label{v}
\end{equation}
where $v_{bot}(x; y)$ is the meridional velocity adjacent to the bathymetry, a single value for each fluid column (indexed by zonal and meridional coordinates $x$ and $y$, and independent of depth $z$). $v_{th}(x,z;y)$ represents the density or thermal wind contribution to the velocity, zero adjacent to the bottom and defined to be in thermal wind balance with the east-west gradient of the in-situ density, $\rho(x,z; y)$. $v_{Ekm}(x,z;y)$ represents the Ekman or wind-stress contribution to the velocity, found within upper layers of the ocean.

The total estimated overturning streamfunction $\tilde{\Psi}(z; y)$, equivalent to the total estimated cumulative MOC transport up to depth $z$, calculated from the decomposition diagnostic, will be written in the form 
\begin{equation}
\tilde{\Psi}(z; y) = \Psi_{bot}(z; y) + \Psi_{th}(z; y) + \Psi_{Ekm}(z; y),
\end{equation}
where the terms correspond to contributions from bottom velocities, boundary densities, and surface wind-stresses. As will be shown below, $\Psi_{th}$ can further be expressed in terms of the eastern and western boundary density contributions. We use the notation $\tilde{\Psi}$ for the total estimated overturning streamfunction from the boundary decomposition, in order to distinguish it from the (expected total) overturning streamfunction $\Psi$ calculated directly using Equation \ref{T}. 

For derivation of boundary components contributing to the overturning streamfunction, we assume a local Cartesian coordinate system, for simplicity of presentation only. Actual calculations, discussed in Section \ref{S_App_DD}, exploit the NEMO Arakawa ``C'' model grid, incorporating appropriate metric terms. The depth-independent flow component of the overturning streamfunction, $\Psi_{bot}$, can be simplified to 
\begin{eqnarray}
\Psi_{bot}(z; y) &=& \int_{W(z; y)}^{E(z; y)} \int_{-H(x; y)}^{z} v_{bot}(x; y) dz' dx \\ 
&=& \int_{W(z; y)}^{E(z; y)} \left[ \int_{-H(x; y)}^{z} dz'\right]v_{bot}(x; y)  dx \\
&=& \int_{W(z; y)}^{E(z; y)} Z_{bot}(x,z; y) v_{bot}(x; y) dx, 
\label{Q_{bot}}
\end{eqnarray}
where $W(z; y)$ and $E(z; y)$ represent the western and eastern boundaries at each $z$ for a given $y$, and $Z_{bot}(x,z; y)$ is the depth of the water below the height $z$,
\begin{align}
Z_{bot}(x,z; y) = & H(x; y) + z,  & z > -H(x; y), \\
Z_{bot}(x,z; y) = & 0,  & z \leq -H(x; y) ,
\label{Z_{bot}}
\end{align}
where $H(x; y)$ is the height of the fluid column, i.e. within the fluid column $Z_{bot}=H+z$, and outside the fluid column $Z_{bot} =0$. The depth-independent flow component can be thought of as a compensatory component which introduces a geostrophic reference velocity into the calculation, accounting for pressure variations within the interior of the ocean across the basin.

The thermal wind component $\Psi_{th}(z; y)$ of the overturning streamfunction is calculated using integration by parts, noting that $v_{th}(x,z; y) = 0$ when $z = -H(x; y)$; i.e. the thermal wind velocity is zero at the bathymetry. Integration by parts gives
\begin{equation}
\begin{split}
\Psi_{th}(z; y) = & \int_{W(z; y)}^{E(z; y)} \int_{-H(x; y)}^{z} v_{th}(x,z'; y) dz'dx  \\
= & \int_{W(z; y)}^{E(z; y)} \left( [z'v_{th}(x,z'; y)]_{-H(x; y)}^z - \int_{-H(x; y)}^{z} z' \frac{\partial v_{th}}{\partial z'} (x,z'; y) dz' \right) dx .
\end{split}
\label{Tc_1}
\end{equation}

Since $v_{th}(x,z'; y) = 0$ when $z' = -H(x; y)$, we see that $[z'v_{th}(x,z'; y)]_{-H(x; y)}^z = zv_{th}(x,z; y)$. Further, $zv_{th}(x,z; y)$ can be written in terms of the vertical gradient in velocity
\begin{equation}
zv_{th}(x,z; y) = z\int_{-H(x; y)}^{z} \frac{\partial v_{th}}{\partial z'} (x,z'; y) dz',
\label{e_zvth}
\end{equation}
which in turn can be related to the horizontal density gradient using the thermal wind relationship (Equation \ref{TW}), as follows. Using Equation \ref{TW}, and substituting the value of $zv_{th}(x,z; y)$ from Equation \ref{e_zvth} into Equation \ref{Tc_1},
\begin{align}
\label{Tc_2}
&\hspace{-20pt} \Psi_{th}(z; y)\\ \nonumber
&= \int_{W(z; y)}^{E(z; y)} \left( z\int_{-H(x; y)}^{z} \frac{\partial v_{th}}{\partial z'}(x,z'; y) dz' - \int_{-H(x; y)}^{z} z' \frac{\partial v_{th}}{\partial z'}(x,z'; y) dz' \right) dx \\ \nonumber
&= \int_{W(z; y)}^{E(z; y)} \int_{-H(x; y)}^{z} (z-z') \frac{\partial v_{th}}{\partial z'}(x,z'; y) dz' dx \\ \nonumber
&= - \frac{g}{f(y) \rho_0} \int_{W(z; y)}^{E(z; y)} \int_{-H(x; y)}^{z} (z-z') \frac{\partial \rho }{\partial x} (x,z'; y) \mathrm{d} z' \mathrm{d} x    \\ \nonumber
&= - \frac{g}{f(y) \rho_0} \int_{-H_{\max}(y)}^{z} (z-z') \left[ \int_{W(z; y)}^{E(z; y)} \frac{\partial \rho }{\partial x} (x,z'; y) \mathrm{d} x \right]  \mathrm{d} z' \\ \nonumber
&= - \frac{g}{f(y) \rho_0} \int_{-H_{\max}(y)}^{z} ( z - z' ) \left[ \sum_{d=1}^{n_D(z;y)} ( \rho_{E_d}(z'; y)  - \rho_{W_d}(z'; y) ) \right] \mathrm{d} z' .
\end{align}

\begin{figure*}[ht!]
	\centerline{\includegraphics[width=\textwidth]{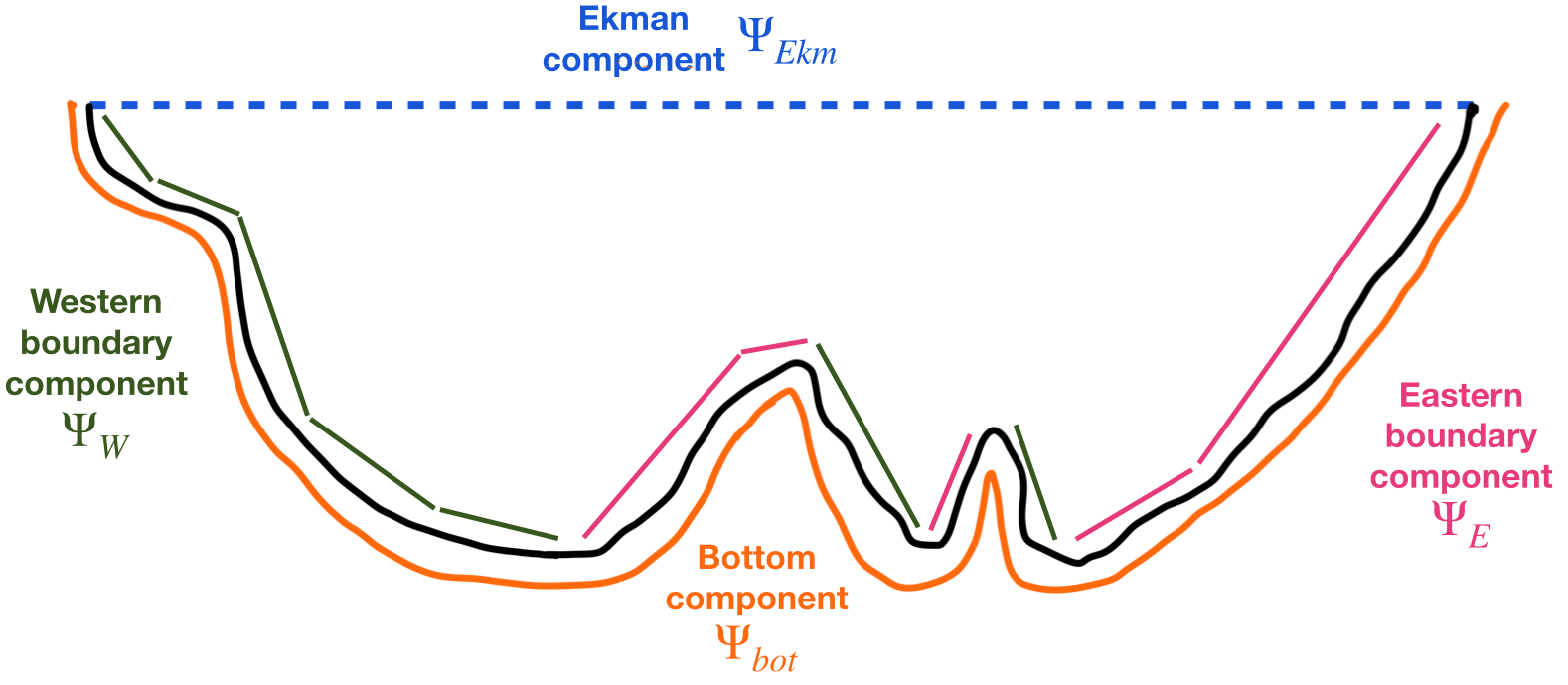}}
	\caption[Schematic of boundary decomposition and contributing components to the AMOC overturning streamfunction.]{Schematic of the boundary decomposition and the contributing components to the AMOC overturning streamfunction. We note in this illustration that multiple pairs of western and eastern boundaries are present at depth, whereas only one pair is present at the surface.}
	\label{F_bdryAMOC}
\end{figure*}

Equation \ref{Tc_2} states that the thermal wind contribution to the overturning streamfunction can be written as the sum of western and eastern boundary density contributions. In the third line of the equation we use the thermal wind relationship (Equation \ref{TW}) to introduce density $\rho(x,z; y)$ and the latitude-dependent Coriolis parameter $f(y)$. In the fourth line the order of integration is reversed, noting that $H_{\max}(y) = \max_x H(x; y)$, the maximum basin depth at meridional coordinate $y$. In the fifth line we introduce boundary densities $\rho_E(z; y) = \rho(E(z; y),z; y)$ and $\rho_W(z; y) = \rho(W(z; y),z; y)$. Here, we are careful to acknowledge the possible presence of $n_D \ge 1$ pairs of western and eastern boundaries for $z$ and $y$, with zonal coordinate intervals $[W_d(z,y),E_d(z,y)]$, $d=1,2,...,n_D(z;y)$, as illustrated in Figure \ref{F_bdryAMOC}. 

\begin{figure*}[ht!]
	\centerline{\includegraphics[width=\textwidth]{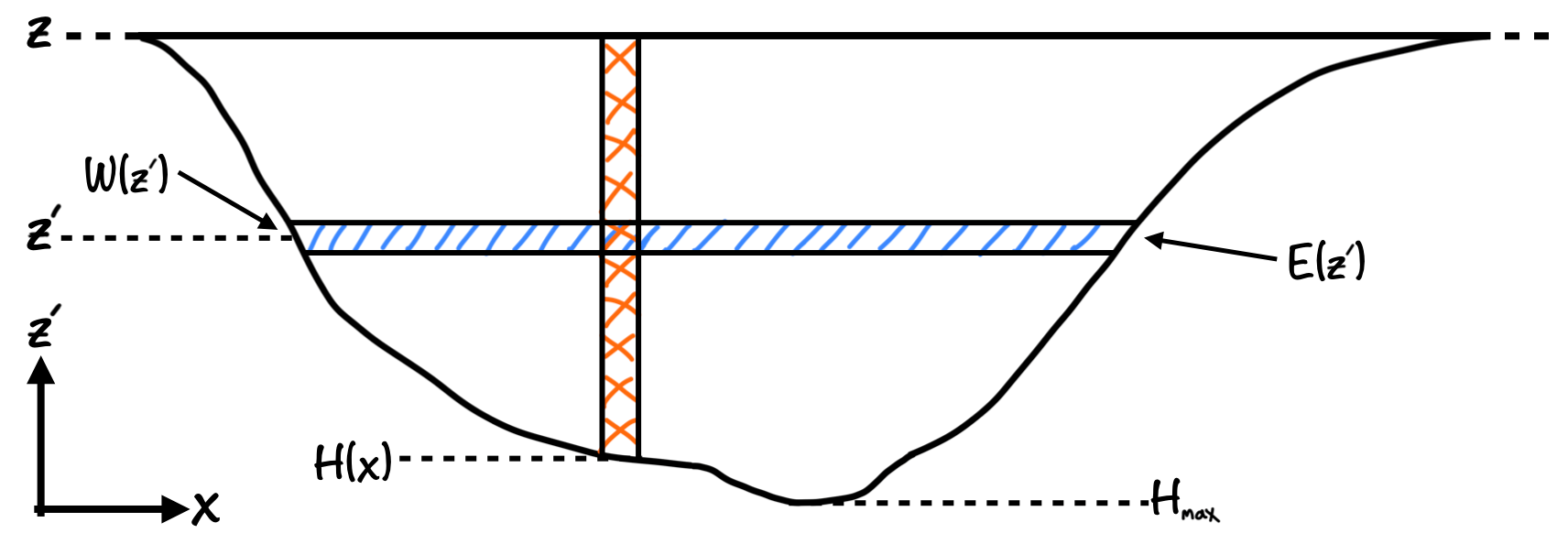}}
	\caption[Schematic of vertical and horizontal integration taking place in calculation of overturning streamfunction up to depth $z$ in Equation \ref{Tc_2}.]{Vertical and horizontal integration taking place in calculation of overturning streamfunction up to depth $z$ in Equation \ref{Tc_2}.}
	\label{f_Integral}
\end{figure*}

Intuitively Equation \ref{Tc_2} is explained as follows, and illustrated in Figure \ref{f_Integral}. For each $y$, the original integral requires a ``vertical'' integration of $\partial \rho / \partial x$ from the bathymetry to depth $z$ (shaded orange in Figure \ref{f_Integral}), followed by ``horizontal'' integration from the western to eastern boundaries. After reversal of the order of integration, the integral involves horizontal integration for each $z'$ from the western to eastern boundaries (shaded blue in Figure \ref{f_Integral}), followed by vertical integration from the maximum water depth to $z$; importantly the result of the horizontal integration can be written purely in terms of western and eastern boundary densities.

The individual western and eastern boundary density contributions from Equation \ref{Tc_2} are
\begin{equation}
\Psi_W(z; y) = \frac{g}{f(y) \rho_0}  \int_{-H_{\max}(y)}^{z} ( z - z' ) \left[ \sum_{d=1}^{n_D(z;y)} \rho_{W_d} (z'; y) \right] \mathrm{d} z',
\end{equation}
\begin{equation}
\Psi_E(z; y) = - \frac{g}{f(y) \rho_0}  \int_{-H_{\max}(y)}^{z} ( z - z' ) \left[ \sum_{d=1}^{n_D(z;y)} \rho_{E_d}(z'; y) \right] \mathrm{d} z',
\end{equation}
so that $\Psi_{th}(z; y) = \Psi_W(z; y) + \Psi_E(z; y)$.

The Ekman contribution to the overturning streamfunction is estimated using surface wind stress data. The total contribution of the wind stress for all $x$ at the given $y$ is calculated and then evenly distributed over the upper $50$m of the ocean, or the depth of the ocean if shallower than $50$m. We recognise the depth of the wind-mixed Ekman layer varies with latitude, longitude and time. However, given our interest lies in the MOC at greater depth, the choice of $50$m has minimal influence on our results. Therefore the Ekman contribution to the overturning streamfunction is defined as
\begin{eqnarray}
\Psi_{Ekm}(z; y) &=& - \int_{W(z; y)}^{E(z; y)} \int_{-h(x; y)}^{z} \frac{1}{h(x; y)}\frac{\tau_S^x(x; y)}{\rho_0 f(y)} \mathrm{d}z'\mathrm{d}x \nonumber\\
&=& - \int_{W(z; y)}^{E(z; y)} \frac{(h(x;y)+z)}{h(x; y)}\frac{\tau_S^x(x; y)}{\rho_0 f(y)} \mathrm{d}x,
\label{Tek}
\end{eqnarray}
where $h(x; y)$ is the depth of the Ekman layer given by
\begin{align}
h(x; y) = & H(x; y),  & H(x; y) \leq 50m, \nonumber \\
h(x; y) = & 50m,  & H(x; y) > 50m ,
\label{e_h}
\end{align}
and $\tau_S^x(x; y)$ is the wind stress on the ocean surface in the zonal direction. The decomposition diagnostic for the overturning streamfunction at a given $y$ up to a given $z$ is therefore 
\begin{equation}
\tilde{\Psi}(z; y) = \Psi_W(z; y) + \Psi_E(z; y) + \Psi_{bot}(z; y) + \Psi_{Ekm}(z; y),
\label{T_smp}
\end{equation}
which equates to
\begin{equation}
\begin{multlined}
\tilde{\Psi}(z; y) =  \frac{g}{\rho_0 f(y)}  \int_{-H_{\max}(y)}^{z} ( z - z' ) \left[ \sum_{d=1}^{n_D(z';y)} \rho_{W_d}(z'; y) \right] \mathrm{d} z' \\
- \frac{g}{f(y) \rho_0}  \int_{-H_{\max}(y)}^{z} ( z - z' ) \left[ \sum_{d=1}^{n_D(z';y)} \rho_{E_d}(z'; y) \right] \mathrm{d} z' \\
+ \int_{W(z; y)}^{E(z; y)} Z_{bot}(x,z; y) v_{bot}(x; y) dx -  \frac{1}{\rho_0 f(y)} \int_{W(z; y)}^{E(z; y)} \int_{-h(x; y)}^{z} \frac{\tau_S^x(x; y)}{h(x; y)} \mathrm{d}z'\mathrm{d}x.
\end{multlined}
\label{T_bdry}
\end{equation}

We aim to investigate and further understand the contribution of these boundary components in relation to: (a) the time-mean overturning streamfunction, (b) temporal variability of the overturning streamfunction and (c) possible driving mechanisms for the observed variability of the overturning streamfunction at different timescales. In order to fulfil these objectives, we diagnose the boundary components using data from a GCM.

\section{Numerical model description}
\label{S3_ModDes}
In this section the general circulation model (GCM) used to investigate the overturning streamfunction decomposition is introduced. An outline of the GCM configuration is given first, followed by a detailed description of the $3$-D lattice structure used for GCM computation. Then we describe the calculation of the overturning streamfunction decomposition components in light of the lattice.

The GCM model configuration used to diagnose the overturning streamfunction into its boundary components is the UK Met Office coupled HadGEM3-GC3.1 model (described by \citealt{Williams2018}, \citealt{Roberts2019}). The GCM incorporates oceanic, atmospheric, land and sea-ice components. The oceanic component of the model (Met Office Global Ocean $6$, $GO6$) is based upon the Nucleus for European Modelling of the Ocean (NEMO $3.6$) model with $75$ vertical levels (\citealt{Storkey2018}). The thickness (vertically) of tracer (``T'') cells increases from 1m at the surface to $\approx250$m at depth. There are 3 model spatial resolutions of particular interest based upon the ``ORCA'' family within NEMO: ORCA12, ORCA025 and ORCA1 at $1/12^\circ$, $1/4^\circ$ and $1^\circ$ zonal grid spacing, respectively, at the equator. The isotropic Mercator grid used leads to a reduction in the meridional grid spacing together with reducing zonal grid spacing when approaching the poles. The ocean model is based upon a tripolar grid with quasi-isotropic grid spacing in the Northern Hemisphere, accommodating two poles in Siberia and Canada. NEMO GCM models incorporate a free surface by means of a ``terrain-following'' vertical s- (or z*-) coordinate system. The atmospheric component of the model consists of a Met Office Unified Model Global Atmosphere, of resolution $N216$ ($\approx60$km, \citealt{Walters2019}) with $85$ vertical levels. Finally, a GSI$8.1$ sea-ice component (\citealt{Ridley2018}) is used at the same resolution as the ocean component. 

Short-term temporal variability of the overturning streamfunction is relatively well understood due to the availability of observational data. The main body of research reported here involves analysis of longer-term trends using annual mean data from a GCM with $1/4^\circ$ oceanic spatial resolution, and a $N216$ atmospheric model using 1950s forcing. This $657$-year control run is one of the longest available. A long control run is essential when investigating temporal variability of the overturning streamfunction on annual and longer timescales. For testing of the AMOC decomposition diagnostics and comparison purposes, the $1^\circ$ and $1/12^\circ$ oceanic spatial resolutions with identical $N216$ atmospheric components are also considered. Each model run is initially spun-up for 30 years using 1950s greenhouse gas forcing before commencement of the control run. All three models form part of the High Resolution Model Intercomparison Project (HighResMIP) for CMIP6 (\citealt{Haarsma2016}). The simulation lengths for the $1^\circ$ and $1/12^\circ$ models are, respectively, $104$ and $176$ years.

\subsection{Application of AMOC decomposition diagnostics within HadGEM3-GC3.1 global climate model}
\label{S_App_DD}

\begin{figure*}[ht!]
	\centerline{\includegraphics[width=12cm]{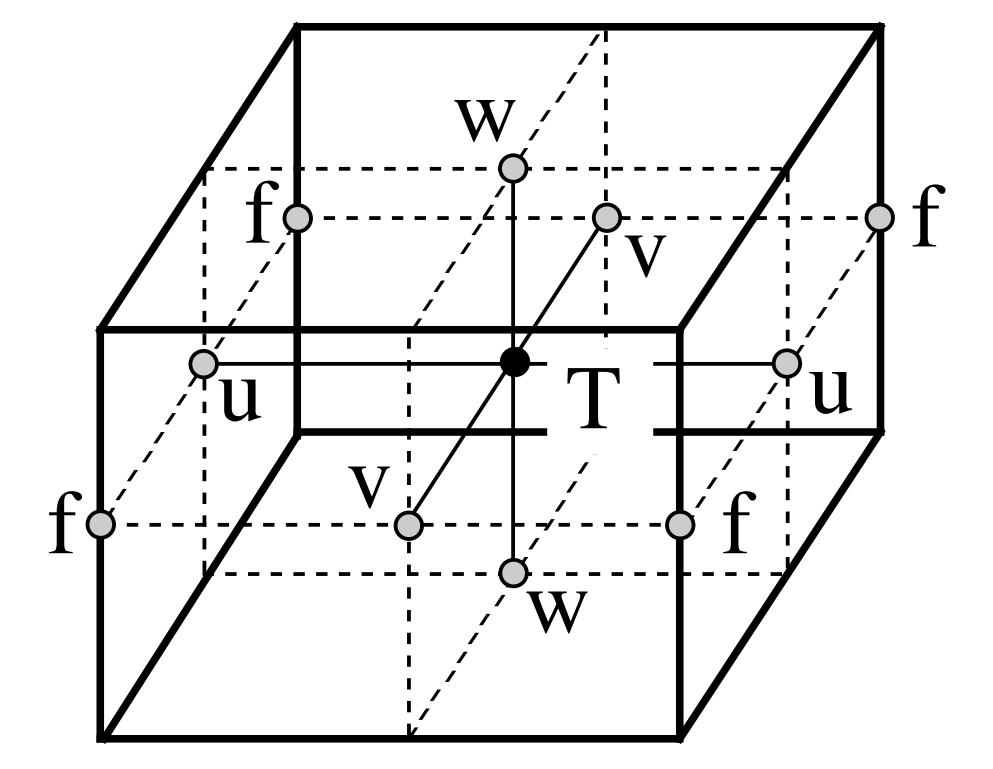}}
	\caption[Arrangement of Arakawa C grid variables (reproduced from \citealt{Madec2016}).]{ Arrangement of Arakawa C grid variables. $T$ indicates a scalar point where temperature, salinity, density and pressure are defined. $(u,v,w)$ indicates vector points, and $f$ indicates vorticity points where both relative and planetary vorticities are defined (reproduced from \citealt{Madec2016}).}
	\label{NEMO_grid}
\end{figure*}

Data fields in NEMO are located on an Arakawa ``C'' grid as shown in Figure \ref{NEMO_grid}, where the centre of each T-cell denotes the location of potential temperature $\theta$, salinity $S$ and pressure $P$, with Cartesian velocity components $u$, $v$, $w$ located at the centre of (and normal to) each cell face. Planetary and relative vorticities, both denoted by $f$ here, are located at the corner of each T-cell at the same vertical level as $u$ and $v$. The grid indices $i$, $j$, $k$ used for the decomposition diagnostic are not aligned with longitude and latitude everywhere; we find at high latitudes the alignment changes, therefore calculations are performed using the model grid rather than lines of constant latitude. In the following discussion, for ease of description, we nevertheless refer to the $i$ and $j$ indices, and the corresponding grid locations, as ``longitude'' and ``latitude''.

\begin{figure*}[ht!]
	\centerline{\includegraphics[width=\textwidth, angle = 0]{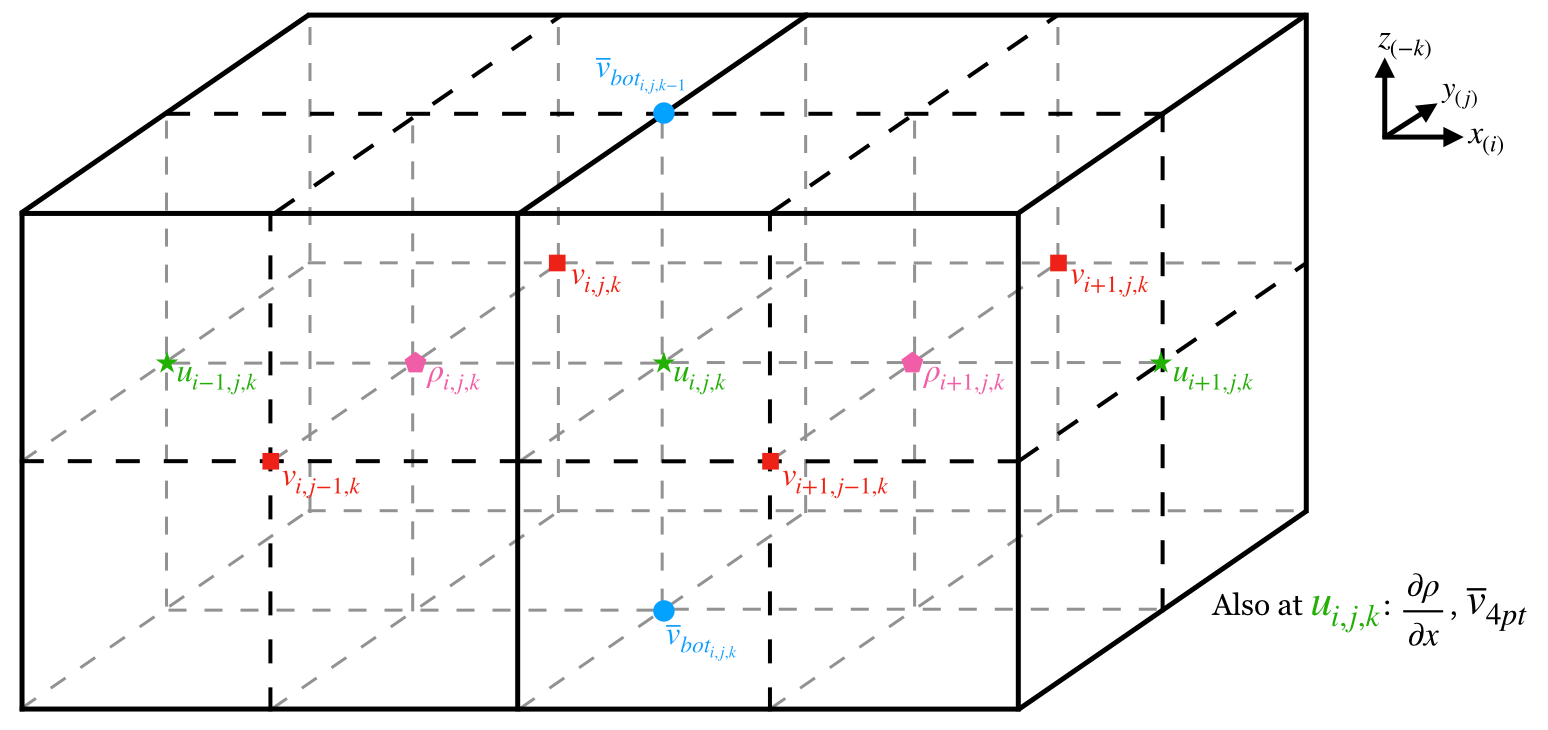}}
	\caption[Locations of variables used within the decomposition diagnostics.]{Locations of variables (for two T-cells) used within the decomposition diagnostics, using $i,j,k$ notation to indicate horizontal $(i,j)$ and vertical ($k$) indices. }
	\label{var_grid}
\end{figure*}

The geostrophic relationship (Equation \ref{GB}), on which the thermal wind relationship (Equation \ref{TW}) of relevance here is based, relates to the $x$-component of the momentum equation. For this reason we centre calculations at the $u$-point, in order to obtain the best representation of the geostrophic relationship within the NEMO framework. Centering calculations at the $u$-point also allows us to compare our calculations with relevant momentum diagnostics output by the model. However, calculation of the geostrophic and thermal wind relationships at the $u$-point causes some difficulties: (a) the calculation requires averaging of terms such as the planetary vorticity $f$ and northward velocity $v$ to calculate the Coriolis acceleration $fv$; (b) density points are located at the centre of the T-cell (Figure \ref{var_grid}) rather than at the $u$ or $v$-points and therefore, at a boundary, the closest density point is half a grid cell away from the boundary; and (c) partial cells present at the bottom bathymetry have differing depths; thus averaging across two cells can cause bias. Issues arising due to these sidewall cells (incomplete cells adjacent to eastern and western boundaries) and partial cells (incomplete cells adjacent to bottom bathymetry) are discussed further in Section \ref{S_TM_est}.

Using time-mean potential temperature and practical salinity fields, the in-situ density is calculated at the centre of each T-cell using the $1980$ equation of state ($eos80$ seawater toolbox, \citealt{Millero1980}), in line with the NEMO GCMs.

The boundary contributions to the overturning streamfunction, $\Psi_W$, $\Psi_E$, $\Psi_{bot}$ and $\Psi_{Ekm}$, described in Equation \ref{T_bdry}, are adapted for diagnostic calculation from the GCM data as follows. We assume a locally Cartesian lattice of locations $x_i$, $y_j$, $z_k$ on which $\rho$ is defined for each year indexed by time $t_{\ell}$. Then the western and eastern boundary density terms are calculated relative to a mean depth-dependent basin density $\overline{\rho}(z_k)$, with mean taken over longitude, latitude and time, 
\begin{equation}
\rho^*(x_i, y_j, z_k, t_{\ell}) = \rho(x_i, y_j, z_k, t_{\ell}) - \overline{\rho}(z_k).
\end{equation}
In cases where there are obstacles such as islands or ridges along the sea floor, there will be multiple pairs of western and eastern boundaries, and the sum of their contributions relative to $\overline{\rho}(z_k)$ is taken, as specified by the sum in e.g. Equation \ref{T_W_math} for $\Psi_W$.

Our interest lies in the depth structure of the boundary components contributing to the overturning streamfunction. We calculate this by first calculating the transport at each depth, and then summing the transport from the sea floor to the depth of interest. Equation \ref{Tc_2} is replicated within the model by 
\begin{equation}
\Psi_{W}(z_k, t_{\ell}; y_j) = -\frac{g}{f(y_j) \rho_{0}} \sum_{m=k}^{M}\left(z_{k-1 / 2}-z_{m}\right) \left[ \sum_{d=1}^{n_D(z_m;y_j)} \rho^*_{W_d}(z_m, t_{\ell}; y_j) \right] \Delta z_{m},
\label{T_W_math}
\end{equation}
\begin{equation}
\Psi_{E}(z_k, t_{\ell}; y_j) = + \frac{g}{f(y_j) \rho_{0}} \sum_{m=k}^{M}\left(z_{k-1 / 2}-z_{m}\right) \left[ \sum_{d=1}^{n_D(z_m;y_j)} \rho^*_{E_d}(z_m, t_{\ell}; y_j) \right] \Delta z_{m},
\label{T_E_math}
\end{equation}
where $k$ indexes the centre of a T-cell, with values running from $k=M$ at the deepest cell to $k=1$ at the surface cell. $\Delta z_m = z_{m-1/2} - z_{m+1/2}$ is the thickness of the $m^{\text{th}}$ level and $f(y_j)$ is the Coriolis parameter at latitude $y_j$. $\rho^*_{W}(z_m, t_{\ell}; y_j)$ and $\rho^*_{E}(z_m, t_{\ell}; y_j)$ are the western and eastern boundary anomaly densities at the $m^{\text{th}}$ level at latitude $y_j$ for year $t_{\ell}$. 

At each latitude, the bottom component of the overturning streamfunction  $\Psi_{bot}$ is calculated using the velocity $\overline{v}_{bot}$ at the bottom of the deepest full cell adjacent to the bathymetry, half a grid point below the lowest $u$-point considered, for each longitude. This is then multiplied by the cell thickness and width for each longitude, integrated over longitude, and integrated vertically upwards through the fluid column to the depth of interest. $\overline{v}_{bot}$ is calculated by taking the $4$-point average of the local velocities onto the $u$-point, and using the zonal density gradient across the bottom cell to obtain, via thermal wind, the vertical gradient in meridional velocity. These terms are summed to give the velocity at the bottom of the deepest cell, $\overline{v}_{bot}$. Therefore, mathematically,
%
%
\begin{equation}
\Psi_{bot}(z_k, t_{\ell}; y_j) = \sum_{i\in \mathcal{I}(z_k; y_j)}  (H(x_i; y_j) + z_{k-1/2}) \overline{v}_{bot}(x_i, t_{\ell}; y_j)  \Delta x_i
\label{T_b_math}
\end{equation}
where $\mathcal{I}(z_k; y_j)$ is an index set for longitudes between the western and eastern boundaries $W(z_k; y_j)$ and $E(z_k; y_j)$ at depth $z_k$ for latitude $y_j$, and $\Delta x_i$ is the width of the cell $\Delta x_i = x_{i+1/2} - x_{i-1/2}$.

The Ekman component of the overturning streamfunction is estimated by calculating the total Ekman transport for each latitudinal section from the wind, and then redistributing this transport as currents over the upper 50m, before integrating vertically and zonally. It was confirmed that the initially-calculated total Ekman transport was equal to the eventual Ekman cumulative transport up to the surface (or overturning streamfunction contribution at the surface) from the redistributed currents. The resulting expression for $\Psi_{Ekm}$ is
%
%
\begin{equation}
\Psi_{Ekm}(z_k, t_{\ell}; y_j) = \frac{1}{\rho_0 f(y_j)} \sum_{i\in \mathcal{I}(z_k; y_j)}  \frac{(h(x_i; y_j) + z_k)}{h(x_i; y_j)} \tau_S^x(x_i, t_{\ell}; y_j)  \Delta x_i
\label{T_Ekm_math}
\end{equation}
where $h(x_i; y_j)$ is defined in Equations \ref{e_h}, and $\tau_S^x(x_i, t_{\ell}; y_j)$ is the zonal surface wind stress as before.

Using the overturning streamfunction components defined above, the estimated overturning streamfunction $\tilde{\Psi}$ can be calculated using 
\begin{eqnarray}
\tilde{\Psi}(z_k, t_{\ell}; y_j) =\Psi_{W}(z_k, t_{\ell}; y_j) + \Psi_{E}(z_k, t_{\ell}; y_j) + \Psi_{bot}(z_k, t_{\ell}; y_j) + \Psi_{Ekm}(z_k, t_{\ell}; y_j) .
\label{e_T_tilda_math}
\end{eqnarray}
In order to assess the accuracy of this decomposition, the MOC overturning streamfunction is also calculated by vertically and zonally integrating the meridional velocity (Equation \ref{T}). For this calculation, we use a $4$-point average velocity $\overline{v}$ at each $u$-point, and a 2-point average velocity for cells adjacent to the bathymetry (for which the 4-point average is not available, discussed further in Section \ref{S_TM_est}). The ``true'' or ``expected'' overturning streamfunction $\Psi$ is therefore calculated using
%
%
\begin{equation}
\Psi(z_k, t_{\ell}; y_j) =  \sum_{m=k}^{M+1} \sum_{i\in \mathcal{I}(z_m; y_j)} \overline{v}(x_i,z_m, t_{\ell}; y_j) \Delta x_i \Delta z_m 
\label{e_T_math}
\end{equation}
where $M$ is again the vertical cell index for the bottommost interior cell in the basin, and level $M+1$ indicates an additional partial cell. The resulting overturning streamfunction $\Psi(z_k, t_{\ell}; y_j)$ can thus be compared directly to the total estimated overturning streamfunction $\tilde{\Psi}(z_k, t_{\ell}; y_j)$ calculated as the sum of the boundary components (Equation \ref{T_smp}). 

\section{Time-mean characteristics of the overturning streamfunction}
\label{S_TM_est}

In this section the estimated time-mean overturning streamfunction, calculated as the sum of boundary components, is compared to the expected overturning streamfunction calculated directly from meridional velocities. Contributions made by boundary components to the time-mean overturning streamfunction are also analysed. Conservation of volume is applied at each latitude in the form of a compensatory term, and the role of additional cells, not included in the decomposition diagnostic, is considered.

The time-mean overturning streamfunction $\Psi(z_k; y_j)$ at depth $z_k$ for latitude $y_j$, is calculated using the temporal average
\begin{equation}
\Psi(z_k; y_j) =  \frac{1}{n_Y}\sum_{\ell = 1}^{n_Y} \Psi(z_k, t_{\ell}; y_j)
\label{e_T_avg}
\end{equation}
where $n_Y$ is the number of years in the model run. Time-mean $\tilde{\Psi}(z_k; y_j)$ (Equation \ref{e_T_tilda_math}) and the corresponding time-mean boundary components are calculated in a similar fashion. 

At each latitude $y_j$ individually throughout the Atlantic basin, for latitude interval ($34^\circ$S, $67^\circ$N), we calculate the overturning streamfunction to the surface, $\Psi(0; y_j)$. To conserve volume we expect that this value should be near to zero. However, it is known that a small volume inflow occurs through the Bering Strait, estimated by \cite{McDonagh2015} to be of order 1.1Sv. In addition, freshwater via melting sea ice and river outflow provide other sources of water into the Atlantic. For simplicity, we assume basin volume is conserved for each latitude and therefore impose no net meridional transport at each latitude ($\Psi(0;y_j) = 0$). 

At the RAPID monitoring array (\citealt{Cunningham2007}, \citealt{McCarthy2015}), depth-independent interior flows, other than those within the western boundary wedge, are represented using a uniformly distributed volume conservation term. In our work, depth-independent interior flows are primarily allocated to the bottom component of the decomposition, with volume compensation being used subsequently to close any remaining imbalances in the volume budget. We impose volume conservation throughout the basin by redistributing $\Psi(0; y_j)$ uniformly as a velocity $v_c$ in the opposite direction across the longitude-depth cross-section for each individual latitude. 

For any individual overturning streamfunction component $\Psi_{\eta}$, the compensated component $\Psi_{\eta}^c$ is given by
\begin{eqnarray}
\Psi_{\eta}^c(z_k, t_{\ell}; y_j) = \Psi_{\eta}(z_k, t_{\ell}; y_j) - C_{\eta}(z_k, t_{\ell}; y_j) , \label{e_etaC_T}
\end{eqnarray}
where
\begin{eqnarray}
C_{\eta}(z_k, t_{\ell}; y_j) = \sum_{m=k}^{M}\left( E(z_m; y_j) - W(z_m; y_j) \right) v_c (t_{\ell}; y_j) \Delta z_m  
\label{e_etaC_C}
\end{eqnarray}
and the compensation velocity in this case is
\begin{eqnarray}
v_c(t_{\ell}; y_j) = \frac{\Psi_{\eta}(0; y_j)}{\sum_{m=1}^{M} \left( E(z_m; y_j) - W(z_m; y_j) \right) \Delta z_m } .
\label{e_etaC_vc}
\end{eqnarray}

In Equation \ref{e_etaC_vc}, the denominator represents the total basin cross-sectional area. This compensation is applied independently at each latitude for each individual overturning streamfunction component, and the expected $\Psi$ and estimated $\tilde{\Psi}$ overturning streamfunction. Note that the estimated compensated overturning streamfunction is related to the individual compensated boundary components, as expected, by
\begin{eqnarray}
\tilde{\Psi}^c(z_k, t_{\ell}; y_j) =\Psi_{W}^c(z_k, t_{\ell}; y_j) + \Psi_{E}^c(z_k, t_{\ell}; y_j) + \Psi_{bot}^c(z_k, t_{\ell}; y_j) + \Psi_{Ekm}^c(z_k, t_{\ell}; y_j) .
\label{e_etaC_Ttilda}
\end{eqnarray}

\begin{figure*}[ht!]
	\centerline{\includegraphics[width=\textwidth]{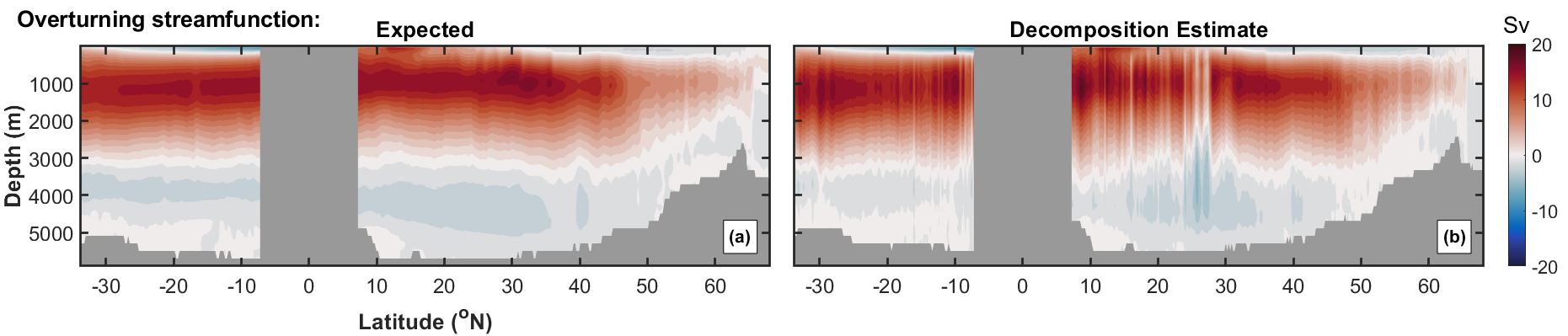}}
	\caption[Comparison of expected and estimated time-mean overturning streamfunction for the $1/4^\circ$ model.]{Comparison of time-mean overturning streamfunction for the $657$ year-long $1/4^\circ$ dataset calculated directly from meridional velocities ($\Psi^c$, Equation \ref{e_T_math}) and the estimated overturning streamfunction $\tilde{\Psi}^c$ reconstructed from the boundary components (Equation \ref{e_etaC_Ttilda}).}
	\label{F_TM_strm}
\end{figure*}

The compensation term corresponding to the expected overturning streamfunction $\Psi$ (Equation \ref{e_T_avg}) is denoted by $\Psi_{vol}$, and is considered further in Chapter \ref{TJ_Var}. Compensated overturning streamfunctions $\Psi^c$ and $\tilde{\Psi}^c$ at each latitude are shown in Figure \ref{F_TM_strm}, with $\Psi^c = \tilde{\Psi}^c = 0$ at the surface. Hereafter, when we discuss the expected and estimated overturning streamfunctions, we are in fact referring to the estimated and expected ``compensated'' overturning streamfunctions ($\Psi^c$, $\tilde{\Psi}^c$).

In Figure \ref{F_TM_strm}, the gradient of the overturning streamfunction at a particular depth indicates the magnitude and direction of volume transport. Flow around positive maxima (deep red) is clockwise, whereas flow around negative minima (deep blue) is anticlockwise. Thus, the figure indicates northward flow near the surface and a southward flow at roughly $1000$-$4000$m. Latitudes near to the equator between $7^\circ$S and $7^\circ$N are omitted since the Coriolis parameter $f$ (present as $1/f$ in $\Psi_W$ and $\Psi_E$) vanishes at the equator, and thus geostrophic balance is no longer a good approximation.

The reconstructed Atlantic overturning streamfunction $\tilde{\Psi}^c$ (Figure \ref{F_TM_strm}(b)) compares reasonably well at leading order to the expected overturning streamfunction $\Psi^c$ (Panel (a)). In terms of spatial variance (explained in Appendix \ref{App_SpatVar}), $\tilde{\Psi}^c$ explains $95.8\%$ of the spatial structure of $\Psi^c$. However, there are clear differences between the two streamfunctions. Potential reasons for these differences will be discussed in the remainder of the chapter. 

\begin{figure*}[ht!]
	\centerline{\includegraphics[width=16cm]{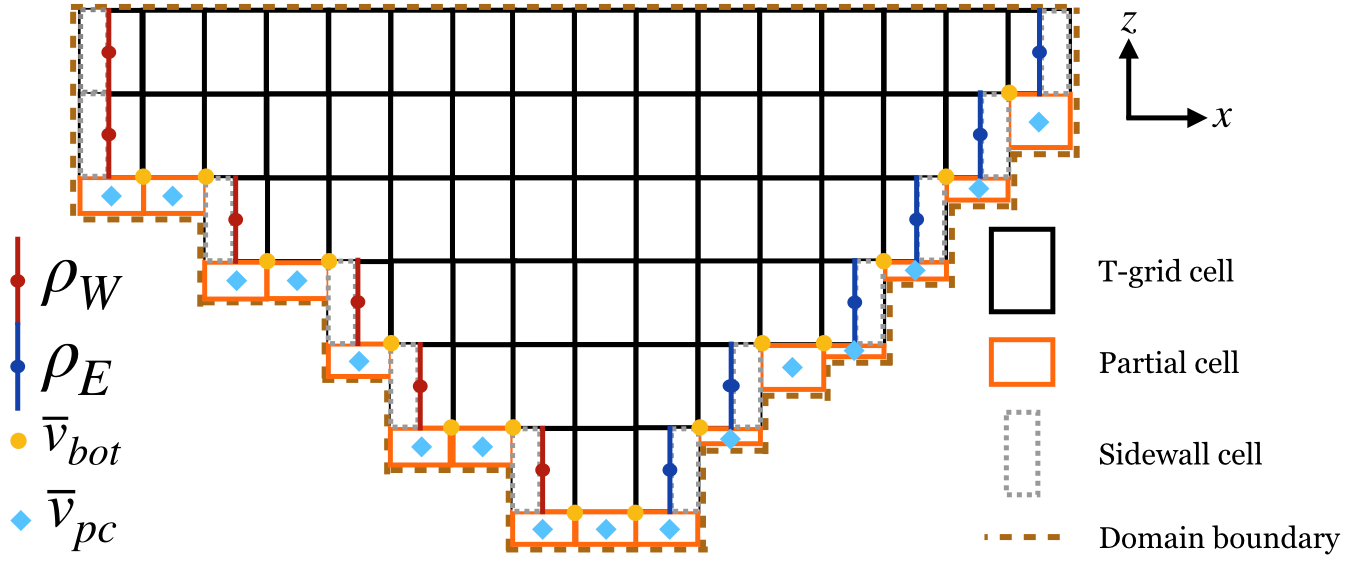}}
	\caption[A longitude-depth cross-section of cells, including interior, partial and sidewall cells.]{A longitude-depth cross-section of cells, including interior, partial and sidewall cells. Sidewall cell contributions are calculated using $\overline{v}_n$ located at the centre of the full T-cell adjacent to boundary (same location as $\rho_W$ and $\rho_E$ in figure). Partial cell contributions are calculated using the northward velocity in the partial cell, $\overline{v}_{pc}$.}
	\label{F_Sch_Cells}
\end{figure*}

A region of clear difference between $\tilde{\Psi}^c$ and $\Psi^c$ is the latitude range $25^\circ$N to $30^\circ$N. Here strong western boundary currents are present, and a possible reason for weaker transport within $\tilde{\Psi}^c$ is model formulation. As discussed in Section \ref{S3_ModDes}, primary variables (e.g. $\rho$, $\theta$, $S$) are located at the centre of a T-cell. Cells containing full field information are known as ``complete'' cells or interior cells. Near boundaries and bathymetry, additional cells are present, referred to as ``sidewall'' and ``partial'' cells, respectively. The latter cells replicate the original boundary as closely as possible given model resolution. Unfortunately, partial cells are of varying vertical thickness relative to adjacent interior cells at the same $k$-level. It is therefore challenging to calculate unambiguously the density gradient across a $u$-cell within a partial cell. Hence the decomposition calculation cannot be replicated routinely using densities for these cells. This is an unavoidable limitation of our approach. Sidewall and partial cells are illustrated in Figure \ref{F_Sch_Cells}.

\begin{figure*}[ht!]
	\centerline{\includegraphics[width=\textwidth]{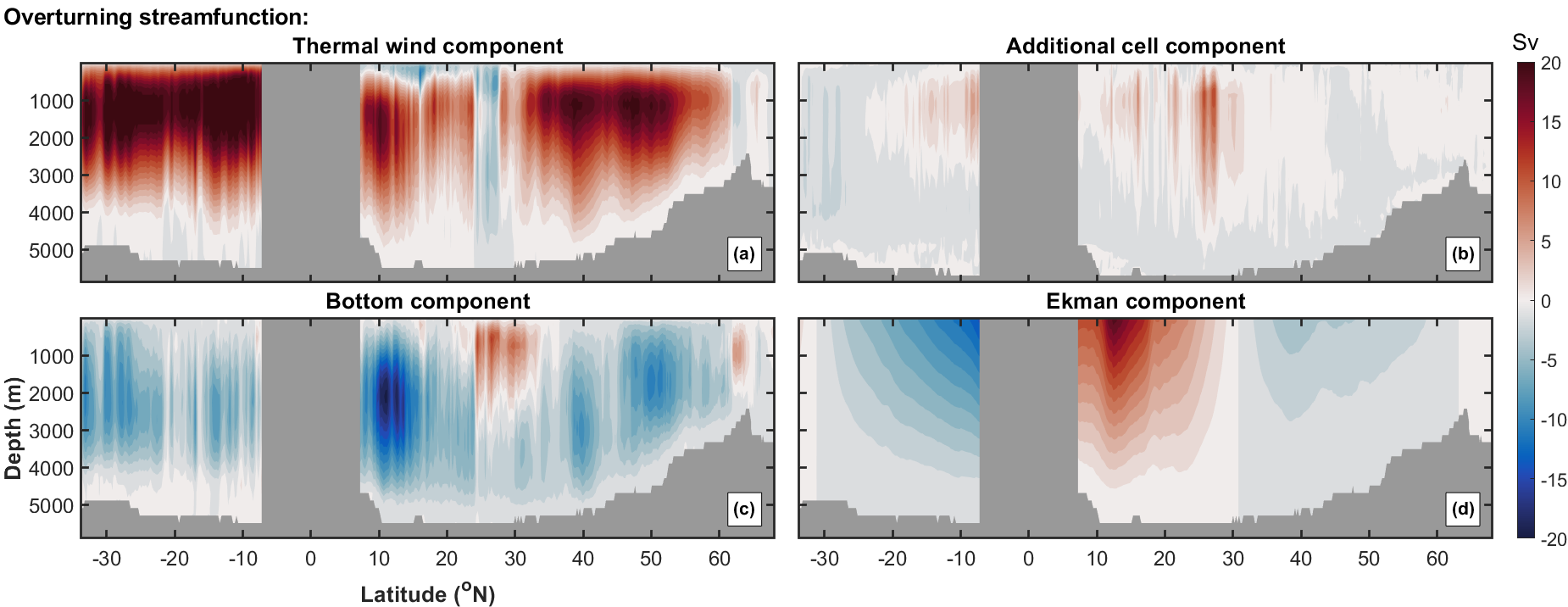}}
	\caption[Time-mean contributions from boundary components to the estimated overturning streamfunction $\tilde{\Psi}^c$ for $1/4^\circ$ model.]{Time-mean contributions from boundary components to the estimated overturning streamfunction $\tilde{\Psi}^c$ shown in Figure \ref{F_TM_strm} for $1/4^\circ$ model. Panels are; (a) thermal wind contribution, $\Psi_W^c+\Psi_E^c$, (b) additional cell contribution $\Psi_{AC}^c$ from partial and sidewall cells, (c) depth-independent flow contribution $\Psi_{bot}^c$ and (d) Ekman contribution $\Psi_{Ekm}^c$.}
	\label{F_TM_BC}
\end{figure*}

However, using northward velocities present in these incomplete cells, we are able to estimate their compensated overturning streamfunction contributions, referred to as $\Psi_{AC}^c(z_k, t_{\ell}; y_j)$. This calculation is similar to that performed for the expected overturning streamfunction $\Psi$ (Equation \ref{e_T_math}, without compensation applied), but for additional cells (i.e. sidewall and partial cells) only. Figure \ref{F_TM_BC} shows boundary component and additional cell contributions to the total estimated compensated overturning streamfunction $\tilde{\Psi}^c$.

The thermal wind component is equivalent to the aggregated contribution $\Psi_W^c+\Psi_E^c$ of western and eastern boundary density components. $\Psi_W^c$ and $\Psi_E^c$ have similar large magnitudes but opposite signs. Very little is gained from visual inspection of $\Psi_W^c$ and $\Psi_E^c$ individually. The thermal wind component dominates the majority of northward transport within the basin, except for in the surface region between $7^\circ$N and $25^\circ$N where strong easterly winds result in northward Ekman transport, and $25^\circ$N to $30^\circ$N where strong boundary currents through the Florida Straits result in a large depth-independent flow contribution, $\Psi_{bot}^c$. \cite{Sime2006} also observe a strong contribution of the ``external mode'' component, similar to our depth-independent component, in HadCM$3$ data.

The contribution of additional cells to the time-mean overturning streamfunction is relatively large in magnitude. At the current spatial resolution of $1/4^\circ$, the width of sidewall cells near the coastline can vary with longitude and latitude. Similarly, the depths of partial cells can vary from a few metres to approximately $260$m deep. Unsurprisingly, in regions of strong western boundary currents and bottom flows, the contribution made by these cells to the overturning streamfunction throughout the water column is significant and cannot be ignored. We find that by adding $\Psi_{AC}^c$ to the decomposition estimate $\tilde{\Psi}^c$, the spatial variance of ${\Psi}^c$ explained by $\tilde{\Psi}^c + \Psi_{AC}^c$ rises from $95.8\%$ to $97.9\%$. Further analysis of the additional cells and how to minimise their neglected contribution is discussed in Section \ref{S3_ResCmp}.

\section{Impact of GCM model resolution on $\Psi_{AC}^c$}
\label{S3_ResCmp}
Issues related to sidewall and partial cells arise because of the finite resolution of the NEMO computational grid. For coarse spatial resolution, the area corresponding to additional cells in a longitude-depth cross-section is relatively large. Therefore, increasing grid resolution should reduce the contribution of additional cells. However, as spatial resolution increases, the number of additional cells also increases to describe the bathymetry more precisely, and hence the total area corresponding to partial and sidewall cells may reduce relatively slowly. Therefore, it is interesting to consider the effect of grid resolution on the importance of additional cell contributions to the overturning streamfunction. 
\begin{figure*}[ht!]
	\centerline{\includegraphics[width=\textwidth]{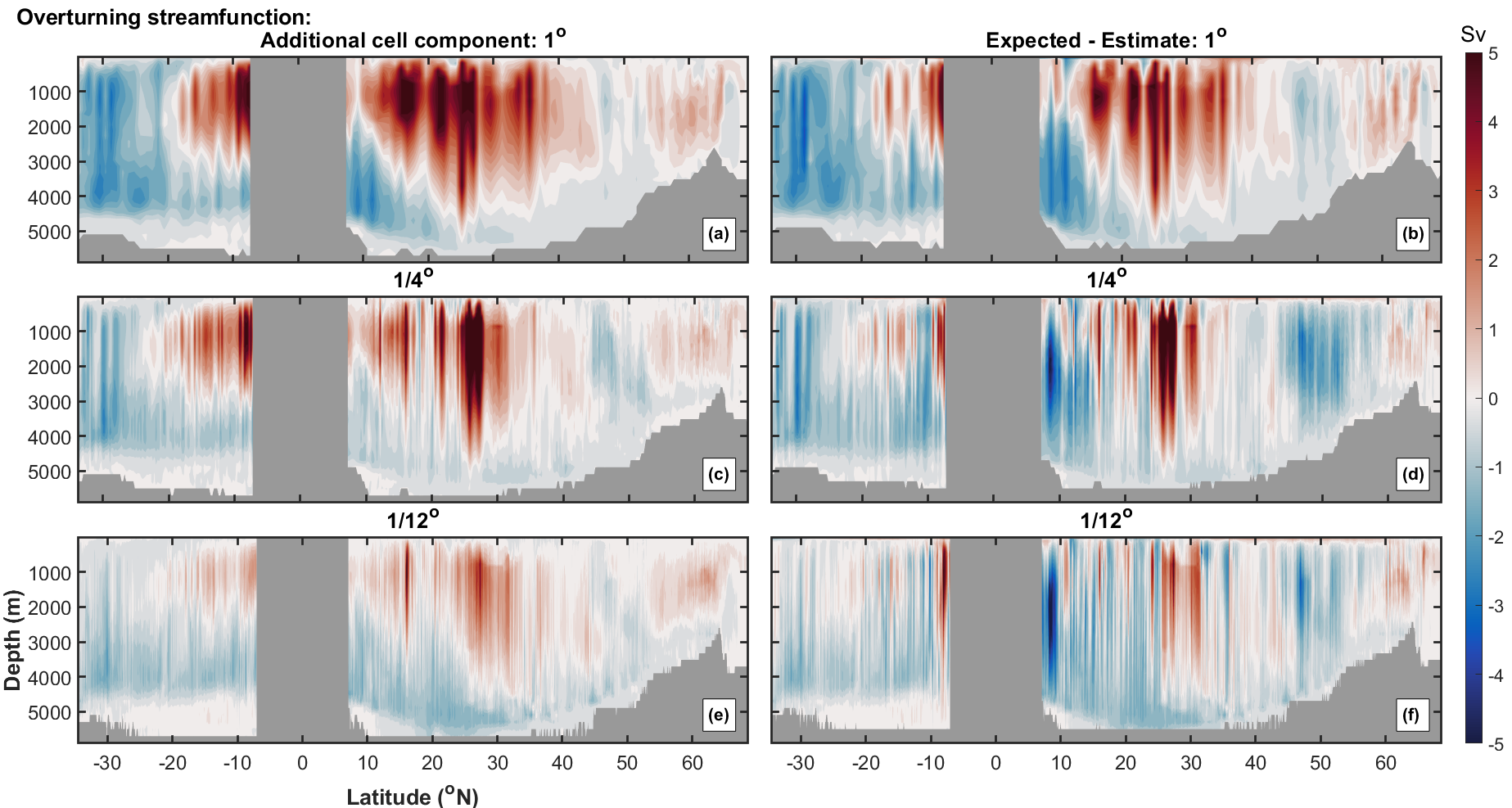}}
	\caption[Comparing time-mean contribution from additional cells and the difference between expected $\Psi$ and decomposition estimate $\tilde{\Psi}^c$ of the overturning streamfunction.]{Left: time-mean contribution from additional cells ($\Psi_{AC}^c$) for $1^\circ$ (a), $1/4^\circ$ (c) and $1/12^\circ$ (e) grid resolutions. Right: difference between expected $\Psi^c$ and decomposition estimate $\tilde{\Psi}^c$ (b,d and f).}
	\label{F_AC_c}
\end{figure*}

To test the role of ocean model grid resolution in determining the additional cell contribution, the decomposition diagnostic is applied to GCM data corresponding to $1^\circ$ and $1/12^\circ$ spatial resolutions, and compared with the decomposition using $1/4^\circ$ data discussed previously. Figure \ref{F_AC_c} shows the contribution $\Psi_{AC}^c$ of additional cells in comparison to the difference $\Psi^c-\tilde{\Psi}^c$ between the estimated $\tilde{\Psi}^c$ and expected overturning streamfunction $\Psi^c$. The figure suggests that the additional cell contribution is at least partly responsible for the structure of $\Psi^c-\tilde{\Psi}^c$; this will shortly be quantified. Further, increasing spatial resolution reduces the additional cell contribution in some areas. However increasing resolution does not bring better agreement between $\Psi^c$ and $\tilde{\Psi^c}$ at all latitudes for the resolutions considered; we still find regions poorly explained by $\tilde{\Psi^c}$ even when additional cells are considered.

For the same underlying initial settings (e.g. atmospheric resolution, spin-up period) applied to the three spatial resolutions, we find that for volume compensated overturning streamfunctions, the spatial variance (Appendix \ref{App_SpatVar}, Table \ref{Tab_SV}) of $\Psi^c$ explained by $\tilde{\Psi}^c$ (without additional cells) increases with increasing resolution. Adding the additional cell contribution $\Psi_{AC}^c$ into the decomposition estimate $\tilde{\Psi}^c$ for all resolutions in turn results in a significant improvement in spatial variance explained for the $1^\circ$ and $1/4^\circ$ models, and a smaller improvement for the $1/12^\circ$ model.

\begin{table}[h!]
	\centering
	\begin{tabular}{ |c||c|c|c| }
		\hline
		\multicolumn{4}{|c|}{Spatial variance of $\Psi^c$ explained} \\
		\hline
		& $1^\circ$ &$1/4^\circ$&$1/12^\circ$\\
		\hline
		$\tilde{\Psi}^c$  & $93.1\%$    &$95.8\%$& $98.3\%$\\
		$\tilde{\Psi}^c+\Psi_{AC}^c$ &   $98.3\%$  & $97.9\%$   &$98.6\%$\\
		\hline
	\end{tabular}
	\caption[Quantifying the role of additional cell contributions to the overturning streamfunction, using spatial variation of $\Psi^c$ explained by $\tilde{\Psi}^c$ and $\tilde{\Psi}^c+\Psi_{AC}^c$.]{Investigating the role of additional cell contributions to the overturning streamfunction: spatial variation of $\Psi^c$ explained by $\tilde{\Psi}^c$ and $\tilde{\Psi}^c+\Psi_{AC}^c$.  }
	\label{Tab_SV}
\end{table}

We note that for the $1/4^\circ$ model, as shown in Appendix \ref{App_SpatVar}, $\tilde{\Psi}^c$ provides a nearly unbiased estimate for the expected overturning streamfunction $\Psi^c$. It is interesting, however, that $\tilde{\Psi}^c+\Psi_{AC}^c$ alone also provides an estimate for $\Psi^c$ with increased bias but considerably reduced uncertainty. We conclude that inclusion of additional cells introduces considerable extra information into the expected overturning streamfunction relative to our estimate based on boundary quantities, at the expense of increased bias (seen by comparing the value of $\sigma$ with $\sigma^*$, and $\mu$ with $\mu^*$ in Equation \ref{e_A_BiasVar}).

Previous work by \cite{Allison2009} and \cite{Burton2010} used similar decomposition methods to investigate the variability of the ACC through the Drake Passage, and the overturning in the North Atlantic, respectively. Both studies were able to capture the temporal variability well, but capturing the time-mean value of the ACC or AMOC overturning streamfunction was more problematic. \cite{Burton2010} found her estimates for the time-mean overturning streamfunction for a specific depth and latitude were double the value expected from direct application of Equation \ref{T}. No explanation for this discrepancy, similar to that encountered by \cite{Allison2009}, was provided. 

The ability of the decomposition diagnostic to capture the spatial structure of $\Psi^c$ is promising. Discrepancies shown in Figure \ref{F_TM_strm} are partially explained by the contribution of the additional cells, but further improvement might be possible. We explore potential sources of discrepancies between estimated and expected overturning streamfunctions in subsequent sections. Direct comparison of the observed overturning streamfunction timeseries at RAPID and SAMBA with corresponding GCM-based timeseries is made in Section \ref{S3_RS}.  

We note that contributions from additional cells are henceforth incorporated into the definition of the estimated overturning streamfunction, namely 
\begin{eqnarray}
	\tilde{\Psi}^c(z_k, t_{\ell}; y_j) &=& \Psi_{W}^c(z_k, t_{\ell}; y_j) + \Psi_{E}^c(z_k, t_{\ell}; y_j) + \Psi_{bot}^c(z_k, t_{\ell}; y_j) \nonumber \\ 
	&+& \Psi_{Ekm}^c(z_k, t_{\ell}; y_j) + \Psi_{AC}^c(z_k, t_{\ell}; y_j). \quad \quad 
	\label{e_tilde_TcF}
\end{eqnarray}
\section{Exploring the discrepancy between ${\Psi}^c$ and $\tilde{\Psi}^c$}
One motivation for performing decomposition diagnostic calculations at $u$-points on the model grid is that equivalent momentum trends and their constituents for these exact same locations can be accessed directly. 

The nature of discrepancies between $\Psi^c$ and $\tilde{\Psi}^c$ (Figure \ref{F_AC_c}) raises questions regarding possible theoretical oversights in constructing the decomposition diagnostic. In this section we will show that the reconstructed time-mean overturning streamfunction calculated using underlying momentum contributions is better than that achieved using the decomposition diagnostic. We will show that the main reason for the discrepancy is the non-linear dependence of in-situ density on potential temperature, and the use of annual-mean potential temperature fields within the in-situ density calculation. By adopting a correctly time-averaged density, we will demonstrate a marked improvement in the performance of the decomposition diagnostic.

First, we will discuss the contribution of the Coriolis acceleration and bottom drag to the differences between decomposition-based $\tilde{\Psi}^c$ and expected overturning streamfunction $\Psi^c$.

\subsection{The effect of the Coriolis acceleration calculation}

Analysis of the uncompensated overturning streamfunctions shows a depth-dependent difference between the estimated $\tilde{\Psi}^c$ and expected $\Psi^c$ streamfunctions for numerous latitudes (not shown), which increases as the surface is approached. This suggests the possibility that small errors near the bottom of the fluid column are propagated upwards due to vertical integration in overturning streamfunction calculations.

Possible sources of error could be the simplification of the Coriolis acceleration ($fv$) calculation near bathymetry, or lack of consideration of bottom friction and drag within the diagnostic framework. An investigation, reported in Appendix \ref{App_Coriolis}, suggests that the greater number of calculation locations within the Energy and Enstrophy (EEN) scheme used to calculate the Coriolis acceleration ($fv$) by NEMO improves the resulting overturning streamfunction estimates, reducing errors caused by bathymetry in places where the simplified 4-point velocity method used in the decomposition performs poorly. Improvement in estimated overturning streamfunction is found especially in regions of rough bathymetry; these improvements are, however, small in comparison to the differences $\Psi^c - \tilde{\Psi}^c$ in time-mean overturning streamfunction.

\subsection{Direct calculation of the overturning streamfunction using momentum trend terms}

We now examine differences between momentum trend terms output directly by the NEMO model and equivalent trends calculated using information (e.g. $\theta$, $S$) exploited in the construction of the decomposition diagnostics. Each momentum term indicates the contribution made by a different physical process to the $u$-component of fluid momentum (Equation \ref{GB}). 

Annual-mean momentum trend terms will be considered for a single year, output after a $30$-year spin-up (for $1.4^\circ$ model run u-bx$950$). For fair comparison, the compensated overturning streamfunction using the decomposition diagnostics is also recalculated using data from the same run. In addition to further understanding the contributions made by various momentum trend terms and explaining the lack of closure of the decomposition diagnostics, we hope to quantify the optimal theoretical performance of our decomposition diagnostics using only those momentum terms which relate to processes accounted for within the decomposition. 

\begin{figure*}[ht!]
	\centerline{\includegraphics[width=\textwidth]{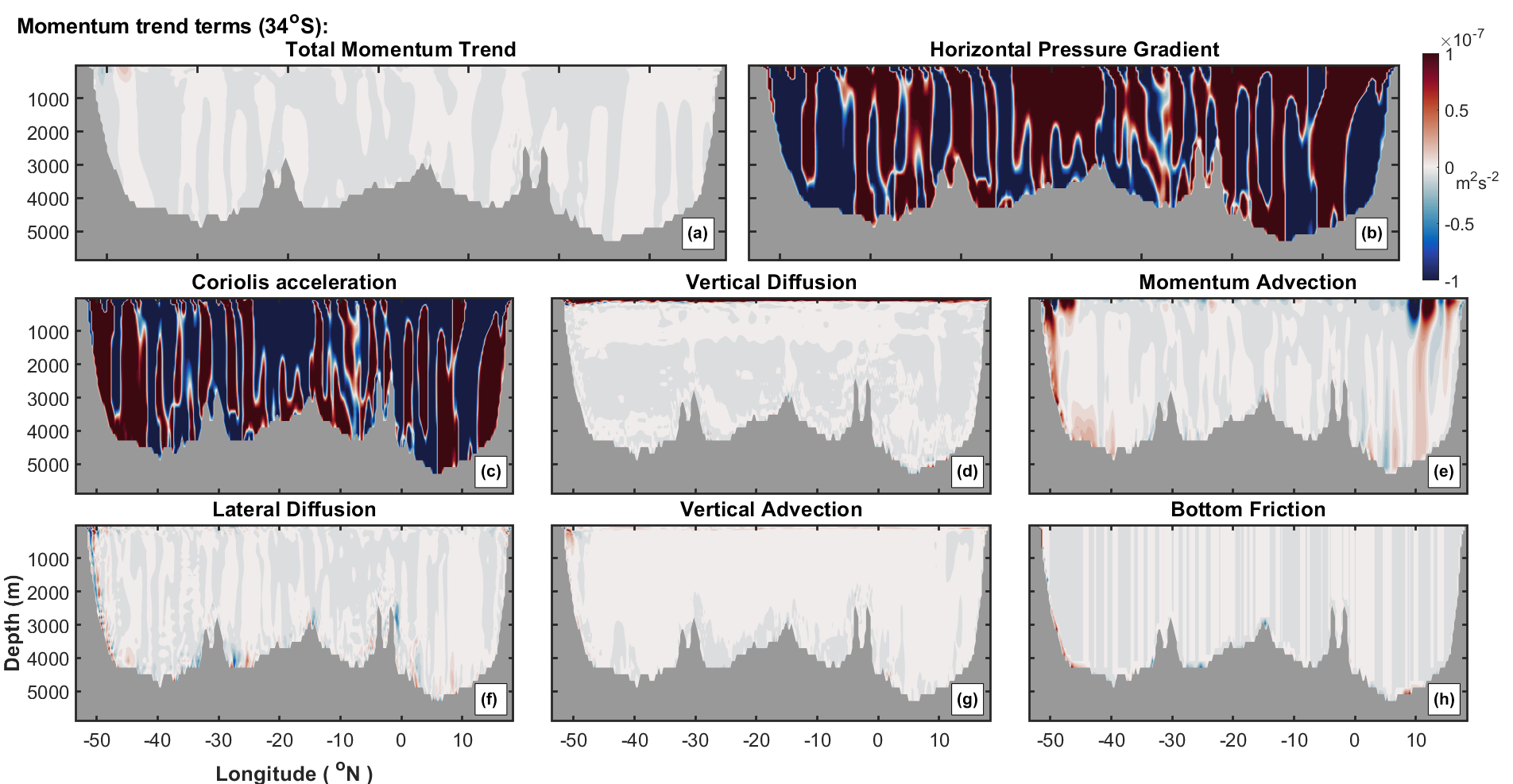}}
	\caption[One year average contribution made by each of the momentum trend terms for a single latitude ($34^\circ$S).]{One year average contribution made by each of the momentum trend terms for a single latitude ($34^\circ$S). (a) total momentum trend, (b) horizontal pressure gradient trend, (c) Coriolis acceleration or $fv$, (d) vertical diffusion, (e) momentum advection, or sum of relative vorticity and kinetic energy gradient, (f) lateral diffusion, (g) vertical advection and (h) bottom friction. Colour scale chosen in order to see features in momentum trends at the expense of saturation found in Panels (b) and (c).}
	\label{MomDiag_all}
\end{figure*}

Momentum trend terms output by the NEMO model (shown in Figure \ref{MomDiag_all}) are the: horizontal pressure gradient, planetary vorticity (Coriolis acceleration), relative vorticity (vortex force), vertical diffusion, lateral diffusion, vertical advection, bottom friction, gradient of the kinetic energy and total momentum trend. The sum of all momentum trend contributions is equal to the ``total momentum'' trend (Panel (a)), which at longer timescales tends to zero (i.e. ${\partial u}/{\partial t} \rightarrow 0 $). Visible in Figure \ref{MomDiag_all} are the dominant magnitudes of the horizontal pressure gradient (b) and the Coriolis acceleration (c) terms.

The decomposition diagnostic, in its thermal wind term $\Psi_W^c + \Psi_E^c$, takes into account the role of the depth-dependent part of the pressure gradient force, encapsulated by the horizontal pressure gradient momentum term. In addition, the depth-independent flow term $\Psi_{bot}^c$ can be viewed as incorporating the depth-independent component of the pressure gradient force (when bottom friction is negligible and geostrophy accurate); this is equivalent to the Coriolis acceleration term $fv$. The Ekman term $\Psi_{Ek}^c$ is equivalent to the vertical diffusion momentum trend term. The optimal theoretical capability of the decomposition diagnostic is therefore equivalent to the overturning streamfunction calculated using the sum of these three momentum trend terms only. Additional, more localised, processes such as momentum advection (i.e. the sum of the gradient of kinetic energy, and relative vorticity terms), lateral diffusion and bottom friction are neglected within the decomposition diagnostics; to describe the overturning streamfunction fully, these additional contributions would need to be included.

\begin{figure*}[ht!]
	\centerline{\includegraphics[width=\textwidth]{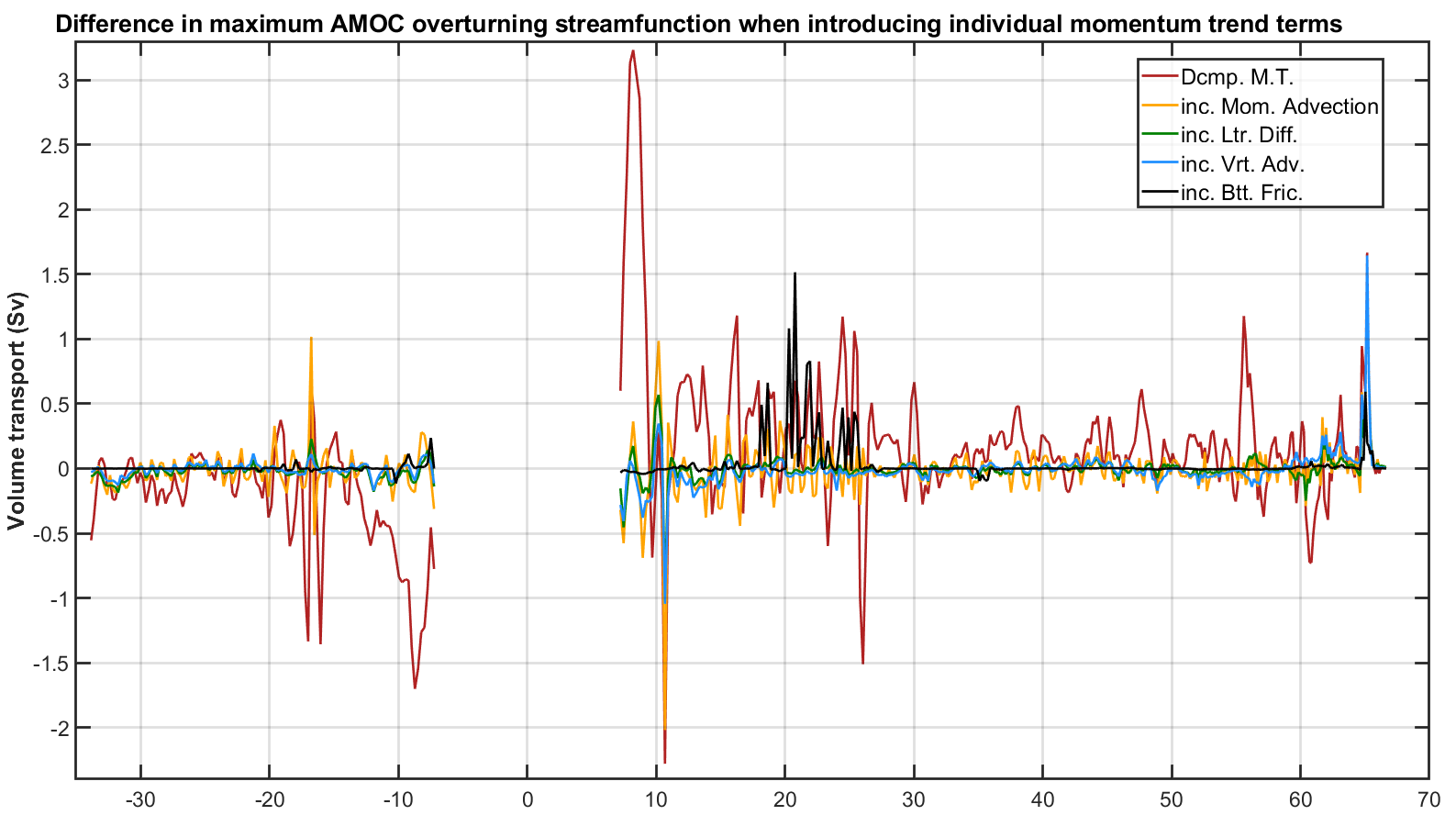}}
	\caption[Investigating role of momentum trend terms, not contributing to the decomposition diagnostic, in explaining the total overturning streamfunction calculated using all momentum trend terms.]{ Investigation into the role of momentum trend terms, not contributing to the decomposition diagnostic, in explaining the total overturning streamfunction calculated using all momentum trends, for a yearly average. The figure shows differences in maximum overturning streamfunction as a function of latitude, between the expected overturning streamfunction and streamfunctions estimated using (a) only momentum trends used in boundary decomposition (Dcmp M.T., red), (b) momentum trends used in boundary decomposition and momentum advection (orange), (c) momentum trends used in boundary decomposition, momentum advection and lateral diffusion (green), (d) momentum trends used in boundary decomposition, momentum advection, lateral diffusion and vertical advection (blue) and (e) all momentum trends (black).}
	\label{MD_RMS_Trsp}
\end{figure*} 

Figure \ref{MD_RMS_Trsp} emphasises the role of each of the additional momentum trend contributions not taken into account within the decomposition diagnostic. Improvement at certain latitudes would be found if additional information such as the momentum advection contribution were to be included within the decomposition diagnostic. However the inclusion of additional terms would require a radically different approach to decomposition, detracting from the relative simplicity of the current diagnostics. Further, the added complexity of an amended decomposition would require information from the ocean interior, inconsistent with the aim of understanding and quantifying the overturning streamfunction using boundary information only. We note that the total momentum trend line (black) should display smallest discrepancy values in the figure, but surprisingly we find large values between $20^\circ$N and $30^\circ$N. We speculate that this feature is due to the large number of additional cells at these latitudes due to numerous islands and ridges, exacerbating inaccuracies in representing bottom friction effects within this model run.   

\begin{figure*}[ht!]
	\centerline{\includegraphics[width=\textwidth]{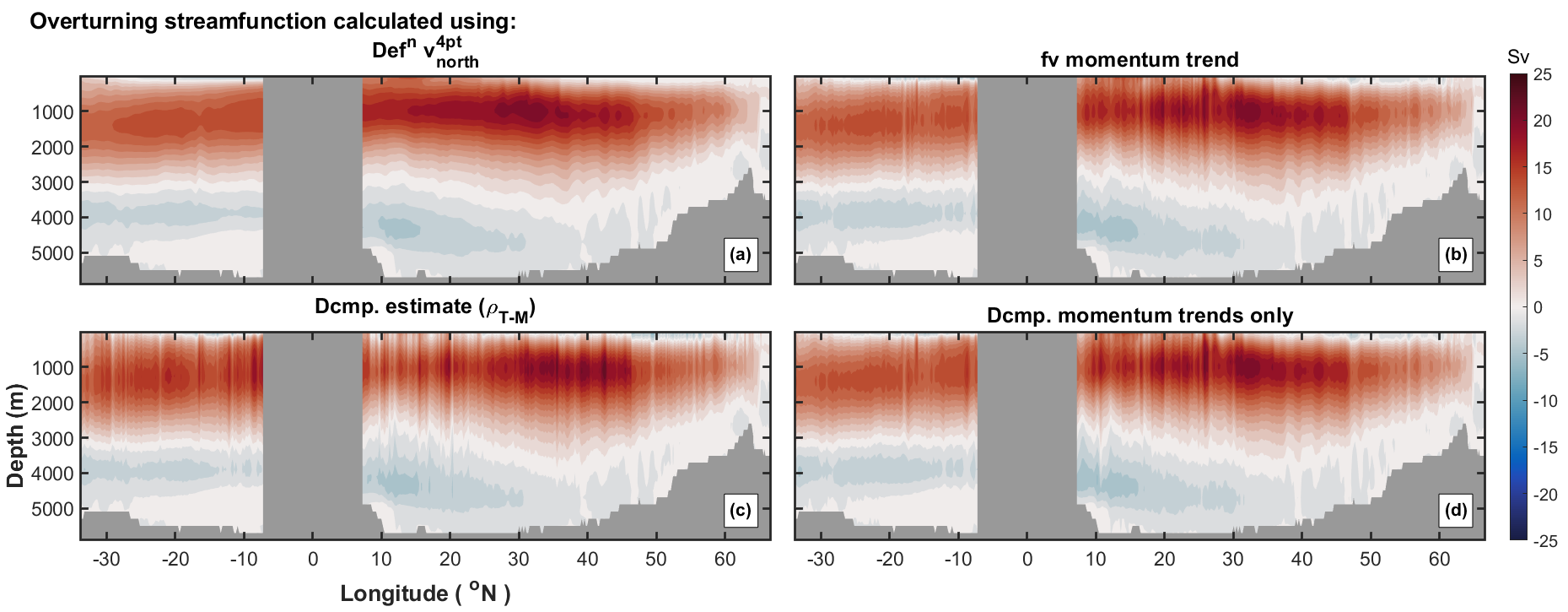}}
	\caption[Comparison of overturning streamfunctions calculated using definition, decomposition diagnostic and momentum trend terms.]{Comparison of overturning streamfunctions: (a) using meridional velocities ($v^{4pt}_{north}, $Equation \ref{T}) (b) using Coriolis acceleration $fv$ momentum trend (and Equation \ref{T}) (c) estimate from the decomposition diagnostic ($\tilde{\Psi}^c$) and (d) using only momentum trend terms accounted for in the decomposition diagnostic.}
	\label{MD_AMOC_Trsp}
\end{figure*}

\begin{figure*}[ht!]
	\centerline{\includegraphics[width=\textwidth]{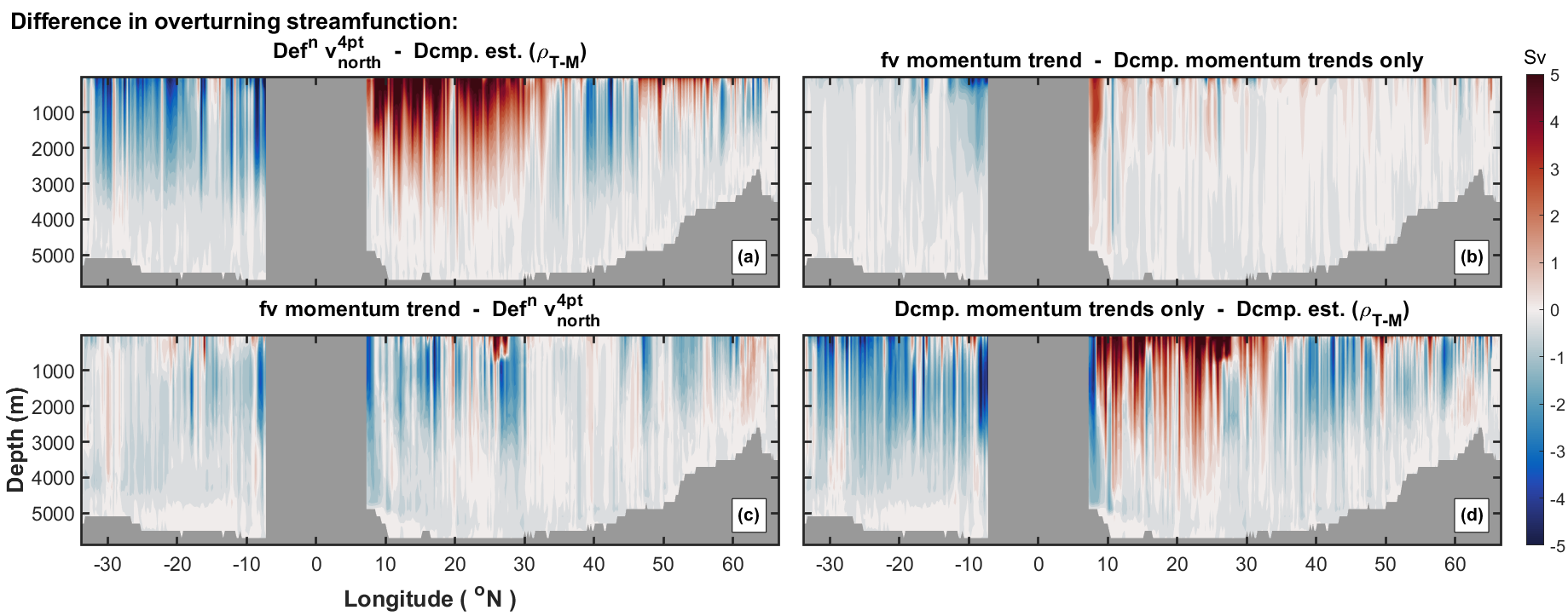}}
	\caption[Comparison of the differences between overturning streamfunctions shown in Figure \ref{MD_AMOC_Trsp}.]{Comparison of the differences between overturning streamfunctions shown in Figure \ref{MD_AMOC_Trsp}: (a) using meridional velocities - estimate from the decomposition diagnostic ($\tilde{\Psi}^c$), (b)  using Coriolis acceleration $fv$ momentum trend - using momentum trends contributing to the decomposition diagnostic only, (c) using momentum trends contributing to the decomposition diagnostic only -  estimate from the decomposition diagnostic ($\tilde{\Psi}^c$) and (d) using Coriolis acceleration $fv$ momentum trend - using meridional velocities.}
	\label{MD_Trsp_Diff}
\end{figure*}

In Figure \ref{MD_AMOC_Trsp}, we compare different estimates for the total compensated overturning streamfunction: (a) $\Psi^c$, using meridional velocities from the model (Equation \ref{T}), (b) the equivalent overturning streamfunction calculated using the $fv$ momentum trend term, by dividing by $f$ and using Equation \ref{T} (c) $\tilde{\Psi}^c$, using the decomposition diagnostic, and (d) the overturning streamfunction calculated using only momentum trend terms accounted for in the decomposition diagnostic. Figure \ref{MD_Trsp_Diff} shows the differences between the estimates of total overturning streamfunction in Figure \ref{MD_AMOC_Trsp}. Comparing Panels (a), (b), (c) and (d) of Figure \ref{MD_Trsp_Diff} reveals that the decomposition diagnostic (Figure \ref{MD_AMOC_Trsp} (c)) performs less well than would be expected from consideration of its constituent momentum trend terms. This is particularly evident in Figure \ref{MD_Trsp_Diff}(c), where large differences are found between the decomposition estimate ($\tilde{\Psi}^c$) and the equivalent calculated using only momentum trends accounted for in the decomposition diagnostic. Figure \ref{MD_AMOC_Trsp}(a,b) and Figure \ref{MD_Trsp_Diff}(c) show the overturning streamfunction (Equation \ref{T}) calculated using meridional velocities and the Coriolis acceleration $fv$ momentum trend are similar.

Detailed analysis of the main decomposition contributions ($\Psi_W+\Psi_E$, $\Psi_{bot}$ and $\Psi_{Ek}$), compared with the equivalent contributions calculated using momentum trends, localises the source of error to the density term, and the thermal wind component $\Psi_W+\Psi_E$ (not shown). Specifically, for a simple ocean box away from any bathymetry, there is a clear difference between the $\Psi_W+\Psi_E$ term calculated from density and the equivalent calculated using momentum trend terms; the latter is calculated using the horizontal pressure gradient momentum trend with bottom pressure removed from each fluid column, and then integrated vertically and zonally.  


Similar comparisons of the vertical diffusion trend with the Ekman component $\Psi_{Ek}$, and of the equivalent $fv$ terms using the (a) EEN scheme, (b) $4$-point (4pt) averaging method $\Psi_{bot}$ and (c) the Coriolis acceleration $fv$ momentum trend, show little to no differences. This suggests strongly that the source of transport discrepancy lies in the boundary density terms $\Psi_W$ and $\Psi_E$.

\subsection{Influence of time-averaging input fields upon the horizontal pressure gradient momentum trend}

\subsubsection*{Motivating investigations}

One possible reason for the differences found when comparing the main decomposition components and their momentum trend equivalents could be the assumption made within the decomposition of a rigid lid, and hence the absence of a free surface. Using the rigid lid approximation, T-cell depths are constant with latitude and longitude since surface waves are neglected. A free surface and hence s- rather than z-coordinate system, was therefore introduced for this investigation, to improve the correspondence between the diagnostic model framework and NEMO. An s-coordinate (or $\sigma$-coordinate) system (\citealt{SONG1994228}, \citealt{Song1998}) can be though of as a terrain-following coordinate system. Its definition exploits the ratio of the pressure at any point in the ocean to the pressure at the bottom of the ocean. 

Further, the horizontal pressure gradient (hpg) trend calculated using input fields used for the decomposition diagnostic was compared directly with the horizontal pressure gradient trend output for a single year from the u-bx$950$ run. This reduces the number of integrations (avoids vertical integration) and the possibility for error propagation throughout the calculation. Initial attempts at recalculating the horizontal pressure gradient trend using annually averaged input fields gave poor agreement throughout the basin and worsened with depth. The significant depth-dependence of the difference suggested possible issues with accuracy of the depths used for the centre of T-cells, and W-levels (water depth at top of each T-cell) within the calculation. 

In an attempt to obtain full agreement for the horizontal pressure gradient trend, the explicit NEMO method for calculating the in-situ density anomaly $\rho_d$,  
\begin{equation}
	\rho_d = \frac{\rho - \rho_0}{\rho_0},
	\label{rhd}
\end{equation}
was also introduced. Here $\rho_0$ is the nominal density of seawater ($= 1026 \text{kgm}^{-3}$) and $\rho$ is the in-situ density. However, agreement between outputted momentum trend fields from NEMO and the equivalent calculated using annually time-average input fields (e.g. potential temperature, salinity and cell depths)  was again poor. The same calculation based on monthly data also gave poor agreement, possibly attributed to internal gravity waves. Agreement is only found when instantaneous input fields are used. 

\subsubsection*{The role of time-averaging}
\label{TM_TA}
\begin{figure*}[ht!]
	\centerline{\includegraphics[width=\textwidth]{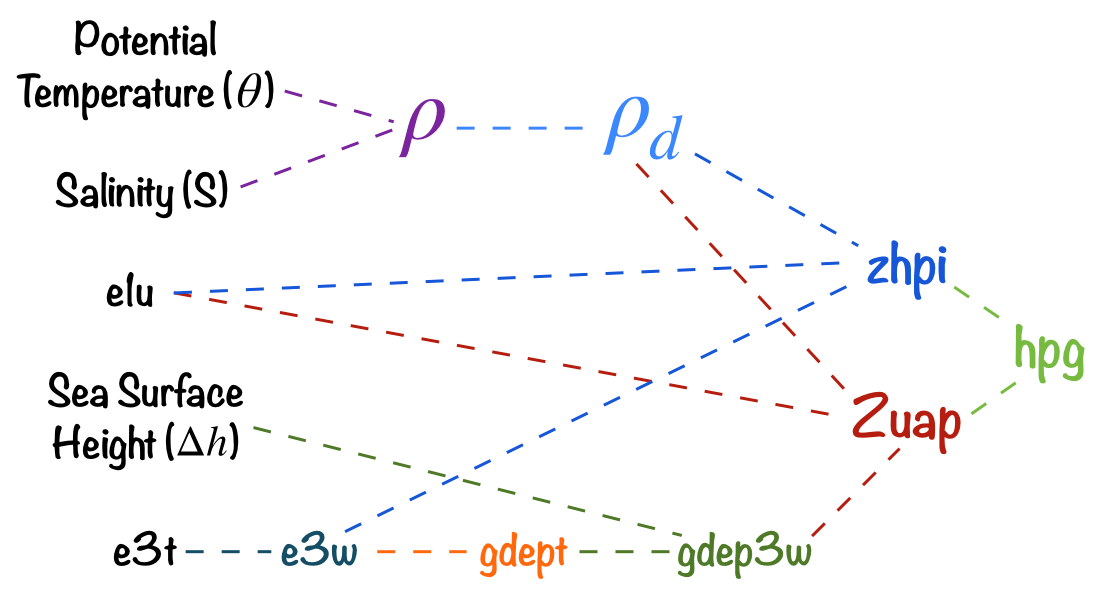}}
	\caption[Schematic of terms used to calculate horizontal pressure gradient ($hpg$) momentum trend term.]{Schematic of the terms used to calculate the horizontal pressure gradient ($hpg$) momentum trend. Percentage difference between the two ways of calculating each term (Equation \ref{e_PrcDff}) is calculated to investigate the role of time-averaging. Terms include the potential temperature ($\theta$), salinity ($S$), width of u-cells ($e1u$), sea surface height ($\Delta h$), depth of T-cells ($e3t$), in-situ density ($\rho$),  depth of W-levels ($e3w$), depth of T-cells with respect to geoid ($gdept$), in-situ density anomaly ($\rho_d$),   depth of W-levels with respect to geoid ($gdept3w$), hydrostatic pressure gradient along s-surfaces ($zhpi$) and s-coordinate pressure gradient correction ($zuap$).}
	\label{hpg_terms}
\end{figure*}

In the NEMO model, the horizontal pressure gradient trend is calculated from instantaneous $\theta$, $S$ and other fields (shown in Figure \ref{hpg_terms}). In the overturning streamfunction decomposition diagnostic, we use a time-average fields for $\theta$, $S$ etc. 

The influence of time-averaging horizontal pressure gradient trends is analysed using monthly averaged data for June $1996$ and for December $1996$, from the $1/4^\circ$ u-bj$748$ dataset, illustrated here for location $55.21^\circ$S, $38.75^\circ$W. The horizontal pressure gradient trend is first calculated for each of the two months separately, as a function of four main input parameters $\theta_i, S_i, SSH_i, e3t_i$, namely potential temperature, salinity, sea surface height and the depth of T-cell fields for the $i^{th}$ month ($i=1,2$). This yields $hpg(\theta_1, S_1, SSH_1, e3t_1)$ and $hpg(\theta_2, S_2, SSH_2, e3t_2)$ for the individual months. Then the time-average horizontal pressure gradient trend over the two months is calculated using

\begin{equation}
\overline{hpg}_{AA} = \frac{hpg(\theta_1, S_1, SSH_1, e3t_1) + hpg(\theta_2, S_2, SSH_2, e3t_2)}{2}.
\label{TM_c}
\end{equation}

The influence of time-averaging is quantified by comparing $\overline{hpg}_{AA}$ with the equivalent horizontal pressure gradient trend $\overline{hpg}_{AB}$ calculated using time-average input fields (averaged over the two months)
\begin{equation}
\overline{hpg}_{AB}= hpg(\overline{\theta}, \overline{S}, \overline{SSH}, \overline{e3t}),
\label{TM_ic}
\end{equation}
where, for example, $\overline{\theta}$ is the average potential temperature over June and December 1996. In simple terms, there are two methods for calculating the time-mean horizontal pressure gradient trend. Both are dependent on time-averaging, but the point at which the average is taken is crucial to obtaining correct results for time-average horizontal pressure gradient. The accurate method is to calculate the horizontal pressure gradient trend initially for each month independently and then time-average (Equation \ref{TM_c}). We denote this estimate as $hpg_{AA}$, with subscript referring to ``averaging after''. The second method estimates $hpg_{AB}$ using time-average inputs, with subscript referring to ``averaging before''; this is the method used in the decomposition diagnostic.

We find that $hpg_{AB} \neq hpg_{AA}$. According to Jensen's inequality (\citealt{Jensen1906}), equality between $hpg_{AB}$ and  $hpg_{AA}$ occurs in general when the function $hpg$ is linear. The observed inequality suggests that the horizontal pressure gradient trend may be non-linear with respect to at least one of its input fields.

One method to quantify the influence of time-averaging of input fields is to calculate the percentage difference between estimates of the horizontal pressure gradient trend computed using the two methods:
\begin{equation}
\text{Percentage Difference} = \frac{\overline{hpg}_{AA} - \overline{hpg}_{AB}}{0.5 \times (\overline{hpg}_{AA} + \overline{hpg}_{AB})} \times 100\% .
\label{e_PrcDff}
\end{equation}

This metric is used to assess the non-linearity of all input arguments to the horizontal pressure gradient trend calculation, described graphically in more detail in Figure \ref{hpg_terms}. From the figure, we see that the horizontal pressure gradient trend is calculated as the sum of two terms $zuap$ and $zhpi$, explained further in the figure caption. These terms are themselves functions of other variables including $\rho_d$ as described in the figure. Ultimately, we see that the horizontal pressure gradient trend is a function of all the inputs on the left hand side of the figure. 

\begin{figure*}[ht!]
	\centerline{\includegraphics[width=\textwidth]{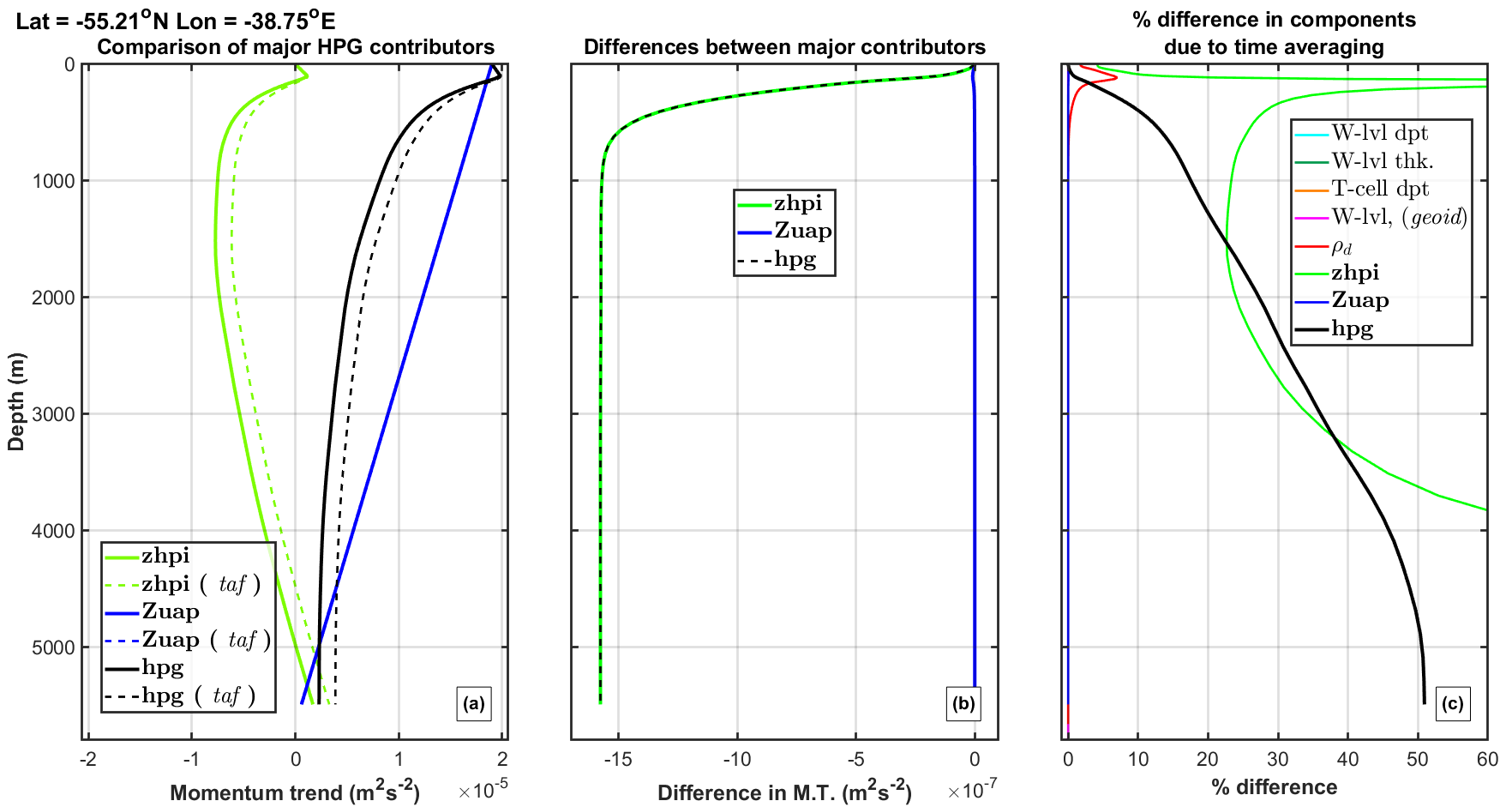}}
	\caption[Investigating influence of time-averaging on main components of horizontal pressure gradient momentum trend term calculation.]{Exploring the role of time-averaging. (a) main components of horizontal pressure gradient momentum trend calculation, horizontal pressure gradient momentum trend ($hpg$, black), hydrostatic pressure gradient along s-surfaces ($zhpi$, green) and s-coordinate pressure gradient correction ($zuap$, blue). Solid lines represent averages of each individual monthly contribution. Dashed lines represent contributions calculated using time-average input fields. (b) difference in the main components of horizontal pressure gradient momentum trend calculation (shown in (a)). (c) percentage difference between time-averaging approaches for different input field variables of horizontal pressure gradient momentum trend.}
	\label{Mhpg}
\end{figure*}

Figure \ref{Mhpg}(a) explores $h_{AA}$ and $h_{AB}$ for $h=hpg, zuap$ and $zhpi$, as a function of water depth. Panel (b) of Figure \ref{Mhpg} then shows the difference $h_{AA} - h_{AB}$ with depth. Finally, Panel (c) gives the percentage difference (Equation \ref{e_PrcDff}) with depth. The figure shows clearly that the source of the difference found in the horizontal pressure gradient trend can be attributed to the $zhpi$ term, which itself is a function of $\rho_d$, shown in red in Panel (c) of Figure \ref{Mhpg}.
Figures \ref{hpg_terms} and \ref{Mhpg} indicate a key time-averaging issue stemming from the calculation of the in-situ density anomaly. The method of time-averaging of other input terms, including sea surface height, depth of T-cells etc. has a negligible effect. It is interesting to note, however, that the difference and percentage difference between the two calculations for the s-coordinate pressure gradient correction term $zuap$ is negligible, even though it is nominally $\rho_d$-dependent. 

Given the importance of in-situ density $\rho_d$ to estimation of the horizontal pressure gradient trend and hence to our overturning decomposition, we proceed to investigate the main time-dependent input components contributing to its calculation. Referring to Figure \ref{hpg_terms}, these are the potential temperature $\theta$ and salinity $S$. The contributions of these terms are now individually analysed by averaging them (e.g. $\overline{\theta}$, Equation \ref{e_ATh}) whilst allowing each other input field in the horizontal pressure gradient trend calculation to vary with time, then calculating the time-average horizontal pressure gradient trend between both months using 
\begin{equation}
\overline{hpg}_{\theta AB}=  \frac{hpg(\overline{\theta}, S_1, SSH_1, e3t_1) + hpg(\overline{\theta}, S_2, SSH_2, e3t_2)}{2}
\label{e_ATh}
\end{equation}
for potential temperature, and
\begin{equation}
\overline{hpg}_{SAB} =  \frac{hpg(\theta_1, \overline{S}, SSH_1, e3t_1) + hpg(\theta_2, \overline{S}, SSH_2, e3t_2)}{2}
\label{e_AS}
\end{equation}
for salinity.
\begin{figure*}[ht!]
	\includegraphics[width=\textwidth]{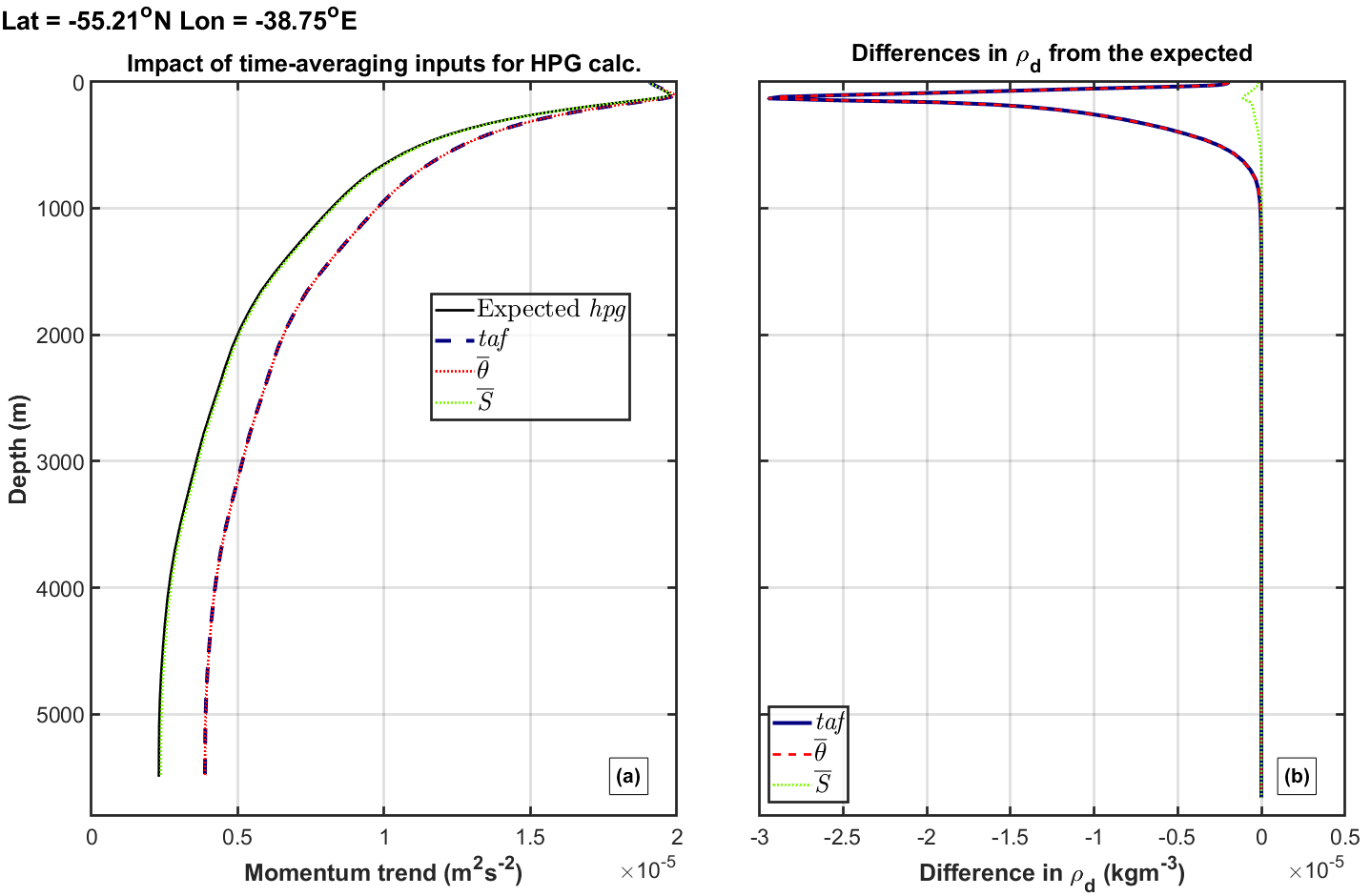}
	\caption[Investigating role of time-average potential temperature and salinity on horizontal pressure gradient trend term.]{Left: horizontal pressure gradient momentum trend for $\overline{hpg}_{AA}$ (black, Equation \ref{TM_c}), $\overline{hpg}_{AB}$ (dashed navy, Equation \ref{TM_ic}, denoted by time-average after or $taf$), $\overline{hpg}_{\theta AB}$ (dotted red, Equation \ref{e_ATh}) and $\overline{hpg}_{SAB}$ (dotted green, Equation \ref{e_AS}). $\overline{hpg}_{SAB}$ overlies $\overline{hpg}_{AA}$, $\overline{hpg}_{\theta AB}$ overlies $\overline{hpg}_{AB}$. Right: differences in in-situ density $\rho_d$ from the expected $\overline{\rho_d}_{AA}$ for each experiment. Navy denotes $\overline{\rho_d}_{AA}-\overline{\rho_d}_{AB}$, dotted red $\overline{\rho_d}_{AA}-\overline{\rho_d}_{\theta AB}$ and dotted green $\overline{\rho_d}_{AA}-\overline{\rho_d}_{SAB}$.}
	\label{TS_hpg}
\end{figure*}

Figure \ref{TS_hpg} indicates that the dependence of $\rho_d$ on potential temperature is problematic for horizontal pressure gradient and overturning calculations using time-average input. This is shown by the almost identical behaviour of the $\overline{hpg}_{AB}$ (dashed navy) and $\overline{hpg}_{\theta AB}$ (dotted red) curves in Panel (a) of Figure \ref{TS_hpg}, and the corresponding curves in Panel (b). The influence of time-average salinity, SSH and depths of T-cells on the horizontal pressure gradient trend, whilst not zero, is negligible in comparison. 

\begin{figure*}[ht!]
	\centerline{\includegraphics[width=\textwidth]{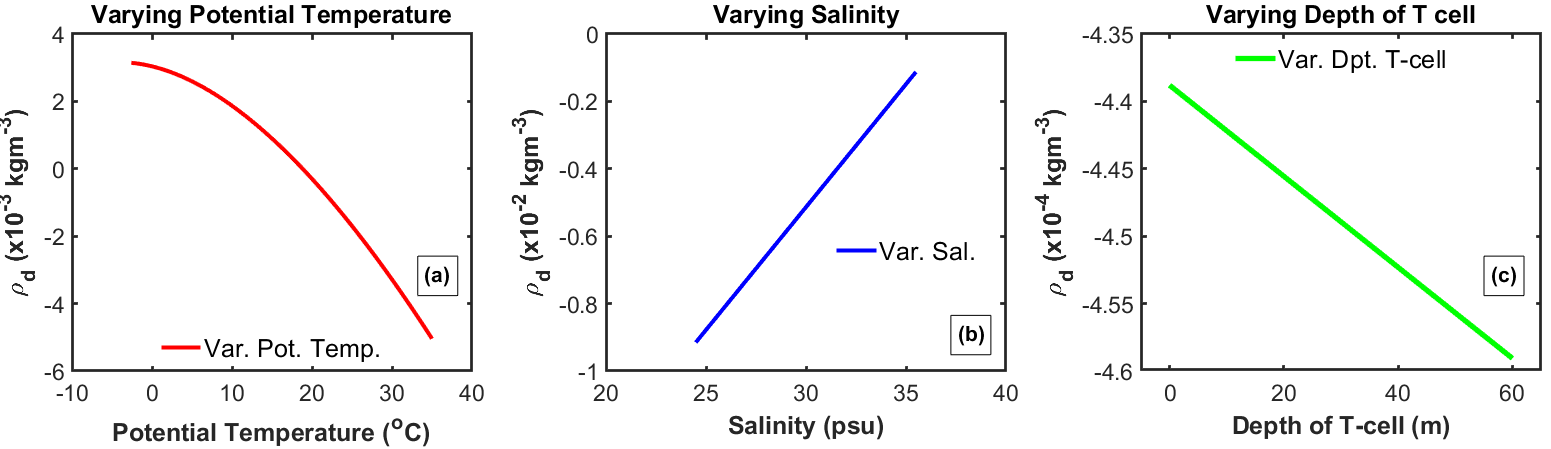}}
	\caption[Behaviour of in-situ density with varying potential temperature, salinity and depth of T-cell.]{The behaviour of in-situ density with varying potential temperature, salinity and depth of T-cell.}
	\label{Lin_TS}
\end{figure*}

In fact it can be shown  (Figure \ref{Lin_TS}, Panel (a)) that for fixed values of salinity $S$ and T-cell depth $e3t$, in-situ density anomaly $\rho_d$ is a non-linear function of potential temperature $\theta$. In comparison, the variation of $\rho_d$ with $S$ (for fixed $\theta$ and $e3t$), and of $\rho_d$ with $e3t$ (for fixed $\theta$ and $S$) is effectively linear. However, at low (polar) temperatures the influence of salinity on setting the density structure is greater than that of temperature. Hence, here it may be the case that incorrectly time-averaged salinities may also play a larger role. More generally, we might expect regional variations in the relative importance of salinity and potential temperature to the discrepancy in density calculation when using time-averaged inputs. 

A further experiment was conducted to examine the usefulness of a time-average $\rho$ given by 
\begin{equation}
\overline{\rho}_{AA} = \frac{\rho(\theta_1, S_1) + \rho(\theta_2, S_2)}{2},
\end{equation}
calculated as the average of $\rho$ for different instantaneous input fields, in the calculation of the horizontal pressure gradient momentum trend. It is found that a properly time-average density field $\overline{\rho}_{AA}$ provides a much improved horizontal pressure gradient momentum trend calculation, as discussed further in Section \ref{S3_Ndens}. The agreement between the horizontal pressure gradient trend calculated using instantaneous fields and that obtained using a correctly time-average density field shows promise as a route to a more accurate MOC overturning streamfunction decomposition into boundary components.

In summary, in the standard overturning streamfunction decomposition diagnostic we estimate the in-situ density $\rho$ using time-average quantities involving $\overline{\theta}$ and $\overline{S}$, rather than accumulating the time-average value of $\rho$ from instantaneous density fields. This approach has been shown to be inaccurate, since $\rho$ is a non-linear function of $\theta$. In order to account correctly for the contribution of $\rho$ to volume transport, a time-average estimate of $\rho$ based on direct accumulation of instantaneous $\rho$ fields is required.

\subsection{Overturning streamfunction estimated using correct NEMO time-mean density}
\label{S3_Ndens}

Now we explore the usefulness of an improved decomposition diagnostic incorporating a correctly time-average density field $\overline{\rho}_{NEMO}$ output directly from NEMO, but unfortunately currently unavailable as a standard output. For a trial case, a single year of correctly time-average instantaneous density fields is output from run u-bx950. This time-average density field is used within the overturning streamfunction decomposition diagnostics in place of, and to compare with, the original approach using time-average $\theta$ and $S$ to calculate in-situ density; in the discussion below, we adopt the shorthand ``T-M'' (time-mean) for brevity, particularly in tables and figures. As an initial step, we compare the performance of $\overline{\rho}_{NEMO}$ with other methods for calculating the horizontal pressure gradient momentum trend.

\begin{table}[h!]
\centering
\begin{tabular}{ |p{3.75cm}||p{3.3cm}|p{3.3cm}|p{3.3cm}| }
	\hline
	\multicolumn{4}{|c|}{Quantifying improvement using $\overline{\rho}_{NEMO}$ as part of hpg momentum trend calculation } \\
	\hline
	& (a) NEMO output in-situ density $hpg_1(\overline{\rho}_{NEMO})$ & (b)T-M fields and NEMO code $hpg_2(\overline{\theta}, \overline{S}, ...)$ & (c)T-M fields with eos80 tools $hpg_3(\overline{\theta}, \overline{S}, ...)$ \\
	\hline
	Mean squared error $(\text{m}^2\text{s}^{\text{-2}})^2$  & $3.142 \times 10^{-31}$    &$1.700 \times10^{-29}$& $1.804\times10^{-29}$ \\
	Mean absolute difference $(\text{m}^2\text{s}^{\text{-2}})$&   $2.040\times10^{-16}$  & $7.253\times10^{-16}$   &$7.687\times10^{-16}$ \\
	\hline
\end{tabular}
\caption[Performance of different methods for calculating the horizontal pressure gradient trend (using the horizontal pressure gradient trend output from NEMO as truth) globally for all latitudes, longitudes and depths.]{Performance of different methods for calculating horizontal pressure gradient momentum trend, using horizontal pressure gradient momentum trend from NEMO as truth, globally for all latitudes, longitudes and depths. Methods evaluated are (a) $hpg_1(\overline{\rho}_{NEMO})$, using correctly time-average density from NEMO, (b) $hpg_2(\overline{\theta}, \overline{S}, ...)$. using time-mean input fields alongside explicit NEMO formulation and (c) $hpg_3(\overline{\theta}, \overline{S}, ...)$ using time-mean input fields alongside $eos80$ to calculate in-situ density. All cells are volume weighted before calculating the resulting mean squared error and the mean absolute difference for all depths, longitudes and latitudes.}
\label{Tab_TMhpg}
\end{table}

Table \ref{Tab_TMhpg} quantifies the performance of the three methods: (a) using NEMO output time-average in-situ density $\overline{\rho}_{NEMO}$, (b) using time-average $\theta$ and $S$ alongside replicated NEMO code to calculate horizontal pressure gradient trend, and (c) using the $eos80$ toolbox and time-average $\theta$ and $S$ to estimate the in-situ density and horizontal pressure gradient trend. 

Performance in Table \ref{Tab_TMhpg} is quantified in terms of volume weighted mean squared error and the mean absolute difference between each of the three methods and the true horizontal pressure gradient trend for the whole ocean. This assessment therefore takes account of all ocean cells, providing a global view of performance. The improvement found when using a correctly time-averaged in-situ density (a) rather than time-average input fields (b,c) to estimate the in-situ density is clear. In terms of mean absolute difference, use of correctly time-averaged NEMO density output (a) is around $3.5$ times better than use of time-mean input fields (b,c). In terms of mean squared error, use of correctly time-averaged NEMO density output is around 60 times better than use of time-mean input fields. The difference between methods (b, NEMO) and (c, eos80), which both use time-average input fields to estimate the in-situ density is somewhat surprising, given they should be the same; however the difference is small.

\begin{figure*}[ht!]
	\centerline{\includegraphics[width=\textwidth]{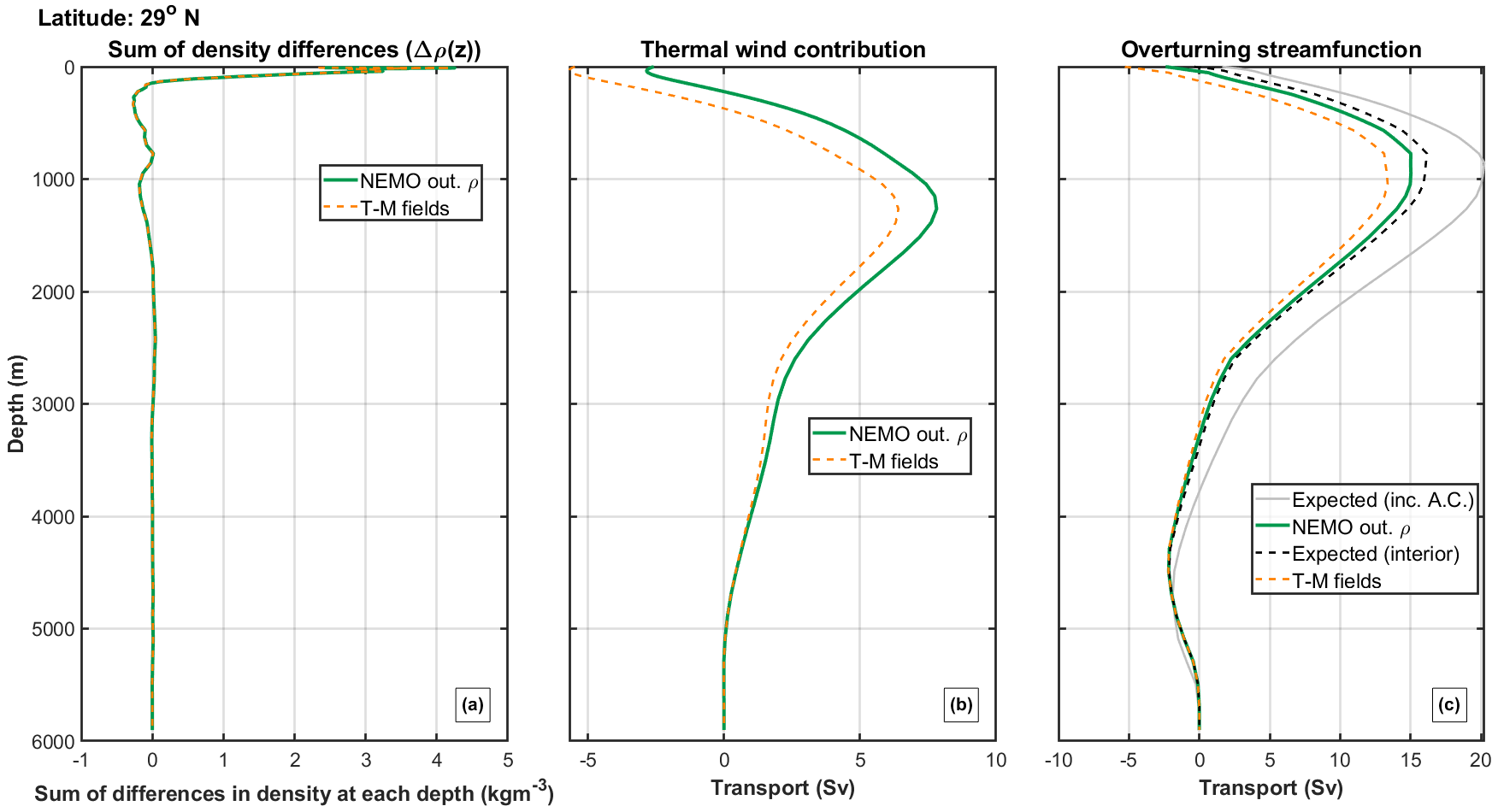}}
	\caption[Assessment of impact of using density calculated from time-average fields using $eos80$ versus time-average density output directly by NEMO calculated using instantaneous fields at $29^\circ$N.] {Assessment of impact of using density calculated from time-average fields using $eos80$ (orange) versus time-average density output directly by NEMO (green) calculated using instantaneous fields at $29^\circ$N, Left: sum of density differences for each depth between the western and eastern boundaries, as explained in the text. Centre: resulting volume-uncompensated thermal wind component of the overturning. Right: resulting volume-uncompensated interior overturning streamfunction (ignoring the additional cell contribution) compared with the expected interior overturning streamfunction calculated using the definition of the overturning streamfunction (Equation \ref{T}) with (light grey) and without (dashed black) additional cell contributions.}
	\label{Trsp_Dens}
\end{figure*}

In-depth analysis of the resulting overturning streamfunction calculated using the $eos80$ toolbox and $\overline{\rho}_{NEMO}$ highlights the improvement when a correctly time-average density is adopted. The improvement is found to be latitude dependent (and note that results for a single latitude are shown in Figure \ref{Trsp_Dens}), emphasising the role of localised physical mechanisms at different latitudes as discussed previously.

The zonal transect at any given depth may consist of multiple intervals of water, each with its own western and eastern boundary. Therefore, in Figure \ref{Trsp_Dens}, Panel (a) shows the sum of the corresponding density differences over all such intervals, for the complete ocean transect.

The difference between profiles for time-average input field calculations (orange) and profiles using time-average density fields directly output from NEMO (green) is small in Panel (a), but becomes noticeable in the upper $1000$m for both the geostrophic overturning streamfunction (i.e. thermal wind component, Panel (b)) and the interior overturning streamfunction (Figure \ref{Trsp_Dens}, Panel (c)).

The estimated interior overturning streamfunction in Panel (c) of Figure \ref{Trsp_Dens} includes contributions from depth-independent flow and Ekman components. We can therefore compare these interior overturning streamfunction estimates with the expected overturning streamfunction with (light grey) and without (dashed black) additional cell contributions. For the latitude $29^\circ$N considered, and in general, a clear improvement is found from incorporating time-average density into the decomposition diagnostic. For certain latitudes no improvement is found, highlighting the importance of additional contributions from small scale processes which we are unable to capture utilising only boundary information. 

The overall performance of the interior overturning streamfunction (using $\overline{\rho}_{NEMO}$) from the decomposition, in estimating the expected interior overturning streamfunction, is given in Table \ref{Tab_TrspCmpDns} over all latitudes and depths in the Atlantic basin. Relative to the original decomposition diagnostic (which uses time-mean input fields and $eos80$), the performance of the $\overline{\rho}_{NEMO}$-based estimate shows improvement.
\begin{table}[h!]
	\centering
	\begin{tabular}{ |p{5cm}||p{4.4cm}|p{4.4cm}| }
		\hline
		\multicolumn{3}{|c|}{Quantifying improvement using $\overline{\rho}_{NEMO}$ as part of interior streamfunction calculation } \\
		\hline
		& (a) NEMO output density $\tilde{\Psi}_1(\overline{\rho}_{NEMO})$ & (b) T-M fields with eos80 tools $\tilde{\Psi}_2(\overline{\theta}, \overline{S}, ...)$\\
		\hline
		Mean squared error $(\text{Sv}^2)$  & $4.46 \times 10^{-3}$    &$7.87 \times 10^{-3}$\\
		Mean absolute difference $(\text{Sv})$&   $1.35 \times 10^{-2}$  & $1.90 \times 10^{-2}$\\
		\hline
	\end{tabular}
	\caption[Performance of different methods for calculating interior overturning streamfunction, compared to the expected interior overturning streamfunction calculated using Equation \ref{T} and meridional velocities, over all latitudes and depths in the Atlantic basin.]{Performance of different methods for calculating interior overturning streamfunction, compared to the expected interior overturning streamfunction calculated using Equation \ref{T} and northward velocities, over all latitudes and depths in the Atlantic basin. Methods evaluated are (a) $\tilde{\Psi}_1(\overline{\rho}_{NEMO})$, using correctly time-average density from NEMO, and (b) $\tilde{\Psi}_2(\overline{\theta}, \overline{S}, ...)$ using time-average $\theta$ and $S$ fields alongside $eos80$ to calculate in-situ density. All cells are depth-weighted before calculating the resulting mean squared error and the mean absolute difference for all depths and latitudes.}
	\label{Tab_TrspCmpDns}
\end{table}

These results demonstrate that including correctly time-average densities in the calculation of the overturning streamfunction decomposition diagnostic improves its performance in reproducing time-averaged characteristics of the overturning streamfunction.  However, the unavailability of a direct time-average in-situ density $\overline{\rho}_{NEMO}$ output from the HadGEM3-GC3.1 model hinders further use of time-average densities. 

\section{Variation of time-mean overturning with latitude}
\label{s_var_tm_ovr_lat}
\begin{figure*}[ht!]
	\centerline{\includegraphics[width=\textwidth]{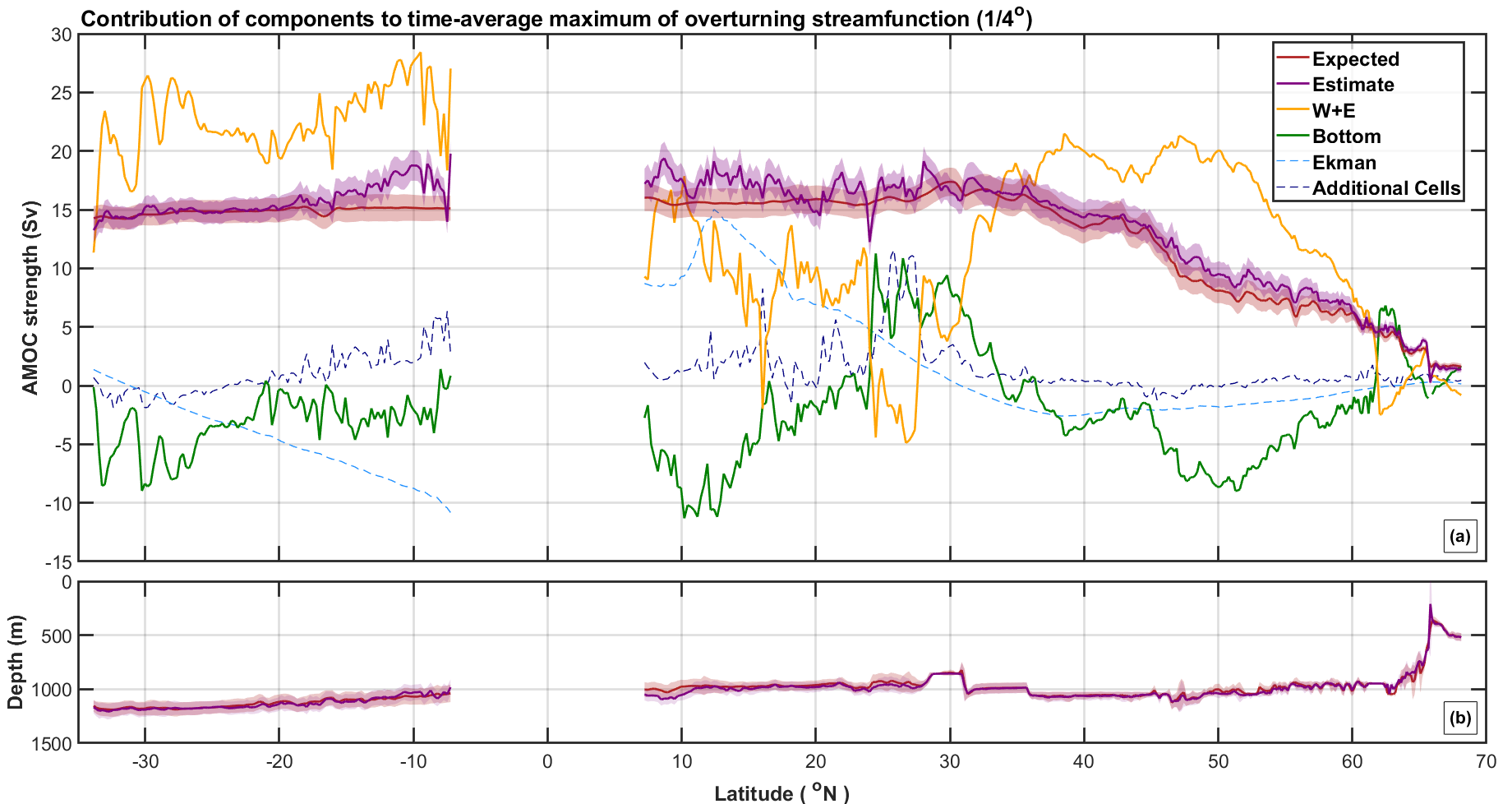}}
	\caption[Contribution of boundary components to time-mean maximum overturning streamfunction for $1/4^\circ$ model.]{Contributions to the time-average maximum overturning streamfunction for the $1/4^\circ$ model. Panel (a) shows time-average maximum overturning streamfunctions $\Psi_{\max}^c$ (red) and $\tilde{\Psi}_{\max}^c$ (purple) with latitude, for all years in model run. It also shows corresponding contributions of boundary components to $\tilde{\Psi}^c$, namely the sum of western and eastern boundary components (Thermal wind, orange), bottom component (green), Ekman component (dashed light blue) and additional cell contribution (dashed navy). Shaded regions indicate $\pm1$ temporal standard deviation of the maximum overturning streamfunction. Panel (b) shows the time-average depths $d_{\max}^c(y)$ and $\tilde{d}_{\max}^c$ of the maximum overturning streamfunction for $\Psi_{\max}^c$ (red) and $\tilde{\Psi}_{\max}^c$ (purple).}
	\label{p_Qtd_MxStrmLtt}
\end{figure*}

We now consider the full latitudinal extent of the AMOC, examining the maximum (with respect to $z$) of the overturning streamfunction, for each latitude and year. The maximum overturning streamfunction may be defined either in terms of the expected $\Psi^c$ or the estimated $\tilde{\Psi}^c$, e.g. Figure \ref{F_TM_strm}. For each latitude-year combination, we also find the corresponding depth of the maximum streamfunction. We then calculate time-average maxima, illustrated in Figure \ref{p_Qtd_MxStrmLtt}, for the $1/4^o$ model (outline in Section \ref{S3_ModDes}). That is, for example in the case of $\Psi^c$,
\begin{eqnarray}
	\Psi_{\max}^c(y) &=& \underset{t}{\text{mean}} \max_{z} \{\Psi^c(z,t;y)\}, \\
	d_{\max}^c(y) &=& \underset{t}{\text{mean}} \ \underset{z}{\arg\!\max} \{\Psi^c(z,t;y)\} ,
	\label{e_Psi_max}
\end{eqnarray}
with similar definitions for $\tilde{\Psi}^c$-based values. Averages are taken over the entire length of each model control run. Contributions of boundary components shown in Figure \ref{p_Qtd_MxStrmLtt} are taken at the same depth $\tilde{d}_{\max}^c$ as those of the maximum estimated overturning $\tilde{\Psi}^c$ and then averaged over the entire model run. In this way the maximum estimated overturning streamfunction can therefore also be decomposed into its boundary components.

Figure \ref{p_Qtd_MxStrmLtt}(a) illustrates the variation with latitude of contributions made by each boundary component, highlighting (i) the dominance of the thermal wind (W+E) contribution in the southern hemisphere, (ii) compensating variations between the components for latitudes from $7^\circ$N to $35^\circ$N, and large changes in boundary contributions with latitude, (iii) significant Ekman contribution in equatorial regions, and (iv) the reversal of density and bottom components at high northern latitudes and $24^\circ$N to $28^\circ$N. 

Further, Figure \ref{p_Qtd_MxStrmLtt} indicates good agreement between $\Psi_{\max}^c$  and $\tilde{\Psi}_{\max}^c$ for the majority of the basin. $\Psi_{\max}^c$  and $\tilde{\Psi}_{\max}^c$ vary smoothly with latitude, with the latter showing a greater level of variability in equatorial regions. The relatively constant magnitude (15Sv) of both $\Psi_{\max}^c$  and $\tilde{\Psi}_{\max}^c$ at all latitudes up to $35^\circ$N can be attributed to conservation of volume between latitudes. Interestingly, the relative stability of both $\Psi_{\max}^c$  and $\tilde{\Psi}_{\max}^c$ is not replicated in the constituent boundary components, especially in the northern hemisphere subtropics. Between $10^\circ$N and $20^\circ$N we find the reduction in $\tilde{\Psi}_{th_{\max}}^c$ is compensated by a large $\tilde{\Psi}_{Ekm_{\max}}^c$ contribution, attributed to easterly trade winds in the region. Further northwards between $25^\circ$N and $30^\circ$N, the reversal of $\tilde{\Psi}_{th_{\max}}^c$ and $\tilde{\Psi}_{bot_{\max}}^c$ contributions to the maximum overturning streamfunction coincides with strong western boundary currents near the Florida Strait. For the same reason we find a large $\tilde{\Psi}_{AC_{\max}}^c$ contribution for this latitude range.

This same analysis is extended to both $1^\circ$ and $1/12^\circ$ models to assess the impact of spatial resolution on the latitudinal variation of $\Psi_{\max}^c$  and $\tilde{\Psi}_{\max}^c$; results are shown in Figures \ref{p_RC_MxStrmLtt} and \ref{p_RC_BdryMxStrmLtt} below. In the $1^\circ$ model, we find that at low latitudes the Ekman component dominates, possibly due to shallow sub-tropical cells, leading to a strong surface contribution. Consequently, we find that the depth of maximum $d_{\max}^c$ and $\tilde{d}_{\max}^c$ for these latitudes lie in the upper few hundred metres; depths which are not our main focus. In order to acquire an appropriate depth for the maximum overturning, we constrain the calculation to only use values corresponding to depths of maxima greater than $500$m. This constraint is found to have little or no effect when applied to the $1/4^\circ$ and $1/12^\circ$ models.

Panel (a) of Figure \ref{p_RC_MxStrmLtt} shows $\Psi_{\max}^c$ and $\tilde{\Psi}_{\max}^c$ for the three model resolutions, and indicates a clear mean state difference between the three models throughout the basin, up to $35^\circ$N. Following the separation of the Gulf Stream, we notice a significant drop in the maximum overturning streamfunction in both higher resolution models, compared with a more gradual weakening for the $1^\circ$ model. Panel (b) shows the corresponding average depths $d_{\max}^c$ and $\tilde{d}_{\max}^c$, corresponding to $\Psi_{\max}^c$ and $\tilde{\Psi}_{\max}^c$ for each latitude within the Atlantic basin. Surprisingly, we find the $1^\circ$ model exhibits the deepest overturning maximum in contrast to shallower maxima for both higher resolution models. We might expect higher resolution models to yield more accurate and realistic results, therefore, it's surprising that higher resolution models have shallower maxima. 
\begin{figure*}[ht!]
	\centerline{\includegraphics[width=\textwidth]{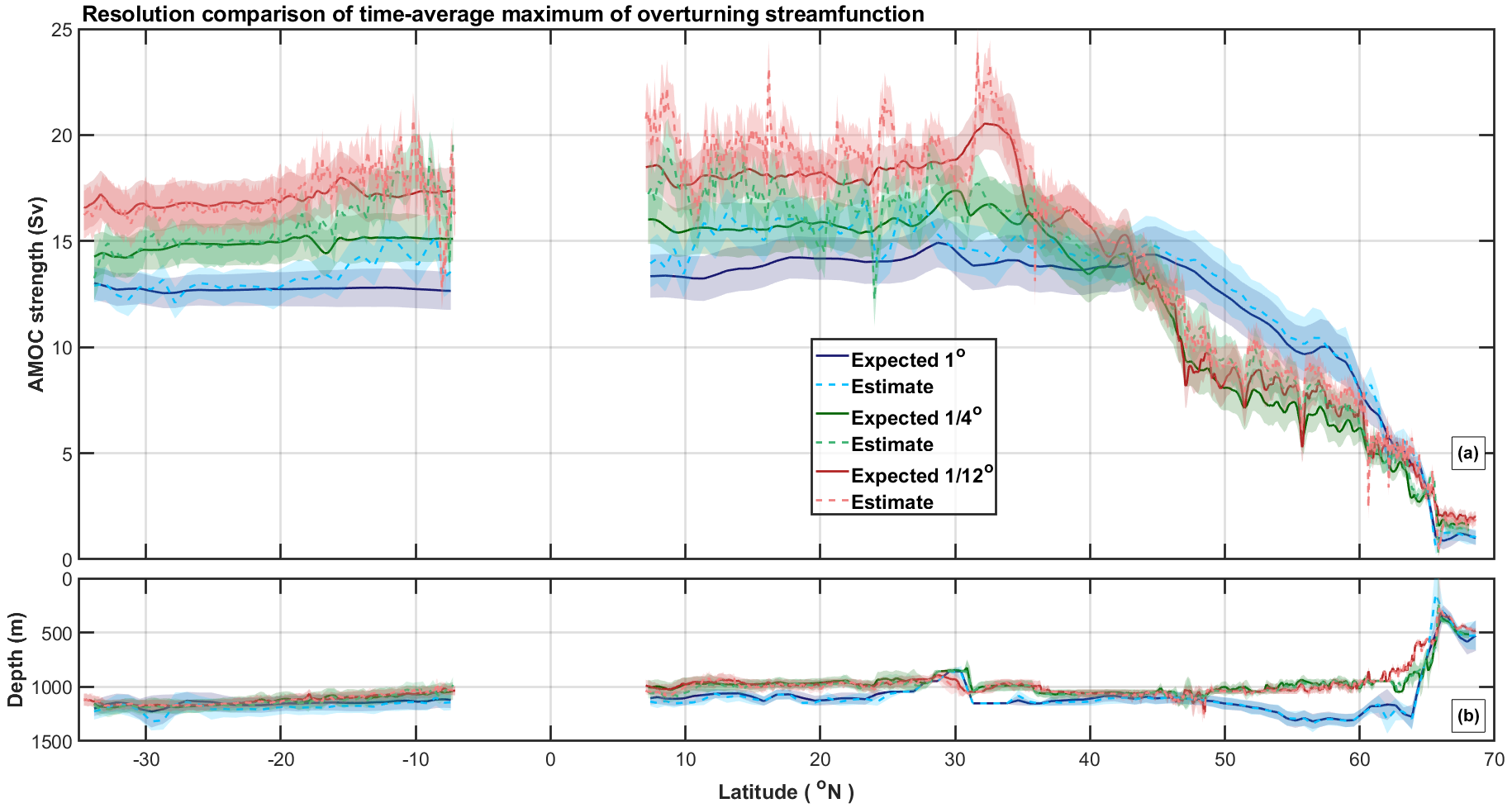}}
	\caption[Spatial resolution comparison of time-mean maximum overturning streamfunction with latitude.]{Contributions to the time-average maximum overturning streamfunction for different model spatial resolutions. Panel (a) shows the time-mean maximum overturning streamfunction, $\Psi_{\max}^c$ and $\tilde{\Psi}_{\max}^c$, as a function of latitude for $1^\circ$ (navy, dashed light blue), $1/4^\circ$ (dark green, dashed light green) and $1/12^\circ$ (red, dashed red) model resolutions. The time-average is taken over the length of the run in each case. Shaded regions indicate $\pm1$ temporal standard deviation for each quantity. Panel (b) shows the corresponding average depths $d_{\max}^c$ and $\tilde{d}_{\max}^c$ corresponding to $\Psi_{\max}^c$ and $\tilde{\Psi}_{\max}^c$ for each latitude of the Atlantic basin.}
	\label{p_RC_MxStrmLtt}
\end{figure*}

\cite{Hirschi2020} discuss the impact of model resolution upon the overturning circulation, and show that eddy-resolving models tend to have a stronger time-average maximum streamfunction (see the $1/12^\circ$ results in Figure \ref{p_RC_MxStrmLtt}). They also comment on a significant weakening in depth-space overturning north of $30^\circ$N, also present in the figure here. They attribute the enhanced weakening at higher resolution to the pathways taken by the AMOC at higher latitudes; at higher resolution we observe a more realistic subpolar gyre and therefore a more realistic northward pathway. This pathway, which includes the North Atlantic Current, impacts the representation of the overturning strength, especially in depth coordinates. Due to gyre circulation, we find the northward and southward transport cancel when using depth coordinates, leading to an apparent weakening of the overturning. \cite{Hirschi2020} show further that the use of density-coordinate-based diagnostics would improve our ability to capture the overturning strength at high northern latitudes. This is especially true for high spatial resolution models, due to greater horizontal circulation. 

\begin{figure*}[ht!]
	\centerline{\includegraphics[width=\textwidth]{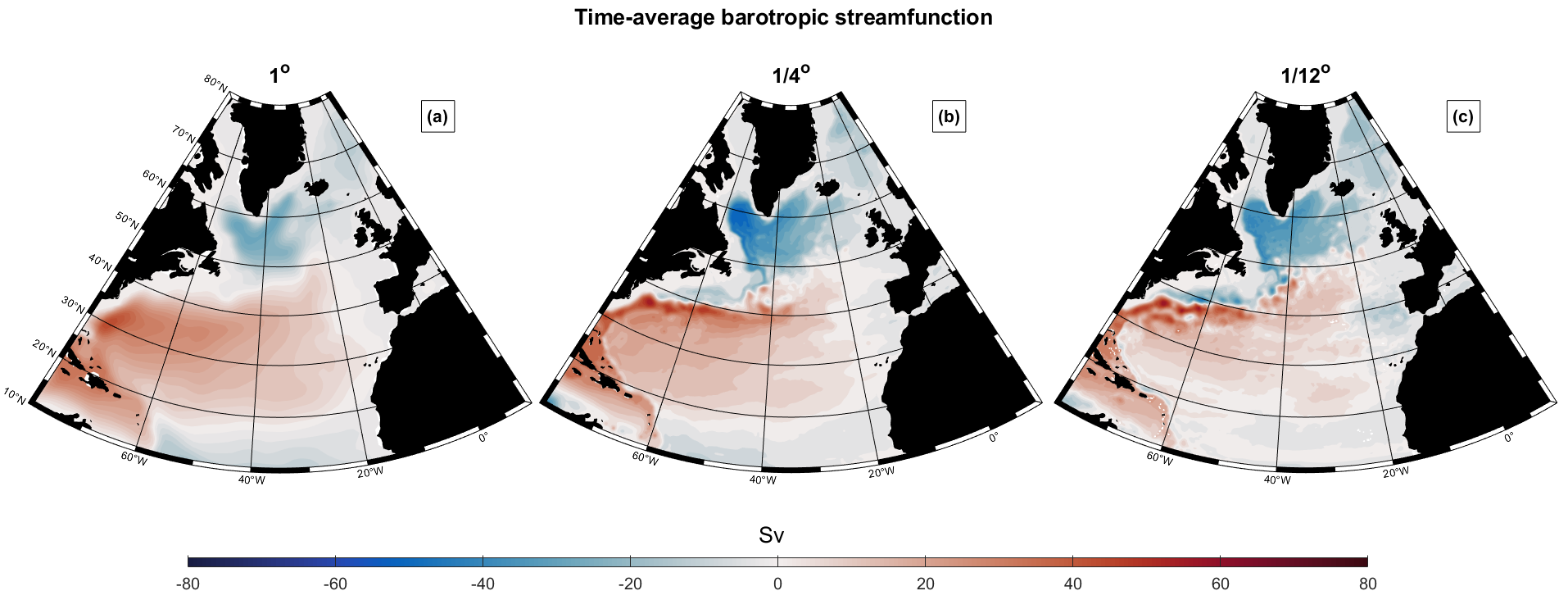}}
	\caption[Spatial resolution comparison of time-mean barotropic streamfunction in the North Atlantic over all years of model runs.]{Time-average barotropic streamfunction $\Psi_v$ for the (a)  $1^\circ$, (b)  $1/4^\circ$ and  $1/12^\circ$ models within the North Atlantic. The time-average is taken over the length of the run in each case.}
	\label{p_RC_BrtStrm}
\end{figure*}

Motivated by \cite{Hirschi2020}, to show the more realistic subpolar gyre within both higher resolution models, the time-average barotropic streamfunction, 
\begin{equation}
	\Psi_v(x,y) = \int_{W(y)}^{x}\int_{-H(x,y)}^{0}v_N(x',y,z) \ \mathrm{d} z \ \mathrm{d} x',
	\label{e_psi_v}
\end{equation}
across all years for the North Atlantic is shown in Figure \ref{p_RC_BrtStrm}. $\Psi_v$ is the cumulative sum from the western boundary at a given $y$ to a $x$ of interest, of the vertical sum of meridional velocities from the ocean floor to the surface. We choose to take the integral from the western boundary eastward at a given $y$. Further, a constant is added to the values of $\Psi_v$ for all longitudes and latitudes, so that $\Psi_v$ for the north-westernmost grid location is zero.

Figure \ref{p_RC_BrtStrm} shows that the subpolar and subtropical gyres within the higher resolution models are stronger, resulting in greater horizontal circulation, which at subpolar latitudes is responsible for much of the water mass transformation, reducing the role of depth-space overturning. This leads to the significant weakening of $\Psi_{\max}^c$ and $\tilde{\Psi}_{\max}^c$ at high northern latitudes in Figure \ref{p_RC_MxStrmLtt}.

\subsection{Boundary contributions to $\tilde{\Psi}_{\max}^c$} 
\begin{figure*}[ht!]
	\centerline{\includegraphics[width=\textwidth]{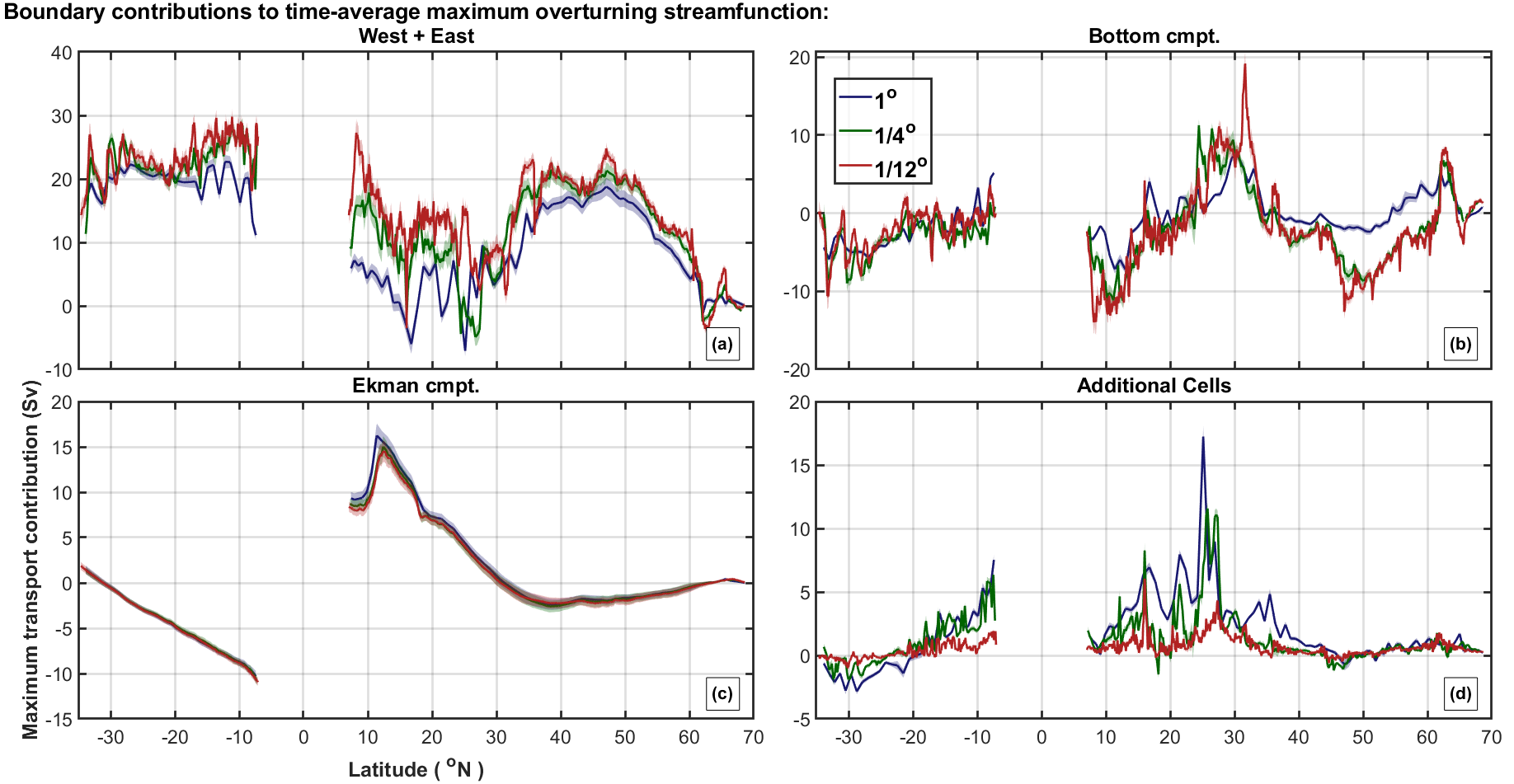}}
	\caption[Boundary contributions to the time-mean estimated maximum overturning streamfunction for the $1^\circ$, $1/4^\circ$ and $1/12^\circ$ model resolutions.]{Time-mean maximum overturning streamfunction boundary contributions to $\tilde{\Psi}_{\max}^c$ for $1^o$ (navy), $1/4^o$ (green) and $1/12^o$ (red) model resolutions. Panel contents are: (a) $\Psi^c_{{th}_{\max}}$, (b)  $\Psi^c_{{bot}_{\max}}$, (c)  $\Psi^c_{{Ekm}_{\max}}$ and (d)  $\Psi^c_{{AC}_{\max}}$. Time-average is taken over the length of the run in each case. Shaded regions indicate $\pm1$ temporal standard deviation of the annual-mean data for the corresponding component. We note model run length differs with resolution. }
	\label{p_RC_BdryMxStrmLtt}
\end{figure*}

Figure \ref{p_RC_BdryMxStrmLtt} illustrates the boundary contributions to the estimates for $\tilde{\Psi}_{\max}^c$ in Figure \ref{p_RC_MxStrmLtt} made by the density (western and eastern), bottom, Ekman and additional cell components for each model resolution. For brevity, we choose to refer to these quantities as $\Psi^c_{{th}_{\max}}$, $\Psi^c_{{bot}_{\max}}$, $\Psi^c_{{Ekm}_{\max}}$ and $\Psi^c_{{AC}_{\max}}$ respectively.

Panel (c) shows little variation in  $\Psi^c_{{Ekm}_{\max}}$ between model resolutions; this is no surprise since common initial atmospheric conditions and atmospheric resolution are used for each model. In Panel (d), it is promising to see that  $\Psi^c_{{AC}_{\max}}$ reduces with increasing model resolution. Large peaks between $20^\circ$N and $30^\circ$N in the coarser models are attributed to shallow, narrow channels between islands along the eastern coast of the U.S; these narrow channels are found to be smaller than the computational grid size. For example, the large peak in the $1^\circ$ model (blue line) at $25^\circ$N is due to the narrow channel between Florida and the Bahamas. This channel is narrower than the $\approx100$km grid box size corresponding to a $1^\circ$ grid cell, and hence the contribution of this region is attributed to  $\Psi^c_{{AC}_{\max}}$ instead of $\Psi^c_{{th}_{\max}}$ and  $\Psi^c_{{bot}_{\max}}$.

\begin{figure*}[ht!]
	\centerline{\includegraphics[width=\textwidth]{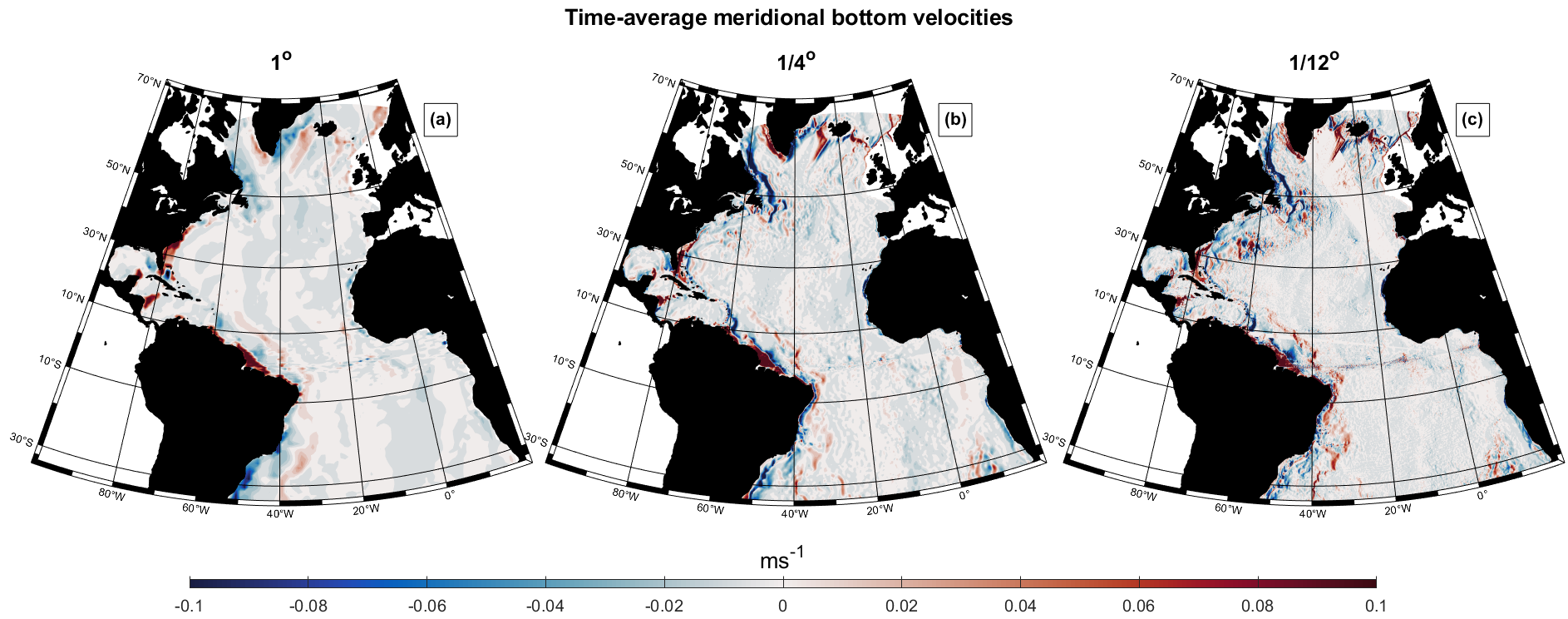}}
	\caption[Time-mean meridional bottom velocities of the Atlantic basin for the (a) $1^\circ$, (b) $1/4^\circ$ and (c) $1/12^\circ$ model resolutions.]{Time-mean meridional bottom velocities for the (a) $1^\circ$, (b) $1/4^\circ$, and (c) $1/12^\circ$ model resolutions. The time-average is taken over the length of the run in each case.}
	\label{p_RC_Bvel}
\end{figure*}

Panel (a) of Figure \ref{p_RC_BdryMxStrmLtt} illustrates the combined contribution to $\tilde{\Psi}_{\max}^c$ of eastern and western boundary densities ($\Psi^c_{{th}_{\max}}$), throughout the Atlantic basin. Noticeable is the weaker contribution in the $1^\circ$ model from $10^\circ$S to $25^\circ$N, attributed to an increase in densities on the western boundary (or decrease in the east), and to coarser model bathymetry. The coarser resolution results in a greater proportion of the contribution being attributed to $\Psi^c_{{AC}_{\max}}$, due to the greater area attributed to additional cell. Similarly, in Panel (b) we find a weaker  $\Psi^c_{{bot}_{\max}}$ contribution to $\tilde{\Psi}_{\max}^c$ for the $1^\circ$ model at latitudes $40^\circ-60^\circ$N, clear from inspection of bottom velocities illustrated in Figure \ref{p_RC_Bvel}. These latitudes correspond to the presence of a weaker deep western boundary current (DWBC) for that model resolution (Figure \ref{p_RC_Bvel}). The $1/4^\circ$ and $1/12^\circ$ models show comparable values for the strength of the DWBC, and hence similar $\Psi^c_{{bot}_{\max}}$ contributions to  $\tilde{\Psi}_{\max}^c$. Regions of strong (positive and negative) $\Psi^c_{{bot}_{\max}}$ in Figure \ref{p_RC_BdryMxStrmLtt} relate closely to boundary currents such as the Gulf Stream ($25^\circ$ to $35^\circ$N) and the southward flows around the eastern shores of the Caribbean islands ($5^\circ$ to $15^\circ$N) and Labrador Sea ($45^\circ$ to $60^\circ$N).

\begin{figure*}[ht!]
	\centerline{\includegraphics[width=\textwidth]{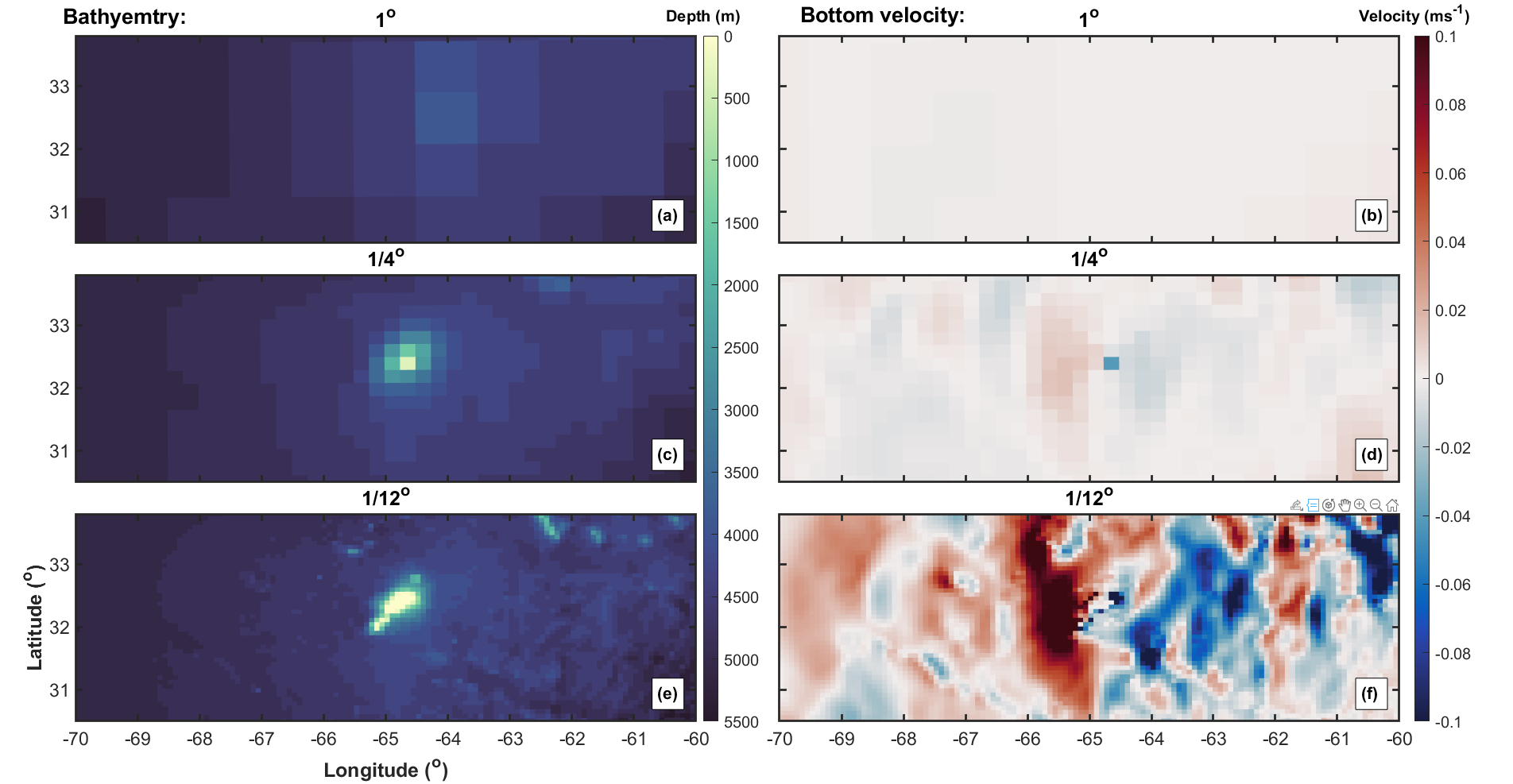}}
	\caption[Bathymetry and time-mean meridional bottom velocities for Bermuda region in the $1^\circ$, $1/4^\circ$ and $1/12^\circ$ models.]{Bathymetry and time-mean meridional bottom velocities for Bermuda region. ($70^\circ$ to $60^\circ$W, $30.5^\circ$ to $33.8^\circ$N) for (a,b) $1^\circ$, (c,d) $1/4^\circ$ and (e,f) $1/12^\circ$ model resolution. The time-average is taken over the length of the run in each case.}
	\label{p_RC_Bermuda}
\end{figure*}

In Figure \ref{p_RC_MxStrmLtt}(a), a large peak for $\Psi_{\max}^c$ and $\tilde{\Psi}_{\max}^c$ in the $1/12^\circ$ model is present at $32^\circ$N, caused by a strong contribution from $\Psi^c_{{bot}_{\max}}$ (Figure \ref{p_RC_BdryMxStrmLtt}(b)) there. This sudden increase in $\Psi^c_{{bot}_{\max}}$ is due to the changes in bathymetry around the island of Bermuda, which is resolved by the $1/12^\circ$ model but not the other resolutions. Hence bottom flows present around the island are different in the $1/12^\circ$ model compared with coarser resolution models (Figure \ref{p_RC_Bermuda}).

Only in the $1/12^\circ$ model is Bermuda represented as an island (i.e. land at $z=0$). In both $1/4^\circ$ and $1^\circ$ models, Bermuda appears as a seamount which does not penetrate the surface. Panels (b,d,f) of Figure \ref{p_RC_Bermuda} illustrate a clear difference in the local time-average bottom velocities across model resolutions. The $1/12^\circ$ model exhibits strong northward velocities to the west of Bermuda, resulting in the strong northward contribution from $\Psi^c_{{bot}_{\max}}$ for this region (Figure \ref{p_RC_MxStrmLtt}(a) and \ref{p_RC_BdryMxStrmLtt}(b)). 

\subsection{Variation in thermal wind component, $\Psi^c_{{th}_{\max}}$, with latitude}
\label{s_TW_Dcmp}

Appendix \ref{App_MaxWestEast} illustrates the difficulty of explaining variation in $\Psi^c_{{th}_{\max}}$ in terms of its constituent eastern and western boundary contributions. The close coupling of $\Psi^c_{{W}_{\max}}$ and $\Psi^c_{{E}_{\max}}$ is to be expected, however, due to a number of considerations outlined here. First, both eastern and western densities are defined relative to a depth-dependent basin average density, introducing correlation between the values of $\Psi^c_{{W}_{\max}}$ and $\Psi^c_{{E}_{\max}}$. Next, $\Psi^c_{{th}_{\max}}$ is estimated using multiple pairs of eastern and western boundaries, whose contributions are summed at each latitude. However, due, e.g., to the presence of islands, the number of pairs of eastern and western boundaries is latitude-dependent. This leads potentially to larger variability in the magnitudes of $\Psi^c_{{W}_{\max}}$ and $\Psi^c_{{E}_{\max}}$ with latitude, e.g. at Caribbean latitudes in Figure \ref{p_MaxStrm_WE}(a,b) of the Appendix. Finally, the maximum ocean depth varies with latitude. Therefore, $\Psi^c_{{W}_{\max}}$ and $\Psi^c_{{E}_{\max}}$ are estimated by integration over different depths as a function of latitude. This introduces further correlation between $\Psi^c_{{W}_{\max}}$ and $\Psi^c_{{E}_{\max}}$ with latitude. Implementation of a local reference density, normalisation by the number of boundary pairs and integration from $2429$m to the surface within the $1^\circ$ model does slightly improve our ability to decouple the boundary components (not shown), but the underlying issue of coupled  western and eastern boundary components remains. A similar issue is encountered by \cite{Waldman2021}.

\begin{figure*}[ht!]
	\centerline{\includegraphics[width=14cm]{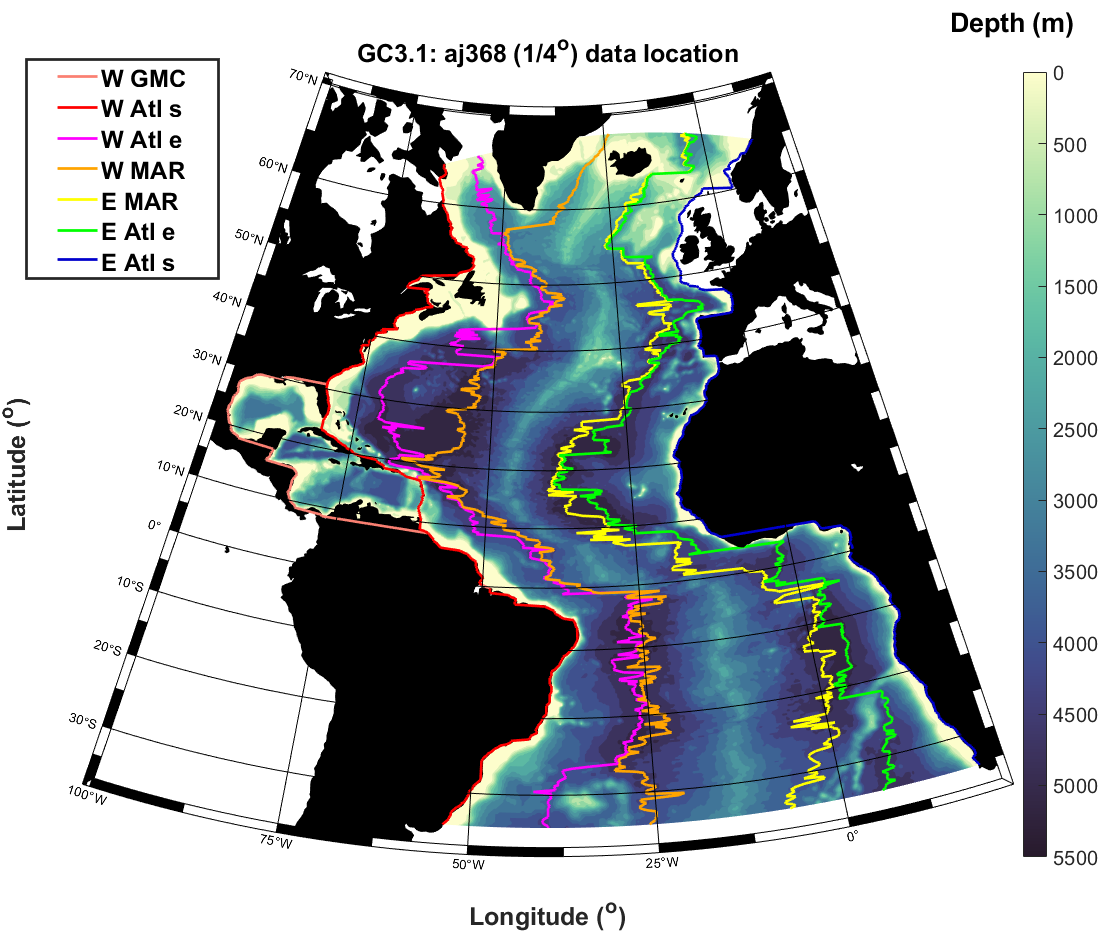}}
	\caption[Bathymetry of the Atlantic basin within $1/4^\circ$ model, including sub-domains for further regional decomposition.]{Bathymetry of the Atlantic basin within the $1/4^\circ$ model. Coloured lines indicate boundaries used for regional decomposition of thermal wind contributions to the overturning streamfunction. Going from left to right, red: start of western Atlantic region (and eastern boundary of GMC where this exists); magenta: end of western Atlantic region; orange: start of MAR region; yellow: end of MAR region; green: start of eastern Atlantic region; blue: end of eastern Atlantic region.}
	\label{p_Bath_sp}
\end{figure*}

To improve our understanding of the boundary density components and to locate the boundary densities which are important to $\tilde{\Psi}^c_{\max}$, a further regional decomposition is applied to the $1/4^\circ$ model data. $\Psi^c_{{W}_{\max}}$ and $\Psi^c_{{E}_{\max}}$ contributions are split into 4 domains, namely: (a) Mid Atlantic Ridge ($\Psi^{c,MAR}_{{W}_{\max}}$, $\Psi^{c,MAR}_{{E}_{\max}}$), (b) Gulf of Mexico and Caribbean Sea ($\Psi^{c,GMC}_{{W}_{\max}}$, $\Psi^{c,GMC}_{{E}_{\max}}$), (c) Atlantic boundary ($\Psi^{c,AB}_{{W}_{\max}}$, $\Psi^{c,AB}_{{E}_{\max}}$) and (d) remainder unaccounted for by the other sub-components ($\Psi^{c,R}_{{W}_{\max}}$, $\Psi^{c,R}_{{E}_{\max}}$). Figure \ref{p_Bath_sp} shows how the regions are defined within the Atlantic basin.

For each latitude, the regions in (a)-(c) above are defined by inspection of the bathymetry. Within the southern hemisphere, the western Atlantic region extends from the longitude corresponding to the South American coastline (red) to the nearest longitude corresponding to the deepest bathymetry (magenta, between the South American coastline and the MAR) at that latitude. The MAR region is defined in a similar manner relative to the longitude of the shallowest MAR bathymetry (between the orange and yellow lines). Finally, the eastern Atlantic region extends to the longitude of the African coastline (blue) from the nearest longitude corresponding to the deepest bathymetry (green, between the MAR and African coastline). Regions in the northern hemisphere are defined similarly, apart from at latitudes corresponding to the Gulf of Mexico and Caribbean Sea (GMC). Here, the western Atlantic boundary is assumed to follow the Caribbean islands, from Trinidad and Tobago northwards along the eastern coast of the Dominican Republic and Cuba to Florida. A separate GMC region is then defined. The eastern boundary of the GMC region (red) is coincident with the western boundary of the Atlantic region where no islands are present, and traces the western boundary of the Caribbean islands otherwise. For each latitude, longitudes not included in any of the Atlantic, MAR and GMC regions are considered to belong to the residual category. 

\begin{figure*}[ht!]
	\centerline{\includegraphics[width=\textwidth]{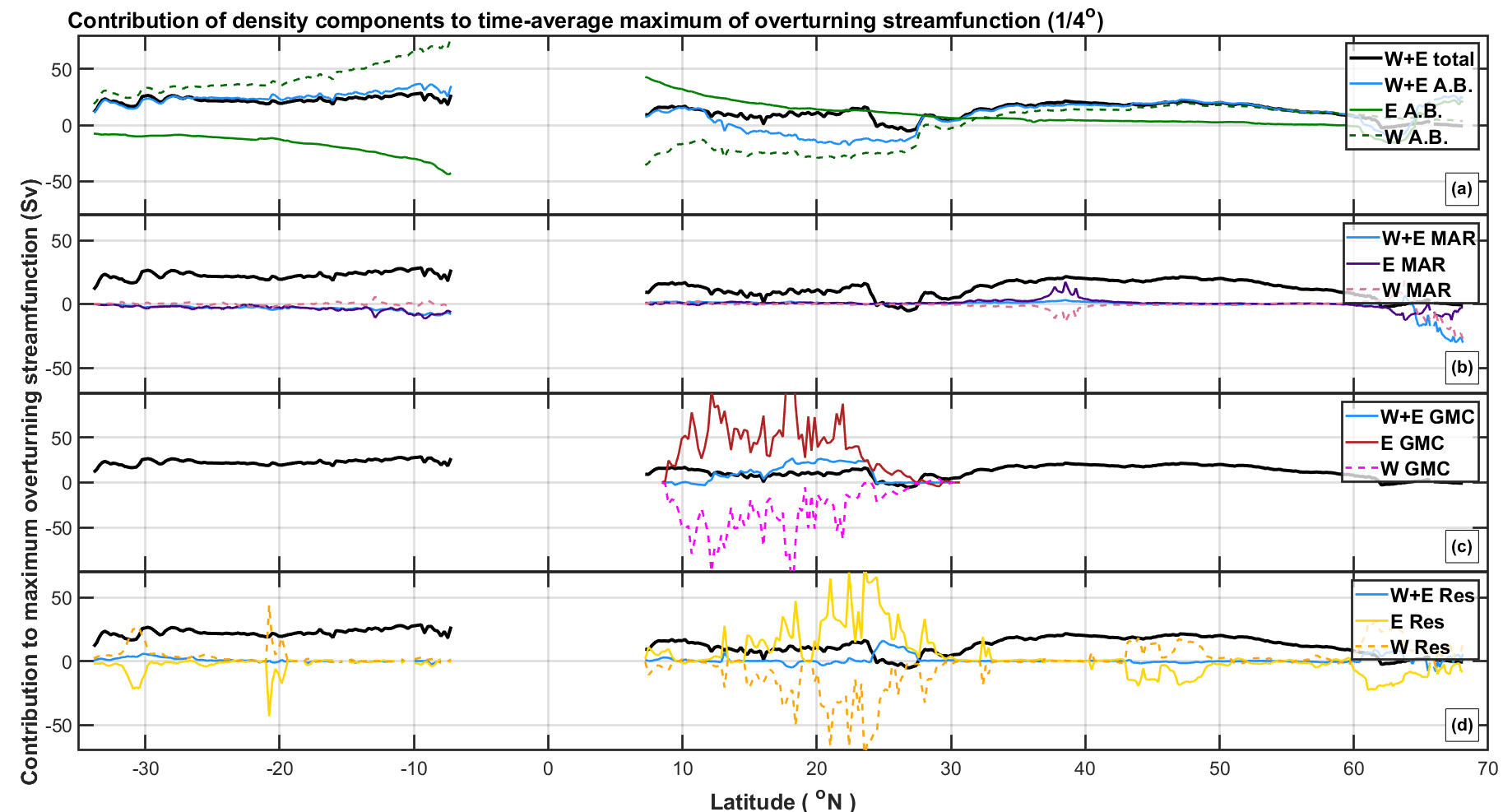}}
	\caption[Contribution made by regionally decomposed sub-components to $\Psi^{c}_{\max}$]{Further decomposition of $\Psi^c_{{W}_{\max}}$ and $\Psi^c_{{E}_{\max}}$ by zonal sub-domains (see Figure \ref{p_Bath_sp}). (a) the contribution made by the Atlantic boundaries (where Gulf of Mexico and Caribbean Sea is neglected, $\Psi^{c,AB}_{{W}_{\max}}$,  $\Psi^{c,AB}_{{E}_{\max}}$), (b) the contribution made by the Mid Atlantic Ridge ($\Psi^{c,MAR}_{{W}_{\max}}$,  $\Psi^{c,MAR}_{{E}_{\max}}$), (c) the contribution of the Gulf of Mexico and Caribbean Sea ($\Psi^{c,GMC}_{{W}_{\max}}$, $\Psi^{c,GMC}_{{E}_{\max}}$) and (d) the remaining contribution not accounted for by the other components ($\Psi^{c,R}_{{W}_{\max}}$, $\Psi^{c,R}_{{E}_{\max}}$). Black line shows $\Psi^c_{{th}_{\max}}$ and blue is the sum of the eastern and western sub-components (e.g. (b) shows the sum of eastern and western contribution made by the MAR, $\Psi^{c,MAR}_{{th}_{\max}}$). For each panel, solid lines indicate eastern and dashed western contributions respectively. }
	\label{p_MaxStrm_WE4_inv}
\end{figure*}

Figure \ref{p_MaxStrm_WE4_inv} shows the contribution made by the eastern and western boundaries of each domain. In each panel, the black line indicates $\Psi^c_{{th}_{\max}}$ and the blue line the sum of the western and eastern contributions for that particular sub-component (e.g. Panel (a) shows the sum $\Psi^{c,AB}_{{th}_{\max}}$ of western and eastern contributions $\Psi^{c,AB}_{{W}_{\max}}+\Psi^{c,AB}_{{E}_{\max}}$ from the Atlantic boundary, with analogous notation used for the other sub-components).  

Panel (a) indicates that the majority of $\Psi^c_{{th}_{\max}}$ (black) in the southern hemisphere is accounted for by  $\Psi^{c,AB}_{{th}_{\max}}$ (blue), as would be expected. Moving northwards, the change in sign of Coriolis parameter $f$ results in a sign change for the sub-components. The negative $\Psi^{c,AB}_{{th}_{\max}}$ (blue) at low northern latitudes is due to the dominant western boundary contribution (dashed green), somewhat surprisingly suggesting southward upper-ocean flow due to these boundaries. We find a large change in $\Psi^{c,AB}_{{W}_{\max}}$ near $27^\circ$N where western boundary densities change across the Gulf Stream as it separates from the boundary, leading to a change in sign of $\Psi^{c,AB}_{{W}_{\max}}$. Conversely $\Psi^{c,AB}_{{E}_{\max}}$ (solid green) is relatively stable, showing smooth decline northwards from the equator, possibly attributed to the presence of relatively flat eastern along boundary isopycnals (discussed further in Chapter \ref{TJ_Bdry}). This steady decline in conjunction with the sudden increase in $\Psi^{c,AB}_{{W}_{\max}}$ near $27^\circ$N results in a strengthening of $\Psi^{c,AB}_{{th}_{\max}}$ (blue) and $\Psi^{c}_{{th}_{\max}}$ (black).  Panel (a) clearly indicates the majority of the geostrophic contribution to the Atlantic's maximum overturning streamfunction is governed by $\Psi^{c,AB}_{{th}_{\max}}$, and $\Psi^{c,AB}_{{W}_{\max}}$ shows greater latitude-dependence than that of the eastern boundary. Further analysis of Atlantic boundary densities and the physical mechanisms leading to the density structure observed are investigated and discussed in Chapter \ref{TJ_Bdry}.   

For the MAR region, Panel (b) indicates a very weak contribution made by boundary density components $\Psi^{c,MAR}_{{W}_{\max}}$ (dashed purple) and $\Psi^{c,MAR}_{{E}_{\max}}$ (solid purple) to $\Psi^c_{{th}_{\max}}$ (black). The sum $\Psi^{c,MAR}_{{th}_{\max}}$ of boundary contributions is small except for the very highest northern latitudes. We find a slight increase in the magnitude of boundary sub-components near $39^\circ$N, likely to be attributable to the Azores, and a decrease at high northern latitudes attributed to boundary densities around eastern Iceland. 

For the GMC, Panel (c) shows that the eastern boundary contribution $\Psi^{c,GMC}_{{E}_{\max}}$ (solid red) is the dominant contributor. There appears to be a gradual weakening of the magnitude of $\Psi^{c,GMC}_{{W}_{\max}}$ (dashed magenta) with latitude relative to $\Psi^{c,GMC}_{{E}_{\max}}$, resulting in a northward contribution to geostrophic part of the maximum overturning streamfunction $\Psi^{c,GMC}_{{th}_{\max}}$ for GMC (solid blue). 

Finally Panel (d) shows the maximum overturning streamfunction contributions unaccounted for by the other 3 sub-components. These contributions can be attributed to the presence of small islands and ridges within the Atlantic basin. The peaks in the southern hemisphere can be explained by the presence of the Malvinas islands ($30^\circ$S) and the juncture of the Walvis ridge with the African continent ($20^\circ$S). Similarly, the contributions between $15^\circ$N and $29^\circ$N are attributed to islands such as the Bahamas, Turks and Caicos and Canary islands, all of which fall outside the other sub-domains (Figure \ref{p_Bath_sp}).

\section{Summary}

In this chapter, the theory for decomposing the meridional overturning streamfunction into contributions related to variables on the ocean boundaries has been outlined, and its application to output of HadGEM3-GC3.1 models at different spatial resolutions was studied. The decomposition diagnostic relies on (a) boundary densities within the nearest full cells to coastal boundaries, (b) ocean surface wind stress, and (c) meridional velocity in the bottom cell of each fluid column.

Reconstructed estimates of the total overturning streamfunction obtained by summing these three components have been shown to perform generally well throughout the Atlantic basin, and are able to capture the time-mean overturning streamfunction well when all components are volume-compensated. Large contributions to the time-mean overturning streamfunction are made by boundary density and depth-independent terms within the interior of the ocean, and the Ekman component dominates the surface boundary layer. The time-mean estimate of the total overturning streamfunction calculated from model data at $1/4^\circ$ spatial resolution, is able to capture $95\%$ of the spatial variation in the expected overturning streamfunction. Using data from higher-resolution models increases the spatial variance explained. Additional incomplete cells adjacent to bathymetry make a sizeable contribution to the time-mean overturning streamfunction, and its spatial variability. An estimate for the additional cell contribution is made using meridional velocities in these additional cells.

Discrepancies have been found between the estimated overturning streamfunction (from the decomposition) and the expected overturning streamfunction (calculated using meridional velocities) at certain latitudes within the Atlantic basin. Investigation of the various terms contributing to the GCM's momentum budget indicated that the theoretical capability of the diagnostic framework was greater than that achieved. This prompted an investigation of the influence of time-average input fields on the decomposition, and the contributions of localised processes (e.g. momentum advection). The non-linear dependence of in-situ density on potential temperature $\theta$ leads to errors if annually-averaged potential temperature data is used for the diagnostic calculations. Time-averaging of salinity and the inclusion of varying cell thickness (via the model's $e3t$ term) have also been shown to make a small contribution to the observed discrepancy, but these effects are insignificant in comparison to errors induced by time-averaging potential temperature. Calculations using densities averaged over every time-step have been shown to improve the resulting horizontal pressure gradient trend and overturning streamfunction estimate. 

The time-mean maximum overturning streamfunction (in depth space) is found to be relatively constant with latitude, for latitudes south of $35^\circ$N, and thereafter reduces with increasing latitude, for all model resolutions considered. There is good agreement between the expected time-mean maximum $\Psi_{\max}^c$ and the decomposition estimate $\tilde{\Psi}_{\max}^c$ for all latitudes. The boundary contributions to $\tilde{\Psi}_{\max}^c$ show considerably larger variation with latitude, of the order of $20$Sv. Differences in boundary contributions at different model resolutions can be attributed to better characterisation of bathymetry and bottom currents in the higher resolution models. At all model resolutions, the sum of western and eastern density contributions dominates  $\tilde{\Psi}_{\max}^c$ in the southern hemisphere; the northern hemisphere exhibits more complex behaviour. 

An investigation into the relative importance of western and eastern boundary density contributions to the total thermal wind contribution ($\Psi^c_{{th}_{\max}}$) of the estimated maximum overturning streamfunction is complicated by the strong correlation between these eastern and western boundary components. 
%
%
Further decomposition of the thermal wind contribution into sub-domains (-components) for the mid-Atlantic ridge (MAR), Gulf of Mexico and Caribbean Sea (GMC), Atlantic boundary and remaining contributions (Rest) reveal (a) a significant positive (or northward) contribution within the GMC due to eastern GMC densities, (b) a weak overall contribution made by the MAR, (c) a gradual weakening of eastern Atlantic boundary density contribution with latitude, and large changes in the Atlantic western boundary density contribution near $27^\circ$N. We also find (d) a strong effect of the Bahamas and surrounding islands on density-driven contributions to the maximum overturning streamfunction at approximately $26^\circ$N.
 
This work motivates the next chapter, where we investigate the contribution of boundary components to temporal variability of the Atlantic overturning circulation. \clearpage

\chapter{Variability of AMOC boundary components}  \label{TJ_Var}

In recent decades, significant progress has been made in understanding the underlying physical mechanisms of the AMOC. Characterising the natural variability of the overturning is critical to identifying and quantifying the potential impacts of a changing climate. Data from cross-basin monitoring arrays, buoys, floats and high resolution models have suggested possible drivers of variability at short timescales and demonstrated the apparent lack of coherence in AMOC variability with latitude. In this chapter, we seek to improve our understanding of the spatial and temporal variability of the AMOC and the factors contributing to it. We aim to improve our understanding of the AMOC's variability at decadal timescales and longer. We consider the temporal variability of the maximum overturning streamfunction timeseries at RAPID ($26^\circ$N) and SAMBA ($34^\circ$S) latitudes, and the temporal variation of the contributing boundary components. We investigate the correlation between the thermal wind and bottom components, and the source of large oscillations found in their timeseries at SAMBA for the $1/4^\circ$ HadGEM-GC3.1 model. The basin-wide spatial variability of in-situ densities and bottom velocities, used in the overturning streamfunction decomposition outlined in Chapter \ref{TJ_TM}, is summarised for different timescales; this analysis leads to the implementation of a multiple linear regression model incorporating Correlation Adjusted coRrelation scores (MLR-CAR, \citealt{Zuber2010}) to quantify the contributions of boundary components to the total overturning streamfunction variability at different depths, latitudes and timescales. We also briefly consider the influence of model resolution upon inferences from MLR-CAR analysis.

\section{Overview: AMOC temporal and spatial variability}

The AMOC varies on timescales from days to millennia. The relatively short timeseries available from the RAPID AMOC observing programme has shed some light on short-term AMOC variability (Sections \ref{I_obsAMOC} and \ref{I_var_obs}). On seasonal timescales, observations and GCM output both show that wind stress on the ocean's surface dominates AMOC variability (e.g. \citealt{Hirschi2007}) via processes such as Ekman transport and coastal upwelling, resulting in large seasonal fluctuations in the strength of the overturning. 

GCMs and observations agree in general (\citealt{Buckley2016}, Section \ref{I_var_mod}) that: (a) intra-annual timescales are dominated by wind stress via Ekman transport; (b) longer inter-annual timescales, the thermal wind (or geostrophic) component dominates, implying the density structure of the basin is very important; (c) the AMOC is not coherent between the subtropical and subpolar gyres on inter-annual timescales. Interestingly the sub-tropical gyre is dominated by inter-annual variability whereas the sub-polar gyre is dominated by variability with decadal periods (e.g. \citealt{Buckley2016}). Models have shown at decadal and longer timescales that there is some meridional coherence within the AMOC (e.g. \citealt{Delworth1993}). Further: (d) upper ocean western boundary density anomalies are important, playing a key role in AMOC variability. On decadal timescales it has been shown that the western boundary buoyancy anomalies are meridionally coherent, originating from the subpolar gyre or along the subtropical-subpolar gyre boundary (e.g. \citealt{Robson2012a}, \citealt{Buckley2012}). Interestingly, \cite{Frajka-Williams2018} find a lack of coherence when comparing fluctuations in transport at $16^\circ$N and $26^\circ$N between 2004 and 2017. With numerous years of data from several monitoring arrays throughout the Atlantic basin (RAPID, MOVE, OSNAP and SAMBA), they suggest the average strength and variability of the overturning streamfunction varies with latitude, with greater variability found in the South Atlantic (\cite{Frajka-Williams2019}).

Numerous model investigations have shown that inter-annual to decadal variability is strongly linked to regions of deep water formation at high latitudes, particularly the convective mixing occurring in the Labrador Sea (e.g. \citealt{Kuhlbrodt2007}, \cite{Gelderloos2012}, Section \ref{I_var_mod}). However, no conclusive observational evidence for a link between Labrador Sea deep water formation and AMOC variability has been shown, and greater variability is in fact found at OSNAP East (\citealt{Lozier2019}). Convective mixing is strongly tied to the North Atlantic Oscillation resulting in fluctuations in convective mixing rates. Periods (e.g. in 1972) of total shut-down in convective mixing within the Labrador Sea have been observed (e.g. \citealt{Gelderloos2012}). 

Salinity anomalies become important at centennial to millennial timescales within the HadCM3 model. Since salinity drives overall density fluctuations at high latitudes, this leads to variations in the AMOC (e.g. \citealt{Jackson2013}). 

The first dynamical decomposition of the overturning circulation was performed by \cite{Lee1998}  to investigate the time-mean and variability of the Indian overturning circulation. Recently, \cite{Waldman2021} decomposed the Atlantic overturning circulation into multiple combinations of components, including Ekman, barotropic, baroclinic, thermal, haline and western and eastern boundary contributions, to investigate centennial variability of the AMOC within the CNRM-CM6 model. The thermal wind component explains over 80\% of the low-frequency AMOC transport variance for all latitudes. They show that this variability is driven by southward propagating western boundary temperature anomalies at depths between 500m and 1500m, originating in the western subpolar gyre. Some variability can be attributed to the contributions made by salinity anomalies along the eastern boundary of the South Atlantic. They find it difficult to interpret variability of individual boundary contributions, due to the close correlation between boundary components (similar to Section \ref{s_TW_Dcmp}). This allows for density-compensating thermohaline variations and covarying densities at both boundaries, especially in the South Atlantic. The sources of mid-depth density variation found, are attributed to deep convection in the Labrador Sea and dense water overflow in the Davis Strait.

Many studies have highlighted a possible slowing of the AMOC (e.g. \citealt{Rahmstorf2015}, \citealt{Caesar2018,Caesar2021}, Section \ref{I_AMOCrole}), emphasizing the need for improved understanding of the inherent natural variability shown by the AMOC through time and space. 

\section{Evaluation of decomposition at RAPID $\&$ SAMBA}
\label{S3_RS}
Having understood the limitations of the overturning decomposition diagnostic in Chapter \ref{TJ_TM}, this section investigates its performance in comparison to published results for the RAPID and SAMBA arrays, located in the North and South Atlantic, respectively. The decomposition diagnostic is used in its final form (Equation \ref{e_tilde_TcF}) for output of HadGEM-GC3.1 simulations at $1^\circ$, $1/4^\circ$ and $1/12^\circ$ spatial resolutions, where time-average input fields $\theta$ and $S$ are used alongside the $eos80$ seawater library (\citealt{Millero1980}) to calculate the thermal wind component of the overturning streamfunction. The depth-independent, Ekman and additional cell components are calculated as in Section \ref{S_App_DD}.

\subsection{RAPID array} \label{S3_RS_Rpd}
The RAPID array (introduced in Section \ref{I_oAMOC}) monitors the basin-wide meridional volume transport from Florida across to Tenerife at $26.5^\circ$N. Recent estimates of $17.0 \pm 3.3 $Sv are quoted by \cite{Frajka-Williams2019} for the maximum of the overturning streamfunction, at $\approx 1100$m depth. 

\begin{figure*}[ht!]
	\centerline{\includegraphics[width=\textwidth]{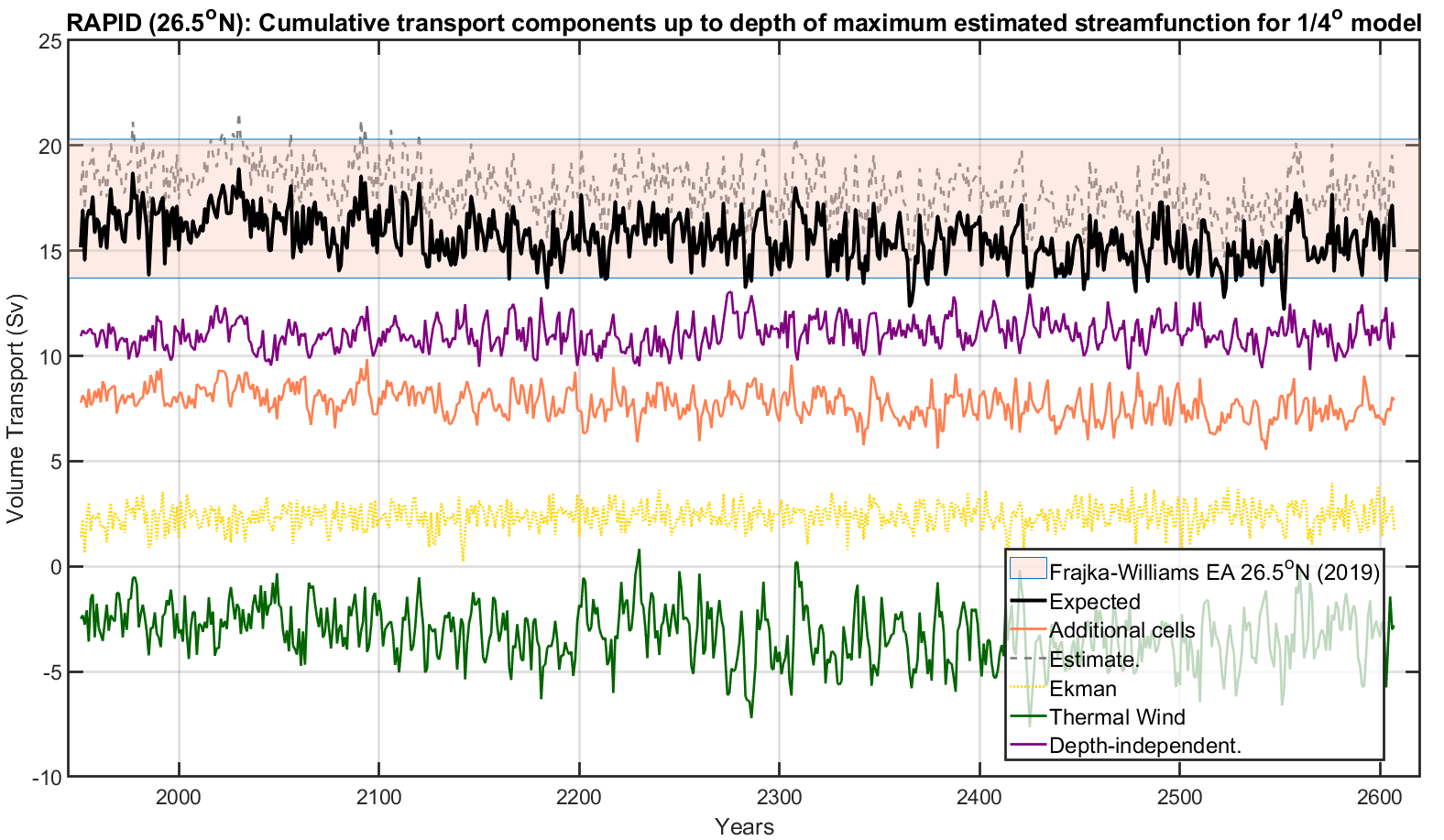}}
	\caption[Timeseries of maximum overturning streamfunction at $26.5^\circ$N for the $1/4^\circ$ model resolution, and contributing components estimated using decomposition diagnostic.]{Timeseries of the maximum overturning streamfunction at $26.5^\circ$N for the $1/4^\circ$ model resolution, and contributing components estimated using the decomposition diagnostic. Shaded red region indicates the $\pm$ one standard deviation interval for the RAPID-observed streamfunction as quoted by \cite{Frajka-Williams2019}. The volume-compensated total expected streamfunction $\Psi^c_{\max}$ is shown in black. Other coloured lines indicate the volume-compensated estimated streamfunction ($\tilde{\Psi}_{\max}^c$, dashed grey), depth-independent flow contribution ($\Psi_{bot_{\max}}^c$, purple), additional cell contribution ($\Psi_{AC_{\max}}^c$, orange), Ekman transport ($\Psi_{Ekm_{\max}}^c$, yellow) and the thermal wind component ($\Psi_{W_{\max}}^c+\Psi_{E_{\max}}^c$, dark green).}
	\label{Rpd_ts}
\end{figure*}

The measurement methodology at RAPID uses a telephone cable to estimate the transport through the Florida Strait, and density moorings elsewhere along the boundaries of the open ocean. In contrast, the decomposition diagnostic relies on boundary densities and velocities for the whole section from Florida to the African coastline. Therefore, small differences in overturning streamfunction estimates between the two approaches are to be expected. Figure \ref{Rpd_ts} shows a timeseries of the maximum compensated overturning streamfunction at $26.5^\circ$N for $1/4^\circ$ model resolution, and contributing components estimated using the decomposition diagnostic, calculated using the methodology from Section \ref{s_var_tm_ovr_lat} and Equation \ref{e_Psi_max}. Specifically, the boundary component contributions are taken at the depth of the maximum of the compensated estimate for the overturning streamfunction $\tilde{\Psi}_{\max}^c$, hereafter, for brevity, known as the maximum estimated overturning streamfunction.

The estimated maximum overturning streamfunction ($\tilde{\Psi}_{\max}^c$, light grey, i.e. the cumulative meridional transport from the bottom up to the depth of the maximum) is approximately $1$ to $2$Sv larger than the expected maximum ($\Psi^c_{\max}$, black) overturning streamfunction calculated directly using meridional velocities (Equation \ref{T}). Possible explanations for the offset were discussed earlier in Chapter \ref{TJ_TM}, involving the quality of decomposition estimates for the thermal wind component $\Psi_{W_{\max}}^c+\Psi_{E_{\max}}^c$. For the majority of the timeseries, both the expected and estimated maximum overturning streamfunctions are within the one standard deviation uncertainty band for the observed maximum overturning streamfunction found by \cite{Frajka-Williams2019}. This suggests that the decomposition diagnostic is performing relatively well, and by implication suggest the $1/4^\circ$ HadGEM-GC3.1 model is performing well. Figure \ref{Rpd_ts} is further supported by Table \ref{Tab_SVRpd}, which gives the percentage temporal variation of $\Psi_{\max}^c$ explained by $\tilde{\Psi}_{\max}^c$ and $\tilde{\Psi}^c_{\max} - \Psi_{AC_{\max}}^c$ (which excludes additional cells). 

For this analysis, increasing HadGEM3-GC3.1 model resolution to $1/12^\circ$ appears to improve the ability of the overturning diagnostic to explain the temporal variation of expected maximum overturning streamfunction. Moreover, at coarser model resolutions, inclusion of additional cell contribution $\Psi_{AC_{\max}}^c$ provides clear improvement in diagnostic skill. The poor results found for the $1/4^\circ$ model, ignoring additional cells, can be attributed to the
models ability to resolve the bathymetry of the western boundary around the Bahamas, and flow through the Florida Strait.

\begin{table}[h!]
	\centering
	\begin{tabular}{ |P{2.7cm}||P{2.7cm}|P{2.7cm}|P{2.7cm}| }
		\hline
		\multicolumn{4}{|c|}{Temporal variance of $\Psi^c_{\max}$ explained at RAPID ($26.5^\circ$N)} \\
		\hline 
		& $1^\circ$ & $1/4^\circ$ & $1/12^\circ$ \\
		\hline
		$\tilde{\Psi}^c_{\max}-\Psi_{AC_{\max}}^c$ & $75.9\%$    &$56.0\%$ & $95.8\%$\\
		$\tilde{\Psi}^c_{\max}$ &   $99.8\%$  & $97.9\%$ & $97.7\%$\\
		\hline
	\end{tabular}
	\caption[Role of additional cell contribution to the temporal variation of the maximum expected overturning streamfunction at the RAPID array, $26.5^\circ$N.]{Temporal variation and additional cell contribution to the maximum expected overturning streamfunction at the RAPID array ($26.5^\circ$N), calculated at the depth of maximum estimated overturning streamfunction: temporal variation of $\Psi^c_{\max}$ explained by $\tilde{\Psi}^c_{\max}-\Psi_{AC_{\max}}^c$ and $\tilde{\Psi}^c_{\max}$ calculated using Equation \ref{e_PrcVarExp} in the Appendix.}
	\label{Tab_SVRpd}
\end{table}

Whereas comparison of total overturning streamfunction estimates from RAPID observations and the decomposition diagnostic is possible, comparison of boundary contributions apart from the Ekman component $\Psi_{Ekm_{\max}}^c$ is not possible. At RAPID, the magnitude of the Ekman component has been observed to be of the order of $5$Sv (\citealt{Team2021}), whereas the decomposition diagnostic provides estimates of the order of $3$Sv, shown in Figure \ref{Rpd_ts}. Other observed values (\citealt{Team2021}) are provided for observed Gulf Stream ($31$Sv) and Upper-mid ocean (-$18$Sv) contributions. The upper mid-ocean transport per day, based on the RAPID mooring data, is the vertical integral of the transport per unit depth down to the deepest northward velocity (at approximately $1100$m).

Large additional cell $\Psi_{AC_{\max}}^c$ and depth-independent flow $\Psi_{bot_{\max}}^c$ contributions, due to strong currents through the Florida Strait, dominate the estimated maximum overturning streamfunction. In the real ocean these contributions to the overturning streamfunction at $26.5^\circ$N are mainly detected by telephone cable rather than mooring data. It is interesting to note a small negative $\Psi_{th_{\max}}^c$ contribution to the maximum estimated overturning streamfunction, suggesting a weak dependence of AMOC strength on geostrophic density gradients at RAPID.

\begin{figure*}[ht!]
	\centerline{\includegraphics[width=\textwidth]{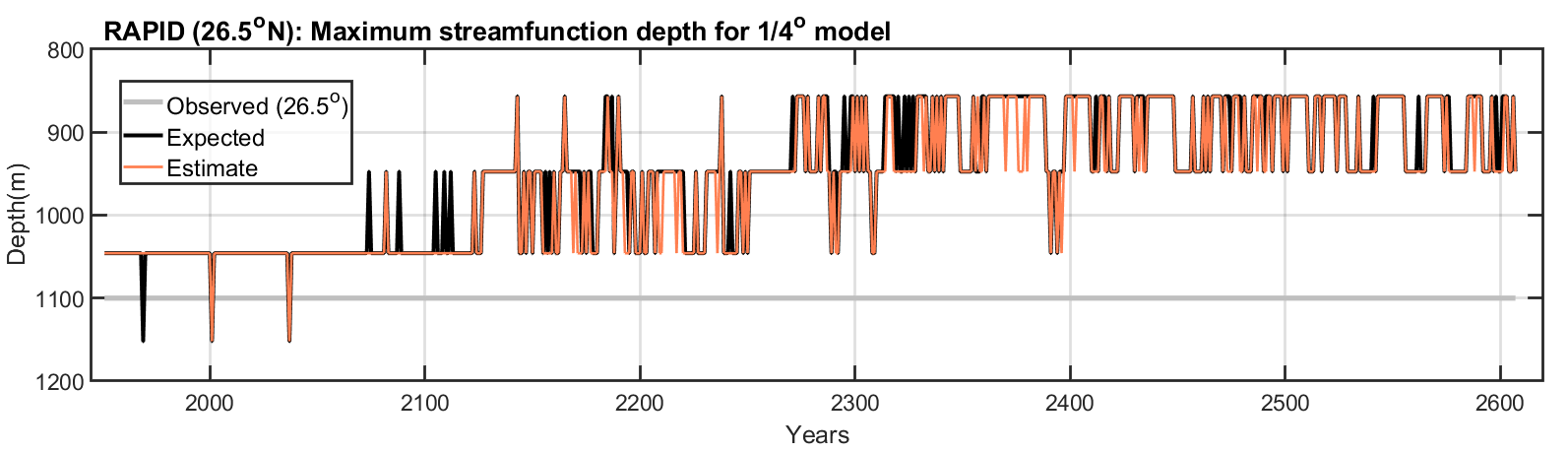}}
	\caption[Depth of maximum overturning streamfunction for $26.5^\circ$N at $1/4^\circ$ model resolution.]{Depth of maximum overturning streamfunction for $26.5^\circ$N at $1/4^\circ$ model resolution, using the volume-compensated total expected overturning streamfunction $\Psi^c_{\max}$ (black) and the volume-compensated estimated overturning streamfunction ($\tilde{\Psi}_{\max}^c$, orange). Value from RAPID observations relative to a level of no motion at $4820$dbar in grey.}
	\label{p_Rpd_ts_dpt}
\end{figure*}

The depth of the maximum overturning streamfunction becomes shallower in the $1/4^\circ$ model with time, as illustrated in Figure \ref{p_Rpd_ts_dpt}. The depth initially is around $\approx 1050$m, slightly shallower than that reported for the RAPID array. But, at the end of the $1/4^\circ$ model run, the depth of the maxima has reduced to only $\approx900$m. This is a shift of approximately 1 to 2 vertical cells, considering the gridded structure of the model. A shallower AMOC maximum streamfunction could indicate a weakening of the overturning circulation, with less dense return flows from high latitudes, or alternatively a less dense lower AMOC cell due to less dense AABW water, leading to a shallower upper AMOC cell. The expected maximum overturning streamfunction does show some signs of weakening through time, with weak trends found in the thermal wind and additional cell contributions. The reduction in thermal wind contribution to the maximum estimated overturning streamfunction would suggest either denser western boundary densities or lighter eastern boundary densities at depth, with the latter being most likely (see Figure \ref{p_LT_Bden} for trends in boundary densities).

\subsection{SAMBA} \label{S3_RS_Smb}
\begin{figure*}[ht!]
	\centerline{\includegraphics[width=\textwidth]{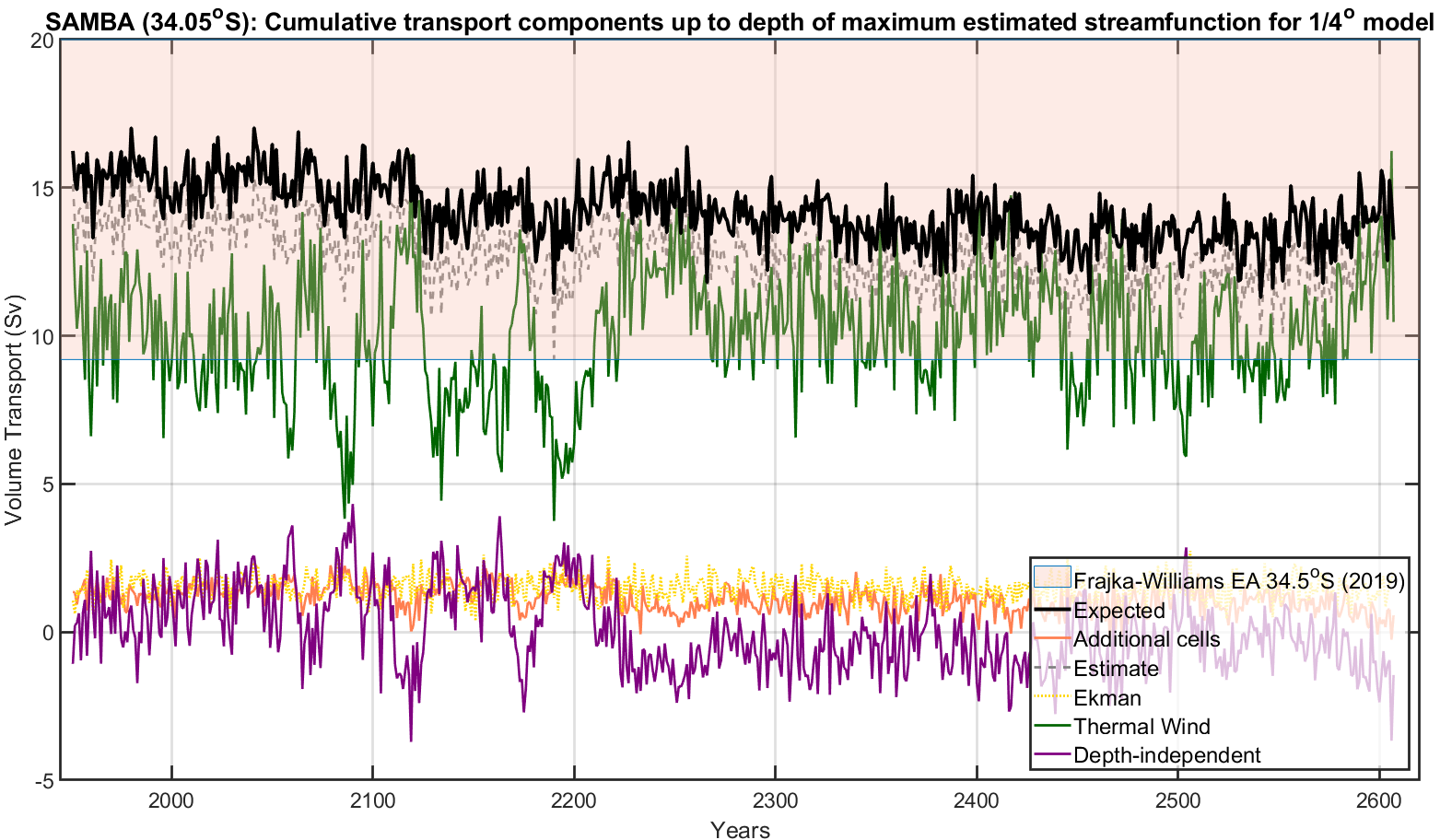}}
	\caption[Maximum overturning streamfunction timeseries for $34.05^\circ$S at $1/4^\circ$ model resolution, and contributing boundary components estimated using the decomposition diagnostic.]{Maximum overturning streamfunction timeseries for $34.05^\circ$S at $1/4^\circ$ model resolution, and contributing boundary components estimated using the decomposition diagnostic. Shaded red region indicates the $\pm$ one standard deviation interval for the SAMBA-observed maximum overturning streamfunction as quoted by \cite{Frajka-Williams2019}. The volume-compensated total expected maximum overturning streamfunction $\Psi^c_{\max}$ is shown in black. Other coloured lines indicate the volume-compensated estimated maximum streamfunction ($\tilde{\Psi}_{\max}^c$, dashed grey), depth-independent flow contribution ($\Psi_{bot_{\max}}^c$, purple), additional cell contribution ($\Psi_{AC_{\max}}^c$, orange), Ekman contribution ($\Psi_{Ekm_{\max}}^c$, yellow) and the thermal wind component ($\Psi_{W_{\max}}^c+\Psi_{E_{\max}}^c$, dark green).}
	\label{Smb_ts}
\end{figure*}

The SAMBA array (introduced in Section \ref{I_oAMOC}) is situated at approximately $34.5^\circ$S in the South Atlantic, to investigate the South Atlantic MOC, including features such as the eastern Aghulas leakage and the western Malvinas current. In contrast to RAPID, no telephone cable measurements are used; the monitoring array relies on density profiles, PIES (pressure inverted echo sounders) and CPIES (``sea'' PIES) providing baroclinic and barotropic estimates of the temporal variability of the overturning streamfunction. The use of PIES and CPIES at $1350$dbar renders the zero net flow assumption obsolete; no volume compensation is applied at SAMBA. However there is an issue with PIES and CPIES measurements due to sensor drift. Capturing the time-average barotropic (depth-independent) component is difficult and unreliable. As a work-around, a time-mean reference velocity is acquired from the OFES (Ocean For the Earth Simulator) model, which is also used to provide an estimate of the meridional volume transport in regions inshore of the $1,350$dbar isobath (see \citealt{Meinen2018} for more information). The monitoring array estimates a maximum overturning streamfunction of the order of $14.6 \pm 5.4$Sv, shown by the shaded pink region in Figure \ref{Smb_ts}. Similarly to RAPID, comparison of boundary components except for the Ekman component is not possible at SAMBA. Moreover, capturing the time-mean overturning streamfunction at SAMBA is difficult due to sensor drift. The focus of research at SAMBA tends therefore to be short-term variability, quantified in terms of transport anomalies as opposed to ``absolute'' values.

Applying the decomposition diagnostic requires the existence of both western and eastern boundaries. Due to variations in coastline locations at different model resolution we find the eastern boundary near South Africa is absent in the $1/4^\circ$ and $1^\circ$ model at latitude $34.5^\circ$S; therefore exact replication of the monitoring array is not possible. The closest possible latitudes for which the eastern boundary is present ($34.05^\circ$S and $33.84^\circ$S for both $1/4^\circ$ and $1^\circ$ resolutions respectively) are used, along with $34.53^\circ$S for the $1/12^\circ$ model.

Figure \ref{Smb_ts} indicates that the reconstructed estimate $\tilde{\Psi}_{\max}^c$ is comfortably within the one standard deviation limits for observations at SAMBA. The density component $\Psi_{th_{\max}}^c$ is considerably larger than any other contributor to  $\tilde{\Psi}_{\max}^c$. There is a strong negative relationship between the density component  $\Psi_{th_{\max}}^c$ of the maximum overturning streamfunction and the depth-independent flow $\Psi_{bot_{\max}}^c$ component, especially for years $100$ to $200$, indicating deviations from the average contributions expected, but having little to no effect upon the total estimated maximum overturning streamfunction. Some compensation between $\Psi_{bot_{\max}}^c$ and $\Psi_{th_{\max}}^c$ is to be expected due to the formulation of the decomposition diagnostic and the volume compensation applied (within the decomposition diagnostic) to ensure net zero flow. At this latitude, the contribution $\Psi_{AC_{\max}}^c$ of additional cells is smaller due to (a) the presence of fewer additional cells due to simpler bathymetry in comparison to the RAPID section (especially on the western boundary, no Florida Strait etc.), and (b) the presence of a weaker Malvinas current along the western boundary (compared with the Gulf Stream found near the RAPID array). The Ekman contribution $\Psi_{Ekm_{\max}}^c$ is stable around $1.5$Sv; this contrasts with estimates for $\Psi_{Ekm_{\max}}^c$ from \cite{Meinen2018} which show greater variation. This difference may be due to the higher resolution temporal data used in \cite{Meinen2018} compared with the annual mean data used here. A general weakening of both the maximum estimated and expected overturning streamfunction is observed in time, in conjunction with a shallowing of the maximum, especially in the first 150 years (not shown). It is not clear to which component this weakening might be attributed; Table \ref{Tab_CrSmb} in Appendix \ref{App_SmbRap}, shows correlations between boundary components and the total estimate, suggesting that $\Psi_{th_{\max}}^c$ is a candidate. Interestingly, at both SAMBA and RAPID, we find an increase in maximum expected and estimated overturning streamfunction in the final 50 years of this 657 year run, with $\Psi_{th_{\max}}^c$ showing similar tendencies.

Similarly to the RAPID timeseries (Table \ref{Tab_SVRpd}), the total estimated maximum streamfunction $\tilde{\Psi}^c_{\max}$ for SAMBA captures the temporal variation of the expected maximum overturning streamfunction $\Psi^c_{\max}$ well (Appendix \ref{App_SmbRap}, Table \ref{Tab_SVSmb}). Generally, we find the percentage variance explained is higher than for the RAPID timeseries, except for particularly poor results found for the $1/4^\circ$ simulation.

\subsection{Further analysis at SAMBA} \label{S3_RS_SmbMore}

Closer investigation of Figure \ref{Smb_ts} shows high negative correlation between $\Psi^c_{th_{\max}}$ and $\Psi^c_{bot_{\max}}$ evaluated at the depth of maximum estimated overturning streamfunction $\tilde{\Psi}_{\max}^c$. For the $1/4^\circ$ model this relationship is particularly evident for the period $2060$ to $2240$, where large oscillations are present in both components. These oscillations are found to be present for a number of latitudes up to $30^\circ$S. Investigating correlations of all the components at the maximum streamfunction depth (Appendix \ref{App_SmbRap} Table \ref{Tab_CrSmb}) for all three resolutions shows that  this relationship exists in both the $1/4^\circ$ and $1/12^\circ$ models. In this section, we investigate the relationship between $\Psi^c_{th_{\max}}$ and $\Psi^c_{bot_{\max}}$ for the $1/4^\circ$ model further. 

\subsubsection{Maxima and minima of $\Psi^c_{bot_{\max}}$}
\begin{figure*}[ht!]
	\centerline{\includegraphics[width=\textwidth]{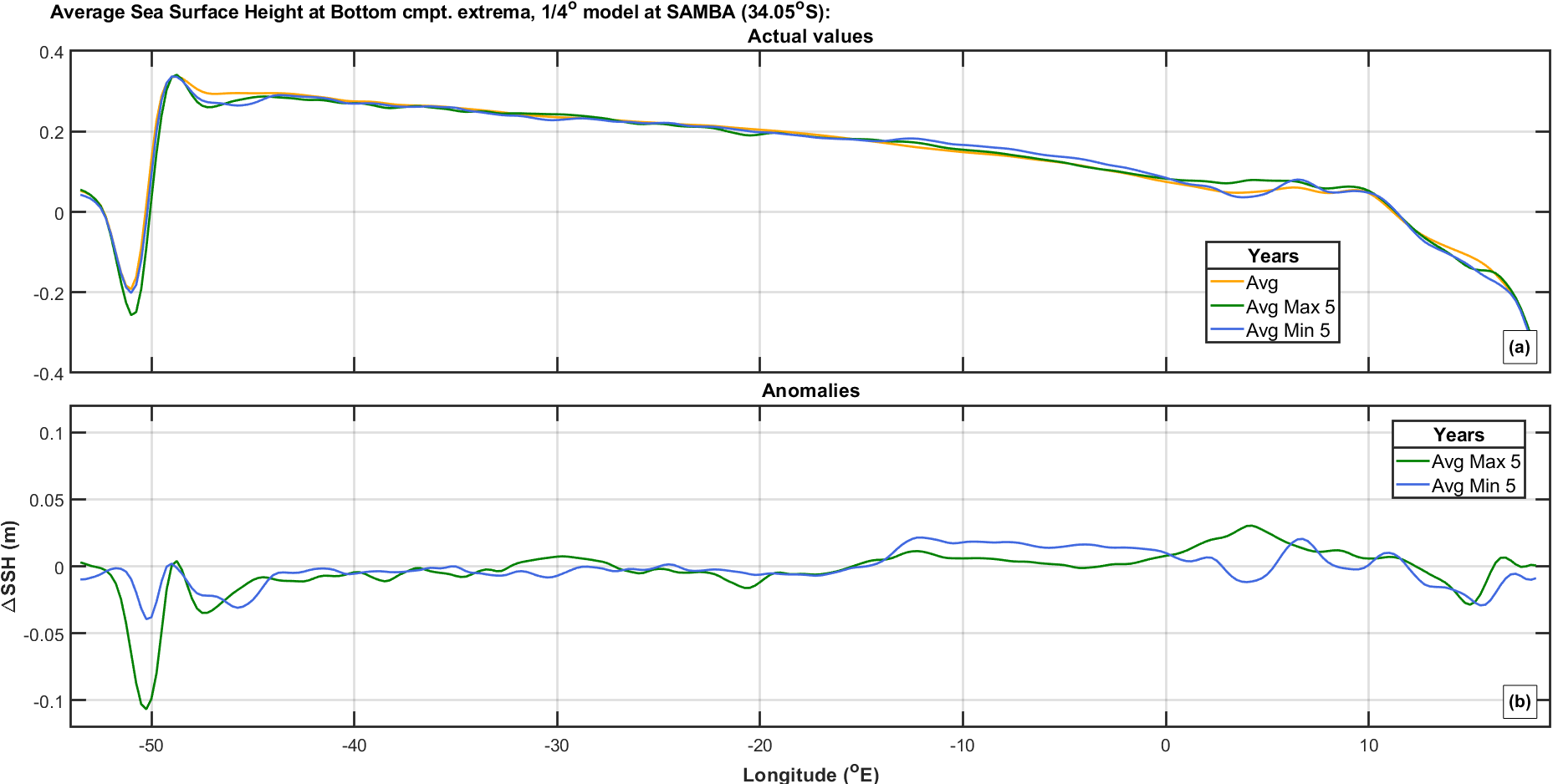}}
	\caption[Average sea surface height over the $5$ years yielding local maximum and minimum of the $\Psi^c_{bot_{\max}}$ contribution to the maximum estimated streamfunction at SAMBA, with longitude.]{Sea surface height for years of maximum and minimum $\Psi^c_{bot_{\max}}$. Panel (a): average SSH over the $5$ years yielding local maximum (green) and minimum (blue) of the $\Psi^c_{bot_{\max}}$ contribution to the maximum overturning streamfunction, plotted against longitude along $34.05^\circ$S. The average of SSH over the entire simulations is also shown in orange. Panel (b) shows corresponding anomalies (i.e. differences from the temporal average over entire simulation) with longitude.}
	\label{p_MxMn_SSH}
\end{figure*}

\begin{figure*}[ht!]
	\centerline{\includegraphics[width=\textwidth]{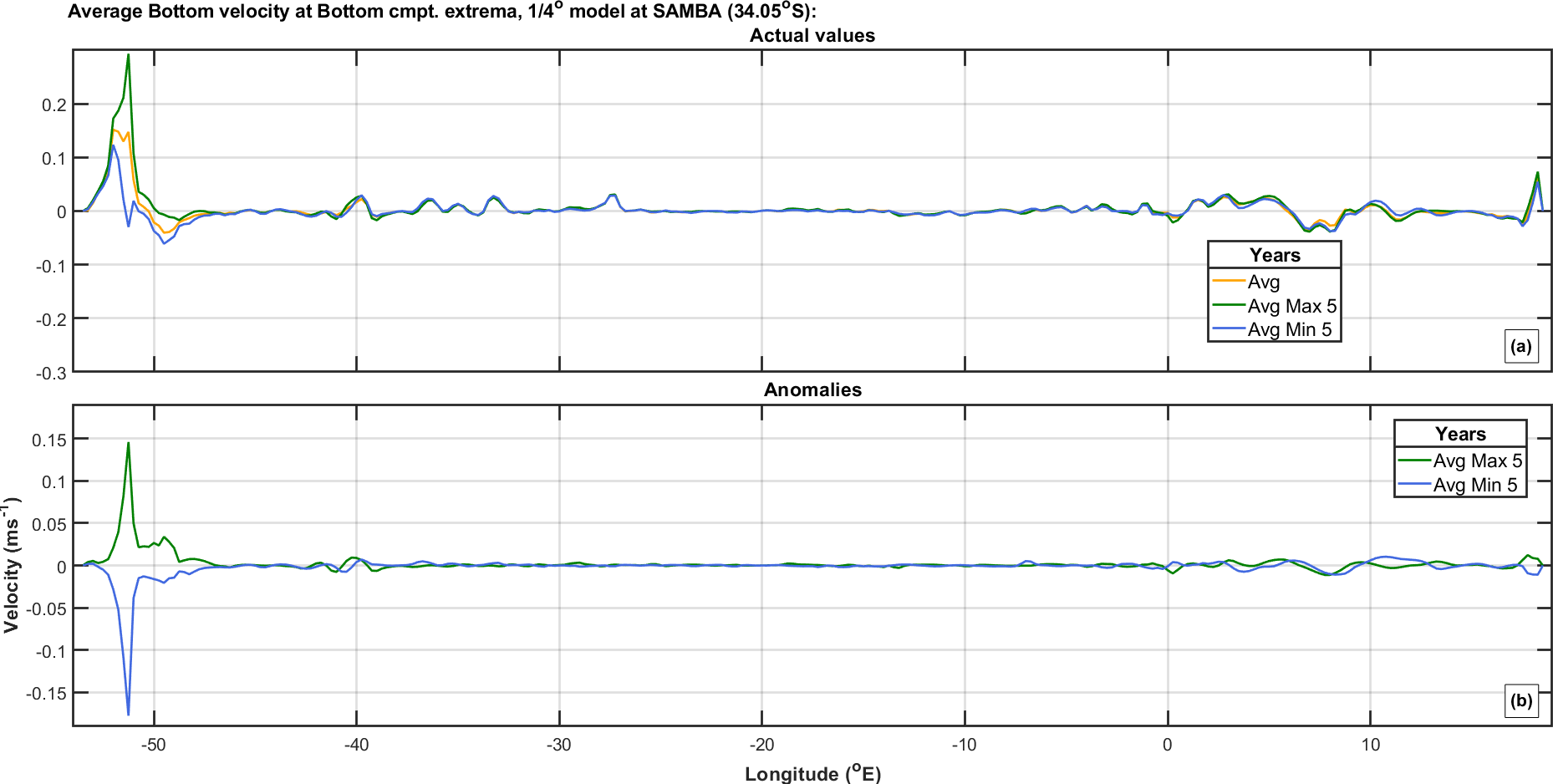}}
	\caption[Average bottom velocities over the $5$ years yielding local maximum and minimum of the $\Psi^c_{bot_{\max}}$ contribution to the maximum overturning streamfunction at SAMBA, plotted as a function of longitude at $34.05^\circ$S.]{Bottom velocities for years of maximum and minimum $\Psi^c_{bot_{\max}}$. Panel (a): average bottom velocities over the $5$ years yielding local maximum (green) and minimum (blue) of the $\Psi^c_{bot_{\max}}$ contribution to the maximum overturning streamfunction, plotted as a function of  longitude at $34.05^\circ$S. The overall temporal average of bottom velocities is also shown in orange. Panel (b) shows corresponding anomalies (i.e. differences from the temporal average).}
	\label{p_MxMn_Bv}
\end{figure*}

First, we isolate the years for which the $\Psi^c_{bot_{\max}}$ contribution (see Figure \ref{Smb_ts}) is a local temporal maximum or minimum. Then, we isolate the sea surface height (SSH), bottom densities and bottom velocities for each longitude of the SAMBA section corresponding to these years, and calculate average values for years of $\Psi^c_{bot_{\max}}$ maxima and minima. Figures \ref{p_MxMn_SSH} and \ref{p_MxMn_Bv} show the resulting composite SSH and bottom velocity properties with respect to longitude, for latitude $34.05^\circ$S.

Panel (a) of Figures \ref{p_MxMn_SSH} and \ref{p_MxMn_Bv} shows the average SSH and bottom velocity (orange) over the model run, and the average SSH and bottom velocity corresponding to the 5 years with local maxima (green) and minima (blue) of the $\Psi^c_{bot_{\max}}$ contribution to the maximum estimated overturning streamfunction. Panel (b) shows corresponding anomalies relative to the average over the whole simulation (orange in Panel (a)).

Figure \ref{p_MxMn_SSH}(a) indicates a large change in SSH with longitude near $50^\circ$W, in the vicinity of the Brazil-Malvinas Confluence (BMC), where the colder northward Malvinas current (centred at $52^\circ$W) meets the warmer southward Brazil current (centred at $49^\circ$W). Figure \ref{p_MxMn_SSH}(b) emphasises the decrease in SSH anomaly for $\Psi^c_{bot_{\max}}$ maxima years near $51^\circ$W, associated with colder denser waters. Variation in the SSH anomaly for $\Psi^c_{bot_{\max}}$ minima years (blue) in the same region is also present, but is not of the same magnitude. 

Figure \ref{p_MxMn_Bv}(a) indicates larger bottom velocities between $48^\circ$W and $54^\circ$W, again in the region of the BMC. Figure \ref{p_MxMn_Bv}(b) emphasises the increase in bottom velocity anomaly for $\Psi^c_{bot_{\max}}$ maxima years for these longitudes, and a corresponding decrease in bottom velocity anomaly for $\Psi^c_{bot_{\max}}$ minima years (blue) of approximately the same magnitude. 

During years of strong $\Psi^c_{bot_{\max}}$ contribution, Figure \ref{p_MxMn_Bv} suggests a stronger northward Malvinas current (centred at $52^\circ$W) but a weaker southward Brazil current (centred at $49^\circ$W). In contrast, in years of weak $\Psi^c_{bot_{\max}}$, a weaker northward Malvinas current and stronger southward Brazil current are observed.

Figures (not shown) for bottom density, analogous to Figures \ref{p_MxMn_SSH} and \ref{p_MxMn_Bv}, reveal anomalously light densities near $53^\circ$W for $\Psi^c_{bot_{\max}}$ minima years, but heavier densities at longitudes in the region of $53^\circ$W for $\Psi^c_{bot_{\max}}$ maxima years. The most significant features for each of SSH, bottom velocity and bottom density are found at longitudes corresponding to the Brazilian continental shelf, with little difference over the rest of the section.

Figure \ref{Smb_ts} indicates that in the time period 2060 to 2240, there are clear intervals corresponding to systematic weakening and strengthening of the bottom component $\Psi^c_{bot_{\max}}$ (purple) contributing to the maximum estimated overturning streamfunction. In the same period, the opposite temporal trends are observed in the thermal wind component $\Psi^c_{th_{\max}}$ (green). To further investigate the relationship between the contributions of $\Psi^c_{bot_{\max}}$ and $\Psi^c_{th_{\max}}$ to the maximum estimated overturning streamfunction at SAMBA, we consider three time intervals, $2065$ to $2090$, $2090$ to $2119$ and $2163$ to $2175$. For each of the three time intervals, we investigate the changes in SSH, bottom velocity and bottom density in terms of (a) longitude-time sections of anomalies, and (b) temporal trend in anomaly as a function of longitude. 

\subsubsection{Analysis of time intervals $2065-2090$, $2090-2119$ and $2163-2175$}

\begin{figure*}[ht!]
	\centerline{\includegraphics[width=\textwidth]{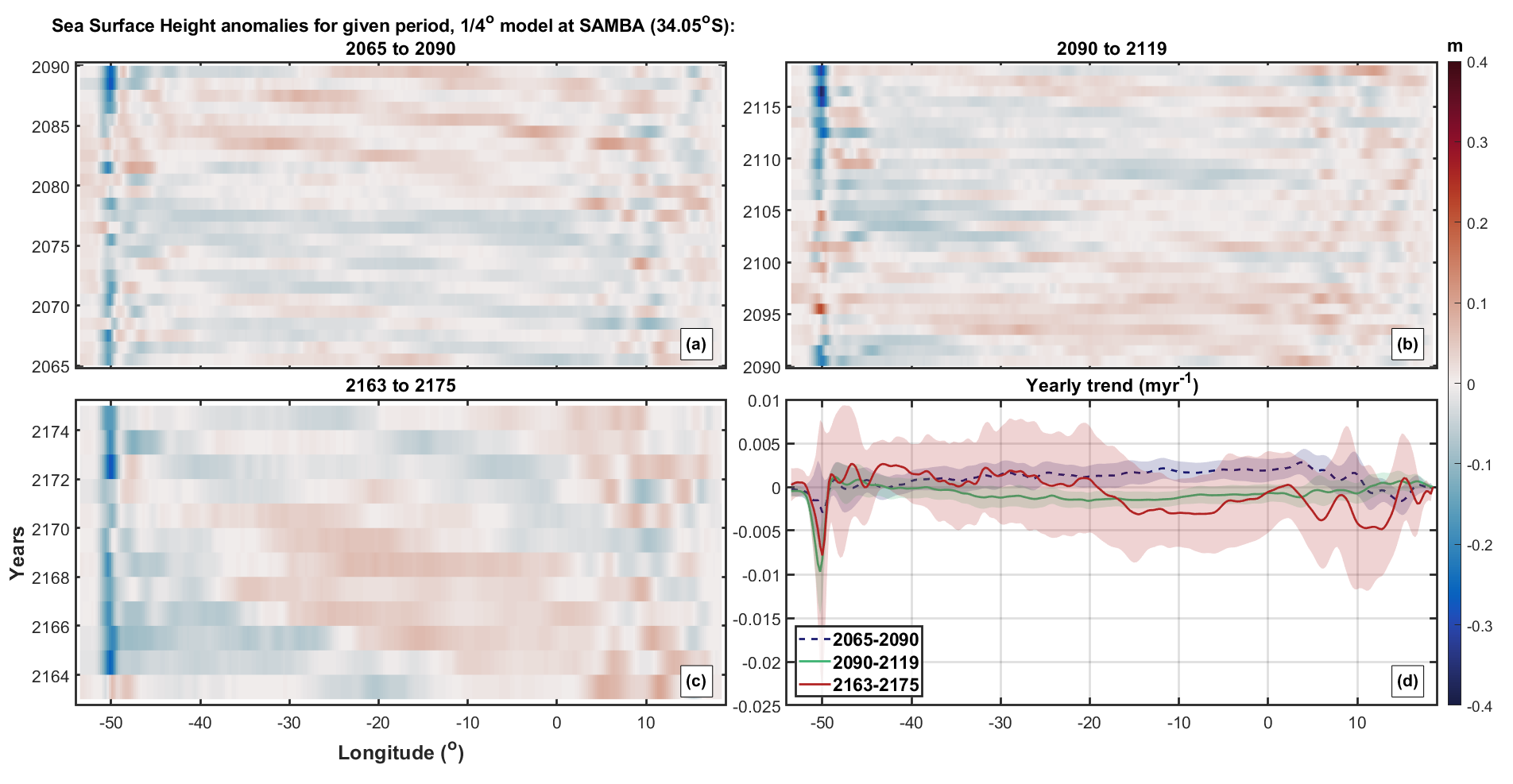}}
	\caption[Sea surface height anomalies (m) with time and longitude for three selected time intervals at SAMBA array for $1/4^\circ$ model.]{Panels (a)-(c) show sea surface height anomalies (m) with time and longitude for three selected time intervals (a) $2065-2090$, (b) $2090-2119$ and (c) $2163-2175$. Panel (d) shows the trend in SSH at each longitude for each of the three time intervals, $2065-2090$ (dashed blue), $2090-2119$ (green) and $2163-2175$ (red). Dashed lines in Panel (d) indicate a time interval corresponding to increasing $\Psi^c_{bot_{\max}}$, and solid lines time intervals of decreasing $\Psi^c_{bot_{\max}}$. Shaded regions indicate $95$\% confidence intervals for the estimated trend.}
	\label{p_Anl_SSH}
\end{figure*}

\begin{figure*}[ht!]
	\centerline{\includegraphics[width=\textwidth]{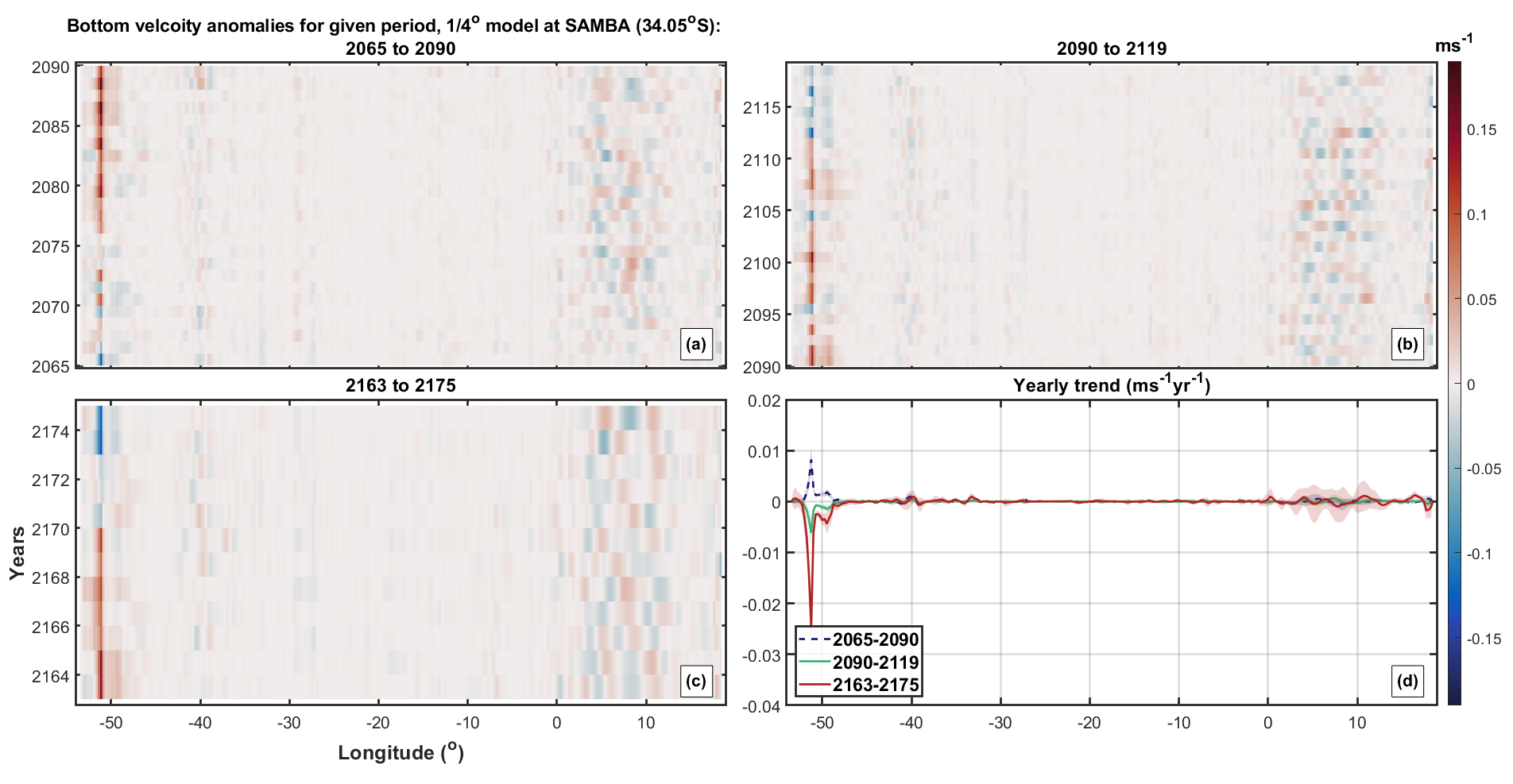}}
	\caption[Meridional bottom velocity anomalies (ms$^\text{-1}$) with time and longitude for three selected time intervals at SAMBA array for $1/4^\circ$ model.]{Panels (a)-(c) show meridional bottom velocity anomalies (ms$^\text{-1}$) with time and longitude for three selected time intervals (a) $2065-2090$, (b) $2090-2119$ and (c) $2163-2175$. Panel (d) shows the trend in bottom velocity at each longitude for each of the three time intervals, $2065-2090$ (dashed blue), $2090-2119$ (green) and $2163-2175$ (red). Dashed lines in Panel (d) indicate a time interval corresponding to increasing $\Psi^c_{bot_{\max}}$, and solid lines time intervals of decreasing $\Psi^c_{bot_{\max}}$. Shaded regions indicate $95$\% confidence intervals for the estimated trend.}
	\label{p_Anl_Bv}
\end{figure*}

\begin{figure*}[ht!]
	\centerline{\includegraphics[width=\textwidth]{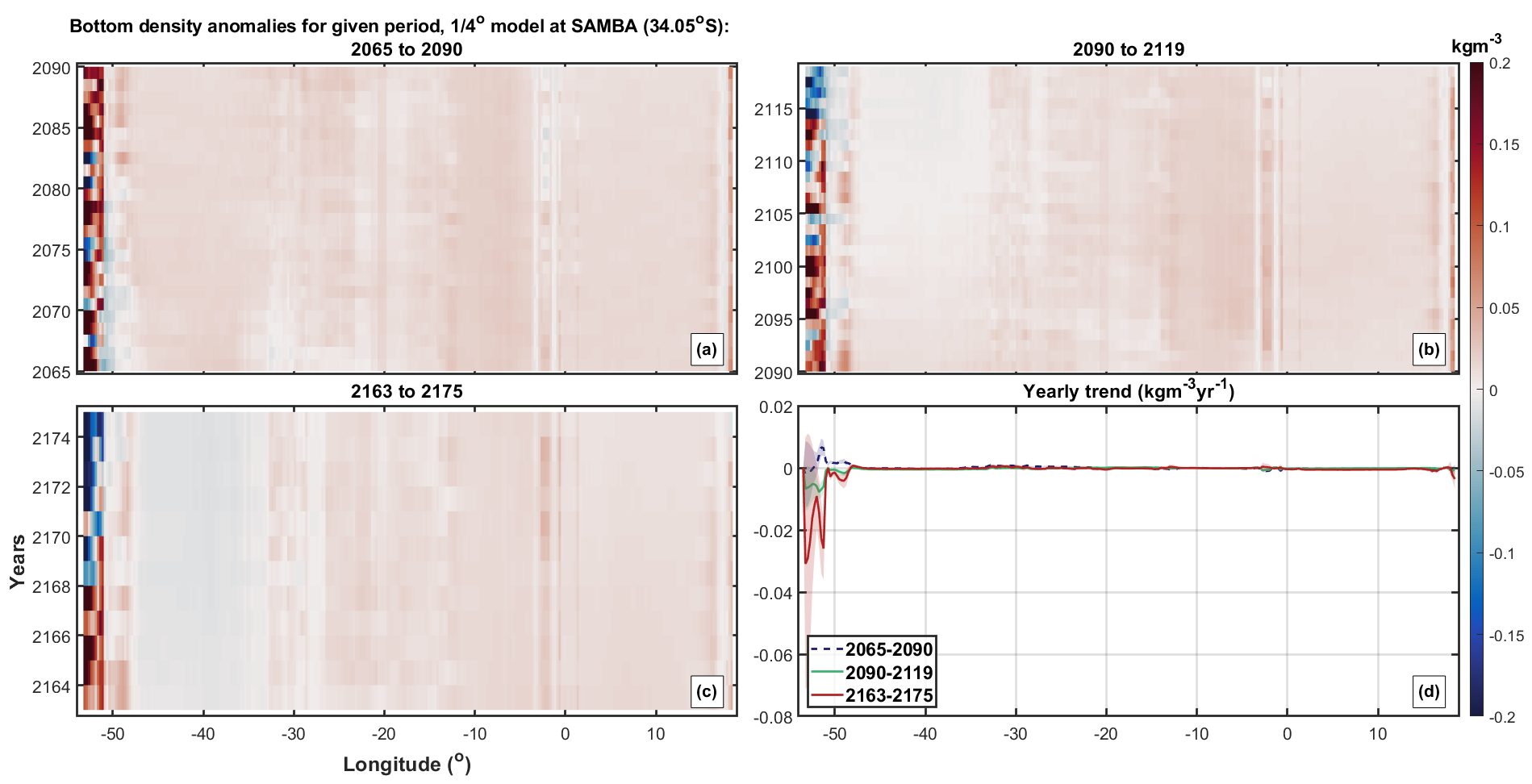}}
	\caption[Bottom density anomalies (kgm$^\text{-3}$) with time and longitude for three selected time intervals at SAMBA array for $1/4^\circ$ model.]{Panels (a)-(c) show bottom density anomalies (kgm$^\text{-3}$) with time and longitude for three selected time intervals (a) $2065-2090$, (b) $2090-2119$ and (c) $2163-2175$. Panel (d) shows the trend in bottom density at each longitude for each of the three time intervals, $2065-2090$ (dashed blue), $2090-2119$ (green) and $2163-2175$ (red). Dashed lines in Panel (d) indicate a time interval corresponding to increasing $\Psi^c_{bot_{\max}}$, and solid lines time intervals of decreasing $\Psi^c_{bot_{\max}}$. Shaded regions indicate $95$\% confidence intervals for the estimated trend.}
	\label{p_Anl_Bdens}
\end{figure*}

Panels (a)-(c) of Figures \ref{p_Anl_SSH}, \ref{p_Anl_Bv}, and \ref{p_Anl_Bdens} show the anomalies in SSH, bottom velocity and bottom densities with respect to time and longitude for each of the three time periods $2065$ to $2090$, $2090$ to $2119$ and $2163$ to $2175$. For each quantity of interest, the anomaly is calculated by subtracting the temporal mean (over all years of model run) from the value of the quantity in a given year. In each figure, Panel (a) refers to a period where $\Psi^c_{bot_{\max}}$ increases from a minimum to a maximum, whereas both Panels (b) and (c) start at a maximum of $\Psi^c_{bot_{\max}}$ and end at a minimum.
 
In Figure \ref{p_Anl_SSH}(a)-(c), anomalies are present for all longitudes and years but the largest magnitude anomalies appear along the western side of the basin. Otherwise, there are regions of each section which suggest local persistent positive (or negative) anomalies with time and longitude, e.g. the region of positive anomaly centred near year 2168 and longitude $20^\circ$W in Figure \ref{p_Anl_SSH}(c). There appears to be westward propagation of SSH anomalies in this region, possibly attributed to Rossby waves. Panel (d) shows large negative trends in SSH near $51^\circ$W for time intervals $2090$ to $2119$ and $2163$ to $2175$ respectively. The uncertainty in the trend for time interval $2163$ to $2175$ (red) is relatively large, due to the relatively short length of the time interval.

Figure \ref{p_Anl_Bv}(a)-(c) presents corresponding anomalies in bottom velocities. In contrast to SSH, bottom velocity anomalies are concentrated in two bands of longitudes, around 51$^\circ$W and 8$^\circ$E . Despite large anomalies in the eastern section, Panel (d) shows no clear trends here. In contrast, the western part of the section shows a significant trend in anomalies in a narrow band around 51$^\circ$W, positive for the time interval in Panel (a), and negative for the other two time intervals. For the latter time intervals, this suggests a weakening of bottom velocities in conjunction with the decreasing $\Psi^c_{bot_{\max}}$ contribution to the maximum estimated streamfunction. For the first time interval (in Panel (a)) we observe strengthening of bottom velocities with increasing $\Psi^c_{bot_{\max}}$.

Finally, Figure \ref{p_Anl_Bdens} shows anomalies in bottom densities. Anomalies with large magnitudes are concentrated at westernmost longitudes, in a band west of approximately 50$^\circ$W. It is interesting that this band is concentrated further onshore for the western coastline relative to bottom velocity and SSH anomalies. Panel (d) indicates a significant positive temporal trend in bottom density for the first time interval of increasing $\Psi^c_{bot_{\max}}$; i.e. bottom water is becoming denser. In contrast, Panel (d) also suggests generally significant negative trends for the other time intervals (of decreasing $\Psi^c_{bot_{\max}}$), corresponding to lighter bottom waters. The lightening of western boundary bottom densities through time (e.g. red colour in Panel (c)) correspond to an increasing west-east density gradient and therefore, an increase in $\Psi^c_{th_{\max}}$ contribution.

Summarising, changes in the $\Psi^c_{bot_{\max}}$ contribution to the maximum overturning streamfunction occur predominantly in the western part of the section. Investigation of three time intervals of significant change in $\Psi^c_{bot_{\max}}$ suggests a weakening of $\Psi^c_{bot_{\max}}$ is coupled to variability of the Brazil-Malvinas Confluence (BMC). We find (e.g. Figure \ref{p_MxMn_Bv}(a)) during years of strong $\Psi^c_{bot_{\max}}$, a strong northward Malvinas current and a weaker southward Brazil current; in years of weaker $\Psi^c_{bot_{\max}}$, a weaker northward Malvinas current and stronger Brazil current is present. Prominent features in the longitude-time sections of each of SSH, bottom velocity and bottom density near the Brazilian continental shelf suggest denser waters are present during years of strong $\Psi^c_{bot_{\max}}$, possibly due to a stronger Malvinas current drawing colder ACC waters northward. 

\subsubsection{Brazil-Malvinas Confluence (BMC)}
\label{S_BMC}

We hypothesise that a driving factor in the fluctuations found in the $\Psi^c_{bot_{\max}}$ contribution to the estimated maximum overturning streamfunction $\tilde{\Psi}^c_{\max}$ at SAMBA in this model is the BMC, and in particular changes in the relative strengths of the northward Malvinas and southward Brazil currents. In this section we investigate the location of the BMC with respect to latitude. Previous work (\citealt{Combes2014}) has shown that the location of the BMC is near $38^\circ$S, and that this location exhibits strong seasonal and inter-annual variability of some $1^\circ$ to $6^\circ$ latitude (\citealt{Goni2011}).

\cite{Goni2011} used satellite observations of SSH and SST to locate the latitude of the BMC. The latitude of maximum gradient of SSH or SST along the 1000m isobath is considered to identify the location of the BMC. Both SSH and SST data show comparable results for the latitude of the BMC.

\begin{figure*}[ht!]
	\centerline{\includegraphics[width=\textwidth]{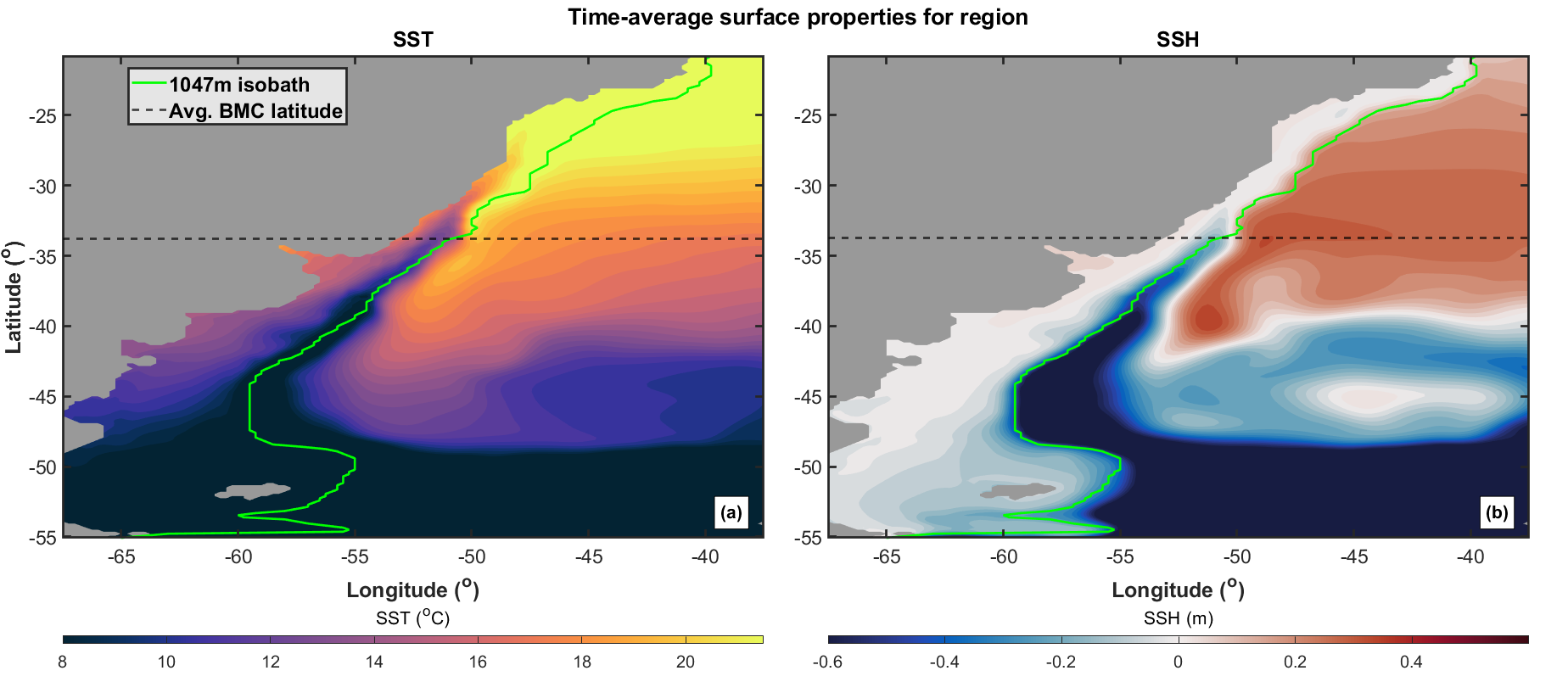}}
	\caption[Time-average sea surface temperature and sea surface height above the geoid for the western South Atlantic in the $1/4^\circ$ model.]{Time-average sea surface temperature (SST, Panel (a)) and sea surface height (SSH, Panel (b)) above the geoid for the western South Atlantic. Time-average taken over entire $1/4^\circ$ model control run. The green line denotes the location of the $1047$m isobath, while dashed black lines show the average latitude of the BMC using the corresponding SST or SSH data. The average BMC latitude from SST data is $33.76^\circ$S and $33.74^\circ$S from SSH.}
	\label{p_Avg_SS}
\end{figure*}

Panels (a) and (b) of Figure \ref{p_Avg_SS} show the time-average SST and SSH taken over the entire $1/4^\circ$ model control run, respectively. The green line denotes the location of the 1047m isobath, the closest isobath depth to 1000m within the model configuration. The dashed black lines indicate the time-average latitude of the BMC calculated using the methodology of \cite{Goni2011}, based on SST (Panel (a)) and SSH (Panel (b)). The latitude of the BMC is near $34^\circ$S when estimated from model SSH and SST, almost $4^\circ$ further north than the expected location based on observations (\citealt{Combes2014}). We also note that seasonality alone cannot be a driver, since we are examining annual average data. We find the BMC latitude to be near $38^\circ$S in the $1^\circ$ model.  

This rather surprising estimate for the location of BMC in the $1/4^\circ$ model warrants further investigation. Figure \ref{p_Avg_SS} suggests that a combination of strong Malvinas current and weak Brazil current is present in the model from both time-average SST and SSH data. Therefore, the cold water of the Malvinas current penetrates further north, resulting in a more northerly BMC than expected within the $1/4^\circ$ model.

\begin{figure*}[ht!]
	\centerline{\includegraphics[width=12cm]{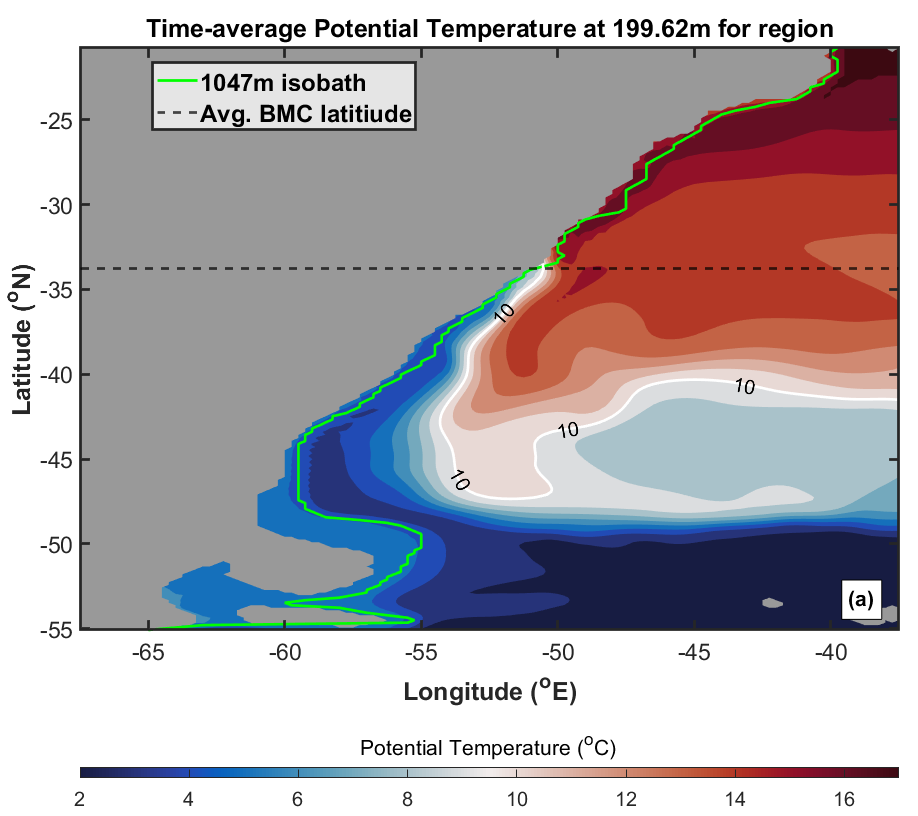}}
	\caption[Time-average potential temperature at 200m for $1/4^\circ$ model, and resulting BMC latitude.]{Time-average potential temperature ($\theta$) at 200m. The green line denotes the 1047m isobath and the white line shows the $10^\circ$C isotherm. Time-average taken over entire $1/4^\circ$ model control run. Using the approach of \cite{Garzoli1987}, the location of BMC is denoted by the latitude of intersection of the isotherm and isobath. The average latitude of the BMC is denoted by the black dashed line is $33.79^\circ$S.}
	\label{p_Avg_PT200}
\end{figure*}

Prior to the satellite-based analysis of \cite{Goni2011}, \cite{Garzoli1987} determined the location of the BMC as the latitude where the $10^\circ$C isotherm at 200m intersects with the 1000m isobath. To validate our estimates for BMC location from SSH and SST data, we also calculate the BMC latitude using this approach. The 10$^\circ$C isotherm is shown in Figure \ref{p_Avg_PT200} as a white line on the potential temperature ($\theta$) plot at depth 200m. The 1047m isobath is shown as a green line. The time-average latitude of the BMC over the entire $1/4^\circ$ model run is then shown by the black dashed line.

\begin{figure*}[ht!]
	\centerline{\includegraphics[width=\textwidth]{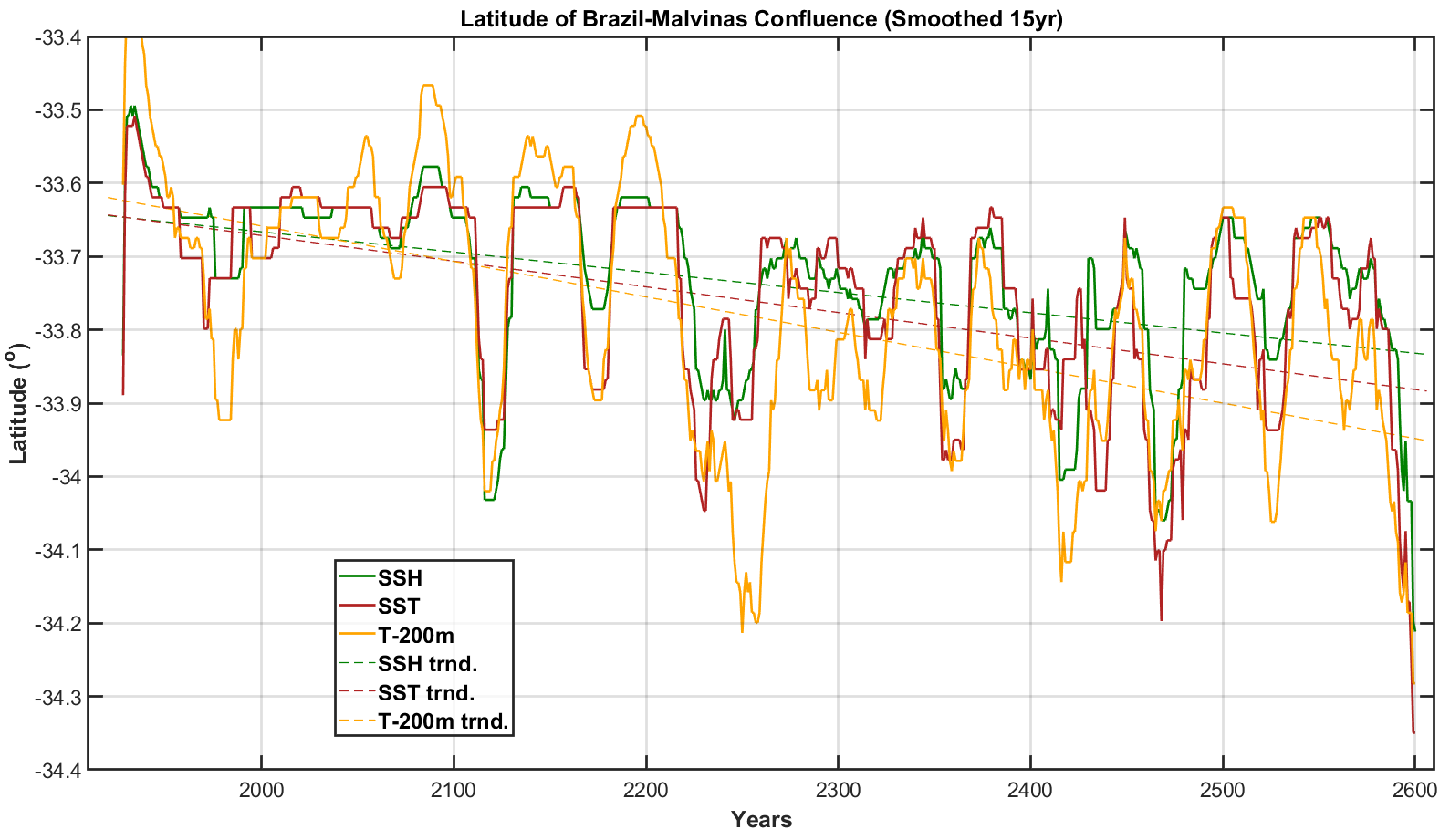}}
	\caption[Timeseries of Brazil-Malvinas confluence latitude calculated using SSH, SST and $10^\circ$C $\theta$ isotherm for the $1/4^\circ$ model.]{Timeseries of Brazil-Malvinas confluence latitude calculated using SSH (green), SST(red) and $10^\circ$C isotherm (orange) for the $1/4^\circ$ model. Coloured dashed lines represent the linear trend shown in each respective timeseries.}
	\label{p_mgr_3}
\end{figure*}

Figure \ref{p_mgr_3} shows the smoothed timeseries of the BMC latitude throughout the $1/4^\circ$ model for all three BMC latitude calculations, using SSH (green), SST(red) and the $10^\circ$C isotherm (orange). The spin-up phase of the control run is included in the figure, between 1920 and 1950. A $\pm7$ year running mean is used to remove short-term variability. All three methods used show excellent agreement for the location of the BMC, and moreover show similar temporal variability throughout, with slight differences in the extent of excursions. Further, all estimates show a southward trend in the location of the confluence in time, with corresponding linear fits shown by the coloured dashed lines. 

\begin{figure*}[ht!]
	\centerline{\includegraphics[width=\textwidth]{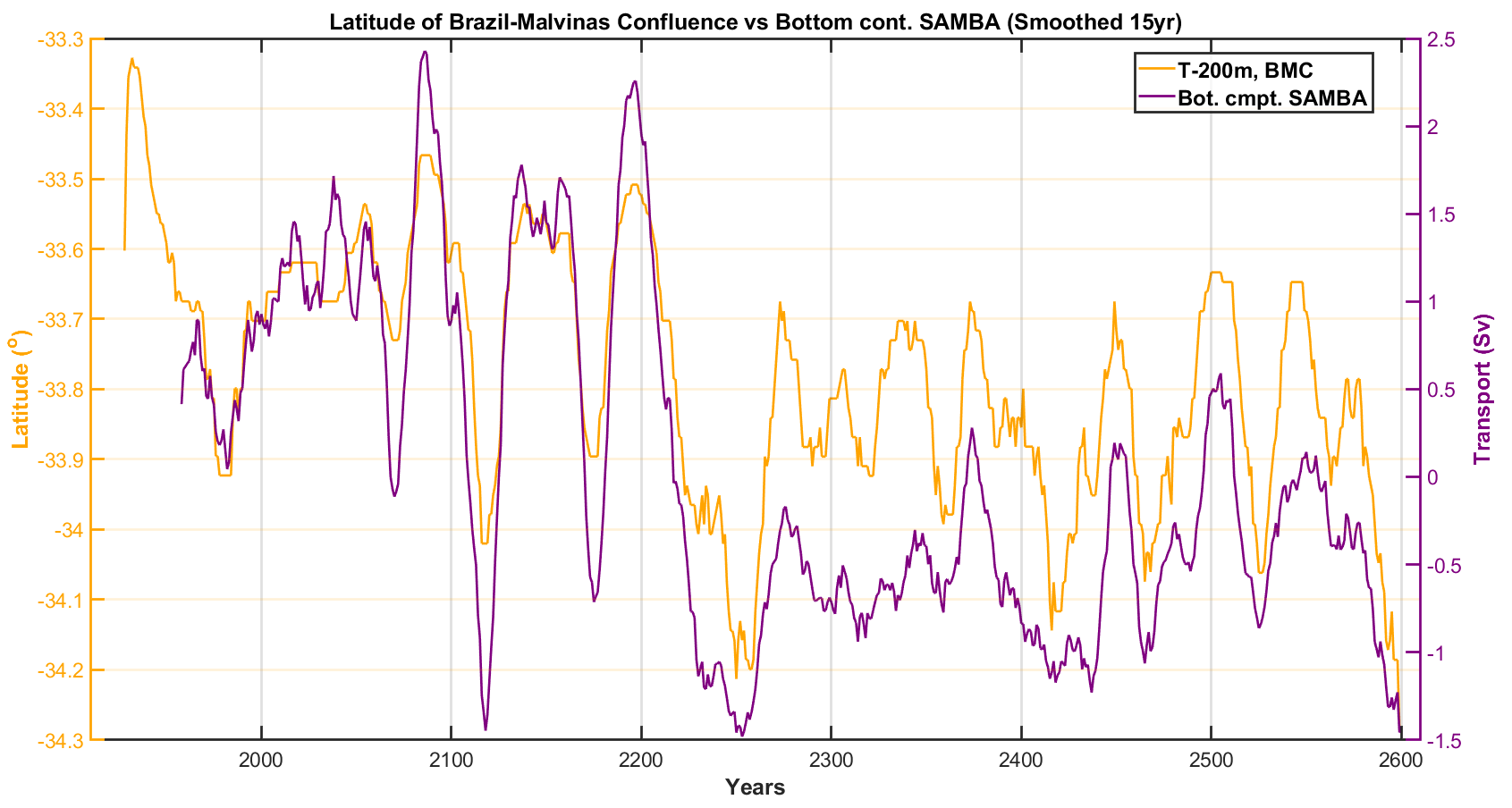}}
	\caption[Comparison of $\Psi^c_{bot_{\max}}$ contribution to the maximum estimated overturning streamfunction and location of Brazil Malvinas confluence.]{Timeseries of the $\Psi^c_{bot_{\max}}$ contribution (purple) to the maximum estimated overturning streamfunction  $\tilde{\Psi}_{\max}^c$ at SAMBA ($34.05^\circ$S) on the right hand y-axis, and the latitude of the Brazil-Malvinas confluence (orange) on the left hand axis. The $R^2$ value comparing the timeseries is 0.66.}
	\label{p_Tgr_Btt}
\end{figure*}

Figure \ref{p_Tgr_Btt} compares timeseries of the $\Psi^c_{bot_{\max}}$ (depth-independent, bottom) contribution (purple) to the maximum estimated overturning streamfunction $\tilde{\Psi}_{\max}^c$ at SAMBA and the latitudinal fluctuations of the BMC (orange) using the \cite{Garzoli1987} methodology. During periods of a northward (southward) shifting BMC, the figure shows increased (decreased) $\Psi^c_{bot_{\max}}$. A similar comparison between $\Psi^c_{th_{\max}}$ at SAMBA and the latitudinal fluctuations of the BMC shows strong  negative correlation (not shown). This supports the inference that fluctuations found in $\Psi^c_{bot_{\max}}$ and $\Psi^c_{th_{\max}}$ in Figure \ref{Smb_ts} due to changes in bottom velocities and densities are strongly associated with the corresponding fluctuations in the latitude of the Brazil-Malvinas confluence. Preliminary investigations of the relationship between temporal variation of BMC latitude with the meridional wind stress, zonal wind stress and curl of the wind stress did not indicate strong trends. 

\section{Standard deviation of bottom properties}

We now turn our attention to the spatial and temporal variability of boundary properties relevant to the decomposition diagnostic, for the whole Atlantic basin. 

\subsection{Decadal and long term standard deviation (sd.)} \label{Std:Thr}

The standard deviation of a variable is one measure of the variation of the variable about its mean. It is a useful way of quantifying whether, for example, the basin-wide mean overturning streamfunction over a $p$-year period (e.g. $p=10$ for decadal mean overturning streamfunction) at some location changes over the full period of data available (itself dependent on model resolution). For a sample $\{x_i\}_{i=1}^n$ of n values of random variable $X$, the sample standard deviation used here is defined by
\begin{equation}
	sd\left[\left\{x_i\right\}_{i=1}^n\right] = \sqrt{\frac{1}{n-1}\sum_{i=1}^{n}\left(x_i - \overline{x}\right)^2}.
\end{equation}
with sample mean $\overline{x}=(1/n)\sum_{i=1}^n x_i$.

Using the $1/4^\circ$ model dataset as an example, we have access to annual data for the overturning streamfunction $\Psi^c(t_{\ell},y_j,z_k)$, and each of its contributing boundary components. In general, $\Psi^c$ is a function of time $t_{\ell}$ ($\ell = 1,...,n_t = 657$), latitude $y_j$ ($j=1,...,n_y = 491$) and depth $z_k$ ($k=1,...,n_z = 75$), but the focus here is on temporal characteristics, so we write $\Psi(t_{\ell})$ only, and perform separate analyses for each individual basin latitude and depth.

The $p$-year average of $\Psi^c$, written $A_p$, is estimated and its variability examined using the standard deviation $sd$. The $p$-year average of $\Psi^c$ estimated using years $h$ to $p-1+h$ is given by
\begin{equation}
	A_{ph} = \frac{1}{p} \sum_{\ell=h}^{p-1+h}\Psi^c_{\ell}
\end{equation} 
for $h=1,2,...,(n_t+1-p)$ where $\Psi^c_{\ell}=\Psi^c(t_{\ell})$. The standard deviation for $A_p$ over all $n_t$ years, for initial year $q$ of calculation, is found using
\begin{equation}
	s_{pq} = sd\left[ A_{pq}, A_{p (q+p)}, A_{p (q+2p)}, ... ,  A_{p (q + \lfloor \frac{n_t+1-p}{p}\rfloor p)} \right]
\end{equation}
for $q =1,2,...,p$, the $\lfloor\bullet\rfloor$ ``floor'' symbol indicates the largest integer less than or equal to $\frac{n_t+1-p}{p}$. 

Here, $s_{pq}$ represents the standard deviation of the $A_p$, assuming that the $p$-year average is defined with respect to starting year $q$. Since we are concerned that estimates for $A_p$ may be sensitive to the choice of starting year, for each latitude and depth, we explicitly quantify variation in $A_p$ independent of the choice of starting year. Specifically, we can use the average of the standard deviations $s_{pq}$ over $q$ to give
\begin{equation}
	\overline{s}_p = \frac{1}{p} \sum_{q=1}^{p} s_{pq}
	\label{e_sp_bar}
\end{equation}
which is an estimate of the standard deviation of $A_p$ over the $n_t$ years of data, not dependent on the starting year for the calculation. Further, we can use the standard deviation of $s_{pq}$ ($q=1,2,...,p$) given by
\begin{equation}
	\sigma_{sp}=sd\left[ \left[ s_{pq} \right]_{q=1}^p \right]
\end{equation}
as a diagnostic to assess the extent to which the starting year affects the calculation of the standard deviation of $A_p$. Even when $\sigma_{sp}$ is small, we can be confident that the arbitrary choice of starting year does not impact the estimate for the standard deviation of $A_p$. When $\sigma_{sp}$ is large, $\overline{s}_p$ is still the best choice of estimate for the standard deviation of the $p$-year average of $\Psi^c$, $A_p$, but we are somewhat more careful with our interpretation.

As a heuristic to aid visual assessment of $\overline{s}_p$ and $\sigma_{sp}$ over latitude and depth, we calculate the ratio
\begin{equation}
	\sigma_{sp}^* = \frac{\sigma_{sp}(y_j, z_k)}{\max_{j',k'}(\overline{s}_p(y_{j'},z_{k'}))}
	\label{e_sigma_sp}
\end{equation}
of $\sigma_{sp}$ per latitude-depth combination with respect to the maximum value of $\overline{s}_p$ (over all latitudes and depths).

\subsection{Linear trend in bottom densities}
\label{s_LR_D}

\begin{figure*}[ht!]
	\centerline{\includegraphics[width=\textwidth]{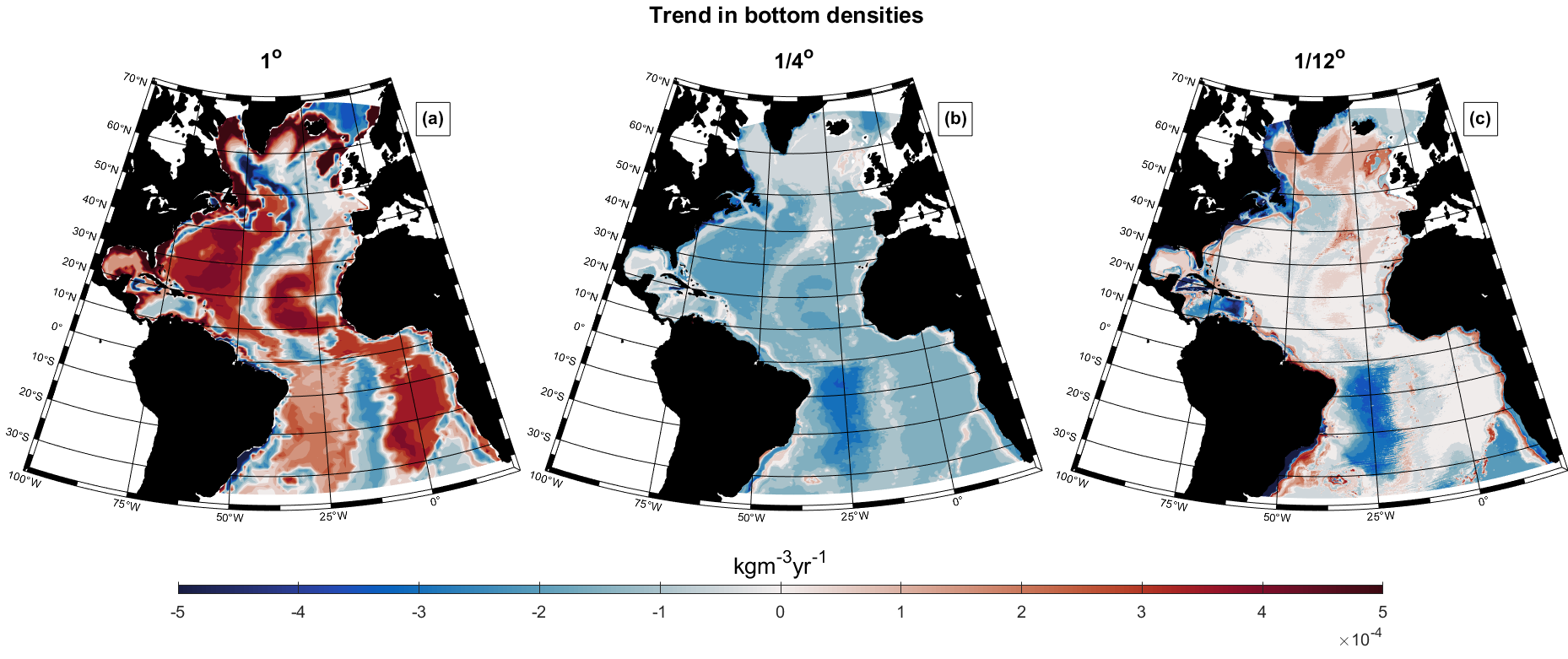}}
	\caption[Temporal trend in in-situ bottom densities for the (a) $1^\circ$, (b) $1/4^\circ$ and (c) $1/12^\circ$ model datasets in the Atlantic basin.]{Temporal trend in in-situ bottom densities over the entire model control run for the (a) $1^\circ$, (b) $1/4^\circ$ and (c) $1/12^\circ$ model datasets in the Atlantic basin.}
	\label{p_LT_Bden}
\end{figure*}

In-situ bottom densities at all model resolutions exhibit linear trends in time, illustrated in Figure \ref{p_LT_Bden}. These were estimated as the slope term from linear regression of bottom density, with time as the covariate over the full length of each model control run.
The trend in the $1^\circ$ model (Figure \ref{p_LT_Bden}(a)) is larger than that in both higher resolution models. For the majority of the basin we see a general positive trend in the $1^\circ$ model, indicating increasing densities in time, except along the Mid Atlantic Ridge (densities over shallower bathymetry getting lighter) and in the Labrador and Nordic Seas. In contrast, the $1/4^\circ$ model exhibits a negative trend throughout the basin, suggesting a lightening of densities everywhere with time. The highest resolution model (Figure \ref{p_LT_Bden}(c)) shows large regions of little to no trend. However, for the southern hemisphere there is a negative trend within the Brazil and Argentine basins, also present in the $1/4^\circ$ model. In the northern subpolar region of the $1/12^\circ$ model we find a positive trend within the Labrador Sea and south of Greenland, suggesting densities increasing in time here. 

The trends in bottom densities for higher resolution models are likely to be attributed to model spin-up of the deep ocean. The shorter length $1^\circ$ model could be showing some signs of longer term variability; however, the trends here are also likely to be model spin-up. The lightening of densities in the $1/4^\circ$ model Argentine-Brazil basin is further discussed in Section \ref{ACC_BdryD} and could be caused by a reduction in Weddell Sea deep water formation, as a result of large biases in Southern Ocean and ACC characteristics in the $1/4^\circ$ model.

We calculate $\overline{s}_p$, $\sigma_{sp}$, $\sigma_{sp}^*$ here having eliminated the linear trend in time, using the residuals of bottom densities found with the fitted linear regression model. However, removal of the trends shown in Figure \ref{p_LT_Bden} is found to have negligible influence on the estimates of $\overline{s}_p$, $\sigma_{sp}$, $\sigma_{sp}^*$. This is not surprising given the magnitudes of the  trends reported in the figure, relative to the typical magnitude of in-situ density.

\subsection{Averaged standard deviations, \textbf{$\overline{s}_p$}}

\begin{figure}[ht!]
	\centerline{\includegraphics[width=\textwidth]{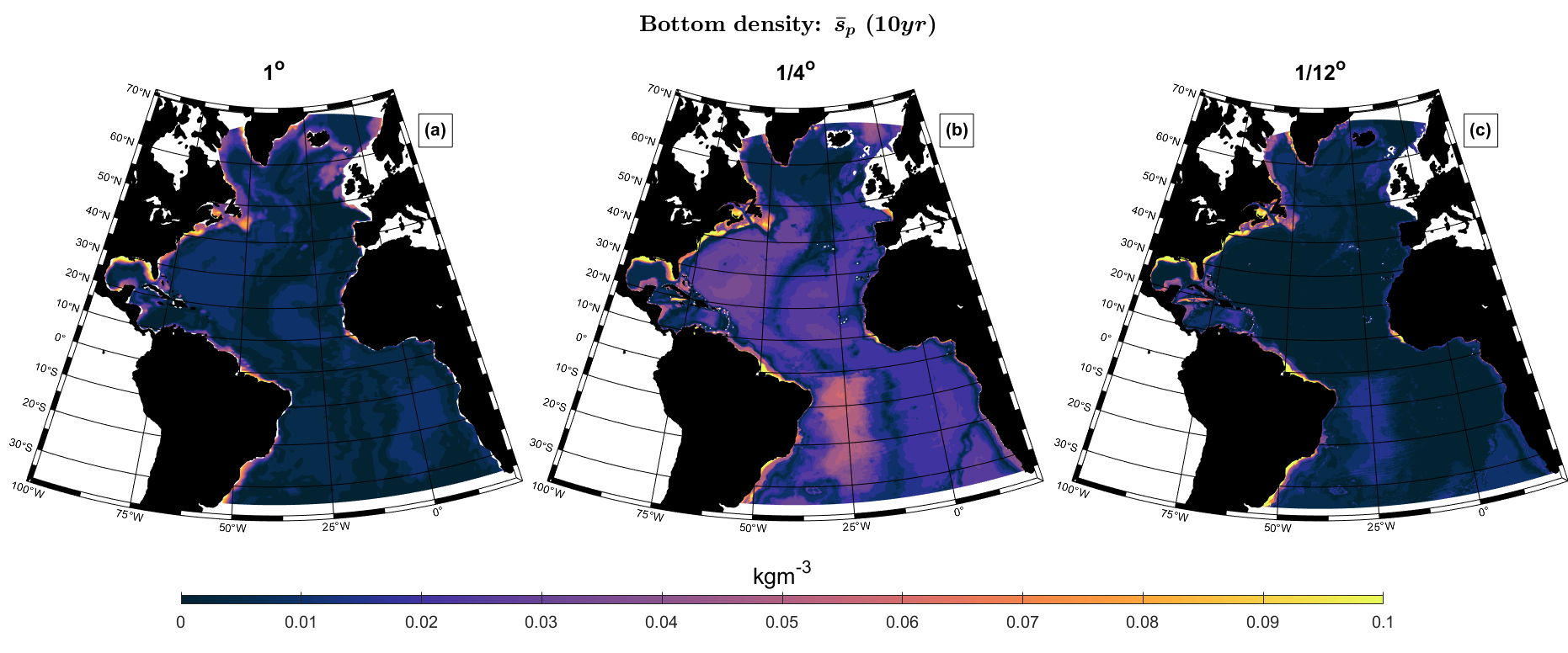}} 
	\caption[Averaged standard deviation $\overline{s}_{10}$ estimated using $p=10$ year observation periods of in-situ bottom density for $1^\circ$, $1/4^\circ$ and (c) $1/12^\circ$ model resolution within the Atlantic basin.]{Averaged standard deviation $\overline{s}_{10}$ estimated using $p=10$ year observation periods for in-situ bottom density. Panel (a) $1^\circ$, (b) $1/4^\circ$ and (c) $1/12^\circ$ show $\overline{s}_{10}$ for each model resolution within the Atlantic basin.}
	\label{p_sb_p10}
\end{figure}

\begin{figure}[ht!]
	\centerline{\includegraphics[width=11cm]{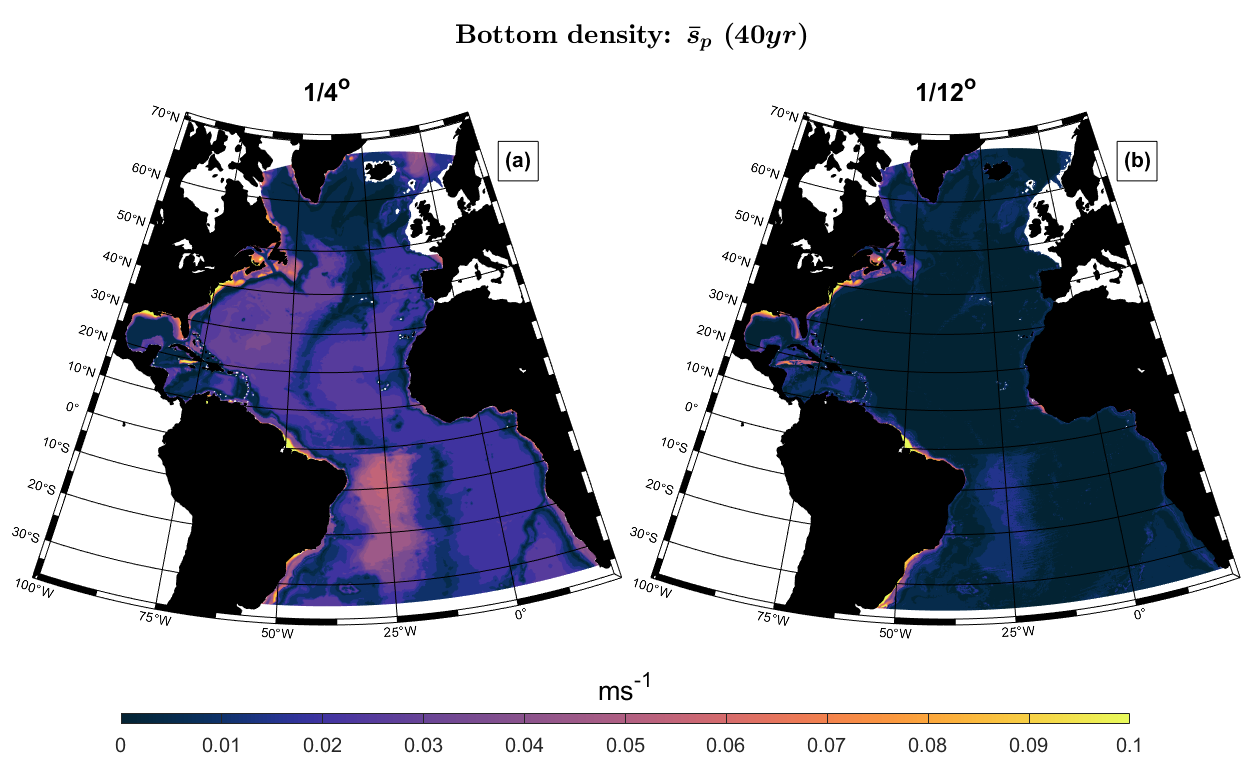}} 
	\caption[Averaged standard deviation $\overline{s}_{40}$ estimated using $p=40$ year observation periods of in-situ bottom density for $1^\circ$, $1/4^\circ$ and (c) $1/12^\circ$ model resolution within the Atlantic basin.]{Averaged standard deviation $\overline{s}_{40}$ estimated using $p=40$ year observation periods for in-situ bottom density. Panel (a) $1/4^\circ$ and (b) $1/12^\circ$ show $\overline{s}_{40}$ for each model resolution within the Atlantic basin.}
	\label{p_sb_p40}
\end{figure}

In this section we investigate the averaged standard deviation $\overline{s}_p$ (Equation \ref{e_sp_bar}). Our primary interest is in longer timescale variability. Therefore, we concentrate our investigation of bottom properties on periods $p=10$ and $40$ years shown in Figures \ref{p_sb_p10} and \ref{p_sb_p40}. We also mention results for $p=1$.

The $1/4^\circ$ model (Figure \ref{p_sb_p10}) exhibits sizeable decadal variation ($\overline{s}_{10}$) for the interior, especially prominent within the Brazil and Argentine basins. All three model resolutions show the largest variation in decadal means ($\overline{s}_{10}$) along the western boundary, which is slightly smaller than that exhibited by the $\overline{s}_1$ case. Interestingly, on closer inspection, decadal variation near the western boundary seems to increase with increasing resolution; this is particularly evident near (a) $30^\circ$N on the western boundary, (b) $10^\circ$N on the eastern boundary, and (c) $20-30^\circ$S on the western boundary. All of these locations coincide with meandering western or eastern boundary currents, and the increase in decadal variation with higher resolution is likely due to the greater ability of higher resolution models to resolve these boundary currents. At high northern latitudes the $1^\circ$ model exhibits greater variability near the Greenland-Scotland ridge overflow regions, than both higher resolution models. This could be related to the weaker subpolar gyre and DWBC found in the $1^\circ$ model and changes in deep water properties and its southward transport.

For $p=1$ (not shown), estimates of $\overline{s}_1$ (average yearly standard deviation) show similar characteristics at each resolution. There is a significant $\overline{s}_1$ value near the western boundary throughout the Atlantic basin, due to the presence of western boundary currents. The only major difference between $\overline{s}_1$ values for different model resolutions is found within the $1/4^\circ$ model. Here we find a larger $\overline{s}_1$ within the majority of the interior and stronger signal in the Brazil-Argentine basin, very similar to that found within Figure \ref{p_sb_p10}(b) for $\overline{s}_{10}$, suggesting changes in bottom water densities in the basin. Figure \ref{p_sb_p10} shows greater variation between resolutions for $\overline{s}_{10}$ compared to $\overline{s}_1$. 

For $p=40$ years, we cannot estimate $\overline{s}_{40}$ reliably for the $1^\circ$ model due to its short simulation length of 104 years. Figure \ref{p_sb_p40} for $p=40$ years at $1/4^\circ$ and $1/12^\circ$ model resolutions shows relatively similar structure to that shown in Figure \ref{p_sb_p10}. Comparing the $1/4^\circ$ and $1/12^\circ$ estimates for $\overline{s}_{40}$, there is agreement except for (a) the Cayman Trough near the Caribbean Sea, and (b) along the eastern boundary in the southern hemisphere. Generally, there is a greater multi-decadal variability ($\overline{s}_{40}$) within the interior for the $1/4^\circ$ model as mentioned previously. Differences for the Cayman trench could be attributed to the ability of the model to accurately resolve the trough itself.

The largest effect of increasing $p$ from 10 to 40 years is found in the $1/12^\circ$ model. Here, we see weaker $\overline{s}_{40}$ values near (a) $30^\circ$N on the western boundary, and (b) the northern section of the Gulf of Mexico. More generally, there is somewhat less variability near the western boundary, suggesting lesser variability in bottom densities and western boundary currents at longer multi-decadal timescales, in the Gulf of Mexico and Caribbean Sea. Cayman Trough densities in the $1/12^\circ$ model show similar variability for both $p=$ 10 and 40 years, whereas shallower regions in the Gulf do not show variability at longer timescales. The lack of change in variability between averages over 10 and 40 years for the $1/4^\circ$ model in the Argentine-Brazil basin and over a large part of the basin suggests long term fluctuations, unaffected by the removal of model linear trend. Another possibility could be a non-linear trend in bottom densities, with largest changes at the start of the model run. Section \ref{ACC_BdryD} will discuss discrepancies in ACC transport through the Drake Passage, Antarctic boundary densities, and the magnitude of Weddell Gyre, which could possibly lead to long-term variability or shift in Atlantic bottom densities, via changes in Antarctic deep water formation. 

\subsection{Bottom velocities}

\begin{figure*}[ht!]
	\centerline{\includegraphics[width=\textwidth]{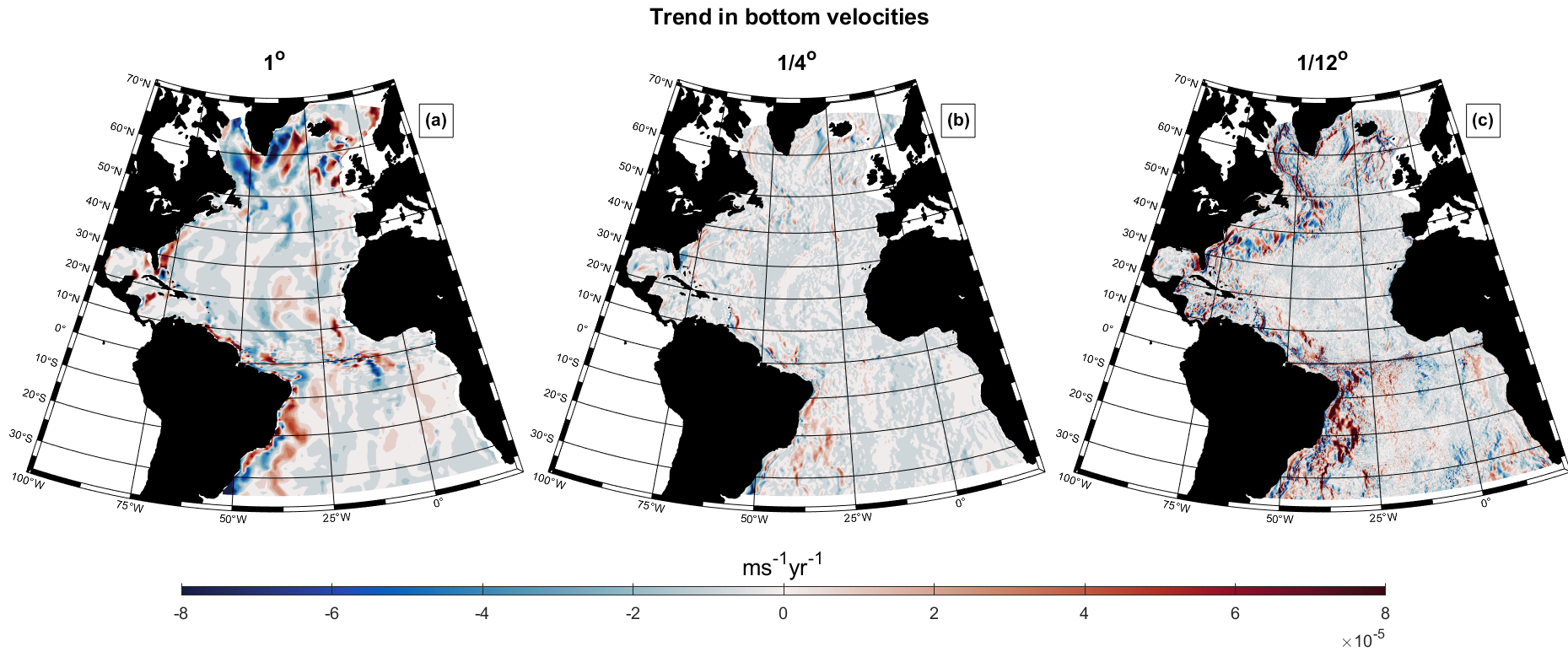}}
	\caption[Linear temporal trend in meridional bottom velocities across entire length of model runs for the $1^\circ$, $1/4^\circ$ and $1/12^\circ$ model datasets in the Atlantic basin.]{Linear temporal trend in meridional bottom velocities for the (a) $1^\circ$, (b) $1/4^\circ$ and (c) $1/12^\circ$ model datasets in the Atlantic basin.}
	\label{p_LT_Bvel}
\end{figure*}

The bottom component ($\Psi^c_{bot}$) of the overturning streamfunction is calculated from bottom velocities within the basin. We find relatively weak linear trends in the bottom velocities at each model resolution, illustrated in Figure \ref{p_LT_Bvel}. From Figure \ref{p_RC_Bvel}, the largest average values of bottom velocities are $\pm$0.1ms$^{\text{-1}}$, large compared to the magnitudes of linear trends in Figure \ref{p_LT_Bvel}; therefore, there is little concern about the impact of model drift.

Linear trends are strongest in the $1^\circ$ and $1/12^\circ$ models (Figure \ref{p_LT_Bvel}(a)(c)) especially on the western side of the basin, to be expected since the strongest boundary currents occur here. Recalling that the $1/4^\circ$ timeseries is considerably longer than that of the other two model simulations, this may have some influence on the relative size of trends observed. The $1^\circ$ model exhibits larger trends in the northern subpolar regions, coinciding with the significantly weaker DWBC found (Figure \ref{p_RC_Bvel}) and weaker subpolar gyre (Figure \ref{p_RC_BrtStrm}). Interestingly, we note alternating positive and negative trends for the $1/12^\circ$ model, where the DWBC flows southwards near the western coast of the U.S., suggesting strong influence of bathymetry on bottom flows in these regions.
\begin{figure}[ht!]
	\centerline{\includegraphics[width=\textwidth]{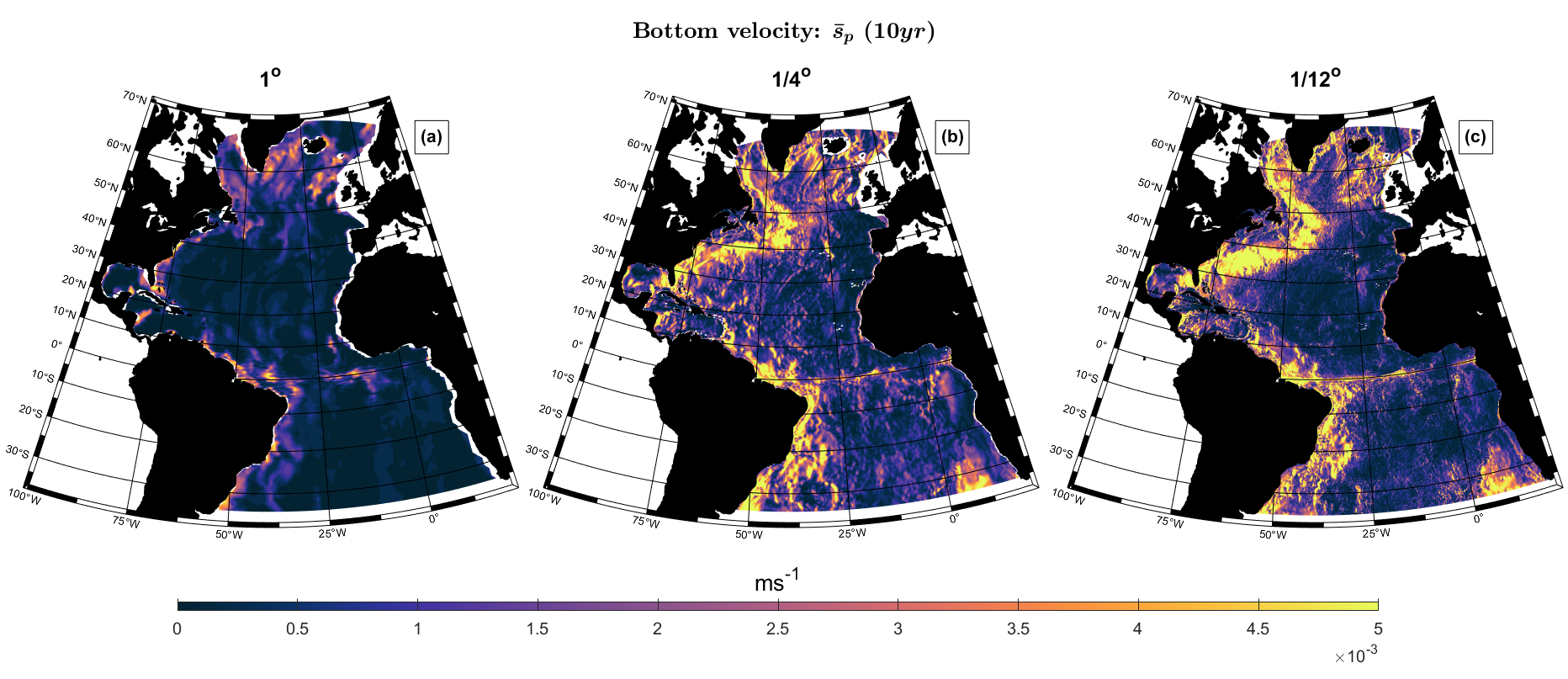}} 
	\caption[Averaged standard deviation $\overline{s}_{10}$ estimated using $p=10$ year observation periods of meridional bottom velocity for $1^\circ$, $1/4^\circ$ and (c) $1/12^\circ$ model resolutions within the Atlantic basin.]{Averaged standard deviation $\overline{s}_{10}$ estimated using $p=10$ year observation periods for bottom velocity. Panel (a) $1^\circ$, (b) $1/4^\circ$ and (c) $1/12^\circ$ shows $\overline{s}_{10}$ for each model resolution within the Atlantic basin.}
	\label{p_sbv_p10}
\end{figure}

Figure \ref{p_sbv_p10} shows the averaged standard deviation, $\overline{s}_{p=10}$ for bottom velocities at each model resolution. Larger decadal variation is observed for both higher resolution models (Figure \ref{p_sbv_p10}(b)(c)), especially along the western boundary. This is likely due to greater variability within surface and deep western boundary currents, and improved resolution of the bottom currents along the shelf in higher resolution models. Output of all three model resolutions shows a relatively large $\overline{s}_{10}$ near the equator, even away from coastlines. This effect could possibly be attributed to tropical waves travelling along the equator. For the longer period variation  ($\overline{s}_{p=40}$, not shown), we find the $1/4^\circ$ and $1/12^\circ$ models to again be very similar to each other and to the corresponding $\overline{s}_{p=10}$, but with smaller magnitudes and therefore less variation at longer timescales. We find a greater extent of large bottom velocity $\overline{s}_{10}$ (Figure \ref{p_sbv_p10}) in comparison to regions of large bottom density $\overline{s}_{10}$ (\ref{p_sb_p10}). The normalised variation in standard deviation, \textbf{$\sigma_{sp}^*$} for bottom densities and velocities is discussed in Appendix \ref{Var_BD_Nrm}. 

\section{Multiple Linear Regression and Correlation Adjusted coRrelation}

In this section we assess the extent to which the variation of the expected compensated overturning streamfunction $\Psi^c$ can be explained in terms of the variation of boundary component contributions. That is, we seek to quantify the relative importance of boundary components in explaining $\Psi^c$ variations using statistical modelling. There is a large body of relevant statistics literature, including \cite{George2000} and \cite{Davison2003}. Here, we will make use of straightforward statistical techniques such as multiple linear regression (MLR), and so-called correlation-adjusted correlations (CAR).

\begin{figure*}[ht!]
	\centerline{\includegraphics[width=\textwidth]{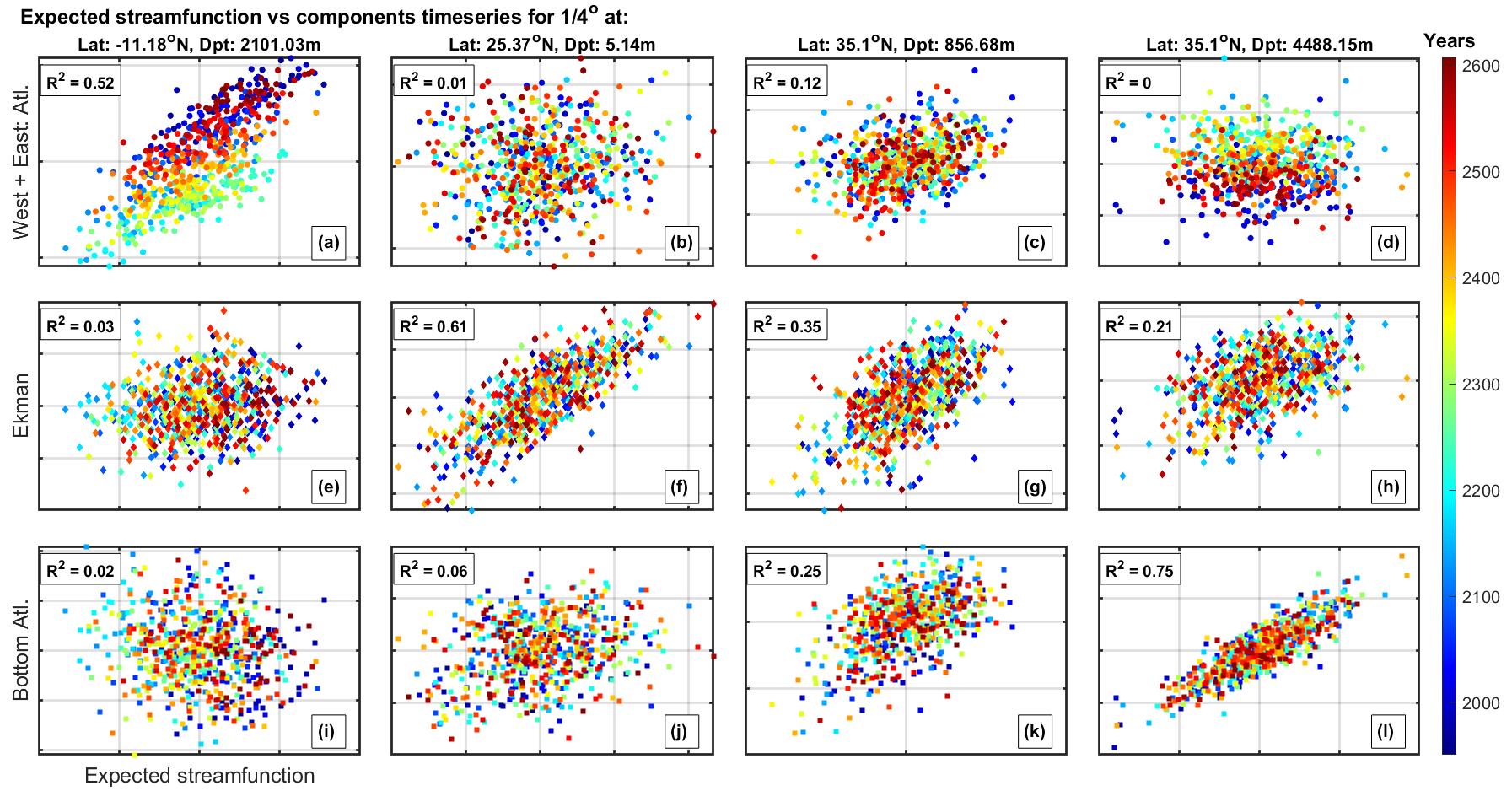}}
	\caption[Scatter plots of detrended timeseries (centred to zero time-mean) of West + East Atlantic densities, Ekman and bottom Atlantic (GMC not included) components against the expected compensated overturning streamfunction $\Psi^c$ for 4 latitude-depth combinations.]{Scatter plots of detrended and time-mean-removed timeseries of West + East Atlantic densities (Panels (a)(b)(c)(d)), Ekman (Panels (e)(f)(g)(h)) and Bottom Atlantic (Panels (i)(j)(k)(l)) components against the expected compensated overturning streamfunction $\Psi^c$ for 4 latitude-depth combinations. Locations for Panels are as follows: (a)(e)(i) are latitude $11.18^\circ$S and depth $2102$m,  (b)(f)(j) are latitude $25.37^\circ$N and depth $5$m, (c)(g)(k) are latitude $35.1^\circ$N and depth $857$m and (d)(h)(l) are latitude $35.1^\circ$N and depth $4488$m. The squared correlation $R^2$ values are shown in the top left corner of each Panel.}
	\label{f_Corr}
\end{figure*}

To begin, Figure \ref{f_Corr} shows scatter plots from timeseries of detrended $\Psi^c$ (centred so that its time-mean is zero) and the main boundary components, for 4 latitude-depth combinations within the Atlantic basin. Panels (a), (e) and (i) show the relationship between $\Psi^c$ (x-axis) and the Atlantic western and eastern densities ($\Psi^{c,AB}_{W}$ + $\Psi^{c,AB}_{E}$), Ekman ($\Psi^c_{Ekm}$) and Atlantic bottom ($\Psi^{c,AB}_{bot}$, excluding bottom velocity contribution from GMC) components respectively, at latitude $11.18^\circ$S and depth $2102$m. $\Psi^{c,AB}_{bot}$ and $\Psi^{c,GMC}_{bot}$ refer to the depth-independent contribution made by Atlantic and Gulf of Mexico and Caribbean Sea basins, as shown in the decomposition of the basin into sub-domains in Figure \ref{p_Bath_sp}. The value of squared correlation between the components and the expected streamfunction is also shown in the top left of each panel. For this latitude-depth combination, we observe that Panel (a) exhibits the strongest correlation ($R^2=0.52$), suggesting the variability in annual mean $\Psi^c$ is more related to variability in densities along the eastern and western boundaries of the Atlantic section, than to $\Psi^c_{Ekm}$ or $\Psi^{c,AB}_{bot}$.

At latitude $35.1^\circ$N and depth $4488$m (Panels (d)(h)(l)) we find $\Psi^{c,AB}_{bot}$ (Panel (l)) shows the greatest correlation with expected streamfunction $\Psi^c$ ($R^2=0.75$). In a similar fashion, $\Psi^c_{Ekm}$ appears to be relatively influential at latitude $25.37^\circ$N and depth $5$m (Panel (f)), confirming the view that the Ekman contribution ($\Psi^c_{Ekm}$) is important near the surface, and bottom velocities ($\Psi^{c,AB}_{bot}$) are important at depth.

Another common approach to assess the relationship between two variables is partial correlations. The method involves estimating the residuals of two multiple linear regressions on a set of so-called controlling variables, and then calculating the correlation between the regression residuals. In this sense, only that correlation between variables which cannot be explained by the controlling variables is accounted for in the partial correlation. Thus, for example, if we were interested in the contribution made by $\Psi^{c,AB}_{bot}$ to $\Psi^c$, firstly we would individually regress $\Psi^{c,AB}_{bot}$ and $\Psi^c$ onto all the remaining boundary components. Then we would calculate the correlation between residuals from both these regression models as an estimate of the partial correlation between $\Psi^{c,AB}_{bot}$ and $\Psi^c$. The partial correlation lies between $-1$ and $1$, with a value near $\pm1$ implying a strong partial correlation. A weakness of both correlation analysis and partial correlation is the lack of physical interpretability. The ability to quantify the contribution is important to understanding the contributions made by the boundary components, and the physical implications thereof. Therefore, instead of partial correlations we use correlation-adjusted correlation (CAR) to assess the importance of boundary components in explaining variation of $\Psi^c$.

\subsection{CAR methodology}
The following discussion draws on the presentation of CAR scores by \cite{Zuber2010} and linear regression modelling from \cite{Davison2003}. Consider fitting a linear regression to a sample $\{y_i\}_{i=1}^n$ of size $n$ for response $Y$ in terms of a sample $\{x_j\}_{i=1,j=1}^{n,p}$ of $p$ explanatory variables $\{X_j\}_{j=1}^p$, using a linear model. For each observation, we fit
\begin{eqnarray}
	y_i = \sum x_{ij} \beta_j + \epsilon_i \label{Rgr1}
\end{eqnarray}
where $\{\epsilon_i\}_{i=1}^n$ are Gaussian noise random variables with zero mean and some unknown variance $\sigma^2$, and $\{\beta\}_{j=1}^p$ is a set of regression coefficients we want to estimate. We can write Equation~\ref{Rgr1} in matrix terms as 
\begin{eqnarray}
	\mathbf{y} = \mathbf{X} \mathbf{\beta} + \mathbf{\epsilon} \label{Rgr2}
\end{eqnarray}
where $\mathbf{y}$ is a $n \times 1$ column vector of response values, $\mathbf{\beta}$ is a $p \times 1$ column vector of regression parameters and $\mathbf{X}$ is a $n \times p$ matrix of values for the explanatory variables. The least squares and maximum likelihood estimate $\hat{\mathbf{\beta}}$ for $\mathbf{\beta}$ is then
\begin{eqnarray}
	\hat{\mathbf{\beta}} = (\mathbf{X}' \mathbf{X})^{-1}\mathbf{X}'\mathbf{y}, \label{Rgr3}
\end{eqnarray}
and the fitted values of $Y$ are
\begin{eqnarray}
	\hat{\mathbf{y}} = \mathbf{X} \hat{\mathbf{\beta}}. \label{Rgr4}
\end{eqnarray}

\begin{figure*}[ht!]
	\centerline{\includegraphics[width=\textwidth]{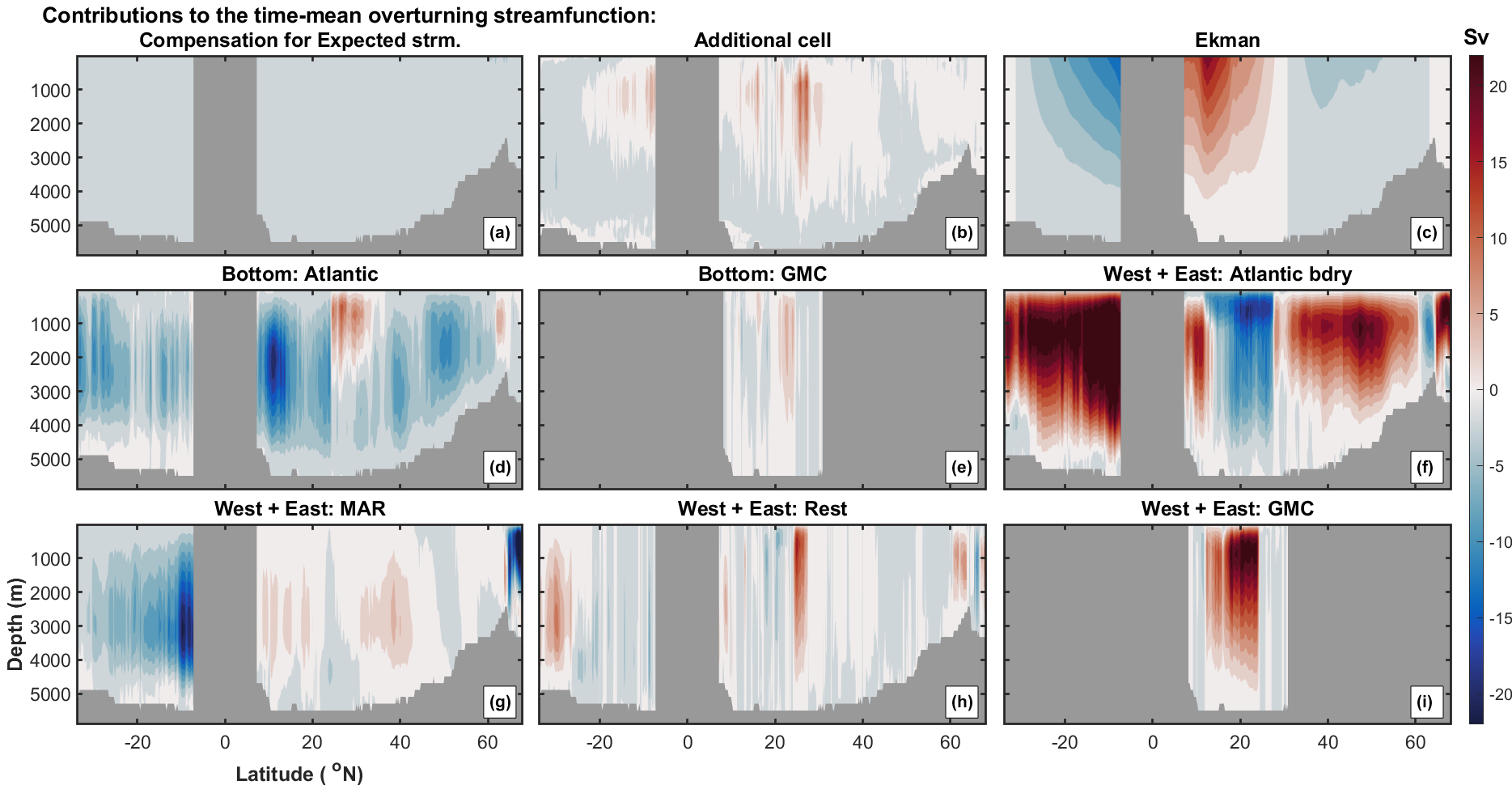}}
	\caption[Contribution to the time-mean overturning streamfunction for $1/4^\circ$ model.]{Contributions to the time-mean overturning streamfunction made by (a) compensation to $\Psi^c$ ($\Psi_{vol}$, volume imbalance), (b) additional cell contribution ($\Psi^c_{AC}$), (c) Ekman component ($\Psi^c_{Ekm}$), (d) bottom component for the Atlantic boundary ($\Psi^{c,AB}_{{bot}}$), (e) bottom component within the GMC ($\Psi^{c,GMC}_{{bot}}$), (f) eastern and western density contribution of the Atlantic boundary ($\Psi^{c,AB}_{W}$, $\Psi^{c,AB}_{E}$), (g) eastern and western density contribution of MAR ($\Psi^{c,MAR}_{W}$, $\Psi^{c,MAR}_{E}$), (h) the remaining unaccounted for eastern and western density contribution ($\Psi^{c,R}_{W}$, $\Psi^{c,R}_{E}$) and (i) eastern and western density contribution of the GMC ($\Psi^{c,GMC}_{W}$, $\Psi^{c,GMC}_{E}$). The same component will be used as explanatory variables in the MLR-CAR analysis.}
	\label{p_TM_CAR}
\end{figure*}

For the case of our AMOC decomposition, the response $Y$ is the total expected AMOC overturning streamfunction ($\Psi^c$), and the explanatory variables $\{X_j\}_{j=1}^{13}$ are: (a) Ekman, (b) additional cell, (c) Atlantic bottom, (d) GMC bottom, (e) compensation included for $\Psi^c$ (volume imbalance term), and the eastern and western density contributions of (f) Atlantic, (g) GMC, (h) MAR and (i) remainder components (time-mean contributors for which are illustrated in Figure \ref{p_TM_CAR}). It can be shown that the variance of the estimated parameter vector is given by
\begin{eqnarray}
	\mathbf{var}(\hat{\mathbf{\beta}}) = \sigma^2 (\mathbf{X}'\mathbf{X})^{-1}. \label{Rgr5}
\end{eqnarray}
The goodness of fit of the linear model is often quantified using the $R^2$ statistic, which is given by
\begin{eqnarray}
	(\mathbf{y}'\mathbf{y}) (1-R^2) &=& (\mathbf{y}-\hat{\mathbf{y}})' (\mathbf{y}-\hat{\mathbf{y}}) \nonumber \\
	&=& (\mathbf{y}-\mathbf{X} (\mathbf{X}'\mathbf{X})^{-1} \mathbf{X}' \mathbf{y})' (\mathbf{y}-\mathbf{X} (\mathbf{X}'\mathbf{X})^{-1} \mathbf{X}' \mathbf{y})' \nonumber \\
	&=& \mathbf{y}' \mathbf{y} - \mathbf{y}' \mathbf{X} (\mathbf{X}'\mathbf{X})^{-1} \mathbf{X}' \mathbf{y}. \label{Rgr6}
\end{eqnarray}
Equation~\ref{Rgr5} suggests that, in order to decorrelate the estimated regression vector $\hat{\mathbf{\beta}}$, we need to pre-multiply it by a quantity proportional to the square root of its variance. This motivates the definition of a CAR vector given by
\begin{eqnarray}
	\mathbf{c} = (\mathbf{y}' \mathbf{y})^{-1} (\mathbf{X}' \mathbf{X})^{-1/2}\mathbf{X}'\mathbf{y}. \label{Rgr7}
\end{eqnarray}
With this definition, Equation~\ref{Rgr6} becomes 
\begin{eqnarray}
	(\mathbf{y}'\mathbf{y}) (1-R^2) = \mathbf{y}' \mathbf{y} (1 - \mathbf{c}' \mathbf{c} ) \label{Rgr8}
\end{eqnarray}
or
\begin{eqnarray}
	R^2 = \mathbf{c}' \mathbf{c}. \label{Rgr9}
\end{eqnarray} 
That is, the squares of the elements $\{c_j\}_{j=1}^p$ of the CAR vector sum to $R^2$, and quantify the contribution of each explanatory variable to the quality of fit of the regression model. If we define the CAR scores $\{\phi_j\}_{j=1}^p$ by
\begin{eqnarray}
	\phi_j = c_j^2,
\end{eqnarray}
then $\phi_j$ quantifies the value (in terms of $R^2$) of explanatory variable $X_j$ in the regression model. Similarly, we see that the part of the sum of squares $\mathbf{y}'\mathbf{y}$ described by explanatory variable $X_j$ is given by $(\mathbf{y}'\mathbf{y})\phi_j$. It is therefore also useful to define the scaled CAR scores using
\begin{eqnarray}
	\phi_j^* = (\mathbf{y}'\mathbf{y}) c_j^2. \label{Rgr10}
\end{eqnarray}

Both CAR scores $\{\phi_j\}$ and scaled CAR scores $\{\phi_j^*\}$ provide useful measures of how different explanatory variables explain the response variable. By mean-centering the response and explanatory variables before CAR analysis, the CAR scores can be expressed in terms of the variance of the response.

In summary, CAR scores supplement a standard regression analysis, providing a clearer description of how explanatory variables correlate with the response variable. In a standard regression analysis, the estimated regression coefficients are correlated. It is therefore possible to explain the variability in the response almost equally well using different choices of regression variables. This means that it is difficult to uniquely allocate variation of the response to a particular explanatory variable (and this is one motivation for considering partial correlations). The CAR vector is created by rotating the estimated regression variables such that they become uncorrelated and independent, but such that rotated regression variables explain the variation of the response exactly as well as the original regression variables. This means that the CAR vector (and scores) gives an  ``importance'' measure for the contribution of each explanatory variable in explaining the response. CAR scores therefore also provide a means of ranking the importance of contributions from different explanatory variables. It is important to note that the quality of fit of the MLR-CAR model is identical to that of the MLR model. However, the MLR-CAR scores aid the physical interpretation of the role of the covariates. Further details are provided by \cite{Zuber2010}, with applications, extensions and discussion including \cite{Bocinsky2014}, \cite{Teisseyre2016}, \cite{Kessy2018}, \cite{Setiawan2018}, \cite{Welchowski2019}, \cite{Bommert2021}, \cite{Mimi2021}.

The time-mean contributions of the decomposition components to the overturning streamfunction for the $1/4^\circ$ model are shown in Figure \ref{p_TM_CAR}, where eastern and western density contributions are summed together into a single thermal wind term for each region for conciseness. The components contributing to $\Psi^c$ are: (a) compensation to $\Psi^c$ ($\Psi_{vol}$, volume imbalance); (b) additional cell contribution ($\Psi^c_{AC}$); (c) Ekman component ($\Psi^c_{Ekm}$); (d) bottom component for the Atlantic boundary ($\Psi^{c,AB}_{{bot}}$); (e) bottom component within the GMC ($\Psi^{c,GMC}_{{bot}}$); (f) eastern and western boundary density contribution of the Atlantic boundary ($\Psi^{c,AB}_{W}$; $\Psi^{c,AB}_{E}$); (g) eastern and western boundary density contribution of the MAR ($\Psi^{c,MAR}_{W}$, $\Psi^{c,MAR}_{E}$); (h) eastern and western boundary density contribution of the GMC ($\Psi^{c,GMC}_{W}$, $\Psi^{c,GMC}_{E}$) and (i) the remaining unaccounted for eastern and western boundary density contribution ($\Psi^{c,R}_{W}$, $\Psi^{c,R}_{E}$). To be clear, the eastern and western boundary density contributions for each region are accounted for as separate explanatory variables within the MLR-CAR score analysis, but for brevity are usually displayed as the sum of components. 

Figure \ref{p_TM_CAR} shows the dominant contribution made by Atlantic boundary density components (Panel (f)) to the time-mean overturning streamfunction $\Psi^c$ throughout the basin. It is interesting to note, however, that densities in the Gulf of Mexico and Caribbean Sea (Panel (i)) are responsible for a large proportion of the northward surface transport for latitudes near $20^\circ$N. As discussed in Chapter \ref{TJ_TM}, the Atlantic bottom (Panel (d)) and additional cell (Panel (b)) components make large contributions to the overturning streamfunction near $28^\circ$N, due to boundary currents. The large contribution in the same region from the ``remainder'' or ``Rest'' density term (Panel (h)), is attributed to intermediatory islands, e.g. the Bahamas. Other regions of strong density contributions that do not come from the GMC, MAR or main Atlantic boundaries are at depth in the southern hemisphere, due to the Malvinas islands and Walvis ridge. Panel (g) shows MAR densities make a large contribution to time-mean $\Psi^c$ at southern hemisphere low latitudes and high northern latitudes due to the Reykjanes ridge.

Now we estimate the MLR-CAR scores of boundary component contributions to the temporal variation of $\Psi^c$ with respect to latitude and depth for the $1/4^\circ$ model. Specifically, for each latitude-depth combination we use compensated boundary component contributions which have been temporally linearly detrended, with time-mean removed. The analysis is performed for various timescales by first filtering the data using a 6th order Butterworth band-pass filter. 

The CAR vector estimating the contribution of each boundary component is closely related to the $R^2$ value of the MLR model. Usually, $R^2<1$ since the regression model is not perfect. That is, there is variation in $\Psi^c$ unexplained by boundary components for each latitude-depth pair. It is useful to also visualise this unexplained variation, shown under the title ``Unexplained in MLR'' in the figures that follow. We find the total $R^2$ value of the MLR-CAR model is comfortably above an acceptable threshold (95$\%$) except for cells adjacent to the ocean's surface. Here, the overturning streamfunction is zero so shows no variation (by construction, to conserve volume), and therefore we cannot reasonably expect the MLR-CAR model to perform well.

\subsection{Results}
\begin{figure*}[ht!]
	\centerline{\includegraphics[width=\textwidth]{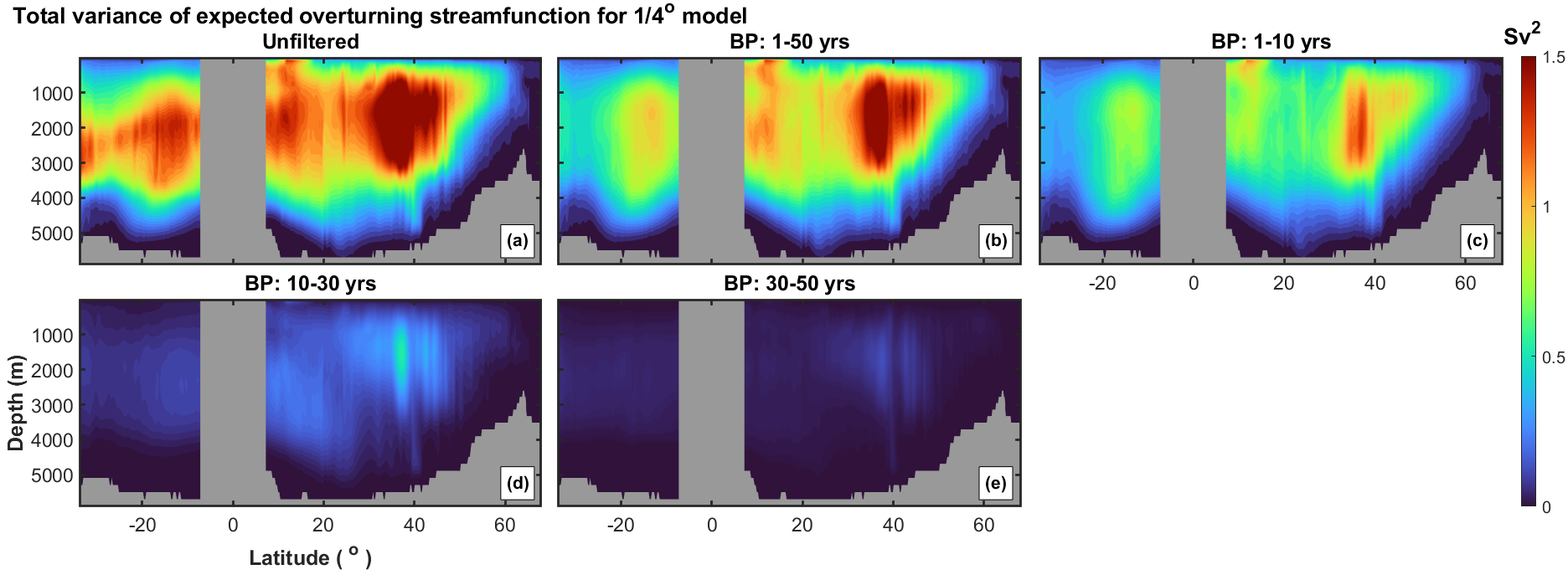}}
	\caption[Total variance of $\Psi^c$ for increasing timescales in the $1/4^\circ$ model.]{Total variance of $\Psi^c$ for (a) unfiltered annual-mean data, and for data that has been band-pass filtered (b) 1-50 years, (c) 1-10 years, (d) 10-30 years and (e) 30-50 years in the $1/4^\circ$ model.  Estimation is not possible in shaded grey regions due to bathymetry and latitudes near the equator.}
	\label{f_v_CAR_T}
\end{figure*}

Figure \ref{f_v_CAR_T} shows the total variance of $\Psi^c$ for the different band-pass timescales under consideration. By construction (Equation \ref{Rgr8}), the total variance of $\Psi^c$ is equal to the sum of scaled CAR scores $\phi_i^*$ over components $i$, plus the unexplained variance for the chosen timescale band. 

We see that the total variance of $\Psi^c$ reduces systematically, with larger variation at shorter timescales (and high frequencies). The largest variance is found between $30^\circ$N and $45^\circ$N for all timescales. This supports the results in Section \ref{Std:Thr}, where large variation was found in bottom velocities and densities at short and long periods for this region, south of Nova Scotia. The greater variance can be attributed to the DWBC flowing over and around a number of seamounts (New England and Corner Rise), resulting in large variation in boundary properties.

The MLR-CAR analysis below investigates the contributions made by each boundary component to the total variance of $\Psi^c$. Results are presented as latitude-depth plots for the whole Atlantic basin in two formats, (a) showing scaled CAR scores (Equation~\ref{Rgr10}), the contribution of each component to the total variance of $\Psi^c$ (in $\text{Sv}^2$), and (b) the percentage contribution of the scaled CAR score of each component (such that the total percentage contribution per latitude-depth pair is 100\%). Unfiltered (annual mean) data is presented first, then results for a range of different band-pass time filter intervals are discussed.

\subsubsection{Unfiltered data}

\begin{figure*}[ht!]
	\centerline{\includegraphics[width=\textwidth]{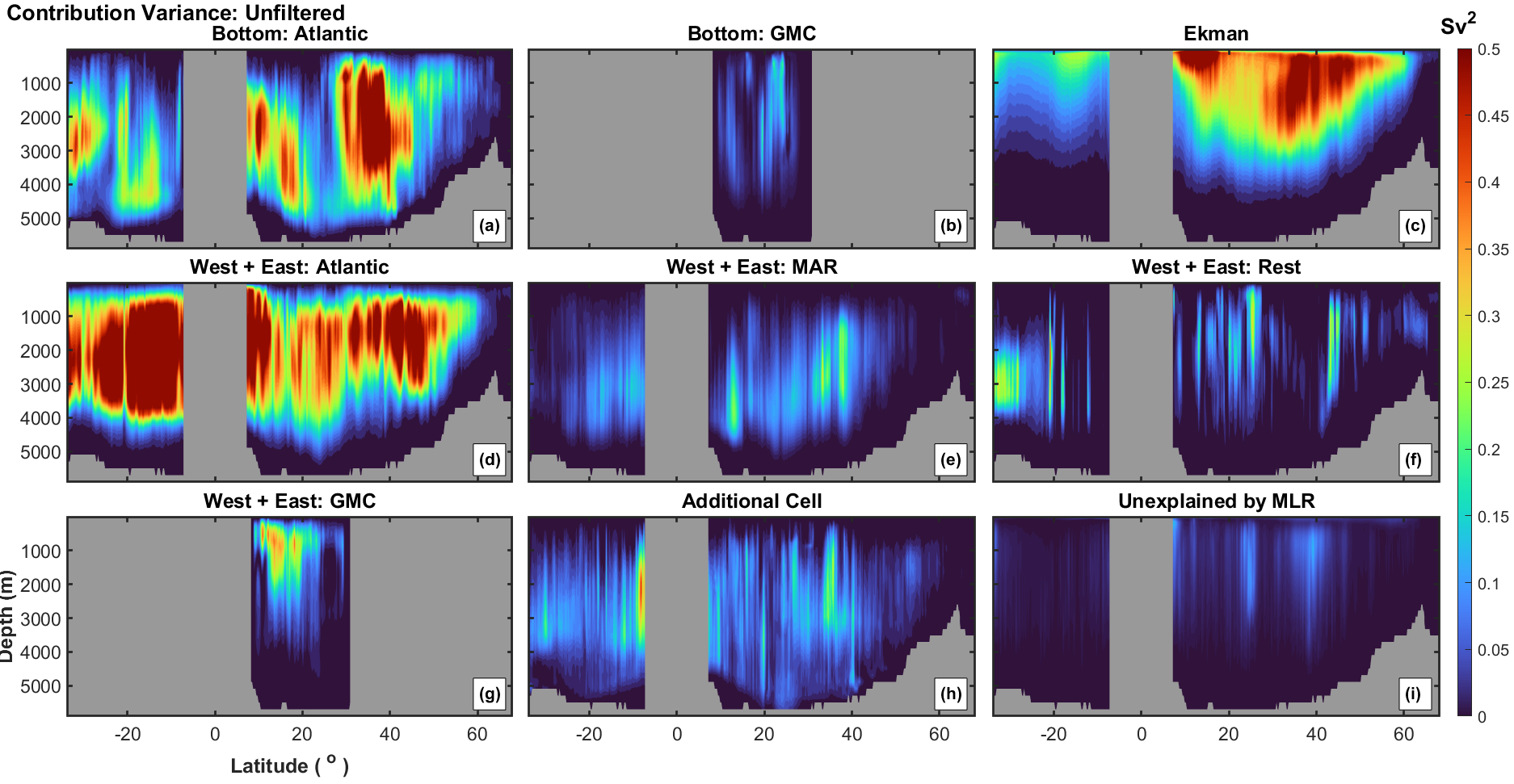}}
	\caption[Scaled CAR score contributions of boundary components to the variance of $\Psi^c$ for unfiltered annual-mean data.]{Scaled CAR score contributions of boundary components to the variance of $\Psi^c$ for unfiltered annual-mean data. Panels show the contribution made by (a) $\Psi^{c,AB}_{{bot}}$, (b) $\Psi^{c,GMC}_{{bot}}$, (c) $\Psi^c_{Ekm}$, (d) $\Psi^{c,AB}_{W}$ + $\Psi^{c,AB}_{E}$, (e) $\Psi^{c,MAR}_{W}$ + $\Psi^{c,MAR}_{E}$,  (f) $\Psi^{c,R}_{W}$ + $\Psi^{c,R}_{E}$, (g) $\Psi^{c,GMC}_{W}$ + $\Psi^{c,GMC}_{E}$, (h) $\Psi^c_{AC}$ and (i) variance unexplained by MLR-CAR. Estimation is not possible in shaded grey regions corresponding to one or more of bathymetry, latitudes near the equator, and latitudes not relevant for GMC components.}
	\label{f_v_CAR_1}
\end{figure*}

Figure \ref{f_v_CAR_1} shows the contribution of each component (using the scaled CAR score $\phi_i^*$ per component $i$) to the total variance exhibited by $\Psi^c$ (Figure \ref{f_v_CAR_T}(a)), for the full annual-mean dataset with no filter applied. In the figure, western and eastern boundary contributions are summed for convenience (thermal wind contribution) but will be considered individually later, and Panel (i) shows the unexplained variance from the MLR-CAR model. Further, the contribution from the $\Psi^c$ compensation term ($\Psi_{vol}$, Panel (a) Figure \ref{p_TM_CAR}) is not shown, since it was found to be negligible.

\begin{figure*}[ht!]
	\centerline{\includegraphics[width=\textwidth]{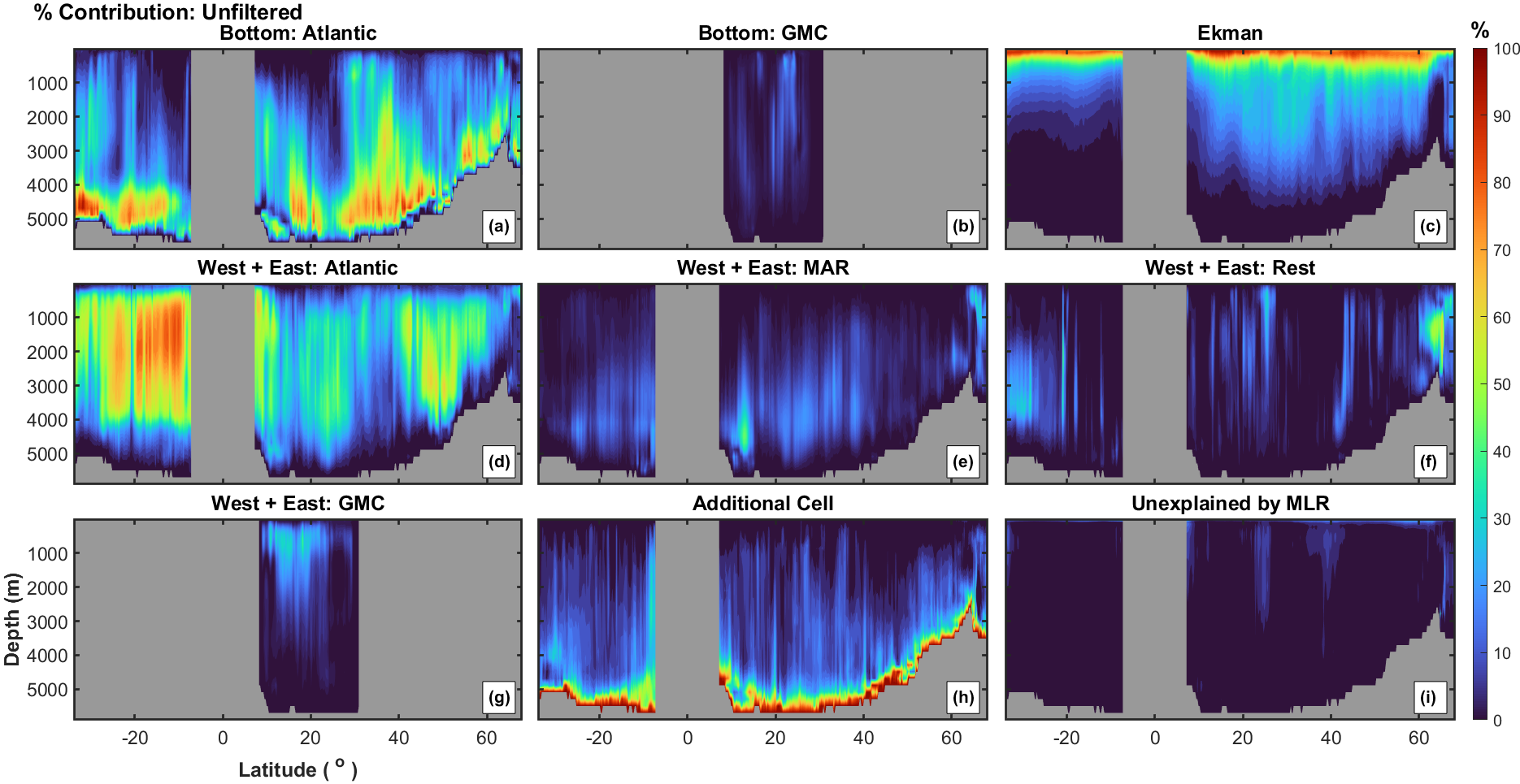}}
	\caption[Percentage scaled CAR score contributions of boundary components to the variance of $\Psi^c$ for unfiltered annual-mean data.]{Percentage scaled CAR score contributions of boundary components to the variance of $\Psi^c$ for unfiltered annual-mean data. Panel details are given in Figure \ref{f_v_CAR_1} caption.}
	\label{f_p_CAR_1}
\end{figure*}

We find a large $\Psi^{c,AB}_{{bot}}$ variance contribution (Panel (a)) at mid-depths between $30^\circ$N and $45^\circ$N, in addition to small pockets at mid-depths in the subtropical and South Atlantic regions. We find the largest $\Psi^{c}_{{Ekm}}$ contributions (Panel (c)) within the upper 2500m in the northern hemisphere, particularly between $10^\circ$N and $20^\circ$N, and $35^\circ$N and $50^\circ$N, at the locations of surface easterlies and westerlies. The density contribution to variance is large especially from the Atlantic boundaries (Panel(d), $\Psi^{c,AB}_{W}$ + $\Psi^{c,AB}_{E}$) and the GMC, (Panel (g), $\Psi^{c,GMC}_{W}$ + $\Psi^{c,GMC}_{E}$). The largest contributor to the variance within the southern hemisphere at mid-depth is $\Psi^{c,AB}_{W}$ + $\Psi^{c,AB}_{E}$, with little contribution from $\Psi^{c,AB}_{{bot}}$ near latitudes of the Malvinas islands and current. $\Psi^{c,R}_{W}$ + $\Psi^{c,R}_{E}$ (Panel (f)) exhibits a relatively strong contribution at 3500m near $30^\circ$S, potentially attributable to the Malvinas islands or the Walvis ridge; analysis of the individual components (not shown) reveals that this contribution is due to $\Psi^{c,R}_{W}$ (eastern side of the Walvis ridge or eastern side of the Malvinas islands). Scaled CAR scores for the other components are small. Summarising, Figure \ref{f_v_CAR_1} provides general context regarding the contributions made by different boundary components on a basin-wide scale: large contributions from Atlantic boundary densities ($\Psi^{c,AB}_{W}$ + $\Psi^{c,AB}_{E}$), wind-stresses ($\Psi^{c}_{{Ekm}}$) due to westerly and trade winds, and the Atlantic bottom velocity tern ($\Psi^{c,AB}_{{bot}}$) south of Newfoundland, attributable to the DWBC and presence of seamounts. 

Figure \ref{f_p_CAR_1} shows the corresponding percentage local scaled CAR score for each boundary component. That is, for each latitude-depth pair, the figure shows the value of $100 \times \phi_i^*/\sum_j \phi_j^*$, where the indices $i$ and $j$ refer to the 13 components considered for this decomposition. Figure \ref{f_p_CAR_1} illustrates that the CAR score contributions are consistent with physical intuition. From Panel (c) we see that $\Psi^{c}_{{Ekm}}$ dominates near the ocean's surface, with some residual influence down to 3000m depth in the northern hemisphere. Panel (a) shows that $\Psi^{c,AB}_{{bot}}$ dominates the abyss, but with sizeable influence at depths as shallow as 2000m for latitudes between $20^\circ$N and $40^\circ$N. Panel (h) shows that additional cells provide important contributions to variance near bathymetry ($\Psi^c_{AC}$). From Panel (d) we see that $\Psi^{c,AB}_{W}$ + $\Psi^{c,AB}_{E}$ variation is an important explanatory factor in the ocean's interior, particularly in the southern hemisphere. The influence of Greenland boundary densities is visible in Panel (f) ($\Psi^{c,R}_{W}$ + $\Psi^{c,R}_{E}$) at high northern latitudes, as is the contribution from boundary densities within the Gulf of Mexico and Caribbean Sea (in Panel (g), $\Psi^{c,GMC}_{W}$ + $\Psi^{c,GMC}_{E}$). 

Figures \ref{f_v_CAR_1} and \ref{f_p_CAR_1} suggest that $\Psi^c_{AC}$ (Panels (h)) has a large role in explaining the variance (of small absolute magnitude) near bathymetry, but in a basin-wide context this contribution is small. In contrast, the scaled CAR scores for the Atlantic bottom velocity term $\Psi^{c,AB}_{{bot}}$ between $30^\circ$N and $45^\circ$N is large, but the corresponding local percentage contribution is relatively small in shallower water, and locally dominant at greater depths. The scaled CAR scores for Ekman, Atlantic bottom velocity and Atlantic boundary density contributions between $30^\circ$N and $45^\circ$N are all large, reflecting the large variance of $\Psi^c$ here. $\Psi^{c}_{{Ekm}}$ appears to be dominant down to almost 3000m in Figure \ref{f_v_CAR_1} Panel (c) but is confined to shallow waters in Figure \ref{f_p_CAR_1}, suggesting greater local influence of the Atlantic boundary densities and Atlantic bottom velocity components. 

\subsubsection{Band-pass filtering admitting cycles with periods between 1 and 50 years}

The total length of the $1/4^\circ$ model simulation analysed here is 657 years. In simple terms, to estimate periodic behaviour in the overturning streamfunction requires that the data length corresponds to at least 10 cycles. Therefore, in this analysis we can only expect to be able to identify cycles with periods of at most approximately 50 years with any confidence. With that caveat in mind, initially we apply a band-pass filter to eliminate cyclic variation with periods greater than 50 years.

\begin{figure*}[ht]
	\centerline{\includegraphics[width=\textwidth]{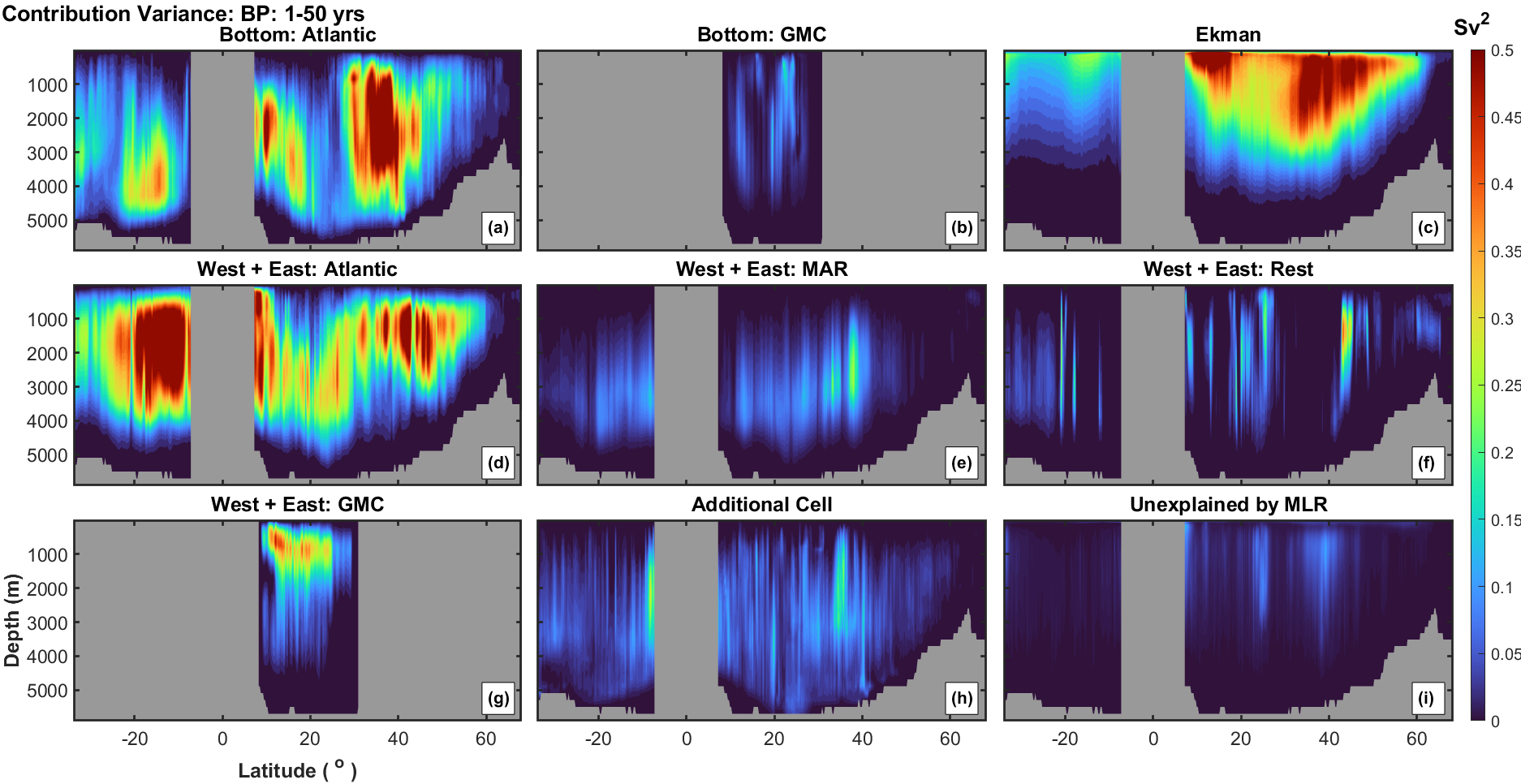}}
	\caption[Scaled CAR score contributions of boundary components to the variance of $\Psi^c$ for band-pass filtered annual-mean compensated overturning streamfunction data, with variation corresponding to periods greater than 50 years removed.]{Scaled CAR score contributions of boundary components to the variance of $\Psi^c$ for band-pass filtered annual-mean compensated overturning streamfunction data, with variation corresponding to periods greater than 50 years removed. Panels show the contribution made by (a) $\Psi^{c,AB}_{{bot}}$, (b) $\Psi^{c,GMC}_{{bot}}$, (c) $\Psi^c_{Ekm}$, (d) $\Psi^{c,AB}_{W}$ + $\Psi^{c,AB}_{E}$, (e) $\Psi^{c,MAR}_{W}$ + $\Psi^{c,MAR}_{E}$,  (f) $\Psi^{c,R}_{W}$ + $\Psi^{c,R}_{E}$, (g) $\Psi^{c,GMC}_{W}$ + $\Psi^{c,GMC}_{E}$, (h) $\Psi^c_{AC}$ and (i) variance unexplained by MLR-CAR. Estimation is not possible in shaded grey regions corresponding to one or more of bathymetry, latitudes near the equator, and latitudes not relevant for GMC components.}
	\label{f_v_CAR_1_50}
\end{figure*}

With contributions to the total variance of $\Psi^c$ at longer timescales (greater than 50 years) removed, we find in Figure \ref{f_v_CAR_1_50} a reduction in $\phi_i^*$ for $\Psi^{c,AB}_{W}$ and $\Psi^{c,AB}_{E}$ (Panel (d); c.f. Figure~\ref{f_v_CAR_1}). The reduction is particularly evident in the northern hemisphere tropical region, and also south of $18^\circ$S, suggesting longer term variation or spin-up in these boundary densities. Other plots (not shown) suggest that the southern hemisphere weakening can be attributed to a weaker contribution by $\Psi^{c,AB}_{W}$, whereas in the northern hemisphere we find a reduction of $\Psi^{c,AB}_{E}$. The contribution of $\Psi^{c,AB}_{{bot}}$ (Panel (a)) shows a slight reduction at mid-depths near $30^\circ$S and $17^\circ$N. However, $\Psi^{c,AB}_{{bot}}$ near $35^\circ$N is still a large factor in $\Psi^c$ variation. Variation of the Atlantic bottom velocity component $\Psi^{c,AB}_{{bot}}$ in this region is likely due to the separation of the Gulf Stream from the U.S. coast and its interaction with the southward flowing DWBC. We also note the disappearance of the $\Psi^{c,R}_{W}$ + $\Psi^{c,R}_{E}$ contribution (Panel (f)) at depths in the southern hemisphere near $30^\circ$S. The percentage contribution of components is very similar to that shown in Figure \ref{f_p_CAR_1} (for unfiltered data), with the contribution of $\Psi^{c}_{{Ekm}}$ penetrating deeper, and a stronger $\Psi^{c,AB}_{{bot}}$ at depths to compensate for the reduction in $\Psi^{c,AB}_{W}$ and $\Psi^{c,AB}_{E}$ in both hemispheres.

\subsubsection{Band-pass filtering admitting cycles with periods between 1 and 10 years}

\begin{figure*}[ht!]
	\centerline{\includegraphics[width=\textwidth]{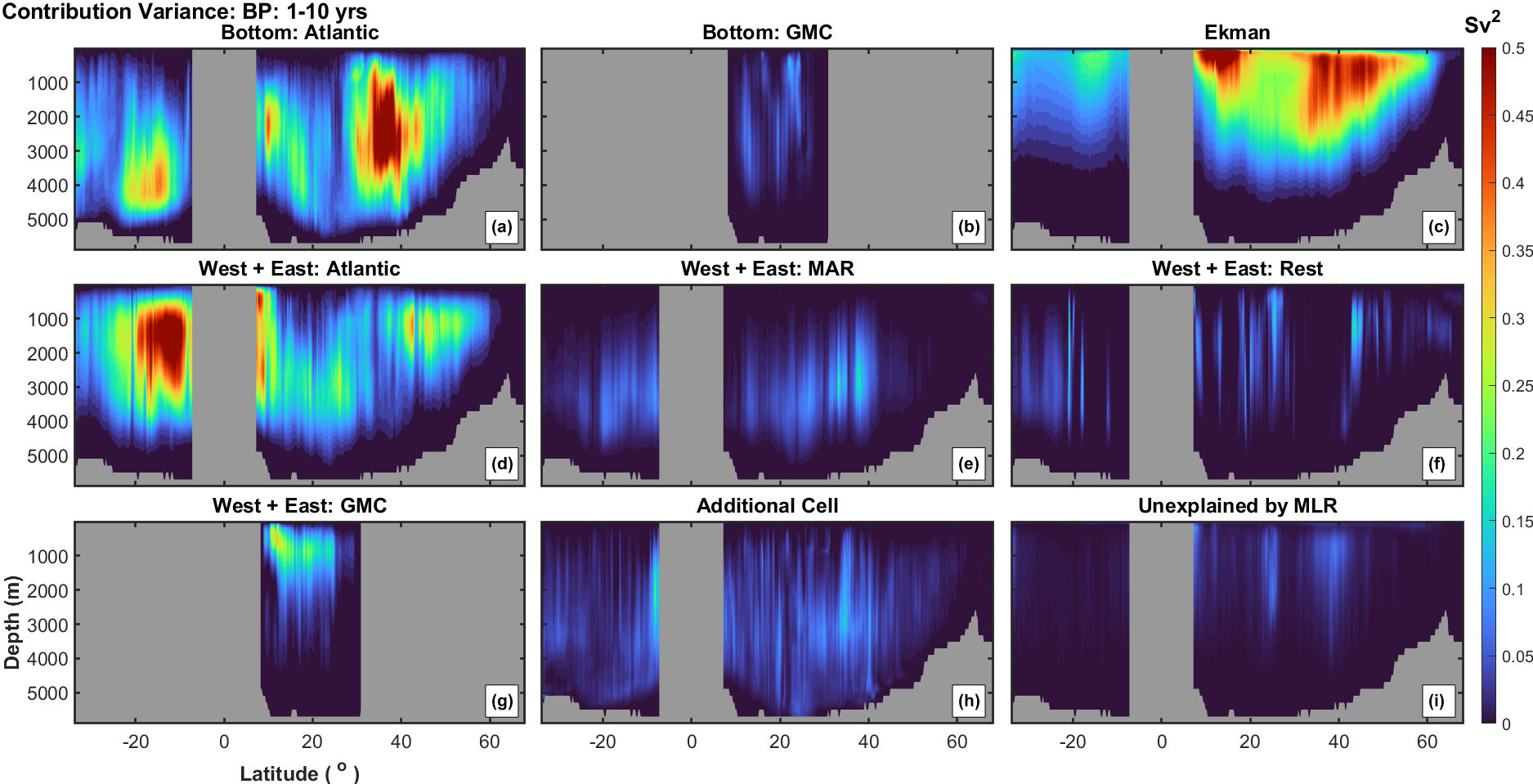}}
	\caption[Scaled CAR score contributions of boundary components to the variance of $\Psi^c$ for band-pass filtered annual-mean compensated overturning streamfunction data, with variation corresponding to periods greater than 10 years removed.]{Scaled CAR score contributions of boundary components to the variance of $\Psi^c$ for band-pass filtered annual-mean compensated overturning streamfunction data, with variation corresponding to periods greater than 10 years removed. Panels show the contribution made by (a) $\Psi^{c,AB}_{{bot}}$, (b) $\Psi^{c,GMC}_{{bot}}$, (c) $\Psi^c_{Ekm}$, (d) $\Psi^{c,AB}_{W}$ + $\Psi^{c,AB}_{E}$, (e) $\Psi^{c,MAR}_{W}$ + $\Psi^{c,MAR}_{E}$,  (f) $\Psi^{c,R}_{W}$ + $\Psi^{c,R}_{E}$, (g) $\Psi^{c,GMC}_{W}$ + $\Psi^{c,GMC}_{E}$, (h) $\Psi^c_{AC}$ and (i) variance unexplained by MLR-CAR. Estimation is not possible in shaded grey regions corresponding to one or more of bathymetry, latitudes near the equator, and latitudes not relevant for GMC components.}
	\label{f_v_CAR_1_10}
\end{figure*}

Figure \ref{f_v_CAR_1_10} shows the scaled CAR scores $\phi_i^*$ of each component $i$ for band-pass filtered annual-mean overturning streamfunction data, such that all variation corresponding to periods greater than 10 years is removed, illustrating contributions to the variability of $\Psi^c$ on short sub-decadal timescales. 

Figure \ref{f_v_CAR_1_10} exhibits similar characteristics to those in Figures \ref{f_v_CAR_1} and \ref{f_v_CAR_1_50}, with weakening of each of the major contributions, most notably for $\Psi^{c,AB}_{W}$ and $\Psi^{c,AB}_{E}$. At these short timescales we see weaker Atlantic boundary density contributions ($\Psi^{c,AB}_{W}$, $\Psi^{c,AB}_{E}$) in the northern hemisphere, but with the contribution between $10^\circ$S and $20^\circ$S remaining relatively strong. The GMC density contribution shows a reduction, but remains important for shallow waters in the GMC region. The similarity of results using 1-50 year and 1-10 year band-pass filters indicates that variation on short timescales dominates within the Atlantic basin, as previously noted (e.g. \citealt{Buckley2016}).

\begin{figure*}[ht!]
	\centerline{\includegraphics[width=\textwidth]{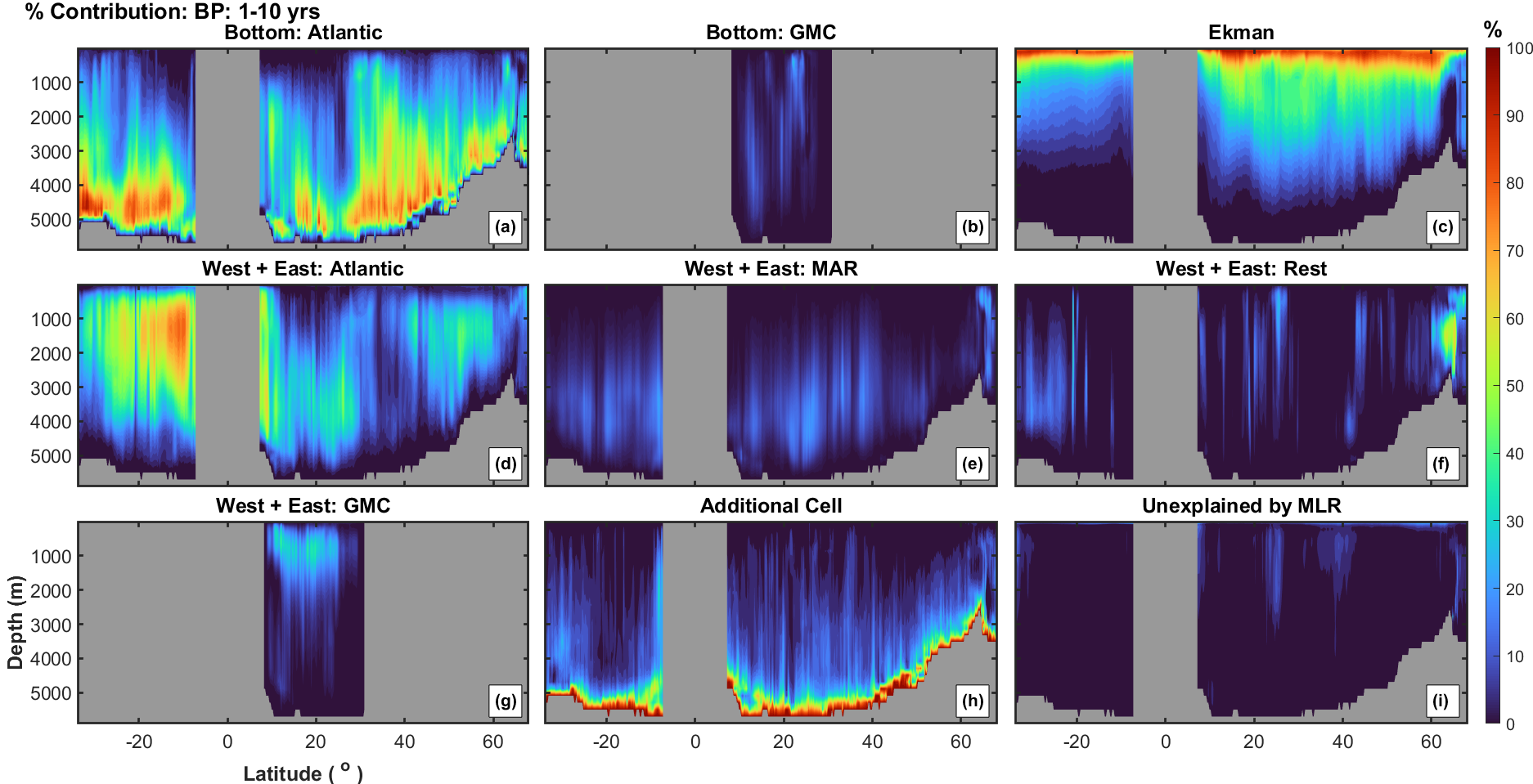}}
	\caption[Percentage scaled CAR score contributions of boundary components to the variance of $\Psi^c$ for band-pass filtered annual-mean compensated overturning streamfunction data, with variation corresponding to periods greater than 10 years removed.]{Percentage scaled CAR score contributions of boundary components to the variance of $\Psi^c$ for band-pass filtered annual-mean compensated overturning streamfunction data, with variation corresponding to periods greater than 10 years removed. Panel details are given in Figure \ref{f_v_CAR_1_10} caption.}
	\label{f_p_CAR_1_10}
\end{figure*}

Figure \ref{f_p_CAR_1_10} shows the corresponding percentage $100 \times \phi_i^*/\sum_j \phi_j^*$ for band-pass filtered annual-mean overturning streamfunction data, with all variation corresponding to periods greater than 10 years removed. The characteristics of the figure are similar to those of the unfiltered data in Figure \ref{f_p_CAR_1} and of the 1-50 year band-pass filtered data (not shown). The most notable difference (also seen in the 1-50 year band-pass filtered data) is a weakened relative contribution from $\Psi^{c,AB}_{W}$ + $\Psi^{c,AB}_{E}$ in the northern hemisphere (Panel (d)), compensated for by somewhat increased $\Psi^{c,AB}_{{bot}}$ and $\Psi^{c}_{{Ekm}}$ contributions (Panels (a) and (c)). In the southern hemisphere, we find a greater $\Psi^{c,AB}_{{bot}}$ contribution at depth, compensating for a weaker Atlantic boundary density component at depth, and at southerly latitudes.

\begin{figure*}[ht!]
	\centerline{\includegraphics[width=\textwidth]{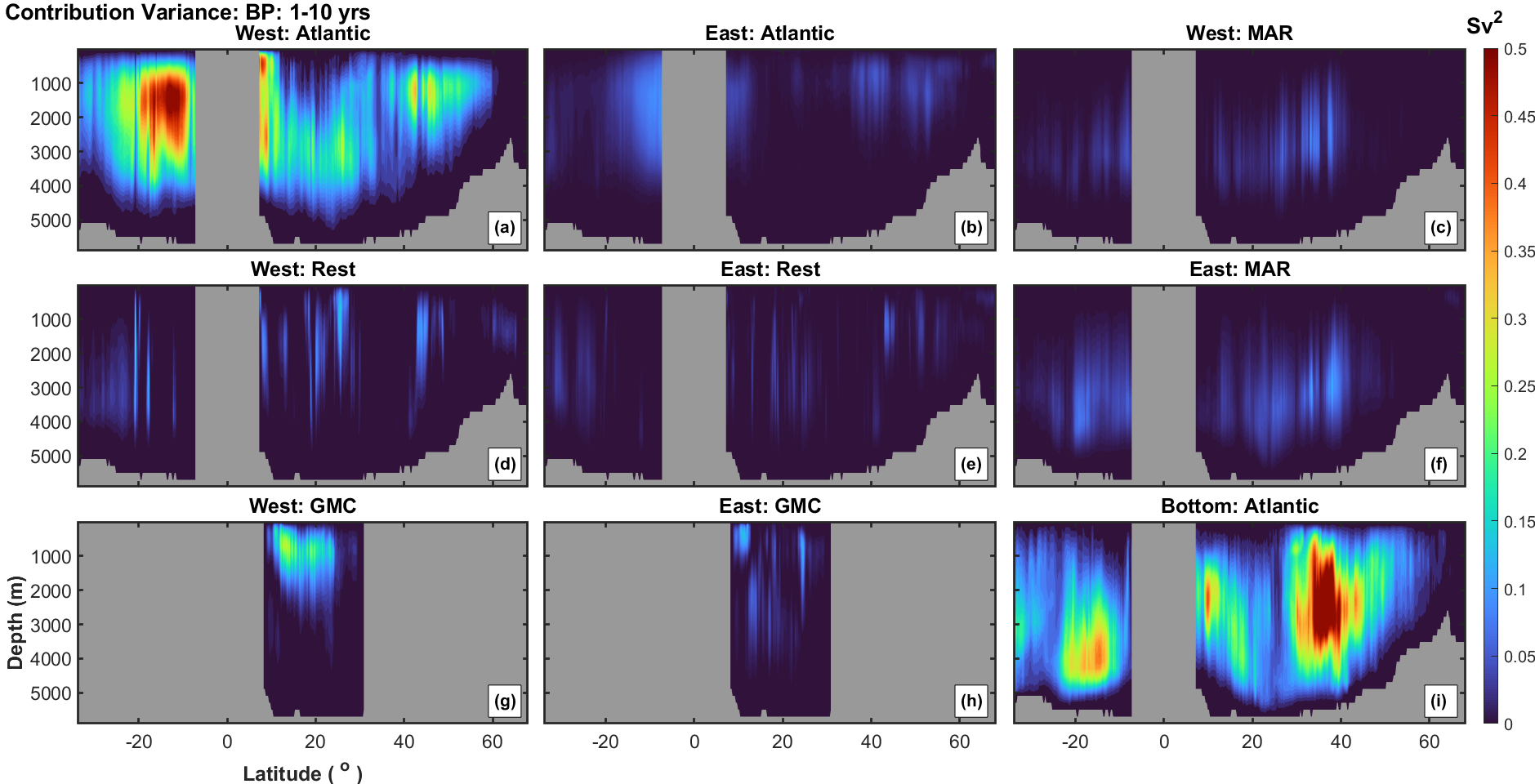}}
	\caption[Scaled CAR score contributions of density driven boundary components and $\Psi^{c,AB}_{{bot}}$ to the variance of $\Psi^c$ for band-pass filtered annual-mean compensated overturning streamfunction data, with variation corresponding to periods greater than 10 years removed.]{Scaled CAR score contributions of density driven boundary components and $\Psi^{c,AB}_{{bot}}$ to the variance of $\Psi^c$ for band-pass filtered annual-mean compensated overturning streamfunction data, with variation corresponding to periods greater than 10 years removed. Panels show the contribution made by (a) $\Psi^{c,AB}_{W}$, (b) $\Psi^{c,AB}_{E}$, (c) $\Psi^{c,MAR}_{W}$, (d) $\Psi^{c,R}_{W}$, (e) $\Psi^{c,R}_{E}$, (f) $\Psi^{c,MAR}_{E}$, (g) $\Psi^{c,GMC}_{W}$, (h) $\Psi^{c,GMC}_{E}$ and (i) $\Psi^{c,AB}_{{bot}}$. Estimation is not possible in shaded grey regions corresponding to one or more of bathymetry, latitudes near the equator, and latitudes not relevant for GMC components.}
	\label{f_v_CAR_1_10d}
\end{figure*}

Figure \ref{f_v_CAR_1_10d} illustrates the scaled CAR scores $\phi_i^*$ for individual components $i$ corresponding to eastern and western boundary densities separately, in order to compare the relative contributions of those boundaries for 1-10 year band-pass filtered data. The figure shows that western boundary density variation dominates, in the Atlantic ($\Psi^{c,AB}_{W}$, Panel (a)) and GMC ($\Psi^{c,GMC}_{W}$, Panel (g)) for these short timescales; eastern boundary density variations do not contribute much to overturning variability. Similar spatial features are seen for the percentage contribution of the density boundary components (not shown); we note a relatively large contribution (50\%) from eastern MAR densities ($\Psi^{c,MAR}_{W}$) at high northern latitudes near Greenland. Inter-annual variation in the Brazil current could be the cause of the large contribution made by western Atlantic boundary densities near $10^\circ$S and $20^\circ$S. In the same region we find a relatively large contribution by $\Psi^{c,AB}_{{bot}}$, suggesting variability in local boundary currents. 

\subsubsection{Band-pass filtering admitting cycles with periods between 10 and 30 years}

\begin{figure*}[ht!]
	\centerline{\includegraphics[width=\textwidth]{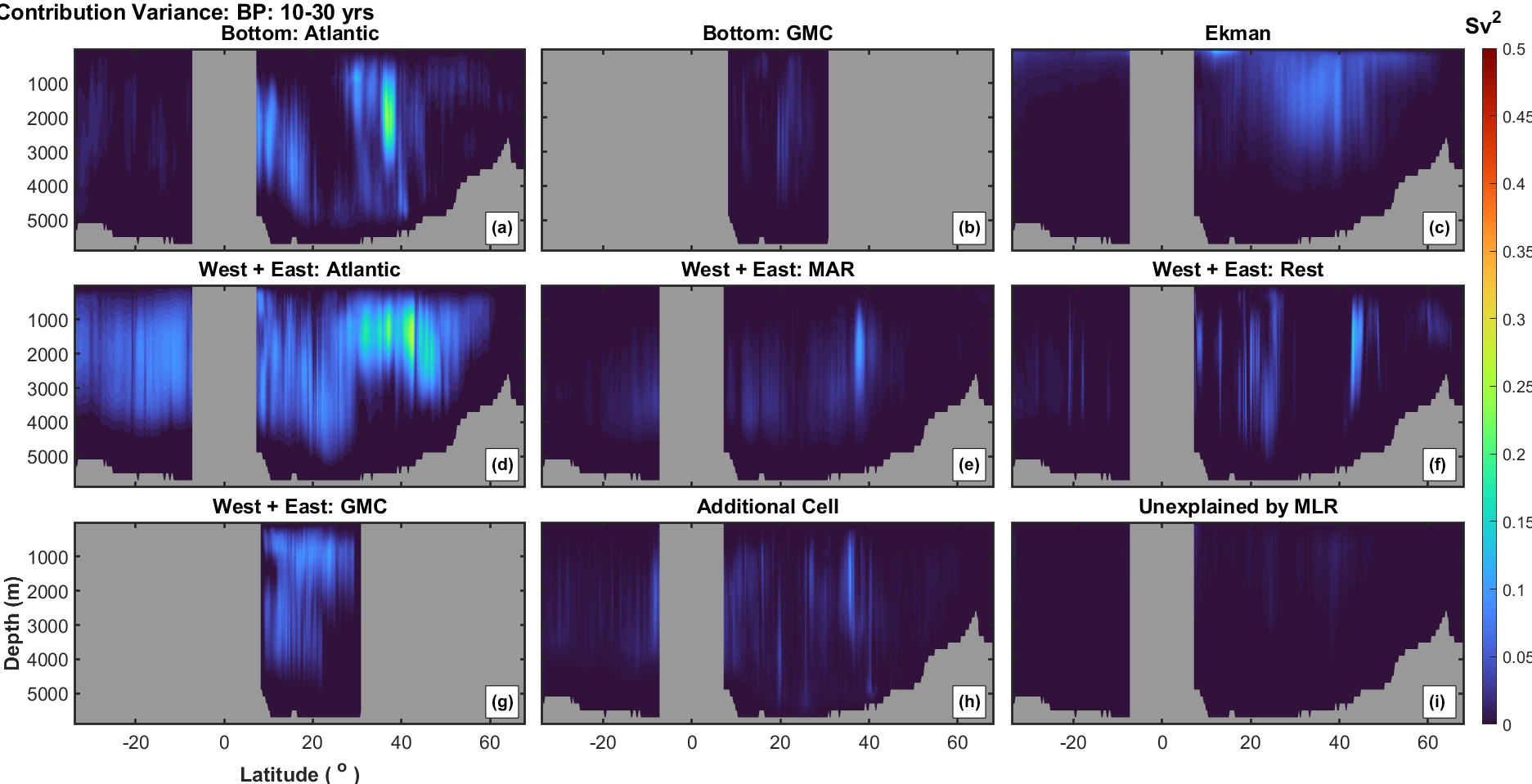}}
	\caption[Scaled CAR score contributions of boundary components to the variance of $\Psi^c$ for band-pass filtered annual-mean compensated overturning streamfunction data, with variation corresponding to periods less than 10 years, or greater than 30 years removed.]{Scaled CAR score contributions of boundary components to the variance of $\Psi^c$ for band-pass filtered annual-mean compensated overturning streamfunction data, with variation corresponding to periods less than 10 years, or greater than 30 years removed. Panels show the contribution made by (a) $\Psi^{c,AB}_{{bot}}$, (b) $\Psi^{c,GMC}_{{bot}}$, (c) $\Psi^c_{Ekm}$, (d) $\Psi^{c,AB}_{W}$ + $\Psi^{c,AB}_{E}$, (e) $\Psi^{c,MAR}_{W}$ + $\Psi^{c,MAR}_{E}$,  (f) $\Psi^{c,R}_{W}$ + $\Psi^{c,R}_{E}$, (g) $\Psi^{c,GMC}_{W}$ + $\Psi^{c,GMC}_{E}$, (h) $\Psi^c_{AC}$ and (i) variance unexplained by MLR-CAR. Estimation is not possible in shaded grey regions corresponding to one or more of bathymetry, latitudes near the equator, and latitudes not relevant for GMC components.}
	\label{f_v_CAR_10_30}
\end{figure*}

\begin{figure*}[ht!]
	\centerline{\includegraphics[width=\textwidth]{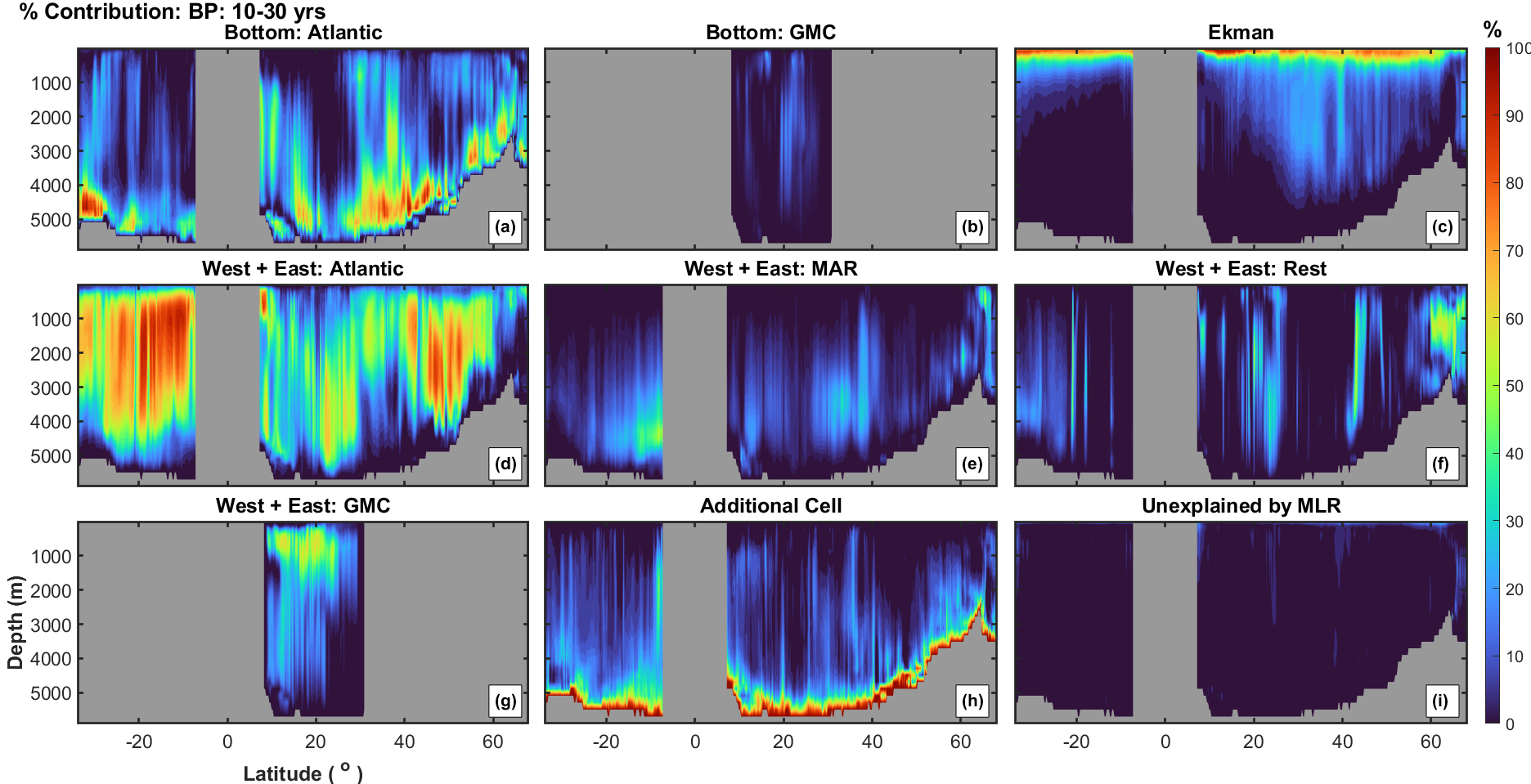}}
	\caption[Percentage scaled CAR score contributions of boundary components to the variance of $\Psi^c$ for band-pass filtered annual-mean compensated overturning streamfunction data, with variation corresponding to periods less than 10 years, or greater than 30 years removed.]{Percentage scaled CAR score contributions of boundary components to the variance of $\Psi^c$ for band-pass filtered annual-mean compensated overturning streamfunction data, with variation corresponding to periods less than 10 years, or greater than 30 years removed. Panel details are given in Figure \ref{f_v_CAR_10_30} caption.}
	\label{f_p_CAR_10_30}
\end{figure*}

Figures \ref{f_v_CAR_10_30} and \ref{f_p_CAR_10_30} show scaled CAR scores for boundary components, and their corresponding percentage contributions, for band-pass filtered annual-mean data with all variation corresponding to periods less than 10 years, or greater than 30 years, eliminated. The figure therefore illustrates the influence of each boundary component on the overturning streamfunction on timescales between 10 and 30 years.

For these timescales, in comparison with the 1-10 year case in Figure \ref{f_v_CAR_1_10}, we see markedly reduced contributions to the total variance of $\Psi^c$ from every component, due to the reduced variance in $\Psi^c$ (Figure \ref{f_v_CAR_T}); the structural features of Figures \ref{f_v_CAR_1_10} and \ref{f_v_CAR_10_30} are otherwise generally similar. Surprisingly, the contribution from $\Psi^{c,AB}_{W}$ + $\Psi^{c,AB}_{E}$ at mid to high northern latitudes remains relatively large, compared with reduced contributions in the subtropics and particularly in the southern hemisphere between $10^\circ$S and $15^\circ$S. 

The reduced contributions to total variance (Figure \ref{f_v_CAR_10_30}) impact the resulting percentage contribution made by each component shown in Figure \ref{f_p_CAR_10_30}. Relative to the analysis of 1-10 year band-pass filtered data in Figure~\ref{f_p_CAR_1_10}, we observe a weakening of the contributions from $\Psi^{c}_{{Ekm}}$ and $\Psi^{c,AB}_{{bot}}$ (Panels (c) and (a)). This is compensated by an increase of boundary density contributions from the Atlantic and GMC (Panels (d) and (g)), supporting the view that the geostrophic component is more important to decadal variability of the overturning ($\Psi^c$) rather than its interannual variations. Atlantic bottom component ($\Psi^{c,AB}_{{bot}}$) remains influential for latitudes between $30^\circ$N and $50^\circ$N, and between $25^\circ$S and $35^\circ$S, where strong boundary currents are present. 


\begin{figure*}[ht!]
	\centerline{\includegraphics[width=\textwidth]{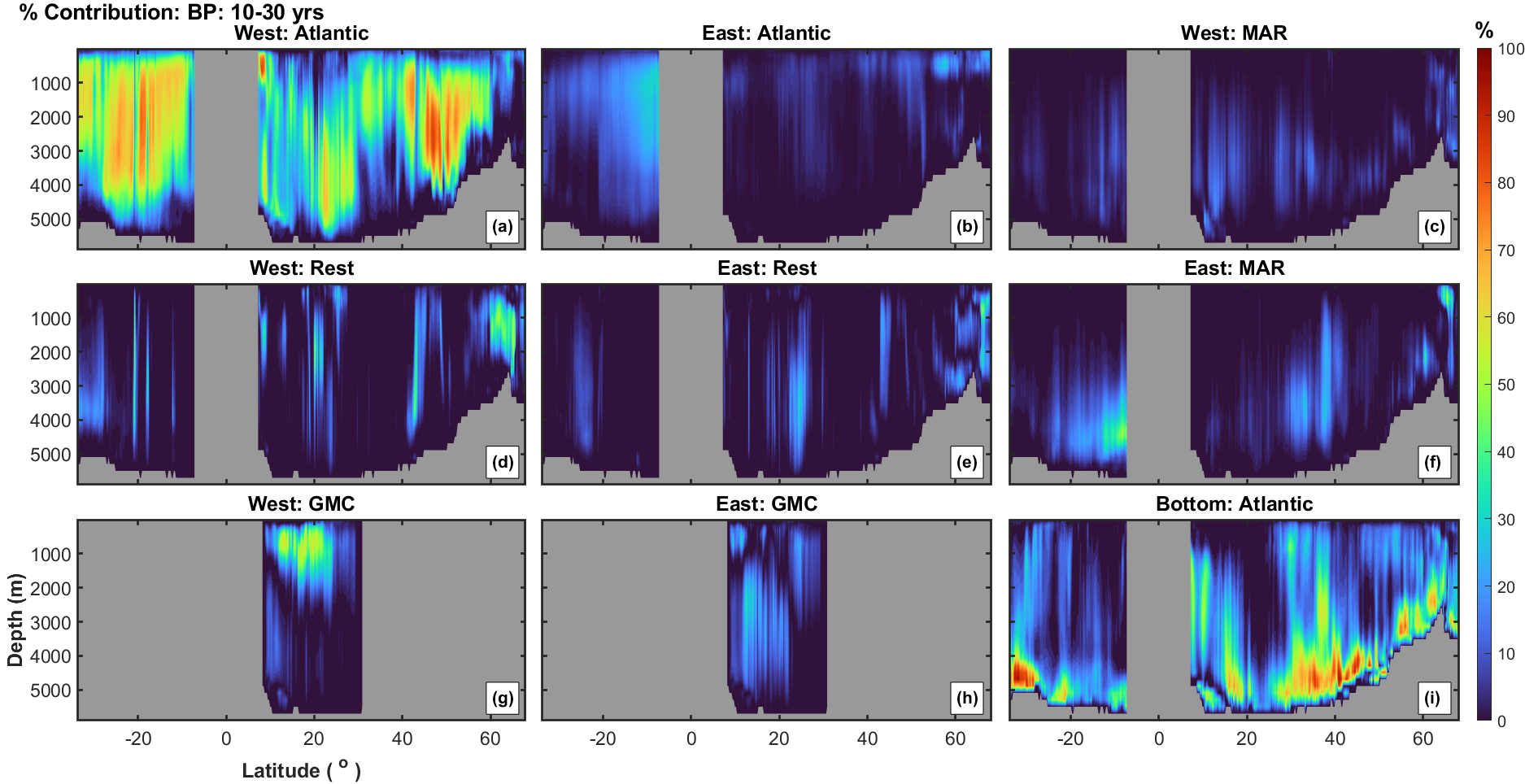}}
	\caption[Percentage scaled CAR score contributions of boundary density components and $\Psi^{c,AB}_{{bot}}$ to the variance of $\Psi^c$ for band-pass filtered annual-mean compensated overturning streamfunction data, with variation corresponding to periods less than 10 years, or greater than 30 years, removed.]{Percentage scaled CAR score contributions of boundary density components and $\Psi^{c,AB}_{{bot}}$ to the variance of $\Psi^c$ for band-pass filtered annual-mean compensated overturning streamfunction data, with variation corresponding to periods less than 10 years, or greater than 30 years, removed. Panels show the contribution made by (a) $\Psi^{c,AB}_{W}$, (b) $\Psi^{c,AB}_{E}$, (c) $\Psi^{c,MAR}_{W}$, (d) $\Psi^{c,R}_{W}$, (e) $\Psi^{c,R}_{E}$, (f) $\Psi^{c,MAR}_{E}$, (g) $\Psi^{c,GMC}_{W}$, (h) $\Psi^{c,GMC}_{E}$ and (i) $\Psi^{c,AB}_{{bot}}$. Estimation is not possible in shaded grey regions corresponding to one or more of bathymetry, latitudes near the equator, and latitudes not relevant for GMC components.}
	\label{f_p_CAR_10_30d}
\end{figure*}
There is evidence for increased importance of the MAR density terms in the southern hemisphere in Panel (e) of Figure \ref{f_p_CAR_10_30}, barely visible in the corresponding panel of Figure \ref{f_p_CAR_1_10}. On closer analysis of constituent boundary density terms in Figure \ref{f_p_CAR_10_30d}, the MAR signal is seen to stem from the contribution of densities on the western side of the ridge, $\Psi^{c,MAR}_{E}$ (Figure \ref{f_p_CAR_10_30d} Panel (f)). 

The decomposition shown in Figure \ref{f_p_CAR_10_30d} indicates the dominant percentage contributions of western boundary densities to the variance of 10-30 year band-pass filtered data, particularly within the interior of the basin; the contribution of $\Psi^{c,AB}_{W}$ dominates within the majority of the basin, $\Psi^{c,GMC}_{W}$ near the surface and $\Psi^{c,R}_{W}$ at high northern latitudes along the coast of Greenland. Panel (b) shows evidence for an increased contribution made by $\Psi^{c,AB}_{E}$ in the southern hemisphere, although it is of small magnitude. In general, the eastern boundary densities play a small role at these timescales too. 

\subsubsection{Band-pass filtering admitting cycles with periods between 30 and 50 years}

\begin{figure*}[ht!]
	\centerline{\includegraphics[width=\textwidth]{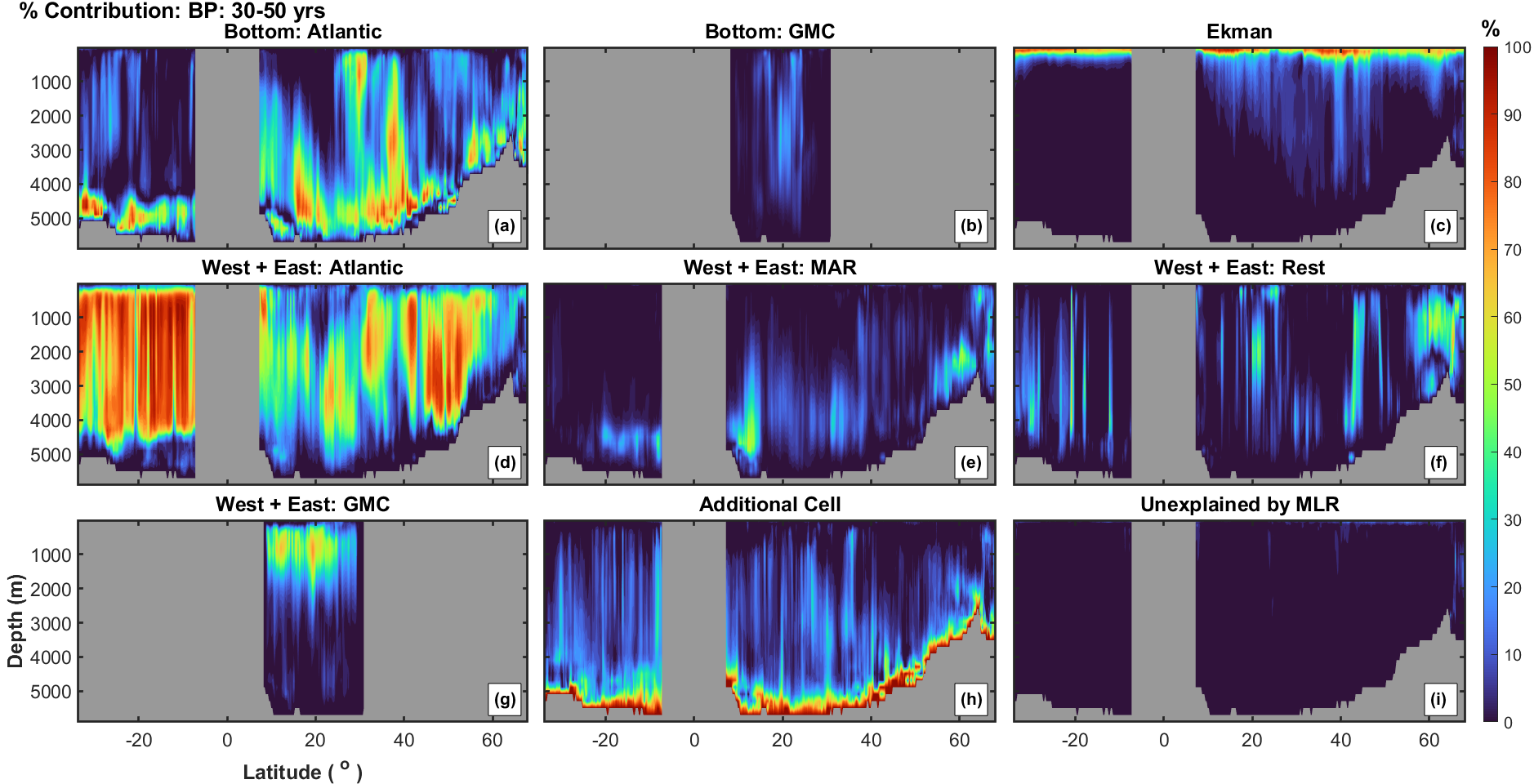}}
	\caption[Percentage scaled CAR score contributions of boundary components to the variance of $\Psi^c$ for band-pass filtered annual-mean compensated overturning streamfunction data, with variation corresponding to periods less than 30 years, or greater than 50 years removed.]{Percentage scaled CAR score contributions of boundary components to the variance of $\Psi^c$ for band-pass filtered annual-mean compensated overturning streamfunction data, with variation corresponding to periods less than 30 years, or greater than 50 years removed. Panels show the contribution made by (a) $\Psi^{c,AB}_{{bot}}$, (b) $\Psi^{c,GMC}_{{bot}}$, (c) $\Psi^c_{Ekm}$, (d) $\Psi^{c,AB}_{W}$ + $\Psi^{c,AB}_{E}$, (e) $\Psi^{c,MAR}_{W}$ + $\Psi^{c,MAR}_{E}$,  (f) $\Psi^{c,R}_{W}$ + $\Psi^{c,R}_{E}$, (g) $\Psi^{c,GMC}_{W}$ + $\Psi^{c,GMC}_{E}$, (h) $\Psi^c_{AC}$ and (i) variance unexplained by MLR-CAR. Estimation is not possible in shaded grey regions corresponding to one or more of bathymetry, latitudes near the equator, and latitudes not relevant for GMC components}
	\label{f_p_CAR_30_50}
\end{figure*}

Figure \ref{f_p_CAR_30_50} shows percentage scaled CAR scores for the boundary components, for band-pass filtered annual-mean data, with all variation corresponding to periods less than 30 years and greater than 50 years eliminated. As previously mentioned, at increased timescale we find weaker total variance, and hence weaker boundary contributions to the total variance (not shown). We find $\Psi^{c,AB}_{{bot}}$ and $\Psi^{c,AB}_{W}$ + $\Psi^{c,AB}_{E}$ scaled CAR scores at 30-50 year timescales (not shown), exhibit a similar structure to that found in the corresponding panels of Figure \ref{f_v_CAR_10_30} (for 10-30 year band-pass filter).  

Compared to Figure \ref{f_p_CAR_10_30}, we note an increase of boundary density contributions (Panels (d) and (g)) at the expense of $\Psi^{c}_{{Ekm}}$ (Panels (c)) and $\Psi^{c,AB}_{{bot}}$ (Panel (a)). Further analysis (not shown) indicates that the boundary density contributions are dominated by $\Psi^{c,AB}_{W}$, with a slight intensification of the $\Psi^{c,AB}_{E}$ contribution between $7^\circ$S and $20^\circ$S in the upper 2000m. The $\Psi^{c,AB}_{W}$ increased contribution at decadal and longer timescales near $45^\circ$N would suggest density variations in DWBC properties and dense waters formed in the Labrador Sea dominate the local variation in the overturning streamfunction in this model simulation. Similarly multi-decadal variability in the southern hemisphere is almost solely dependent on western boundary densities. We find that with increasing timescales, the relative role of $\Psi^{c,AB}_{W}$ + $\Psi^{c,AB}_{E}$ increases within the basin, suggesting that if longer timescales were available, we'd find some agreement with \cite{Waldman2021}, who saw that at centennial timescales, AMOC overturning streamfunction variability in the CNRM-CM6 model is dominated by the thermal wind component (greater than 80\%), and the Ekman component is negligible.

We note an area of increased $\Psi^{c,AB}_{{bot}}$ contribution in the upper 1000m at $30^\circ$N. Interestingly we find that the contribution of southern hemisphere MAR densities at depth in Panel (e) is weaker for these timescales; in contrast, a stronger contribution for latitudes between $10^\circ$N and $20^\circ$N is observed. Further analysis (not shown) indicates that this is due to reduction in $\Psi^{c,MAR}_{E}$ (relative to that observed in the 10-30 year band-pass timescale) in the southern hemisphere, and increase in $\Psi^{c,MAR}_{W}$ at depth near $15^\circ$N at timescales between 30 and 50 years.   

\subsection{Summary of MLR-CAR score analysis for 1$^\circ$ and 1/12$^\circ$ models}

The MLR-CAR score analysis was repeated for the $1^\circ$ and $1/12^\circ$ models, and inferences are summarised here. The analysis procedure is similar to that reported above with one exception. Density contributions for the $1^\circ$ and $1/12^\circ$ models are now assigned to western and eastern basin boundary terms only, in contrast to the regional separation (to Atlantic, MAR and GMC) applied to the $1/4^\circ$ model data. This gives a somewhat less detailed description of the boundary density contributions to the overturning variability. Further, we note that for the $1^\circ$ and $1/12^\circ$ simulations the timeseries are shorter than for the $1/4^\circ$ simulation. As a result, the uncertainty in our estimates increases, especially for the longer band-pass timescales considered.

For percentage scaled CAR scores, the impact of model resolution is minimal; in general, similar cross-basin features are present for component contributions at all resolutions. The relative importance of the Atlantic bottom component $\Psi^c_{bot}$ appears to increase with increased resolution, especially in the interval $30^\circ$N to $45^\circ$N due to increased fidelity and strength of bottom velocities. $\Psi^c_E$, representing all eastern boundary density contributions, increases for the $1^\circ$ model at 10-30 year timescale and the $1/12^\circ$ model at 30-50 year timescale, for depths around 1000-3000m between $7^\circ$S and $25^\circ$S. However, the relatively short lengths of the timeseries from the $1^\circ$ and $1/12^\circ$ models suggests that we should interpret our findings more cautiously, relative to the $1/4^\circ$ model.

\begin{figure*}[ht!]
	\centerline{\includegraphics[width=\textwidth]{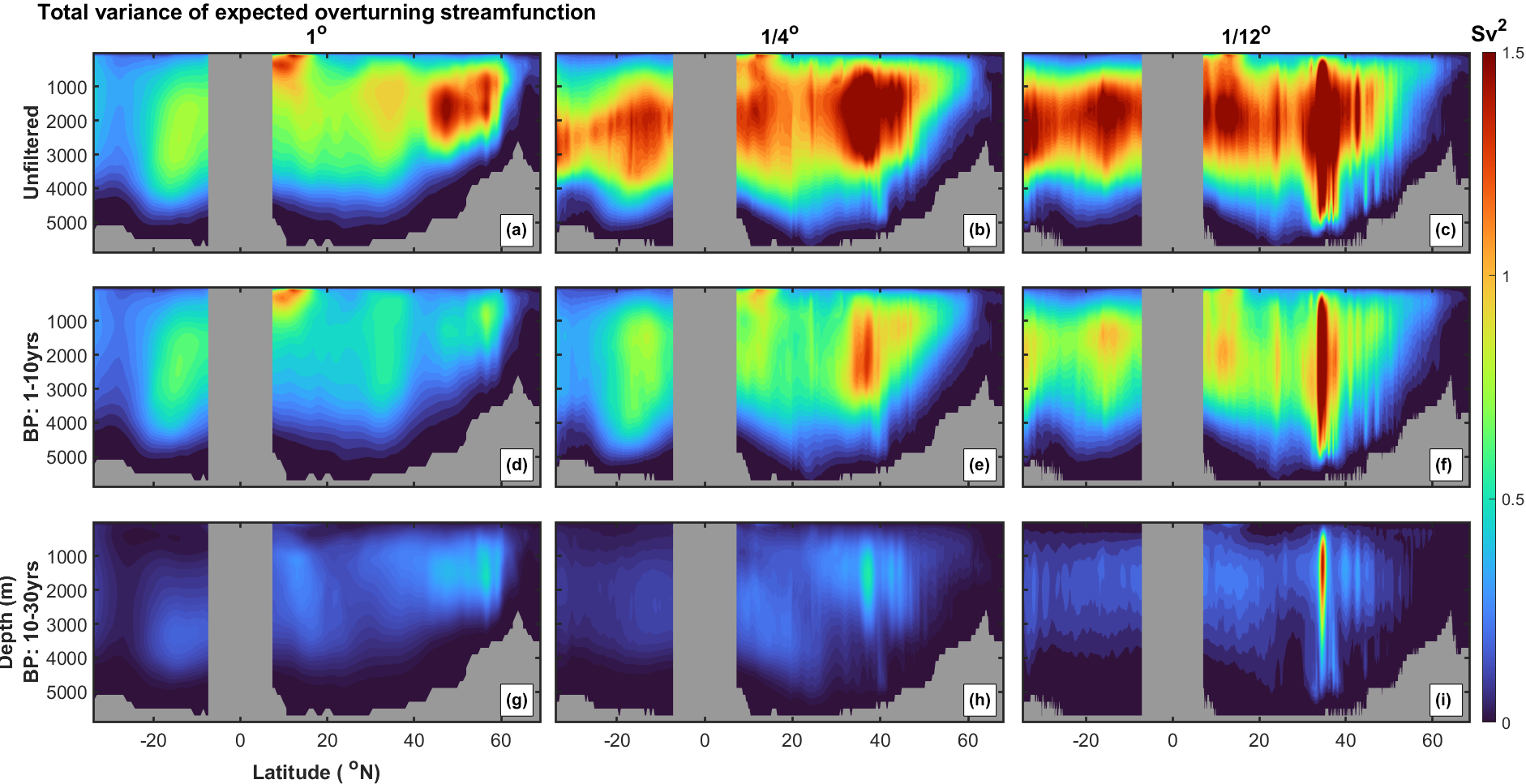}}
	\caption[Total variance of $\Psi^c$ for different model resolutions and band-pass filtered periods.]{Total variance of $\Psi^c$ for different model resolutions and band-pass filters. (a) unfiltered $1^\circ$, (b) unfiltered $1/4^\circ$, (c) unfiltered $1/12^\circ$, (d) band-pass filtered 1-10 years $1^\circ$, (e) band-pass filtered 1-10 years $1/4^\circ$, (f) band-pass filtered 1-10 years $1/12^\circ$, (g) band-pass filtered 10-30 years $1^\circ$, (h) band-pass filtered 10-30 years $1/4^\circ$ and band-pass filtered 10-30 years $1/12^\circ$. Estimation is not possible in shaded grey regions corresponding to one or more of bathymetry and latitudes near the equator.}
	\label{f_vRes_CAR}
\end{figure*}

The total variance of the overturning streamfunction $\Psi^c$ is shown in Figure \ref{f_vRes_CAR}, for all model resolutions and three band-pass timescales (including unfiltered annual-mean data). For unfiltered data in particular, the total variance of $\Psi^c$ is considerably greater for $1/4^\circ$ and $1/12^\circ$ model resolutions. Interestingly, the region of high variance around $37^\circ$N seen in the $1/4^\circ$ and $1/12^\circ$ models is not present at $1^\circ$, regardless of the band-pass timescale.

In terms of scaled CAR score contribution to the total variance of $\Psi^c$, we find for short timescales (band-pass filtered 1-10 years, not shown) a weaker contribution from $\Psi^c_W$ + $\Psi^c_{E}$ for the $1^\circ$ model especially in the southern hemisphere, reflecting the lower total variance of $\Psi^c$ at this coarsest resolution (Figure \ref{f_vRes_CAR}). 

Scaled CAR score boundary contributions of the $1/12^\circ$ model show similar features to those of the $1/4^\circ$ model. The influence of $\Psi^c_{bot}$ penetrates to even shallower water near $37^\circ$N. At longer timescales we find that contributions decrease in a similar fashion to the $1/4^\circ$ model, except for the contribution of $\Psi^c_{bot}$ in a small region between 500-4000m at $37^\circ$N, which is uncharacteristically strong for both 10-30 year and 30-50 year band-pass timescales (reflected in Panels (f) and (i) of Figure \ref{f_vRes_CAR}). The unexplained variance is also relatively large within the upper 1000m at $37^\circ$N. Figure \ref{p_RC_Bvel} of Chapter \ref{TJ_TM} shows large positive and negative meridional velocities near $37^\circ$N for the $1/12^\circ$ model. These changes in velocities are likely to be due to the New England or Corner Rise seamounts, two chains of extinct volcanic mountains, off the coast of Massachusetts. These seamounts do not appear in the $1^\circ$ model bathymetric or bottom velocity data. However, they are partially resolved in the $1/4^\circ$ model, but not to the same extent as the $1/12^\circ$ model. This suggests the ability of the $1/12^\circ$ model to resolve smaller-scale features results in greater variability in this region.

\section{Summary}

In this chapter, we have investigated the spatial and temporal variability of the AMOC and factors contributing to it. 

The decomposition diagnostic has been shown to give good quantification of the maximum overturning streamfunction at $26.5^\circ$N and $34.5^\circ$S, compared with observations taken from the RAPID and SAMBA arrays, and the expected maximum of the compensated overturning streamfunction $\Psi^c_{\max}$ calculated directly from meridional velocities. The maximum of the compensated estimated overturning streamfunction ($\tilde{\Psi}^c_{\max}$) explains a large proportion of the temporal variability in the maximum compensated expected overturning streamfunctions. We find high correlation between thermal wind component ($\Psi^c_{th_{\max}} = \Psi^c_{W_{\max}}+\Psi^c_{E_{\max}}$, sum of western and eastern boundary density components) and the depth-independent component ($\Psi^c_{bot_{\max}}$, bottom) at SAMBA, especially within high resolution models, clearly visible as large oscillations in the $1/4^\circ$ model. 

We considered the role of SSH, bottom velocities and bottom densities for (a) years with maxima and minima in $\Psi^c_{bot_{\max}}$ and (b) specific time-intervals of increasing or decreasing $\Psi^c_{bot_{\max}}$. We find that north-south variation of the latitude of the Brazil-Malvinas confluence impacts western boundary bottom velocities and densities resulting in fluctuation in depth-independent (bottom) and thermal wind (density) contributions to the estimated maximum compensated overturning streamfunction at SAMBA.

Standard deviations of bottom densities at increasing timescales reveal large variations along the western boundary (regions of strong currents) within both higher resolution models, showing larger values of 10-year variability $\overline{s}_{10}$ and 40-year variability $\overline{s}_{40}$ in the Brazil-Argentine basins. Bottom velocities also show large 10-year variability $\overline{s}_{10}$ at high resolutions especially in regions of strong currents (such as the North Brazil, Gulf Stream, North Equatorial currents).

A multiple linear regression model incorporating Correlation Adjusted coRrelation scores (MLR-CAR, \citealt{Zuber2010}) was used to estimate the contributions of boundary components to the total variance of the expected compensated overturning streamfunction $\Psi^c$, using timeseries subject to different band-pass filters, to expose the influence of different timescales. At all timescales, large scaled CAR score contributions to the total variance of $\Psi^c$ are estimated for (a) the Ekman contribution, especially in the upper 2500m for the northern hemisphere, (b) the Atlantic boundary depth-independent (bottom velocity, $\Psi^{c,AB}_{{bot}}$) contribution at mid-depths between $30^\circ$N and $45^\circ$N, and (c) the western Atlantic boundary density component within the majority of the ocean's interior. When the influence of longer timescales are removed from the data (e.g. using a 1-10 year band-pass filter), western (mainly southern hemisphere) and eastern (northern hemisphere) Atlantic boundary density components show a reduced contribution to the total variance. The western Gulf of Mexico and Caribbean Sea boundary density contribution is particularly large at shallow depths, suggesting that surface variance of the total expected overturning streamfunction is dictated by densities in the Gulf of Mexico and Caribbean Sea at these latitudes.

MLR-CAR results based on band-pass filtered data, for different band-pass intervals, suggest that the largest contributions to total variance of the expected overturning streamfunction occur at short timescales (annual, multi-year). Total variance of the expected compensated overturning streamfunction within the interior of the southern hemisphere is (western) boundary-density-dominated, with an influential Atlantic boundary depth-independent contribution at depth also important, and Ekman contribution dominant near the surface. Large variation is found due to the Brazil current, visible within the western Atlantic boundary density and Atlantic bottom velocity terms. In the northern hemisphere, contributions are evenly distributed between the Ekman, western Atlantic boundary density and Atlantic boundary depth-independent contributions, with the latter prevalent throughout the fluid column near $35^\circ$N, possibly due to the interaction of the DWBC and Gulf Stream. The contribution of additional cells is small on a basin-wide scale, but locally crucial in explaining the variance of the expected compensated overturning streamfunction near bathymetry. At longer timescales, the relative importance of the western Atlantic boundary densities increase in the southern hemisphere interior. Interesting localised contributions from eastern and western mid Atlantic ridge boundary density contributions are observed near the equator, possibly connected to Antarctic bottom water moving northward through fracture zones. There is some evidence for important boundary density contributions along the Greenland coast, shown via the boundary density remainder component. 

Estimates of the total variance of the expected overturning streamfunction are considerably larger for the $1/4^\circ$ and $1/12^\circ$ model resolutions than for the $1^\circ$ model. However, the impact of model resolution is found to be minimal for percentage scaled CAR score contributions of boundary components. 
 \clearpage

\chapter{Hydrographic properties on ocean boundaries}  \label{TJ_Bdry}

This chapter investigates along-boundary properties for the Atlantic basin and surrounding coastlines.  We characterise the spatial and temporal structure of the boundary densities underpinning the decomposition diagnostic for the overturning streamfunction, applied to the $1^\circ$, $1/4^\circ$ and $1/12^\circ$ HadGEM-GC3.1 model simulations in Chapters \ref{TJ_TM} and \ref{TJ_Var}. We assess the extent to which isopycnals are flat along the (sloping) eastern boundary, as is commonly assumed within reduced-gravity models.

We introduce a boundary mapping algorithm, motivated by the work of \cite{Hughes2018}, to map large sections of along-shore continental boundaries across multiple depths. We use the mapping algorithm to explore spatio-temporal variations in potential temperature, salinity, neutral and potential density for (HadGEM-GC3.1) model , (GloSea5) reanalysis and (Gouretski and Kolterman) climatological datasets, centred on the Atlantic basin.

\section{Background}

For a meridional pressure gradient to be sustained on the eastern boundary, a velocity into the boundary is required. Since the boundary is impenetrable, coastally-trapped boundary waves propagate and distribute any eastern boundary anomaly evenly along the boundary, and therefore flat eastern boundary isopycnals are likely (Section \ref{I_adjAMOC}). Flat eastern boundary isopycnals imply that stratification of the water column does not vary with latitude. Hence, given that Rossby waves transmit eastern boundary information into the ocean's interior, density stratification is quasi-uniform over much of the domain. This raises the possibility of using properties from only a few stratification profiles to constrain the overturning at latitudes where no data is available, and quantifying eastern boundary contributions to the overturning streamfunction using just a small number of profiles.

The assumption of flat isopycnals along the eastern boundary has been widely used within idealised models (e.g. \citealt{Johnson2002}, \citealt{Cessi2010}, \citealt{Nikurashin2011}, \citealt{Marshall2017}, \citealt{Nieves2018}). However, previous work (e.g. \citealt{Johnson2002}, \citealt{Sun2016}) has shown that the assumption of flat isopycnals is only valid away from the surface, since here Ekman processes dominate, and geostrophy breaks down. Further, \cite{Cessi2013a} argue that instability processes along the eastern boundary act to erode the density structure. They highlight the importance of agesotrophic processes in locations where adiabatic laminar thermocline theories (\citealt{Luyten1983}, \citealt{Pedlosky1983}, \citealt{Gill1985}) predict a singularity as a result of opposing needs for (1) a meridional gradient in buoyancy at the surface and (2) the boundary condition of no normal flow. The relative flatness of eastern boundary isopycnals in basins to the north of the circumpolar Southern Ocean (e.g. Atlantic basin, \citealt{Gnanadesikan1999}, \citealt{Johnson2007}, \citealt{Samelson2009}, \citealt{Radko2011}, \citealt{Nikurashin2011}, \citealt{Shakespeare2012}, \citealt{Marshall2013}, \citealt{Sun2016}, \citealt{Marshall2017}) has been exploited within diagnostic models; the success of this approach may be due to the contrasting roles of the eastern and western boundaries in establishing the hydrographic structure and circulation found in ocean basins, as shown by \cite{Johnson2002}. See Section \ref{I_adjAMOC} for further information.

An important area of research in this thesis is the role that boundary characteristics play in describing the AMOC and its variability. In particular, this requires understanding and exploitation of the density structures on the eastern and western boundaries, which underpin the decomposition diagnostic.

There have been no previous attempts to map density, potential temperature or salinity properties along sloping boundaries of ocean basins, or to characterise their spatial variation. Our ability to map boundary properties by direct observation is limited due to the huge distances involved, and the expense of hydrographic studies. The present work to improve our understanding of the along-boundary properties is motivated in part by the work of \cite{Hughes2018} in mapping bottom pressures.  

\section{Preliminary analysis of boundary densities}
\label{Bdry_exp} \label{Bdry_EW}
The objective of this section is to provide an initial visual inspection of eastern and western boundary potential densities for the Atlantic basin, and to motivate further investigation of along-boundary variation.  We explore the time-mean boundary potential densities (referenced to the surface) for HadGEM-GC3.1 models at three different spatial resolutions, as used in the decomposition diagnostic. Findings are illustrated as latitude-depth contour plots along the sloping eastern and western boundaries.

In estimating a latitude-depth plot of boundary potential density, eastern and western boundaries are defined, at each latitude, as the easternmost and westernmost water boundaries available for the Atlantic basin at the required depth. This mimics the approach used in the development of the overturning streamfunction decomposition diagnostic in Section \ref{S_App_DD}, with intermediate boundaries ignored. Therefore, for each latitude-depth pair, there is a unique longitude corresponding to each of the eastern and western boundaries, and unique eastern and western boundary density values.

%
Figures \ref{F_sBdry_12_300}-\ref{F_sBdry_12_5900} show time-mean potential densities (relative to the surface and averaged over the full time periods of $1/12^\circ$ model output) for western and eastern boundaries, and west-east difference.

\subsection{General features of time-average boundary potential densities at $1/12^\circ$ resolution} 
\begin{figure*}[ht!]
	\centerline{\includegraphics[width=\textwidth]{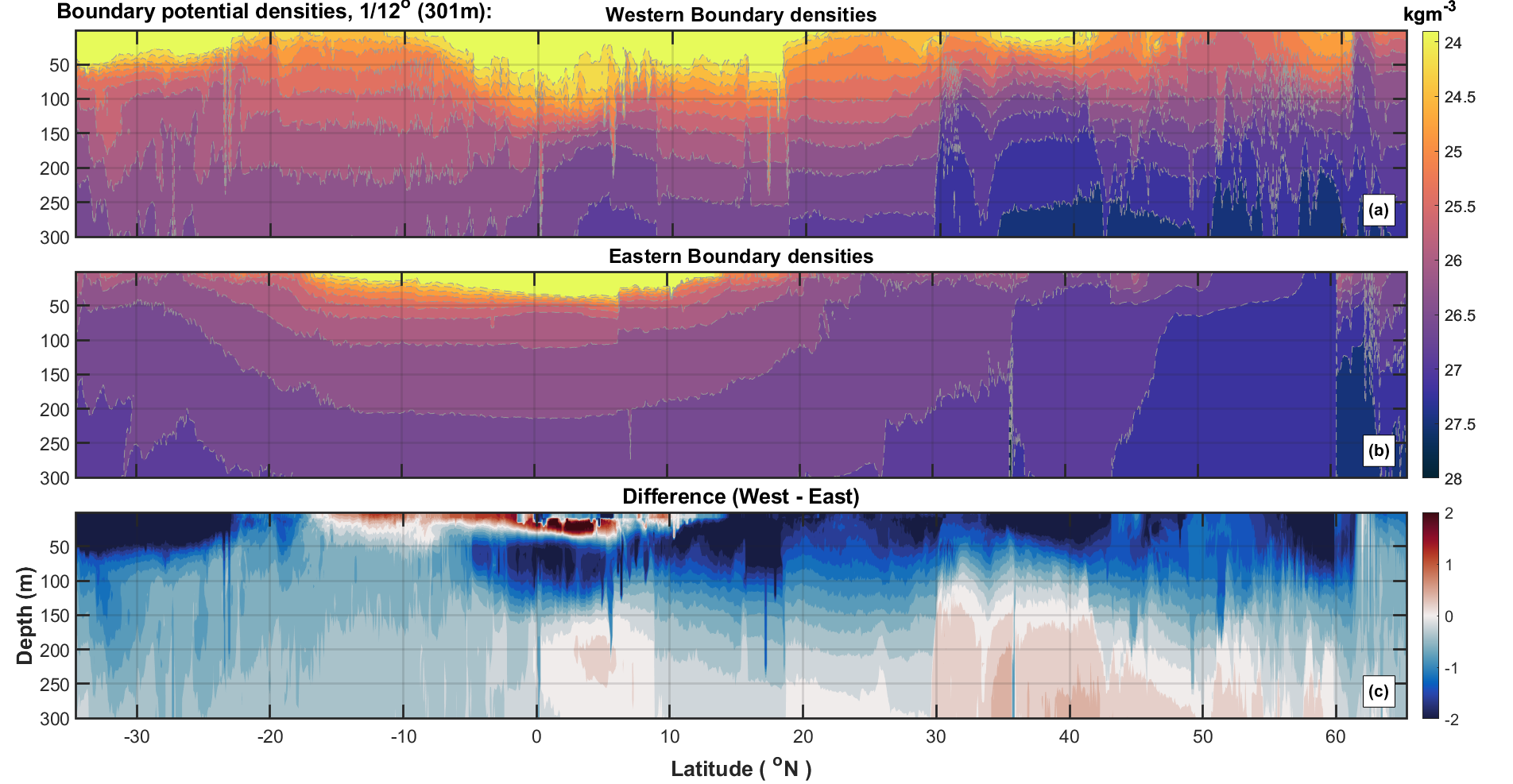}}
	\caption[Western and eastern Atlantic boundary potential densities for the upper $301$m in the $1/12^\circ$ model.]{Western (a) and eastern (b) Atlantic boundary potential densities (referenced to the surface) for the upper $301$m in the $1/12^\circ$ model. Panel (c) indicates the difference between western (a) and eastern (b) densities. Time-mean taken over the 176-year control run.}
	\label{F_sBdry_12_300}
\end{figure*}

Figure \ref{F_sBdry_12_300} for the upper $300$m shows considerable spatial variation in boundary potential density, and suggests that the hypothesis of flat eastern boundary isopycnals is not valid in the mixed layer, as we might expect. However, Figure \ref{F_sBdry_12_300}(b) shows that for latitudes between $15^\circ$S and $10^\circ$N there is some evidence of approximately constant boundary density with latitude on the eastern boundary. We find surface densities near the equator to be lightest due to warming from the sun (surface heat fluxes). Along the western boundary we find that lighter surface densities extend further towards the poles due to western boundary currents (e.g. Gulf Stream and North Brazil current) redistributing warmer conditions poleward.

At high northern latitudes in Figure \ref{F_sBdry_12_300}(a,b), there is evidence for outcropping of isopycnals on both boundaries. We find some similar tendencies on the eastern boundary in the southern hemisphere. The apparent discontinuity in Figure \ref{F_sBdry_12_300}(a) at approximately $30^\circ$N is because the western boundary shifts here from the Gulf of Mexico to the Florida Straits, highlighting a weakness of the current approach to boundary definition, and motivating the improved approach introduced in Section \ref{CntMpBnd}. 

Figure \ref{F_sBdry_12_300}(c) illustrates that, in the upper $100$m of the Atlantic, western boundary waters are generally less dense than the corresponding eastern boundary waters. The reverse trend at 300m for latitudes $30^\circ-50^\circ$N is attributed to lighter outflow waters from the Mediterranean Sea, also leading to a discontinuity in eastern boundary isopycnals at these latitudes.

\begin{figure*}[ht!]
	\centerline{\includegraphics[width=\textwidth]{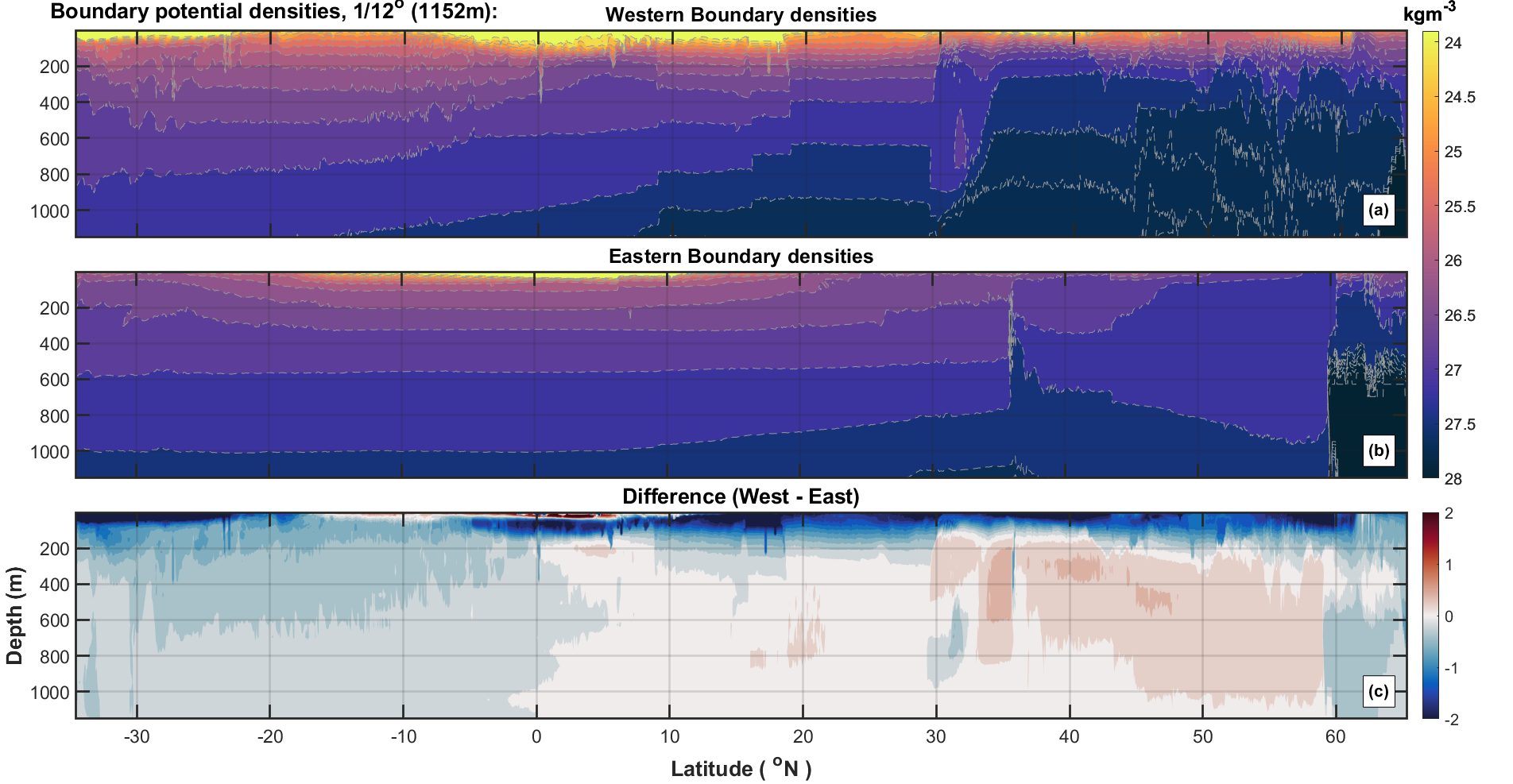}}
	\caption[Time-average western and eastern Atlantic boundary potential densities for the upper $1152$m in the $1/12^\circ$ model.]{Western (a) and eastern (b) Atlantic boundary potential densities (referenced to the surface) for the upper $1152$m in the $1/12^\circ$ model. Panel (c) indicates the difference between the eastern (b) and western (a) densities. Time-mean taken over the 176-year control run.}
	\label{F_sBdry_12_1100}
\end{figure*}

Figure \ref{F_sBdry_12_1100}, the corresponding latitude-depth density contour plot for the upper $1100$m, shows many of the same features as Figure \ref{F_sBdry_12_300}. Panel (a) suggests that western isopycnals vary slowly and approximately linearly (slowly rising northwards) with latitude south of $20^\circ$N, at depths outside the mixed layer. In contrast, Panel (b) shows eastern boundary isopycnal depths vary little with latitude, south of $35^\circ$N and outside the mixed layer. North of $35^\circ$N, the lack of continuity in densities along the eastern boundary suggests a discontinuous boundary or the action of ageostrophic processes, resulting in a breakdown of the flat isopycnal hypothesis. Variation in isopycnal depth near $35^\circ$N can be attributed to the mixing of warmer saltier Mediterranean waters with those of the Atlantic near the Strait of Gibraltar. Above $60^\circ$N on the eastern boundary, the fingerprint of denser waters formed at higher northern latitudes is clear at depths greater than 600m. On the western boundary, at high northern latitudes, denser water masses are introduced due to the southward flowing DWBC and Labrador Current.

\begin{figure*}[ht!]
	\centerline{\includegraphics[width=\textwidth]{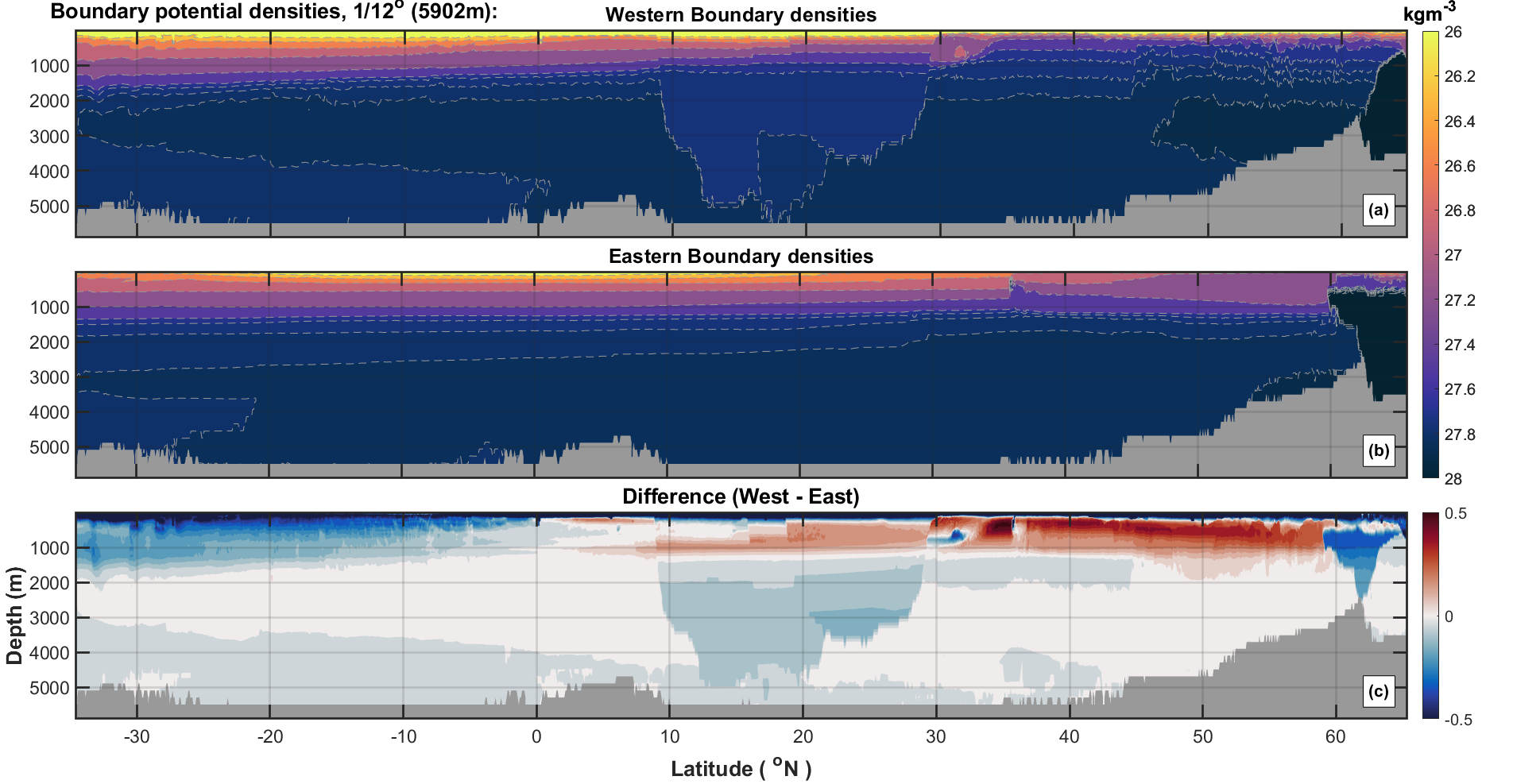}}
	\caption[Western and eastern Atlantic boundary densities for the upper $5902$m in the $1/12^\circ$ model.]{Western (a) and eastern (b) Atlantic boundary potential densities (referenced to the surface) for the upper $5902$m in the $1/12^\circ$ model. Panel (c) indicates the difference between the eastern (b) and western (a) densities. Time-mean taken over the 176-year control run.}
	\label{F_sBdry_12_5900}
\end{figure*}

Figures \ref{F_sBdry_12_1100}(c) and \ref{F_sBdry_12_5900}(c) shows the difference between eastern and western boundary densities, and are consistent with a resulting northward transport in both hemispheres, in the upper 2000m and outside the mixed layer. In the southern hemisphere, denser eastern waters and a negative Coriolis parameter lead to a general northward volume transport. This is inferred by using the geostrophic relationship, denser eastern waters result in a negative pressure gradient across the basin, combined with a negative Coriolis parameter, gives positive meridional (northward) velocities. In contrast, in the northern hemisphere, denser western boundary waters combine with a positive Coriolis parameter, again leading to northward transport. At approximately $62^\circ$N, we find a large region of denser eastern boundary water, likely to be southward flowing from the Norwegian Sea. Figures \ref{F_sBdry_12_5900}(a) and (b) show the presence of denser boundary waters at depth for the highest northern latitudes, attributed to denser waters pooling north of the Greenland-Scotland ridge, with Denmark Strait and Iceland-Scotland overflows resulting in regions of denser waters south of the ridge. At latitudes greater than $63^\circ$N, signs of dense waters are clear on both the eastern and western boundaries; Figures \ref{F_sBdry_12_5900}(a) and (b) show the NADW tongue formed at high latitudes flowing southwards, extending further on the western boundary. In the southern hemisphere of Panel (b), above 2500m, isopcynals exhibit approximately linear structure. Note that the discontinuity on the western boundary in Figure \ref{F_sBdry_12_5900}(a) at around $10^\circ-30^\circ$N is again an artefact of the simplistic approach adopted here to identify boundaries, which can be eliminated by a wiser choice of continuous boundary based on depth contours, as explored in Section \ref{CntMpBnd}.

\subsection*{Impact of model resolution on time-average boundary potential densities}
Plots equivalent to Figures \ref{F_sBdry_12_300}-\ref{F_sBdry_12_5900} at $1/4^\circ$ and $1^\circ$ spatial resolutions (not shown), reveal the general isopycnal structure is very similar across model resolution. We find the isopycnal structure with latitude becomes increasingly smooth, with decreasing model resolution.

Discontinuities are again apparent at approximately $30^\circ$N on the western boundary and $35^\circ$N on the eastern boundary, but less obvious at $1^\circ$, and denser southward flowing waters at high northern latitudes can be seen. We note that the eastern boundary densities at $1^\circ$ spatial resolution do not show obvious signs of Mediterranean outflow. Eastern boundary isopycnals are again approximately flat south of $35^\circ$N, and western boundary isopycnals show an approximately linear isopycnal structure (slowly rising northwards) with latitude, with some discontinuities attributed to contributions from the western Gulf of Mexico for shallower depths, and eastern boundary of the Caribbean Islands at depth. When interest lies in the general time-average characteristics of eastern and western boundary densities with latitude and depth, there appears to be little to gain from increased model resolution. However, we expect that local spatial features are more adequately represented using a higher-resolution model. 

\subsection*{Reconstruction of isopycnal structure using depth profiles from a limited number of locations}
\label{Bdry_MCMC}

There is evidence that the isopycnal structure on an eastern boundary of an ocean basin is less complex, as a function of latitude and depth, than on the western boundary. For example, on the eastern boundary of the Atlantic, we hypothesise that density profiles with depth over a large range of latitudes can be approximated by just a small number of depth profile measurements at well-chosen latitudes. Using just these along slope density profile measurements, linear interpolation is then adequate to estimate density profiles at other latitudes of interest, along both eastern and western boundaries. This would yield a simple useful low-order reconstruction of eastern boundary density in particular. In Appendix \ref{Bdry_MCMC_App}, we quantify the extent to which this is the case on both eastern and western boundaries of the Atlantic. We develop a piecewise linear model for time-mean potential density, using the minimum number of linear pieces to reconstruct along-boundary density structure.

The analysis presented in Appendix \ref{Bdry_MCMC_App} provides a means of automatically selecting locations for depth profile measurements of density along both eastern and western boundaries, in order to provide optimal reconstruction of the density structure (with latitude and depth) for the whole basin. Results indicate that the number of profiles required to reconstruct the eastern boundary density structure is approximately half that required for the western boundary. The methodology raises the possibility of further simplifying the description of boundary density contributions to the AMOC decomposition diagnostic for the whole Atlantic basin, and could be particularly valuable for paleooceanographic applications (discussed in Section \ref{Smm_Dsc}).

\section{Densities along continuous boundaries}
\label{CntMpBnd}

\subsection{Motivating an improved boundary representation}
\label{Bdry_NM_mot}
The results in Section \ref{Bdry_EW} above support the hypothesis of flat Atlantic eastern boundary isopycnals beneath the mixed-layer, south of $35^\circ$N. Moreover, western boundary isopycnals appear to vary approximately linearly with latitude (slowly rising northwards) throughout the basin, but particularly south of $35^\circ$N. However, the approach taken to identify boundaries so far is simplistic, and leads to discontinuities in boundary properties. Specifically, for each latitute-depth pair, a unique longitude is identified corresponding to each of the eastern and western boundaries. Discontinuities in latitude-depth plots appear when adjacent boundary locations are at similar latitudes but different longitudes, e.g. in the Gulf of Mexico, Florida Straits and the zonal boundary of the southern coast of west Africa. 

We now develop a more sophisticated boundary mapping algorithm, able to map basin boundaries for the whole planet continuously at any depth of interest. We expect that this mapping algorithm will provide more reliable estimates of the structure of eastern and western boundary isopycnals for the Atlantic and other basins. The mapping procedure quantified the continuous boundary in terms of a set of discrete locations. Multiple adjacent locations at the same latitude are possible, in contrast to the method used in Section \ref{Bdry_exp}.

The procedure motivated by \cite{Hughes2018} consists of two main stages. In the first (described in Section \ref{CntMpBnd.1}), the Atlantic boundary location is estimated for a given fixed depth. In the second (Section \ref{CntMpBnd.2}), the boundary location corresponding to any ``remote'' second depth is registered (or paired) with the original reference boundary. The final boundary set is constructed by registering (or pairing) the boundary locations at all depths with the reference boundary. We note that the procedure is performed using depth indices ($k$, denoting vertical levels within a model) in place of actual depth.

\subsection{Mapping the boundary at a given depth}
\label{CntMpBnd.1}

We seek a pathway in longitude and latitude which estimates a bathymetric contour around the Atlantic basin at a specified depth. At that depth, we create a binary mask on the model longitude-latitude grid, partitioning ``ocean'' from ``land''. The pathway is created by moving (through mask locations corresponding to ocean only) from a starting point to an end point along the coastline, always keeping the land (i.e. mask locations corresponding to land) to the left in the direction of travel. 

Where necessary, we specify extra artificial land barriers to prevent pathways into regions not of interest in this study. For example, a barrier at latitude 76$^\circ$N across the Labrador and Nordic Seas avoids pathways entering the Arctic region. The steps of the algorithm are as follows.

\textbf{Identify pathway start and end points:} Longitude and latitude indices for initial start and initial end points in the ocean are specified, along with the ``depth'' $k$-level of interest. When the initial start and end points do not lie adjacent to land, the initial start and end points are iterated in one of four pre-specified cardinal directions until land is found (at the specified depth). The resulting points are adopted as the start and end points for the pathway. 

\begin{figure*}[ht!]
	\centerline{\includegraphics[width=\textwidth]{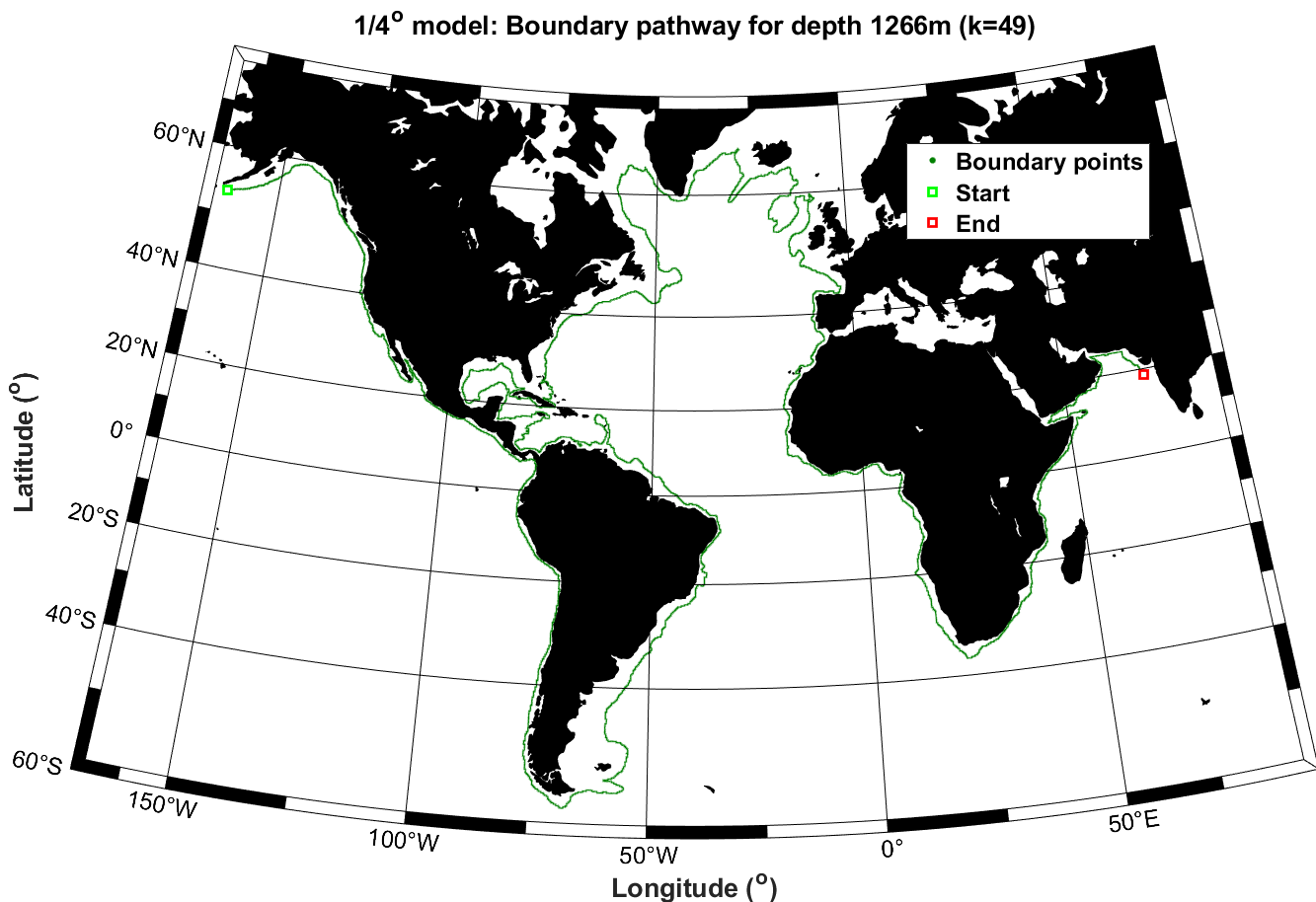}}
	\caption[Boundary pathway at depth level $k=49$ (corresponding to 1266m), for the $1/4^\circ$ model, running from the coast of Alaska to India.]{Boundary pathway at depth level $k=49$ (corresponding to 1266m, in green), for the $1/4^\circ$ model, running from the coast of Alaska to India. The start and end points of the pathway are denoted by green and red squares, respectively.}
	\label{F_mapBdry_k49}
\end{figure*}

\textbf{Iterate around coastline:} From the given starting point, we iterate (or progress) around the coastline (at the specified depth) keeping the land on the left hand side in the direction of travel. This is achieved by attempting iterations in strict sequence relative to the direction of travel, taking the first available option: (1) to the left, (2) forward, (3) to the right, (4) backward (\citealt{Pavlidis1982}, Moore-neighbour tracing algorithm). We note that the backward step is always possible (since it is already a point on the pathway); in practice, for the Atlantic basin mapping, the backward step is never required. Hence, with starting point near the coast of Alaska, the pathway evolves ``hugging'' the Alaskan coast southwards, down the American coast and around the tip of South America into the Atlantic basin. The pathway then follows the east coast of South America northwards through the Caribbean Sea and Gulf of Mexico up to the Labrador Sea. Here, an artificial barrier prevents entry into the Arctic Ocean, resulting in the pathway incorporating the coastline of Greenland, before continuing eastwards and southwards around the west of the UK, continental Europe and Africa. The pathway ends at the end point (off the coast of north-west India). The resulting pathway for $k=49$ in the $1/4^\circ$ model is illustrated in Figure \ref{F_mapBdry_k49}.

\textbf{Estimate the contour by smoothing the pathway:} The pathway found consists of straight line segments between locations on the model longitude-latitude grid. We obtain a more continuous contour (at the specified depth) by smoothing the pathway. Smoothing is particularly important to ensure that estimates of distances along the contour are not inflated (due to the potential ``zig-zag'' nature of a pathway on the discrete longitude-latitude grid). Smoothing is performed using a $\pm n_{\text{HW}}$ point vector moving-average (with respect to angles of longitude and latitude) to yield the contour estimate. Great circle distances between points on the contour (i.e. the smoothed pathway) are then calculated, so that distances between arbitrary points on the contour can be determined. In the current work, we find that $n_{\text{HW}}=3$ provides reasonable estimates for contour distances. 

\subsection{Pairing any other depth contour to the reference contour}
\label{CntMpBnd.2}

To analyse density properties across multiple contours at different depths, it is advantageous to have a common measure of distance along all contours. To achieve this, we choose a ``reference'' contour, to which every other depth contour will be ``registered''; in the present analysis of the Atlantic basin, we choose the $2262$m ($k=56$) depth contour as reference. Here, we describe how the registration of other depth (``remote'') contours onto the reference contour is performed.

\textbf{Identify nodes on the reference contour}: We identify equally-spaced locations along the reference contour, and call these reference nodes (\citealt{Hughes2018}). In the example discussed below, an inter-node interval of 400km is used. For definiteness, we assume we find $N+1$ reference nodes $P_0, P_1, ..., P_N$.

\textbf{Identify remote nodes}: For each reference node, we find the closest point (in terms of great circle distance) on the remote depth contour, and call this the ``remote node''; on each remote contour, there will hence also be $N+1$ remote nodes labelled $Q_0, Q_1, ..., Q_N$. 


\textbf{Define segment pairs on the reference and remote contours}: We next define a set of reference segments $\{A_i\}_{i=1}^N$ and remote segments $\{B_i\}_{i=1}^N$ having consecutive reference and remote nodes as start and end points, as follows:
\begin{eqnarray}
	A_i=(P_{i-1},P_i] \text{ for } i=1,2,...,N, \\
	B_i=(Q_{i-1},Q_i] \text{ for } i=1,2,...,N  .
\end{eqnarray}

Thus, $A_i$ (and $B_i$) consists of the boundary points on the reference (and remote) contour between consecutive nodes $P_{i-1}$ and $P_i$ (and $Q_{i-1}$ and $Q_i$). We refer to $(A_i, B_i)$ as a segment pair.

We want to find a unique mapping between points on the reference and remote contours. However, the procedure above to identify segment pairs does not guarantee this. It is possible that a point on the remote contour can occur in more than one remote segment and hence in more than one reference segment. This means that the remote point could be linked to more than one point on the reference contour; this would obviously not produce a unique pairing. Therefore, we proceed iteratively to accept remote segments into the final boundary mapping, favouring segment pairs with the smallest separation, whilst avoiding non-unique pairs of remote - reference boundary points (specifically by eliminating overlapping remote segments). 

\textbf{Calculate the horizontal distance between a segment pair}: The distance $D_i>0$ between segments $A_i$ and $B_i$ is calculated by first finding $n_{\text{Prt}}$ within-segment locations along each segment, which are approximately equally-spaced in terms of the number of points on the pathway (or ``boundary index'') for that segment. We then calculate the great circle distance between the corresponding $n_{\text{Prt}}$ points in each segment, and use the average distance as a measure $D_i$ of the distance between segments.  In the current work, we find that $n_{\text{Prt}}=10$ is a suitable choice, but note that in general this choice should consider the internode distance on reference and remote contours. 

\begin{figure*}[ht!]
	\centerline{\includegraphics[width=\textwidth]{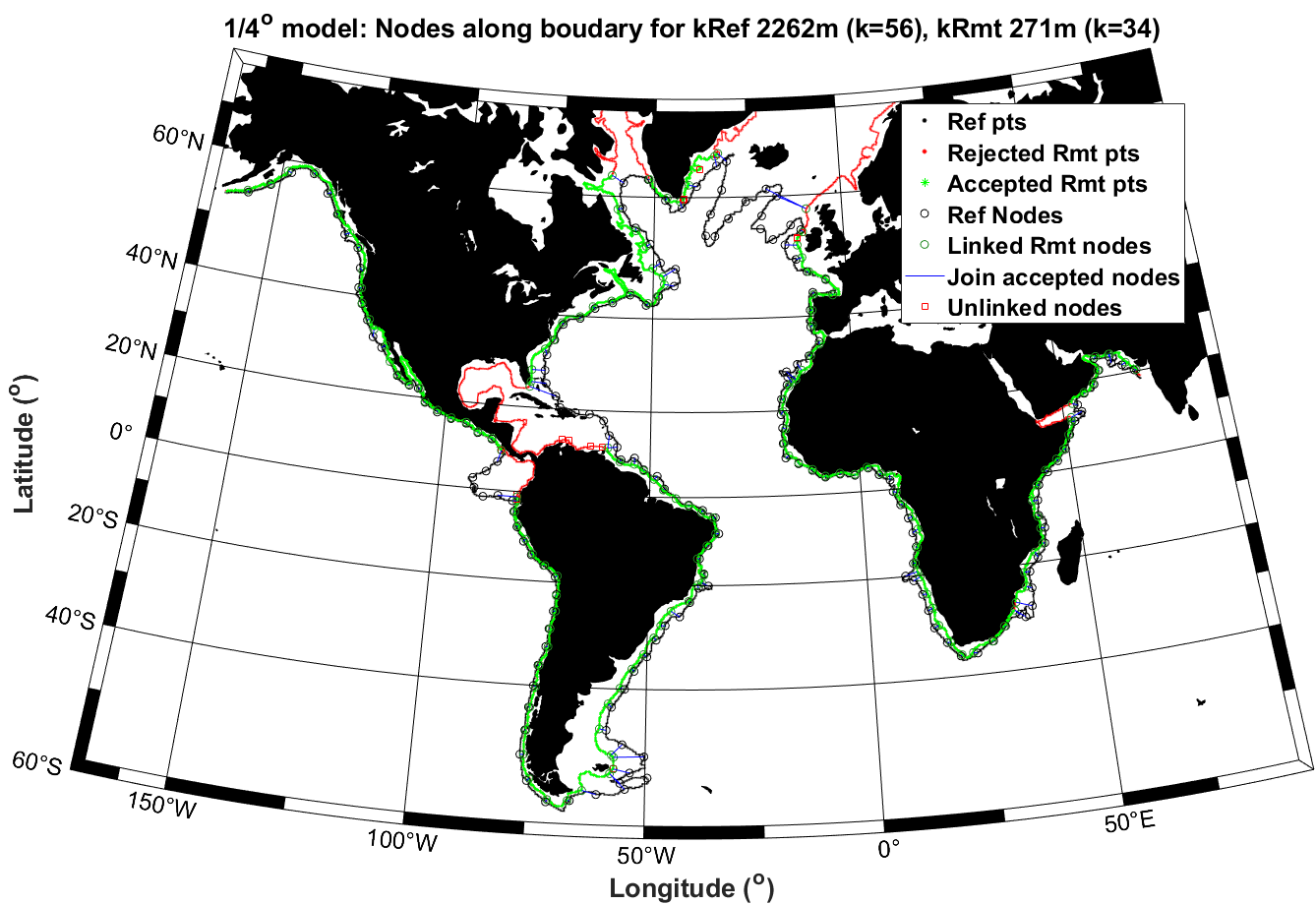}}
	\caption[Pairing of reference contour for depth-level $k=56$ (2262m) and remote contour at depth-level $k=34$ (271m).]{Pairing of reference contour (black dots) for depth-level $k=56$ (2262m) and remote contour at depth-level $k=34$ (271m). Accepted points on the remote contour are shown in green, and rejected points in red. Nodes spaced every 400km along the reference contour are shown as black circles. Corresponding nodes on the remote contour are shown as green circles (if linked) and red squares (if not linked). Linked reference-remote node pairs are joined by a blue line (e.g. visible around the Malvinas Islands). The pathway shown runs from the coast of Alaska to India for the $1/4^\circ$ model.}
	\label{F_mapNod_k56k34}
\end{figure*}
 
\textbf{Create final boundary mapping by iteratively accepting/rejecting segment pairs}:  We start (a) by finding the segment pair with smallest overall separation distance, and admit it to the final boundary mapping. Then (b) we consider the candidate segment pair with next smallest overall separation distance. We only admit the segment pair if its remote segment does not overlap with any of the remote segments already admitted to the final boundary mapping. If overlapping occurs, the candidate segment pair is rejected (since the candidate remote segment is effectively already used in the final boundary mapping, with a preferred, smaller separation distance). We continue (c) by iterating over all remaining segment pairs (ordered in terms of increasing separation distance), until all segment pairs have either been accepted or rejected. In the example discussed here, segment pairs with distance greater than 600km are not considered realistic pairings, and hence are rejected. Densities for points on the remote contour which are not mapped in the final boundary mapping are allocated NaN values. 

Once the mapping is complete, we have a consistent measure of distance along contours at all depths, allowing boundary characteristics at a given distance to be fairly compared over all depths. We can therefore visualise e.g. boundary densities in terms of depth-distance plots, analogous to the depth-latitude plots shown in Section \ref{Bdry_exp}, but avoiding the latter's discontinuities.

Figure \ref{F_mapNod_k56k34} shows the pairing of reference contour $k=56$ (2262m) and remote contour $k=34$ (271m). Rejected points (within rejected segments), shown as red dots, can be seen in the Gulf of Mexico and Caribbean Sea where the reference contour follows the outline of the Caribbean islands, whereas the shallower remote contour follows the Central American shoreline; hence the distance between segment pairs on reference and remote contours is large, and the segment pairings rejected. Similar behaviour is observed at high northern latitudes, e.g. on the Norwegian coast.

\subsection{Calculation of boundary densities}
\label{CntMpBnd.3}
We extract potential temperature $\theta$, salinity $S$, T-cell depth, longitude and latitude for each point along the boundary for all depth contours available. We then calculate potential and neutral densities using the Gibbs Seawater Oceanographic toolbox utilising the $TEOS$-$10$ algorithms (\citealt{McDougall2020}). Specifically, for calculation of potential density, the conservative temperature and absolute salinity fields are first calculated using the NEMO EOS-80 potential temperature and salinity fields, and then the potential density field is calculated in line with the $TEOS$-$10$ methodology.

Investigations in Section \ref{TM_TA} demonstrated discrepancies in density calculation using time-average input fields, especially on secondary calculations such as the overturning streamfunction. For our current purpose of calculating the boundary density, we must accept the accuracy of densities from time-average input fields. Nevertheless, we again highlight the absence of instantaneous outputs from the HadGEM-GC3.1 model, or equivalently of correctly time-average densities.

From Section \ref{CntMpBnd.2}, the along-contour distance $D$ is known for contours at all depths. Hence we can interpolate boundary densities (calculated independently for each depth contour) onto a common distance for visualisation.

\subsection{Datasets analysed}
\label{Bdry_Mod}
The along-boundary mapping algorithm outlined in Sections \ref{CntMpBnd.1}-\ref{CntMpBnd.3} is applied to HadGEM-GC3.1 model simulations (at three different oceanic and atmospheric spatial resolutions), GloSea5 reanalysis and \cite{GouretskiViktorandKoltermann2004} (GK) climatology. 

For the HadGEM-GC3.1 model (Section \ref{S3_ModDes}, \citealt{Williams2018} and \citealt{Roberts2019}) at $1^\circ$ oceanic spatial resolution, we use two model simulations which have different atmospheric resolutions. The coarser N96 atmospheric resolution is denoted by $LL$ (low ocean, low atmosphere) and the ``medium'' N216 atmospheric resolution by $LM$ (low ocean resolution, medium atmosphere resolution). The $1^\circ$-$LL$ model run has a length of 399 years, and the $1^\circ$-$LM$ a length of 104 years. In addition, we use the $1/4^\circ$ HadGEM-GC3.1 model (Chapters \ref{TJ_TM} and \ref{TJ_Var}), with run length of 657 years. All HadGEM-GC3.1 models considered are control runs with fixed 1950s forcing, where a 30 year spin-up was used prior, using EN4 dataset for initial conditions. We extract annual-mean properties for visualisation. 

The GloSea5 reanalysis dataset (\citealt{MacLachlan2015}) is used for seasonal forecasting at N216 atmospheric and $1/4^\circ$ oceanic spatial resolutions, similar therefore to the HadGEM-GC3.1 $1/4^\circ$ model. The vertical resolution is also the same as that of the HadGEM-GC3.1 models. The oceanic and sea-ice components of the model are based on NEMO's three-dimensional variational ocean data assimilation, which itself is based on the multi-institution NEMOVAR project (\citealt{Mogensen2009}, \citealt{Mogensen2012}).  We use monthly data available for 23 years from 1995 to 2018.

The \cite{GouretskiViktorandKoltermann2004} (GK) climatology dataset is produced at $1/2^\circ$ horizontal resolution for 44 vertical levels, with vertical spacings ranging from 10m at the surface to 250m at depth. The climatology is isopycnally averaged in order to minimise bias due to formation of artificial water masses. It incorporates 1,059,535 hydrographic profiles from the World Ocean Circulation Experiment (WOCE), German Oceanographic Data Centre, Alfred-Wegener-Institute, Arctic and Antarctic Research Institute and the French Oceanographic Data Centre.

\section{Characteristics of the time-average boundary densities in an extended Atlantic basin}

\begin{figure*}[ht!]
	\centerline{\includegraphics[width=\textwidth]{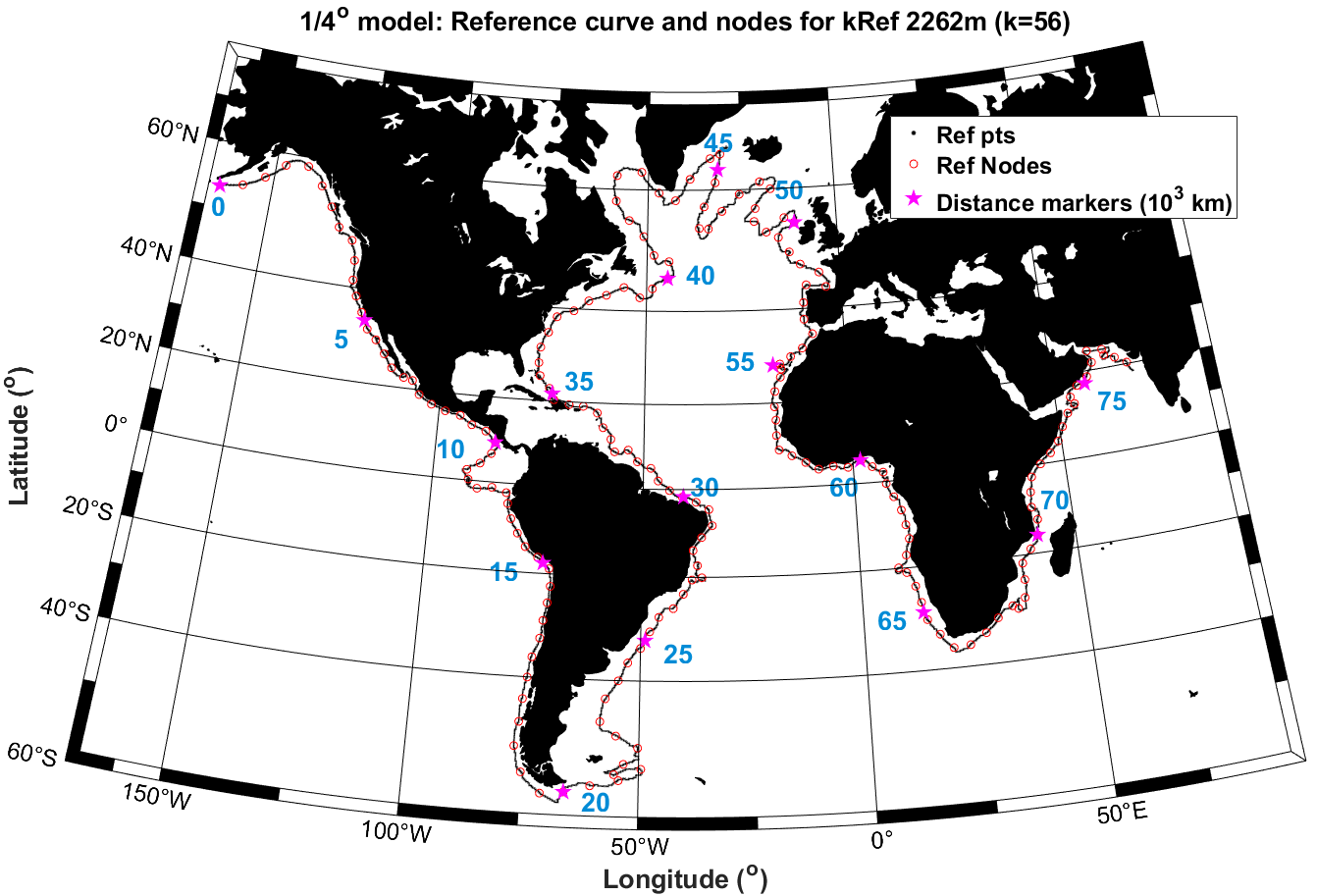}}
	\caption[Reference contour at $k=56$ (2262m) for the extended Atlantic basin with nodes, distance markers and cumulative distance.]{Reference contour (black dots) at $k=56$ (2262m) for the extended Atlantic basin with nodes (red circles), distance markers (pink star, every $5 \times 10^3$km) and cumulative distance (blue numbers, $\times 10^3$km). Remote contours at all depths are referenced to this reference contour.}
	\label{F_mapRef_k56}
\end{figure*}

We first discuss the general features of the Atlantic along-boundary neutral densities for the $1/4^\circ$ HadGEM-GC3.1 model. We then proceed to investigate the characteristics of boundary isopycnals in the $1^\circ$-$LL$ and $1^\circ$-$LM$ HadGEM-GC3.1 models, GloSea5 reanalysis and GK climatology. To reduce potential issues with model drift, only the first 100 years of output for each model are considered in the time-average. GloSea5 time-averaging is performed for all 23 years of data available. Figure \ref{F_mapRef_k56} shows the reference contour at depth level $k=56$ (2262m), together with the common along-boundary distance scale used to reference boundary contours at all depth levels $k$ (blue numbers, $\times 10^3$km).

\subsection{General features of boundary isopycnals in the $1/4^\circ$ HadGEM-GC3.1 model}

\begin{figure*}[ht!]
	\centerline{\includegraphics[width=\textwidth]{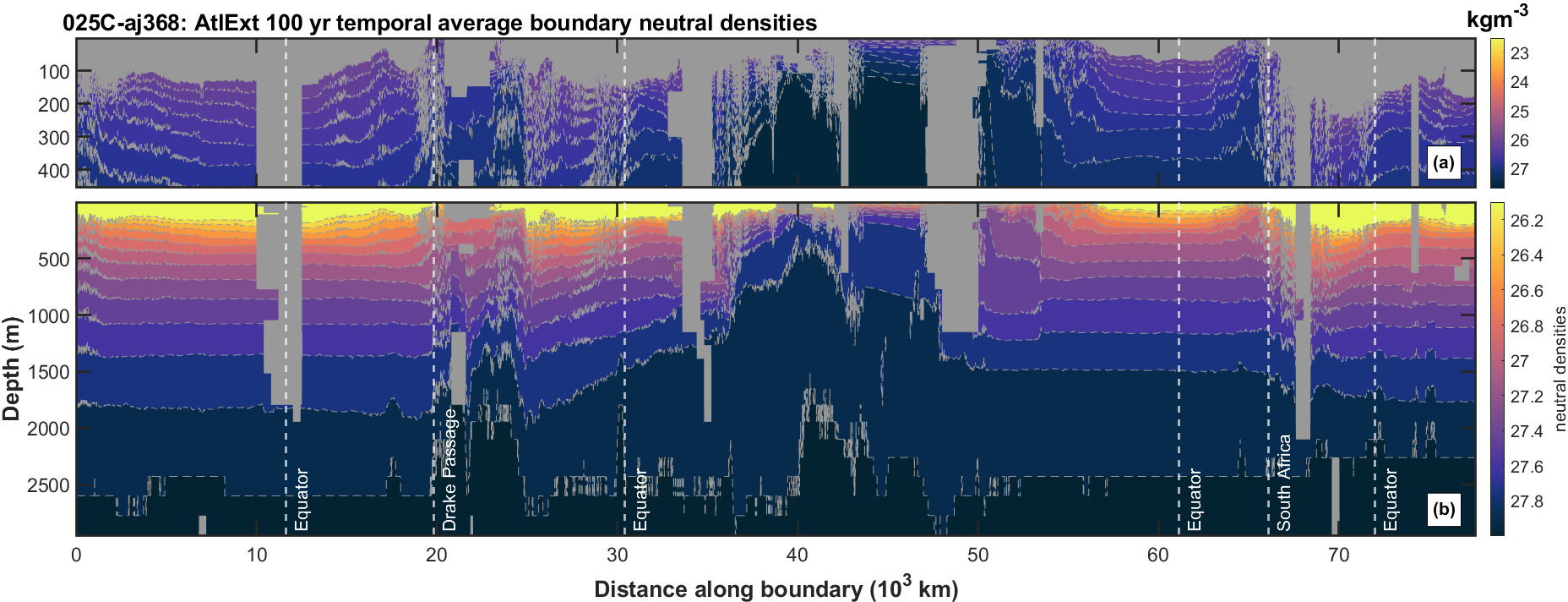}}
	\caption[Time-average neutral densities for the first 100 years of the $1/4^\circ$ model for the upper 400m and for all depths.]{Time-average neutral densities for the first 100 years of the $1/4^\circ$ model for (a) the upper 400m and (b) all depths. Dashed white lines indicate locations of Equator, Drake Passage and southern tip of South Africa. }
	\label{F_BD_n025M}
\end{figure*}

Figure \ref{F_BD_n025M} shows along-boundary neutral densities for the extended Atlantic region, corresponding to the pathway shown in Figure \ref{F_mapRef_k56}. Panel (a) concentrates on the along-boundary isopycnal structure for the upper 400m, whereas Panel (b) uses a different colour scale to emphasise isopycnal structure on the boundary at depth. Grey regions in this figure, and subsequent figures in this chapter, indicate depth-distance combinations rejected during boundary mapping. This occurs when remote and reference contour segment pairs do not meet the criteria outlined in Section \ref{CntMpBnd.2}.  

Figure \ref{F_BD_n025M}(a) shows large variability in isopycnal structure along the boundary near the surface. Typically, within the mixed-layer where wind-stress variability dominates, the hypothesis of flat eastern boundary isopycnals breaks down. However, eastern boundary regions corresponding to distances between $1$-$19\times 10^3$km (Pacific) and $55$-$65\times 10^3$km (Atlantic) are seen to have relatively flat isopycnals even within the upper 400m. At greater depths, Panel (b) also suggests the presence of flat isopycnals along eastern boundaries.

\begin{figure*}[ht!]
	\centerline{\includegraphics[width=\textwidth]{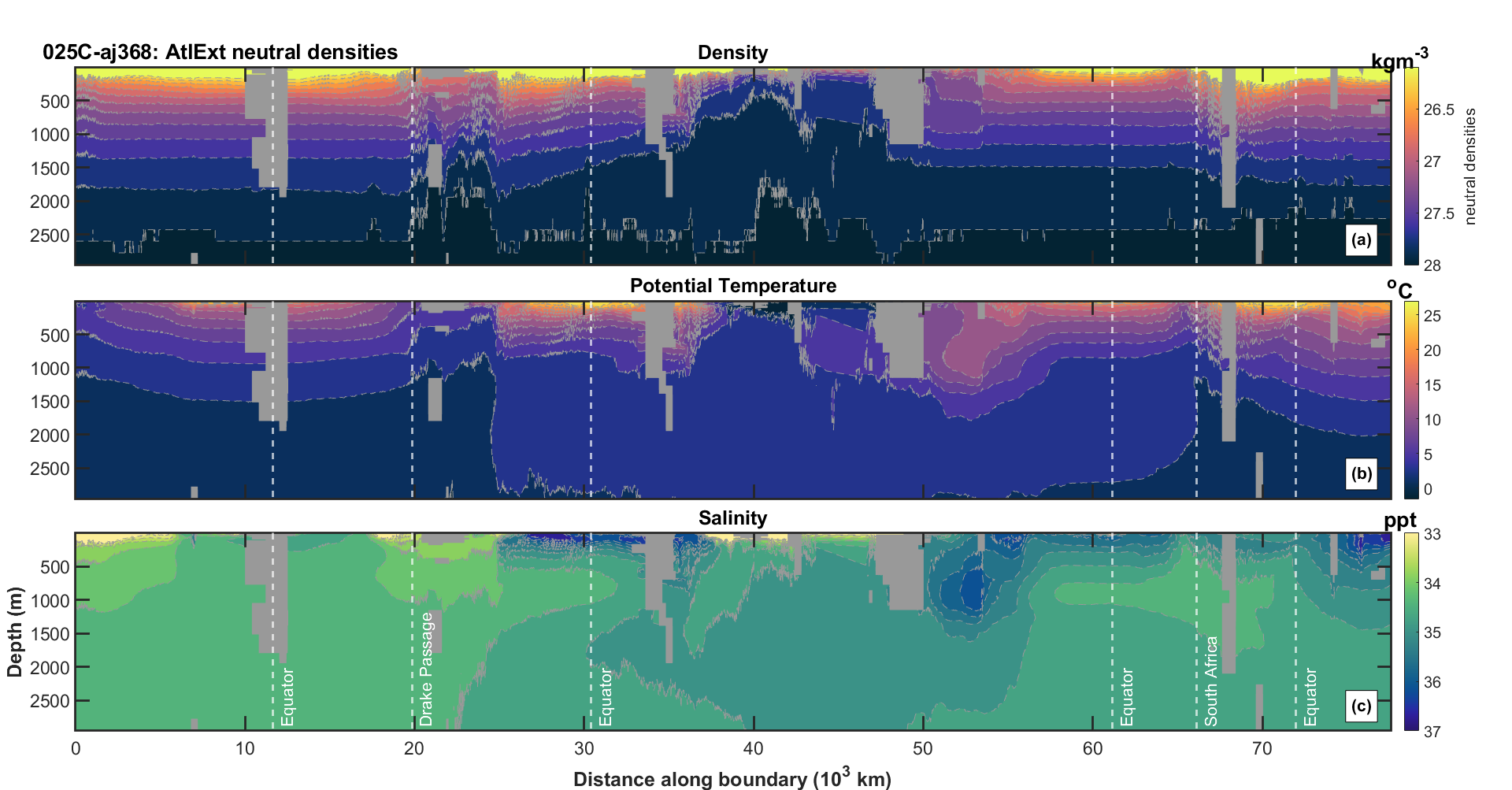}}
	\caption[Depth-distance plots for neutral density, potential temperature and salinity, time-average for the first 100 years of $1/4^\circ$ model.]{Depth-distance plots for (a) neutral density, (b) potential temperature and (c) salinity, time-average for the first 100 years of $1/4^\circ$ model. Dashed white lines indicate locations of the Equator, Drake Passage and southern tip of South Africa. }
	\label{F_BD_dps025M}
\end{figure*}

With a starting point offshore of Alaska, the first $19\times 10^3$km follows the eastern boundary of the Pacific. Along-boundary isopycnals are remarkably flat here below 400m. This uniform structure is also reflected by the corresponding isotherms shown in Figure \ref{F_BD_dps025M}(b). Locations near the Equator at distance $12\times 10^3$km are rejected by the boundary mapping algorithm in the region of the Galapagos Islands, due to excessive distances between remote and reference contours. At $20\times 10^3$km, a breakdown in flat isopycnals occurs at the Drake Passage. The following $3\times 10^3$km exhibits highly-variable isopycnal structure, persisting around the Malvinas islands until we return to the Atlantic's western boundary near southern Brazil. Here, the steep gradients in isopycnals at depth can be attributed to colder denser Antarctic Bottom Waters (AABW) moving northward from the Weddell Sea. Similarly, denser surface waters can be attributed to the Malvinas current. 

Northwards from southern Brazil ($24\times 10^3$km), for $10\times 10^3$km towards the Caribbean islands, gently upward-sloping isopycnals are observed at depths from 350m down to 2000m. A number of shallower points are rejected in the Gulf of Mexico. Further north, isopycnals continue to slope upwards, and we note a large shallowing of deeper isopycnals (depths 500-1500m) at around $36\times 10^3$km, likely attributed to the separation of the Gulf Stream from the U.S. coast. At $39.5\times 10^3$km, near Nova Scotia and Newfoundland, another steep tilt in deep isopycnals (400-800m) is witnessed, associated with cold southward-bound Labrador Current and DWBC waters shown by the time-average meridional bottom velocities in Figure \ref{p_RC_Bvel}.

The densest boundary surface waters are observed near Newfoundland ($40\times 10^3$km). This location corresponds to the end of upward-sloping isopycnals along the sloping western boundary; entering the Labrador Sea, more variable isopycnals occur due to strong air-sea interaction and deep convection. Once more, shallower remote contours at $43\times 10^3$km are rejected due to the large distances between reference and remote contours entering into Hudson Bay and the Arctic Ocean. In this region, we constrain boundary contours from entering Baffin bay, forcing them to return along the west Greenland coast. Isopycnals along eastern Greenland ($44$-$45\times 10^3$km) and along the western side of the Reykjanes ridge are relatively smooth, with a gentle downward slope. Gradients in surface isopycnals (upper 200m) are smaller than those of deeper isopycnals, resulting in a large pool of waters (300-900m) with density of approximately 27.7kgm$^{\text{-3}}$ at a distance of $45\times 10^3$km, possibly attributed to Iceland-Scotland Overflow Waters flowing around the Reykjanes ridge.

Near Iceland the contour path heads southward and then northward along the eastern side of the Reykjanes ridge, crossing from the western to the eastern side of the Atlantic basin. Local bathymetry leads to rejection of points (eastern side of Reykjanes ridge) shown in Figure \ref{F_BD_n025M} prior to $50\times 10^3$km. In this region boundary paths for different depths have separated, leading to shallower contours being rejected. Variable surface isopycnals occur on the west coast of Ireland (distance $50\times10^3$km) due to ageostrophic processes, before an interval of flat isopycnals along the eastern Atlantic boundary, corresponding to the western coast of France, Portugal and north-east Africa. 

A perturbation in the isopycnal structure is found near the Mediterranean outflow at $53\times 10^3$km in Figures \ref{F_BD_dps025M}(b,c); the Strait of Gibraltar appears as a thin vertical column of rejected points in shallow waters. At this location, there is a clear ``step'' in isopycnal depths (at intermediate depths, Figure \ref{F_BD_n025M}). Above 750m, isopycnal levels drop suddenly indicating the presence of lighter waters, whereas at depths between 900m and 1400m isopycnals become shallower, indicating the presence of denser (warmer, saltier) Mediterranean Overflow Waters (Figure \ref{F_BD_dps025M}(b,c)).

The eastern Atlantic boundary ($51$-$66\times 10^3$km) is similar to its Pacific counterpart, with flat boundary isopycnals. Crossing the Equator ($61\times 10^3$km), a region of light water is found near the surface, caused by warming and evaporation from surface buoyancy fluxes. Flat eastern boundary isopycnals persist on the sloping boundary for the majority of the East-African coastline (until $66\times 10^3$km). As the boundary path passes the Cape of Good Hope into the Indian Ocean, isopycnals slump due to the presence of warmer and saltier surface waters from the Madagascar and Agulhas currents.

\begin{figure*}[ht!]
	\centerline{\includegraphics[width=\textwidth]{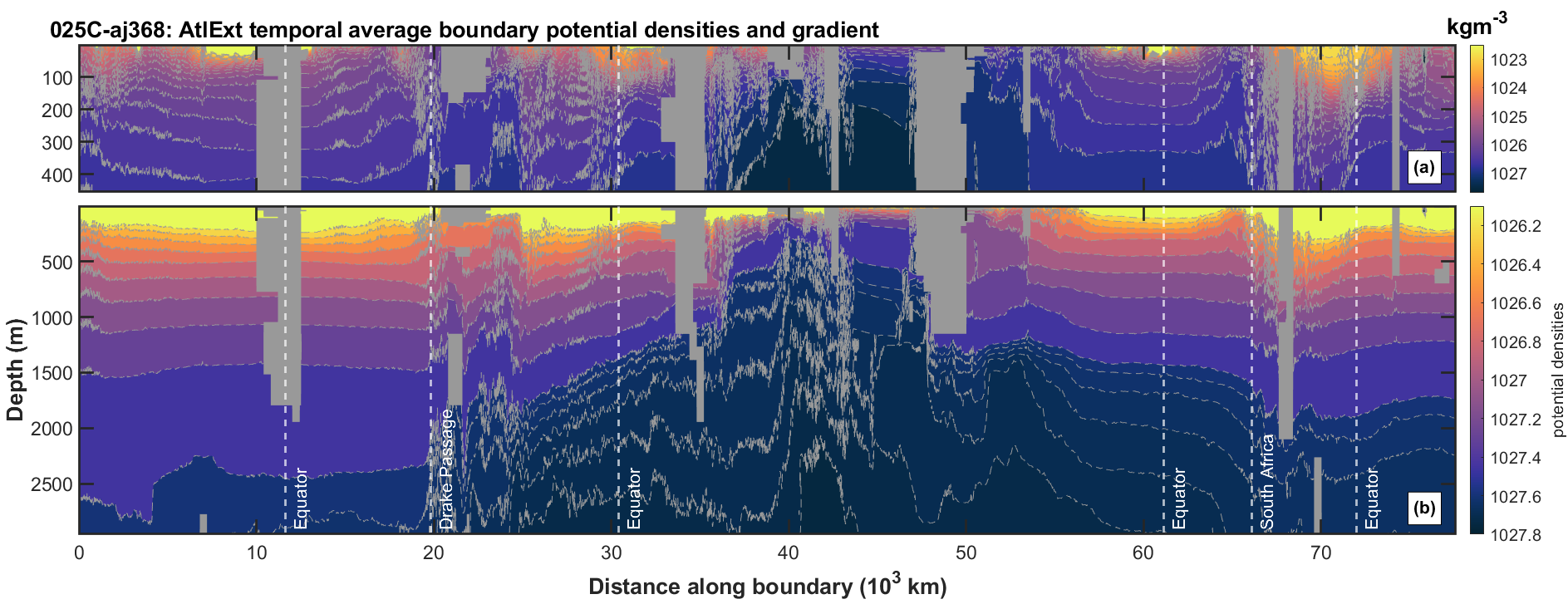}}
	\caption[Time-average boundary  potential densities for the first 100 years of the $1/4^\circ$ model for the upper 400m and all depths.]{Time-average boundary potential densities (referenced to the surface) for the first 100 years of the $1/4^\circ$ model for (a) the upper 400m and (b) all depths. Dashed white lines indicate the locations of the Equator, Drake Passage and southern tip of South Africa.}
	\label{F_BD_p025M}
\end{figure*}

At approximately $68\times 10^3$km, a region of rejected points down to 2000m is caused by a sudden change in the reference contour pathway off the coast of Durban, South Africa. Surprisingly, boundary isopycnals along the east coast of Africa (western Indian Ocean) are relatively flat, passing through the Mozambique channel and northward past Ethiopia and the Gulf of Aden ($74\times 10^3$km). This flatness of isopycnals along the western coast of the Indian Ocean is interesting, and is addressed further in Section \ref{Bdry_WMec}.

To complement the along-boundary investigation of neutral densities above, we briefly consider potential density, referenced to the surface. Figure \ref{F_BD_p025M} shows the potential density equivalent of Figure \ref{F_BD_n025M}. Here again, there is evidence for flat isopycnals on eastern boundaries. However, isopycnals are generally less stable in comparison to the neutral density isopycnals, especially at depth and in the northern Atlantic ($30$-$50\times 10^3$km). For example, near the Mediterranean outflow ($53\times 10^3$km), Figure \ref{F_BD_p025M}(b) shows dense water up to 1500m, leading to a steep gradient in deep isopycnals. 

\subsection{Boundary densities from model at different resolutions, reanalysis and climatology} 

Having characterised the boundary density structure for the extended Atlantic section in the $1/4^\circ$ model, we now explore the equivalent sections for various other datasets for comparison. 

\begin{figure*}[ht!]
	\centering
	\includegraphics[width=\textwidth]{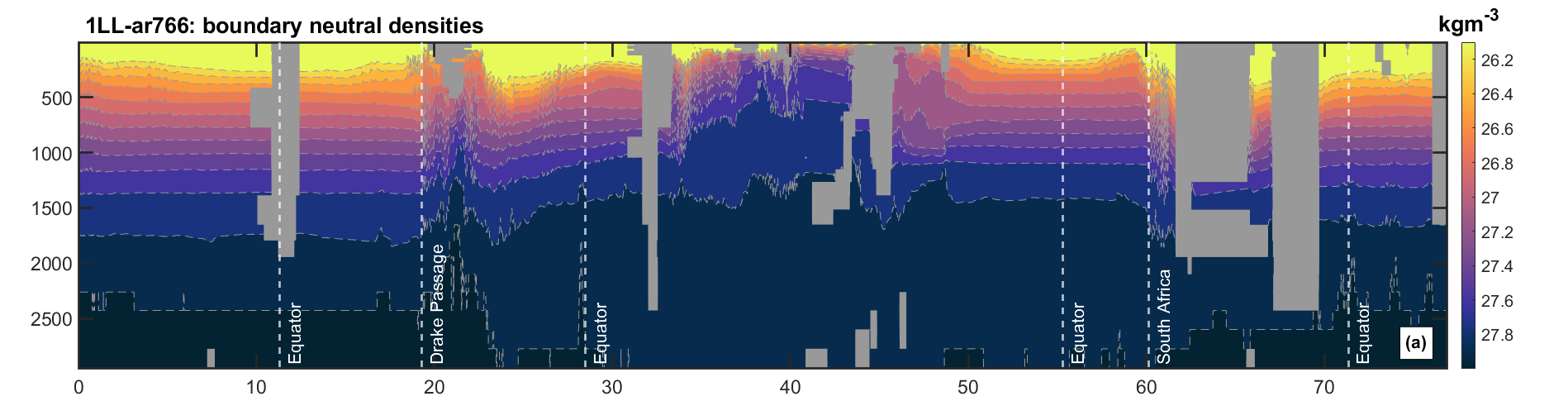}\\
	\includegraphics[width=\textwidth]{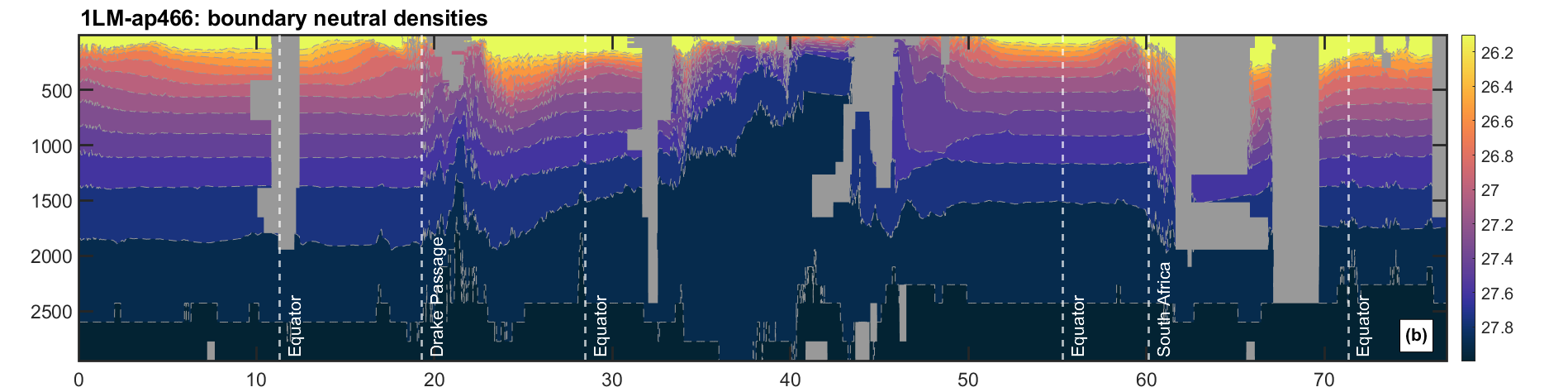}\\
	\includegraphics[width=\textwidth]{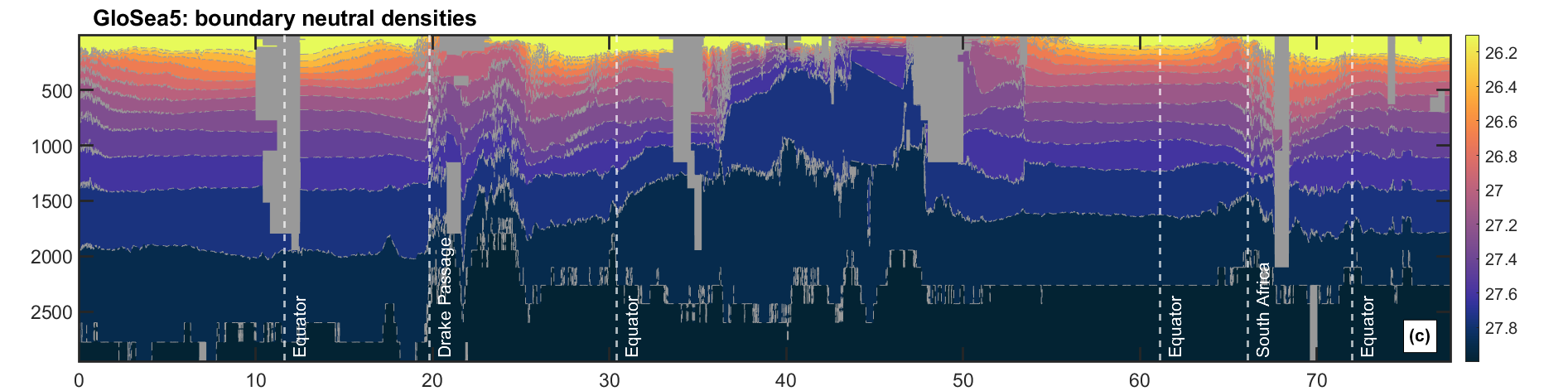}\\
	\includegraphics[width=\textwidth]{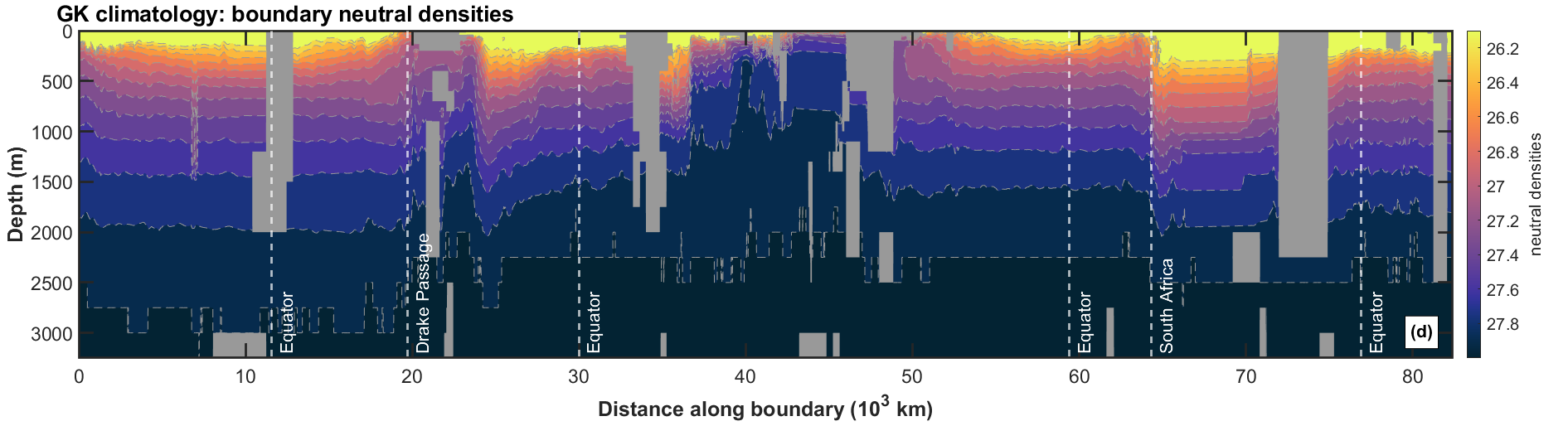}
	\caption[Neutral densities along sloping Atlantic boundary for the (a) $1^\circ$-$LL$ model, (b) $1^\circ$-$LM$ model, (c) GK climatology and (d) GloSea5 reanalysis for all depths.]{Neutral densities along sloping Atlantic boundary for the (a) $1^\circ$-$LL$ model, (b) $1^\circ$-$LM$ model, (c) GK climatology and (d) GloSea5 reanalysis for all depths. Model outputs are averaged over the first 100 years and GloSea5 over 23 years of run. Dashed white lines indicate locations of the Equator, Drake Passage and southern tip of South Africa.}
	\label{F_BD_Rest}
\end{figure*}

Figure \ref{F_BD_Rest} shows along-boundary neutral densities for (a) low ($LL$, N96) and (b) medium ($LM$, N216) atmospheric resolutions of the $1^\circ$ model over the first 100 years. Panels (c) and (d) show the isopycnal structure for the GloSea5 reanalysis and GK climatology respectively. We note for the $1^\circ$ models and GK climatology, the pathway taken by the reference contour deviates near Madagascar ($65\times 10^3$km). In the $1/4^\circ$ model and GloSea5 datasets it follows the African coastline through the Mozambique channel, whereas in coarser models and climatology it follows the east coast of Madagascar instead, resulting in a slightly longer section. Elsewhere the pathways taken by the reference contours are almost identical.

All panels of Figure \ref{F_BD_Rest}, with common colour scale, show similar along-boundary isopycnal structures. Two main differences are observed between the $1^\circ$ models (Panels (a,b)). First, the coarser resolution atmospheric model ($1^\circ$-$LL$, (a)) shows lighter surface density conditions throughout the basin. Secondly, the $1^\circ$-$LL$ also exhibits less dense waters near the surface in the North Atlantic (centred at $38\times 10^3$km at 1000m depth). For the $1/4^\circ$ and $1^\circ$-$LM$ models, and to a certain extend the GK climatology, deep isopycnals (depth 500-1500m near $36$-$45\times 10^3$km) slope upwards with increasing distance along the western Atlantic boundary; the $1^\circ$-$LL$ model and GloSea5 output show flatter isopycnals in this region. The latter suggest less dense waters near Newfoundland ($40\times 10^3$km), possibly attributed to changes in deep water formation and resulting Labrador Current and DWBC.

Isopycnal structure for the GloSea5 dataset shown in Figure \ref{F_BD_Rest}(c) is most similar to the $1^\circ$-$LL$ model. In general however, at great depths, GloSea5 shows signs of denser AABW (following the Drake Passage at around $24\times 10^3$km), penetrating further up the slope than for any other dataset considered; the $1^\circ$-$LL$ model exhibits the lightest deep waters (2000m and deeper). However, generally all datasets considered show remarkably similar isopycnal structures.

An investigation of the influence of seasonality on along-boundary isopycnal structure was conducted using monthly GloSea5 data, averaged into seasons of three consecutive months starting in December (i.e. DJF, MAM, JJA and SON). Results (not shown) indicated little to no seasonal effects, except near the east coast of Greenland where the 27.6kgm$^{\text{-3}}$ isopycnal (located near 500m depth) was 100-200m shallower in atmospheric summer months; this variation is attributed to changes in the East Greenland Coastal Current and mixed layer depths.   

Appendix \ref{App_Bdry_Tmp} investigates the variation of neutral boundary densities over different timescales. For the $1^\circ$ models, temporal variability outside the surface layer is largest in the North Atlantic, specifically locations downstream of areas of deep water formation. Greater variability at depth in the $1^\circ$-$LM$ model than the $1^\circ$-$LL$ model, can likely be attributed to greater variability in mixed-layer depth due to higher atmospheric resolution. The $1/4^\circ$ model shows interesting variability at 1200m depth, along both Atlantic boundaries, for all timescales considered. The along-boundary extent of these features might be attributed to internal or boundary waves. Remnants of this increased variability are propagated into the Pacific and Indian Oceans. In general, little difference in variability is found when  comparing eastern and western boundary sections.

\subsection{Mechanisms setting isopycnal slopes on western boundaries}
\label{Bdry_WMec}
The previous section establishes general consensus regarding along-boundary isopycnal structure across models, reanalysis and climatology for the Atlantic and surrounding boundaries. The presence of flat eastern boundary isopycnals is demonstrated, which can be explained by the action of boundary waves produced in response to anomalies in density (present due to e.g. propagation around the basin, or wind- or local-buoyancy-forcing). The condition of no flow through the boundary prevents a pressure gradient forming; therefore, any anomaly is quickly spread along the boundary by coastally-trapped waves, leading to flat isopycnals. However, on western boundaries, isopycnal structures along the Atlantic and Indian basins show contrasting behaviours, considered further here.

On the western boundary, flows are not geostrophic due to friction and non-linearities, therefore flat isopcynals are not to be expected here. However, in the presence of flat Atlantic eastern boundary isopycnals, a Coriolis parameter increasing with distance from the equator, and assuming the thermal wind relationship dominates the MOC, western boundary isopycnals must slope upward towards the north to sustain an overturning circulation. Using the thermal wind relation (Equation \ref{TW}), an increasing Coriolis parameter with latitude requires that the zonal density gradient ($\partial \rho$/$\partial x$) increases in magnitude ($\rho_W>\rho_E$) with latitude in the northern hemisphere to maintain the thermal wind shear ($\partial v$/$\partial z$). Therefore, upward-sloping western boundary isopycnals are necessary to increase the East-West density gradient, and maintain the overturning circulation. However, from this argument, we cannot infer a causal relationship between the isopycnal structure and overturning circulation.

The sloping western boundary isopycnals observed, are consistent with the results outlined by \cite{MacCready1994}, who discuss the frictional decay of abyssal boundary currents (e.g. DWBC), and present a theory for the observed longevity of these boundary currents along a sloping bathymetry. The majority of the current's energy is stored as potential energy via the upturn of isopycnals (normal to the boundary) on its slope. However, the rate of decay of the energy of the abyssal boundary current is dictated by the current's kinetic energy, which is considerably smaller than its potential energy. They develop a simple ($1\frac{1}{2}$-layer) mathematical model with a diffusion equation governing the DWBC vertical thickness, to predict the change in cross-sectional shape of the abyssal current during ``spin-down'' along the boundary. They find, as the current spins-down, kinetic energy is dissipated and replenished from a large pool of available potential energy, leading to the current's upper tip exhibiting down-slope movement, whilst the bottom of the current becomes thicker and wider. The resulting along-slope transport of the abyssal current is found to remain stable, even as its energy decreases. For a typical DWBC case, such an abyssal current may travel thousands of kilometers before being drained due to Ekman pumping. The Atlantic DWBC is known to flow southward from the Nordic Seas along the western side of the Atlantic down to the Southern Ocean, where it becomes a part of the ACC. Motivated by the findings of \cite{MacCready1994}, it is possible that the DWBC slowly migrates down-slope as it flows southward; if the density signature is maintained, this would result in a gradual deepening of isopycnals (southwards) along the boundary, and hence generate a gradient in along-boundary isopycnals. 

Along the Indian Ocean's western boundary, in marked contrast to its Atlantic counterpart, flat isopycnals are present within a basin exhibiting a predominately wind-driven overturning circulation (\citealt{Lee1998}), for which neither large East-West density gradients nor sloping western boundary isopycnals are necessary. Further, we note that the Indian Basin exhibits no DWBC; for a DWBC to be formed, a source of surface dense water is required at high latitudes in the basin (such as the Nordic Seas in the case of the Atlantic); however, no such source exists in the Indian Ocean (\citealt{Tamsitt2019}). The densest Indian Deep Water is formed diffusively within the interior from diapycnal mixing of abyssal waters (\citealt{Talley2013}). It is interesting that flat western boundary isopycnals coincide with the absence of a DWBC within the Indian Basin, in line with \cite{MacCready1994}.

A further point of interest is the presence of deep eastern boundary currents (DEBC) within the southern hemisphere of the Pacific, Atlantic and Indian Oceans. Their presence at depths between 2000m and 4000m would likely influence the along-boundary density structure. The dynamics of these DEBCs are not very well understood (\citealt{Tamsitt2019}), and current GCMs do not represent them well (\citealt{Yang2020a}). They are smaller than their western counterparts, but still play an important role in the overturning circulation (\citealt{Wunsch1983}, \citealt{Arhan2003}, \citealt{Tamsitt2017}). \cite{Yang2020a} proposes mechanisms for the dynamics of the Pacific DEBC using GCMs and idealised models. They show that DEBCs do not behave like typical boundary currents, or their stronger DWBC counterparts. They propose a framework to explain the Pacific DEBC as a manifestation of topographic stretching plus a dynamical mode, that decays away from the eastern boundary. This occurs only when temperature, diffusion, viscosity and stratification effects are considered together. However, their linearised framework does not work well for the Atlantic DEBC, due to its non-linear dynamics.

\section{Summary}

In this chapter, we have investigated along-boundary properties of an extended Atlantic basin. Latitude-depth contour plots for eastern and western boundary densities for the $1^\circ$, $1/4^\circ$ and $1/12^\circ$ HadGEM-GC3.1 models reveal that density variation on the eastern boundary is less complex, with intervals of flat isopycnals outside the surface layer, particularly for latitudes south of $35^\circ$N. In contrast, western boundary isopycnals slope approximately linearly with latitude, above 2500m and south of $35^\circ$N. The regularity of density variation with latitude and depth is disrupted by features such as the Mediterranean Outflow (on the eastern boundary), and Gulf Stream separation (on the western boundary). Lighter surface densities are found on the western boundary due to strong boundary currents there. For depths between 200m and 1500m, eastern boundary densities are heavier than their western counterparts in the southern hemisphere, whereas the reverse is true in the northern hemisphere. These findings do not appear to be dependent on model spatial resolution. 

We have introduced a boundary mapping algorithm which determines the location of sloping continental boundaries. The algorithm allows us to map and project the ocean's sloping boundaries in a two-dimensional (along-boundary distance, and depth) coordinate system. We have used the mapping algorithm to explore spatio-temporal variations in potential temperature, salinity, neutral and potential densities for (HadGEM-GC3.1) model simulations, (GloSea5) reanalysis and (Gouretski and Kolterman) climatology, centred on the Atlantic basin. The general consensus across these datasets is that the assumption of flat isopycnals along the eastern boundary is valid. Quantifying whether isopycnal slope at a location is zero or not is a challenging problem. Slope estimates are uncertain because locally along the boundary we only have limited data to estimate them. Therefore, estimates of isopycnal gradients are either very noisy, or require a lot of smoothing using non-local data, casting doubt on the estimate of the local isopycnal gradient. 

Along the eastern Pacific boundary, isopycnals are flat. At the Drake Passage and near the Malvinas islands, northward flow of colder denser Malvinas current and deep AABW result in a shallowing of isopycnals. Along the western Atlantic boundary, isopycnals slope gently upwards along the boundary in a northward direction. Discrepancies between boundary densities in different models are greatest at depth (approximately 1500m) in the North Atlantic, and may be related to specifics of deep water formation at high latitude within each model. 

To the north of the Mediterranean outflow, the lack of continuous boundary between the British Isles and continental Europe, and ageostrophic eastern boundary flows result in non-flat isopcynals. For the eastern Atlantic boundary at latitudes south of the Mediterranean outflow, flat isopycnals occur. These are disturbed temporarily near the Cape of Good Hope, where isopycnals deepen due to the input of warm Indian Ocean waters. However, beyond South Africa on the western boundary of the Indian Ocean, isopycnals remain flat, in contrast to their behaviour on the western Atlantic boundary. 

One explanation for the observed linear slope in Atlantic western boundary isopycnals is the presence of an overturning circulation in the basin together with flat eastern boundary isopycnals and a Coriolis parameter increasing with distance from the equator. Thermal wind balance would indicate western boundary isopycnals must slope upwards to the north to maintain the overturning circulation. These results are consistent with the work of \cite{MacCready1994}, where a southward-propagating DWBC migrates normally down the western continental slope with along-boundary distance (becoming thicker and wider at depth), leading to a slope in isopcynals. 

We hypothesise that flat isopycnals on the western boundary of the Indian Ocean are explained by the mainly wind-driven overturning and weak geostrophic component (\citealt{Lee1998}) there, making sloping (western boundary) isopycnals unnecessary. Further, the absence of a clear DWBC in the Indian basin, together with the presence of flat western boundary isopycnals there, is again  consistent with the view set out by \cite{MacCready1994}. The observed flat isopycnal structure along the eastern Atlantic and Pacific boundaries is attributed to the no normal flow through the boundary condition, resulting in any anomalies on the eastern boundary being propagated quickly and distributed evenly along the boundary by coastally-trapped waves.

 \clearpage

\chapter{Investigating bias in the Antarctic Circumpolar Current within HadGEM-GC3.1 models}  \label{TJ_ACC}

Inspection of Antarctic Circumpolar Current (ACC) transport in HadGEM-GC3.1 models at different spatial resolutions reveals large differences in estimates for total transport through the Drake Passage. In this chapter, the overturning streamfunction decomposition diagnostic developed for the AMOC in Chapter \ref{TJ_TM} is modified for decomposition of the zonal ACC transport at the Drake Passage, the point of narrowest constriction of the ACC. ACC transport is decomposed into the sum of terms involving only boundary properties, and an additional $\beta$ term arising from a meridionally-varying Coriolis parameter. Decomposing the cumulative ACC transport in this way helps us to investigate the role of model spatial resolution, and identify possible physical mechanisms which may give rise to resolution dependence. The boundary densities along the Antarctic coastline are mapped using the algorithm from Chapter \ref{TJ_Bdry} and spatial characteristics discussed for various models, reanalysis and climatology.

\section{Background} \label{Sct_ACC_Bck}

The Southern Ocean is a crucial part of the global ocean circulation. Its unique bathymetry and lack of meridional boundary (i.e. zonally-unblocked latitudes), in conjunction with an equator-to-pole temperature gradient and strong westerly winds, results in an eastward flowing ACC. The ACC is the strongest ocean current in the world ($173.3\pm10.7$Sv, \citealt{Donohue2016}), vital in transporting heat, carbon and nutrients between the major ocean basins (see Section \ref{I_SO}). 

The ACC is strongly associated with steeply meridionally-tilted isopycnals in the Southern Ocean, with isopycnals of lighter water masses outcropping at lower latitudes. Using geostrophy (Equation \ref{GB}), the isopycnal slope can be related to the perpendicular transport: a steep negative gradient from south to north results in eastward flow. The slope of the isopycnals is also related to the strength of the westerly winds encircling Antarctica. These winds result in large-scale coastal Ekman upwelling, with surface waters moving northward and a geostrophic return flow at depth. In the absence of a zonal land boundary, the zonal pressure gradient is only supported at great depths, resulting in a relatively deep geostrophic return flow (see Section \ref{I_dynL}). NADW waters are found to upwell before the Antarctic shelf to replace the northward-moving surface waters. The upwelling not only steepens isopycnals, but returns old, nutrient-rich waters to the surface. This cycle in the Southern Ocean is commonly known as the Deacon cell, (\citealt{Doos1994}). In general, in regions of strong currents (such as the ACC, Gulf Stream and Kuroshio), baroclinic instabilities are prominent, giving rise to eddies which act to flatten the isopycnals, opposing the steepening from the wind-driven Deacon cell. 

Early estimates of ACC transport covered a wide range of values, in part due to the use of different reference depths for hydrographic calculations, and lack of knowledge of near-bottom current characteristics. \cite{Bryden1977} showed a level of no-motion near the sea-floor would not be appropriate, due to bottom currents through the Drake Passage. In the late 1970s and early 1980s, the International Southern Ocean Studies (ISOS) programme (\citealt{Whitworth1982}, \citealt{Whitworth1983}, \citealt{Whitworth1985}) led to significantly improved understanding of the Southern Ocean. \cite{Cunningham2003} reanalysed the ISOS data and estimated a mean ACC transport of $134$Sv with uncertainty of between $15$Sv and $27$Sv.

\cite{Meredith2011} review hydrographic estimates for baroclinic transport through the Drake Passage ($SR1b$ section) for the period $1993-2009$ (including data from the World Ocean Circulation Experiment). The baroclinic transport through the Drake Passage is found to be $136.7\pm6.9\text{Sv}$ ($\pm$ indicates standard deviation of transport), a relatively consistent value throughout the period of observation. Using a combination of CPIES and current meters at a location slightly west of $SR1b$ ($cDrake$ experiment), \cite{Chidichimo2014} estimate the average baroclinic transport over a 4-year period ($2007-2011$) through the Drake Passage, to be $127.7\pm8.1\text{Sv}$. \cite{Donohue2016}, quotes an additional depth-independent (or barotropic) transport component of $45.6$Sv, calculated using bottom current recorders, as a part of the $cDrake$ experiment. This updated combined (baroclinic plus barotropic) estimate of $173.3\pm10.7$Sv for the transport through the Drake Passage, represents an increase of approximately $30\%$ over the benchmark interval of $130$ to $140$Sv, typically expected for ACC transport in climate models (see Section \ref{I_obsACC} for further information regarding observing the ACC). 

\cite{Allison2009} decomposed the ACC into its boundary components within the pre-industrial control integration of the GFDL CM2.1 model. She found a relatively stable ACC of $130$Sv using zonal velocities; however, her estimate of ACC transport, using boundary components and a $\beta$ term, fell 20Sv short of this value. We speculate that the discrepancy might be accounted for by contributions from additional cells. The dominant terms contributing to the time-mean ACC in \cite{Allison2009} are the northern density (83Sv) and depth-independent components (30Sv).    

The transport accumulated from the bottom to the surface, calculated using zonal velocities (see Equation \ref{e_ACC_def} below) at $66.5^\circ$W, is shown in Figure \ref{F_Tru_Trsp} for the three HadGEM-GC3.1 models (with $1^\circ$, $1/4^\circ$ and $1/12^\circ$ resolutions) considered in this chapter. The timeseries shown include the initial 30-year ``spin-up'' phase for each model. Model resolution has a dramatic effect on the corresponding strength of the ACC. The figure also shows estimates for $T_{ACC}$ from observations, given by \cite{Meredith2011} and \cite{Donohue2016}; the $1/4^\circ$ model resolution in particular underestimates the observed ACC transport.  

The work conducted in this chapter is part of a wider effort to understand the ACC, and Southern Ocean characteristics, in the current generation of UK Met Office GCMs.
\begin{figure}[ht!]
	\centerline{\includegraphics[width=\textwidth]{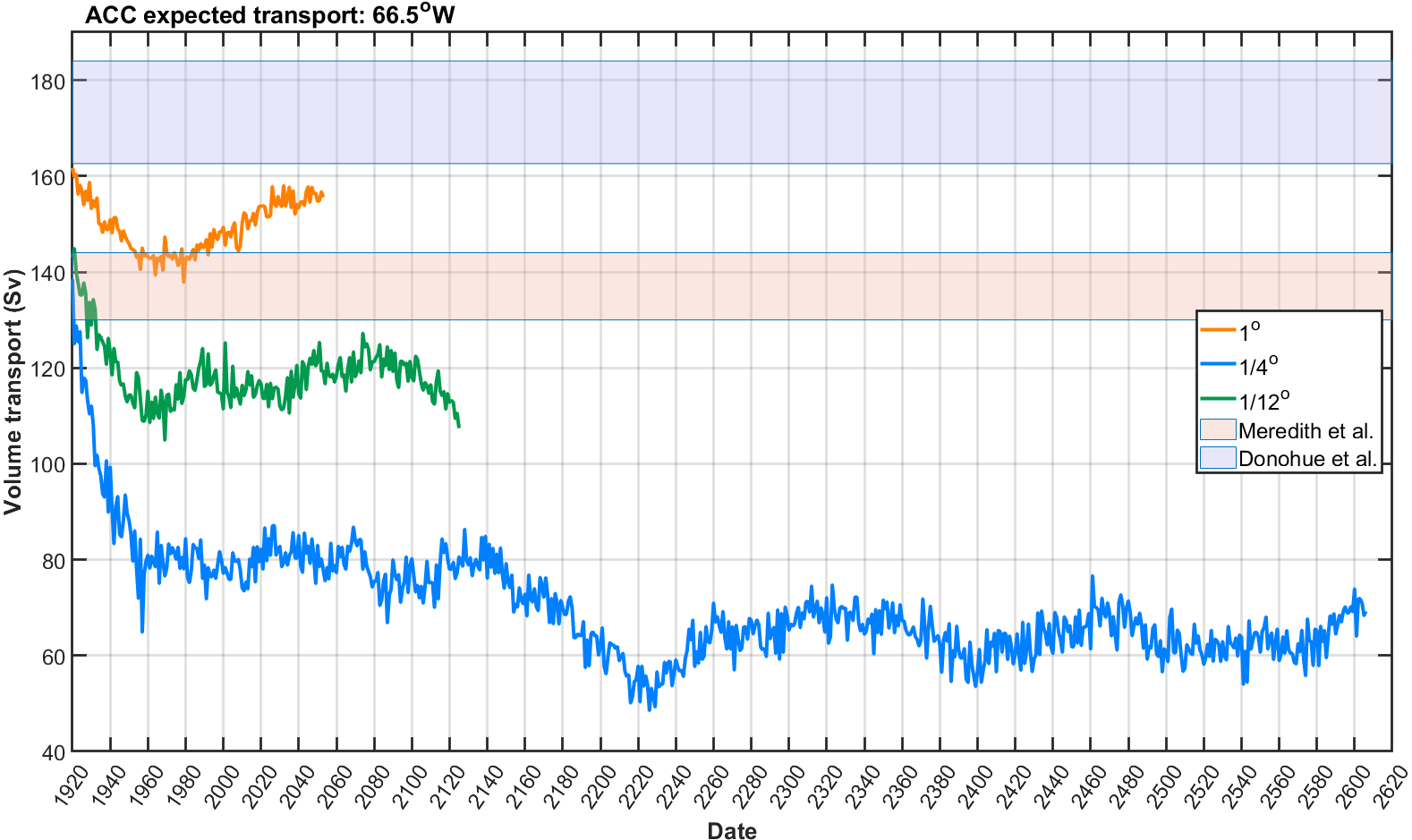}}
	\caption[Timeseries of volume transport $T_{ACC}$ integrated up to the surface through the Drake Passage (Section: $66.5^\circ$W) calculated using Equation \ref{e_T_ACC_math} for the $1^\circ$, $1/4^\circ$ and $1/12^\circ$ HadGEMGC3.1 spatial resolutions.]{Timeseries of volume transport $T_{ACC}$ integrated up to the surface through the Drake Passage (Section: $66.5^\circ$W) calculated using Equation \ref{e_T_ACC_math} for the $1^\circ$, $1/4^\circ$ and $1/12^\circ$ HadGEMGC3.1 spatial resolutions. Shaded red region indicates the baroclinic transport estimate using hydrographic data from \cite{Meredith2011} (without bottom flow), blue region indicates estimate quoted by \cite{Donohue2016} using cDrake data (including depth-independent transport). }
	\label{F_Tru_Trsp}
\end{figure}

\section{Method} \label{Sct_ACC_Mth}

Following the method derived for decomposing the AMOC's overturning streamfunction into boundary components in Chapter \ref{TJ_TM}, we now outline the analogous decomposition of the ACC zonal cumulative transport into Ekman, depth-independent (bottom), northern and southern boundary density and $\beta$ terms. The $\beta$ term is introduced to account for the contribution made by a varying Coriolis parameter across the meridional section (south to north). Much of the derivation here is similar to that presented in detail in Chapter \ref{TJ_TM}; for this reason, we present the current derivation concisely, referring the reader to Chapter \ref{TJ_TM} for details as necessary.

The zonal volume transport (with east as positive) below a given depth z through the meridional section corresponding to the Drake Passage is defined as the meridional and vertical integral of zonal velocities
\begin{equation}
T_{ACC}(z;x) = \int_{S(z;x)}^{N(z;x)} \int_{-H(y;x)}^{z} u_E(y,z';x) \mathrm{d} z' \mathrm{d} y
\label{e_ACC_def}
\end{equation} 
where $S(z;x)$ and $N(z;x)$ are the southern and northern boundaries respectively at depth $z$ for zonal coordinate $x$. $H(y;x)$ is the maximum depth of the fluid column at meridional coordinate $y$. The zonal velocity $u_E$ can be partitioned into the sum of (a) thermal wind ($u_{th}$), (b) depth-independent ($u_{bot}$) and (c) Ekman ($u_{Ekm}$) components, in an analogous manner to that shown in Equation \ref{v} for the AMOC. Hence the corresponding cumulative transport components can be calculated.

The depth-independent cumulative transport component $T_{bot}(z;x)$ can be thought of as introducing a geostrophic reference into the calculation. We write



\begin{equation}
T_{bot}(z;x) =  \int_{S(z;x)}^{N(z;x)} Z_{bot}(y,z;x) u_{bot}(y;x) \mathrm{d} y
\label{e_ACC_int_1}
\end{equation} 
for bottom zonal velocity $u_{bot}(y;x)$, where bottom depth $Z_{bot}(y,z;x)$ is given by
\begin{align}
Z_{bot}(y,z; x) = & H(y; x) + z,  & z > -H(y; x), \\
Z_{bot}(y,z; x) = & 0,  & z \leq -H(y; x) .
\label{e_Z_{bot}_ACC}
\end{align}

\begin{figure*}[ht!]
	\centerline{\includegraphics[width=\textwidth]{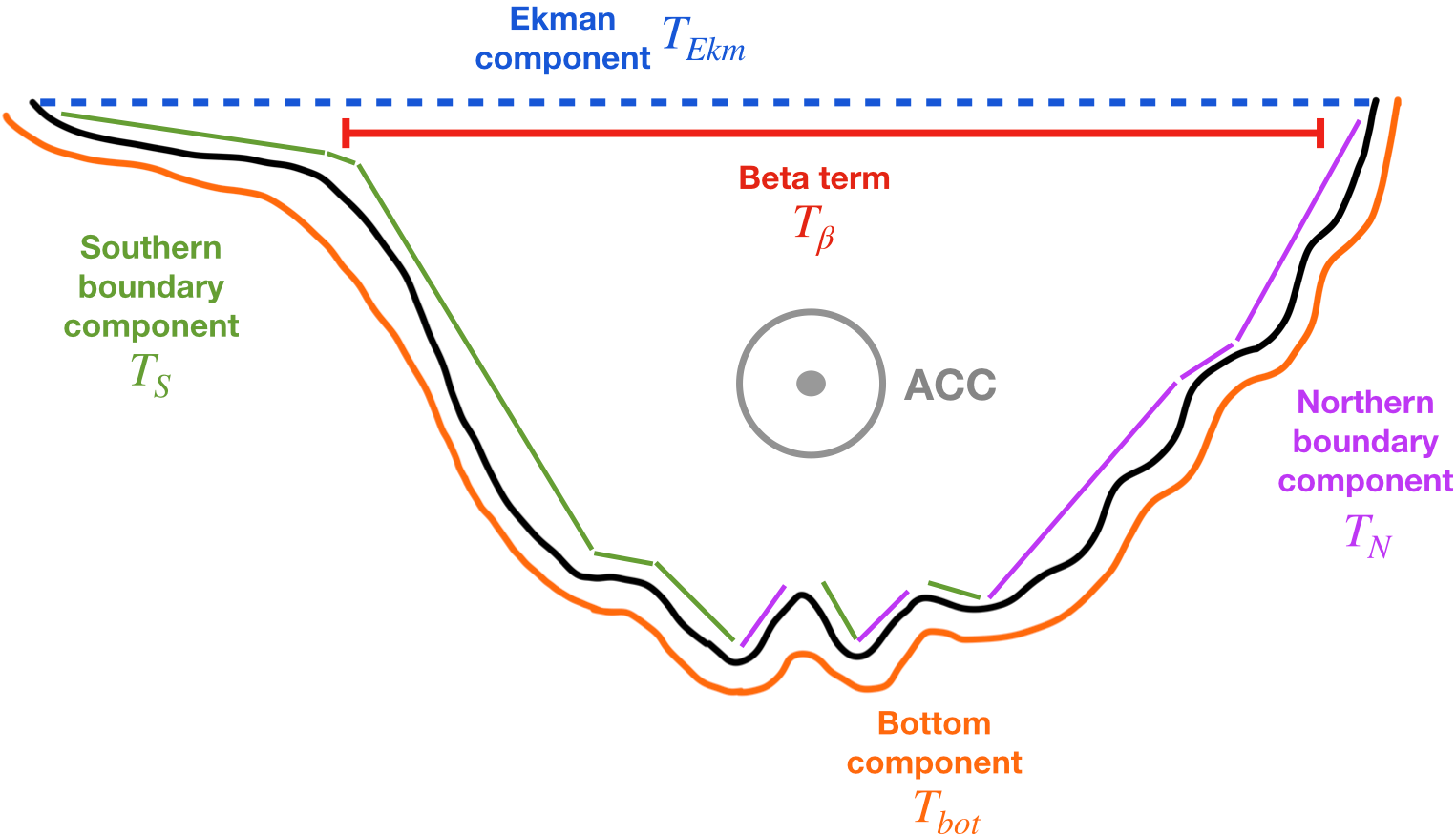}}
	\caption[Schematic of the boundary decomposition and the contributing components to the ACC transport through the Drake Passage.]{Schematic of the boundary decomposition and the contributing components to the ACC transport through the Drake Passage. We note in this illustration that multiple pairs of southern and northern boundaries are present at depth, whereas only one pair is present at the surface. The grey circle indicates eastward-flowing ACC.}
	\label{F_bdryACC}
\end{figure*}

The thermal wind cumulative transport component $T_{th}(z;x)$ is calculated (following a rationale similar to that outlined in Equations \ref{Tc_1}-\ref{Tc_2} for the AMOC) as a meridional and vertical integral of the zonal thermal wind velocity $u_{th}(y,z; x)$. The integral is simplified using integration by parts, taking $u_{th}(y,z; x) = 0$ when $z = -H(y; x)$. In marked contrast to the calculation of meridional transport for the AMOC, the Coriolis parameter ($f(y)$, introduced via the thermal wind relationship, Equation \ref{TW}) is not constant with latitude in the meridional integral for zonal ACC transport; hence we cannot simply move $f(y)$ ``outside'' the south-north integral in the current case. The resulting expression for $T_{th}(z;x)$ is
\begin{equation}
T_{th}(z;x) = \frac{g}{\rho_0} \left(\int_{S(z;x)}^{N(z;x)} \left[ z\int_{-H(y;x)}^{z}  \frac{1}{f(y)}\frac{\partial \rho}{\partial y}(y,z';x)dz' - \int_{-H(y;x)}^{z} z' \frac{1}{f(y)}\frac{\partial \rho}{\partial y}(y,z';x) dz' \right] \mathrm{d}y \right),
\label{e_ACC_TW_1}
\end{equation}
where $\rho$ is neutral density, and $\rho_0$ is the nominal density of seawater. Equation \ref{e_ACC_TW_1} is re-arranged in a manner analogous to Equation \ref{Tc_2}, to obtain
\begin{align}
	\label{e_ACC_TW_2}
	T_{th}(z; x) &= \frac{g}{\rho_0} \left[ \int_{-H_{max}(x)}^{z} (z-z') \left[ \sum_{d=1}^{n_D(z;x)} \left( \frac{\rho_{N_d}(z';x)}{f_{N_d}(x)} - \frac{\rho_{S_d}(z';x)}{f_{S_d}(x)} \right) \right] \mathrm{d}z' \right. \nonumber \\ 
	&+ \left. \int_{-H_{max}(x)}^{z} (z-z') \int_{S(z;x)}^{N(z;x)} \frac{\beta(y) \rho(y,z';x)}{f(y)^2} \mathrm{d}y \mathrm{d}z' \right],
\end{align}
where $\beta(y)=\frac{\partial f}{\partial y}(y)$, and $\rho_S(z;x)$ and $\rho_N(z;x)$ are the southern and northern boundary densities at each $z$ for a given $x$. Further, $f_S(x)$ and $f_N(x)$ are Coriolis parameter values at the southern and northern boundary density points for a given $x$. $H_{\max}(x)$ is the maximum basin depth at zonal coordinate $x$, and we are careful to acknowledge the possible presence of $n_D \ge 1$ pairs of southern and northern boundaries for depth $z$ and zonal coordinate $x$, with meridional intervals $[S_d(z,x),N_d(z,x)]$, $d=1,2,...,n_D(z;x)$, as illustrated in Figure \ref{F_bdryACC}. 

It can be seen from Equation \ref{e_ACC_TW_2} that the resulting expression for the zonal ACC thermal wind decomposition takes a similar form to that of the meridional AMOC thermal wind decomposition, with the addition of a $\beta$ term. The thermal wind contribution of the ACC transport can therefore be written as the sum of southern and northern boundary density terms  $T_S(z;x)$ and $T_N(z;x)$, and a $\beta$ term ($T_{\beta}$) accounting for the role of a varying Coriolis parameter across the section. In the presence of $n_D(z;x) \ge 1$ pairs of southern and northern boundaries, the boundary terms consist of sums over the individual independent meridional intervals
\begin{equation}
	T_{N}(z; x) =  \frac{g}{\rho_0} \int_{-H_{max}(x)}^{z} (z-z') \left[ \sum_{d=1}^{n_D(z;x)} \left( \frac{\sigma_{N_d}(z;x)}{f_{N_d}(x)} \right) \right] \mathrm{d}z'
	\label{T_Nrt}
\end{equation}

\begin{equation}
	T_{S}(z; x) =  - \frac{g}{\rho_0} \int_{-H_{max}(x)}^{z} (z-z') \left[ \sum_{d=1}^{n_D(z;x)} \left( \frac{\sigma_{S_d}(z;x)}{f_{S_d}(x)} \right)  \right] \mathrm{d}z'
	\label{T_Sth}
\end{equation}

\begin{equation}
	T_{\beta}(z; x) = + \int_{-H_{max}(x)}^{z} (z-z') \int_{S(z;x)}^{N(z;x)} \frac{\beta(y) \sigma(y,z';x)}{f(y)^2} \mathrm{d}y \mathrm{d}z'.
	\label{T_Beta}
\end{equation}

The difference between the sloping boundary density components is what sets the thermal wind condition. Therefore, any constant can be added to the boundary density components in these equations. Hence, in Equations \ref{T_Nrt}-\ref{T_Beta}, to better distinguish the properties of the northern and southern density components, the neutral density of the deepest point $\rho_{deep}$ is subtracted from the neutral densities at the boundaries, yielding a referenced density or neutral density anomaly $\sigma$. Thus, for the northern density term corresponding to the $d^\text{th}$ boundary, the anomaly with respect to the density at the deepest point is
\begin{equation}
\sigma_{N_d}(z;x) = \rho_{N_d}(z;x,\mathcal{S}, \theta, P=0) - \rho_{deep}(z;x,\mathcal{S}, \theta, P=0) 
\end{equation}
with a similar definition for $\sigma_{S_d}(z;x)$ and $\sigma(y,z;x)$, for salinity $\mathcal{S}$, potential temperature $\theta$ and pressure $P$. We note that $\rho_{deep}$ is time-dependent, varying from year to year. The calculation for $T_{\beta}$ requires interior information for neutral densities within the section, and therefore the ACC decomposition is no longer solely a function of boundary information. Theoretically, in-situ densities should be used within the thermal wind relationship, as is done in Chapters \ref{TJ_TM} and \ref{TJ_Var}. However, for improved visualisation of the local density structure, neutral densities are used for the decomposition calculation, this has a negligible impact on the resulting ACC transport. We calculate the neutral densities for the section using the method outlined by \cite{Jackett1997} and \cite{McDougall2020}. 

The final, Ekman component of the cumulative ACC transport decomposition is estimated using the meridional wind stresses ($\tau^y$), in a similar fashion to that described in Equations \ref{Tek} and \ref{e_h} for the AMOC. The Ekman contribution to the cumulative ACC transport is given by
\begin{eqnarray}
T_{Ekm}(z; x) &=& - \frac{1}{\rho_0}\int_{S(z; x)}^{N(z; x)} \int_{-h(y; x)}^{z} \frac{1}{h(y; x)}\frac{\tau_s^y(y; x)}{f(y)} \mathrm{d}z'\mathrm{d}y \nonumber \\
&=& - \frac{1}{\rho_0}\int_{S(z; x)}^{N(z; x)} \frac{(h(y;x)+z)}{h(y; x)}\frac{\tau_s^y(y; x)}{f(y)} \mathrm{d}y
\label{e_ACC_Tek}
\end{eqnarray}
where $h(y; x)$ is the depth of the Ekman layer given by
\begin{align}
h(y; x) = & H(y; x),  & H(y; x) \leq 50m, \nonumber \\
h(y; x) = & 50m,  & H(y; x) > 50m,
\label{e_ACC_h}
\end{align}
and $\tau_s^y(y; x)$ is the meridional wind stress on the ocean's surface.

Therefore the full decomposition diagnostic of the zonal cumulative volume transport for a given $x$ with respect to depth, illustrated in Figure \ref{F_bdryACC}, is given by
\begin{equation}
\tilde{T}_{ACC}(z; x) = T_N(z; x) + T_S(z; x) + T_{\beta}(z;x) + T_{bot}(z; x) + T_{Ekm}(z; x)
\label{T_smp_ACC}
\end{equation}
or
\begin{equation}
\begin{multlined}
\tilde{T}_{ACC}(z; x) =  \frac{g}{\rho_0} \int_{-H_{max}(x)}^{z} (z-z') \left[ \sum_{d=1}^{n_D(z;x)} \left( \frac{\sigma_{N_d}(z;x)}{f_{N_d}(x)} \right) \right] \mathrm{d}z'\\ - \frac{g}{\rho_0} \int_{-H_{max}(x)}^{z} (z-z') \left[ \sum_{d=1}^{n_D(z;x)} \left( \frac{\sigma_{S_d}(z;x)}{f_{S_d}(x)} \right)  \right] \mathrm{d}z' \\ 
+ \int_{-H_{max}(x)}^{z} (z-z') \int_{S(z;x)}^{N(z;x)} \frac{\beta(y) \sigma(y,z';x)}{f(y)^2} \mathrm{d}y \mathrm{d}z' +  \int_{S(z;x)}^{N(z;x)} Z_{Bot} u_{bot}(y;x) \mathrm{d} y \\
- \frac{1}{\rho_0}\int_{S(z; x)}^{N(z; x)} \frac{(h(y;x)+z)}{h(y; x)}\frac{\tau_s^y(y; x)}{f(y)} \mathrm{d}y .
\end{multlined}
\label{e_tilde_T_ACC}
\end{equation}

\section{Decomposition of the ACC within the HadGEM3-GC3.1 general circulation model}

The decomposition is applied to data from the $1^\circ$, $1/4^\circ$ and $1/12^\circ$ oceanic spatial resolution control runs of the HadGEM-GC3.1 HighResMIP project, with HadGEM-GC3.1 model configurations as described in Chapter \ref{TJ_TM}. We include $30$-year spin-up phases at each model resolution, incorporating 1950s forcing and N216 atmospheric resolution. The spin-up phases are included to investigate the initial temporal trends within the models, where the greatest change in ACC transport is found, as illustrated in Figure \ref{F_Tru_Trsp}. All three model spin-ups begin from the same EN4 conditions.

For the present zonal cumulative transport decomposition, the calculation is centred upon the $v$-points of the NEMO computational grid shown in Figure \ref{NEMO_grid} and \ref{var_grid} (rather than the $u$-points used for meridional transport decomposition of the AMOC). The method outlined in Equations \ref{e_ACC_int_1}, \ref{T_Nrt}, \ref{T_Sth}, \ref{T_Beta} and \ref{e_ACC_Tek} is then applied as follows for interior cells (cells containing full field information, Figure \ref{F_Sch_Cells}) only. In the notation of Chapter \ref{TJ_TM}, the southern and northern boundary density contributions at depth $z_k$, time $t_\ell$ and longitude section $x_i$ take the form
\begin{equation}
	T_{S}(z_k, t_{\ell}; x_i) = -\frac{g}{\rho_{0}} \sum_{m=k}^{M}\left(z_{k-1 / 2}-z_{m}\right) \left[ \sum_{d=1}^{n_D(z_k;x_i)}  \frac{\sigma_{S_d}(z_m, t_{\ell}; x_i) \Delta z_{m}}{f_{S_d}(x_i)} \right] 
	\label{T_S_math}
\end{equation}
and
\begin{equation}
	T_{N}(z_k, t_{\ell}; x_i) = +\frac{g}{\rho_{0}} \sum_{m=k}^{M}\left(z_{k-1 / 2}-z_{m}\right) \left[ \sum_{d=1}^{n_D(z_k;x_i)} \frac{\sigma_{N_d}(z_m, t_{\ell}; x_i) \Delta z_{m}}{f_{N_d}(x_i)} \right]
	\label{T_N_math}
\end{equation}
where $k$ indexes the centre of a T-cell, with values running from $k=M$ at the deepest interior cell to $k=1$ at the surface. $\Delta z_m = z_{m-1/2} - z_{m+1/2}$ is the thickness of the $m^{\text{th}}$ level, and $f_{S_d}(x_i)$ and $f_{N_d}(x_i)$ are the Coriolis parameters for the $d^\text{th}$ southern and northern boundaries at longitude section $x_i$. Further, $\sigma_{S_d}(z_m, t_{\ell}; x_i)$ and $\sigma_{N_d}(z_m, t_{\ell}; x_i)$ are the corresponding neutral density anomalies at the $m^{\text{th}}$ depth level. 

The $\beta$ cumulative transport component $T_{\beta}$ takes the form


\begin{equation}
	T_{\beta}(z_k, t_{\ell}; x_i) = \sum_{m=k}^{M}\left(z_{k-1 / 2}-z_{m}\right) \sum_{j\in \mathcal{I}(z_m; x_i)} \frac{\beta(y_j;x_i) \bar{\sigma}(y_j, z_m, t_{\ell}; x_i) }{f(y_j; x_i)^2}  \Delta y_{j} \Delta z_{m} ,
	\label{T_Beta_math}
\end{equation}
where $\bar{\sigma}(y_j, z_m, t_{\ell}; x_i)$ represents the neutral density anomaly (relative to $\rho_{deep}$) averaged onto the $v$-point, since the density is located at the centre of the T-cell rather than at the $v$-point. $\mathcal{I}(z_k; x_i)$ is the index set for meridional coordinates between the southern and northern boundaries $S(z_k; x_i)$ and $N(z_k; x_i)$, and $\Delta y_j=y_{j-1/2} - y_{j+1/2}$ is the meridional cell width. The bottom component $T_{bot}$ is given by
\begin{equation}
	T_{bot}(z_k, t_{\ell}; x_i) = \sum_{j\in \mathcal{I}(z_k; x_i)}  (H(y_j; x_i) + z_{k-1/2}) \overline{u}_{bot}(y_j, t_{\ell}; x_i)  \Delta y_j ,
	\label{Ta_b_math}
\end{equation}
where $\overline{u}_{bot}(y_j, t_{\ell}; x_i)$ is the sum of (a) the $4$-point average of the local zonal velocities (averaged onto the $v$-point) and (b) the vertical gradient in meridional velocity, obtained using the meridional density gradient across the bottom cell and thermal wind relationship (see Equation \ref{T_b_math} and related text in Chapter \ref{TJ_TM}). The Ekman component takes the form
\begin{equation}
	T_{Ekm}(z_k, t_{\ell}; x_i) = - \frac{1}{\rho_0} \sum_{j\in \mathcal{I}(z_k; x_i)}  \frac{(h(y_j; x_i) + z_k)}{h(y_j; x_i)} \frac{\tau_s^y(y_j, t_{\ell}; x_i)}{f(y_j)}  \Delta y_j ,
	\label{Ta_Ek_math}
\end{equation}
where $h(y_j; x_i)$ is defined in Equation \ref{e_ACC_h}, and $\tau_s^y(y_j, t_{\ell}; x_i)$ is the meridional surface wind stress on the NEMO grid as before.

Near boundaries and bathymetry, additional cells are present, referred to as ``sidewall'' and ``partial'' cells respectively. The latter cells replicate the original boundary as closely as possible given model resolution. Unfortunately, potential temperature and salinity fields are not available in the additional cells, since the width of sidewall cells and depths of partial cells vary; hence the decomposition calculation cannot be replicated reliably for these cells. This is an unavoidable limitation of our approach. Sidewall and partial cells are illustrated in Figure \ref{F_Sch_Cells}. However, using zonal velocities present in these incomplete cells, we are able to estimate their transport contributions, referred to as $T_{AC}(z_k, t_{\ell}; x_i)$. $T_{AC}$ is calculated using Equation \ref{e_ACC_def} by directly integrating zonal velocities vertically and meridionally within the additional cells. Therefore,
\begin{equation}
	T_{AC}(z_k, t_{\ell}; x_i) =  \sum_{m=k}^{M+1} \sum_{j\in \mathcal{I}(z_m; x_i)} \overline{u}_{AC}(y_j,z_m, t_{\ell}; x_i) \Delta y_j \Delta z_m ,
	\label{e_T_AC_math}
\end{equation}
where $M+1$ is the vertical cell index for the additional cell in the basin, and $\overline{u}_{AC}(y_j,z_m, t_{\ell}; x_i)$ is the averaged zonal velocity within depth-latitude combinations $z_m,y_j$ corresponding to additional cells only, and set to zero otherwise.  

Using the transport components defined above, the estimated total zonal transport $\tilde{T}_{ACC}$, for each depth, time and longitude section, is then
\begin{eqnarray}
	\label{e_tilde_T_ACC_math}
	\tilde{T}_{ACC}(z_k, t_{\ell}; x_i) &=& T_{S}(z_k, t_{\ell}; x_i) + T_{N}(z_k, t_{\ell}; x_i) + T_{\beta}(z_k, t_{\ell}; x_i) \\
	&+& T_{bot}(z_k, t_{\ell}; x_i) + T_{Ekm}(z_k, t_{\ell}; x_i) + T_{AC}(z_k, t_{\ell}; x_i)\nonumber .
\end{eqnarray}

To validate this decomposition, the cumulative ACC transport is also calculated directly using Equation \ref{e_ACC_def} by vertically and meridionally integrating the zonal velocity (see Figure \ref{F_Tru_Trsp}). For this calculation, we use a $4$-point average velocity $\overline{u}$ at each NEMO $u$-point. The ``true'' or ``expected'' cumulative transport $T_{ACC}$ is then
\begin{equation}
	T_{ACC}(z_k, t_{\ell}; x_i) =  \sum_{m=k}^{M+1} \sum_{j\in \mathcal{I}(z_m; x_i)} \overline{u}(y_j,z_m, t_{\ell}; x_i) \Delta y_j \Delta z_m ,
	\label{e_T_ACC_math}
\end{equation}
where $M$ is the vertical cell index for the bottommost interior cell in the basin, and level $M+1$ indicates an additional cell (similar to Equation \ref{e_T_math}). The resulting total cumulative transport $T_{ACC}(z_k, t_{\ell}; x_i)$ includes additional cell contributions, and can thus be compared directly to the total estimated transport $\tilde{T}_{ACC}(z_k, t_{\ell}; x_i)$ calculated as the sum of the boundary and $\beta$ components (Equation \ref{e_tilde_T_ACC_math}). 

\section{Interrogating ACC Transport within HadGEMGC3.1 models at the Drake Passage}

In this section, we explore various aspects of cumulative ACC transport characteristics at $66.5^\circ$W. Section \ref{Sct_Inter_TT} compares decomposition estimate $\tilde{T}_{ACC}$ with expected transport $T_{ACC}$ as a function of HadGEMGC3.1 model resolution. Section \ref{Sct_Inter_CS} examines properties along a meridional section including salinity, potential temperature and neutral densities, for the initial model spin-up phase. Section \ref{Sct_Inter_Dcm} describes results for the components of the cumulative transport decomposition analysis. Finally, Section \ref{Sct_Inter_Wdd} considers possible physical rationales for the weakening of the ACC during and after the spin-up phase in the $1/4^\circ$ model. Note that hereafter, ACC transport refers to the ACC volume transport integrated from the sea floor up to the surface.

\subsection{Total transport at $66.5^\circ$W} \label{Sct_Inter_TT}
Disparities in volume transport $T_{ACC}$ through the Drake Passage, calculated directly from zonal velocities using Equation \ref{e_ACC_def}, have already been illustrated in Figure \ref{F_Tru_Trsp}. Disparities between transports $T_{ACC}$ in models with spatial resolutions of $1^\circ$ (orange), $1/4^\circ$ (blue) and $1/12^\circ$ (green) raise questions about the resolution dependence of the ACC within HadGEMGC3.1 models, and indeed whether the models at any resolution provide realistic estimates of the ACC transport. The red and blue shaded regions of Figure \ref{F_Tru_Trsp} indicate recent observations of $T_{ACC}$ by \cite{Meredith2011} and \cite{Donohue2016}, as discussed in Section \ref{Sct_ACC_Bck}. 

These literature values are calculated near or on the $SR1b$ repeat hydrography line, close to $61^\circ$W longitude. However, we choose to perform all our direct transport and decomposition calculations at $66.5^\circ$W, the nearest meridional section for which northern and southern boundaries of the Drake Passage are present for all model resolutions. 

Figure \ref{F_Tru_Trsp} indicates an initial weakening of $T_{ACC}$ for all spatial resolutions. This is especially evident for the $1/4^\circ$ model, exhibiting a reduction of almost $60$Sv within the first $30$-year spin-up phase of model integration; corresponding weakening of around $20$Sv and $30$Sv is observed for $1^\circ$ and $1/12^\circ$ model resolutions, respectively. None of the model-calculated transports lie within the values given by \cite{Donohue2016}. Transport calculated from the $1^\circ$ model stabilises at $\approx 157$Sv, mid-way between the literature estimates. Surprisingly, $T_{ACC}$ calculated from the $1/4^\circ$ model continues to weaken post $30$-year spin-up, eventually stabilising at around $65$Sv. Clearly the behaviour of $T_{ACC}$ during the initial $30$-year spin-up phase for each model run is critical to understanding the ACC weakening.

\begin{figure*}[ht!]
	\centerline{\includegraphics[width=\textwidth]{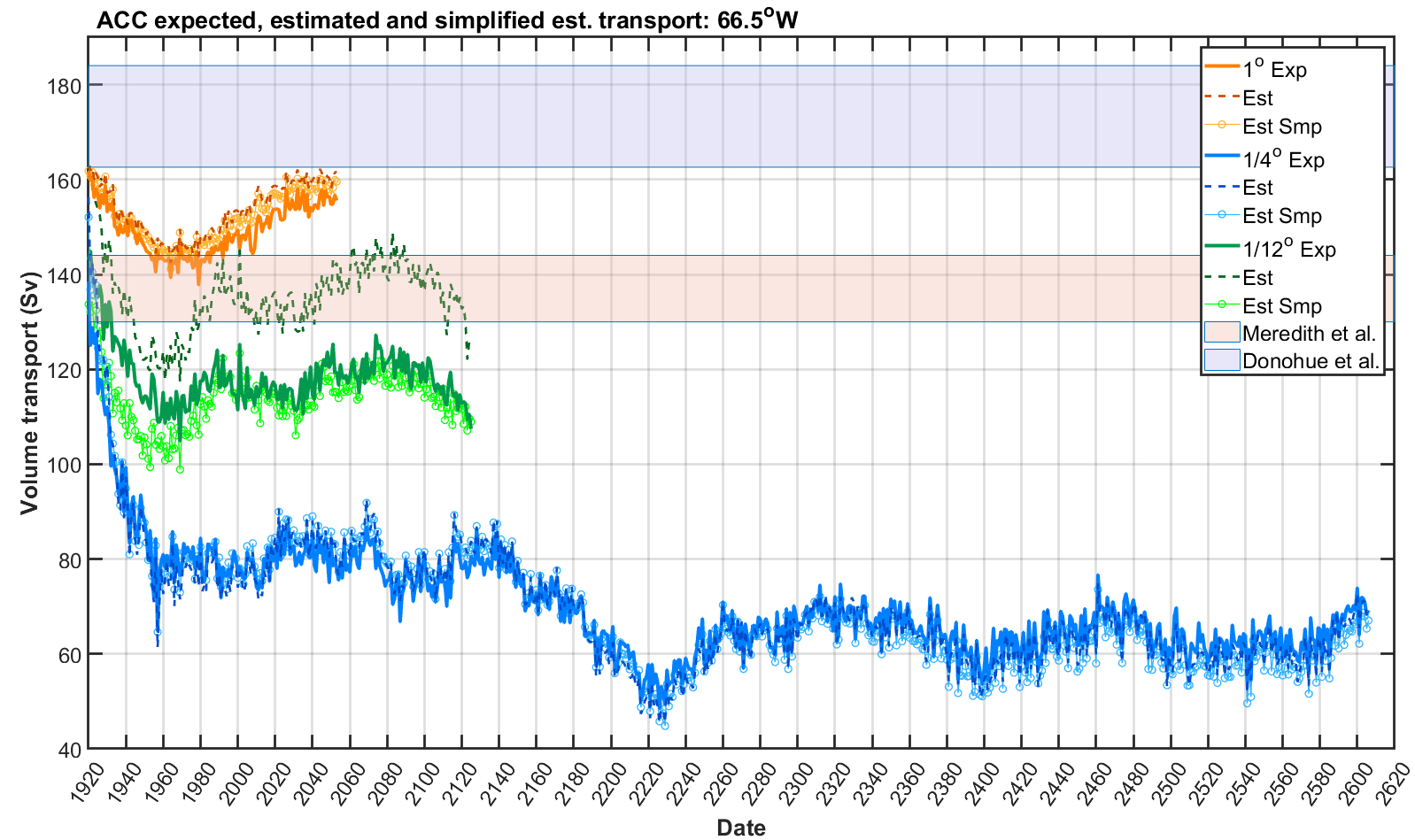}}
	\caption[Timeseries of cumulative volume transports: expected $T_{ACC}$, estimated $\tilde{T}_{ACC}$ and simplified boundary estimate $\tilde{T}_{ACC_{SP}}$ at the surface through the Drake Passage for $1^\circ$, $1/4^\circ$ and $1/12^\circ$ spatial resolution at $66.5^\circ$W.]{Timeseries of cumulative volume transports: expected $T_{ACC}$, estimated $\tilde{T}_{ACC}$ (dashed) and simplified boundary estimate $\tilde{T}_{ACC_{SP}}$ (line with circles) at the surface through the Drake Passage for the $1^\circ$ (orange), $1/4^\circ$ (blue) and $1/12^\circ$ (green) spatial resolution at $66.5^\circ$W. Dashed lines indicate estimated volume transport comprising multiple northern and southern boundary densities, $\beta$, depth-independent flow and Ekman components. Lines with circles indicate the estimated volume transport using a simplified boundary density term, where a single density value is used for each boundary at each depth. The shaded red region indicates the estimate from \cite{Meredith2011}, blue that from \cite{Donohue2016} (including a depth-independent component of transport).}
	\label{F_SB_Trsp}
\end{figure*}

Timeseries of the three ACC decomposition estimates $\tilde{T}_{ACC}$ (Equation \ref{e_tilde_T_ACC}) at $66.5^\circ$W are shown as dashed lines alongside directly-calculated $T_{ACC}$ in Figure \ref{F_SB_Trsp}. The discrepancy between $\tilde{T}_{ACC}$ and ${T}_{ACC}$ for the $1^\circ$ and $1/4^\circ$ models is small, with the $1/4^\circ$ estimate in particular performing extremely well in capturing the expected ACC transport. The corresponding discrepancy for the $1/12^\circ$ model is large. We explain this observation as follows. As shown in e.g. Equations \ref{T_S_math} and \ref{T_N_math}, calculation of the decomposition diagnostic can involve multiple pairs of southern and northern boundaries. The number of pairs is greatest for the $1/12^\circ$ model resolution, which exhibits greatest bathymetric detail, leading to numerous small trenches being incorporated into the calculation. We hypothesise that density representation in these small trenches is poor, resulting in an increased estimate $\tilde{T}_{ACC}$ relative to ${T}_{ACC}$ for this model resolution. 

To explore this effect further, a decomposition using a single pair (of southernmost and northernmost boundaries) at each depth to calculate the boundary density contributions was performed, ignoring the contributions of all intermediate boundaries. The resulting estimated transport $\tilde{T}_{ACC_{SP}}$ for the $1/12^\circ$ model at $66.5^\circ$W gives much improved agreement with ${T}_{ACC}$ as shown also in Figure \ref{F_SB_Trsp} (as lines with circles). The coarser nature of the $1^\circ$ and $1/4^\circ$ model bathymetries leads to little difference between $\tilde{T}_{ACC_{SP}}$  and $\tilde{T}_{ACC}$ for those model resolutions.

The improvement found for the $1/12^\circ$ model at $66.5^\circ$W using the single boundary pair decomposition is not replicated for other meridional sections (e.g. between Antarctica and South Africa, New Zealand, Malvinas Islands) exhibiting large-scale intermediate bathymetry effects (such as islands, large ridges and trenches). This highlights the strong influence of bathymetry upon the diagnostics. In general we find the use of multiple southern and northern boundaries to give the best representation of $\tilde{T}_{ACC}$ across resolutions for varying meridional sections, discussed further later in the chapter. 

\subsection{Cross-sectional properties at the Drake Passage}  \label{Sct_Inter_CS}

We investigate the characteristics of underlying physical quantities, available in HadGEMGC3.1 model output, which may be related to the apparent weakening transport at Drake Passage. Depth-latitude cross-sectional comparisons at $66.5^\circ$W are made for potential temperature $\theta$, salinity $\mathcal{S}$, and neutral densities $\rho_{neut}$ for each of the three model resolutions. Our focus lies on the initial 30-year spin-up phase of the model integrations, where significant changes in transport are found across all resolutions. 
\begin{figure*}[ht!]
	\centerline{\includegraphics[width=\textwidth]{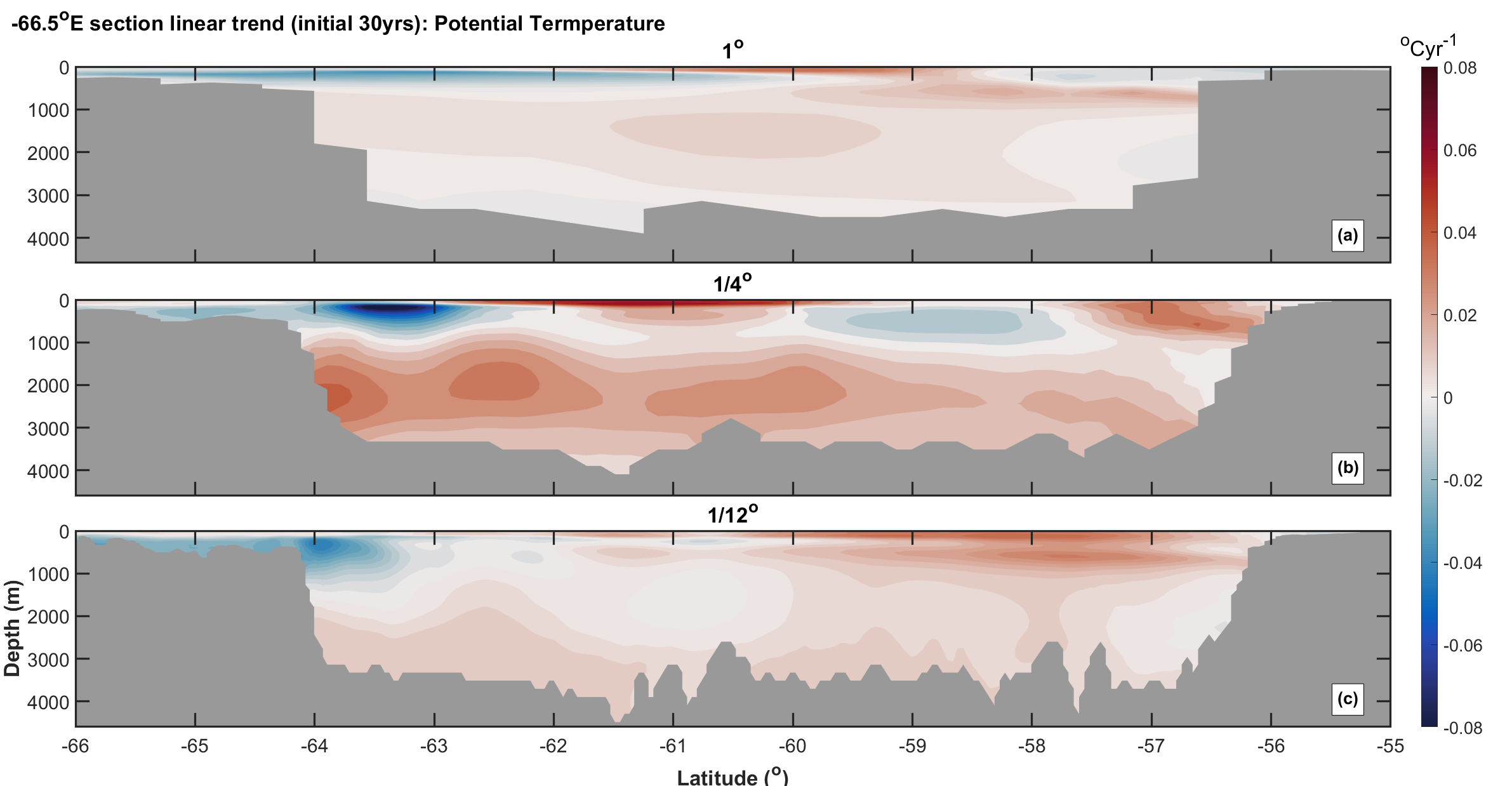}}
	\caption[Cross-sections of trend in potential temperature for the first 30 years (spin-up phase) of model run, estimated using linear regression. Section taken at $66.5^\circ$W at the Drake Passage for the $1^\circ$, $1/4^\circ$ and $1/12^\circ$ spatial resolutions.]{Cross-sections of trend in potential temperature $\theta$ for the first 30 years (spin-up phase) of model run, estimated using linear regression. Section taken at $66.5^\circ$W at the Drake Passage for the (a) $1^\circ$, (b) $1/4^\circ$ and (c) $1/12^\circ$ spatial resolutions. Red indicates increasing potential temperature.}
	\label{F_T_Trnd}
\end{figure*}

\begin{figure*}[ht!]
	\centerline{\includegraphics[width=\textwidth]{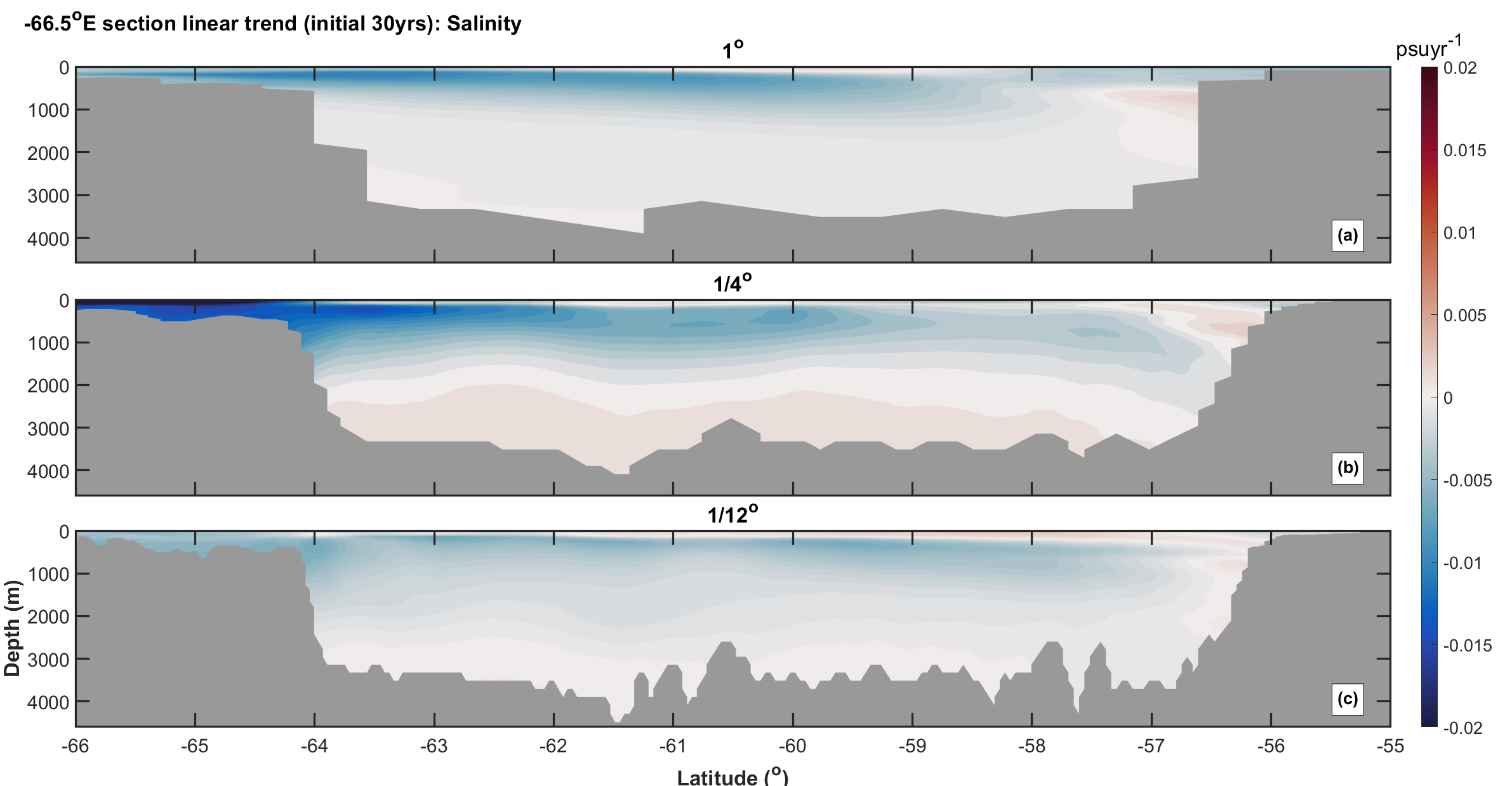}}
	\caption[Cross-sections of the trend in salinity during the first 30 years (spin-up phase) of model run, estimated using linear regression. Section taken at $66.5^\circ$W at the Drake Passage for the $1^\circ$, $1/4^\circ$ and $1/12^\circ$ model resolutions.]{Cross-sections of the trend in salinity $\mathcal{S}$ during the first 30 years (spin-up phase) of model run, estimated using linear regression. Section taken at $66.5^\circ$W at the Drake Passage for the (a) $1^\circ$, (b) $1/4^\circ$ and (c) $1/12^\circ$ model resolutions. Blue indicates decreasing salinity.}
	\label{F_S_Trnd}
\end{figure*}

Figures \ref{F_T_Trnd} and \ref{F_S_Trnd} show the initial trends found for $\theta$ and $\mathcal{S}$ within the spin-up phase. The trend was estimated using linear regression over the first 30 years of output, independently at each depth-latitude pair available. Figure \ref{F_T_Trnd} suggests that temperatures at each model resolution generally exhibits a warming trend within the interior of the section, and a slight cooling near the Antarctic shelf. The $1/4^\circ$ model in Panel (b) shows greater warming at depth and significant cooling in the upper 500m near $63.5^\circ$S. Similar cooling characteristics are found in the $1/12^\circ$ model. The $\theta$ trends in the northern part of the section show a significant warming in both higher resolutions models, indicating a lightening of waters near the northern boundary within the spin-up phase. 

\begin{figure*}[ht!]
	\centerline{\includegraphics[width=\textwidth]{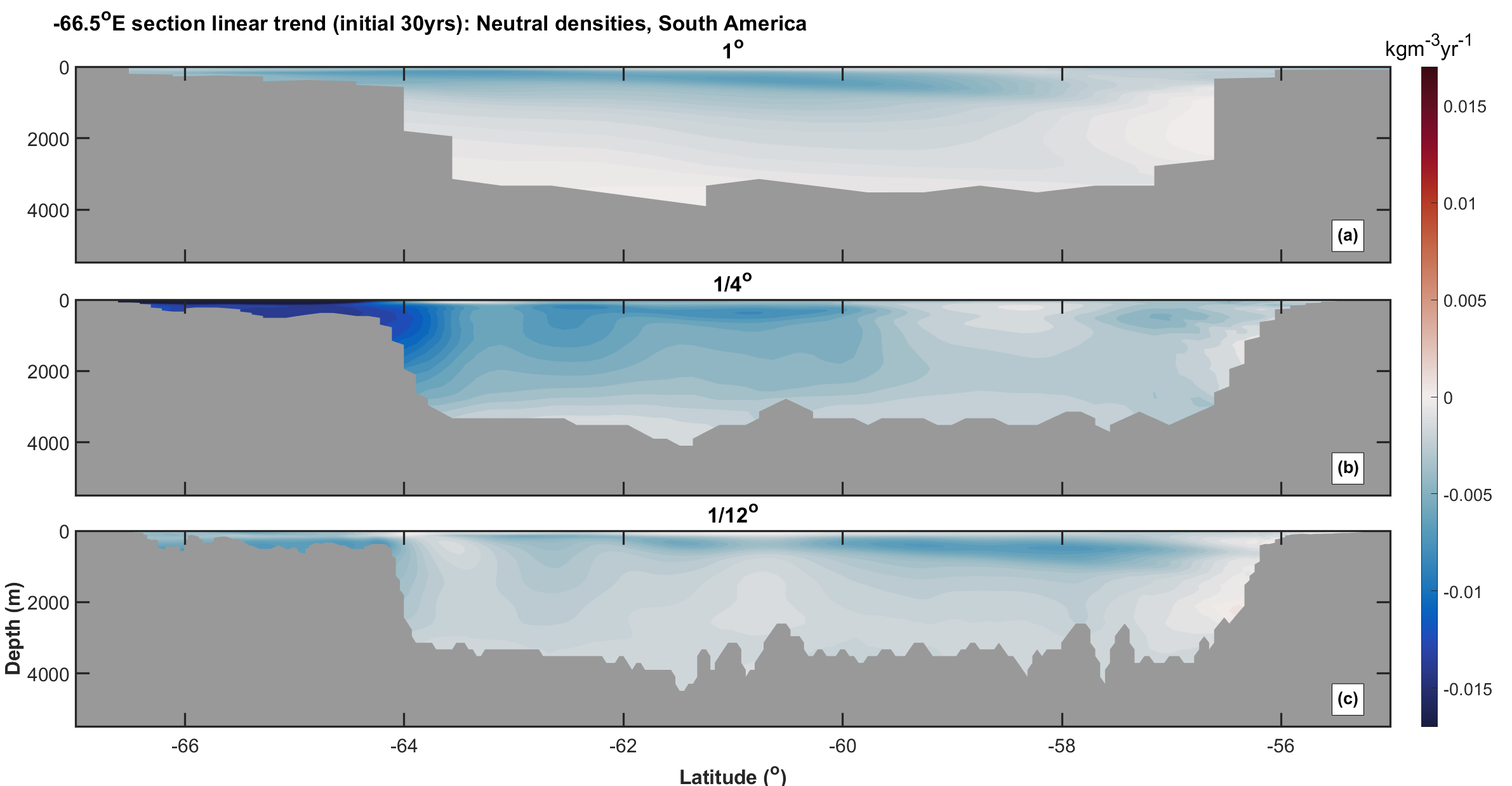}}
	\caption[Cross-sections of trend in neutral densities during the first 30 years (spin-up phase) of model run, estimated using linear regression. Section taken at $66.5^\circ$W at the Drake Passage for the $1^\circ$, $1/4^\circ$ and $1/12^\circ$ model resolutions.]{Cross-sections of trend in neutral densities $\rho_{neut}$ during the first 30 years (spin-up phase) of model run, estimated using linear regression. Section taken at $66.5^\circ$W at the Drake Passage for the (a) $1^\circ$, (b) $1/4^\circ$ and (c) $1/12^\circ$ model resolutions. Blue indicates decreasing neural density.}
	\label{F_ND_Trnd}
\end{figure*}

At high latitudes, density variations are driven mainly by changes in salinity rather than temperature. We find a general freshening or reduction in salinity in Figure \ref{F_S_Trnd} throughout a large portion of the interior, for all 3 model resolutions considered. This general spatial characteristic is enhanced for the $1/4^\circ$ model in Panel (b). The large decrease in salinity observed near the southern boundary would indicate a similar large change in density, possibly due to freshwater input from melting sea-ice or a freshening of the along-coast boundary current. The trends found in $\theta$ and $\mathcal{S}$ (Figures \ref{F_T_Trnd} and \ref{F_S_Trnd}) suggest a general lightening of densities across the section, with greater density reduction along the southern boundary in the $1/4^\circ$ model output as shown in Figure \ref{F_ND_Trnd}.
\begin{figure*}[ht!]
	\centerline{\includegraphics[width=\textwidth]{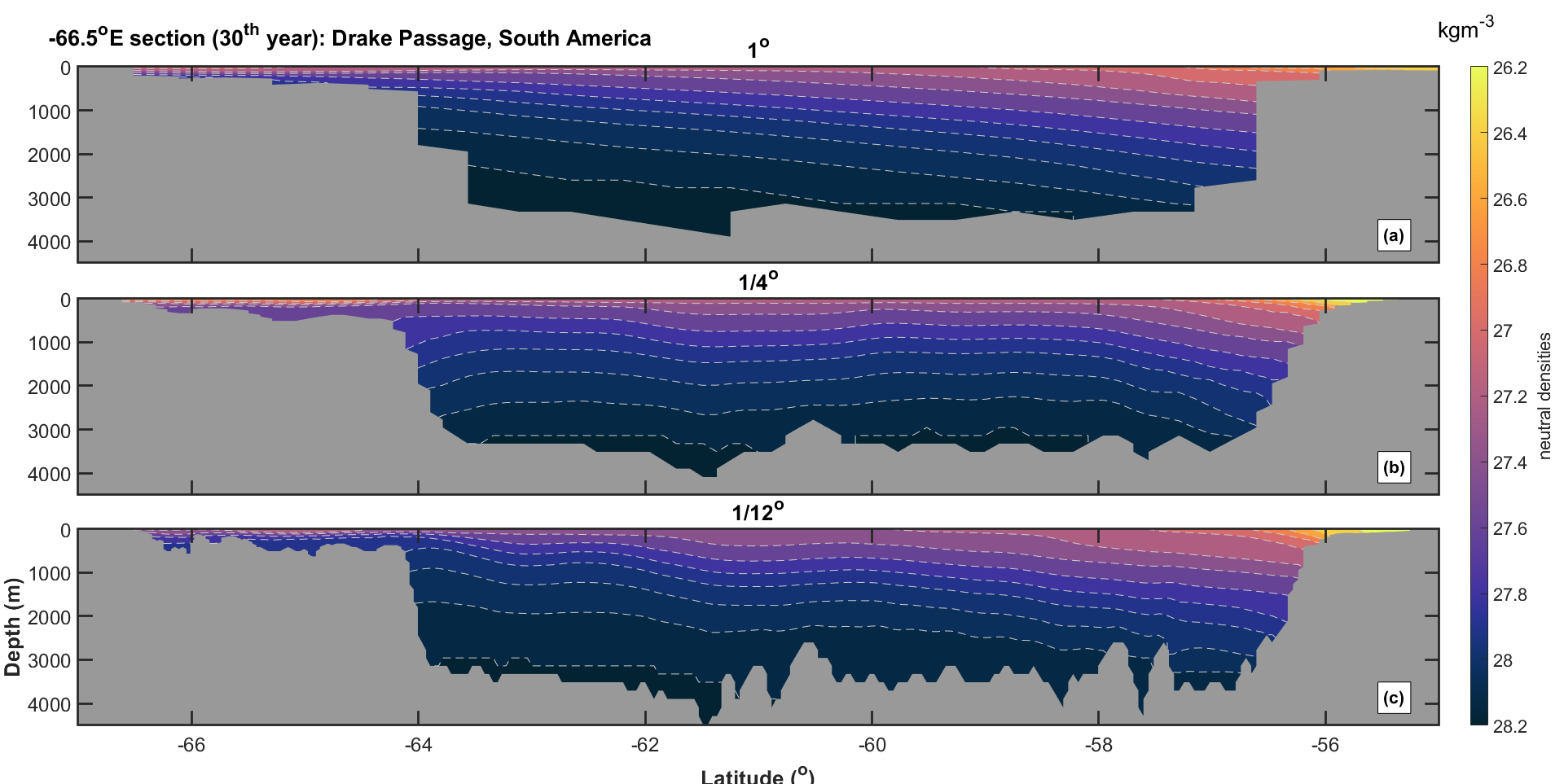}}
	\caption[Cross-sections of neutral densities $\rho_{neut}$ for the 30$^{\text{th}}$ year of model spin-up. Section taken at $66.5^\circ$W at the Drake Passage for the $1^\circ$, $1/4^\circ$ and $1/12^\circ$ model resolutions.]{Cross-sections of neutral densities $\rho_{neut}$ for the 30$^{\text{th}}$ year of model spin-up. Section taken at $66.5^\circ$W longitude at the Drake Passage for the (a) $1^\circ$, (b) $1/4^\circ$ and (c) $1/12^\circ$ model resolutions.}
	\label{F_ND_Avg}
\end{figure*}

\begin{figure*}[ht!]
	\centerline{\includegraphics[width=\textwidth]{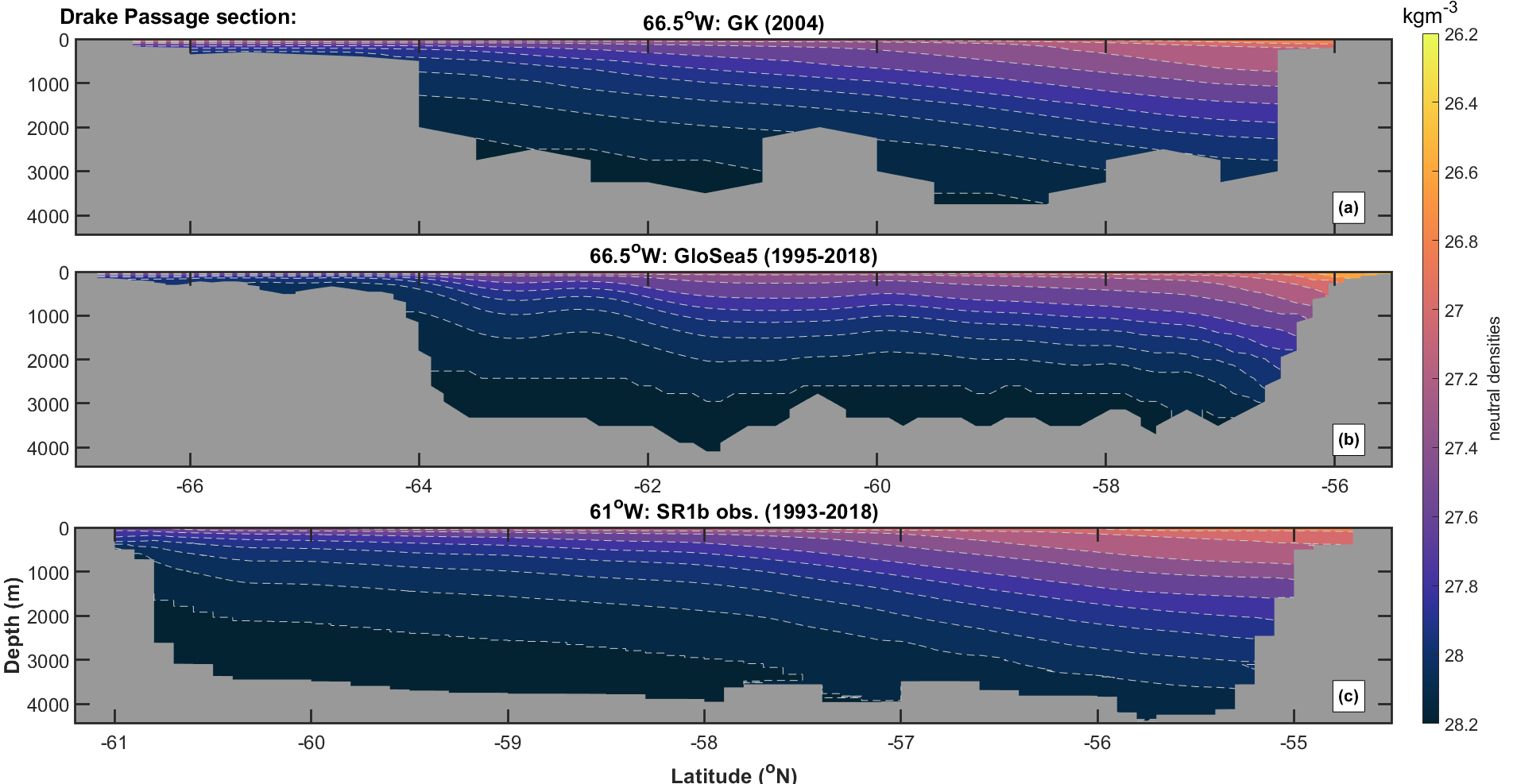}}
	\caption[Cross-sections of average neutral densities $\rho_{neut}$ for the  \cite{GouretskiViktorandKoltermann2004} climatology, GloSea5 reanalysis and SR1b section shipboard observations. Section taken at $66.5^\circ$W at the Drake Passage for GK and GloSea5, observations from SR1b section located close to $61^\circ$W.]{Cross-sections of average neutral densities $\rho_{neut}$ for the (a) \cite{GouretskiViktorandKoltermann2004} climatology, (b) GloSea5 reanalysis and (c) SR1b section shipboard observations. Section taken at $66.5^\circ$W at the Drake Passage for (a) GK and (b) GloSea5. Panel (c) shows observations from the SR1b section located close to $61^\circ$W; note the change in latitudinal extent of the section.}
	\label{F_ND_ObsAvg}
\end{figure*}

The different trends in $\theta$, $\mathcal{S}$ and $\rho_{neut}$ in each model simulation, result in material differences between cross-sectional $\rho_{neut}$ after the first $30$ years. These differences are most prominent near the southern boundary. Panel (a) of Figure \ref{F_ND_Avg} shows $\rho_{neut}$ in the 30$^{\text{th}}$ year of model spin-up for the $1^\circ$ model, in reasonable agreement with the expected isopycnal structure inferred from current climatological, reanalysis and observational data shown in Figure \ref{F_ND_ObsAvg}. Specifically, isopycnals rise from the northern towards the southern boundary, with many outcropping before the southern boundary. $\rho_{neut}$ for the spin-up phase of the $1/12^\circ$ model shown in Panel (c) exhibits similar spatial features to the $1^\circ$ model output, except for a few oscillations in isopycnals near the Antarctic coast, also present in the GloSea5 reanalysis dataset (Figure \ref{F_ND_ObsAvg}(b)). These features could possibly signify re-circulations within the interior of the ocean via standing eddies etc. 

From Figure \ref{F_Tru_Trsp}, we know that significant weakening of ACC transport is observed for the $1/4^\circ$ model simulation. The corresponding $\rho_{neut}$ after the spin-up phase (shown in Figure \ref{F_ND_Avg}(b)), shows slumping of isopycnals near the southern boundary, as opposed to outcropping as expected. We attribute this to the previously observed freshening on the Antarctic shelf. Near the Antarctic coastline, the local north-south isopycnal gradient, in conjunction with geostrophy, implies the presence of a local return flow where isopycnals slump, in contrast to the main eastward ACC flow. A local westward flow along the coastline through the Drake Passage weakens the net transport. There is also evidence of similar slumping of southern boundary isopycnals in the $1/12^\circ$ model, particularly between 800m and 1200m.   

\subsection{Decomposition of ACC transport into boundary and $\beta$ components} \label{Sct_Inter_Dcm}

\begin{figure*}[ht!]
	\centerline{\includegraphics[width=\textwidth]{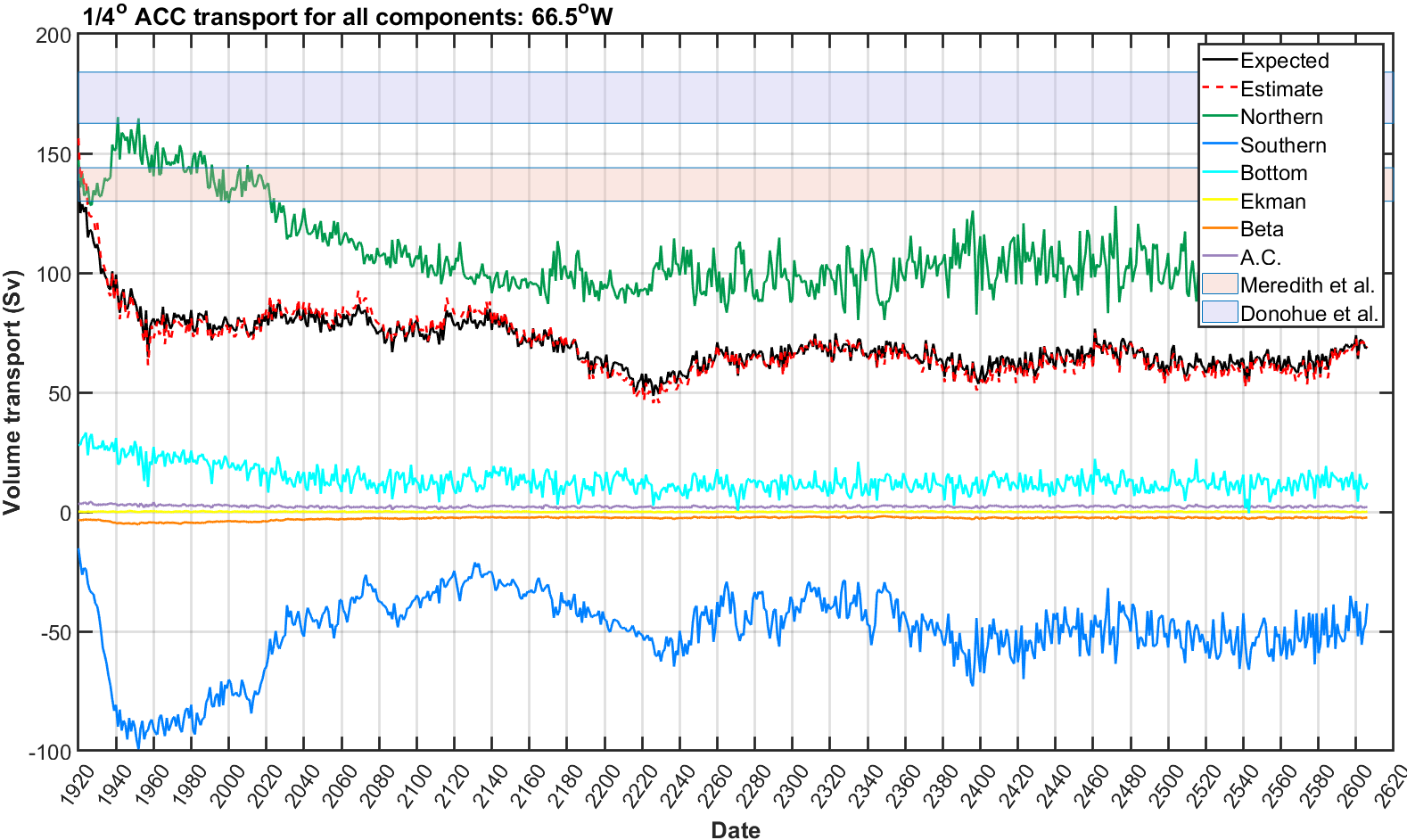}}
	\caption[Timeseries of expected $T_{ACC}$, estimated $\tilde{T}_{ACC}$ and contributing components through the Drake Passage at $66.5^\circ$W for the $1/4^\circ$ model.]{Timeseries of expected $T_{ACC}$ (black), estimated $\tilde{T}_{ACC}$ (red dashed) and contributing component cumulative volume transports at the surface through the Drake Passage at $66.5^\circ$W for the $1/4^\circ$ model. Contributing components are (a) northern boundary density ($T_N$, green), (b) southern boundary density ($T_S$, dark blue), (c) bottom (depth-independent $T_{bot}$, light blue), (d) Ekman ($T_{Ekm}$, yellow), (e) beta ($T_{\beta}$, orange) and (d) additional cell ($T_{AC}$, purple). Shaded red and blue regions indicate estimates from \cite{Meredith2011} and \cite{Donohue2016} respectively.}
	\label{F_Dcmp_All_Qtd}
\end{figure*}

The decomposition estimate $\tilde{T}_{ACC}$, shown as dashed lines in Figure \ref{F_SB_Trsp}, is the sum of six components (Equation \ref{e_tilde_T_ACC_math}). Three components correspond to northern, southern and $\beta$ density contributions. The depth-independent component corresponds to zonal bottom velocities, and the Ekman component to meridional wind stresses. The final component arises from the transport contributions of additional cells. Figure \ref{F_Dcmp_All_Qtd} illustrates the evolution of each of these components in time, for the transport accumulated up to the surface from the $1/4^\circ$ model. For reference, timeseries of ${T}_{ACC}$ and $\tilde{T}_{ACC}$ are also given. 

We observe that $T_{\beta}$, $T_{Ekm}$ and $T_{AC}$ are small throughout the period. The biggest contribution from any of these components, of approximately $-5$Sv, is due to $T_{\beta}$ within the first $50$ years. $T_{\beta}$ subsequently decreases in magnitude and stabilises near $-2$Sv. This pattern of initial increase and subsequent decrease and stabilisation of $T_{\beta}$ is also seen in $1^\circ$ and $1/12^\circ$ model outputs, which stabilise at $-5.5$Sv and $-4$Sv respectively. $T_{Ekm}$ is found to oscillate between $\pm 0.4$Sv throughout the period for all model resolutions (not shown). $T_{AC}$ lies at approximately $2$Sv for both higher-resolution models; for the 1$^\circ$ model, however, its value is near $10$Sv. We attribute the larger value of $T_{AC}$ at 1$^\circ$ to larger model grid cell size and hence greater relative importance of additional cells in the transport calculation. The results for $T_{\beta}$ and $T_{Ekm}$ are similar to those found by \cite{Allison2009}; however, our $T_{AC}$ contribution is smaller than the residual term they find.

\begin{figure*}[ht!]
	\centerline{\includegraphics[width=\textwidth]{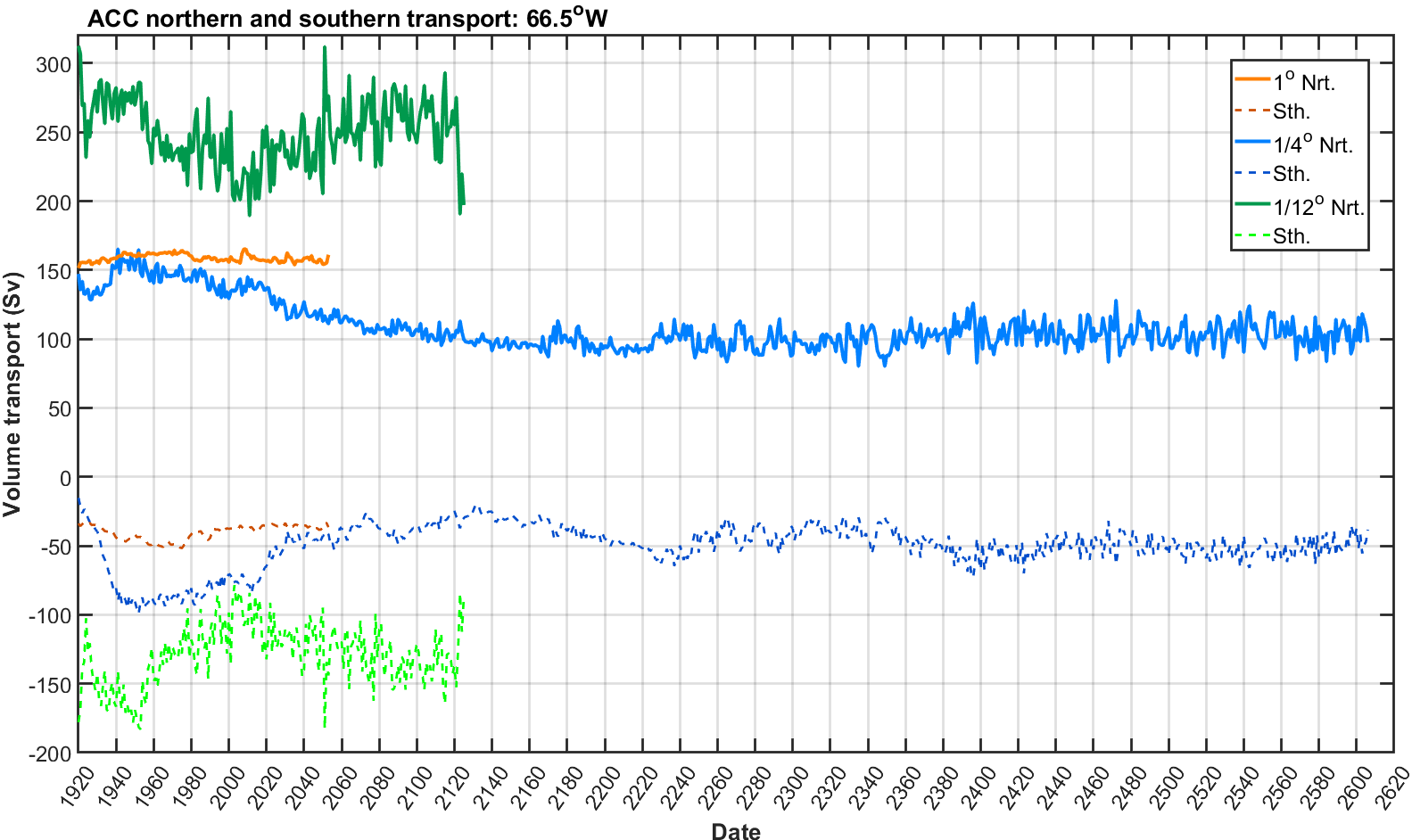}}
	\caption[Timeseries of northern and southern boundary density contributions to the estimated volume transport $\tilde{T}_{ACC}$ through the Drake Passage for the $1^\circ$, $1/4^\circ$ and $1/12^\circ$ spatial resolutions. Section taken at $66.5^\circ$W.]{Timeseries of northern ($T_N$, solid) and southern ($T_S$, dashed) boundary density contributions to the estimated volume transport $\tilde{T}_{ACC}$ integrated up to the surface through the Drake Passage for the $1^\circ$ (orange), $1/4^\circ$ (blue) and $1/12^\circ$ (green) spatial resolutions. Section taken at $66.5^\circ$W.}
	\label{F_Dcmp_NS}
\end{figure*}

The major contributors to $\tilde{T}_{ACC}$ are $T_S$, $T_N$ and $T_{bot}$. Timeseries of $T_S$ and $T_N$ for each model resolution are shown in Figure \ref{F_Dcmp_NS}. $T_S$ and $T_N$ generally show high negative correlation for each model resolution. This negative dependence is similar to that observed for western and eastern boundary contributions in the Atlantic overturning streamfunction. That is, for the ACC, $T_S$ and $T_N$ tend to compensate each other, except for the early years. 

For the $1/4^\circ$ model, the magnitude of $T_S$ increases more quickly than that of $T_N$ within the first 40 years. We have observed from Figures \ref{F_T_Trnd}-\ref{F_ND_Trnd} a large freshening of the southern boundary for the $1/4^\circ$ model; this freshening appears to increase the magnitude of the southern boundary transport contribution by $70$Sv. This increase in southern boundary contribution is partially compensated by an increase in northern contribution; however, southern boundary freshening reduces ACC transport by approximately 60Sv in the first 40 years for the $1/4^\circ$ model. A similar but weaker trend is observed within the $1/12^\circ$ model. In contrast, timeseries of $T_S$ and $T_N$ in the $1^\circ$ model remain relatively stable, with some initial strengthening of both components within the first 50 years. This coincides with slight weakening in $\tilde{T}_{ACC}$ at this time (see Figure \ref{F_SB_Trsp}). Differences in magnitudes of $T_S$ and $T_N$ for the $1/12^\circ$ resolution are attributed to the greater number of boundary pairs found at that resolution. The contribution made by $T_S$ is considerably larger within HadGEM-GC3.1 models than found by \cite{Allison2009}; in their study, $T_S$ has a magnitude of 5Sv only. Of course, we note that the magnitudes of $T_N$ and $T_S$ are arbitrary and dependent on the choice of reference density used.

The initial strengthening of $T_S$ for the $1/4^\circ$ model reverses after the first 40 years. That is, for a period of approximately 150 years following the initial phase, $T_S$ weakens and returns eventually to a value similar to its initial value. During the same 150 year period, $T_N$ exhibits a slow weakening, compensating for the change in $T_S$. Hence, the overall estimated cumulative transport $\tilde{T}_{ACC}$ does not exhibit a strong trend during this 150 year period; for the same period, no strong trend is seen in the expected transport $T_{ACC}$ either. A similar pattern of behaviour is observed for the $1/12^\circ$ model, but with greater interannual temporal variability.

\begin{figure*}[ht!]
	\centerline{\includegraphics[width=\textwidth]{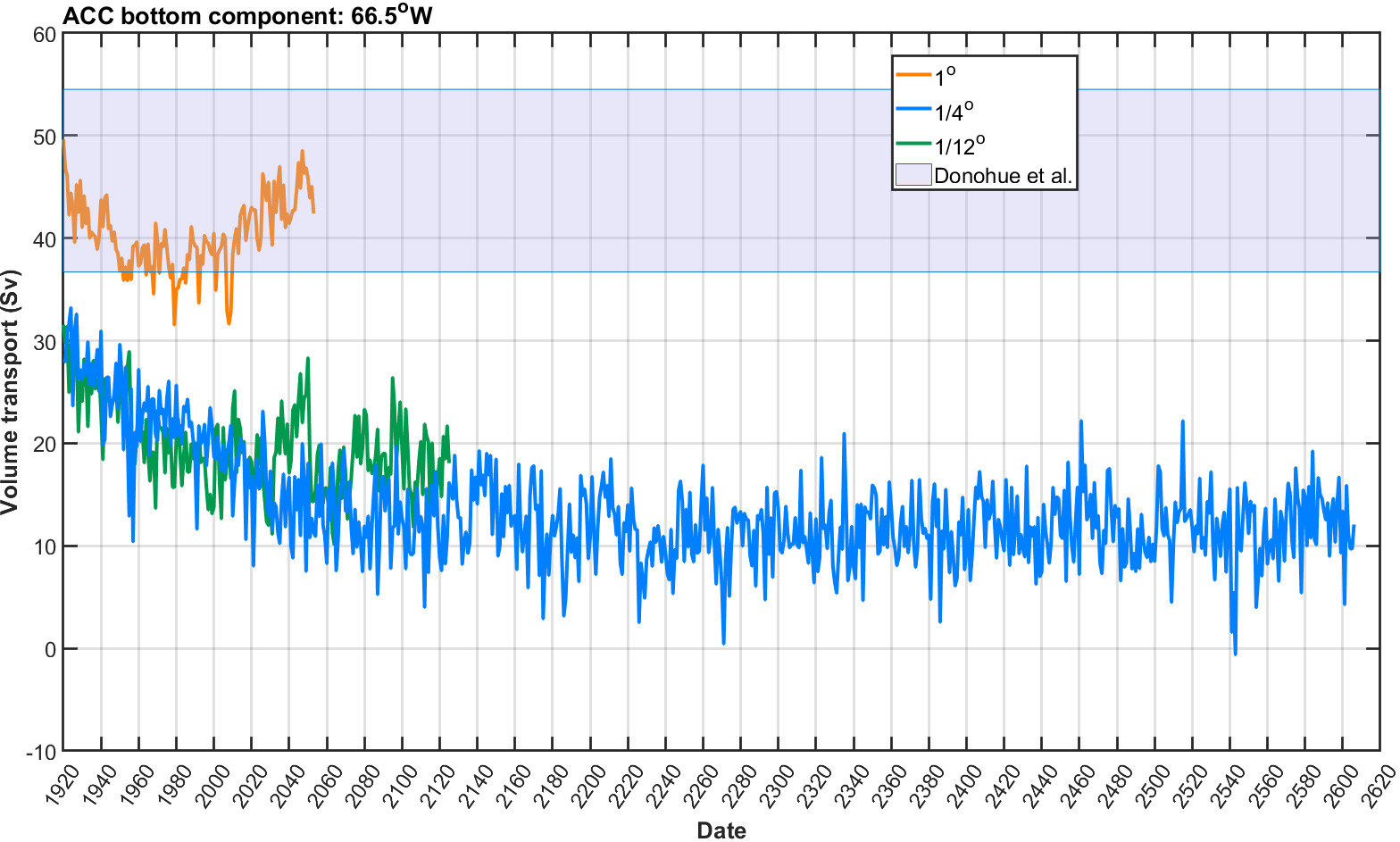}}
	\caption[Timeseries of the depth-independent flow contribution $T_{bot}$ to the estimated cumulative volume transport $\tilde{T}_{ACC}$ the Drake Passage for the $1^\circ$, $1/4^\circ$ and $1/12^\circ$ models. Section taken at $66.5^\circ$W.]{Timeseries of the depth-independent flow contribution $T_{bot}$ to the estimated cumulative volume transport $\tilde{T}_{ACC}$ through the Drake Passage for the $1^\circ$ (orange), $1/4^\circ$ (blue) and $1/12^\circ$ (green) models. Section taken at $66.5^\circ$W. Blue shaded region indicates the range of observed values found by \cite{Donohue2016} using current meters.}
	\label{F_Dcmp_bot}
\end{figure*}

The final large contributor to $\tilde{T}_{ACC}$ is the bottom or depth-independent flow component $T_{bot}$, shown by \cite{Donohue2016} to contribute over a quarter of the observed ACC transport through the Drake Passage. Figure \ref{F_Dcmp_bot} shows the \cite{Donohue2016} estimate as a blue shaded region, alongside the decomposition estimates for the $1^\circ$ (orange), $1/4^\circ$ (blue) and $1/12^\circ$ (green) models. Only the $1^\circ$ model provides estimates for $T_{bot}$ within the range of values suggested by \cite{Donohue2016}.

Higher resolution models exhibit a starting value for $T_{bot}$ which is already some 10Sv smaller than the range of values suggested by \cite{Donohue2016}; the value of $T_{bot}$ subsequently weakens further. The end value of $T_{bot}$ for the $1/4^\circ$ simulation is around $35$Sv weaker than that suggested by \cite{Donohue2016}. The final value for $T_{bot}$ from the $1/12^\circ$ model is also found to be around $25$Sv weaker than expected, accounting for almost half the difference between the literature value and the expected $T_{ACC}$ from that model. 

A possible future investigation to help explain discrepancies in estimates for $T_{bot}$, would compare bottom currents measured by \cite{Donohue2016} with those generated from GCMs, confirming the extent to which GCMs are able to resolve bottom flows within the Drake Passage for different model resolutions. In particular, this may reveal why the coarser resolution model apparently outperforms both higher resolution models here.
 
\subsection{The Weddell gyre} \label{Sct_Inter_Wdd}

Sections \ref{Sct_Inter_TT}-\ref{Sct_Inter_Dcm} suggest a significant change in physical properties along the southern boundary of the Drake Passage within the $1/4^\circ$ model over the spin-up phase, leading to a weakening ACC transport. The freshening of the Antarctic shelf leads to a stronger return flow along the shelf. Here we consider whether these changes might be associated with changes in the characteristics of the Weddell Gyre, formed by wind stresses and Coriolis effects in the Weddell Sea. 

\subsubsection*{Sea surface height in Weddell Sea and the Drake Passage}
\begin{figure*}[ht!]
	\centerline{\includegraphics[width=\textwidth]{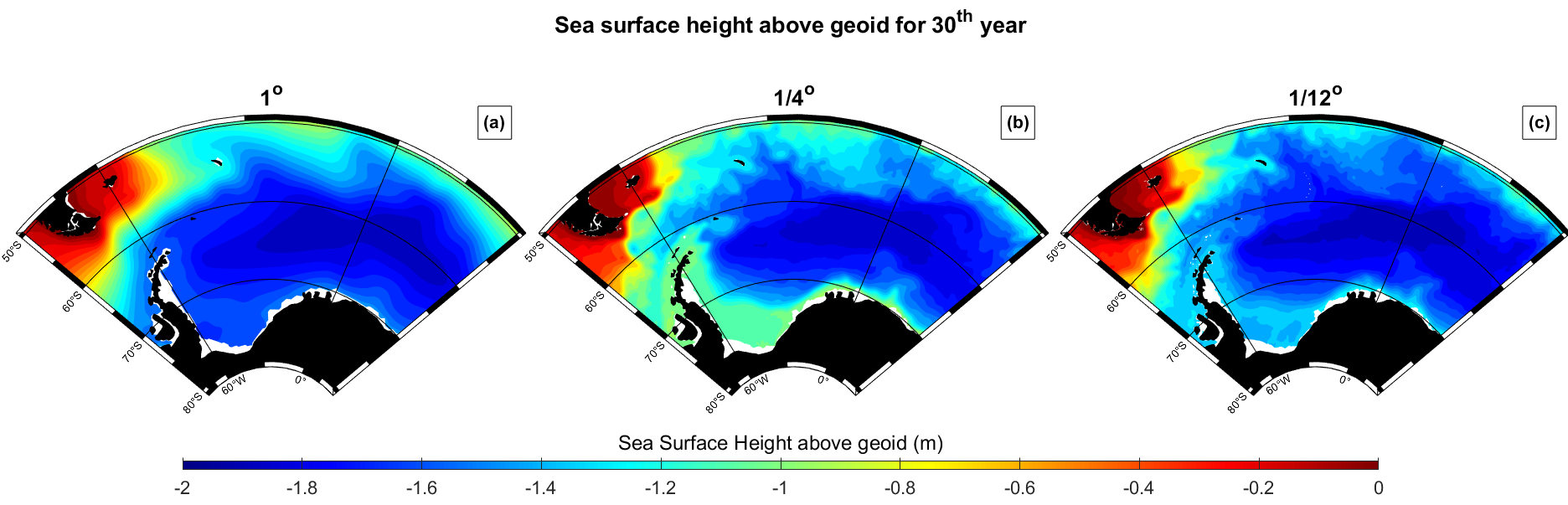}}
	\caption[Weddell Sea surface height above the geoid for $1^\circ$, $1/4^\circ$ and $1/12^\circ$ resolution models for the $30^{\text{th}}$-year of spin-up.]{Weddell Sea surface height above the geoid for (a) $1^\circ$, (b) $1/4^\circ$ and (c) $1/12^\circ$ resolution models for the $30^{\text{th}}$-year of spin-up. Region displayed is from $80^\circ$W to $30^\circ$E and $85^\circ$S to $40^\circ$S.}
	\label{F_SSH_sp}
\end{figure*}
Figure \ref{F_SSH_sp} shows the annual-mean sea surface height (SSH) from the (a) $1^\circ$, (b) $1/4^\circ$ and (c) $1/12^\circ$ resolution models for the 30$^{\text{th}}$ year of model spin-up. We observe large differences in SSH between models at different resolution near the Antarctic shelf region. We note that SSH is generally negative with respect to the geoid. The $1^\circ$ model shows good agreement with the SSH structure observed using satellite altimetry data ($Rio$ MDT dataset from AVISO, not shown, \textit{personal correspondence with Pat Hyder}). The smaller, (less negative) SSH observed in the $1/4^\circ$ (and to a lesser extent $1/12^\circ$) model suggests a fresher and less dense shelf region. Panel (b) shows a tongue-like structure protruding westward into the Drake Passage, suggesting possible changes in the structure of the Weddell gyre in the region. In fact, the smaller SSH observed near the Antarctic coastline in the $1/4^\circ$ model extends all the way around the Antarctic land mass. 

\begin{figure*}[ht!]
	\centerline{\includegraphics[width=\textwidth]{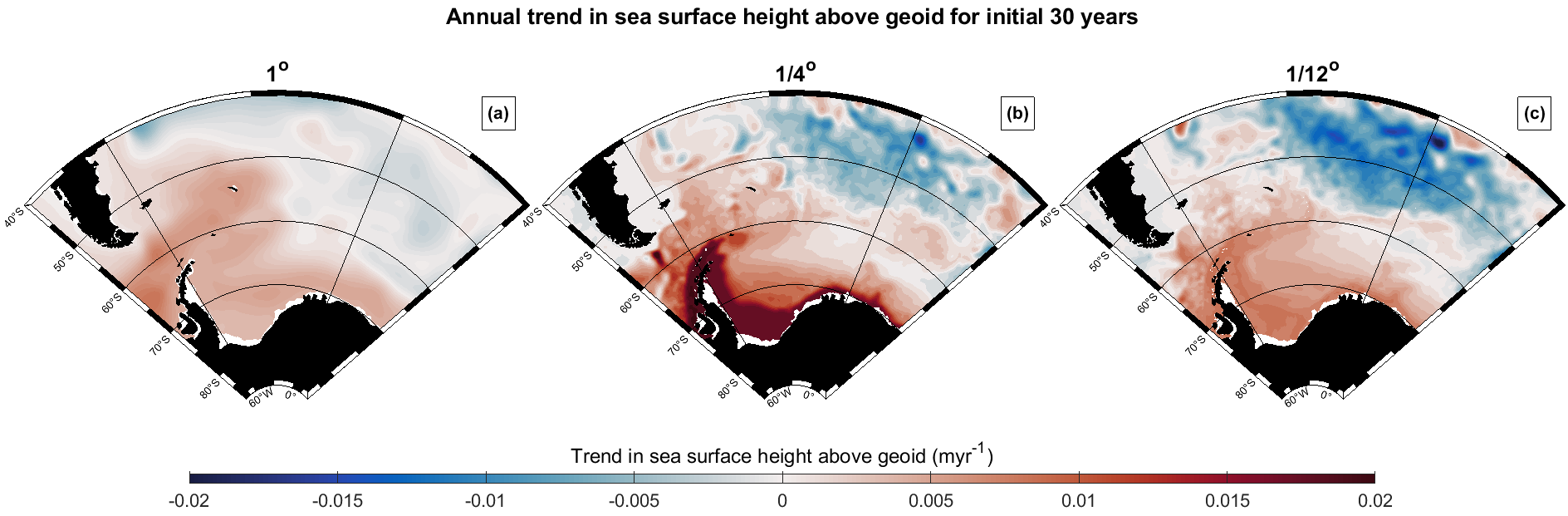}}
	\caption[Trend in Weddell Sea surface height above the geoid for $1^\circ$, $1/4^\circ$ and $1/12^\circ$ resolution models for the initial $30$-year spin-up phase.]{Trend in sea surface height above the geoid for (a) $1^\circ$, (b) $1/4^\circ$ and (c) $1/12^\circ$ resolution models for the initial $30$-year spin-up phase. Region displayed is from $80^\circ$W to $30^\circ$E and $85^\circ$S to $40^\circ$S.}
	\label{F_SSH_spt}
\end{figure*}

There is a mainly positive trend in SSH over the first 30 years, shown in Figure \ref{F_SSH_spt} for all model resolutions. The $1/4^\circ$ model exhibits the strongest trend, especially along the Antarctic shelf, suggesting significant changes to the properties of the shelf current. 

\subsubsection*{Characteristics of the Weddell Gyre}
Two possible mechanisms for a freshening Antarctic shelf in the $1/4^\circ$ model are (a) excessive sea-ice melt around Antarctica, and (b) leakage of the Weddell gyre. \cite{Meijers2016} discuss the strong correlation between the curl of the wind stress integrated over the Weddell Gyre and the water properties of the westward flowing shelf current near Elephant Island. Using observations from hydrographic cruise CTDs and bottom landers, they hypothesise that the westward flowing shelf current is an extension of the Antarctic slope current, with flow across the South Scotia Ridge between the Weddell and Scotia Seas being controlled by changes in the strength of the Weddell Gyre. 

\begin{figure*}[ht!]
	\centerline{\includegraphics[width=\textwidth]{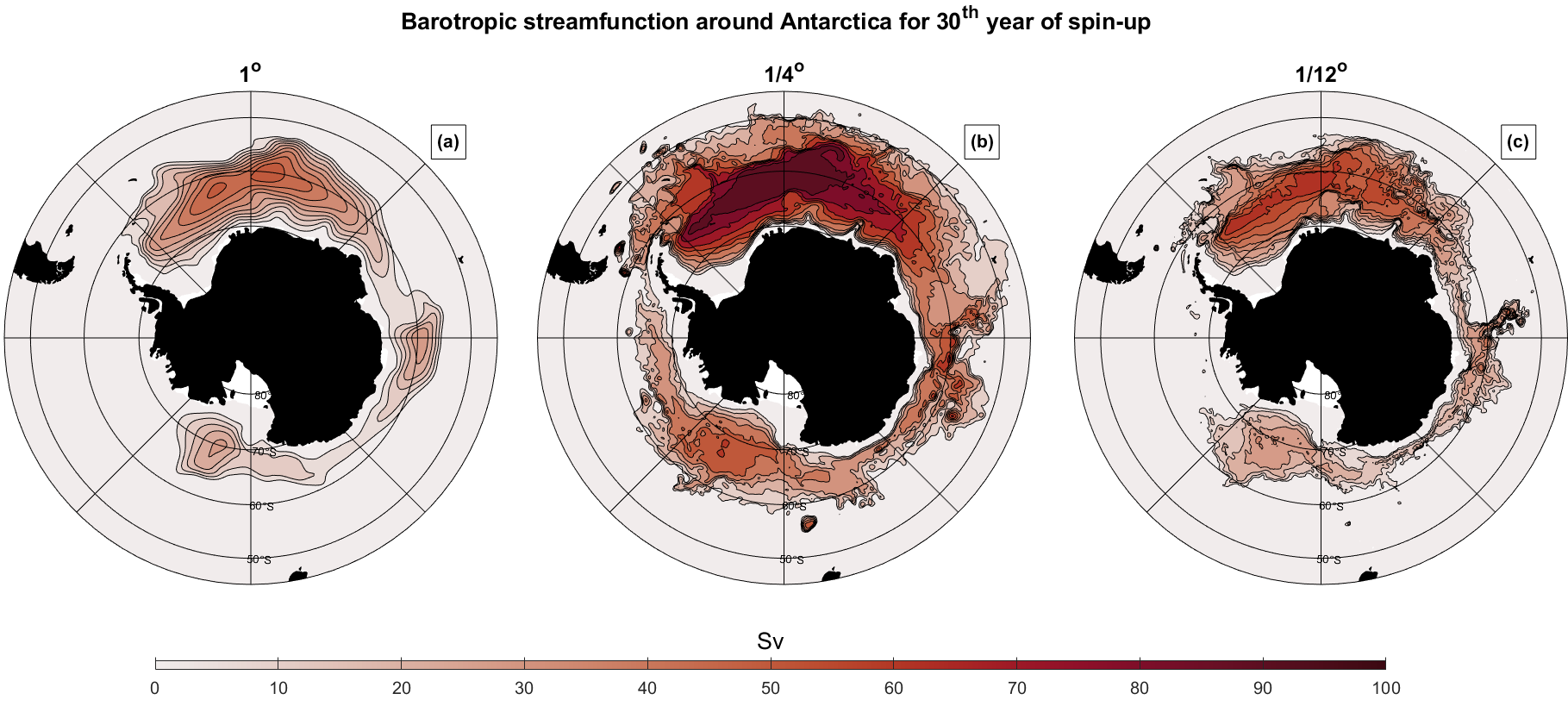}}
	\caption[Barotropic streamfunction $\psi_u(x,y)$ around Antarctica for $1^\circ$, $1/4^\circ$ and $1/12^\circ$ resolution models for the $30^{\text{th}}$-year of model spin-up.]{Barotropic streamfunction $\psi_u(x,y)$ around Antarctica for (a) $1^\circ$, (b) $1/4^\circ$ and (c) $1/12^\circ$ resolution models for the $30^{\text{th}}$-year of model spin-up. Only positive values of streamfunction are shown, to emphasise the location and strength of the Weddell and Ross Sea gyres (in the upper and lower halves of each panel, respectively).}
	\label{F_BrtS_TM}
\end{figure*}
Figure \ref{F_BrtS_TM} shows the barotropic streamfunction $\psi_u$ for the 30$^{\text{th}}$ year at each model resolution, defined by
\begin{equation}
	\psi_u(x,y) = \int_{S(x)}^{y}\int_{-H(x,y)}^{0}u_E(x,y',z) \ \mathrm{d} z \ \mathrm{d} y'
	\label{e_psi_u}
\end{equation}
for zonal velocity $u_E$, southern boundary $S$ and sea floor depth $H$. That is, $\psi_u$ is the cumulative sum from the southern boundary at a given $x$ to a meridional coordinate $y$ of interest, of the vertical sum of zonal velocities from the ocean floor to the surface. We choose to take the integral from the southern boundary northward at a given $x$, since the starting values at the southernmost location for the longitudes of interest correspond to land, and hence $\psi_u(x,S(x))=0$. Figure \ref{F_BrtS_TM} shows stronger Weddell and Ross Sea gyres for the $1/4^\circ$ model. Here, the Weddell Gyre is found to protrude into the Drake Passage, leading to a weaker overall eastward transport through the passage.

\subsection*{Ocean-only GCMs, and $1/4^\circ$ sensitivity experiments}  \label{Sct_Inter_Oth}
In Appendix \ref{Sct_Inter_Oth_A} we summarise preliminary investigations into the impact of model parameterisation, and atmospheric coupling, on the ACC transport at the Drake Passage. We note that further investigations into ACC transport have recently been conducted by the UK MetOffice, showing clear ACC transport sensitivity to characteristics of the Southern Ocean bathymetry.

\subsection{Decomposition analysis for other ACC longitude sections} \label{Sct_LngSct}

\begin{figure*}[ht!]
	\centerline{\includegraphics[width=14cm]{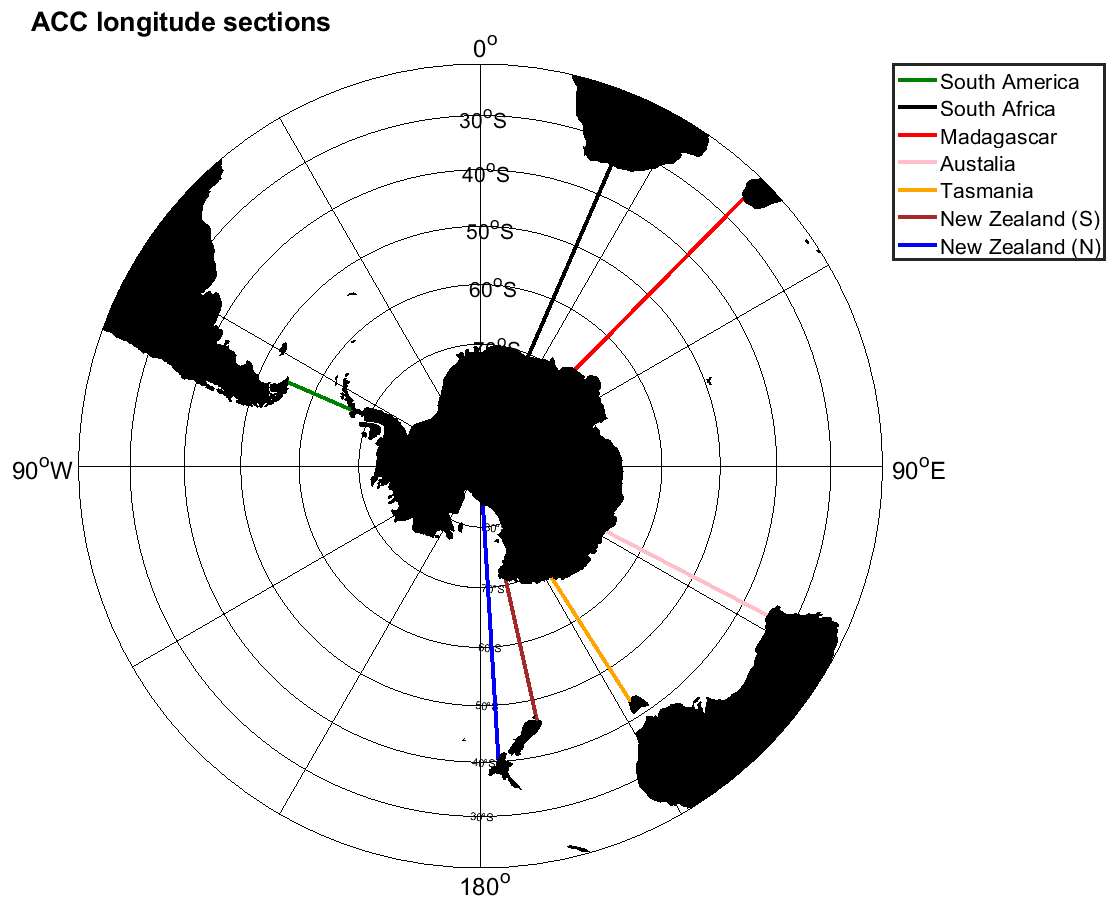}}
	\caption[Locations of seven meridional sections, at which total cumulative ACC transports up to the surface $T_{ACC}$ and $\tilde{T}_{ACC}$ are calculated. ]{Locations of seven meridional sections, at which total cumulative ACC transports $T_{ACC}$ and $\tilde{T}_{ACC}$ are calculated. Meridional sections are from Antarctica to South America (green), South Africa (black), Madagascar (red), western Australia (pink), Tasmania (orange), New Zealand South Island (brown) and New Zealand North Island (blue).}
	\label{F_LngSct}
\end{figure*}

In addition to the decomposition of transport through the Drake Passage reported in Section \ref{Sct_Inter_Dcm}, we also estimate total cumulative transport $T_{ACC}$ at the surface and its decomposition estimate $\tilde{T}_{ACC}$ for 6 other longitude sections, as illustrated in Figure \ref{F_LngSct}. The sections correspond to (a) South Africa - Antartica, (b) Madagascar - Antarctica, (c) western Australia - Antarctica, (d) Tasmania - Antarctica, (e) New Zealand South Island - Antarctica and (f) New Zealand North Island - Antarctica. 

\begin{figure*}[ht!]
	\centerline{\includegraphics[width=\textwidth]{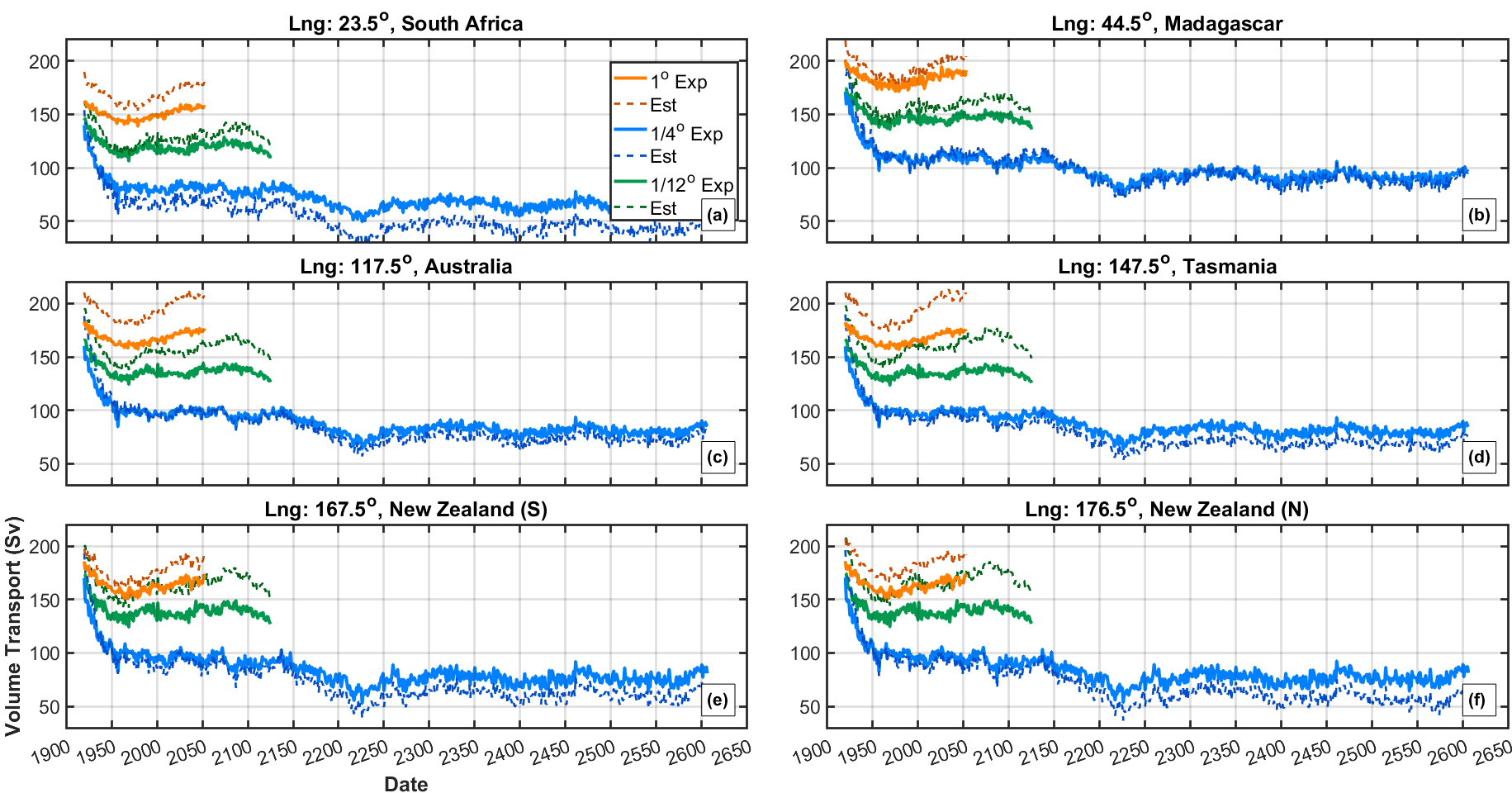}}
	\caption[Timeseries of expected $T_{ACC}$ and estimated $\tilde{T}_{ACC}$ cumulative volume transports up to the surface through each of 6 meridional sections for the $1^\circ$, $1/4^\circ$ and $1/12^\circ$ models.]{Timeseries of expected $T_{ACC}$ (solid) and estimated $\tilde{T}_{ACC}$ (dashed) cumulative volume transports up to the surface through each of 6 meridional sections for the $1^\circ$ (orange), $1/4^\circ$ (blue) and $1/12^\circ$ (green) models.  Meridional sections show are (a) South Africa - Antartica, (b) Madagascar - Antarctica, (c) Australia - Antarctica, (d) Tazmania - Antarctica, (e) New Zealand South Island - Antarctica and (f) New Zealand South Island - Antarctica.}
	\label{F_TE_LngSct}
\end{figure*}
Figure \ref{F_TE_LngSct} gives timeseries of $T_{ACC}$ and $\tilde{T}_{ACC}$ through each section for all model resolutions. The figure suggests that many of the findings of the decomposition analysis at the Drake Passage are common to all meridional sections considered. In particular, the values of transports $T_{ACC}$ and $\tilde{T}_{ACC}$ from GCM outputs at different model resolutions are ordered: values from the $1^\circ$ model tend to be larger than those from the $1/12^\circ$ model, which themselves are larger than those from the $1/4^\circ$ model. Moreover, the values of $T_{ACC}$ for the Madagascar - Antarctica section are typically the largest at each model resolution; we attribute this to the wider section at this longitude and the incorporation of flows which are not part of the ACC at latitudes near Madagascar (e.g. Agulhas Return Flow); similarly, values of $T_{ACC}$ and $\tilde{T}_{ACC}$ at the South America - Antractica (Drake Passage) section are relatively small. We note the weaker transport observed for the South Africa - Antarctica section (Panel (a), across model resolutions) is due to the westward-flowing Agulhas current at the northern edge of this meridional section. We find varying levels of agreement between $T_{ACC}$ and $\tilde{T}_{ACC}$ across model resolution and section; generally, the $1/4^\circ$ model performs best in this respect. For the $1/12^\circ$ model, $\tilde{T}_{ACC}$ is always larger than $T_{ACC}$, with best agreement found for the South Africa and Madagascar sections. The Madagascar section also yields best agreement between $T_{ACC}$ and $\tilde{T}_{ACC}$ for the $1^\circ$ and notably the $1/4^\circ$ model. Together, these results appear to emphasise the important role of local bathymetric characteristics on the quality of estimates of ACC transport decomposition.

Results of the transport decomposition diagnostic applied to these additional meridional sections highlight the importance of $T_{\beta}$ for wider sections (not shown). The increased magnitude of $T_{\beta}$ for wider sections is attributed to greater density variations across those sections, and the incorporation of density contributions (otherwise accounted for by northern and southern boundary densities for narrower sections), also resulting in smaller $T_S$ and $T_N$. Interestingly, the magnitude of $T_{bot}$ from the decomposition analysis, is approximately three times larger for the Drake Passage and Madagascar - Antarctica sections, than for any of the other sections. This is likely to be due to strong bottom currents in these meridional sections, and worthy of further investigation. 

\section{Mapping along-boundary densities for the Antarctic}
\label{ACC_BdryD}
The investigations of ACC transport through the Drake Passage, and meridional cross-sections of potential temperature and salinity there, indicate significant changes in southern boundary properties. Here, we apply the boundary mapping algorithm developed previously for the Atlantic and neighbouring continuous boundaries to the Antarctic continent. The objective is to examine along-boundary properties within the control $1^\circ$ and $1/4^\circ$ HadGEM-GC3.1 models, GloSea5 reanalysis (\citealt{MacLachlan2015}) and GK climatology (\citealt{GouretskiViktorandKoltermann2004}). 

\subsection{Along-boundary mapping}
\begin{figure*}[ht!]
	\centerline{\includegraphics[width=14cm]{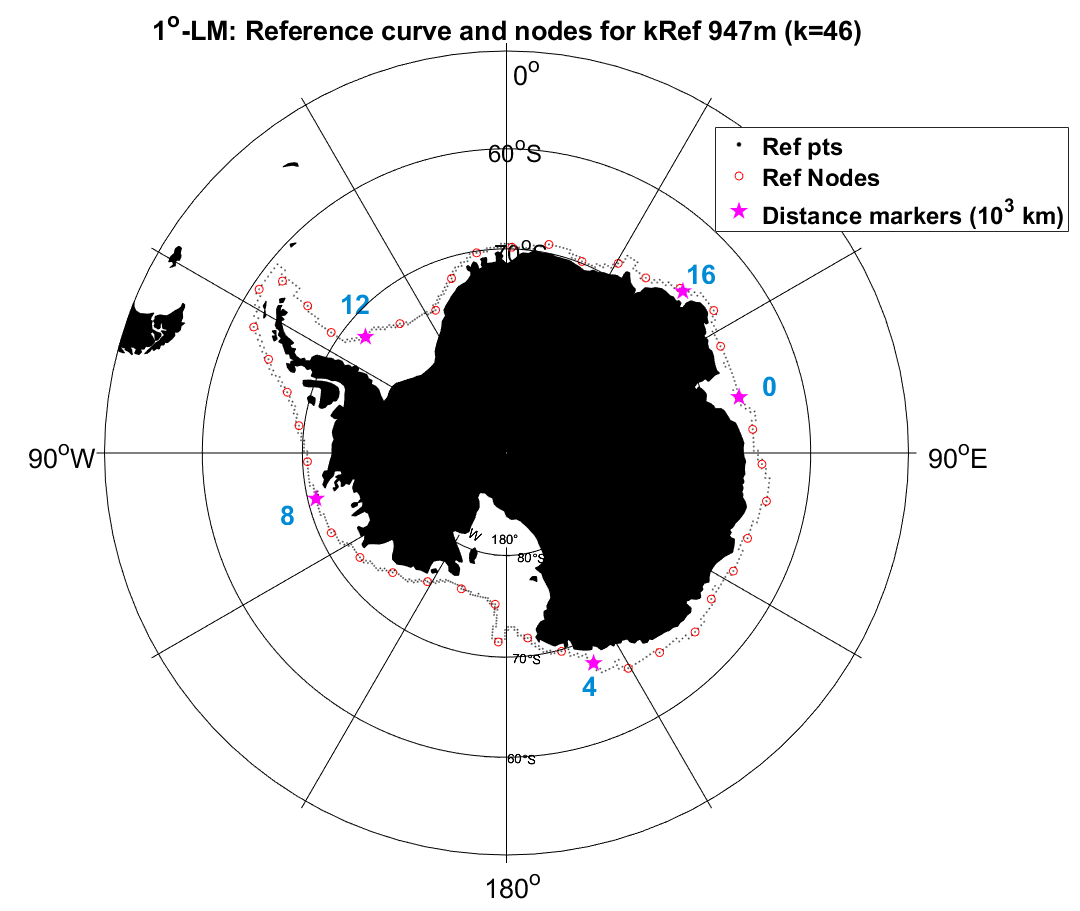}}
	\caption[Reference contour at depth level $k=46$ (947m) for the Antarctic coastline with nodes, distance markers and cumulative distance.]{Reference contour (black dots) at depth level $k=46$ (947m) for the Antarctic coastline with nodes (red circles), distance markers (pink stars, every $4 \times 10^3$km) and cumulative distance (blue numbers, $\times 10^3$km). Contours at all other depths are referenced to this contour.}
	\label{F_MapAnt_1pth}
\end{figure*}

The mapping algorithm is applied as described in Sections \ref{CntMpBnd.1}-\ref{CntMpBnd.2}, with minor modifications. For the Antarctic boundary, we use a reference contour at depth $k=46$ (947m), and an initialisation point offshore of Antarctica near $80^\circ$E. Figure \ref{F_MapAnt_1pth} shows the reference contour for the $1^\circ$ model together with the common along-boundary distance scale used to reference boundary contours at all depth levels. Reference nodes (red circles) are located every 400km, for the clockwise pathway.

\subsection{Boundary neutral densities for the $1^\circ$-$LM$ model}
Section \ref{Sct_Inter_TT} shows that the best estimate for ACC transport is provided by the HadGEM-GC3.1 $1^\circ$ model with medium atmospheric resolution ($LM$). Section \ref{Sct_Inter_CS} demonstrates good agreement of cross-sectional properties at the Drake Passage between the $1^\circ$-$LM$ model and climatological, observational and reanalysis datasets. Therefore, we begin by characterising the time-mean boundary density structure around Antarctica in the $1^\circ$-$LM$ model. To reduce potential issues with model drift, only the first 100 years of output for each GCM model is considered; GloSea5 time-averaging is performed over all 23 years of data available. 

\begin{figure*}[ht!]
	\centerline{\includegraphics[width=\textwidth]{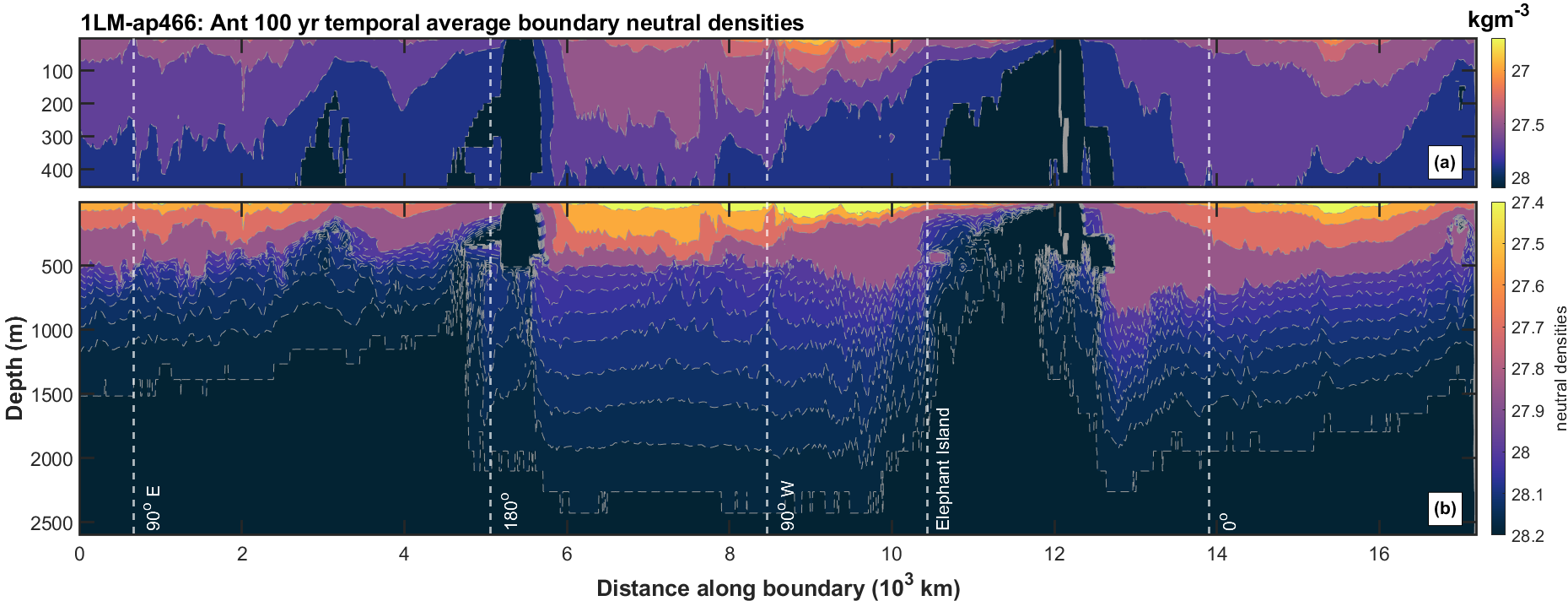}}
	\caption[Along-boundary time-average neutral densities for the first 100 years of the $1^\circ$-$LM$ model for the upper 400m and for all depths.]{Along-boundary time-average neutral densities for the first 100 years of the $1^\circ$-$LM$ model for (a) the upper 400m and (b) all depths. Dashed white lines indicate locations of longitudes $90^\circ$E, $180^\circ$, $90^\circ$W, $0^\circ$ and that of Elephant Island.}
	\label{F_BD_1LM}
\end{figure*}

Figure \ref{F_BD_1LM} shows neutral densities on the sloping boundary around Antarctica for the $1^\circ$-$LM$ model, for (a) the upper 400m and (b) all depths, with an adjusted colour bar to highlight isopycnal structure at depth. 

The upper few hundred metres is dominated by local wind-stress variability, and therefore along-boundary continuity in isopycnals is not found nor expected. However, at depth, we might expect some continuity in density structure, since the ACC is reasonably longitudinally coherent, and any major changes in Antarctic coastline densities would require compensatory effects elsewhere to maintain a constant ACC.

Following the boundary clockwise around Antarctica, from $80^\circ$E we find that isopycnal depths vary with along-boundary distance throughout the fluid column for the first $4.5 \times 10^3$km, with a small upward slope. At approximately $5 \times 10^3$km (near longitude $180^\circ$), on entering the Ross Sea (where deep water formation is known to occur), we observe a large change in isopycnal depths. Near the surface, a region of significantly denser water is found. Below this, there is a sudden deepening of along-boundary isopycnals. Connecting to this feature, at along-boundary distances of $4.5-5.0 \times 10^3$km and a depth of approximately 300m, we find a horizontal tongue of deep water which extends along the boundary.

After the Ross Sea region of dense surface water, we enter the Amundsen and Bellinghausen Seas at distances $6-10\times 10^3$km and find well-stratified boundary densities. Flat isopycnals are observed at depth, along with a gentle upward slope in isopycnals for depths 100-400m. 

At approximately $10\times 10^3$km we pass through the Drake Passage where there is a shallowing of all isopycnals. Passing Elephant Island, the southernmost point of the SR1b hydrographic section, a steep gradient in isopycnals is found for depths between 500m and 2500m, which continues into the Weddell Sea (at $11-13 \times 10^3$km). Like the Ross Sea, this sheltered location is a key region of deep water formation, and is a prominent source of AABW. At $12\times 10^3$km, dense water dominates the entire fluid column, a clear indicator not only of dense water formation near the surface but also of convective sinking of dense waters along the shelf and continental slope. Following the Weddell Sea, we find isopcynals below 500m slope gently upwards along East Antarctica.

Findings from Figure \ref{F_BD_1LM} agree with the \cite{Thompson2018} characterisation of regions of dense water formation in both the Weddell and Ross Seas. However, \citealt{Thompson2018} discuss a third region of dense water formation near the Adelie coast (at $17.6 \times 10^3$km) which is not immediately visible within Figure \ref{F_BD_1LM} for the $1^\circ$-$LM$ model; on close inspection however, a small region of denser water can be found in Panel (b) at around 250m. Section \ref{GS5_S} below considers the Adelie coast region further, and the impact of seasonality on deep water formation.

\subsection{Boundary neutral densities in other datasets}
\begin{figure*}[ht!]
	\centering
	\includegraphics[width=\textwidth]{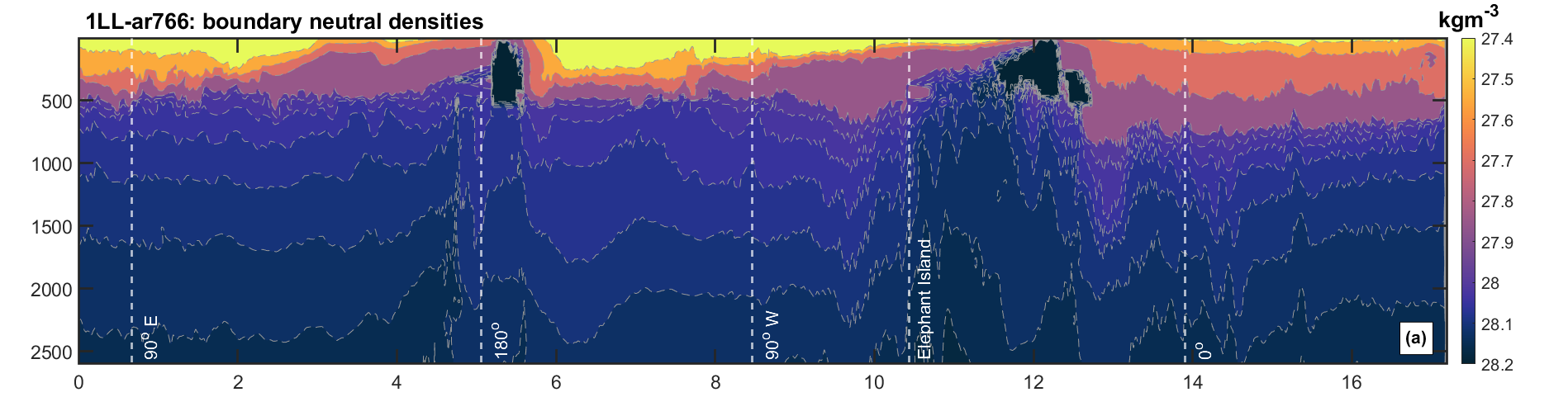}\\
	\includegraphics[width=\textwidth]{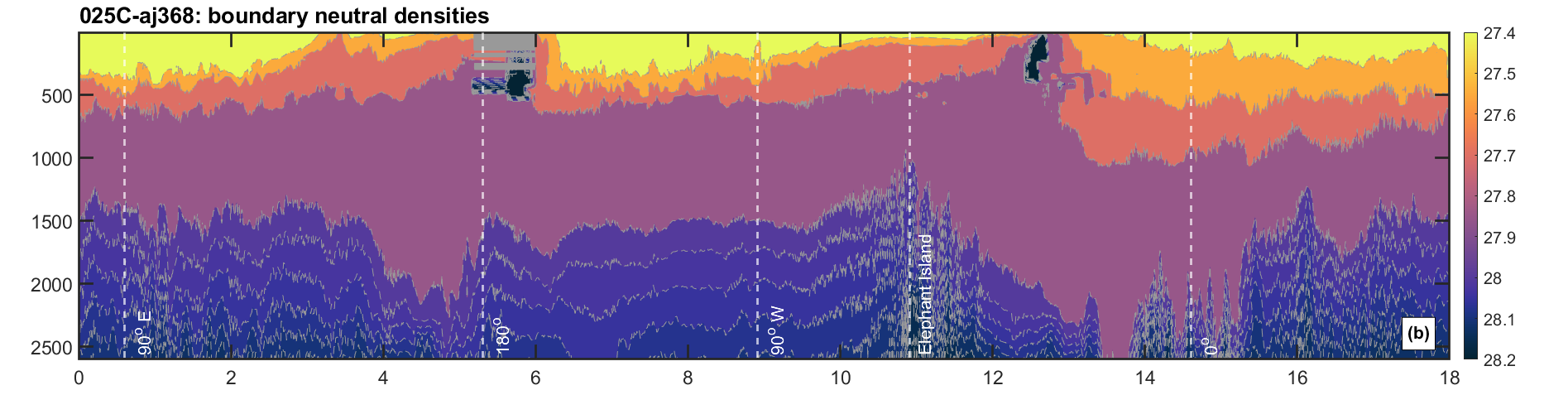}\\
	\includegraphics[width=\textwidth]{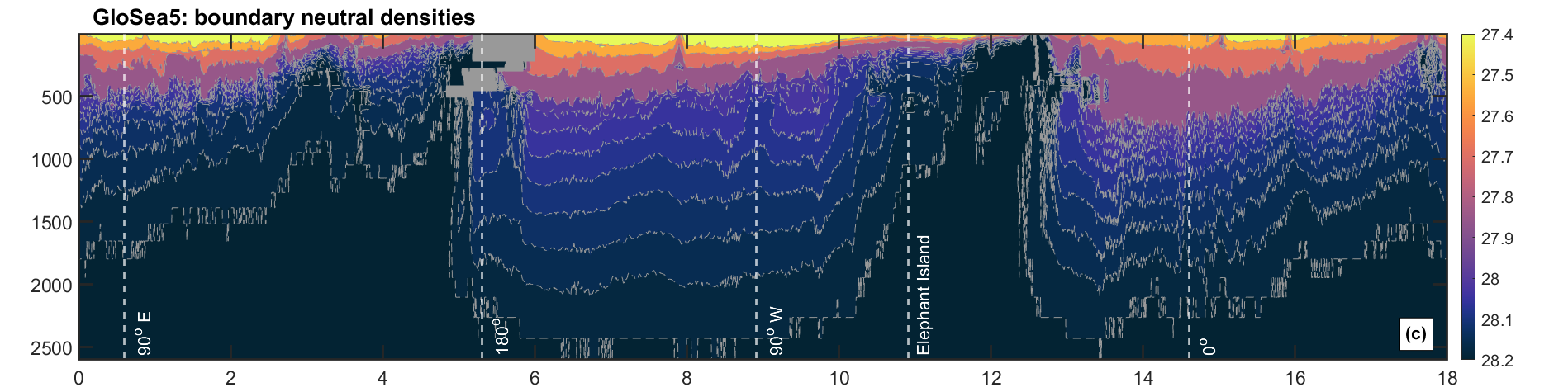}\\
	\includegraphics[width=\textwidth]{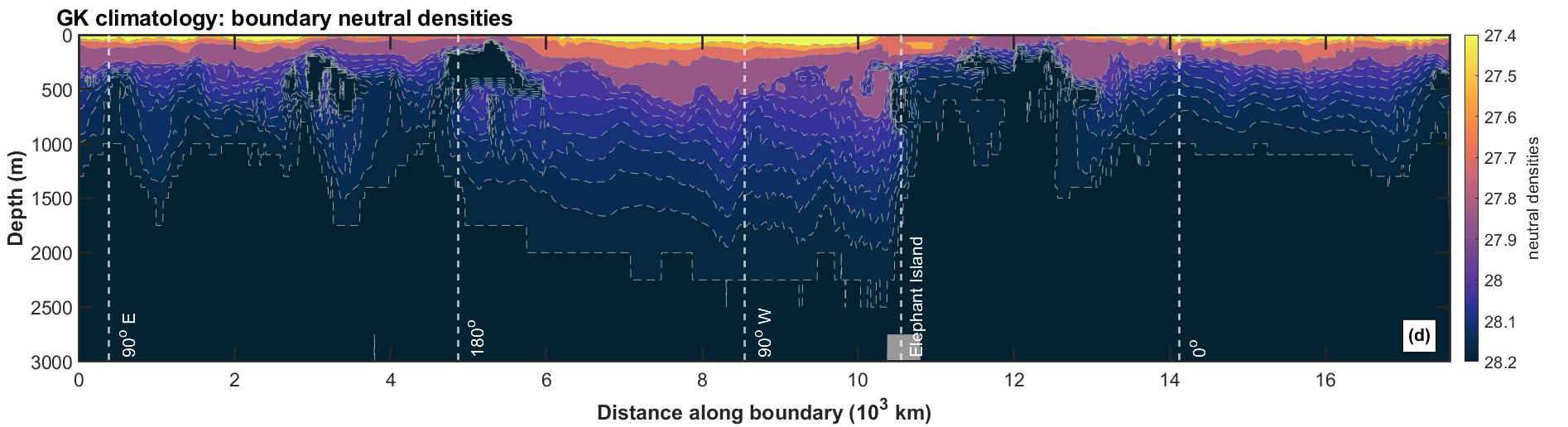}
	\caption[Neutral densities along the sloping Antarctic coastline for the (a) $1^\circ$-$LL$ model, (b) $1/4^\circ$ model, (c) GloSea5 reanalysis and (d) GK climatology, for all depths.]{Neutral densities along the sloping Antarctic boundary for the (a) $1^\circ$-$LL$ model, (b) $1/4^\circ$ model, (c) GloSea5 reanalysis and (d) GK climatology, for all depths. Model outputs are averaged over the first 100 years, and GloSea5 over 23 years of run. Dashed white lines indicate locations of longitudes $90^\circ$E, $180^\circ$, $90^\circ$W, $0^\circ$ and that of Elephant Island.}
	\label{F_BD_Ant_Rest}
\end{figure*}

Figure \ref{F_BD_Ant_Rest} presents the time-average along-boundary neutral densities for the (a) $1^\circ$-$LL$ model, (b) $1/4^\circ$ model, (c) GloSea5 reanalysis and (d) GK climatology. Figure \ref{F_SB_Trsp} shows weaker ACC transport than expected for the $1/4^\circ$ model, suggesting model inaccuracies in representing Southern Ocean dynamics. 

Using a common colour scale for both Figures \ref{F_BD_1LM} and \ref{F_BD_Ant_Rest}(b), we find a stark difference in densities on the boundary for the corresponding models. The 27.8kgm$^{\text{-3}}$ (dark pink) density contour penetrates down to 1500m for the $1/4^\circ$ model, whereas in the $1^\circ$ model its maximum depth is only about 500m. Lighter densities throughout Figure \ref{F_BD_Ant_Rest}(b) indicate clearly warmer and fresher Southern Ocean and Antarctic coastline bottom waters in the $1/4^\circ$ model. Boundary potential temperature and salinity properties (not shown) indicate a generally fresh upper 1500m, leading to colder surface waters. However, at depth, significantly warmer waters are found in the $1/4^\circ$ model. 

The $1/4^\circ$ model does show small regions of dense surface waters in the Weddell and Ross Seas. No connectivity to deeper water is found, suggesting no pathway for dense waters sinking down the shelf and continental slope. Isopycnal structure below 1500m is relatively stable, except for regions prior to the Ross Sea (Figure \ref{F_BD_Ant_Rest}(b), at $5\times 10^3$km) and after the Weddell Sea (at $14\times 10^3$km). These features  suggest the presence of regions of warming or freshening at depth, supported by significant unexplained warming (not shown) found along East Antarctica (at $13.5$-$16 \times 10^3$km) at 2000m depth. 

Figure \ref{F_BD_Ant_Rest}(c) shows time-average along-boundary neutral densities for the GloSea5 reanalysis dataset, for the period $1995$-$2018$. The spatial structure of boundary densities is very similar to that exhibited by the $1^\circ$-$LM$ model in Figure \ref{F_BD_1LM}. Weddell and Ross Sea deep water formation regions are clearly defined, despite a slightly different bathymetry resulting in minor differences in contour location, resulting in some surface data being rejected. Within GloSea5, there is more evidence of deep water formation near the Adelie coast (at $17.6 \times 10^3$km). Relatively flat or slightly tilted isopycnals are found along the western Antarctic Peninsula, between the notable regions of deep water formation in the Ross and Weddell Seas. The slopes of isopycnals along eastern Antarctica (at $12.5-17.5 \times 10^3$km) for GloSea5 are somewhat steeper than those for the $1^\circ$-$LM$. 

Comparison of the coarser atmospheric resolution $1^\circ$-$LL$ GCM (Figure \ref{F_BD_Ant_Rest}(a), discussed in Section \ref{Bdry_Mod}) and the \cite{GouretskiViktorandKoltermann2004} climatology data (Figure \ref{F_BD_Ant_Rest}(d)), again reveals differing boundary density structures. 

For the GK dataset, much denser waters at depth are present, with three clear pockets of dense water near the surface in the Ross Sea (at $3.5$, $5.2\times 10^3$km) and the Weddell Sea (at $12.1\times 10^3$km). Prior to the Ross Sea, isopycnal depths vary significantly with distance, with less clear connectivity between surface and deep dense water in comparison to the Weddell Sea. Between the Ross and Weddell Seas, isopycnals slope downward (in contrast to all other datasets considered), and post Weddell Sea we observe relatively flat isopycnals. Near the Adelie coast, a small region of dense water near 500m is found. In terms of boundary density magnitudes, the GK dataset is closest to the GloSea5 and the $1^\circ$-$LM$ model; however its detailed isopycnal structure differs from all datasets considered. We note that observational data for the Antarctic coastline is sparse. Since the GK methodology uses interpolation in the absence of data, there is some doubt as to the veracity of the detailed density structure estimated.

Density structure for the $1^\circ$-$LL$ model (Figure \ref{F_BD_Ant_Rest}(a)) exhibits localised regions of deep water formation, with no connectivity along the boundary from dense surface water to the deep ocean. $1^\circ$-$LL$ model waters are lighter generally than those of the GloSea5, $1^\circ$-$LM$ and GK datasets, with relatively flat isopycnals below 700m, except for some variability prior to the Ross and Weddell Seas. The general density structure is similar to that of the $1/4^\circ$ model in Figure \ref{F_BD_Ant_Rest}(b), except that waters as somewhat denser, and that dense surface regions are clear but isolated. There is no evidence for dense surface water near the Adelie coast. The lack of connectivity between dense surface and deep water along the boundary in the $1^\circ$-$LL$ and $1/4^\circ$ models would suggest mixing through the interior via open ocean convection, instead of down-slope flow of dense waters.

When comparing first and last years of model runs (not shown), densities become generally lighter for both the $1/4^\circ$ and $1^\circ$-$LL$ models, suggesting model drift. With fresher surface waters and warmer water at intermediate depths, we find isolated regions of dense surface water near the Weddell and Ross Seas for both models; again, no connectivity between the surface and depth is found at these locations.

The large differences in density structure found between model, climatology and reanalysis datasets, emphasise the difficulty in representing Southern Ocean and Antarctic coastline properties, and the importance of direct observation of these remote and hostile waters.

\subsection{Mechanisms underlying observed isopycnal structure}

We hypothesise that the isopycnal structure on the Antarctic boundary for the $1^\circ$-$LM$ model and GloSea5 reanalysis dataset is influenced by the presence of an Antarctic Slope Current (ASC). The ASC is a near-circumpolar current flowing westward along Antarctica's shelf and slope regions, disrupted only by the western Antarctic Peninsula between the Pacific and Atlantic Oceans. There is uncertainty regarding the formation location of the ASC. The ASC behaves like a shelf break current, typically formed by wind stresses aligned parallel to the coastline, acting as a barrier to mixing between shelf and open-ocean waters. The formation area of the ASC spans from the Bellingshausen Sea to the western Ross Sea ($6$-$10 \times 10^3$km along our boundary). \cite{Thompson2018} show the ASC disappears near the western Antarctic Peninsula; here instead, there is an eastward flowing slope current along with the southern edge of the ACC. 

Concentrating on Figures \ref{F_BD_1LM} and \ref{F_BD_Ant_Rest}(c), starting at the Weddell Sea (at along-boundary distance $12 \times 10^3$km), we find dense water near the surface, caused by local bathymetric and atmospheric conditions. Moving further along the boundary, isopycnals are then found to slope upwards to the east along East Antarctica until the Ross Sea. However, following the Ross Sea (at $5 \times 10^3$km, again exhibiting dense water throughout the fluid column), the pattern of gently sloping isopycnals is not repeated as we move along the boundary back to the Weddell Sea. Instead, we find that isopycnals are relatively flat. 

We speculate that this feature may be related to the disconnect between Bellinghausen and Weddell basins, and the absence of ASC here. On the Antarctic boundary, we hypothesise that the down-slope movement of the ASC transports denser surface waters westwards (anti-clockwise) from the Ross Sea along the sloping boundary of East Antarctica to the Weddell Sea, causing a shallow downwards isopycnal slope in the direction of the ASC. In contrast, there is no ASC westwards from the Weddell Sea, through the Drake Passage, along the western Antarctic Peninsula to the Ross Sea. As a result, we observe relatively flat isopycnals here. 

Similarities can be drawn between the Antarctic boundary isopycnal structure in Figures \ref{F_BD_1LM} and \ref{F_BD_Ant_Rest}(c), and that observed along western boundaries of the Atlantic discussed in Section \ref{CntMpBnd}. The sloping isopycnals found between Ross and Weddell Seas along East Antarctica are consistent with a boundary current (the ASC here) propagating (anti-clockwise) around Antarctica and slowly moving down slope to maintain kinetic energy as outlined by \cite{MacCready1994} (see Section \ref{Bdry_WMec} for more detail; inspection of boundary and bottom velocities might reveal the location and characteristics of the ASC and other deep along-boundary return currents). In contrast, isopycnals along the western Antarctic Peninsula behave similarly to those of the western boundary of the Indian Ocean. There is no strong boundary current, and hence no clear pathway for moving dense Weddell Sea water around Elephant Island into the Bellinghausen and Amundsen Seas. 

\subsection{Seasonal features of along-boundary densities from GloSea5 reanalysis}
\label{GS5_S}
\begin{figure*}[ht]
	\centerline{\includegraphics[width=\textwidth]{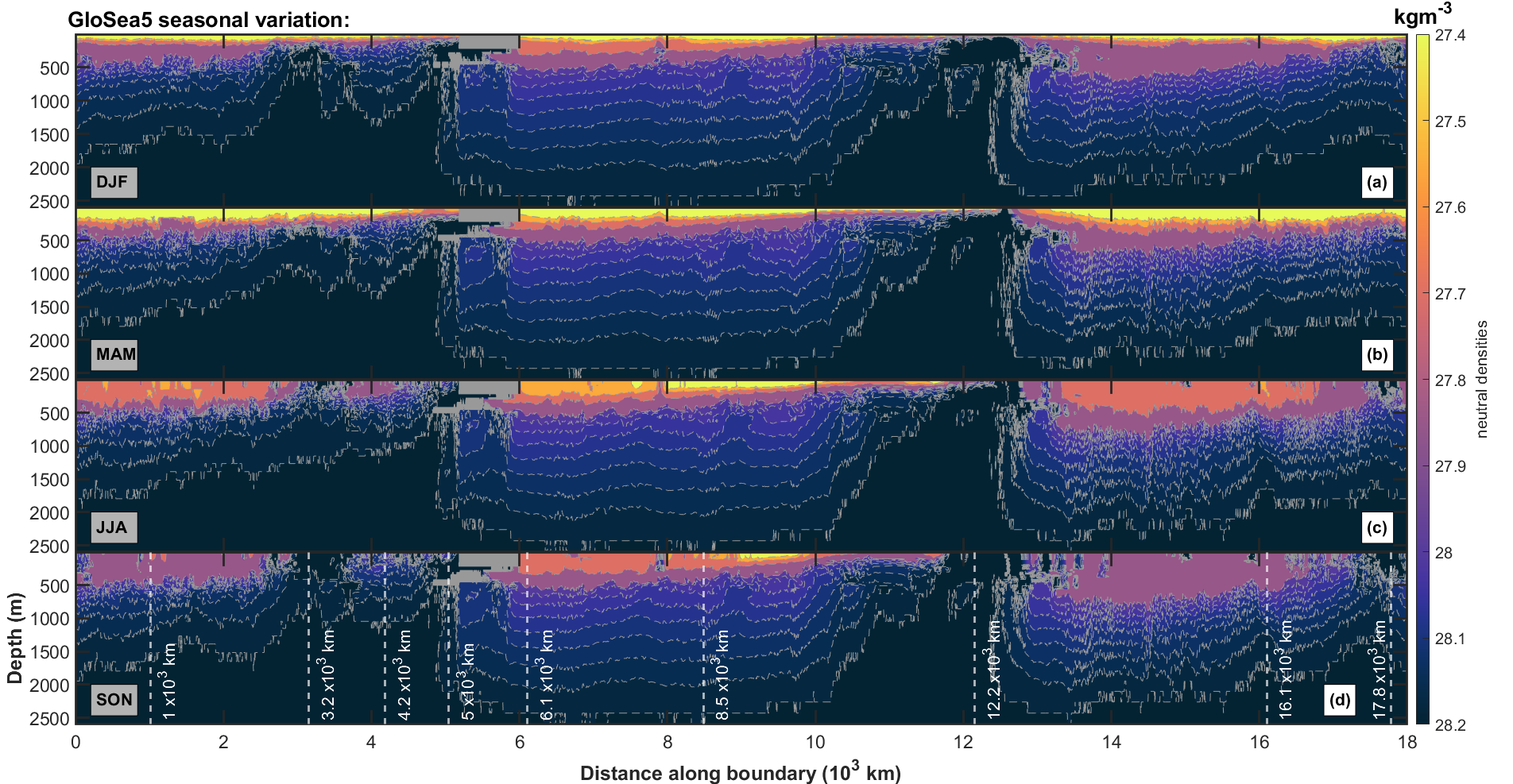}}
	\caption[Seasonally-averaged neutral densities on the Antarctic boundary in the GloSea5 reanalysis dataset for DJF, MAM, JJA and SON. Dashed white lines indicate along-boundary distances at which month-depth illustrations are given in Figure \ref{F_BD_GS5_M}. ]{Seasonally-averaged neutral densities on the Antarctic boundary in the GloSea5 reanalysis dataset for (a) DJF, (b) MAM, (c) JJA and (d) SON. Dashed white lines indicate along-boundary distances at which month-depth illustrations are given in Figure \ref{F_BD_GS5_M}.}
	\label{F_BD_GS5_S}
\end{figure*}

Analysis of the monthly GloSea5 data reveals the impact of seasonality on along-boundary neutral densities, as shown in Figure \ref{F_BD_GS5_S}. Panels correspond to time-average seasonal boundary densities for the three-month periods (a) DJF, (b) MAM, (c) JJA and (d) SON.

\begin{figure*}[ht!]
	\centerline{\includegraphics[width=\textwidth]{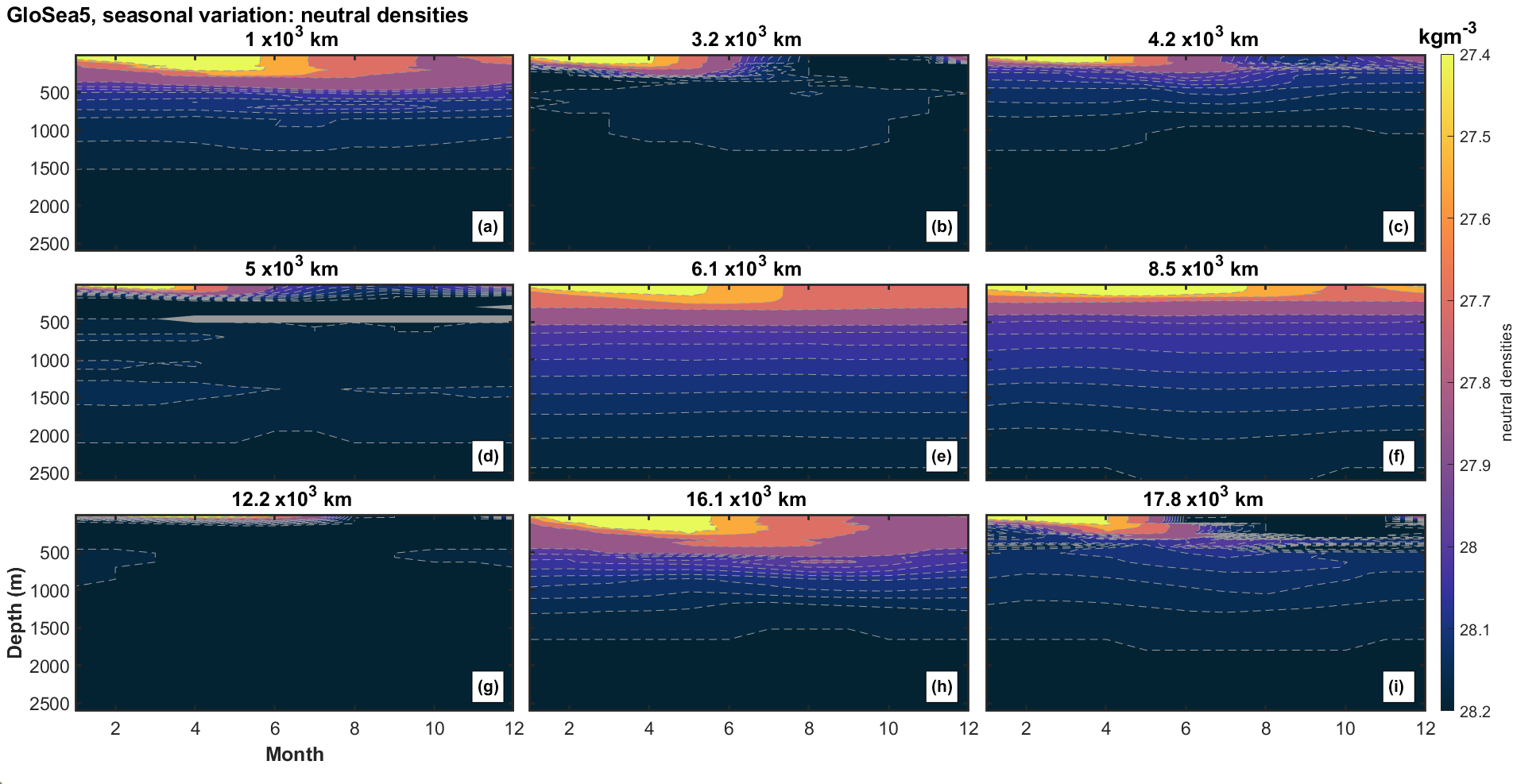}}
	\caption[Monthly-averaged neutral densities on the Antarctic boundary in the GloSea5 reanalysis dataset for nine different along-boundary distances (given in Figure \ref{F_BD_GS5_S}(d)).]{Monthly-averaged neutral densities on the Antarctic boundary in the GloSea5 reanalysis dataset for nine different along-boundary distances (given in Figure \ref{F_BD_GS5_S}(d)). The first month is January.}
	\label{F_BD_GS5_M}
\end{figure*}

Figure \ref{F_BD_GS5_S}(c) and (d) exhibit denser surface waters and deeper mixed-layers, caused by wind-driven mixing and brine rejection during ice formation in preceding stormy winter months. In autumn, Figure \ref{F_BD_GS5_S}(b) shows distinct stratification and lighter fresher near-surface waters due to ice melt throughout the preceding summer months. The lightest and warmest year-round surface waters are found along the western Antarctic Peninsula, in agreement with \cite{Thompson2018}, as seen from Figure \ref{F_BD_GS5_M}(e, f) below.  

Figure \ref{F_BD_GS5_M} shows the monthly-average boundary density variation with depth, at nine along-boundary locations given in Figure \ref{F_BD_GS5_S}(d). In the Weddell Sea (Figure \ref{F_BD_GS5_M}(g)) and western Ross Sea (Figure \ref{F_BD_GS5_M}(d)), dense water is present throughout the year, with very shallow regions of light near-surface water in the first few months of the year. However, there is clear seasonality in deep water formed in the eastern Ross Sea ($3.2 \times 10^3$km) and near the Adelie coast ($17.8 \times 10^3$km), highlighted by Figure \ref{F_BD_GS5_M}(b) and (i). Both regions show lighter surface conditions during the first 4-5 months of the year, proceeding to dense surface waters on the boundary later in the year. 

Seasonality found near the Adelie coast impacts the resulting gradient in isopycnals along the boundary (Fiugure \ref{F_BD_GS5_S}). During periods of deep water formation, more steeply-sloping isopycnals are found, whereas during the rest of the year isopycnals have a gentler along-boundary slope. However, along the western Antarctic Peninsula ($6$-$10 \times ^3$km) we find seasonality has minimal effect on the density structure. As a result, along-boundary isopycnals there remain relatively flat.

\section{Summary}

This chapter has developed the cumulative transport decomposition diagnostic for the Antarctic Circumpolar Current (ACC), attributing total zonal transport through various sections to contributions on ocean boundaries together with a Coriolis-related $\beta$ term. The decomposition diagnostic has been applied to output from the HadGEM3-GC3.1 model at $1^\circ$, $1/4^\circ$ and $1/12^\circ$ spatial resolutions across the Drake Passage and other ACC meridional sections.

Analysis (Section \ref{Sct_Inter_TT}) reveals a large resolution dependence of the expected ACC volume transport $T_{ACC}$. Estimates of $T_{ACC}$ from the $1/4^\circ$ model are poor, stabilising at a mere $60$Sv, compared with an observational estimate of $173$Sv (\citealt{Donohue2016}). Estimates of $T_{ACC}$ from the $1^\circ$ and $1/12^\circ$ models are better, at approximately $150$Sv and 120Sv respectively, but nevertheless underestimate the observed transport. A large weakening in $T_{ACC}$ with time is observed within the initial 30-year ``spin-up'' phase of the model run for all three model resolutions, where the initial ocean conditions are taken from the EN4 dataset. 

Reconstructed estimates $\tilde{T}_{ACC}$ of the total ACC transport has been found to perform well for the Drake Passage within the $1^\circ$ and $1/4^\circ$ models. Discrepancies within the reconstructed $1/12^\circ$ model estimate are attributed to the influence of greater bathymetric detail at this high resolution. Specifically, the presence of a large number of small trenches and ridges near the sea floor, where physical properties are not well represented, leads to an overestimate of $T_{ACC}$. We find that a simplified decomposition, incorporating the northernmost and southernmost boundary contributions only, results in an improved estimate for $\tilde{T}_{ACC}$ from the $1/12^\circ$ model.

We analysed properties of the Drake Passage section (Section \ref{Sct_Inter_CS}), including potential temperature, salinity and neutral density for the initial 30-year spin-up phase of the model run, to investigate possible causes for the large weakening in transport. Trends in these properties include large changes in salinity and neutral density near the southern boundary in the $1/4^\circ$ model. The $1/4^\circ$ model output also indicates slumping rather than outcropping of isopycnals near the southern boundary. In contrast, the neutral density structure of the $1^\circ$ and $1/12^\circ$ models is found to correspond well with climatology (\citealt{GouretskiViktorandKoltermann2004}) and GloSea5 reanalysis.

We used the decomposition diagnostic to calculate the boundary and $\beta$ transport components contributing to the estimated cumulative transport $\tilde{T}_{ACC}$ (Section \ref{Sct_Inter_Dcm}). Large changes in the southern density component $T_S$ for the $1/4^\circ$ simulation explain the initial weakening observed for $\tilde{T}_{ACC}$. Gradual recovery of $T_S$ later in the timeseries is compensated by gradual weakening of the northern density component $T_N$; hence little impact on $\tilde{T}_{ACC}$ is observed after the spin-up phase. Transport contributions of additional cells, Ekman and $\beta$ components are negligible. 

We compared the bottom (or depth-independent) component $T_{bot}$ calculated from the decomposition diagnostic with an estimate for $T_{bot}$ from \cite{Donohue2016}, calculated using observations of bottom velocities. The decomposition-based $T_{bot}$ is in good agreement with observations for the $1^\circ$ model, but found to be smaller than expected for both of the higher resolution models. 

Analysis of sea surface height (SSH) data for the Weddell Sea and Drake Passage (Section \ref{Sct_Inter_Wdd}) reveals smaller SSH near the Antarctic coastline for the $1/4^\circ$ degree model. Indeed, at this model resolution, a strong positive trend in SSH (getting smaller) in time is observed around the entire Antarctic coastline, suggesting changes in the Antarctic Slope Current (ASC) during model spin-up. Large-scale sea-ice melt near the coastline, and changes in the Weddell Gyre leakage due to a stronger Weddell Gyre observed in the $1/4^\circ$ model, are noted as possible explanatory mechanisms.

The transport decomposition analysis has been performed for other meridional sections across the ACC in Section \ref{Sct_LngSct}. Many of the characteristics of the decomposition are common to all meridional sections. Expected ${T}_{ACC}$ and its decomposition estimate $\tilde{T}_{ACC}$ are found to be relatively consistent across sections, except for the wider Madagascar - Antarctica section, where non-ACC currents are also present. We find weaker $T_{bot}$ for all other sections except  Madagascar - Antarctica and the Drake Passage. $T_{\beta}$ is found to increase in magnitude for the wider sections, for which it represents a larger proportion of the overall density contribution.

Investigation of along-boundary neutral densities reveals good agreement between boundary density structures for the HadGEM-GC3.1 $1^\circ$-$LM$ model and the GloSea5 reanalysis dataset. More generally, there are considerable differences in isopycnal structure between different model, climatology and reanalysis datasets. Boundary density values, and locations of surface dense water along the Antarctic coastline, are comparable for the GloSea5, GK and $1^\circ$-$LM$ model datasets. However, GK does not replicate (a) the presence of flat isopycnals between the Weddell and Ross Seas along the western Antarctic Peninsula, and (b) sloping isopycnals along East Antarctica between the Ross and Weddell Seas. In contrast, the $1/4^\circ$ and $1^\circ$-$LL$ models show significantly warmer waters at depth and fresher surface upper layers, resulting in lighter densities throughout. Both $1/4^\circ$ and $1^\circ$-$LL$ models exhibit weaker isolated regions of dense water near the surface, with a lack of connectivity along the boundary to the deep ocean. The contrasting density features of model, climatology and reanalysis datasets emphasise the difficulty in characterising along-boundary density, and the physical mechanisms underlying its structure, especially given sparse direct observational data from this remote region.

We hypothesise that the presence of an ASC is important in setting the  sloping isopycnal structure between the Ross and Weddell Seas along East Antarctica. The absence of such a current leads to flatter isopycnals along the western Antarctic Peninsula. Investigation of GloSea5 data reveals that regions of deep water formation in the Weddell and western Ross Seas exhibit dense surface water on the boundary throughout the year. However, deep water formation regions of the eastern Ross Sea and Adelie coast show seasonal characteristics, with dense water throughout the fluid column only from June to December.

 \clearpage

\chapter{Summary and conclusions}  \label{TJ_Smm}
The Meridional Overturning Circulation (MOC) is a system of surface and deep ocean currents, eddy-driven circulations and regions of water mass transformation. The Atlantic MOC (AMOC) exchanges fluid (and hence heat, salt, nutrients, carbon, etc.) vertically and horizontally throughout the Atlantic basin, strongly influencing the climate of the northern hemisphere. Climate models suggest that changes to the AMOC are indicators and drivers of climate shifts of considerable societal concern. The Antarctic Circumpolar Current (ACC) is a dominant feature of the MOC, connecting all of the Earth's major ocean basins. Long-term variation of the ACC affects the rate of heat and carbon exchange between atmosphere and ocean in the Southern Ocean, with further implications for climate change. 

The AMOC and ACC generated by general circulation models (GCMs) are often difficult to interpret, involving many interacting complex physical processes. Simple theoretical diagnostic models are therefore useful in explaining observed behaviour in terms of elementary mechanisms including thermal wind and geostrophy, and the characteristics of fundamental quantities such as density on the ocean boundary.

This thesis has considered ocean boundary properties and their impact on the AMOC and the ACC. The aim of the thesis has been to determine the extent to which large-scale ocean circulation features such as the AMOC and ACC can be reconstructed and understood in terms of properties measured on ocean boundaries. Specific goals are four-fold: (a) Development and application of a basin-wide AMOC decomposition diagnostic explaining the total meridional overturning streamfunction in space and time in terms of its constituent boundary components. (b) Quantification of basin-wide spatial and temporal structure and variability of boundary components, for the Atlantic basin.  (c) Exploration of ACC transport and contributing components within a hierarchy of HadGEM-GC3.1 models using a decomposition diagnostic modified for zonal flow. (d) Characterisation of density, potential temperature and salinity along the sloping continental boundaries around the Atlantic and Antarctic continent, using a boundary mapping algorithm.

This chapter summarises the findings of the thesis in a broader oceanographic context, and outlines opportunities for further study. Due to the increasing prevalence of models with $1/4^\circ$ spatial resolution within the latest GCMs, we draw conclusions for the HadGEM-GC3.1 $1/4^\circ$ model in particular from across the thesis.

\section{Thesis summary}

Chapter \ref{TJ_TM} develops the MOC decomposition diagnostic. In the Atlantic, this diagnostic allows us to decompose the AMOC overturning streamfunction for a given latitude into five constituent boundary components:  western and eastern boundary densities, ocean surface wind stress, meridional velocity in bottom cells, and meridional velocities in additional side-wall and partial cells. The decomposition diagnostic is evaluated using the HadGEM-GC3.1 model at $1^\circ$, $1/4^\circ$ and $1/12^\circ$ spatial resolutions. Estimates of the total overturning streamfunction in terms of the five components generally perform well throughout the Atlantic basin, and are able to capture the time-mean AMOC overturning streamfunction when all components are volume-compensated to form individual streamfunctions which close at the surface. Remarkably, boundary information is largely sufficient to reconstruct the time-mean overturning streamfunction and its temporal variability, without any information from the ocean's interior.

Building on the existing literature, we show that western and eastern boundary density components generally dominate the basin-wide time-mean overturning streamfunction within the ocean's interior (c.f. \citealt{Buckley2016}, \citealt{McCarthy2020}). The Ekman (wind stress) term dominates near the surface (\citealt{Ekman1905}, \citealt{Baehr2004}). For latitudes $25-30^\circ$N, boundary currents through the Florida Straits result in a strong bottom (depth-independent) flow contribution (\citealt{Sime2006}).  Contributions from additional incomplete cells adjacent to the bathymetry also make a sizeable contribution in regions of strong boundary currents. As model spatial resolution increases, the size and importance of contributions from additional cells reduces.

The theoretical capacity of the decomposition diagnostic framework is greater than that achieved in practice using time-average quantities which do not preserve non-linear dependencies, such as that between in-situ density and potential temperature (because of Jensen's inequality, \citealt{Jensen1906}). Calculations using instantaneous densities, averaged over time-steps, improve the reconstructed horizontal pressure gradient momentum trend, and hence the overturning streamfunction estimate. We conclude that decomposition of the overturning using time-average quantities should be undertaken with caution, and that models such as NEMO might usefully output correctly time-average computations for densities in the future.  

In agreement with \cite{Hirschi2020}, we find that AMOC strength, quantified in terms of the maximum (with respect to depth) of the overturning streamfunction, increases with model spatial resolution for all latitudes south of $35^\circ$N. We also find that AMOC strength south of $35^\circ$N is approximately constant with latitude. The five constituent boundary components are also approximately independent of latitude in the southern hemisphere, but show large variability with latitude in the northern hemisphere. Boundary density terms no longer dominate for the latitude range $15-30^\circ$N. Further regional decomposition of density terms reveals a number of interesting features, including a large  contribution to the maximum estimated overturning streamfunction from eastern boundary density in the Gulf of Mexico and Caribbean Sea, as well as the eastern boundary of the Bahamas and neighbouring islands. 

Chapter \ref{TJ_Var} examines the spatial and temporal variability of the AMOC and the mechanisms by which boundary components contribute to it. Compared with data from RAPID and SAMBA arrays (\citealt{Frajka-Williams2019}), the decomposition diagnostic gives good estimates for the maximum overturning streamfunction. The overturning reconstructed from boundary properties captures the temporal variability of the full overturning well. 

For the $1/4^\circ$ model, statistical analysis (using MLR-CAR, \citealt{Zuber2010}, combining elements of linear regression and correlation analysis) is used to estimate the contributions of boundary components to the total variance of the expected overturning streamfunction, for different timescales set using band-pass filtering. The largest variation is found at short timescales. At all timescales, we observe large contributions from the Ekman component in the upper 2500m for the northern hemisphere, attributable to variable winds. Contributions from the depth-independent component at mid-depths between $30^\circ$N and $45^\circ$N are large, possibly due to seamounts and the interaction of the Gulf Stream and DWBC. The influence of the western Atlantic boundary density dominates variation at mid-depths. Variation in the upper layers at Gulf of Mexico and Caribbean Sea latitudes is dominated by western boundary densities in those marginal basins. The Ekman contribution to the total variance of the overturning streamfunction dominates near the surface for all latitudes and timescales considered (e.g. \citealt{Srokosz2015}, \citealt{Buckley2016}). Due to the discretised model grid, the contribution of additional cells is always important locally near bathymetry. On shorter timescales (i.e. at higher frequency) in the southern hemisphere, western densities provide important contributions at mid-depths, with an influential Atlantic boundary depth-independent contribution at depth, attributed to variation of the Brazil current. In the northern hemisphere, mid-depth contributions are evenly distributed between the Ekman, western Atlantic boundary density and Atlantic boundary depth-independent contributions, with the latter prevalent throughout the fluid column near $35^\circ$N, possibly due to the separation of the Gulf Stream and its interaction with the DWBC. At longer timescales, we find, in the southern hemisphere, that western Atlantic boundary densities assume increasing importance. The Atlantic boundary depth-independent contribution remains influential at depth. At high northern latitudes, Greenland boundary densities make larger contributions to the total variance. 

Corresponding analysis for the $1^\circ$ and $1/12^\circ$ models reveals similar contributions of boundary components. However, with increasing model resolution, the total variance of the overturning increases. This is particularly evident near $37^\circ$N for the $1/12^\circ$ model due to better resolution of the New England and Corner Rise seamount chains.

In Chapter \ref{TJ_Bdry}, along-boundary properties for the Atlantic and surrounding basins are considered in detail. Using a rudimentary analysis in latitude-depth space for $1^\circ$, $1/4^\circ$ and $1/12^\circ$ HadGEM-GC3.1 model output, flat eastern boundary isopycnals are observed to the south of $35^\circ$N at all model resolutions, as commonly assumed within reduced-gravity models, supporting the idea that fast boundary wave propagation removes density anomalies (\citealt{Johnson2002}). 

Motivated by \cite{Hughes2018}, a boundary tracking algorithm is developed to pinpoint the location of sloping continental boundaries across multiple depths. The algorithm produces novel descriptions of neutral and potential densities, potential temperature and salinity, centred on the Atlantic basin from HadGEM-GC3.1 model, GloSea5 reanalysis and \cite{GouretskiViktorandKoltermann2004} climatology. There is general agreement across all sources for the presence of flat isopycnals along the eastern Pacific boundary, along the eastern Atlantic boundary south of the Mediterranean, and along the western boundary of the Indian Ocean. Isopycnals whose depth varies linearly with along-boundary distance are only found along the western boundary of the Atlantic basin up to the Labrador Sea; we find some intervals of steeper isopycnal gradients with along-boundary distance e.g. in the vicinity of Gulf Stream separation near Cape Hatteras. Discrepancies between the boundary density structure in different models are greatest at depth in the North Atlantic, due potentially to differences in model descriptions of deep water formation at high latitudes. 

Given that eastern boundary isopycnals are flat, that the thermal wind relationship dominates the MOC, and given that the Coriolis parameter $f$ varies with latitude, isopycnals on the western boundary of the Atlantic must slope to maintain an overturning circulation. The linear slope of isopycnals on the western boundary is consistent with down-slope migration of southward-propagating DWBC proposed by \cite{MacCready1994}. The absence of a similar mechanism in the Indian Ocean is a possible cause for flat isopycnals on the western boundary there.

Chapter \ref{TJ_ACC} shows that estimates for the magnitude of ACC transport at the Drake Passage in HadGEM-GC3.1 models at different resolutions do not agree, and all underestimate the $173$Sv observed by \cite{Donohue2016}, which includes a depth-independent contribution of 45Sv. ACC strengths of 150Sv, 60Sv and 120Sv are found in the $1^\circ$, $1/4^\circ$ and $1/12^\circ$ resolution simulations at steady state. Further, in the $1/4^\circ$ model run, the ACC transport declines rapidly in the first 40 years. These issues reflect a wider range of deficiencies in the HadGEM-GC3.1 description of Southern Ocean dynamics. The UK Met Office is interested in understanding and interpreting these deficiencies. As part of this effort, the decomposition diagnostic for the overturning streamfunction, developed in Chapter \ref{TJ_TM} is modified to decompose the zonal cumulative ACC transport at the Drake Passage, reported in Chapter \ref{TJ_ACC}. The total ACC transport is described in terms of boundary properties and an additional $\beta$ term arising from a meridionally-varying Coriolis parameter. Decomposition-based estimates for the total ACC transport perform well for the Drake Passage within the $1^\circ$ and $1/4^\circ$ models. Discrepancies at $1/12^\circ$ resolution are attributed to inaccurate density information within localised trenches, only resolved at this resolution; an improved estimate is obtained using a simplified boundary description at the Drake Passage. 

As the $1/4^\circ$ model spins up, lighter densities and freshening develop near the southern boundary, causing isopycnals to slump towards Antarctica rather than outcrop, strengthening the local along-boundary near-surface ASC return flow. The southern density contribution to the ACC transport increases by 70Sv within the first 40 years, explaining the initial weakening of ACC transport observed. There is an increase in SSH in time, from an initial negative value, around the entire Antarctic coastline, potentially due to large-scale coastal sea-ice melt or changes in Weddell Gyre leakage at this resolution (\citealt{Meijers2016}). 

Density characteristics along the sloping boundary of Antarctica vary with choice of model resolution, reanalysis or climatology dataset, making precise characterisation of along-boundary density structure problematic. Nevertheless, along-boundary neutral densities from the HadGEM-GC3.1 $1^\circ$-$LM$ model (at medium atmospheric resolution) agree with those from the GloSea5 reanalysis dataset. Given that the $1^\circ$-$LM$ model also reproduces a realistic ACC transport, we might infer that these estimates are therefore more reliable than others. Two distinct regions of deep water formation are present in the Weddell and Ross Seas, with dense surface water throughout the year. Regions of deep water formation in the eastern Ross Sea and on the Adelie coast show seasonal variation, with dense water throughout the fluid column from June to December. 

Mapping of Antarctic boundary densities for $1^\circ$-$LM$ and GloSea5 indicates flat isopycnals along the western Antarctic Peninsula, and sloping isopycnals along East Antarctica. The latter could be explained by a westward Antarctic Slope Current migrating down the boundary, preserving kinetic energy at the expense of its potential energy (\citealt{MacCready1994}). Along the western Antarctic Peninsula, no Antarctic Slope Current is present. 

\subsection*{Summary of $1/4^\circ$ HadGEM-GC3.1 model issues}

The HadGEM-GC3.1 $1/4^\circ$ model with medium atmospheric resolution (N216) is part of the next generation of UK Met Office GCMs (HighResMIP, \citealt{Haarsma2015}). The spatial resolution typically used for current CMIP5 and CMIP6 models is $1^\circ$, with $1/4^\circ$ resolution becoming more popular. It is interesting therefore to summarise the findings of this D.Phil. thesis regarding the characteristics of the $1/4^\circ$ model, in studying the AMOC, ACC and continental along-boundary properties. 

The $1^\circ$-$LM$ model provides estimates of both AMOC and ACC strengths which agree relatively well with observations. Moreover, the model provides representations of boundary densities which compare well with GloSea5 reanalysis for the Atlantic and Antarctic coastlines. With increasing model spatial resolution at $1/4^\circ$, we might expect similarly good or even better correspondence.

In the Atlantic, the time-mean latitude-depth structure and magnitude of the overturning streamfunction for the $1/4^\circ$ model shows good agreement with time-mean observations at RAPID and SAMBA arrays. Along-boundary potential temperature, salinity and density for the Atlantic and surrounding boundaries are in good agreement with estimates from models at other spatial resolutions, reanalysis and climatology. However a weaker AABW signal is observed near the Drake Passage, especially later in the model run of 657 years.

Fluctuations in the depth-independent and thermal wind (boundary density) components for the $1/4^\circ$ model at SAMBA, reveal that these components are dependent on the Brazil-Malvinas confluence (BMC). These fluctuations appear since the BMC is located almost $4^\circ$ further north than expected (\citealt{Goni2011}), suggesting either a weaker Brazil or stronger Malvinas current within the model.

However, ACC transport through the Drake Passage shows large weakening throughout the $1/4^\circ$ model run, stabilising at 60Sv, only one third of the expected value reported by \cite{Donohue2016}. Similar results are found for other meridional sections across the ACC. The weakened ACC is due to a freshening of the southern boundary and a slumping of isopycnals, resulting in a stronger reverse flow (ASC) along the boundary. Preliminary analysis of local sea surface height and barotropic streamfunction support fresher shelf waters (less dense) and considerably larger Weddell and Ross Sea gyres, respectively. 

A possible mechanism for the strengthened Malvinas current is the northward shift of the main ACC pathway, caused by a northward shift of the wind-stress curl and westerly winds. These changes in wind-stress patterns could result in stronger spin-up of gyres, and weaken or even reverse the shelf or slope current through the Drake Passage.

Further, along-boundary investigations for the sloping Antarctic coastline reveal considerably lighter densities compared to the GloSea5 reanalysis data and model output at $1^\circ$. Fresher surface water suggests excessive freshwater flux. As a result, little or no surface dense water formation is observed in the Weddell and Ross Seas. This model simulation also lacks along-boundary down-slope flow of dense water. Together, this suggests a lack of realistic deep water formation, and AABW supply is impaired. This effect is also suggested by the Atlantic along-boundary isopycnal structure in Chapter \ref{TJ_Bdry}: AABW signal to the East of the Drake Passage is reduced compared to other models, reanalysis and climatology datasets. A lightening with time of bottom densities in the Brazil-Argentine basin, coupled to increased decadal and multi-decadal variability of bottom densities there, suggests upstream impairment of deep water formation near the Antarctic coast.

Drawing points from the previous paragraphs together, it is interesting that good agreement is found between observations and model-based estimates of AMOC strength for the $1/4^\circ$ model. Moreover, good agreement is also found for the $1^\circ$-$LM$ and $1/12^\circ$ models. These results might be expected for GCMs calibrated to AMOC strength. However, agreement between observations and model-based estimates of ACC transport are poor at $1/4^\circ$ in particular. The model also provides a poor representation of sloping Antarctic boundary densities, a significant freshening of the Antarctic shelf, slumping rather than outcropping of isopycnals and a stronger ASC as a result. Somewhat surprisingly, this suggests that adequate representation of these Southern Ocean circulation characteristics is not necessary for reasonable representation of Atlantic circulation in the $1/4^\circ$ model, and perhaps more generally.

\section{Discussion}
\label{Smm_Dsc}

Results in Chapters \ref{TJ_TM} and \ref{TJ_Var} suggest that the AMOC decomposition diagnostic provides a useful means to quantify the Atlantic overturning streamfunction and its variability, using only information gathered from boundaries. From Chapter \ref{TJ_Bdry} we find that the structure of western and eastern boundary densities can be represented simply in terms of flat or linearly sloping isopycnals with along-boundary distance. Together, these results suggest an improved method for monitoring basin-wide AMOC, using a small set of density profiles at carefully-chosen locations on the western and eastern boundaries. 

Observations at these locations over an extended period would allow the characterisation of basin-wide AMOC variation on longer timescales. Analysis of HadGEM-GC3.1 model output here suggests that, on longer timescales, in accordance with the literature, the geostrophic component becomes ever more important in explaining AMOC variability. Further, on decadal and longer timescales, the MLR-CAR model shows that western boundary densities dominate the variability, but that there is also evidence for an increasing contribution from the eastern boundary.

Other investigations using various forms of AMOC decomposition (e.g. \citealt{Waldman2021}) for long timescales are based on time-average GCM outputs such as potential temperature and salinity. Chapter \ref{TJ_TM} highlights deficiencies in this type of analysis which occur when non-linear relationships between properties, such as density and potential temperature, are not appropriately accounted for. This issue could be resolved in a relatively straightforward manner, were correctly time-averaged densities available as primary outputs from GCMs such as HadGEM-GC3.1. Chapter \ref{TJ_Var} shows that total AMOC variability reduces with increasing timescale, therefore becoming harder to diagnose. As a result, the impact of errors incurred due to incorrect calculation of time-average densities might increase with increasing timescale.

The variation of neutral density with depth and latitude along both eastern and western boundaries might be exploited for paleoclimate studies. Here, along-boundary observations are sparse, typically of density (or surrogate properties) for a limited range of depths along the sloping boundary at a given latitude, as opposed to full depth density-profiles. Nevertheless, such data at multiple locations along the boundary may be sufficient to construct a simple piecewise-linear model for boundary density structure, and provide a basis for AMOC reconstruction (\citealt{Lynch-Stieglitz2008}). The impact of bias and uncertainty associated with paleo-density reconstruction (e.g. using $\delta ^{\text{18}}$O) could be managed within a carefully-formulated statistical model; these uncertainties could possibly be larger than the variability of the estimated AMOC (e.g. \citealt{Hirschi2006}, \citealt{Lynch-Stieglitz2008}). 

The decomposition diagnostic is based on the simple principles of geostrophy and thermal wind. However these simple physical relationships do not hold everywhere in the ocean. For example, ageostrophic processes are important in the wind-driven Ekman layer and boundary currents on both sides of the basin. Further, features such as eddies, gyres and internal waves all play a role in ocean circulation, but do not adhere to the simplified dynamics assumed within our decomposition. Nevertheless, it is clear that the decomposition diagnostic is able to capture the overturning strength and variability, without fully accounting for these effects. 

\cite{Cessi2013a} highlight the role of ageostrophic flow along the Atlantic eastern boundary; they argue that instabilities act to erode the density structure here. Agesotrophic processes are likely to be key to the breakdown of flat eastern boundary isopycnals, especially north of $35^\circ$N in the Atlantic; eastern boundary eddies act to disturb the flat isopycnal structure. 

In addition, further consideration of processes such as along-equator winds and their impact on eastern and western boundary densities in the equatorial regions is needed. Density anomalies at the equator are propagated to high latitudes by fast boundary waves, emphasising the importance of equatorial processes, and any processes forming density anomalies along the boundary. 

Mapping along-boundary densities of the Atlantic and surrounding basins has demonstrated the inter-basin continuity of boundary densities. Given that boundary densities are influential for the overturning circulation, this connectivity between basins emphasises that localised changes in boundary properties may have large-scale or global impact.

In the Southern Ocean, the ACC decomposition diagnostic provides a means to quantify the impact of southern boundary freshening on the volume transport. The Coriolis $\beta$ term, incorporated within the diagnostic to ensure complete representation of the zonal transport, contributes minimally to ACC transport for narrow meridional sections, such as the Drake Passage. Combined with a weak Ekman transport contribution there, this suggests that useful monitoring at the Drake Passage could be achieved using only bottom velocity and boundary density measurements. However, the approach would not be appropriate for wider meridional sections, where the Coriolis $\beta$ term is likely to be more influential. 

The current work suggests that ASC flows along the southern boundary have a significant influence on zonal transport through the Drake Passage for some GCM resolutions. As a result, the net zonal transport there depends on both ACC and ASC transport components. Therefore, using the net zonal transport as the main means to quantify the ACC is inappropriate.

Given the zonally-constant ACC strength found in GCM output, we might expect along-boundary Antarctic densities to show little spatial variability. The large variability actually found, suggests that ageostrophic and other compensatory effects must be at play to maintain a constant ACC strength. This serves to highlight our lack of understanding regarding the three dimensional structure of the ACC, a topic considered further in Section \ref{Frt_Smm}.

\section{Further investigations} \label{Frt_Smm}
This section outlines areas of potential future research related to the topics addressed in the thesis. 

\subsection{Continuity of ACC transport}
In Section \ref{ACC_BdryD} we investigate neutral boundary densities around the Antarctic continent for various model resolutions, reanalysis and climatology. 
Intuitively, we might expect that large variations in neutral density observed along the Antarctic coastline would be accompanied by similar variations in the ACC transport. Therefore, the fact that the ACC strength remains relatively constant around the continent (Section \ref{Sct_LngSct}) seems counter-intuitive, and requires further explanation. 

\begin{figure*}[ht!]
	\centerline{\includegraphics[width=\textwidth]{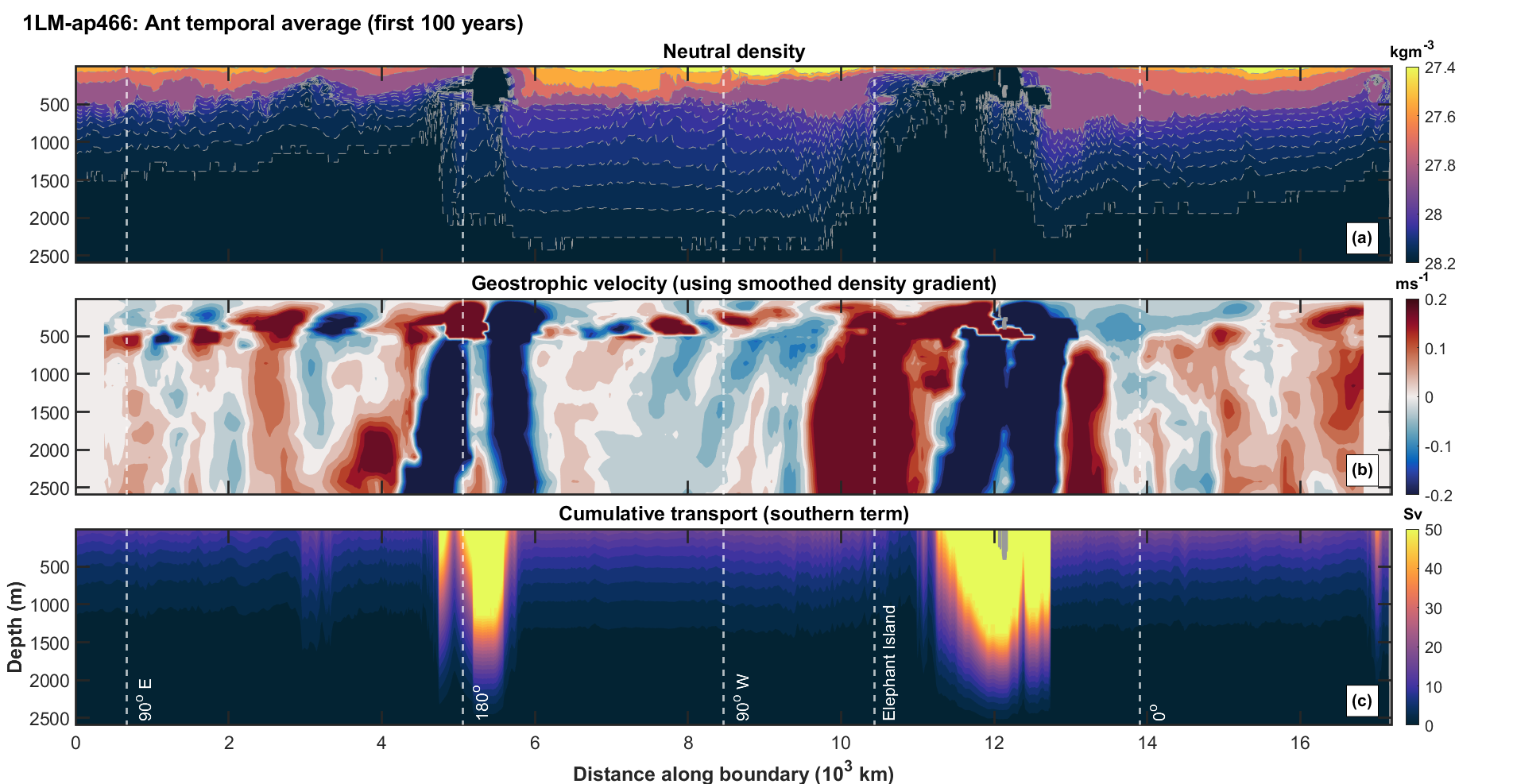}}
	\caption[Along-boundary Antarctic neutral densities for the $1^\circ$-$LM$ model, geostrophic velocity into the boundary and southern boundary cumulative transport contribution to the ACC transport, using diagnostic calculation. Time-average is taken over the first 100 years of model run.]{Panel (a): time-average along-boundary Antarctic neutral densities for the $1^\circ$-$LM$ model. Panel (b) : geostrophic velocity ($v$) into the boundary, calculated from neutral density using a smoothed density gradient. Panel (c) : southern boundary cumulative transport contribution to the ACC, using diagnostic calculation (Chapter \ref{TJ_ACC}). Time-average is taken over the first 100 years of model run. Dashed white lines indicate locations of longitudes $90^\circ$E, $180^\circ$, $90^\circ$W, $0^\circ$ and that of Elephant Island.}
	\label{F_BD_T_1LM}
\end{figure*}

Figure \ref{F_BD_T_1LM} shows (a) neutral boundary densities and (b) resulting geostrophic velocity into the boundary for the $1^\circ$-$LM$ model. Panel (c) is the equivalent southern boundary cumulative transport contribution to the ACC ($T_S$, from the ACC decomposition diagnostic) calculated from boundary density data (shown in Panel (a)), gathered using the boundary mapping algorithm. 

Cumulative transport contributions from southern boundary densities are relatively consistent around Antarctica, with three notable exceptions in the Weddell and Ross Seas, and near the Adelie coast, where dense surface water is found. These regions exhibit a significant increase in southern contribution to the ACC transport. ACC continuity therefore demands compensating changes in the other boundary components.

The $1/4^\circ$ model showed a large weakening in ACC transport, and isolated regions of deep water formation. The corresponding southern boundary cumulative transport contribution (not shown) is found to be markedly different to the $1^\circ$ model (Figure \ref{F_BD_T_1LM}(c)): large contributions are still found near the Weddell and Ross Seas, due to denser surface waters. However, there is greater along-boundary variability of the southern boundary contribution to ACC transport, and generally a larger magnitude of contribution outside regions of deep water formation, due to lighter along-boundary densities (see Figure \ref{F_BD_Ant_Rest}(b)). 
%

A zonally-constant ACC coupled to a variable southern component necessitate variability in other components. Alternatively, ageostrophic properties may contribute in regions of deep water formation, so that the geostrophic principles of the transport decomposition fail. Other possibilities include (a) a greater influence of intermediate boundaries in meridional sections near regions of deep water formation, (b) changes on the northern flank of the ACC to compensate those on the southern boundary, and (c) a greater influence of the $\beta$ term. A considerable obstacle to any investigation would be the difficulty in identifying the northern boundary for the majority of the ACC. Use of an oceanic boundary would be feasible, but compromises the decomposition methodology. 

\subsection{Impact of model resolution on bottom velocities}
The ACC transport is commonly used as an indicator of model performance. The  depth-independent component is a key contributor to the total ACC transport. \cite{Donohue2016}, using bottom current meters through the Drake Passage, estimate the depth-independent component to be of the order of $45$Sv, 30Sv stronger than that simulated within the $1/4^\circ$ and $1/12^\circ$ models (see Figure \ref{F_Dcmp_bot}). Only the $1^\circ$ model performed well in this respect, a somewhat counter-intuitive finding since higher resolution models should resolve bottom flows better.

Further, ACC decompositions at the Drake Passage and elsewhere (Chapter \ref{TJ_ACC}) indicate that the depth-independent component is maximised at the Drake Passage and Madagascar - Antarctica sections, suggesting stronger bottom flows or currents in these regions. Examination of ACC bottom currents through Drake Passage and other meridional sections, for models at different resolution, and a comparison with e.g. cDrake observations, could highlight issues associated with resolving bottom flows in higher-resolution models. An increase in depth-independent component of 30Sv for the $1/12^\circ$ model would bring it into line with the corresponding value for the $1^\circ$ model; as a result, the total ACC transport from the $1/12^\circ$ and $1^\circ$ models would also agree.

\subsection{Location of the Brazil-Malvinas Confluence (BMC)}
Results in Chapter \ref{TJ_Var} indicate that the location of the BMC within the HadGEM-GC3.1 $1/4^\circ$ model is surprisingly $4^\circ$ further north than expected. This might be explained by a stronger Malvinas current driven by a northward shift of the main ACC pathway through the Drake Passage. This itself would require a northward shift of the wind-stress curl and westerly winds. Atmospheric fields could be examined for evidence to support this. Further, a possible relationship between the location of the BMC and the magnitude of the northern density component of the ACC decomposition diagnostic might be sought. 

\subsection{Application and further development of the along-boundary mapping algorithm}
The analysis in Chapter \ref{TJ_Bdry} and Section \ref{ACC_BdryD} demonstrates the utility of the along-boundary mapping algorithm, and the boundary density profiles generated using it. Applying the mapping algorithm more widely (e.g. to CMIP6 models) would offer the opportunity to compare and contrast along-boundary isopycnal structures, and help generate consensus. The sparsity of observational data for the Antarctic coastline makes it difficult to calibrate models, and therefore more likely that boundary characteristics are determined by model-specific dynamics in the Southern Ocean. Examining a broader range of models would provide greater understanding of along-boundary isopycnal structure, and test the influence of the Antarctic Slope Current on the density structure estimated.

The boundary pathway used for the Atlantic and surrounding basins could be extended to accommodate the western Pacific and eastern Indian Oceans, creating a single connected boundary to the north of the Southern Ocean. This analysis would require extra assumptions to accommodate non-continuous boundaries and throughflows between oceans in south-east Asia; preliminary analysis suggests that artificial boundaries in this region might be a useful for some purposes.  

Without observations or model data, the western Pacific along-boundary isopycnal structure is difficult to predict, given the weak Pacific MOC. Logic dictates that isopycnal structure should be relatively similar to that of the East Pacific, with possibly slightly sloping isopycnals to the north, to maintain the PMOC. Initial estimates of along-boundary densities support this thinking. 

Relatively flat isopycnals are predicted for the eastern Indian Ocean as well. Here, the overturning circulation consists of two shallow cells either side of the Equator (e.g. \citealt{Lee1998}). \cite{Lee1998} find that the Ekman component of their decomposition dominates these two shallow overturning cells. Below 500m, they find a weak vertical shear (or geostrophic) contribution (of 2Sv), showing a positive overturning streamfunction throughout the basin. We speculate, based on simple geostrophic calculations, that eastern boundary isopycnals should show signs of a weak upward slope to the north to maintain the positive cell of the overturning streamfunction throughout the basin (similar to the western boundary in the Pacific). Nevertheless, we caution that there are many physical mechanisms at play in this highly seasonal basin, and the weakness of the vertical shear component found by \cite{Lee1998} suggests that along-boundary density structure might not be strongly-coupled to the overturning circulation here. 

\subsection{Reconstruction of along-boundary structure using a limited number of depth profiles}
In Appendix \ref{Bdry_MCMC_App} a piecewise linear model (outlined in Appendix \ref{App_MCMC}) is used to describe the relative complexity of time-mean latitude-depth eastern and western Atlantic boundary densities. The analysis emphasises the greater uniformity of the eastern boundary, and the possibility of estimating along-boundary densities, and overturning streamfunction, with a limited number of density profiles. This provides a quantitative basis to optimise the locations of future density profile observations, even over vast distances across continents, based on GCM output.

The piecewise linear model can be applied routinely to provide a low-dimensional summary of along-boundary isopycnal structure for the Atlantic and Antarctic. In the thesis work, the piecewise linear model is applied to time-mean data. An important but relatively straightforward enhancement would be to extend the algorithm to accommodate time-varying data. Further, if the intention is to simplify the overturning streamfunction calculation, depth-weighting could be introduced to emphasise the important role of densities at depth (Appendix \ref{App_MCMC}).  

\subsection{Similarities between thermal wind and depth-independent components}
Atlantic thermal wind (Chapter \ref{TJ_TM}, Figure \ref{F_TM_strm}(d)) and depth-independent components of the overturning circulation (Figure \ref{F_TM_strm}(c)) show a high degree of negative correlation. Preliminary work (using the UK MetOffice GC2 dataset) suggests it might be feasible to predict the uncompensated depth-independent component using a regression model with uncompensated western boundary, eastern boundary and Ekman components as predictors. If successful, this statistical model would further reduce the complexity of the decomposition diagnostic, making reconstruction of the total overturning streamfunction possible in terms of boundary densities and surface wind-stress only.

\section{Final Word}

The meridional overturning circulation is a fundamental component of the Earth's climate system. This thesis shows that the large-scale four-dimensional overturning circulation can be well-represented by information available only on the ocean's boundaries. New mappings of boundary densities across multiple ocean basins reveal simple structures and large-scale continuity, including regions of flat and linearly-sloping isopcynals extending over thousands of kilometers. The physical mechanisms underlying these simple boundary structures ensure that the effects of local density anomalies are propagated along boundaries, demonstrating the inter-connected nature of the global overturning, and its sensitivity to sources of remote boundary anomalies.

 \clearpage
\appendix
\appendix
\chapter{Appendix}  
\label{TJ_App}

\section{Reliability of decomposition time-mean estimate for AMOC and the role of additional cells} \label{App_SpatVar}

Suppose we have observations $\{\Psi_i\}_{i=1}^n$ of the time-mean total overturning at a depth-latitude combination indexed by $i$, and observations $\{\Psi_i^*\}_{i=1}^n$ of the time-mean interior transport, which ignores contributions from partial and sidewall cells. Suppose we also have estimates $\{\hat{\Psi}_i\}_{i=1}^n$ of the time-mean total overturning from the AMOC decomposition model (also ignoring partial and sidewall cells). We would like to quantify and compare how well our estimate $\hat{\Psi}$ explains $\Psi$ and $\Psi^*$. We can do this by estimating the \textbf{mean} $m$ and \textbf{variance} $v$ of the difference between $\Psi$ or $\Psi^*$ and $\hat{\Psi}$, using the equations 
\begin{eqnarray}
m(x) &=& \frac{1}{n} \sum_{i=1}^{n} x_i \\ 
v(x) &=& \frac{1}{n} \sum_{i=1}^{n} \left( x_i - m(x) \right) ^2
\end{eqnarray}
where $x$ refers to the difference between streamfunction estimates, and the summation is made over all latitudes and depths in the cross-section. We find that for the $1/4^\circ$ model
\begin{eqnarray}
\mu &=& m(\Psi-\hat{\Psi}) = -0.056 \text{ Sv} \nonumber \\
\sigma^2 &=& v(\Psi-\hat{\Psi}) = 1.018 \text{ Sv}^2 \nonumber \\
\mu^* &=& m(\Psi^*-\hat{\Psi}) = -0.161 \text{ Sv} \nonumber \\
\sigma^{*2} &=& v(\Psi^*-\hat{\Psi}) = 0.458 \text{ Sv}^2. \label{e_A_BiasVar} 
\end{eqnarray} 
Results show that the variance of the interior transport error $\Psi^*-\hat{\Psi}$ is almost exactly one quarter the variance of the full transport error $\Psi-\hat{\Psi}$, indicating that additional cells are responsible for increasing the uncertainty of our estimate for $\Psi$. However, the additional contributions also reduce the magnitude of the bias (or mean error) from $-0.161$Sv for $\Psi^*$ to $0.056$ Sv for $\Psi$.

Assuming the true value $\Psi$ follows a Gaussian distribution given $\hat{\Psi}$, we can write
\begin{eqnarray}
\Psi &=& \hat{\Psi} + \mu + \sigma \times \epsilon \nonumber  \\
\Psi^* &=& \hat{\Psi} + \mu^* + \sigma^* \times \epsilon
\end{eqnarray} 
where $\epsilon$ is a random number with a standard Gaussian distribution (with mean zero and variance one). Our data also shows that 
\begin{eqnarray}
v(\Psi) = 24.427 \text{Sv}^2  \nonumber  \\
v(\Psi^*) = 21.466 \text{Sv}^2 .
\end{eqnarray} 
Comparing $v(\Psi)$ with $v(\Psi-\hat{\Psi})$ (or $v(\Psi^*)$ with $v(\Psi^*-\hat{\Psi})$), we see that our estimate explains most of the variability of time-mean transport. For example, the percentage of the variance of $\Psi$ explained by $\hat{\Psi}$ is $(1 -  1.018/24.427) \times 100 = 95.8 \%$. The percentage of the variance of $\Psi^*$ explained by $\hat{\Psi}$ is $(1 - 0.458/21.466 ) \times 100 = 97.9 \%$. 

\subsection*{Percentage variance explained}
In more general terms, the percentage variance explained by estimate $Y$ of variable $X$ is calculated using 
\begin{eqnarray}
1-\frac{\sum_{i=1}^n \left( (x_i - \overline{x}) - (y_i - \overline{y}) \right) ^2}{\sum_{i=1}^n (x_i - \overline{x})^2}
\label{e_PrcVarExp}
\end{eqnarray}
where $\{x_i\}_{i=1}^n$ and $\{y_i\}_{i=1}^n$ are samples of values for $X$ and $Y$ of size $n$, and $\overline{x}$ and $\overline{y}$ are sample means, which might be observed over time or space or both.

\section{Effect of Coriolis acceleration and bottom drag on the estimates of the overturning streamfunction}
\label{App_Coriolis}

Analysis of the overturning streamfunctions in Chapter \ref{TJ_TM} shows a depth-dependent difference for numerous latitudes between the estimated $\tilde{\Psi}^c$ and expected $\Psi^c$ streamfunctions, which increases as the surface is approached. This suggests the possibility that small errors near the bottom of the fluid column are propagated upwards due to vertical integration in overturning streamfunction calculations, leading to large differences near the surface.

Possible sources of error could be the simplification of the Coriolis acceleration ($fv$) calculation near bathymetry, or lack of consideration of bottom friction and drag within the diagnostic framework. These influences can be accounted for as additional terms within the geostrophic balance equation (Equation \ref{HB}), satisfied by the NEMO model. We extend the geostrophic balance equation to include two correction terms $B$ and $C$. $B$ represents a bottom drag correction and $C$ a correction for the simplified treatment of Coriolis acceleration. Informally, the extended equation is
\begin{equation}
	\frac{\partial P}{\partial x} = \rho_0 f(y) v_{4pt} + C + B
	\label{e_HBext}
\end{equation}
where the Boussinesq approximation is applied ($\rho \rightarrow \rho_0$) and $v_{4pt}$ refers to the northward velocity calculated at the $u$-point using a 4-point average of the local northward velocities at the neighbouring $v$-points. 

Although there is a contribution to the overturning streamfunction from bottom drag, we find that the coarse vertical grid structure, especially at depth, renders any inclusion of a bottom drag term within the geostrophic balance equation of negligible value. The contribution of the bottom drag only influences cells adjacent to bathymetry, which are generally additional cells. However, discrepancies between estimated $\tilde{\Psi}^c$ and expected $\Psi^c$ streamfunctions are also found for complete interior cells, which are not influenced by bottom drag. We therefore set $B\approx0$.

Currently, the NEMO model (\citealt{Madec2016}) uses an Energy and Enstrophy conserving scheme (EEN, \ref{EEN_fv}) for the Coriolis term. The Coriolis acceleration term within the \textit{EEN} scheme is calculated using a 12-point average, taking the form of 4 sets of triads (3 points in an ``L'' shape) which together form a weighted average for $fv$ at each $u$-point, according to

\begin{equation}
	\label{EEN_fv}
	(q v^*)_{EEN} = \frac {2}{3} \overline{ ( \overline{q}^x v^*  ) }^{xy}
	+ \frac {2}{3} \left( { \overline{q}^y   \overline{v}^{*xy} } \right)
	- \frac {1}{3} \overline{ \left( { q \overline{v}^{*x} } \right) }^y , \\
\end{equation}
where $v^* = H_v v$ is the flux through a cell face and $H_v$ is the depth of the cell face. Further, $q = (f + \zeta)/H_{\zeta}$, where $\zeta$ is the vorticity and $H_{\zeta}$ is the cell depth at the vorticity point. The $12$ contributions to $(fv)_{EEN}$ in the \textit{EEN} scheme take the form  \textbf{$\frac{f}{H_f}H_v v$} where $H_f$ and $H_v$ are the depths of a grid cell at $f$ and $v$ points respectively. The resulting Coriolis acceleration is equal to the sum of weighted values of $v$ at surrounding $4$ points, with unequal weightings. The Coriolis acceleration correction in Equation \ref{e_HBext} is
\begin{eqnarray}
	C = (fv^*)_{EEN} - fv_{4pt}.
\end{eqnarray} 
The greater number of calculation locations within the EEN scheme improves the resulting overturning streamfunction estimates, reducing errors caused by bathymetry in places where the simplified 4-point velocity method performs poorly. Improvement in estimated overturning streamfunction is found especially in regions of rough bathymetry; these improvements are, however, small in comparison to the overall differences $\Psi^c - \tilde{\Psi}^c$ in time-mean overturning streamfunction present.

\section{Difficulty in decoupling western and eastern contribution to the maximum overturning streamfunction}
\label{App_MaxWestEast}

\begin{figure*}[ht!]
	\centerline{\includegraphics[width=\textwidth]{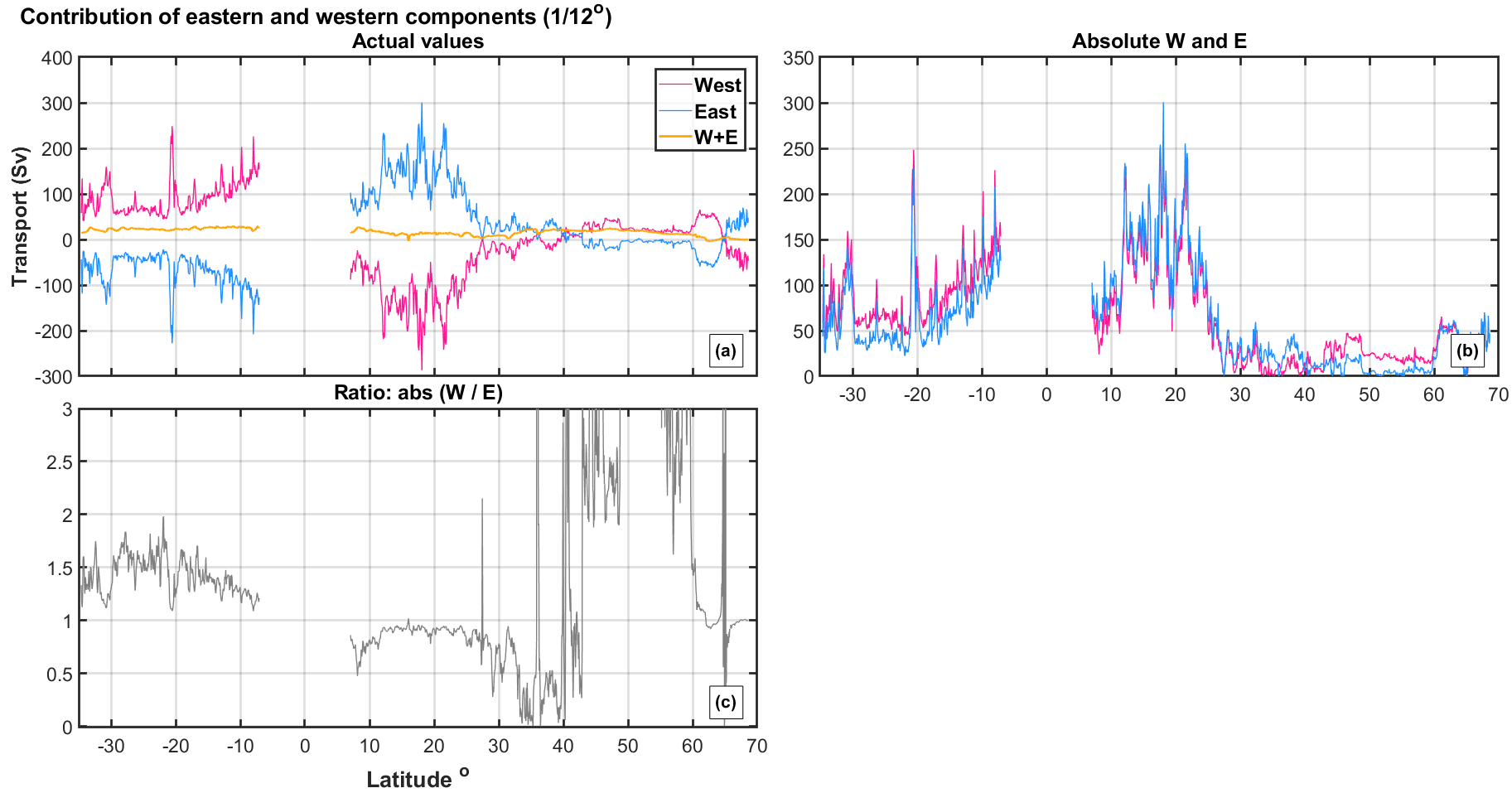}}
	\caption[Western and eastern boundary component contributions to the maximum estimated compensated overturning streamfunction with respect to latitude for the $1/12^\circ$ model.]{Western and eastern boundary component contributions to the maximum estimated overturning streamfunction with respect to latitude for the $1/12^\circ$ model. Panel (a) shows the actual values for the components, (b) the absolute values of $\Psi^c_{{W}_{\max}}$ and $\Psi^c_{{E}_{\max}}$ components and (c) the absolute ratio of $\tilde{\Psi}^c_{\max}$ relative to $\Psi^c_{{E}_{\max}}$. Western boundary component shown in pink, eastern in blue and their sum in orange.}
	\label{p_MaxStrm_WE}
\end{figure*}

In Figures \ref{p_Qtd_MxStrmLtt} and \ref{p_RC_BdryMxStrmLtt}a of Chapter \ref{TJ_TM} we find $\Psi^c_{{th}_{\max}}$ (i.e. the sum of the eastern and western boundary components) exhibits significant variability with latitude. Examination of the individual eastern and western boundary components is difficult due to the magnitude and close coupling found between the components (Figure \ref{p_MaxStrm_WE}). For brevity, we choose to refer to these quantities as $\Psi^c_{E_{\max}}$ and $\Psi^c_{W_{\max}}$, evaluated at depth $\tilde{d}_{\max}^c$ of the maximum estimated streamfunction. 

Figure \ref{p_MaxStrm_WE} highlights the contribution of $\Psi^c_{{W}_{\max}}$, $\Psi^c_{{E}_{\max}}$ and $\Psi^c_{{th}_{\max}}$ components to the maximum estimated overturning streamfunction ($\tilde{\Psi}^c_{\max}$) for each latitude for the $1/12^{\circ}$ model. Panel (a) shows the actual values for the components, (b) the absolute values of $\Psi^c_{{W}_{\max}}$ and $\Psi^c_{{E}_{\max}}$ components and (c) the absolute ratio of  $\Psi^c_{{W}_{\max}}$ relative to $\Psi^c_{{E}_{\max}}$. We find different behaviour in each of three latitude intervals. First, throughout the southern hemisphere, $\Psi^c_{{W}_{\max}}$ is positive, and its magnitude is greater than the (negative-valued) $\Psi^c_{{E}_{\max}}$ (Figure \ref{p_MaxStrm_WE}(a,b)); this leads to a significant northward $\Psi^c_{{th}_{\max}}$ contribution. Next, at low latitudes in the northern hemisphere up to $40^\circ$N, sign reversal of $\Psi^c_{{E}_{\max}}$ and $\Psi^c_{{W}_{\max}}$ components occurs due to the change in Coriolis parameter relative to the southern hemisphere, and now the (positive-valued) $\Psi^c_{{E}_{\max}}$ is of greater magnitude; this again leads to a northward volume transport. Finally, at high northern latitudes, the magnitudes of both $\Psi^c_{{W}_{\max}}$ and $\Psi^c_{{E}_{\max}}$ reduce dramatically. We observe another sign-reversal at around $40^\circ$N, and also a relatively larger $\Psi^c_{{W}_{\max}}$ contribution. Between latitudes $50^\circ$N and  $60^\circ$N, the magnitude of $\Psi^c_{{E}_{\max}}$ is considerably smaller than that of $\Psi^c_{{W}_{\max}}$.

\begin{figure*}[ht!]
	\centerline{\includegraphics[width=\textwidth]{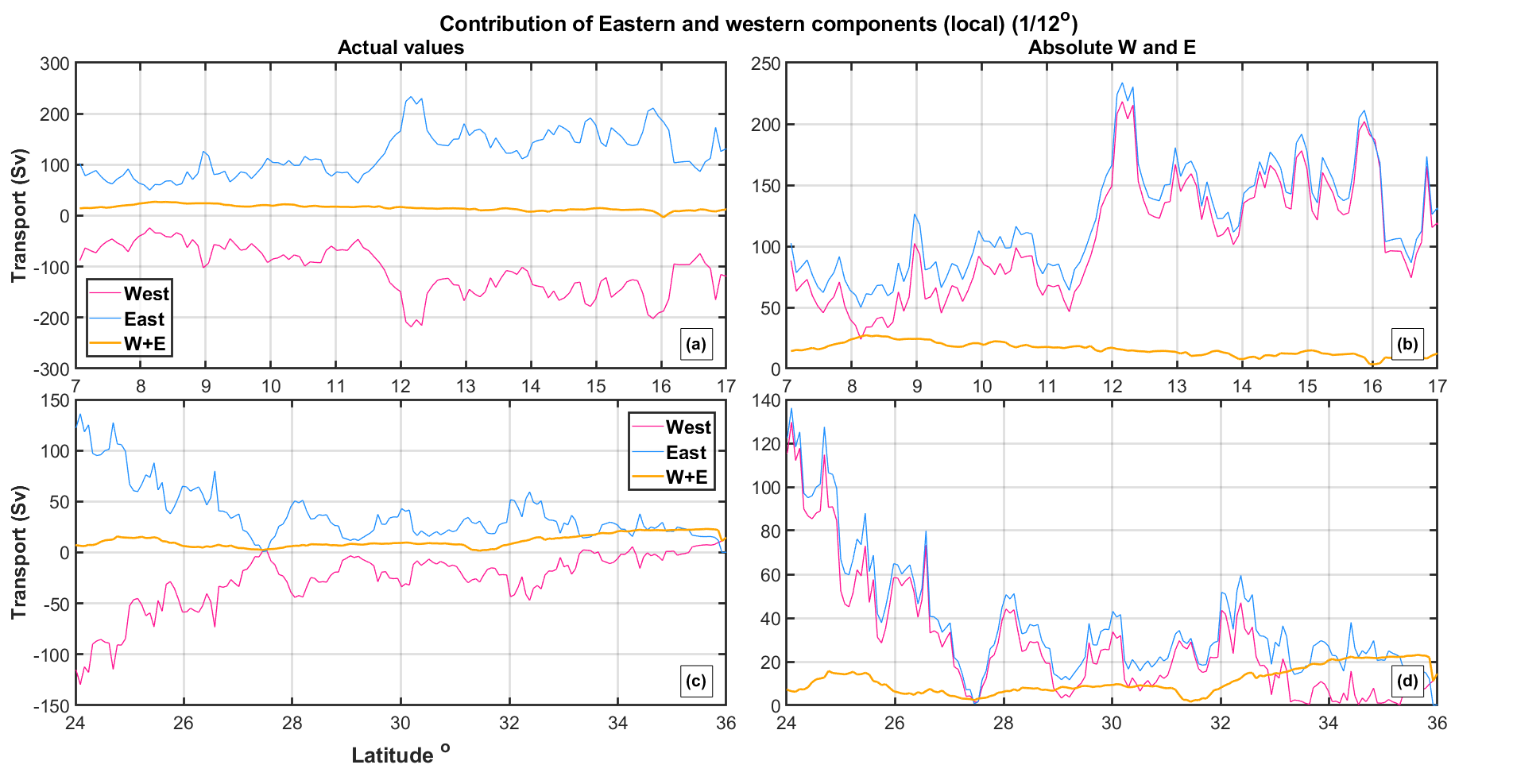}}
	\caption[$\Psi^c_{{E}_{\max}}$ and $\Psi^c_{{W}_{\max}}$ contributions to $\tilde{\Psi}^c_{\max}$ with respect to latitude for the $1/12^\circ$ model.]{$\Psi^c_{{E}_{\max}}$ and $\Psi^c_{{W}_{\max}}$ contributions to $\tilde{\Psi}^c_{\max}$ with respect to latitude for the $1/12^\circ$ model. Panels (a) and (b) refer to latitudes $7^\circ$N to $17^\circ$N  and Panels (c) and (d) refer to latitudes $24^\circ$N to $36^\circ$N. $\Psi^c_{{W}_{\max}}$ is given in pink, and $\Psi^c_{{E}_{\max}}$ in blue. Their sum ($\Psi^c_{{th}_{\max}}$) is given in orange. Panels (b) and (d) show the absolute values of boundary components in Panels (a) and (c).}
	\label{p_MaxStrm_WE_inv}
\end{figure*}

We further investigate the characteristics of the red line in Figure \ref{p_RC_BdryMxStrmLtt}(a) for two latitude bands where a large change in $\Psi^c_{{th}_{\max}}$ is seen, corresponding to latitudes $7^\circ$N to $17^\circ$N and $24^\circ$N to $36^\circ$N. Within these latitude ranges we find large changes in the $\Psi^c_{{th}_{\max}}$ contribution to $\tilde{\Psi}^c_{\max}$. Figure \ref{p_MaxStrm_WE_inv} emphasises the difficulty in decoupling $\Psi^c_{{E}_{\max}}$ and $\Psi^c_{{W}_{\max}}$, and attributing a change in $\Psi^c_{{th}_{\max}}$ to a strengthening or weakening in either of $\Psi^c_{{E}_{\max}}$ or $\Psi^c_{{W}_{\max}}$. The figure shows actual and absolute $\Psi^c_{{E}_{\max}}$ or $\Psi^c_{{W}_{\max}}$ for the two intervals of interest. 

From Figure \ref{p_MaxStrm_WE_inv}(b), we attribute the peak in $\Psi^c_{{th}_{\max}}$ at approximately $8^\circ$N to an apparent relative reduction in the magnitude of $\Psi^c_{{W}_{\max}}$, which is not as pronounced at higher latitudes, resulting in a subsequent reduction in the overall northward $\Psi^c_{{th}_{\max}}$ contribution as latitude increases here. 

For the majority of the northern hemisphere (up to $40^\circ$N), $\Psi^c_{{E}_{\max}}$ has a greater magnitude. In the interval $24^\circ$N to $27^\circ$N, the magnitudes of both $\Psi^c_{{E}_{\max}}$ and  $\Psi^c_{{W}_{\max}}$ reduce significantly. However, the magnitude of $\Psi^c_{{W}_{\max}}$ reduces more quickly at around $25^\circ$N, resulting in an enhanced $\Psi^c_{{th}_{\max}}$ there. In the interval between $25^\circ$N to $27^\circ$N, the magnitudes of $\Psi^c_{{E}_{\max}}$ and  $\Psi^c_{{W}_{\max}}$ contributions reduce in tandem resulting in a weakening $\Psi^c_{{th}_{\max}}$.

Between $30^\circ$N and $35^\circ$N, $\Psi^c_{{E}_{\max}}$ is largest in magnitude. Around $31.5^\circ$N, a relatively strong $\Psi^c_{{W}_{\max}}$ leads to a weakening in $\Psi^c_{{th}_{\max}}$. North of $31.5^\circ$N, $\Psi^c_{{W}_{\max}}$ weakens to a small value, resulting in a relatively large overall northward transport. At near $36^\circ$N, $\Psi^c_{{E}_{\max}}$ becomes negative, but $\Psi^c_{{W}_{\max}}$ positive, again resulting in a small northward transport. 

\section{Decomposition of SAMBA timeseries}
\label{App_SmbRap}

Table \ref{Tab_SVSmb} shows the temporal variation of $\Psi^c_{\max}$ explained by $\tilde{\Psi}^c_{\max} - \Psi_{AC_{\max}}^c$ and $\tilde{\Psi}^c_{\max}$ calculated using Equation \ref{e_PrcVarExp} for the SAMBA array timeseries, discussed in Chapter \ref{TJ_Var}. With $\Psi_{AC}^c$ included, percentage variance explained does not improve materially with model resolution. Without additional cells, the percentage variance explained is better than for RAPID timeseries in general, but particularly poor results are found for the $1/4^\circ$ resolution. 
\begin{table}[h!]
	\centering
	\begin{tabular}{ |P{2.7cm}||P{2.7cm}|P{2.7cm}|P{2.7cm}| }
		\hline
		\multicolumn{4}{|c|}{Temporal variance of $\Psi^c_{\max}$ explained at SAMBA ($34.5^\circ$S)} \\
		\hline
		& $1^\circ$ & $1/4^\circ$ & $1/12^\circ$ \\
		\hline
		$\tilde{\Psi}^c_{\max} - \Psi_{AC_{\max}}^c$  &$97.8\%$ & $62.5\%$ & $88.1\%$\\
		$\tilde{\Psi}^c_{\max}$ &  $99.4\%$ & $91.0\%$ & $92.3\%$\\
		\hline
	\end{tabular}
	\caption{The role of additional cell transport contributions at SAMBA array ($34.5^\circ$S), calculated at the depth of maximum estimated overturning streamfunction: temporal variation of $\Psi^c_{\max}$ explained by $\tilde{\Psi}^c_{\max} - \Psi_{AC_{\max}}^c$ and $\tilde{\Psi}^c_{\max}$ calculated using Equation \ref{e_PrcVarExp}.}
	\label{Tab_SVSmb}
\end{table}

\subsection*{Correlation between components at SAMBA}

Investigating the correlation  between all boundary components at the maximum estimated streamfunction depth, displayed in Figure \ref{Tab_SVSmb}, reveals a strong relationship between $\Psi^c_{th_{\max}}$ and $\Psi^c_{bot_{\max}}$, especially within the $1/4^\circ$ and $1/12^\circ$ models. Results are shown in Table \ref{Tab_CrSmb}. 
\begin{table}[h!]
	\centering
	\begin{tabular}{ |P{3.3cm}||P{2.7cm}|P{2.7cm}|P{2.7cm}| }
		\hline
		\multicolumn{4}{|c|}{Correlation between components at SAMBA ($34.5^\circ$S)} \\
		\hline
		& $1^\circ$ & $1/4^\circ$ & $1/12^\circ$ \\
		\hline
		$\Psi^c_{\max}$ vs. $\tilde{\Psi}^c_{\max}$   &$1.00$ & $0.96$ & $0.96$\\
		$\tilde{\Psi}^c_{\max}$ vs. $\Psi^c_{th_{\max}}$ &  $0.58$ & $0.45$ & $0.62$\\
		$\tilde{\Psi}^c_{\max}$ vs. $\Psi^c_{bot_{\max}}$ &  $0.32$ & $0.01$ & $-0.15$\\
		$\Psi^c_{th_{\max}}$ vs. $\Psi^c_{bot_{\max}}$ &  $-0.46$ & $-0.86$ & $-0.84$\\
		$\Psi^c_{th_{\max}}$ vs. $\Psi^c_{W_{\max}}$ &  $0.17$ & $0.65$ & $0.39$\\
		$\Psi^c_{th_{\max}}$ vs. $\Psi^c_{E_{\max}}$ &  $0.68$ & $-0.03$ & $-0.16$\\
		\hline
	\end{tabular}
	\caption[Correlation between components at SAMBA array.]{Correlation between components at SAMBA array ($34.5^\circ$S), calculated at depth of maximum estimated overturning streamfunction, $\tilde{\Psi}^c_F$ for all model resolutions.}
	\label{Tab_CrSmb}
\end{table}

Lagged correlations between $\Psi^c_{th_{\max}}$ and $\Psi^c_{bot_{\max}}$ are no larger than the values shown in Table \ref{Tab_CrSmb}.

\section{Normalised variation in standard deviation \textbf{$\sigma_{sp}^*$} of Atlantic bottom densities and velocities}
\label{Var_BD_Nrm}
\begin{figure}[ht!]
	\centerline{\includegraphics[width=\textwidth]{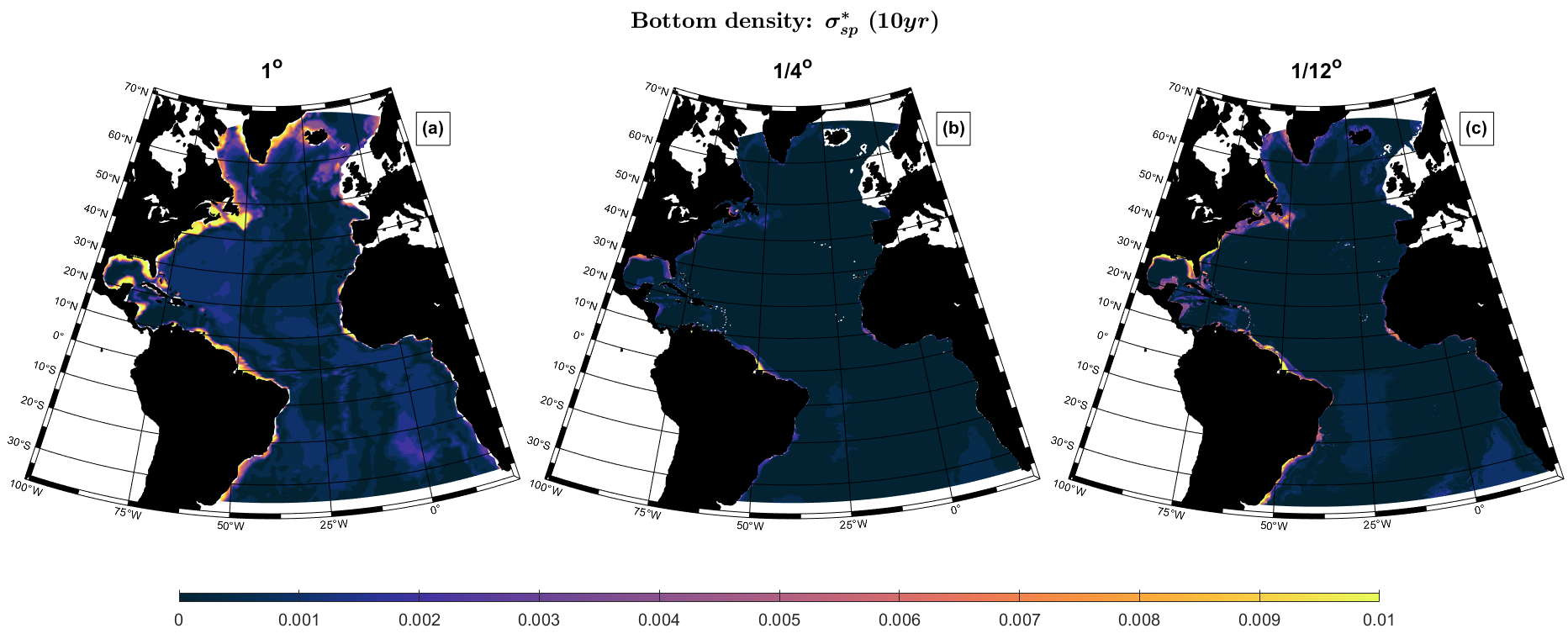}} 
	\caption[Normalised variation in standard deviation, $\sigma_{s10}^*$ for in-situ bottom density for $1^\circ$, $1/4^\circ$ and $1/12^\circ$ models.]{Normalised variation in standard deviation, $\sigma_{s10}^*$ for in-situ bottom density. Panel (a) $1^\circ$, (b) $1/4^\circ$ and (c) $1/12^\circ$ show $\sigma_{s10}^*$ for each model resolution within the Atlantic basin, using observation periods of $p=10$ years.}
	\label{p_SS_p10}
\end{figure}

\begin{figure}[ht!]
	\centerline{\includegraphics[width=12cm]{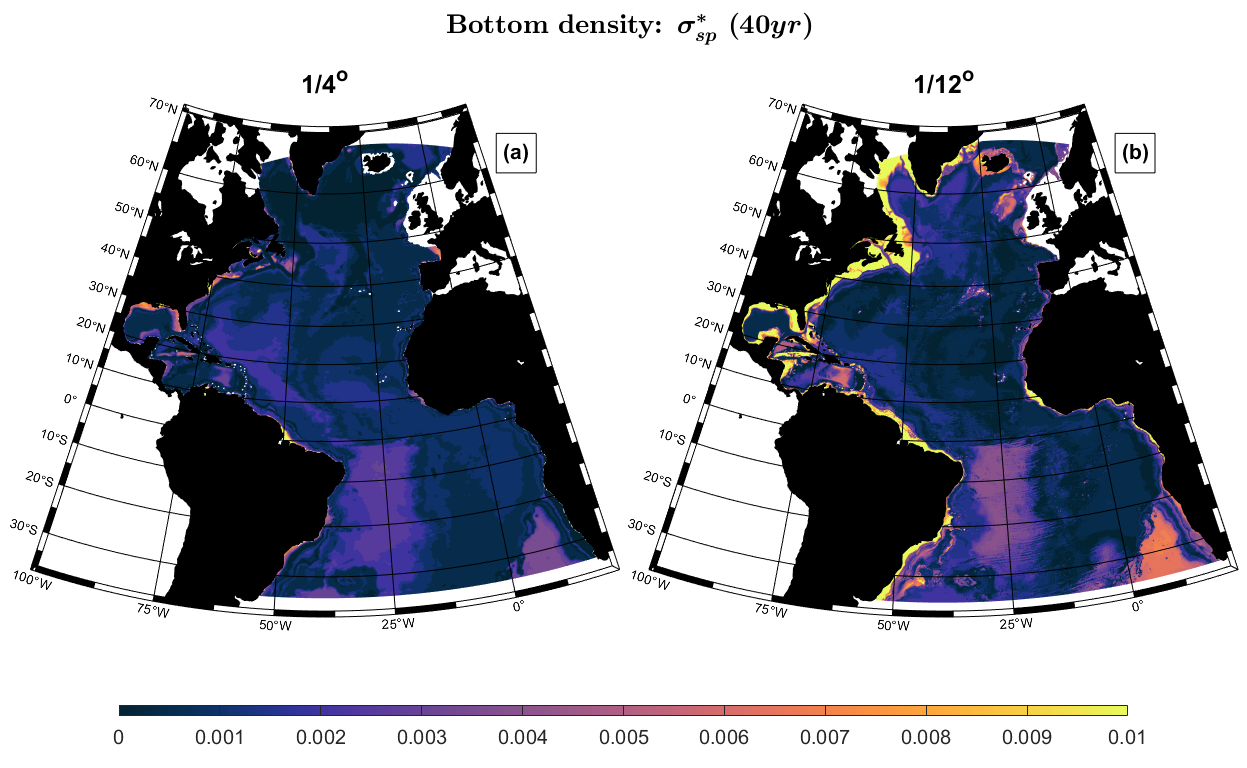}} 
	\caption{Normalised variation in standard deviation $\sigma_{s40}^*$ for bottom densities. Panel (a) $1/4^\circ$ and (b) $1/12^\circ$ show  $\sigma_{s40}^*$ for each model resolution within the Atlantic basin, using observation periods of $p=40$ years.}
	\label{p_SS_p40}
\end{figure}

The normalised variation $\sigma_{sp}^*$ in the standard deviation $\overline{s}_p$ is introduced in Section \ref{Std:Thr}. Using Equation \ref{e_sigma_sp}, we calculate $\sigma_{sp}^*$ for in-situ bottom densities for various periods $p$. $\sigma_{sp}^*$ for $p=10$ and $40$ years is shown in Figures \ref{p_SS_p10} and \ref{p_SS_p40} respectively. Figure \ref{p_SS_p10} indicates large variation between model resolutions. We find that the majority of the interior has a small $\sigma_{s10}^*$ value, for all resolutions, suggesting insensitivity to the starting year. In contrast, near the boundaries, especially in the west, we find large values of $\sigma_{s10}^*$ particularity in the $1^\circ$ and $1/12^\circ$ models, indicating $\overline{s}_{10}$ in western boundary currents are sensitive to the initial starting year.

The $1^\circ$ model shows large values along the majority of the western boundary, including the shores of Greenland; the same cannot be said for the $1/4^\circ$ model. One could argue that a similar structure is found at $1/4^\circ$, but with much smaller magnitudes; regions of strong $\sigma_{s10}^*$ are isolated to (a) northern Brazil and (b) northern Gulf of Mexico. The $1/12^\circ$ model shows large values of $\sigma_{s10}^*$ near (a) Malvinas current, (b) northward Brazilian current, (c) northern Gulf of Mexico and (d) the region of the Gulf Stream. All of these are locations of boundary currents and therefore regions of high variability;  the chosen start year will have greater impact on $\overline{s}_{10}$. 

Figure \ref{p_SS_p40} shows a clear increase in $\sigma_{s40}^*$ for both $1/4^\circ$ and $1/12^\circ$ model resolutions, compared with $\sigma_{s10}^*$. For the $1/4^\circ$ model, increasing $p$ from 10 to 40 results in greater normalised variation in the standard deviation $\sigma_{sp}^*$ off the eastern coast of the United States. This characteristic is also reflected in the $1/12^\circ$ model, for which large values of $\sigma_{s40}^*$ occur along the majority of the western boundary. In contrast to $\sigma_{s10}^*$, we find $\sigma_{s40}^*$ has much larger values within the interior of the southern hemisphere basin for both resolutions, especially evident in the $1/12^\circ$ model (Figure \ref{p_SS_p40}(b)) at (a) south of the Walvis ridge off the south-eastern flank of Africa and (b) the Brazil and Argentine basins.

\begin{figure}[ht!]
	\centerline{\includegraphics[width=\textwidth]{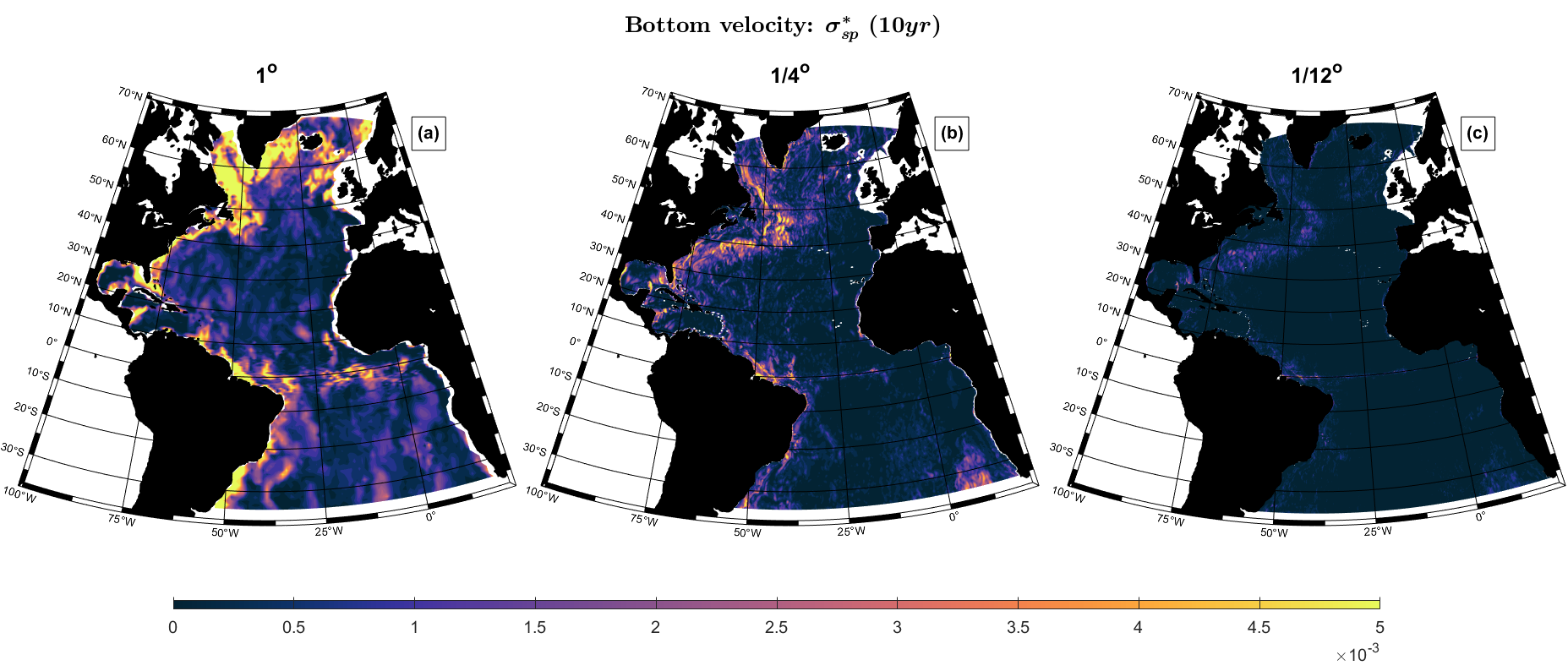}} 
	\caption{Normalised variations in standard deviation $\sigma_{s10}^*$ for bottom velocities. Panel (a) $1^\circ$, (b) $1/4^\circ$ and (c) $1/12^\circ$ for each model resolution within the Atlantic basin, using observation periods of $p=10$ years.}
	\label{p_SSbv_p10}
\end{figure}

Figure \ref{p_SSbv_p10} shows the normalised variation $\sigma_{s10}^*$ in the standard deviation for bottom velocities, with $\sigma_{s10}^*$ larger for coarser resolution models. The $1^\circ$ model in Panel (a) exhibits largest values at high northern latitudes, potentially due to poorer representation of the subpolar gyre and DWBC. Large $\sigma_{s10}^*$ suggest sensitivity to the initial year used for $\overline{s}_{10}$ calculation. Therefore both coarser models suggest larger discrepancies in bottom velocities in the first few years, whereas the $1/12^\circ$ model shows bottom velocities especially in the DWBC are relatively stable for this initial period. The sensitivity of the $1^\circ$ model supports the larger linear trend in the subpolar region, indicating significant changes in DWBC and likely deep water formation in the Labrador and Nordic Seas during the first 10 years. For longer periods ($\sigma_{s40}^*$, not shown), we find a similar spatial structure for both higher resolution models, with the $1/12^\circ$ model showing increased magnitudes of $\sigma_{s40}^*$, suggesting greater sensitivity to the starting year from year 10 to 40.  

\section{Reconstruction of isopycnal structure using depth profiles from a limited number of locations}
\label{Bdry_MCMC_App}

There is evidence in Chapter \ref{TJ_Bdry} that the isopycnal structure on an eastern boundary of an ocean basin is less complex, as a function of latitude and depth, than on the western boundary. For example, on the eastern boundary of the Atlantic, we hypothesise that density profiles with depth over a large range of latitudes can be approximated by just a small number of depth profile measurements at well-chosen locations. Using just these density profile measurements, linear interpolation is then adequate to estimate density profiles at other latitudes of interest. This would yield a simple useful low-order reconstruction of eastern boundary density in particular. In this section we seek to quantify the extent to which this the case on both eastern and western boundaries of the Atlantic. 

Briefly, we estimate a piecewise linear model for time-mean density $\rho$ (over all years of data available) as a function of latitude $y$ and depth $z$, independently for each of the eastern and western boundaries discussed in the Section \ref{Bdry_EW}. We aim to use the minimum number of linear pieces to reconstruct the density structure to some pre-specified quality. We expect that, if the eastern boundary exhibits a simpler isopycnal structure, that the minimum number of linear pieces to explain the eastern boundary will be lower than for the western boundary. The methodology is implemented using Bayesian inference, and is explained in more detail in Appendix \ref{App_MCMC}.

Results below consider reconstruction of time-average eastern and western latitude-depth profiles for the $1^\circ$, $1/4^\circ$ and $1/12^\circ$ models. First we examine the influence of the choice of number of depth profiles on the quality of reconstruction for each model resolution. Next we identify the minimum number of depth profiles required to achieve a specified quality of reconstruction for each model resolution. Then we visualise the actual and reconstructed latitude-depth boundary densities. 

\begin{figure*}[ht!]
	\centerline{\includegraphics[width=12cm]{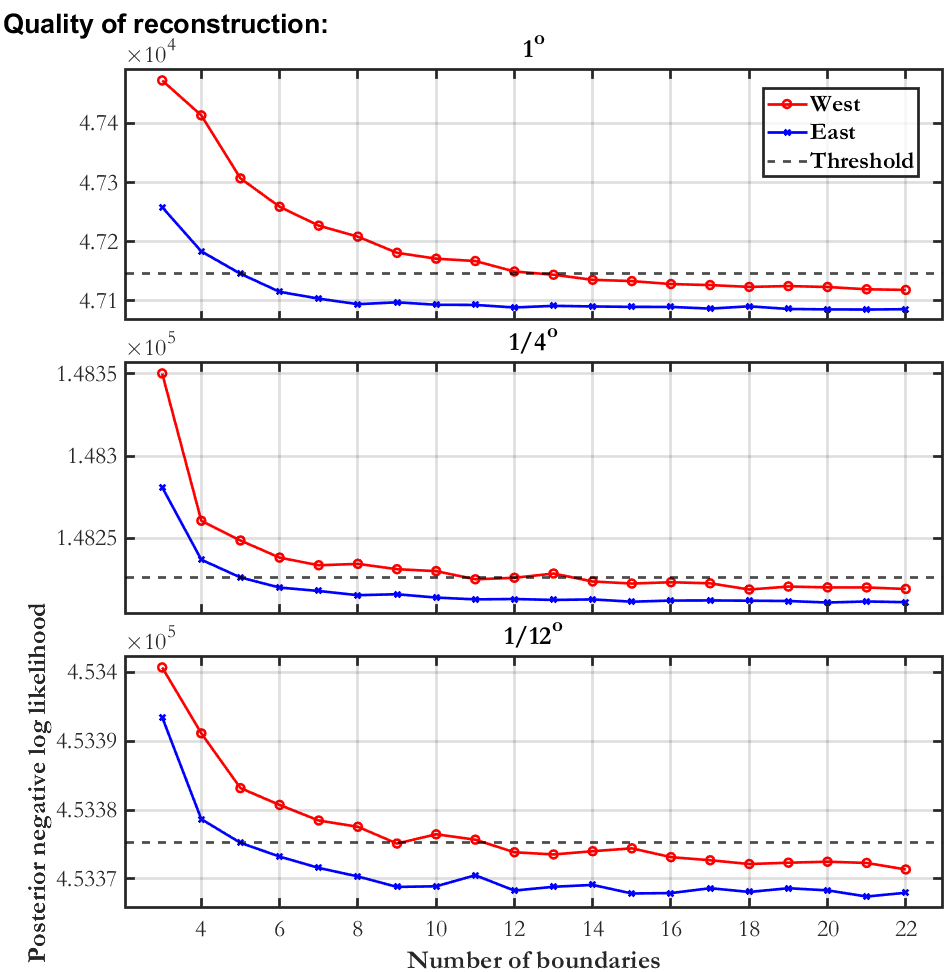}}
	\caption[Reconstruction quality of time-mean latitude-depth density structure for eastern and western boundaries quantified using posterior negative log likelihood, for $1^\circ$, $1/4^\circ$ and $1/12^\circ$ models.]{Reconstruction quality of time-mean latitude-depth density structure for eastern (blue) and western (red) boundaries quantified using posterior negative log likelihood (PNLL), for (a) $1^\circ$, (b) $1/4^\circ$ and (c) $1/12^\circ$ models. Dashed line indicates the PNLL threshold for 5 eastern boundary profiles.}
	\label{F_MCMC_nLL1}
\end{figure*}

Figure \ref{F_MCMC_nLL1} illustrates the quality of reconstruction as a function of the number of depth profiles $n_r$ (see Appendix \ref{App_MCMC}) for the eastern and western boundaries, for each model resolution. Reconstruction quality is quantified in terms of the posterior negative log likelihood (PNLL). PNLL is closely related to the least squares error between the actual and estimated latitude-depth density structure over all latitudes and depths; as PNLL decreases, the quality of reconstruction increases. We see that, to achieve a given value of PNLL requires considerably larger $n_r$ on the western boundary compared to the eastern at all model resolutions. For example, given 5 depth profiles on the eastern boundary, to achieve a comparable quality on the western boundary would require 13 profiles for the $1^\circ$ model.

\begin{figure*}[ht!]
	\centerline{\includegraphics[width=\textwidth]{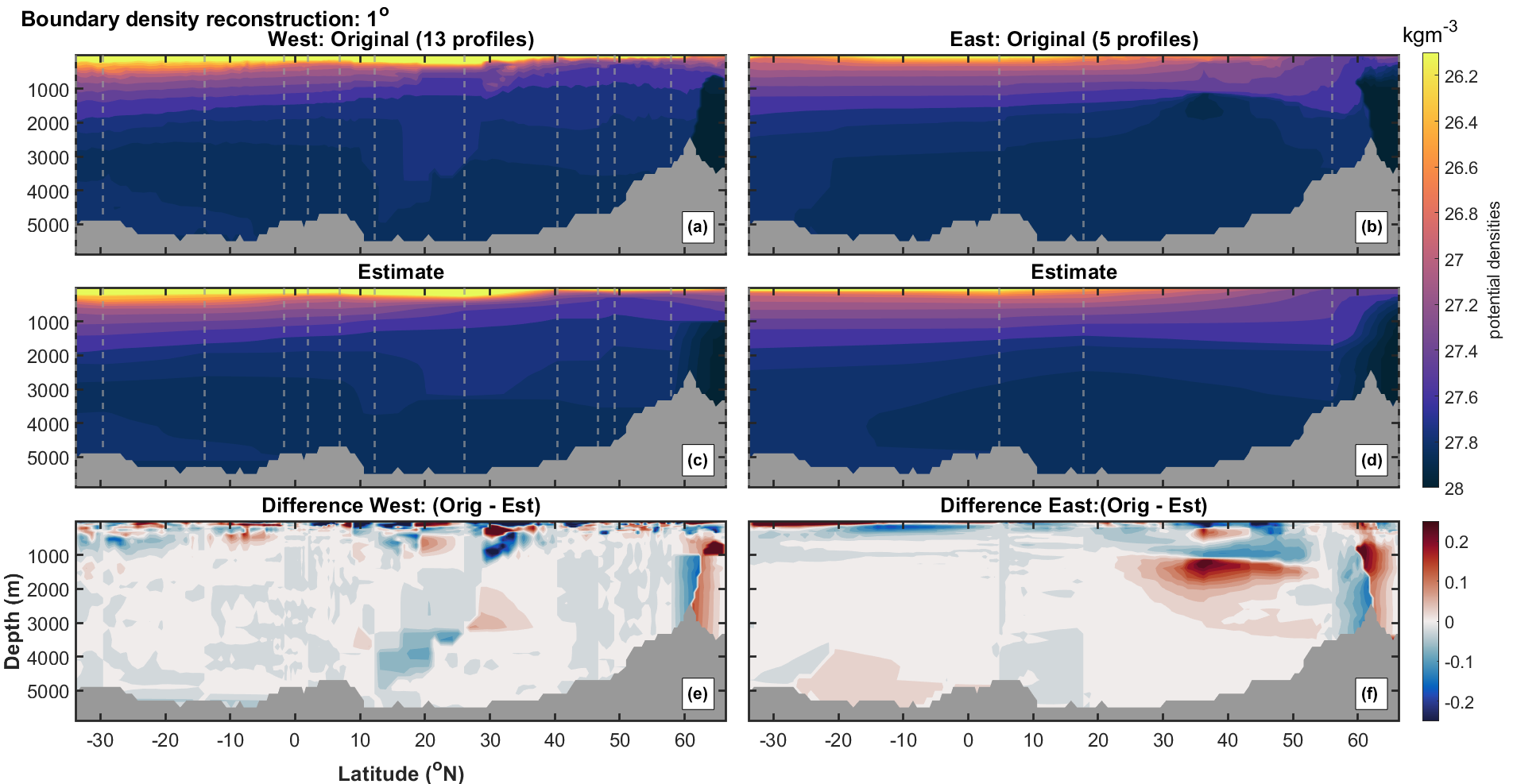}}
	\caption[Original and reconstructed time-mean latitude-depth density structures, and their difference, for the eastern and western boundaries using $n_r=$ 13 and 5 for the West and East respectively, corresponding to equal reconstruction performance in the $1^\circ$ model.]{Original and reconstructed time-mean latitude-depth density structures, and their difference, for the eastern and western boundaries using $n_r=$ 13 and 5 for the West and East respectively, corresponding to equal reconstruction performance in the $1^\circ$ model. Panels are: (a) Original West, (b) Original East, (c) Estimated West, (d) Estimated East, (e) Original-Estimate for West, (f) Original-Estimate for East. Dashed vertical lines indicated the location of depth profiles required for reconstruction.}
	\label{F_MCMC_BdryRecon1}
\end{figure*}

Figure \ref{F_MCMC_BdryRecon1} provides an illustration of actual and reconstructed time-mean latitude-depth density structures, and their difference, for the eastern and western boundaries. The number $n_r$ of depth profiles is chosen so that the quality of reconstruction (over all depths) is the same for the eastern and western boundaries. We note that the number of depth profiles required to explain the eastern boundary is smaller than that for the western boundary. Figure \ref{F_MCMC_BdryRecon1}(e)(f) show the eastern boundary reconstruction struggles at latitudes $35^\circ$N to $50^\circ$N near 2000m, since as discussed previously, the flat eastern boundary approximation breaks down here. The original densities (Panel (a)) shows a region of denser water at 2000m and approximately $37^\circ$N, very close to the Mediterranean outflow. Other notable differences are found near the overflow regions at latitude $60^\circ$N.

The current analysis provides a means of automatically selecting locations for depth profile measurements of density along both eastern and western boundaries, in order to provide optimal reconstruction of the density structure (with latitude and depth) for the whole basin. Results indicate that the number $n_r$ of profiles required to reconstruct the eastern boundary density structure is lower that that required for the western boundary. The current methodology raises the possibility of further simplifying modelling of boundary density contributions to the decomposition diagnostic for the whole Atlantic basin. 

\subsection{Outline of estimation procedure}
\label{App_MCMC}

Here we provide an outline of the statistical analysis performed to assess the complexity of latitude-depth contours for eastern and western boundary densities.

We assume measurements $\{\rho(y_j,z_k)\}_{j=1,k=1}^{n_y,n_z}$ are available for density on a grid of $n_y$ latitudes $\{y_j\}_{i=1}^{n_y}$ and $n_z$ depths $\{z_k\}_{k=1}^{n_z}$ for each of the eastern and western boundaries. The objective of the analysis is to find the smallest subset $\mathcal{R}=\{r_a\}_{a=1}^{n_r}$ of $|\mathcal{R}|=n_r$ latitudes, with $n_r \ll n_y$, which provides an adequate piecewise linear reconstruction of the full latitude-depth cross-sectional density. 

Piecewise linear approximations to functions have been studies for many years (e.g. \citealt{Hamann1994}, \citealt{Muggeo2003}). Writing the piecewise linear estimate for density given $\mathcal{R}$ as $\hat{\rho}(y,z|\mathcal{R})$, we seek to minimise the loss function $L(\mathcal{R})$
\begin{eqnarray}
	L(\mathcal{R}) = \sum_j \sum_k (\rho(y_j,z_k)-\hat{\rho}(y_j,z_k|\mathcal{R}))^2 \label{e_MCMC_1}
\end{eqnarray}
where $\hat{\rho}(y,z|\mathcal{R})$ is the piecewise linear reconstruction given by
\begin{eqnarray}
	\hat{\rho}(y,z|\mathcal{R}) = \frac{\delta_L \rho(r_{a^*},z) + \delta_U \rho(r_{{a^*}+1},z)}{{r_{{a^*}+1}-r_{a^*}}}  \label{e_MCMC_2}
\end{eqnarray}
where $a^*=\argmax{a}\{r_a : r_a<y\}$, $\delta_L=y-r_{a^*}$ and  $\delta_U=r_{a^*+1}-y$. The optimal choice $	\mathcal{R}^*_{n_r}$ of $\mathcal{R}$ for any $n_r$ is then
\begin{eqnarray}
	\mathcal{R}^*_{n_r} = \argmin{\mathcal{R} , |\mathcal{R}|=n_r} L(\mathcal{R}) .  \label{e_MCMC_3}
\end{eqnarray}
Then we find the smallest value of $n_r$ such that the loss using the corresponding $\mathcal{R}^*_{n_r}$ is less than a given value $C$ 
\begin{eqnarray} 
	\mathcal{R}^* = \argmin{\mathcal{R}^*_{n_r}} L(\mathcal{R}^*_{n_r})<C  \label{e_MCMC_4}
\end{eqnarray}
and the corresponding optimal size of $\mathcal{R}$ is $n_r^*=|\mathcal{R}^*|$.

We perform this analysis separately for the eastern and western boundaries. If we compare boundaries fairly, e.g. by adopting a common value for $C$ on both boundaries, we are therefore able to quantify whether one boundary exhibits a more complex density structure, simply by comparing the values of $|\mathcal{R}_E^*|$ and $|\mathcal{R}_W^*|$ for a given $C$.

\subsection{Bayesian inference}
There are many different approaches that could be used to find solutions to Equation \ref{e_MCMC_3}. Here, for each choice of $n_r$, we find the optimal $\mathcal{R}$ using a statistical technique for Bayesian inference called Markov chain Monte Carlo (MCMC; \citealt{Gamerman2006}, \citealt{Gelman2013}). MCMC inference is widely used in many scientific fields, including modelling of ocean density (e.g. \citealt{Economou2019}). For the current work, in precise terms, we seek to find the values of $r_1, r_2, ...r_{n_r}$ which provide the best piecewise linear representation of the latitude-depth density cross-section, by minimising $L(\mathcal{R})$. We proceed as follows.

\textbf{Prior specification:} First, we specify prior distributions $f(r_a)$ for each of the $r_a$, $a=1,2,...,n_r$. Here, we assume that each is uniformly distributed on the latitude domain of interest. The joint prior distribution of all the boundary locations is then $f(\mathcal{R})=\prod_{a=1}^{n_r} f(r_a)$, where each of the terms $f(r_a)$ is a constant defined on the latitude domain. Since the prior is a constant for a given value of $n_r$ regardless of the choices $r_1, r_2, ...r_{n_r}$, the prior plays no further role in the analysis.

\textbf{Likelihood:} Next, we specify a likelihood for any piecewise linear representation given the observed density data. Writing the observed data set as $\{\rho(y_j,z_k)\}_{j=1,k=1}^{n_y,n_z}=D$ for brevity, we assume that this likelihood is Gaussian, of the form
\begin{eqnarray}
	f(D | \mathcal{R}) = \frac{1}{(2 \pi)^{1/2} \kappa} \exp\left[ - \frac{L(\mathcal{R})}{2 \kappa^2}\right]  \label{e_MCMC_5}
\end{eqnarray}
where $L(\mathcal{R})$ is the loss criterion from Equantion~\ref{e_MCMC_1} and $\kappa$ is the measurement uncertainty which we also specify before the analysis. The setting of $\kappa$ is discussed below, in particular to accommodate the effects of varying GCM cell dimensions with latitude and depth.

\textbf{Posterior estimation:} Finally, we estimate the posterior distribution $f(\mathcal{R}|D)$ of the parameters $\mathcal{R}$ by applying Bayes' theorem
\begin{eqnarray}
	f(\mathcal{R}|D) = \frac{f(D | \mathcal{R}) f(\mathcal{R})}{f(D)}  \label{e_MCMC_6}
\end{eqnarray}
where $f(D)$ is called the ``evidence'' and is in general an expensive integral to calculate. $f(\mathcal{R}|D)$ provides the joint posterior distribution of the optimal latitudes $\mathcal{R}$ at which density profile measurements should be made.

\textbf{Gibbs sampling:} Fortunately, using MCMC, we can estimate $f(\mathcal{R}|D)$ without calculating $f(D)$. The procedure we use is called Gibbs sampling, and works by iteratively sampling from each of the following full conditional distributions in turn
\begin{eqnarray}
	&f(r_1 | D, r_2, r_3, ..., r_{n_r})& \label{e_MCMC_7} \\
	&f(r_2 | D, r_1, r_3, ..., r_{n_r})&  \nonumber \\
	&...& \nonumber \\ 
	&f(r_n | D, r_1, r_2, ..., r_{n_r-1})& .  \nonumber
\end{eqnarray}
If a sufficient number of iterations is used, it can be shown that this sampling procedure yields a random sample from the posterior distribution $f(\mathcal{R}|D)$.

\textbf{Metropolis-Hastings sampling:}
Unfortunately, we cannot evaluate the full conditional distributions in Equation~\ref{e_MCMC_7} in closed form. Therefore, we use a Metropolis-Hastings (MH) sampling scheme to sample from each one of them for each iteration of the Gibbs sampler. For example, to sample from $f(r_1 | D, r_2, r_3, ..., r_{n_r})$, suppose we have already generated values $r_1^{i}, r_2^{i}, ..., r_{n_r}^{i}$ at the end of iteration $i$ of the Gibbs sampler, and we are about to start iteration $i+1$. Then we propose a new candidate value $r_1^{i*}$ according to 
\begin{eqnarray}
	r_1^{i*} = r_1^{i} + \epsilon \gamma \label{e_MCMC_8}
\end{eqnarray}
where $\epsilon$ is a random number from the standard Gaussian distribution, and $\gamma$ is a proposal standard deviation specified before hand. This is called a Gaussian random walk proposal. We then accept the candidate $r_1^{i*}$ as $r_1^{i+1}$ with probability $\alpha$ given by
\begin{eqnarray}
	\alpha=\min\left\{ 1, \frac{f(D|r_1^{i*}, r_2^{i}, r_3^{i}, ..., r_{n_r}^{i})}{f(D|r_1^{i},r_2^{i}, r_3^{i}, ..., r_{n_r}^{i})} \right\} . \label{e_MCMC_9}
\end{eqnarray}
If we reject the candidate value, we simply set $r_1^{i+1}=r_1^{i}$. After this accept/reject step for $r_1$, we proceed to perform the same step for $r_2$, $r_3$, $...$ . Again, it can be shown that the MH scheme provides a valid sample from the correct full conditional. It is recommended that the value of $\gamma$ is set so that the acceptance rate of proposals is about $1/4$. It can be seen that the acceptance criterion in Equation \ref{e_MCMC_9} is based on the likelihood ratio of the candidate and current states. This is easily calculated using Equation ~\ref{e_MCMC_5}, and provides for a computationally efficient scheme.

\textbf{Adaptive Metropolis sampling:} After some iterations, the values of $r_1^{i}, r_2^{i}, ..., r_{n_r}^{i}$ already generated typically begin to show correlation. For subsequent iterations, it is therefore advantageous to exploit this correlation structure using the approach of \cite{Roberts2009}. Now we replace the single location proposal in Equation \ref{e_MCMC_8} with a joint proposal for all $n_r$ locations using
\begin{eqnarray}
	\un{r}^{i*} = \un{r}^i + (1-\beta) \mathcal{N}(\un{0}, {2.38^2 \un{\Sigma}_r}/{n_r}) + \beta \mathcal{N}(\un{0}, {0.1^2 \un{I}_{n_r}}/{n_r}) 	\label{e_MCMC_10}
\end{eqnarray}
where $\un{r}^i$ is the full set of locations at iteration $i$, and $\un{r}^{i*}$ is the corresponding candidate vector. $\un{\Sigma}_r$ is an estimate for the covariance structure of the $r$s from previous iterations, and $\un{I}_{n_r}$ is an $n_r \times n_r$ identity matrix. $\beta$ is a parameter which is usually set to 0.05, and $\mathcal{N}$ indicates a Gaussian random variable with given mean and variance.

Using the complete Metropolis-Hasting within Gibbs sampling scheme, we are able to find the joint distributions of optimal locations $\mathcal{R}$ for any choice of $n_r$. In particular, we can take the combination of locations which occurs most often in $f(\mathcal{R}|D)$ as the optimal choice of $\mathcal{R}$ for given $n_r$.

\subsection{Implementation refinements}

Here we describe specific refinements of the method above, needed to apply it reasonably for the eastern and western boundaries of the Atlantic. 

\textbf{Latitude and depth scaling:} We seek reconstructions of the boundary density structures which are equally good at any location in depth-latitude space. Since the depth grid used to calculate densities consists of unevenly spaced depth points $z_k$, it is important to use an appropriate weighting scheme so that the model performs equally well at any choice of depth. Similarly, the latitude grid for the data is uniform: however, this does not correspond to a uniform distance scale. Weighting is also therefore necessary, so that the model performs equally well for any choice of latitude on the boundary. 

Specifically, we prefer to estimate the piecewise linear reconstruction of the boundary density so that the quality of fit is the same per metre of depth, and per metre along the boundary. To achieve this, we modify Equation \ref{e_MCMC_1} to include two weight vectors $\alpha_{\text{Ltt}}$ and $\alpha_{\text{Dpt}}$ so that the modified equation becomes
\begin{eqnarray}
	L(\mathcal{R}) = \sum_j \sum_k \left(\alpha_{\text{Ltt}}(y_j) \ \alpha_\text{Dpt}(z_k) \left(\rho(y_j,z_k)-\hat{\rho}(y_j,z_k|\mathcal{R})\right)\right)^2 \label{e_MCMC_11} .
\end{eqnarray}
Here, $\alpha_{\text{Ltt}}(y_j)$ is the number of kilometres per degree latitude at latitude $y_j$ and similarly $\alpha_\text{Dpt}(z_k)$ is the thickness in metres of the depth grid at $z_k$.

\textbf{Accommodating bathymetry:} The bathymetry of the eastern and western boundaries varies with latitude. Therefore, the value of density at the deepest cells is often missing, because that depth does not actually correspond to ocean. It is possible during the MCMC sampling procedure, that neighbouring locations for depth profiles $\mathcal{R}$ might be selected corresponding locally to relatively shallow bathymetry, with intervening latitudes having deeper bathymetry. In this situation, it is impossible to define the piecewise linear function at depths beyond those available at the two depth profiles in $\mathcal{R}$. When this occurs, we simply use the estimated value of density at the deepest point available, per latitude, to in-fill for deeper cells. This approximation is found to have minimal effect on reconstructions; in any event, the MCMC sampling scheme is able to adjust the locations of depth profiles $\mathcal{R}$ to accommodate the effect.

Further, the locations of neighbouring depth profiles in $\mathcal{R}$ are likely to correspond to different bathymetry depths. When only the deeper depth profile is able to provide a density value, this value of density (at the deeper profile) is assumed for intervening latitudes at that depth. Again, this approximation is found to have minimal effect on reconstructions; in any event, the MCMC sampling scheme is able to adjust the locations of depth profiles $\mathcal{R}$ to accommodate it.

\subsection{Possible extensions}

\textbf{Accommodating boundary profiles for different times:} The description above is easily extended to include optimal reconstruction of boundaries for multiple times. In this case, $\mathcal{L}$ in Equation \ref{e_MCMC_11} takes the form
\begin{eqnarray}
	L(\mathcal{R}) = \sum_j \sum_k \sum_\ell \left(\alpha_{\text{Ltt}}(y_j) \ \alpha_\text{Dpt}(z_k) \left(\rho(y_j,z_k,t_\ell)-\hat{\rho}(y_j,z_k,t_\ell|\mathcal{R})\right)\right)^2 \label{e_MCMC_12}
\end{eqnarray}
where the measurements and reconstructions are now available over latitude, depth and time.

\textbf{Emphasising quality of overturning streamfunction estimates from reconstruction:} The quality of density reconstruction at large depths is of greater importance for estimation of overturning streamfunction. For this reason, it may be advantageous to change the depth weighting vector $\alpha_\text{Dpt}(z_k)$, by increasing weights corresponding to large depths, to improve reconstruction quality at depth. Inspection of the overturning streamfunction (Equation \ref{T}) suggests that a sensible alternative depth weighting $\alpha_\text{Dpt}^\Psi(z_k)$ would be the depth of the centre of the $T$-cell at $z_k$ multiplied by the thickness of the $z_k$ depth cell.

\section{Temporal variability of Atlantic along-boundary neutral densities}
\label{App_Bdry_Tmp}

In this section we investigate the variation of Atlantic boundary neutral densities (discussed in Section \ref{CntMpBnd}) over different timescales. We quantify the variation in terms of standard deviations of the form $\overline{s}_p$ defined in Equation \ref{e_sp_bar} for different time intervals $p$ in years; the calculation performed is the same as that reported in Chapter \ref{TJ_Var}. This analysis is performed using the full timeseries available for each HadGEM-GC3.1 model (as opposed to the first 100 years only), whereas the GK and GloSea5 datasets are not considered due to short dataset lengths. 

\begin{figure*}[ht!]
	\centerline{\includegraphics[width=\textwidth]{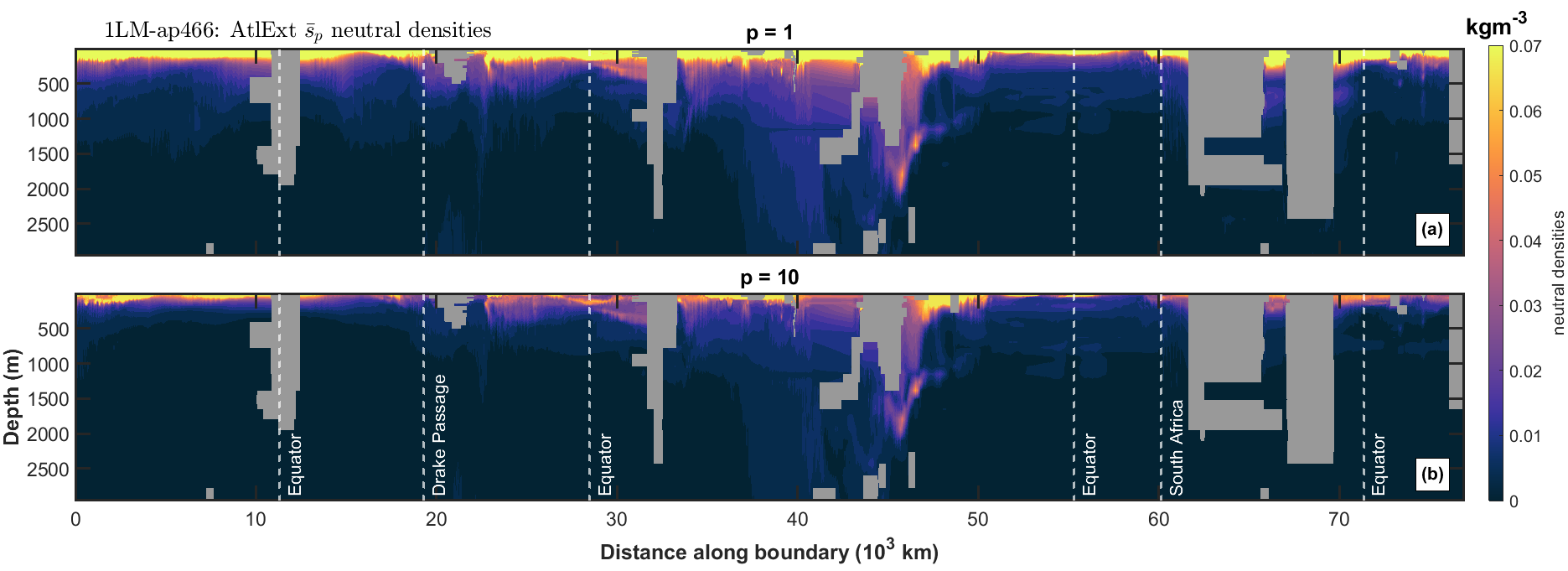}}
	\caption[Average standard deviation $\overline{s}_{p}$ for boundary neutral density in the $1^\circ $-$LM$ model, for various timescales.]{Average standard deviation $\overline{s}_{p}$ for boundary neutral density in the $1^\circ $-$LM$ model. Periods considered are (a) p=1 and (b) p=10 years. White dashed lines indicate locations of the Equator, Drake Passage and southern tip of South Africa.}
	\label{F_BD_nStd_1LM}
\end{figure*}

\begin{figure*}[ht!]
	\centerline{\includegraphics[width=\textwidth]{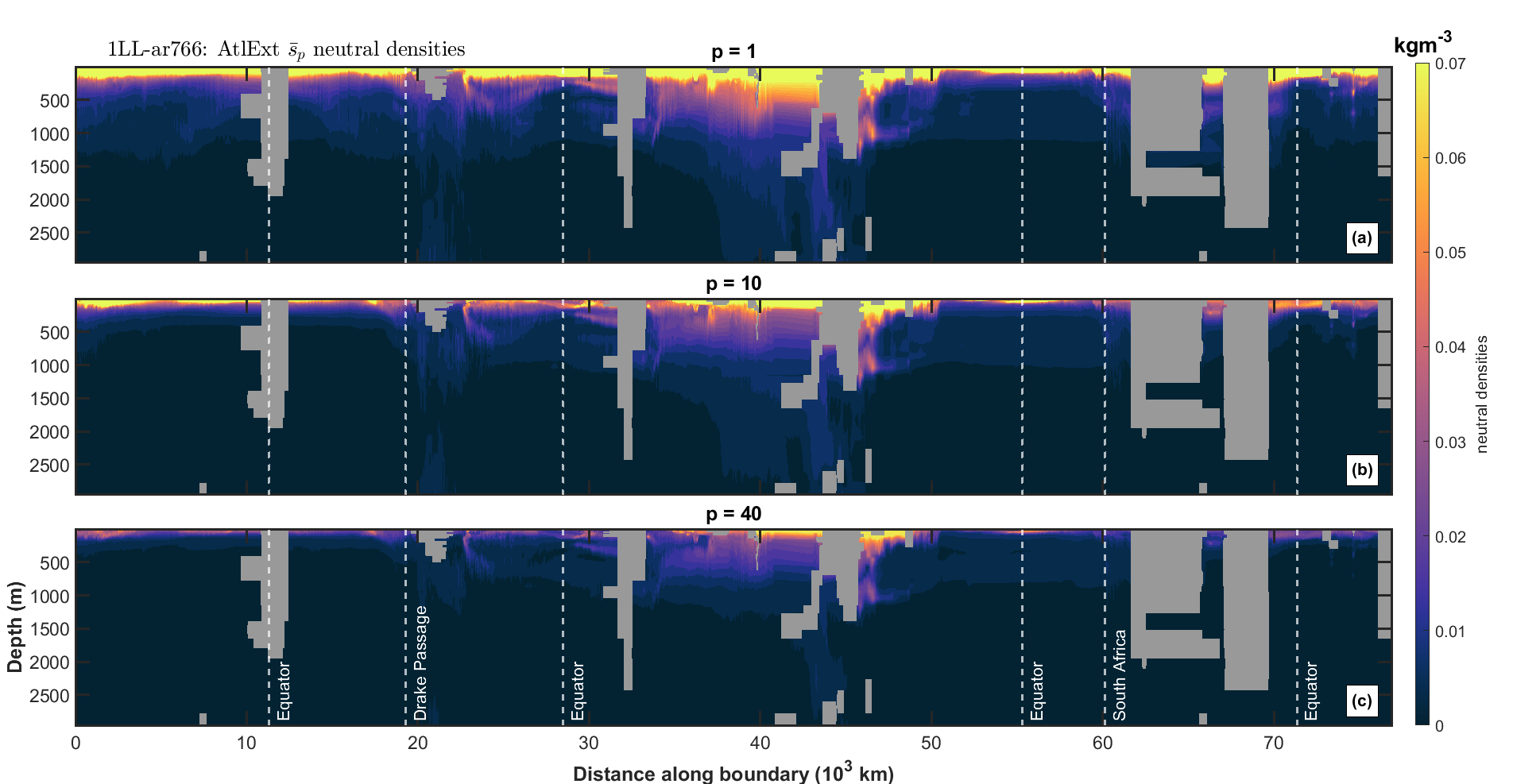}}
	\caption[Average standard deviation $\overline{s}_{p}$ for boundary neutral density in the $1^\circ $-$LL$ model, for various timescales.]{Average standard deviation $\overline{s}_{p}$ for boundary neutral density in the $1^\circ $-$LL$ model. Periods considered are (a) p=1, (b) p=10 and (c) p=40 years. White dashed lines indicate locations of the Equator, Drake Passage and southern tip of South Africa.}
	\label{F_BD_nStd_1LL}
\end{figure*}

Figure \ref{F_BD_nStd_1LM} shows the average standard deviation of boundary densities for time intervals $p=1$ and $p=10$ years. For the upper 300m, $\overline{s}_p$ shows considerable temporal variability due to the influence of wind-stress on both timescales. The variability found reduces with increasing timescale, since Ekman processes tend to dominate on shorter timescales. Away from the surface layers, large values of $\overline{s}_{p}$ indicate considerable variability in boundary densities at depth, near $46\times 10^3$km or the Reykjanes ridge, possibly due to overflows. Considerable variability remains in the surface layers for the 10-year timescales from $47$-$50\times 10^3$km. Variability in this region is replicated in Figure \ref{F_BD_nStd_1LL} for the $1^\circ $-$LL$ model, for $p=$1, 10 and 40 years, with somewhat less pronounced features at depth. Large variability at longer timescales between $35$ and $50\times 10^3$km is attributed to intense air-sea interaction and deep water formation in the Labrador Sea. These results are consistent with that found in Figure \ref{p_sb_p10}, where greater variability at high latitudes and near overflow regions is found in in-situ bottom densities for the $1^\circ$-$LM$ in comparison to both the $1/4^\circ$ and $1/12^\circ$ resolution models.

For both $1^\circ$ models and all timescales considered, temporal variability in neutral density, quantified by $\overline{s}_{p}$, outside the surface layer is largest in the North Atlantic region, specifically locations downstream of areas of deep water formation. Variability near the surface and at depths down to 1500m at locations to the east of the Reykjanes ridge can be attributed to variability in local mixed-layer depth.

\begin{figure*}[ht!]
	\centerline{\includegraphics[width=\textwidth]{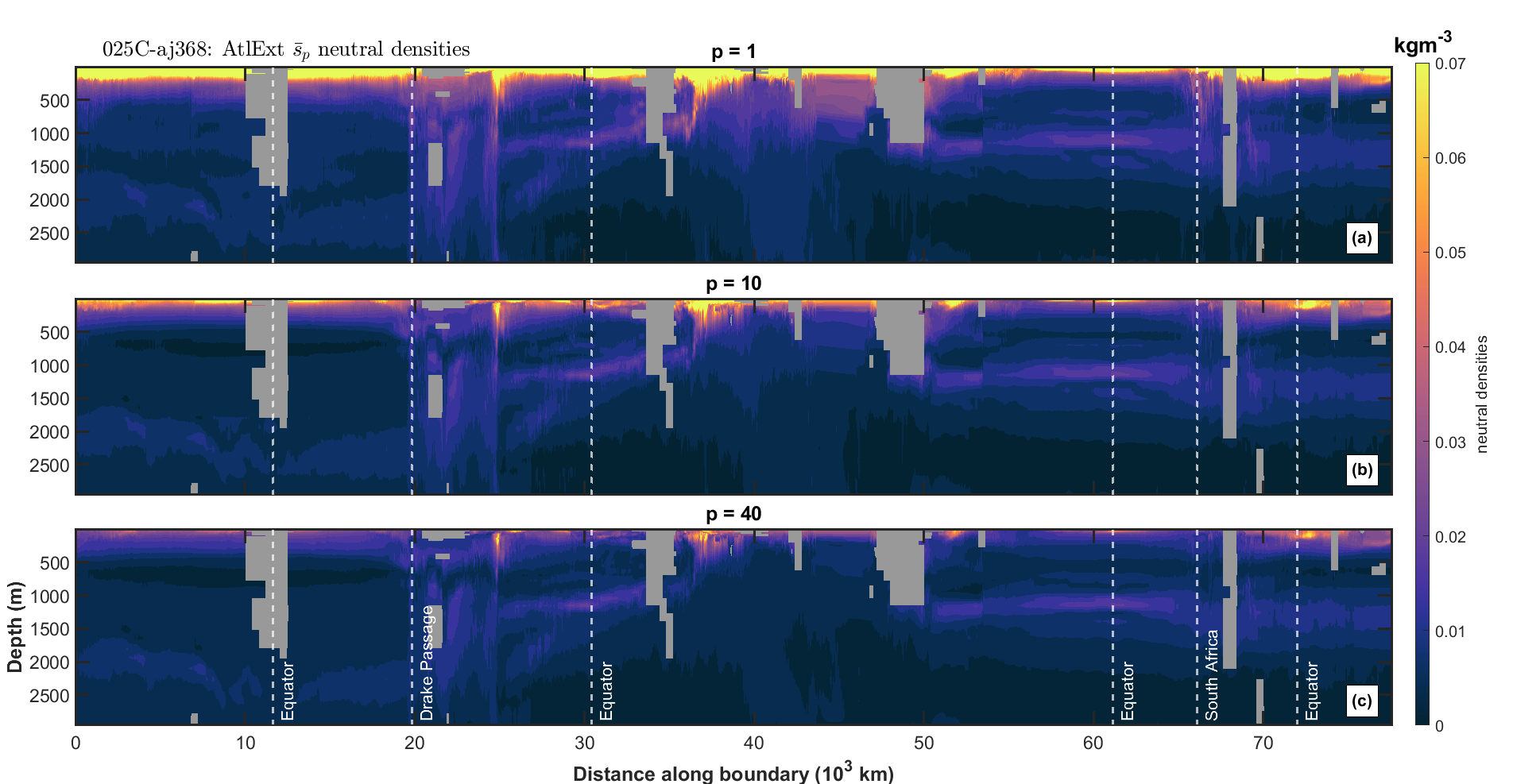}}
	\caption[Average standard deviation, $\overline{s}_{p}$ for boundary neutral density in the $1/4^\circ$ model, for various timescales.]{Average standard deviation, $\overline{s}_{p}$ for boundary neutral density in the $1/4^\circ$ model. Periods considered are (a) p=1, (b) p=10 and (c) p=40 years. White dashed lines indicate locations of equator, Drake Passage and southern tip of South Africa}
	\label{F_BD_nStd_025}
\end{figure*}

Estimates of $\overline{s}_{p}$ for boundary density in the $1/4^\circ$ model at timescales $p=$1, 10 and 40 years exhibit similar surface features to those of the $1^\circ$ models. However, Figure \ref{F_BD_nStd_025} reveals less variability in the North Atlantic and interesting features at all timescales in equatorial regions at depths of approximately 1200m. The latter could be attributed to internal or boundary waves, or possibly depth-variability in the overturning streamfunction. Along the western Atlantic boundary, the $\overline{s}_{p}$ signal appears to increase northward and decrease southward along the boundary from the Equator. The increased variability throughout the fluid column around the Malvinas Islands might be associated with temporal variation of AABW. 

Comparison of $\overline{s}_{p}$ for eastern and western boundary sections across resolutions would suggest somewhat greater variability at depth (300-1000m) on the western boundary.

\section{Ocean-only and sensitivity experiments}  \label{Sct_Inter_Oth_A}
Here we summarise brief investigations into the impact of model parameterisation, and atmospheric coupling, on the ACC transport at the Drake Passage discussed in Section \ref{Sct_Inter_Oth}.

\subsection{$1/4^\circ$ sensitivity experiments} 

In a small number of preliminary sensitivity experiments for the $1/4^\circ$ model, the ACC transport is decomposed into its components, using the methodology outlined in Section \ref{Sct_ACC_Mth}. Model runs of approximately 20 years were considered, with different eddy parameterisation and diffusivity operators starting from EN4 temperature and salinity fields and 1950s fixed forcing. The first model incorporated a weak Gent-McWilliams (GM, \citealt{Gent1990}) parameterisation scheme, and the second a $3\times$ BiLaplacian (3xBiLap, \citealt{Madec2016}) viscosity scheme. A control run was also performed (without GM). Timeseries of expected $T_{ACC}$ and estimated $\tilde{T}_{ACC}$ from both the weak GM and 3xBiLap viscosity iterations were observed to be more constant with time when compared to the control run. Over the 20 years of simulation, $T_{ACC}$ from the 3xBiLap model stabilised after approximately 5 years to 145Sv, compared to stabilisation after a similar period to 135Sv for the weak GM model. In contrast, $T_{ACC}$ from the control run weakens approximately linearly from an initial 150Sv to approximately 105Sv over the period of the simulation, with no stabilisation observed. The reduced weakening found for $\tilde{T}_{ACC}$ was mainly attributed to a stable southern density contribution $T_{S}$ of approximately -10Sv to -20Sv for both weak GM and 3xBiLap runs, in contrast to a strengthening $T_{S}$ for the control run from approximately 0Sv to -90Sv over the period of the simulation. The weak-GM parameterisation scheme acts to dampen the resolved small-scale features such as eddies in the $1/4^\circ$ model. The GM scheme removes available potential energy from the resolved flow by flattening isopycnal slopes. Therefore the $1/4^\circ$ model appears to behave more similarly to the $1^\circ$ model, and hence has an improved ACC transport. The BiLaplacian scheme should also dampen the flow and reduce the variance of the velocity field, by reducing the overall kinetic energy and smoothing out the jets. The BiLaplacian viscosity scheme is more scale-selective in its smoothing compared to a simple increase in viscosity, therefore the scheme will have a tendency to smooth small grid-scale features preferentially.

\subsection{Ocean-only GCM} 

Analysis of forced ocean-only model runs for $1^\circ$, $1/4^\circ$ and $1/12^\circ$ resolutions was conducted to investigate the impact of atmospheric and sea-ice coupling on the ACC transport through the Drake Passage. For a run period of approximately 30 years, ocean-only models showed significantly less weakening of ${T}_{ACC}$ at higher resolutions compared with estimates using a coupled atmosphere. ${T}_{ACC}$ was found to stabilise at $120$Sv and $130$Sv for the $1/4^\circ$ and $1/12^\circ$ models, respectively. In contrast, ${T}_{ACC}$ for the $1^\circ$ model stabilised near $154$Sv. The difference between estimates for different model resolutions is primarily attributable to $T_{bot}$. Notably, the sum of density contributions ($T_S$, $T_N$ and $T_{\beta}$) for the three model resolutions was found to be very similar, reducing from an initial 120Sv to approximately 105Sv for all models over the period of simulation. We conclude that atmospheric and sea-ice coupling contributes to the weakening of the ACC transport in both higher resolution models.

This suggests atmosphere-ocean-sea-ice interactions in the model, near the Antarctic coast, act to cool and freshen surface waters, leading to a slumping rather than outcropping of isopycnals towards the coast, and a reverse flow along the coast. We find freshwater input from ice shelves (basal melt and iceberg calving) is approximately twice as great in the coupled model, compared to the forced ocean-only model. This would explain the freshening observed along the coast, and contribute to gyre spin-up. Local cold easterly wind stress could also play a part in enhancing the cooling of surface water along the Antarctic coastline within coupled models. On a larger scale, changes in winds could act to spin-up gyres.




\pagebreak
\bibliographystyle{apalike}

\addcontentsline{toc}{chapter}{Bibliography}
\bibliography{TJ_bib_Thesis}

\end{document}